\documentclass[a4paper,9pt]{article}
\pdfoutput=1 

\usepackage{jheppub} 
\usepackage{hyperref}
\hypersetup{
    bookmarks=true,         
    unicode=false,          
    pdftoolbar=true,        
    pdfmenubar=true,        
    pdffitwindow=false,     
    pdfstartview={FitH},    
    pdftitle={My title},    
    pdfauthor={Author},     
    pdfsubject={Subject},   
    pdfcreator={Creator},   
    pdfproducer={Producer}, 
    pdfkeywords={keyword1} {key2} {key3}, 
    pdfnewwindow=true,      
    colorlinks=true,       
    linkcolor=red,          
    citecolor=violet,        
    filecolor=magenta,      
    urlcolor=blue,           
    linktocpage=true
}
\usepackage{graphics}
\usepackage{rotating}
\usepackage{subfigure}
\usepackage{pstricks}
\usepackage{color}
\usepackage{amsfonts}
\usepackage{mathrsfs}
\usepackage{epsfig}
\usepackage{amsmath,amssymb,amsthm,graphicx,latexsym}
\usepackage{ulem}
\renewcommand{\arraystretch}{2}

\newcommand{\be}{\begin{equation}}
\newcommand{\ee}{\end{equation}}
\newcommand{\bea}{\begin{eqnarray}}
\newcommand{\eea}{\end{eqnarray}}
\newcommand{\bml}{\begin{subequations}}
\newcommand{\eml}{\end{subequations}}
\newcommand{\bfig}{\begin{figure}}
\newcommand{\efig}{\end{figure}}

\newcommand{\bfx}{\mbox{\boldmath{$x$}}}

\newcommand{\slashed}{\hspace{-1.1ex}/}

\begin{document}
$~~~~~~~~~~~~~~~~~~~~~~~~~~~~~~~~~~~~~~~~~~~~~~~~~~~~~~~~~~~~~~~~~~~~~~~~~~~~~~~~~~~~$\textcolor{red}{\bf TIFR/TH/16-50}
\title{\textsc{\fontsize{25}{17}\selectfont \sffamily \bfseries \textcolor{purple}{COSMOS-$e'$- soft Higgsotic attractors}}}

\author[a]{Sayantan Choudhury
\footnote{\textcolor{purple}{\bf Presently working as a Visiting (Post-Doctoral) fellow at DTP, TIFR, Mumbai, \\$~~~~~$Alternative
 E-mail: sayanphysicsisi@gmail.com}. ${}^{}$}}

\affiliation[a]{Department of Theoretical Physics, Tata Institute of Fundamental Research, Colaba, Mumbai - 400005, India
}

\emailAdd{sayantan@theory.tifr.res.in}

\abstract{In this work, we have developed an elegant algorithm to study the cosmological consequences from a huge class of quantum field theories (i.e. superstring theory, supergravity, extra dimensional theory, modified gravity etc.), which are equivalently described by soft attractors in the effective field theory framework. In this description we have restricted our analysis for two scalar fields - dilaton and Higgsotic fields minimally coupled with Einstein gravity, which can be generalized for any arbitrary number of scalar field contents with generalized non-canonical and non-minimal interactions. We have explicitly used $R^2$ gravity, from which we have studied the attractor and non-attractor phase by exactly computing two point, three point and four point correlation functions from scalar fluctuations using  In-In (Schwinger-Keldysh) and $\delta {\cal N}$ formalism. We have also presented theoretical bounds on the amplitude, tilt and running of the primordial power spectrum, various shapes (equilateral, squeezed, folded kite or counter collinear) of the amplitude as obtained from three and four point scalar functions, which are consistent with observed data. Also the results from two point tensor fluctuations and field excursion formula are explicitly presented for attractor and non-attractor phase. Further, reheating constraints, scale dependent behaviour of the couplings and the dynamical solution for the dilaton and Higgsotic fields are also presented. New sets of consistency relations between two, three and four point observables are also presented, which shows significant deviation from canonical slow roll models. Additionally, three possible theoretical proposals have presented to overcome the tachyonic instability at the time of late time acceleration. Finally, we have also provided the bulk interpretation from the three and four point scalar correlation functions for completeness. 

}

\maketitle
\flushbottom

\section{\textcolor{blue}{Introduction}}

The inflationary paradigm is a theoretical proposal which attempts to solve various long-standing issues with standard Big Bang Cosmology and has been studied earlier in various works \cite{Guth:1980zm,Baumann:2009ds,Senatore:2016aui,Liddle:1999mq,Langlois:2010xc,Riotto:2002yw,Lyth:1998xn,Lyth:2007qh,Weinberg:2003sw,Weinberg:2008hq,Cheung:2007st,Bardeen:1980kt}. But apart from the success of the this theoretical framework it is important to note that there is no single model exists till now using which one can explain the complete evolution history of the universe and also unable to break the degeneracy between various cosmological parameters computed from various models of inflation \cite{Choudhury:2011sq,Choudhury:2011jt,Choudhury:2012yh,Choudhury:2012ib,Choudhury:2013zna,Choudhury:2013jya,Choudhury:2014sxa,Choudhury:2015pqa,Choudhury:2014hja,Choudhury:2016wlj,Choudhury:2015hvr,Bhattacharjee:2014toa,Deshamukhya:2009wc,Kachru:2003sx,Kachru:2003aw,Iizuka:2004ct,Choudhury:2014kma,Choudhury:2013iaa,Choudhury:2013woa,Choudhury:2014sua,Choudhury:2014wsa}. It is important to note that, the vacuum energy contribution generated by the trapped Higgs field in a metastable vacuum state which mimics the role of effective cosmological constant in effective theory. At the later stages of Universe such vacuum contribution dominates over other contents and correspondingly Universe 
expands in a exponential fashion. But using such metastable vacuum state it is not possible to explain the tunneling phenomena and also impossible to explain the end of inflation. To serve both of the purposes shape of the effective potential for inflation should have flat structure. Due to such specific structure effective potential for inflation satisfy flatness or slow-roll condition using which one can easily determine the field value corresponding to the end of inflation. There are various classes of models exists in cosmological literature where one can derive such specific structure of inflation \cite{Choudhury:2011jt,BuenoSanchez:2006rze,Allahverdi:2006iq,Ross:1995dq,Allahverdi:2006rt,Enqvist:2003gh,Allahverdi:2006we}. For an example, the Coleman-Weinberg effective potential serves this purpose \cite{Coleman:1973jx,Barenboim:2013wra}. Now if we consider the finite temperature contributions in the effective potential \cite{Fodor:1994bs,Quiros:1999jp} then such thermal effects need to localize the inflaton field to small expectation values at the beginning of inflation. The flat structure of the effective potential for inflation is such that the scalar inflaton field slowly rolls down in the valley of potential during which the scale factor varies exponentially and then 
infation ends when the scalar inflaton field goes to the non slow-rolling region by violating the flatness condition. At this epoch inflaton field 
evolves to the true minimum very fast and then it couples to the matter content of the Universe and reheats our Universe via subsequent oscillations
about the minimum of the slowly varying effective potential for inflation. These class of models are very successful theoretical probe through which it is possible to explain the characteristic and amplitude of the spectrum of density fluctuations with high statistical accuracy ($2\sigma$ CL from Planck 2015 data \cite{Ade:2015lrj,Ade:2015ava,Ade:2015xua}) and at late times this perturbations act as the seeds for the large scale structure formation, which we observe at the present epoch.
Apart from this huge success of inflationary paradigm in slowly varying regime it is important to mention that, these density fluctuations generated from various class of successful models were unfortunately
large enough to explain the physics of standard Grand Unified Theory (GUT) with well known theoretical frameworks and also it is not possible to explain the observed isotropy of the Cosmic Microwave Background
Radiation (CMBR) at small scales during inflationary epoch. The only physical possibility is that the self interactions of the inflaton field and the associated couplings to other matter field contents would be sufficiently small for which it is possible to satisfy these cosmological and particle physics constraints. But the prime theoretical challenge at this point is that for such setup it is impossible to achieve thermal equilibrium at the end of inflation. Consequently, it is not at all possible to localize the scalar inflaton field near zero Vacuum Expectation Value (VEV), $\langle \phi \rangle =\langle 0|\phi|0\rangle=0,$ where $|0\rangle$ is the corresponding vacuum state in quasi de Sitter space time. Therefore, a sufficient amount of expansion will not be obtained from this prescribed setup. Here it is important to note that, for a broad category of effective potentials the inflaton field evolves with time very slowly compared to the Hubble scale following slow-roll conditions and satisfies all of the observational constraints \cite{Ade:2015lrj,Ade:2015ava,Ade:2015xua} computable from various inflationary observables from this setup. However, apart from the success of slow roll inflationary paradigm the density fluctuations or more precisely the scalar component of the metric perturbations
restricts the coupling parameters to be sufficiently small enough and allows huge fine-tuning in the theoretical set up. This is obviously a not recommendable prescription from model builder's point of view. Additionally, all of these class of models are not ruled out completely by the present observed data (Planck 2015 and other joint data sets \cite{Ade:2015lrj,Ade:2015ava,Ade:2015xua,Ade:2015tva}), as they are degenerate in terms of determination of inflationary observables and associated cosmological parameters in precision cosmology. There are various ideas exist in cosmological literature which can drive inflation. These are appended below:
\begin{itemize}
\item \underline{\textcolor{red}{\bf Category I:}}\\ In this class of models, inflation drives through a field theory which involves a very high energy physics phenomena. Example: string theory and its supergravity extensions \cite{Choudhury:2016wlj,Choudhury:2015hvr,Choudhury:2014sxa,Choudhury:2013zna,Choudhury:2012ib,Choudhury:2012yh,Choudhury:2011sq,Yamaguchi:2011kg,Stewart:1994ts,McAllister:2007bg,Baumann:2009ni,Nilles:1983ge,Linde:1997sj,BasteroGil:2006cm,Choudhury:2011rz,Alishahiha:2004eh,Silverstein:2016ggb,Flauger:2014ana,McAllister:2014mpa,Silverstein:2013wua,Silverstein:2008sg,Panda:2010uq,Panda:2007ie,Mazumdar:2001mm,Choudhury:2002xu,Choudhury:2003vr,Deshamukhya:2009wc,Choudhury:2015baa,Choudhury:2015fzb,Choudhury:2016rtp,Maharana:1997cz,Headrick:2004hz,Minwalla:2003hj,David:2001vm,Gopakumar:2000rw,Sen:2000kd,Rastelli:2000hv,Sen:2002qa,Sen:2002in,Sen:2002an,Sen:1999xm,Sen:2002nu,Mandal:2003tj}, various supersymmetric models \cite{Choudhury:2011jt,BuenoSanchez:2006rze,Allahverdi:2006iq,Ross:1995dq,Allahverdi:2006rt,Enqvist:2003gh,Allahverdi:2006we} etc. 

\item \underline{\textcolor{red}{\bf Category II:}}\\ In this case, inflation is driven by changing the mathematical structure of the gravitational sector. This can be done using the following ways:
\begin{enumerate}
\item Introducing higher derivative terms of the form of $f(R)$, where $R$ is the Ricci scalar \cite{Starobinsky:1979ty,Sebastiani:2015kfa,DeFelice:2010aj,Sotiriou:2008rp}. Example: Starobinsky inflationary framework which is governed by the model \cite{Starobinsky:1979ty}, $f(R)=aR+bR^2$, where the coefficients $a$ and $b$ are given by, $a=M^2_p$ and $b=1/6M^2$. If we set $a=0$ and $b=1/6M^2=\alpha$ then we can get back the theory of scale free gravity in this context. In this paper will we explore the cosmological consequences from scale free theory of gravity. 

\item Introducing higher derivative terms of the form of Gauss Bonnet gravity coupled with scalar field in non-minimal fashion, where the contribution in the effective action can be expressed as \cite{Kanti:2015pda,Nozari:2008ny}, \be S_{GB}=\int d^4 x\sqrt{-g}~f(\phi)\left[R_{\mu\nu\alpha\beta}R^{\mu\nu\alpha\beta}-4R_{\mu\nu}R^{\mu\nu}+R^2\right],\ee where $f(\phi)$ is the inflaton dependent coupling which can be treated as the non-minimal coupling in the present context. This is also an interesting possibility which we have not explored in this paper. Here one cannot consider the Gauss Bonnet term in the gravity sector in 4D without coupling to other matter fields as in 4D Gauss Bonnet term is topological surface term. 

\item Another possibility is to incorporate the effect of non-minimal coupling of inflaton field and the gravity sector \cite{Bezrukov:2007ep,Hertzberg:2010dc,Pallis:2010wt}. The simplest example is, $f(\phi)R$ gravity theory. For Higgs inflation \cite{Bezrukov:2007ep},$ f(\phi)=\left(1+\xi \phi^2\right)$, where $\xi$ is the non-minimal coupling of the Higgs field. Here one can consider more complicated possibility as well by considering a non-canonical interaction between inflaton and $f(R)$ gravity by allowing $f(\phi)f(R)$ term in the 4D effective action \cite{Budhi:2017yjd}. For the construction of effective potential we have considered this possibility.

\item One can also consider the other possibility, where higher derivative non-local terms can be incorporated in the gravity sector \cite{Modesto:2011kw,Biswas:2011ar,Biswas:2012ka,Biswas:2012bp,Biswas:2013cha,Talaganis:2014ida,Biswas:2014tua,Modesto:2014lga,Dona:2015tra,Koshelev:2016xqb}. For example one can consider the possibilities, 
 $Rf_{1}(\Box)R$, 
$R_{\mu\nu}f_{2}(\Box)R^{\mu\nu}$, 
$R_{\mu\nu\alpha\beta}f_{3}(\Box)R^{\mu\nu\alpha\beta}$,
,$Rf_{4}(\Box)\nabla_{\mu}\nabla_{\nu}\nabla_{\alpha}\nabla_{\beta}R^{\mu\nu\alpha\beta}$,
$R_{\mu}^{\nu\alpha\beta}f_{5}(\Box)\nabla_{\alpha}\nabla_{\beta}\nabla_{\nu}\nabla_{\rho}\nabla_{\lambda}\nabla_{\gamma}R^{\mu\rho\lambda\gamma}$,
$R^{\mu\nu\alpha\beta}f_{6}(\Box)\nabla_{\alpha}\nabla_{\beta}\nabla_{\nu}\nabla_{\mu}\nabla_{\lambda}\nabla_{\gamma}\nabla_{\eta}\nabla_{\xi}R^{\lambda\gamma\eta\xi}$,
 where $\Box$ is defined as,
 $\Box=\frac{1}{\sqrt{-g}}\partial_{\mu}\left[\sqrt{-g}~g^{\mu\nu}\partial_{\nu}\right]$ is the d'Alembertian operator in 4D and the  $f_{i}(\Box)\forall i=1,2,\cdots,6$ are analytic entire functions containing higher derivatives up to infinite order. This is itself a very complicated possibility which we have not explored in this paper.
\end{enumerate}
\item \underline{\textcolor{red}{\bf Category III:}}\\
In this case, inflation is driven by changing the mathematical structure of both the gravitational and matter sector of the effective theory. One of the examples is to use Jordan-Brans-Dicke (JBD) gravity theory \cite{Brans:1961sx,Brans:1962zz} along with extended inflationary models which includes non-canonical interactions. By adjusting the value of Brans-Dicke parameters one can study the
observational consequences from this setup. Instead of Jordan-Brans-Dicke (JBD) gravity theory one can also use non local gravity or many other complicated possibilities.
\end{itemize} 
In this paper, we consider the possibility of soft inflationary paradigm in Einstein frame, where a chaotic Higgsotic potential is coupled to a dilaton via exponential type of potential, which is appearing through the conformal transformation from Jordan to Einstein frame in the metric within the framework of scale free $\alpha R^2$ gravity. Here it is important to mention that, in case of soft inflationary model, the dilaton exponential potential is multiplied by an coupling constant of the Higgsotic theory which mimics the role of an effective
coupling constant and its value always decreases with the field value. One can generalize this idea for any arbitrary matter interactions which is also described by generalized $P(X,\phi)$ theory \cite{Chen:2006nt,Khoury:2010gb}(see appendix \ref{df1} for more details). In this context of discussion also it is important to specify that, one can treat the field dependent couplings in the simple effective potentials or may be in a generalized $P(X,\phi)$ functionals contains a decaying behaviour with dilaton field value as it contains an overall exponential factor which is coming from the dilaton potential itself in Einstein frame. This is a very  interesting feature from the point of view of RG flow in QFT as the field dependent coupling in Einstein frame captures the effect of field flow (energy flow). In this context instead of solving directly the RGE for the effective coupling, we solve the dynamical equations for the fields and the effective coupling for power law and exponential attractors. Due to the similarities in both of the techniques here one can arrive at the conclusion that in cosmology solving a dynamical attractor problem in presence of effective coupling in Einstein frame mimics the role of solving RGE in QFT. Thus due to the exponential suppression in the effective coupling in Einstein frame it is naturally expected from the prescribed framework that for suitable choices of the model parameters soft cosmological constraints can be obtained \cite{Berkin:1990ju,Berkin:1991nm}. As in this prescribed framework dilaton exponential coupling plays very significant role, one can ask a very crucial question about its theoretical origin. Obviously there are various sources exist from which one can derive exponential effective couplings or more precisely the effective potentials for dilaton. These possibilities are appended bellow:
\begin{itemize}
\item \textcolor{red}{\underline{\bf Source~I}:}\\ One of the source for dilaton exponential potential is string theory, which is appearing in the \textcolor{red}{\underline{\bf Category~I}}. Specifically, superstring theory and low energy supergravity models are the theoretical possibilities in string theory \cite{Bars:1992sr,Callan:1985ia,Bars:1990rb,Choudhury:2013yg,Choudhury:2013aqa,Choudhury:2014hna,Choudhury:2015wfa,Derendinger:1985kk,Ellis:1986zt,Pilch:2000ue} where dilaton exponential potential is appearing in the gravity part of the action in Jordan frame and after conformal transformation in Einstein frame such dilaton effective potential is coupled with the matter sector. The most important example is $\alpha$-attractor which mimics a class of inflationary models in ${\cal N}=1$ supergravity in 4D. For details see ref.~ \cite{Kallosh:2013yoa,Kallosh:2014rga,Galante:2014ifa,Kallosh:2015lwa,Linde:2015uga,Carrasco:2015uma,Carrasco:2015rva,Carrasco:2015pla,Kallosh:2016gqp,Kallosh:2016sej,Ferrara:2016fwe}.

\item \textcolor{red}{\underline{\bf Source~II}:}\\ Another possible source of dilaton exponential potential is coming from modified gravity theory framework such as, $f(R)$ gravity \cite{Starobinsky:1979ty,Sebastiani:2015kfa,DeFelice:2010aj,Sotiriou:2008rp}, $f(\phi)f(R)$ gravity \cite{Bezrukov:2007ep,Hertzberg:2010dc,Pallis:2010wt,Budhi:2017yjd} and Jordan-Brans-Dicke
theory \cite{Brans:1961sx,Brans:1962zz} in Jordan frame, which are appearing in the \textcolor{red}{\underline{\bf Category~II~(1\& 3)}} and \textcolor{red}{\underline{\bf Category~III}}. After transforming the theory in the Einstein frame via conformal transformation one can derive dilaton exponential potential.

\end{itemize}
 
\begin{figure*}[htb]
                       \centering
                       \subfigure[Diagrammatic representation of attractor phase of soft Higgsotic inflation. In this representative diagram we have shown the steps followed during the computation.]{
                           \includegraphics[width=7.6cm,height=5cm] {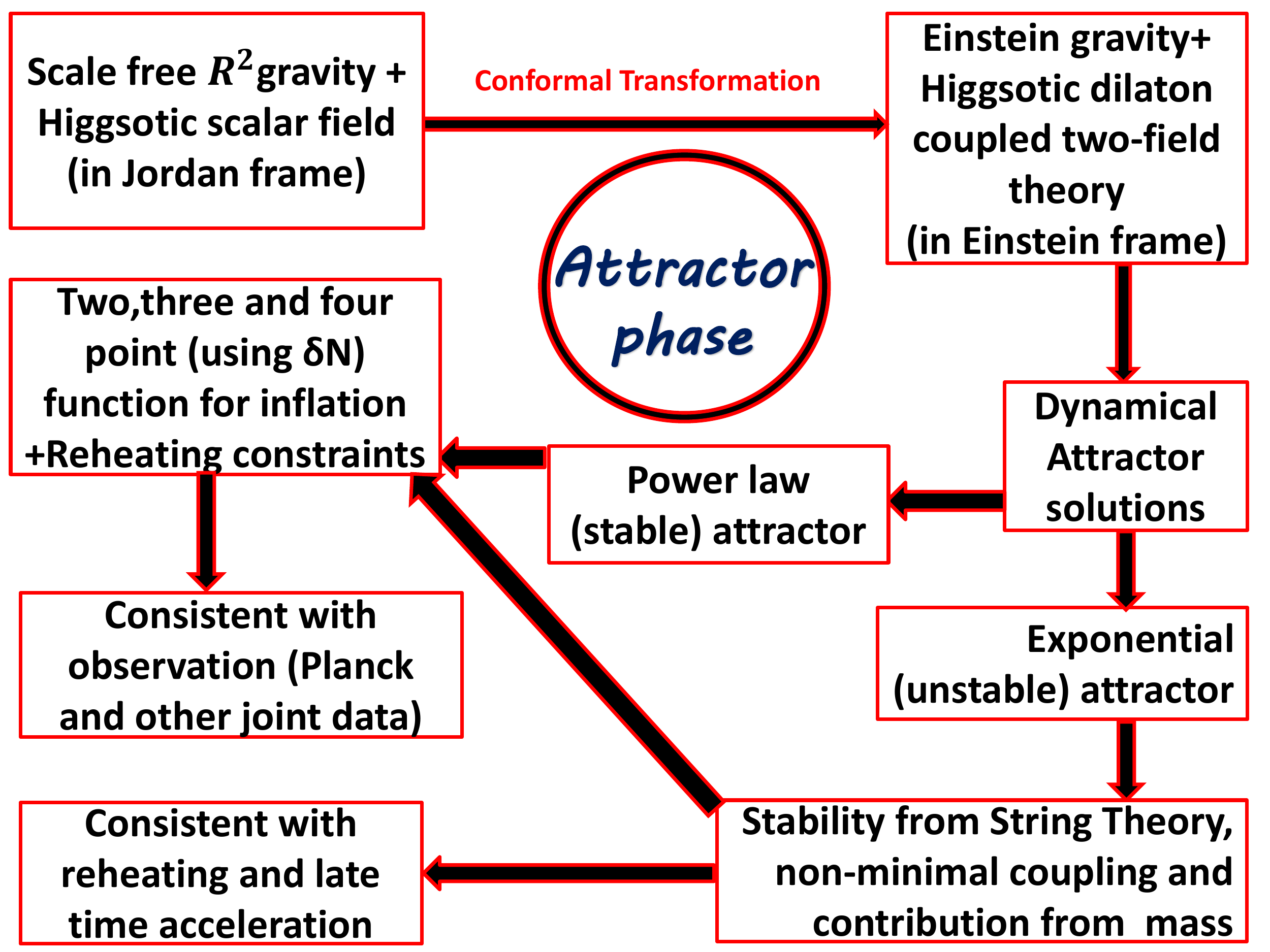}
                           \label{figg1}
                       }
                       \subfigure[Diagrammatic representation of non-attractor phase of soft Higgsotic inflation. In this representative diagram we have shown the steps followed during the computation.]{
                           \includegraphics[width=7.6cm,height=5cm] {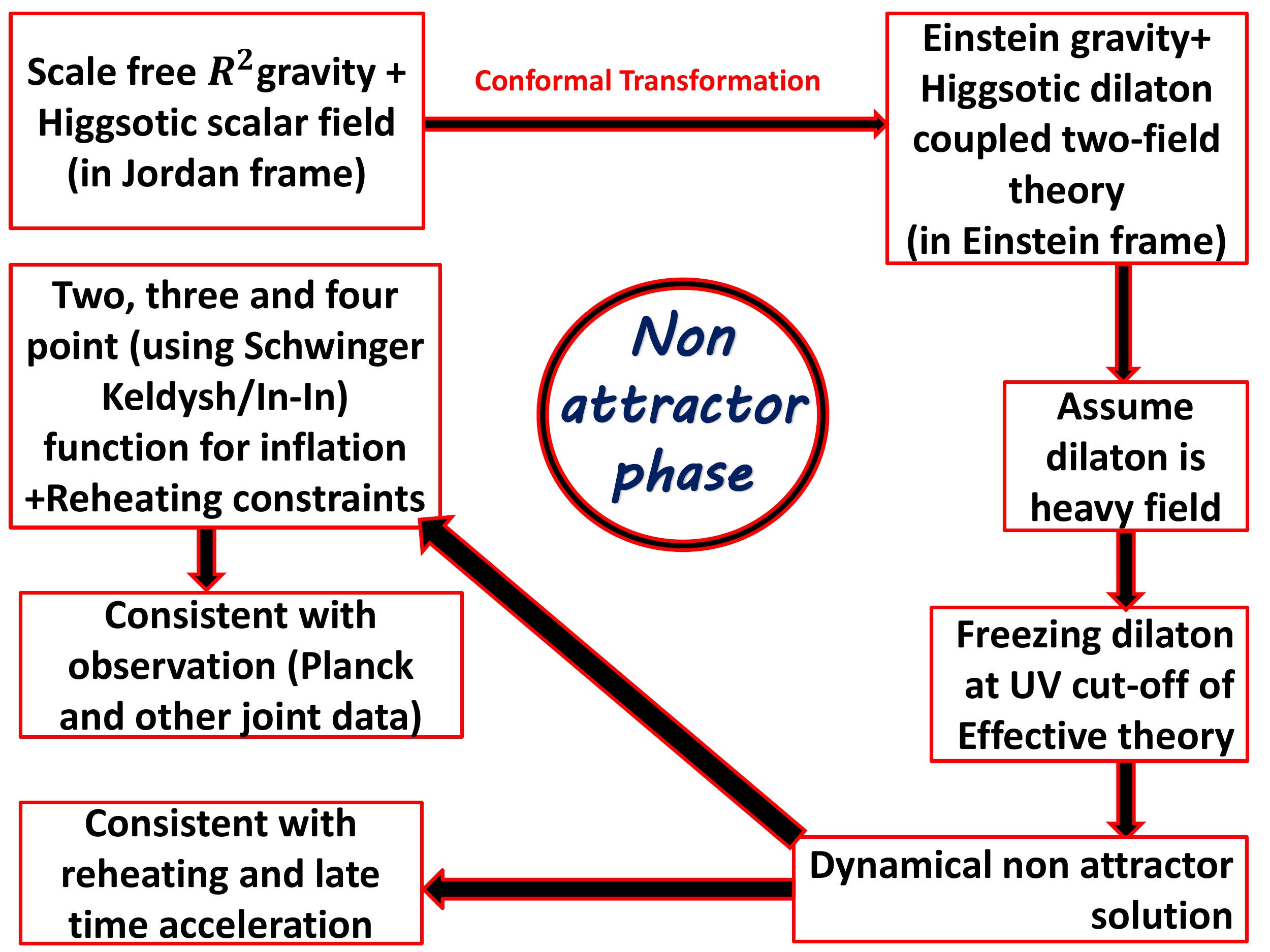}
                           \label{figg2}
                       }
                       \subfigure[Diagrammatic representation of attractor and non-attractor dynamical phase of soft Higgsotic inflation which is coupled with dilaton in Einstein frame.]{
                                \includegraphics[width=9.6cm,height=5cm] {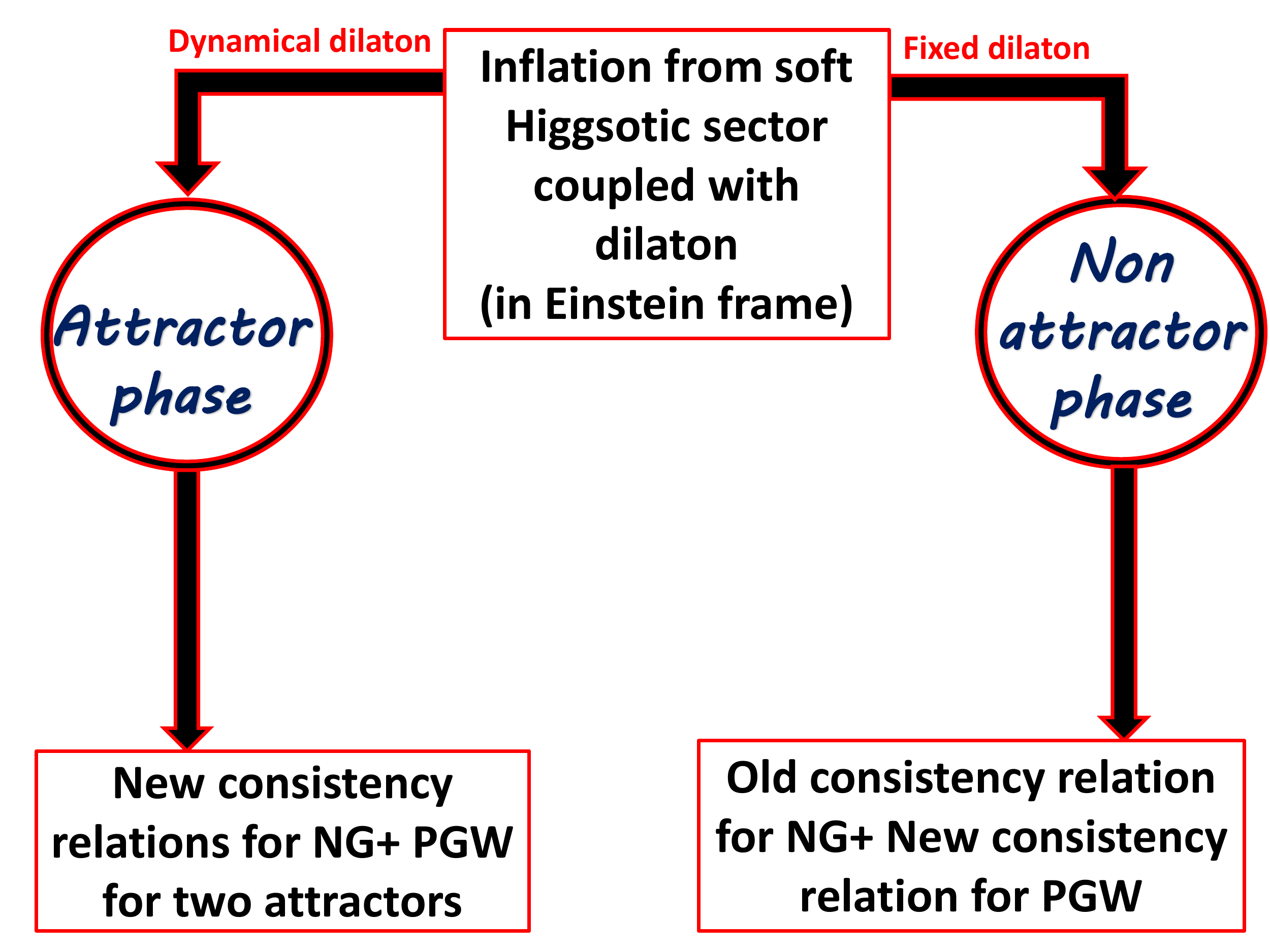}
                                \label{figg3}
                            }
                       \caption[Optional caption for list of figures]{Representative scematic diagram of attractor and non-attractor phase of soft Higgsotic inflation.}
                       \label{sch}
                       \end{figure*}

In fig~(\ref{figg1}), fig~(\ref{figg2}) and fig~(\ref{figg3}), we have shown the diagrammatic representation of attractor and non-attractor phase of soft Higgsotic inflation. In these representative diagrams we have shown the steps followed during the computation.
In this work we have addressed the following important points through which it is possible to understand the underlying cosmological consequences from the proposed setup. These issues are:
\begin{itemize}
 \item Transition from scale free gravity to scale dependent gravity have discussed and its impact on the solutions in the attractor and non-attractor regime of inflation have also discussed. 
 
 \item Explicit calculation of $\delta{\cal N}$ formalism is presented by considering the effect up to second order perturbation in the solution of the field equation in attractor regime. Additionally deviation in the consistency relation between the non-Gaussian amplitude for four point and three point scalar correlation function {\it aka}  Suyama Yamaguchi relation is presented to explicitly show the consequences from attractor and non-attractor phase.
 
 \item Additionally, new sets of consistency relations are presented in attractor and non-attractor phase of inflation to explicitly show the deviation from the results obtained from canonical single field slow roll model.  
 
 \item Detailed numerical estimations are given for all the inflationary observables for attractor and non-attractor phase of inflation which confronts well Planck 2015 data. Additionally, constraints on reheating is also presented for attractor and non-attractor phase.
 
 \item Bulk interpretation are given in terms of $S$, $T$ and $U$ chanel contribution for all the individual terms obtained from three and four point correlation function.
 
 \item Scale dependent behaviour of the non-minimal coupling between inflaton field and additional dilaton field are given in Einstein frame for power law and exponential type of attractor. 
 
 \item Three possible theoretical proposals have presented to overcome the tachyonic instability \cite{Felder:2001kt,Felder:2002jk,Gumrukcuoglu:2016jbh,Saitou:2011hv,Tsujikawa:2010zza} at the time of late time acceleration in Jordan frame and due to this fact the structure of the effective potentials changes in Einstein frame as well. These proposals are inspired from:
 \begin{itemize}
 \item \textcolor{red}{\bf I.} Non-BPS D-brane in superstring theory \cite{Choudhury:2015hvr,Sen:1999mg,Frau:1999qs,Eyras:2000my,Bergshoeff:2000dq,Eyras:2000ig,Brax:2000cf},
 \item \textcolor{red}{\bf II.} An alternative situation where we switch on the effects of additional quadratic mass term in the effective potential,
 \item  \textcolor{red}{\bf III.} Also we have considered a third option where we switch on the effect of non-minimal coupling between scale free $\alpha R^2$ gravity and the inflaton field.
 \end{itemize}

\end{itemize}
Now before going to the further technical details let us clearly mention the underlying assumptions to understand the background physical setup of this paper:
\begin{enumerate}
 \item We have restricted our analysis up to monomial $\phi^4$ model and due to the structural similarity with Higgs potential at the scale of inflation we have identified monomial $\phi^4$ model as Higgsotic model in the present context. 
 
 \item To investigate the role of scale free theory of gravity, as an example we have used $\alpha R^2$ gravity. But the present analysis can be generalized to any class of $f(R)$ gravity models.
 
 \item In the matter sector we allow only simplest canonical kinetic term which are minimally coupled with $\alpha R^2$ gravity sector. For such canonical slow roll models the effective sound speed $c_S=1$. But for more completeness one can consider most generalized version of $P(X,\phi)$ models, where $X=-\frac{1}{2}g^{\mu\nu}\partial_{\mu}\phi\partial_{\nu}\phi$ and the effective sound speed $c_S<1$ for such models. For an example one can consider following  structure \cite{Alishahiha:2004eh,Chen:2006nt}:
 \be P(X,\phi)=-\frac{1}{f(\phi)}\sqrt{1-2Xf(\phi)}+\frac{1}{f(\phi)}-V(\phi),\ee
 which is exactly similar to DBI model. But here one can implement our effective Higgsotic models in $V(\phi)$ instead of choosing the fixed structure of the DBI potential in UV and IR regime. Here one can choose \cite{Alishahiha:2004eh}, $f(\phi)\approx \frac{g}{\phi^4}$, which is known as throat factor in string theory. In string theory $g$ is the parameter which depends on the flux number. But other choices for $f(\phi)$ are also allowed for general class of $P(X,\phi)$ theories which follows the above structure. Similarly one can consider the 
 following structure of $P(X,\phi)$ which are given for tachyon and Gtachyon models given by \cite{Choudhury:2015hvr,Sen:2004nf}: \bea \textcolor{red}{\bf For~Tachyon :}~~~~~~~~~~P(X,\phi)&=&-V(\phi)\sqrt{1-2X\alpha^{'}},\\
 \textcolor{red}{\bf For~GTachyon :}~~~~~~~~~~P(X,\phi)&=&-V(\phi)\left(1-2X\alpha^{'}\right)^{q}~~~~~~~(1/2<q<2),~~~~\eea
 where $\alpha^{'}$ is the Regge slope. Here we consider the most simplest canonical form, $ P(X,\phi)=X-V(\phi)$, where $V(\phi)$ is the effective potential for monomial $\phi^4$ model considered here for our computation.
 
 \item As a choice of initial condition or precisely as the choice of vaccum state we restrict our analysis using Bunch Davies vacuum. If we relax this assumption, then one can generalize the results for $\alpha$ vacua as well.
 
 \item During our computation we have restricted upto the minimal interaction between the $\alpha R^2$ gravity and matter sector. Here one can consider the possibility of non-minimal interaction between $\alpha R^2$ gravity and matter sector. 
 
 \item During the implementation of In-In formalism \cite{Baumann:2009ds} to compute three and four point correlation function we have use the fact that the additional dilaton field $\Psi$ is fixed at Planckian field
 value to get the non attrator behavior of the present setup. One can relax this assumption and can redo the analysis of In-In formalism to compute three and four point correlation function without freezing the dilaton field $\Psi$ and also use the attractor behaviour of the model to simplify the results. 
 
 \item During the computation of correlation functions using semi classical method, via $\delta {\cal N}$ formalism \cite{Choudhury:2015hvr,Sugiyama:2012tj,Choudhury:2014uxa,Lee:2005bb,Domenech:2016zxn,Chen:2013eea}, we have restricted up to second order contributions in the solution of the field equation in FLRW background and also neglected the contributions from the back reaction for all type of effective Higgsotic models derived in Einstein frame. For more completeness, one can relax these assumptions and redo the analysis by taking care of all such contributions.
 
 \item In this work we have neglected the contribution from the loop effects (radiative corrections) in all of the effective Higgsotic interactions (specifically in the self couplings) derived in the Einstein frame. After switching on all such effects one can investigate the numerical contribution of such terms and comment on the effects of such terms in precision cosmology measurement.
 
 \item We have also neglected the interactions between gauge fields and Higgsotic scalar field in this paper. One can consider such interactions by breaking conformal invariance of the $U(1)$ gauge field in presence of time dependent coupling $f(\phi(\eta))$ to study the features of primordial magnetic field through inflationary magnetogenesis \cite{Choudhury:2015jaa,Choudhury:2014hua,Subramanian:2015lua}.
\end{enumerate}

The plan of this paper is as follows:
\begin{itemize}
\item In \underline{\textcolor{purple}{sec}~\ref{s1}}, we start our discussion with $f(R)=\alpha R^2$ gravity where a scalar field is minimally coupled with the gravity sector and contains only canonical kinetic term. Next in the matter sector we choose a very simple monomial model of potential, $V(\phi)=\frac{\lambda}{4}\phi^4$, which can be treated as a Higgs like potential as at the scale of inflation, contribution from the VEV of Higgs almost negligible. 

\item Further, in \underline{\textcolor{purple}{sec}~\ref{s2}}, we provide the field equations in Jordan frame written in spatially flat FLRW background. Next, we perform a conformal transformation in the metric to the Einstein frame and introduce a new dilaton field. Further, we derive the field equations in Einsein frame and try to solve them for two dynamical attractor features as given by-Power law solution, and Exponential solution. However, the second case give rise to tachyonic behaviour which can be resolved by considering- I. non-BPS D-brane in superstring theory, II. via switching on the effect of quadratic term in the effective potential and III. by introducing a non-minimal coupling between matter and $\alpha R^2$ gravity sector. 
 
\item Next, in \underline{\textcolor{purple}{sec}~\ref{s3}}, using two dynamical attractors, Power law and Exponential solution we study the cosmological constraints in presence of two fields. We study the constraints from primordial density perturbation, by deriving the expressions for two point function and the present inflationary observables in \underline{\textcolor{purple}{sec}~\ref{s3b}}. Further, we repeat the analysis for tensor modes and also comment on the future observables-amplitude of the tensor fluctuations and tensor-to-scalar ratio in \underline{\textcolor{purple}{sec}~\ref{s3c}}. Additionally, in \underline{\textcolor{purple}{sec}~\ref{s3d}}, we study the constraint for reheating temperature. Finally, in \underline{\textcolor{purple}{sec}~\ref{s4a}} and \underline{\textcolor{purple}{sec}~\ref{s4b}}, we derive the expression for inflaton and the non minimal coupling at horizon crossing, during reheating and at the onset of inflation for two above mentioned dynamical cosmological attractors. 

\item Further, in \underline{\textcolor{purple}{sec}~\ref{s5}}, we have explored the cosmological solutions beyond attractor regime. Here we restrict ourselves at spatially flat FLRW background and made cosmological predictions from this setup in \underline{\textcolor{purple}{sec}~\ref{s6a}}. To serve this purpose we have used ADM formalism using which we compute two point function and associated present observables using Bunch Davies initial condition for scalar fluctuations in \underline{\textcolor{purple}{sec}~\ref{s6b1}} and \underline{\textcolor{purple}{sec}~\ref{s6b2}}. Further, in \underline{\textcolor{purple}{sec}~\ref{s6c1}} and \underline{\textcolor{purple}{sec}~\ref{s6c2}}, we repeat the procedure for tensor fluctuations as well where we have compute two point function and the associated future observables. We also derive few sets of consistency relations in this context which are different from the usual single field slow roll models. Further, in \underline{\textcolor{purple}{sec}~\ref{s6d}}, we derive the constraints on reheating temperature in terms of observables and number of e-foldings. 

\item Next, in \underline{\textcolor{purple}{sec}~\ref{s7a1}} and \underline{\textcolor{purple}{sec}~\ref{s7a2}}, as a future probe we compute the expression for three point function and the bispectrum of scalar fluctuations using In-In formalism for non attractor case and $\delta {\cal N}$ formalism for the attractor case. Further, we derive the result for non-Gaussian amplitude $f^{loc}_{NL}$ for equilateral and squeezed limit triangular shape configuration. Also we give a bulk interpretation of each of the momentum dependent terms appearing in the expression for the three point scalar correlation function in terms of $S$, $T$ and $U$ channel contributions. Further, for the consistency check we freeze the additional field $\Psi$ in Planck scale and redo the analysis of $\delta {\cal N}$ formalism. Here we show that the expression for the three point non-Gaussian amplititude is slightly different as expected for single field case. Further, in \underline{\textcolor{purple}{sec}~\ref{s7a1}} and \underline{\textcolor{purple}{sec}~\ref{s7a2}}, we compare the results obtained from In-In formalism and $\delta{\cal N}$ formalism for the non attractor phase, where the additional field $\Psi$ is fixed in Planck scale. Finally, we give a theoretical bound on the scalar three point non-Gaussian amplitude.

\item Finally, in \underline{\textcolor{purple}{sec}~\ref{s7b1}} and \underline{\textcolor{purple}{sec}~\ref{s7b2}}, as an additional future probe we have also computed the expression for four point function and the trispectrum of scalar fluctuations using In-In formalism for non attractor case and $\delta {\cal N}$ formalism for the attractor case. Further, we derive the results for non-Gaussian amplitude $g^{loc}_{NL}$ and $\tau^{loc}_{NL}$ for equilateral, counter collinier or folded kite and squeezed limit shape configuration from In-In formalism. Further we give a bulk interpretation of each of the momentum dependent terms appearing in the expression for the four point scalar correlation function in terms of $S$, $T$ and $U$ channel contributions. In the attractor phase following the prescription of $\delta {\cal N}$ formalism we also derive the expressions for the four point non-Gaussian amplitude $g^{loc}_{NL}$ and $\tau^{loc}_{NL}$. Next we have shown that the consistency relation connecting three and four point non-Gaussian amplitude {\it aka} Suyama Yamaguchi relation is modified in attractor phase and further given an estimate of the amount of deviation. Further, for the consistency check we freeze the additional field $\Psi$ in Planck scale and redo the analysis of $\delta {\cal N}$ formalism. Here we show that the four point non-Gaussian amplitude is slightly different as expected for single field case. Finally, we give a theoretical bound on the scalar four point non-Gaussian amplitude.
 \end{itemize}
\section{\textcolor{blue}{\bf Model building from scale free gravity}}
 \label{s1}
 To described the theoretical setup let us start with the total action of $f(R)$ gravity coupled minimally along with a scalar inflaton field $\phi$, as given by:
\bea\label{eq1}
S&=&\int d^{4}x\sqrt{-g}\left[f(R)-\frac{g^{\mu\nu}}{2}
\left(\partial_{\mu}\phi\right)\left(\partial_{\nu}\phi\right)-V(\phi)\right]
\eea
where in general $f(R)$ can be arbitrary function of Ricci scalar $R$. For an example one can choose a generic form given by \cite{Choudhury:2015zlc,Choudhury:2016tzz}:
\be \label{fr} f(R)=\sum^{\infty}_{n=1}a_{n}R^{n},\ee
where $a_{n}\forall n$ are the expansion coefficients for the above mentioned generic expansion. Here one can note down the following features of this generic choice of the expansion:
\begin{enumerate}
\item If we set $a_{1}= M^2_p/2,~ a_{n}= 0 \forall n>1,$ then one can get back well known Einstein Hilbert action (GR) in Joradn frame as given by, $f(R)=M^2_p R/2$. In this particular case Jordan frame and Einstein frame is exactly same because the conformal factor for the frame transformation is unity. This directly implies that no dilaton potential is appearing due to the frame transformation from Jordan to Einstein frame. But since in this paper we are specifically interested in the effects of modified gravity sector, the higher powers of $R$ is more significant in the above mentioned generic expansion of $f(R)$ gravity.  

\item If we set, $a_{1}= a=M^{2}_{p}/2,~ a_{2}= b=\alpha,~ a_{n}= 0 \forall n>2,$ then one can get back the specific structure of the very well known Starobinsky model as given by,
$ f(R)=aR+bR^2=M^2_p R/2+\alpha R^2$.
Here one can treat the $\alpha R^2$ term as an additional quantum correction to the Einstein gravity.

\item One can also set, $ a_{1}= a=M^{2}_{p}/2,~~a_{n}= \alpha \forall n\geq 2,$ then one can get back the following specific structure, $f(R)=M^2_p R/2+\alpha R^n$,
which describes the situation where the Einstein Hilbert gravity action is modified by the monomial powers of $R$. Here also one can treat the $\alpha R^n$ term as an additional quantum correction to the Einstein gravity.

\item In our computation we set, $ a_{1}= a=0,~ a_{2}= b=\alpha,~ a_{n}= 0\forall n>2,$ which is known as scale free gravity in Jordan frame as given by,$f(R)=\alpha R^{2}$,
where $\alpha$ is a dimensionless scale free coefficient. For this type of theory if er perform the conformal transformation from Jordan to Einstein frame then it will induce a constant term in the effective potential and can be interpreted as the 4D cosmological constant using which one can fix the scale of the theory for early and late universe. But in our computation we introduce an additional scalar field in the action in Jordan frame, which we identified to be the inflaton. After conformal transformation in Einstein frame we get an effective potential which is coming from the interaction between the dilaton exponential potential and the inflationary potential as appearing in Jordan frame. We will show that here the two fields- dilaton and inflaton forms dynamical attractors using which one can very easily solve this two field complicated model in the context of cosmology.
\end{enumerate}
Next we will discuss about the structure of the inflational as appearing in Eq~(\ref{eq1}).
Generically in 4D effective theory the effective potential can be expressed as:
\be V(\phi)=\underbrace{V_{ren}(\phi)}_{\bf Renormalizable~part}+\underbrace{\sum^{\infty}_{\delta=5}J_{\delta}(g)\frac{\phi^{\delta}}{M^{\delta-4}_p}}_{\bf Non-renormalizable~part}=\sum^{\infty}_{\delta=0}C_{\delta}(g)\frac{\phi^{\delta}}{M^{\delta-4}_p},\ee
where $J_{\delta}(g)$ and $C_{\delta}(g)$ are the Wilson coeeficients in effective theory. Here $g$ stands for the scale of theory and the dependences of the Wilson coefficients on the scale can be exactly computed for a full UV complete theory using renormalization group equations. In this paper the similar scale dependence on the couplings we will calculate using dynamical attractor method in Einstein frame, which exactly mimics the role of solving renormalization group equations in the context of cosmology. As written here, the total effective potential is made by renormalizable (relevant operators) and non-renormalizable (irrelevant operators) part, which can be obtained by heavy degrees of freedom from a known UV complete theory. In our computation we just concentrate on the renormalizable part of the action, which can be recast as:
\be V(\phi)=\sum^{4}_{\delta=0}C_{\delta}(g)\frac{\phi^{\delta}}{M^{\delta-4}_p},\ee
Next to get the Higgslike monomial structure of the potential we set $C_{3}(g)=0,$ as in this paper our prime motivation is to look into only Higgsotic potentials. Consequently we get:
\be V(\phi)=C_{0}+C_{2}(g)M^2_p\phi^2+C_{4}(g)\phi^4.\ee 
To get the Higgsotic structure of the potential one should set, \bea C_{0}(g)&=&\frac{\lambda}{4}v^4,~~ C_{2}(g)=-\frac{\lambda}{2}v^2,~~   
 C_{4}(g)=\frac{\lambda}{4}.\eea Here $v$ is the VEV of the field $\phi$. Consequently, one can write the potential in the following simplified form:
 \be V(\phi)=\frac{\lambda}{4}(\phi^2-v^2)^2.\ee
 Now we consider a situation where scale of inflation as well as the field value are very very larger than the VEV of the field. This assumption is pretty consistent with inflation with Higgs field. Consequently, in our case the final simplified monomial form of the Higgsotic potential is given by:
\bea\label{pot}
V(\phi)&=&\frac{\lambda}{4}\phi^{4}.\eea
Further varying Eq.~(\ref{eq1}) with respect to the metric and using Eq~(\ref{fr}) and Eq~(\ref{pot})  eqn of motion (modified Einstein eqn) for the $\alpha R^2$ scale free gravity can be written as:
\bea\label{eq2}
\tilde{G}_{\mu\nu}:&=&\alpha \left[\left\{R_{\mu\nu}+2\left(g_{\mu\nu}\Box-\nabla_{\mu}\nabla_{\nu}\right)\right\}+G_{\mu\nu}\right]R=T_{\mu\nu}
\eea
where the D'Alembertian operator is defined as, $\Box=g^{\alpha\beta}\nabla_{\alpha}\nabla_{\beta}=g^{\alpha\beta}\nabla_{\alpha}\partial_{\beta}=\frac{1}{\sqrt{-g}}\partial_{\alpha}\left(\sqrt{-g}
g^{\alpha\beta}\partial_{\beta}\right)$
and the energy-momentum stress tensor can be expressed as:
\bea\label{eq4}
T_{\mu\nu}&=&-\frac{2}{\sqrt{-g}}\frac{\delta\left(\sqrt{-g}{\cal L}_{M}\right)}{\delta g^{\mu\nu}}=\partial_{\mu}\phi\partial_{\nu}\phi-g_{\mu\nu}
\left(\frac{1}{2}g^{\alpha\beta}\partial_{\alpha}\phi\partial_{\beta}\phi+\frac{\lambda}{4}\phi^{4}\right)
\eea
Here it is important to note that the Einstein tensor is defined as, 
$G_{\mu\nu}:=R_{\mu\nu}-\frac{g^{\mu\nu}}{2}R$.
Now after taking the trace of Eq.~(\ref{eq2}) we get,
$R\Box R=\frac{T}{6\alpha},$
where the trace of energy momentum tensor is characterized by the symbol $T=T_{\mu}^{\mu}$.
In this modified gravity picture we have,
$\nabla^{\mu}\tilde{G}_{\mu\nu}=4\alpha\left[\nabla_{\mu},\Box\right]R\neq0$
where we use, 
$\nabla^{\mu}R_{\mu\nu}=\frac{g^{\mu\nu}}{2}\nabla^{\mu}R$, which 
dirctly follows from the Bianchi identity $\nabla^{\mu}G_{\mu\nu}=0$.
Now varying Eq~(\ref{eq1}) with respect to the field $\phi$ we get the following eqn of motion in curved spacetime:
\bea\label{eq7a}
\Box\phi= -V^{'}(\phi)=-\lambda\phi^{3} \Longrightarrow \frac{1}{\sqrt{-g}}\partial_{\alpha}\left(\sqrt{-g}
g^{\alpha\beta}\partial_{\beta}\phi\right)=-\lambda\phi^{3}.
\eea
Further assuming the flat ($k=0$) FLRW background metric the Friedmann Equations can be written from Eq.~(\ref{eq2}) as:
\bea\label{eq8}
H^{2}&=&\left(\frac{\dot{a}}{a}\right)^2=\frac{\rho_{\phi}}{6\alpha R}+\frac{R}{2}-\left(\frac{\dot{R}}{R}\right)H,\\
\label{eq9z}2\dot{H}+3H^{2}&=&2\left(\frac{\ddot{a}}{a}\right)+\left(\frac{\dot{a}}{a}\right)^{2}=-\frac{p_{\phi}}{2\alpha R}
-2\left(\frac{\dot{R}}{R}\right)H-\frac{\ddot{R}}{R}+\frac{R}{4}
\eea
where we have assumed the energy-momentum tensor can be described by perfect fluid as, $T^{\mu}_{\nu}=diag\left(-\rho_{\phi},p_{\phi},p_{\phi},p_{\phi}\right)$
where the energy density $\rho_{\phi}$
and the pressure density $p_{\phi}$ can be expressed for scalar field $\phi$ as:
\bea\label{eq10}
\rho_{\phi}=\frac{\dot{\phi}^2}{2}+\frac{\lambda}{4}\phi^{4},~~~
\label{eq11} p_{\phi}=\frac{\dot{\phi}^2}{2}-\frac{\lambda}{4}\phi^{4}.
\eea
Similarly the field eqn for the scalar field $\phi$ in the flat ($k=0$) FLRW background can be recast as: 
\bea\label{eq11a}
\ddot{\phi}+3H\dot{\phi}+\lambda\phi^{3}&=&0
\eea
In the flat ($k=0$) FLRW background we have the following expressions:
\bea\label{eq12}
R&=&6\left(\dot{H}+2H^2\right),~~
\dot{R}=6\left(\ddot{H}+4H\dot{H}\right),~~
\ddot{R}= 6\left(\dddot{H}+4\dot{H}^2+4H\ddot{H}\right).
\eea
Substituting these results in Eq~(\ref{eq8}) and Eq~(\ref{eq9z}) the Friedmann eqns can be recast in the Jordan frame as:
\bea\label{eq13}
2H\left(\ddot{H}+3H\dot{H}\right)-\dot{H}^2&=&\frac{\rho_{\phi}}{18\alpha},\\
\label{eq14}9\dot{H}\left(\dot{H}+H^{2}\right)+6H\ddot{H}+\dddot{H}&=&-\frac{p_{\phi}}{6\alpha}.
\eea
In the slow-roll regime ($\dot{\phi}^2/2<<\frac{\lambda}{4}\phi^{4}$) the energy density $\rho_{\phi}$
and the pressure density $p_{\phi}$ can be approximated as, $\rho_{\phi}\approx \frac{\lambda}{4}\phi^{4},~
\label{eq16} p_{\phi}\approx -\frac{\lambda}{4}\phi^{4}$.
Consequently Eq~(\ref{eq11a}), Eq~(\ref{eq13}) and Eq~(\ref{eq14}) can be recast as:
\bea\label{eq17xx}
3H\dot{\phi}+\lambda\phi^{3}&\approx& 0,\\
\label{eq17}
2H\left(\ddot{H}+3H\dot{H}\right)-\dot{H}^2&\approx&\frac{V(\phi)}{18\alpha},\\
\label{eq18}9\dot{H}\left(\dot{H}+H^{2}\right)+6H\ddot{H}+\dddot{H}&\approx&-\frac{V(\phi)}{6\alpha}
\eea
where $V(\phi)=\frac{\lambda}{4}\phi^4$.
Further combining Eq~(\ref{eq17}) and Eq~(\ref{eq18}) we get, $\dddot{H}= 3\dot{H}\left(3H^{2}-4\dot{H}\right)$.
For further analysis one can also define following sets of slow-roll parameters in Jordan frame:
\bea\label{eq20}
\epsilon_{H}&=&-\frac{\dot{H}}{H^{2}},~~~
\label{eq21a}\delta_{H}=-\frac{\ddot{H}}{H^{3}}=\left(\frac{\dot{\epsilon_{H}}}{H}-2\epsilon^{2}_{H}\right),~~
\label{eq22}\gamma_{H}=-\frac{\dddot{H}}{H^{4}}=3\epsilon_{H}\left(3+4\epsilon_{H}\right),~~~
\label{eq21} \eta_{H}= -\frac{\ddot{\phi}}{H\dot{\phi}}.\eea
Further using these new sets of parameters Eq~(\ref{eq17}) and Eq~(\ref{eq18}) can be recast into the following simplified form:
\bea\label{eqn22}
2\delta_{H}+\frac{\gamma_{H}}{12}+\frac{21}{4}\epsilon_{H}\approx -\frac{V(\phi)}{18\alpha H^{4}}=-\frac{\lambda\phi^{4}}{72\alpha H^{4}}.\eea
However solving this two-field problem in presence of scale free gravity is itself very complicated for the following reasons:
\begin{itemize}
\item \textcolor{red}{\bf\underline{Complication~I:}}\\ First of all, for a given structure of inflationary potential in Jordan frame (here it is Higgsotic potential as mentioned earlier) it is impossible to solve directly the dynamical equations Eq~(\ref{eq17}), Eq~(\ref{eq18}) and Eq~(\ref{eqn22}) due its complicated coupled structural form.

\item \textcolor{red}{\bf\underline{Complication~II:}}\\ One can use various solution ansatz to get approximated numerical results, but this is also dependent on the structure of the inflaton potential in Jordan frame and how one can able to implement initial condition (starting point) of inflation for arbitrary structure of the effective potential.

\item \textcolor{red}{\bf\underline{Complication~III:}}\\ In connection with the implementation of the initial condition and to check the sufficient condition for inflation in this complicated field theoretical setup one needs to define the expression for number of e-foldings in terms of effective potentials. But this cannot be possible very easily in the present context as the field equations are coupled. 
\end{itemize}
Due to these huge number of difficulties in Jordan frame we transform the total action into the Einstein frame using conformal transformation. After transforming the Jordan frame action into the Einstein frame in the present context we need the solve a two interacting field problem in presence of Einstein gravity. There are several ways one can solve this problem. These possibilities are:
\begin{itemize}
\item \textcolor{red}{\bf\underline{Solution~I:}}\\ The first solution to solve this problem is to follow the well known approach to solve two field models of inflation by following the method of curvature and isocurvature perturbation in the semiclassical $\delta{\cal N}$ formalism. For more accurate results one can also solve directly the Mukhanov-Sasaki equation for this two field model and directly treat fluctuations quantum mechanically. Since this methodology have discussed in various earlier works, we will not discuss this issue in in this paper. See refs.~\cite{Peterson:2010np,Wands:2002bn,GarciaBellido:1995qq,Kaiser:2013sna,Wands:2007bd} fore more details. 

\item \textcolor{red}{\bf\underline{Solution~II:}}\\ Second way of solving this problem is to use dynamical attractor mechanism in the present context where the two fields are connected through specific relations, which can be obtained by solving dynamical field equations in cosmology. This is equivalent to solving renormalization group equations in the context of quantum field theory as the dynamical attractor solution of two fieds captures the effects of all the energy scale. In our computation we explore the possibility of two dynamical attractors:
\begin{enumerate}
\item Power law attractor
\item Exponential attractor
\end{enumerate}
 Here both of them have different cosmological consequences. But they are originated from Higgsotic structure of the effective potential which we will discuss later in the next section in detail. 

\item
  \textcolor{red}{\bf\underline{Solution~III:}}\\
  Final possibility is to freeze the dilaton field in the Planck scale or in the vicinity, so that one can absorb it in the effective couplings in the Higgsotic theory. This is identified as the non-attractor phase in the context of cosmology. The physical justification for such possibilities can also be explained from the UV behaviour of the 4D effective theory, which is known as the UV completion of the effective theory. According to this proposal we have two sectors in the theory:- 
  \begin{enumerate}
  \item  \textcolor{red}{\bf\underline{Hidden sector:}}\\
  Hidden sector is made up of heavy field (in our case dilaton) which lies around the UV cutoff of the effective theory, which is the Planck scale. We can't able to probe directly this sector. But can visualize its imprints on the low energy effective theory.
  
  \item   \textcolor{red}{\bf\underline{Visible sector:}}\\
  Visible sector is made up of lite field (in our case inflaton) which one can probe directly. For present discussion visible sector is important to explain the cosmological evolution.
  
  \end{enumerate}
    Usually in such a prescription one integrate the heavy fields and finally get an effective theory in the visible sector. Here we use the fact that such procedure mimics the role of freezing the heavy dilaton field near the Planck scale. The only difference is, in the case of freezing the dilaton field we only concentrate on the Higgsotic potential. But the integration of heavy field allows all relevant and irrelevant operators. However, by applying the similar argument one can look into only the renormalizable Higgsotic part of the total effective potential. Additionally it is important to note that at late times the dynamical picture is completely opposite where the inflaton field freezes at the vicinity of the Planck scale and the dynamical contribution for late time acceleration comes from dilaton field. In more simpler way one can interpret this physical prescription as the competitive dynamical description of two field. During inflation Higgsotic field wins the game and at late times dilaton serves the same purpose. More precisely, within this prescription dynamic features transfers from dilaton to Higgsotic field (or any scalar inflaton) during inflation and at late times completely opposite situation appears, where the similar transfer takes place from inflaton to dilaton field. 
\end{itemize}
In this paper we explore the possibility of  \textcolor{red}{\bf\underline{Solution~II}} and  \textcolor{red}{\bf\underline{Solution~III}} in detail in the next section. For completeness we briefly review also \textcolor{red}{\bf\underline{Solution~I}} in the appendix.

\section{\textcolor{blue}{Soft attractor: A two field approach}}
\label{s2}
In the present context let us introduce a scale dependent mode $\Psi$, which can be written in terms of a no scale dilaton mode $\Theta$ as:
\bea \label{eq22cx}
\Theta &=& f^{'}(R)M^{-2}_{p}=2\alpha R~M^{-2}_{p}=e^{\sqrt{\frac{2}{3}}\frac{\Psi}{M_p}}
\eea
which mimics the role of a Lagrange multiplier and arises in the Jordan frame without space-time derivatives.
In terms of the newly introduced no scale dilaton mode $\Theta$ the total action of the theory (see Eq~(\ref{eq1})) can be recast as:
\bea
S&=&\int d^{4}x \sqrt{-g}\left\{\frac{M^{2}_{p}}{2}\Theta R-\frac{M^{4}_{p}}{8\alpha}\Theta^{2}-\frac{g^{\mu\nu}}{2}
\left(\partial_{\mu}\phi\right)\left(\partial_{\nu}\phi\right)-\frac{\lambda}{4}\phi^{4}\right\}.
\eea 
To study the behaviour of the proposed $R^2$ theory of gravity here we introduce the following conformal transformation (C.T.) in the metric from Jordan frame to the Einstein frame:
\bea\label{eq23}
g_{\mu\nu}&\xrightarrow{C.T.}&\tilde{g}_{\mu\nu}=\Omega^{2}g_{\mu\nu},~~
\label{eq24}g^{\mu\nu}\xrightarrow{C.T.}\tilde{g}^{\mu\nu}=\Omega^{-2}g^{\mu\nu},~~
\label{24a} \sqrt{-g}\xrightarrow{C.T.}\sqrt{-\tilde{g}}=\Omega^{4}\sqrt{-g}.
\eea
which satisfies the condition,
$g_{\mu\nu}g^{\nu\beta}=\tilde{g}_{\mu\kappa}\tilde{g}^{\kappa\beta}=\delta_{\mu}^{\beta}$.
In the present context conformal factor $\Omega$ is given by:
\bea\label{eq25a}
\Omega&=&\sqrt{\Theta}=e^{\frac{\sqrt{2}}{2\sqrt{3}}\frac{\Psi}{M_p}}.
\eea 
Under this proposed C.T. in the metric the Ricci curvature scalar in the Jordan frame ($R$) related to the Einstein frame ($\tilde{R}$) as:
\bea\label{eq27}
R&=& \Omega^{2}\left[\tilde{R}+6\widetilde{\Box}{\rm ln}~ \Omega-6\tilde{g}^{\mu\nu}\tilde{\partial}_{\mu}{\rm ln}~ \Omega~\tilde{\partial}_{\nu}{\rm ln}~ \Omega\right]
\eea 
where $\tilde{\partial}_{\mu}= \frac{\partial }{\partial \tilde{x}^{\mu}}$ and $\widetilde{\Box}{\rm ln}~ \Omega\equiv\frac{1}{\sqrt{-\tilde{g}}}\partial_{\alpha}\left(\sqrt{-\tilde{g}}
\tilde{g}^{\alpha\beta}\partial_{\beta}{\rm ln}~ \Omega\right)$.
After doing C.T. the total action can be recast in the Einstein frame as~\footnote{Here we apply Gauss's theorem to remove the following
 contribution in the total effective action:
\bea 
\int d^{4}x \sqrt{-\tilde{g}} \sqrt{\frac{3}{2}}M_{p}\tilde{\Box}\Psi&=&\sqrt{\frac{3}{2}}M_{p}\int d^{4}x ~\partial_{\alpha}\left(\sqrt{-\tilde{g}}
\tilde{g}^{\alpha\beta}\partial_{\beta}\Psi\right)=\oint_{\partial {\cal M}} d^{3}x~\left(\sqrt{-\tilde{g}}
\tilde{g}^{\alpha\beta}\partial_{\beta}\Psi\right)n_{\alpha}\equiv 0.
\eea
where $\partial {\cal M}$ represents the boundary of 4-volume and $n_{\alpha}$ be the unit normal.}:
\bea\label{eq311}
S~~\underrightarrow{C.T.}~~ \tilde{S}
&=&\int d^{4}x \sqrt{-\tilde{g}}\left[\frac{M^{2}_{p}}{2}\tilde{R}+\frac{\tilde{g}^{\mu\nu}}{2}\tilde{\partial}_{\mu}\Psi\tilde{\partial}_{\nu}\Psi
+\frac{\tilde{g}^{\mu\nu}}{2}\tilde{\partial}_{\mu}\phi\tilde{\partial}_{\nu}\phi-\tilde{W}(\phi,\Psi)\right]
 \eea
where after applying C.T. the total potential can be recast as:
\bea\label{eq32}
\tilde{W}(\phi,\Psi)&=&\frac{\frac{M^{4}_{p}}{8\alpha}e^{2\sqrt{\frac{2}{3}}\frac{\Psi}{M_p}}+\frac{\lambda}{4}\phi^{4}}{\Omega^4}
=V_0 \left[1+\frac{2\alpha\lambda(\Psi)}{M^{4}_{p}}\phi^{4}\right]
\eea
where $V_0=M^{4}_{p}/8\alpha$
 exactly mimics the role of cosmological constant and the effective matter coupling ($\lambda(\Psi)$) in the potential sector is given by, $\lambda(\Psi)=\frac{\lambda}{\Omega^4}=\lambda e^{-\frac{2\sqrt{2}}{\sqrt{3}}\frac{\Psi}{M_p}}$.
Now varying Eq.~(\ref{eq311}) with respect to the metric the field Eqns can be expressed as:
\bea\label{eq31cf}
\tilde{\cal G}_{\mu\nu}:=\left(\tilde{R}_{\mu\nu}-\frac{\tilde{g}_{\mu\nu}}{2}\tilde{R}\right)=\tilde{T}_{\mu\nu}\left(\phi, \Psi\right)
\eea
where the energy-momentum tensor $\tilde{T}_{\mu\nu}\left(\phi, \Psi\right)$ for the dilaton-inflaton coupled theory can be expressed as:
\bea\label{eq34}
\tilde{T}_{\mu\nu}\left(\phi, \Psi\right)&=&-\frac{2}{\sqrt{-\tilde{g}}}\frac{\delta\left(\sqrt{-\tilde{g}}{\cal L}(\phi,{\Psi})\right)}{\delta \tilde{g}^{\mu\nu}}=\tilde{\partial}_{\mu}\phi\tilde{\partial}_{\nu}\phi+\tilde{\partial}_{\mu}\Psi\tilde{\partial}_{\nu}\Psi-\tilde{g}_{\mu\nu}
\left(\frac{1}{2}\tilde{g}^{\alpha\beta}\tilde{\partial}_{\alpha}\phi\tilde{\partial}_{\beta}\phi \nonumber\right.\\ &&\left. \nonumber
~~~~~~~~~~~~~~~~~~~~~~~~~~~~~~~~~~~~~~~~~~~~~~~~~~~~~~~~~~~~~+\frac{1}{2}\tilde{g}^{\alpha\beta}\tilde{\partial}_{\alpha}\Psi\tilde{\partial}_{\beta}\Psi+\tilde{W}(\phi,\Psi)\right).
\eea
Here for the matter part of the action the following property holds between the Einstein frame and Jordan frame energy-momentum tensor,
$\tilde{T}_{\mu\nu}(\phi,\Psi)\supset\tilde{T}_{\mu\nu}=-\frac{2}{\sqrt{-\tilde{g}}}\frac{\delta\left(\sqrt{-\tilde{g}}{\cal L}_{M}\right)}{\delta \tilde{g}^{\mu\nu}}
=\frac{T_{\mu\nu}}{\Omega^{2}}$
which implies that using the perfect fluid assumption one can write, $
\tilde{T}^{\mu}_{\nu}=diag\left(-\tilde{\rho}_{\phi},\tilde{p}_{\phi},\tilde{p}_{\phi},\tilde{p}_{\phi}\right)=\frac{1}{\Omega^4}
diag\left(-\rho_{\phi},p_{\phi},p_{\phi},p_{\phi}\right)=\frac{T^{\mu}_{\nu}}{\Omega^4}$.
Assuming the flat ($k=0$) FLRW background metric in Einstein frame the Friedmann Equations can be written from Eq~(\ref{eq31cf}) as
~\footnote{It is important to mention here that the time interval in Einstein frame $d\tilde{t}$ is related to the time interval in Jordan frame $dt$ as, $d\tilde{t}=\Omega~ dt.$}:
\bea\label{eq38}
\tilde{H}^2&=&\left(\frac{d\ln a}{d\tilde{t}}\right)^{2}=\frac{\tilde{\rho}}{3M^2_p},\\
\frac{d\tilde{H}}{d\tilde{t}}+\tilde{H}^2&=&\left(\frac{d^{2} a}{d\tilde{t}^{2}}\right)=-\frac{\left(\tilde{\rho}+3\tilde{p}\right)}{6M^{2}_p}
\eea
where the effective energy density ($\tilde{\rho}$) and the effective pressure ($\tilde{p}$) can be written in Einstein frame as:
\bea\label{con1a}
\tilde{\rho}&=&\left(\frac{d\Psi}{d\tilde{t}}\right)^2 +\left(\frac{d\phi}{d\tilde{t}}\right)^2+\tilde{W}(\phi,\Psi)\,,~~~~
\label{con1b}\tilde{p}=\left(\frac{d\Psi}{d\tilde{t}}\right)^2 +\left(\frac{d\phi}{d\tilde{t}}\right)^2-\tilde{W}(\phi,\Psi)\,.
\eea
Additionally, the Hubble parameter in the Einstein frame ($\tilde{H}$) can be expressed its Jordan frame ($H$) counterpart as, $\tilde{H}=\frac{1}{\Omega}\left[H+\frac{1}{2}\frac{d\ln\Omega^2}{dt}\right]=
e^{-\frac{1}{\sqrt{6}}\frac{\Psi}{M_p}}\left\{H+\frac{\dot{\Psi}}{\sqrt{6}M_p}\right\}$.
Also the Klien-Gordon field equations for inflaton field $\phi$ and the new field $\Psi$ can be written in the flat ($k=0$) FLRW background as:
\bea
\label{eqs11}
\frac{d^{2}\phi}{d\tilde{t}^{2}}+3\tilde{H}\frac{d\phi}{d\tilde{t}}+\partial_{\phi}\tilde{W}(\phi,\Psi)&=&0\,\\
\label{eqs21}\frac{d^{2}\Psi}{d\tilde{t}^{2}}+3\tilde{H}\frac{d\Psi}{d\tilde{t}}+\partial_{\Psi}\tilde{W}(\phi,\Psi)&=&0.\,
\eea
Now in the slow-roll regime the field equations are approximated as:
\bea
\label{eqs1n1}
3\tilde{H}\frac{d\phi}{d\tilde{t}}+\lambda(\Psi)\phi^3=0\,\\
\label{eqs2n1}3\tilde{H}\frac{d\Psi}{d\tilde{t}}-\frac{\lambda(\Psi)\phi^4}{\sqrt{6} M_p}=0,\,\\
\label{eqs3n1}\tilde{H}^2=\frac{\tilde{W}(\phi,\Psi)}{3M^2_p}=\frac{V_0}{3M^2_p}\left[1+\frac{2\alpha\lambda(\Psi)}{M^{4}_{p}}\phi^{4}\right],
\eea
 To study the behavior of the proposed model let us consider two cases, where the dynamical features are characterized by:
 \begin{enumerate}
 \item \textcolor{red}{\bf Case I: Power-law solution},
 \item \textcolor{red}{\bf Case II: Exponential solution}.
 \end{enumerate}
 which we discuss in the next subsection.
\subsection{Case I: Power-law solution}
\label{s2a}
We consider here large $\alpha$, small $V_0(\approx 0)$ with $\lambda>0$ with effective potential: \be
\tilde{W}(\phi,\Psi)
\approx \frac{\lambda(\Psi)}{4}\phi^{4}=\frac{\lambda}{4}e^{-\frac{2\sqrt{2}}{\sqrt{3}}\frac{\Psi}{M_p}}\phi^{4}~~~~~  \textcolor{red}{\bf{\bf (for ~~Case ~I)}}.\ee
Consequently the field equations can be recast as:
\bea
\label{eqs1nx}
3\tilde{H}\frac{d\phi}{d\tilde{t}}+\lambda e^{-\frac{2\sqrt{2}}{\sqrt{3}}\frac{\Psi}{M_p}}\phi^3=0,\,\\
\label{eqs2nx}3\tilde{H}\frac{d\Psi}{d\tilde{t}}-\frac{\lambda\phi^4}{\sqrt{6} M_p}e^{-\frac{2\sqrt{2}}{\sqrt{3}}\frac{\Psi}{M_p}}=0,\,\\
\label{eqs3nx}\tilde{H}^2=\frac{\lambda}{12M^2_p}e^{-\frac{2\sqrt{2}}{\sqrt{3}}\frac{\Psi}{M_p}}\phi^{4}.
 \eea
 This is the case where the cosmological constant $V_0$ or more precisely the parameter $\alpha$ will not appear in the final solution.
 The cosmological solutions of Eq.~(\ref{eqs1nx}-\ref{eqs3nx}) are given by~\footnote{Throughout the paper the subscript $`0'$ is used to describe the inflationary epoch.
}:
 \bea\label{po1}
 \textcolor{red}{\bf\underline{\bf Case ~I}}\nonumber\\
\Psi-\Psi_0&\approx&\frac{2\sqrt{2}M_p}{\sqrt{3}}\ln\left(\frac{a}{a_0}\right)=\frac{\sqrt{3}M_p}{\sqrt{2}}\ln\left(\frac{t}{t_0}\right)=-\frac{9}{2\sqrt{6}M_p}\left(\phi^2 -\phi^2_0\right),~~~~~~~~~~~\\
\label{po2aa} a&\approx&a_0\left(\frac{t}{t_0}\right)^{3/4},~~~~~~~~~\\
\label{po3aa} {\cal N}(\phi)-{\cal N}(\phi_0)&=& \frac{1}{M^{2}_{p}}\int^{\phi}_{\phi_0}d\phi \frac{\tilde{V}(\phi)}{\partial_{\phi}\tilde{V}(\phi)}=\frac{\phi^2- \phi^2_0}{8M^{2}_{p}}\approx-\frac{5}{9}\ln\left(\frac{a}{a_0}\right)=-\frac{5}{12}\ln\left(\frac{t}{t_0}\right),~~~~~~
\label{po4}
\eea
 \subsection{Case II: Exponential solution}
 \label{s2b}
 We consider small $\alpha$, large $V_0$ with $\lambda<0$ with effective potential: \be
\tilde{W}(\phi,\Psi)
\approx \frac{M^{4}_{p}}{8\alpha}+\frac{\lambda(\Psi)}{4}\phi^{4}=\frac{M^{4}_{p}}{8\alpha}-\frac{\lambda}{4}e^{-\frac{2\sqrt{2}}{\sqrt{3}}\frac{\Psi}{M_p}}\phi^{4}~~~~~ \textcolor{red}{{\bf (for ~~Case ~II)}}.\ee
Here to aviod any confusion we have taken out the signature of the coupling $\lambda$ outside in the expression for the effective potential for $\lambda<0$ case.

Finally the field equations can be expressed as:
\bea
\label{eqs1nxx}
3\tilde{H}\frac{d\phi}{d\tilde{t}}-\lambda e^{-\frac{2\sqrt{2}}{\sqrt{3}}\frac{\Psi}{M_p}}\phi^3=0\,\\
\label{eqs2nxx}3\tilde{H}\frac{d\Psi}{d\tilde{t}}+\frac{\lambda\phi^4}{\sqrt{6} M_p}e^{-\frac{2\sqrt{2}}{\sqrt{3}}\frac{\Psi}{M_p}}=0,\,\\
\label{eqs3nxx}\tilde{H}^2=\frac{M^{2}_{p}}{24\alpha}-\frac{\lambda}{12M^2_p}e^{-\frac{2\sqrt{2}}{\sqrt{3}}\frac{\Psi}{M_p}}\phi^{4},
\eea 
The cosmological solutions of Eq.~(\ref{eqs1nxx}-\ref{eqs3nxx}) are given by:

\bea
\textcolor{red}{\underline{\bf Case ~II}}\nonumber\\
\label{eq5}\Psi-\Psi_0&\approx&\frac{2\sqrt{2}M_p}{\sqrt{3}}\ln\left(\frac{a}{a_0}\right)=\frac{M^{2}_{p}}{3\sqrt{\alpha}}\left(t-t_0\right)
=-\frac{1}{2\sqrt{6}M_p}\left(\phi^2 -\phi^2_0\right),~~~~~~~~~~~\\
\label{po5} a&\approx&a_0~\exp\left[\frac{M_p}{2\sqrt{6\alpha}}\left(t-t_0\right)\right],~~~~~~~~~\\
\label{po6} {\cal N}(\phi)-{\cal N}(\phi_0)&=& \frac{1}{M^{2}_{p}}\int^{\phi}_{\phi_0}d\phi \frac{\tilde{V}(\phi)}{\partial_{\phi}\tilde{V}(\phi)}=-\frac{M^2_p}{16\alpha\lambda(\Psi)}\left(\frac{1}{\phi^2_0}-\frac{1}{\phi^2}\right)=-\frac{M^2_p}{16\alpha\lambda(\Psi)\phi^2_0}\left[1-\frac{1}{1-\frac{8M^2_p}{\phi^2_0}\ln\left(\frac{a}{a_0}\right)}\right]\nonumber\\
&&~~~~~~~~~~~~~~~~~~~~~~~~~~~~~~~~~~~~~~~~~~~~~~~~~~~~~~~~~~~\approx\frac{M^4_p}{2\alpha\lambda(\Psi)\phi^4_0}\ln\left(\frac{a}{a_0}\right)
\nonumber\\
&&~~~~~~~~~~~~~~~~~~~~~~~~~~~~~~~~~~~~~~~~~~~~~~~~~~~~~~~~~~~\approx\frac{M^5_p}{4\alpha^{3/2}\lambda(\Psi)\sqrt{6} \phi^4_0}\left(t-t_0\right)\eea
\begin{figure*}[htb]
\centering
\subfigure[$\textcolor{blue}{\bf \underline{Case ~I}:}$ Power-law behavior.]{
    \includegraphics[width=7.7cm,height=5cm] {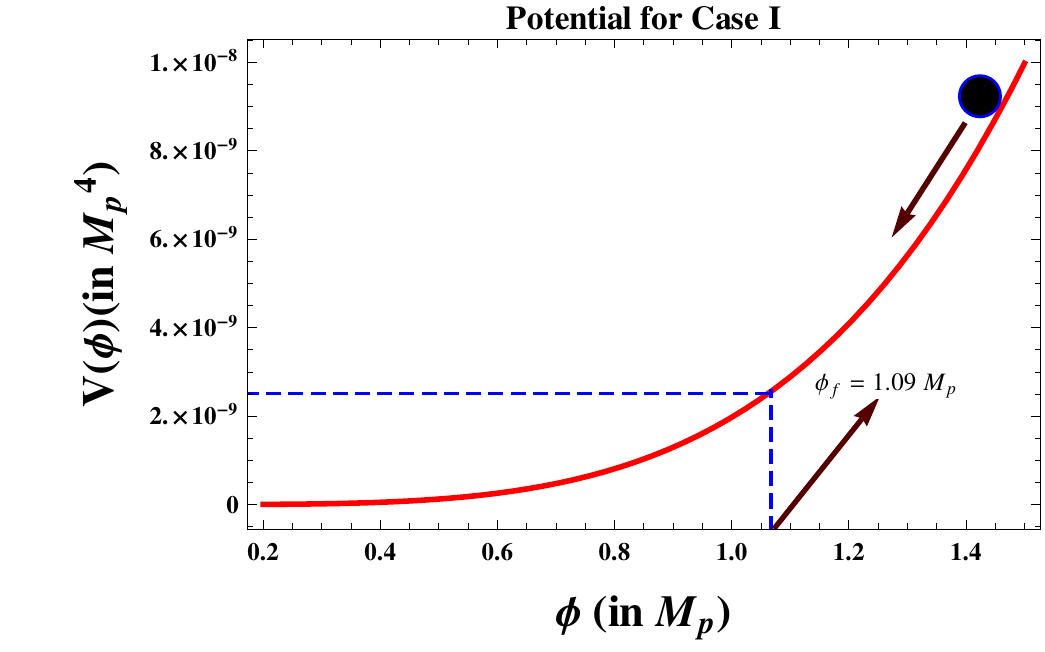}
    \label{fig1}
}
\subfigure[$\textcolor{blue}{\bf \underline{Case ~II}}:$ Tachyonic behaviour.]{
    \includegraphics[width=7.7cm,height=5cm] {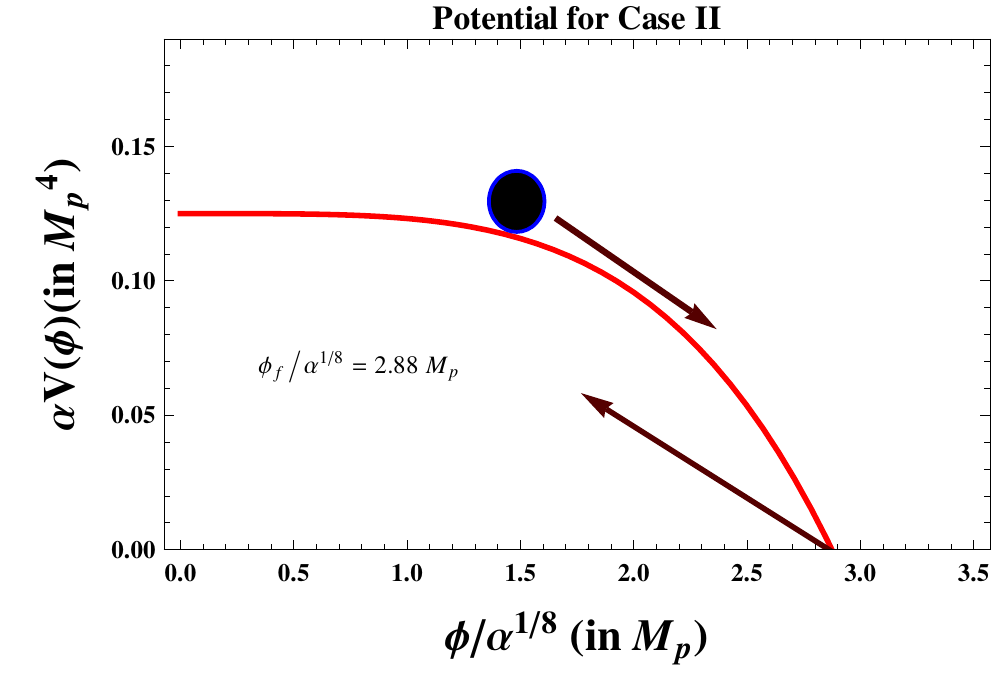}
    \label{fig2}
}
\caption[Optional caption for list of figures]{Behaviour of the inflationary potential for \ref{fig1} $V_0\approx 0$ and $\lambda>0$ ($\textcolor{blue}{\bf {Case ~I}}$) and \ref{fig2}
$V_0\neq0$ and $\lambda<0$ ($\textcolor{blue}{\bf {Case ~II}}$). 
In fig.~\ref{fig1} the inflaton rolls down from a large field value and inflation ends at 
$\phi_{f}\approx 1.09~{ M_p}$. On the other hand in fig.~(\ref{fig2})
the inflaton field rolls down from a small field value and the inflation ends at
the field value $\phi_{f}=2.88~\alpha^{1/8}~M_p$, where the lower bound on the
parameter $\alpha$ is: $\alpha\geq 2.51\times 10^{7}$, which is consistent
with Planck 2015 data \cite{Ade:2015lrj,Ade:2015ava,Ade:2015xua}.} 
\label{fzaa}
\end{figure*}
This is the specific case where the cosmological constant is explicitly appearing in the potential. To end inflation we need to fulfill an extra requirement that $\lambda<0$ and this will finally 
led to massless tachyonic solution. In In fig.~\ref{fig1} and fig.~\ref{fig2} we have shown the behaviour of the inflationary potential for the two cases, 1. $V_0\approx 0$ and $\lambda>0$, 2. $V_0\neq0$ and $\lambda<0$.

 Fig.~\ref{fig1} implies that the inflaton rolls down from a large field value and inflation ends at $\phi_{f}\approx 1.09~{ M_p}$. Also the potential has a global minimum at $\phi=0$,
around which field is start to oscillate and take part in reheating. On the other hand in fig.~(\ref{fig2})
the inflaton field rolls down from a small field value and the inflation ends at the field value $\phi_{f}=2.88~\alpha^{1/8}~M_p$, where the lower bound on the parameter $\alpha$ is,  $ \alpha\geq 2.51\times 10^{7}$, which is consistent
with Planck 2015 data \cite{Ade:2015lrj,Ade:2015ava,Ade:2015xua}. Within this prescription it is possible to completely destroy the effect of cosmological constant at the end of inflationary epoch. But within this setup to explain the particle production during reheating and also explain the late time acceleration of our universe we need additional features in the 
total effective potential in scale free $\alpha R^2$ gravity theory. It is general notion that , the reheating phenomena can only be explained if the effective potential have a local minimum and a remnant contribution (vacuum energy or equivalent to cosmological constant) in the total effective potential finally produce the observed energy density at the present epoch as given by~\footnote{For Einstein gravity one can write the observed energy density at the present epoch in the following form, 
$\rho_{now}\approx 3H^2_0 M^2_p$, where $H_0$ is the Hubble parameter at the present epoch.}
$\rho_{now}\approx 10^{-47}~{\rm GeV }^{4}$, which is necessarily required to explain the late time acceleration of the universe.
Now here one can ask a very relevant question that if we include some additional features to the effective Higgsotic potential, which is also can be treated as a massless tachyonic potential, then how one can interpret the justifiability as well as the behaviour of effective field theory framework around the minimum of the potential which will significantly control  
the dynamical behaviour in the context of cosmology. The most probable answer to this very significant question can be described in various ways. In the present context to get a stable minimum (vacuum) of the derived effective Higgsotic potential in Einstein frame here we discuss few physical possibilities which are appended in following points:
\begin{itemize}
\item\textcolor{red}{\underline{\bf Choice I:}}\\  The first possible solution of the mentioned problem is motivated from non-BPS D-brane in superstring theory. In this prescription  the effective potential have a pair of global extrima at the field value,
$\phi_{extrema}=\phi=\pm \phi_{V}$ for the non-BPS D-brane within the framework of superstring theory \cite{Choudhury:2015hvr,Sen:1999mg,Frau:1999qs,Eyras:2000my,Bergshoeff:2000dq,Eyras:2000ig,Brax:2000cf}. Additionally it is important to note that, here
a one parameter $(\gamma)$ family of global extrima exists at the field value, 
$\phi=\phi_{V}~e^{i\gamma}$ for the brane-antibrane
system. Here $\phi_{V}$ is identified to be the field value where reheating phenomena occurs. At this specified field value of the minimum the brane tension of the D-brane configuration which is exactly
canceled by the negative contribution as appearing in the expression for effective potential in Einstein frame. Here for the sake of simplicity we little bit relax the constraints as appearing exactly in \textcolor{red}{\underline{\bf Case II}}. To explore the behaviour of the derived effective potential here we have allowed both 
of the signatures of the coupling parameter $\lambda$. This directly implies the following constraint condition:
\bea  
-\frac{\lambda}{4}e^{-\frac{2\sqrt{2}}{\sqrt{3}}\frac{\Psi}{M_p}}\phi^{4}_{V}+{\bf \Theta}_{p}&=&0~~~~~~~~~(\textcolor{red}{\bf for ~\lambda<0}),\\
\frac{\lambda}{4}e^{-\frac{2\sqrt{2}}{\sqrt{3}}\frac{\Psi}{M_p}}\phi^{4}_{V}+{\bf \Theta}_{p}&=&0~~~~~~~~~(\textcolor{red}{\bf for ~\lambda>0}),
\eea
where ${\bf \Theta}_{p}$ is the above mentioned additional contribution and in the context of  superstring theory this is given by:
\be\begin{array}{lll}\label{rk9finnsxx}
 \displaystyle {\bf \Theta}_{p}=\left\{\begin{array}{ll}
                    \displaystyle  \sqrt{2}(2\pi)^{-p}g^{-1}_{s} &
 \mbox{\small {\textcolor{red}{\bf for non-BPS Dp-brane}}}  
\\ 
         \displaystyle 2(2\pi)^{-p}g^{-1}_{s} & \mbox{\small {\textcolor{red}{\bf for non-BPS Dp-$\bar{D}$p brane  pair}}}.
          \end{array}
\right.
\end{array}\ee
with string coupling constant $g_{s}$. This implies that the inflaton energy density vanishes at the minimum of the tachyon type of derived effective potential
and in this connection the remnant energy contribution is given by, $V_0=M^4_{p}/8\alpha$ which serves the explicit role of cosmological constant in the context of late time acceleration of the universe. In this case concedering  the additional contribution as mentioned above
the total effective potential can be modified as:
\bea\textcolor{red}{\bf v1:}~~~~
\tilde{W}(\phi,\Psi)
&=&\frac{M^{4}_{p}}{8\alpha}-\frac{\lambda}{4}e^{-\frac{2\sqrt{2}}{\sqrt{3}}\frac{\Psi}{M_p}}\left(\phi^4-\phi^{4}_{V}\right)~~~~~~~~~(\textcolor{red}{\bf for ~\lambda<0}),~~~~\\
\textcolor{red}{\bf v2:}~~~~\tilde{W}(\phi,\Psi)
&=&\frac{M^{4}_{p}}{8\alpha}+\frac{\lambda}{4}e^{-\frac{2\sqrt{2}}{\sqrt{3}}\frac{\Psi}{M_p}}\left(\phi^4-\phi^{4}_{V}\right)~~~~~~~~~(\textcolor{red}{\bf for ~\lambda>0}).~~~~
\eea
Here to aviod any confusion we have taken out the signature of the coupling $\lambda$ outside in the expression for the effective potential for $\lambda<0$ case.

In the present context the field equations can be expressed as:
\bea
\textcolor{red}{\bf For~ v1:}~~~~~~~~~~~~~~~~~~~~~\label{eqs1nxv1}
3\tilde{H}\frac{d\phi}{d\tilde{t}}-\lambda e^{-\frac{2\sqrt{2}}{\sqrt{3}}\frac{\Psi}{M_p}}\phi^3=0,\,\\
\label{eqs2nxv1}3\tilde{H}\frac{d\Psi}{d\tilde{t}}+\frac{\lambda\left(\phi^4-\phi^4_V\right)}{\sqrt{6} M_p}e^{-\frac{2\sqrt{2}}{\sqrt{3}}\frac{\Psi}{M_p}}=0,\,\\
\label{eqs3nxv1}\tilde{H}^2=\frac{M^{2}_{p}}{24\alpha}-\frac{\lambda}{12M^2_p}e^{-\frac{2\sqrt{2}}{\sqrt{3}}\frac{\Psi}{M_p}}\left(\phi^4-\phi^{4}_{V}\right).
\\
\textcolor{red}{\bf For~ v2:}~~~~~~~~~~~~~~~~~~~~~\label{eqs1nxv2}
3\tilde{H}\frac{d\phi}{d\tilde{t}}+\lambda e^{-\frac{2\sqrt{2}}{\sqrt{3}}\frac{\Psi}{M_p}}\phi^3=0,\,\\
\label{eqs2nxv2}3\tilde{H}\frac{d\Psi}{d\tilde{t}}-\frac{\lambda\left(\phi^4-\phi^4_V\right)}{\sqrt{6} M_p}e^{-\frac{2\sqrt{2}}{\sqrt{3}}\frac{\Psi}{M_p}}=0,\,\\
\label{eqs3nxv2}\tilde{H}^2=\frac{M^{2}_{p}}{24\alpha}+\frac{\lambda}{12M^2_p}e^{-\frac{2\sqrt{2}}{\sqrt{3}}\frac{\Psi}{M_p}}\left(\phi^4-\phi^{4}_{V}\right).
 \eea
 The solutions of Eq.~(\ref{eqs1nxv1}-\ref{eqs3nxv2}) are given by:
 \bea\label{po1v1}
\textcolor{red}{\underline{\bf Choice~I(v1)}}~~~\nonumber\\
\Psi-\Psi_0&\approx&\frac{2\sqrt{2}M_p}{\sqrt{3}}\ln\left(\frac{a}{a_0}\right)
=\frac{M^{2}_{p}}{3\sqrt{\alpha}}\left(t-t_0\right) 
 =-\frac{1}{2\sqrt{6}M_p}\left[\left(\phi^2 -\phi^2_0\right)+\phi^4_V\left(\frac{1}{\phi^2}-\frac{1}{\phi^2_0}\right)\right],~~~~~~~~~\\
\label{po2v1} a&\approx&a_0~\exp\left[\frac{M_p}{2\sqrt{6\alpha}}\left(t-t_0\right)\right],~~~~~~~~~\\
\label{po3v1} {\cal N}(\phi)-{\cal N}(\phi_0)&=&\frac{1}{M^{2}_{p}}\int^{\phi}_{\phi_0}d\phi \frac{\tilde{V}(\phi)}{\partial_{\phi}\tilde{V}(\phi)}\approx
-\left(\frac{M^2_{p}}{16\alpha\lambda(\Psi)}+\frac{\phi^4_{V}}{8M^2_p}\right)\left(\frac{1}{\phi^2_0}-\frac{1}{\phi^2}\right) \nonumber\\
&&~~~~~~~~~~~~~~~~~~~~~~~~~\approx\left(\frac{M^4_{p}}{2\alpha\lambda(\Psi)}+\frac{\phi^4_{V}}{\phi^2_0}\right)\ln\left(\frac{a}{a_0}\right),~~~~~~~
\label{po4v1}
\\ \label{po1v2}
\textcolor{red}{\underline{\bf Choice~I(v2)}}~~~\nonumber\\
\Psi-\Psi_0&\approx&\frac{2\sqrt{2}M_p}{\sqrt{3}}\ln\left(\frac{a}{a_0}\right)=\frac{M^{2}_{p}}{3\sqrt{\alpha}}\left(t-t_0\right)=-\frac{1}{2\sqrt{6}M_p}\left[\left(\phi^2 -\phi^2_0\right)+\phi^4_V\left(\frac{1}{\phi^2}-\frac{1}{\phi^2_0}\right)\right],~~~~~~~~~\\
\label{po2v2} a&\approx&a_0~\exp\left[\frac{M_p}{2\sqrt{6\alpha}}\left(t-t_0\right)\right],~~~~~~~~~\\
\label{po3v2} {\cal N}(\phi)-{\cal N}(\phi_0)&=&\frac{1}{M^{2}_{p}}\int^{\phi}_{\phi_0}d\phi \frac{\tilde{V}(\phi)}{\partial_{\phi}\tilde{V}(\phi)}\approx
\left(\frac{M^2_{p}}{16\alpha\lambda(\Psi)}-\frac{\phi^4_{V}}{8M^2_p}\right)\left(\frac{1}{\phi^2_0}-\frac{1}{\phi^2}\right)\nonumber\\
&&~~~~~~~~~~~~~~~~~~~~~~~~~\approx\left(\frac{M^4_{p}}{2\alpha\lambda(\Psi)}-\frac{\phi^4_{V}}{\phi^2_0}\right)\ln\left(\frac{a}{a_0}\right),~~~~~~~~~~~~
\label{po4v2}
\eea
In fig.~(\ref{fig3a}) and Fig.~(\ref{fig3b}) we have shown the variation of the potential with respect to the inflaton field for both the cases. For fig.~(\ref{fig3a}) the inflaton can roll down in both ways. Firstly, this can 
roll down to a global minimum at the field value, $\phi_{V}=0$ from higher to lower field value and take part in particle production procedure during reheating. On the other hand, in the same picture the inflaton can also 
roll down to higher to lower field value in a opposite fashion. In that case the inflaton goes up to the zero energy level of the effective potential and cannot able to explain the
thermal history of the early universe in a proper sense. It is also important to note that, in this picture the position of the 
maximum of the effective potential in Einstein frame is at around the field value, $\phi_{V}=0.42~M_p$.
Fig.~(\ref{fig3b}) is the case where the signature of the coupling $\lambda$ is positive.
Also the behavior of the effective potential is completely opposite compared to the situation arising in fig.~(\ref{fig3a}).
In this case the inflaton field can be able to roll down to higher to lower field value or lower to higher field value. But in both the cases
the inflaton field settle down to a local minimum at, $\phi_{min}=\phi_{V}=0.42~M_p$ and within the vicinity of this point it will produce particles via reheating. 
In both of the situations the lower bound on the
parameter $\alpha$ is fixed at, $\alpha\geq 2.51\times 10^{7}$,  which is perfectly consistent
with Planck 2015 data \cite{Ade:2015lrj,Ade:2015ava,Ade:2015xua}.

\begin{figure*}[htb]
\centering
\subfigure[$\textcolor{blue}{\bf \underline{Choice ~I(v1)}:}$ Modified potential from superstring theory with $\lambda<0$.]{
    \includegraphics[width=7.7cm,height=5cm] {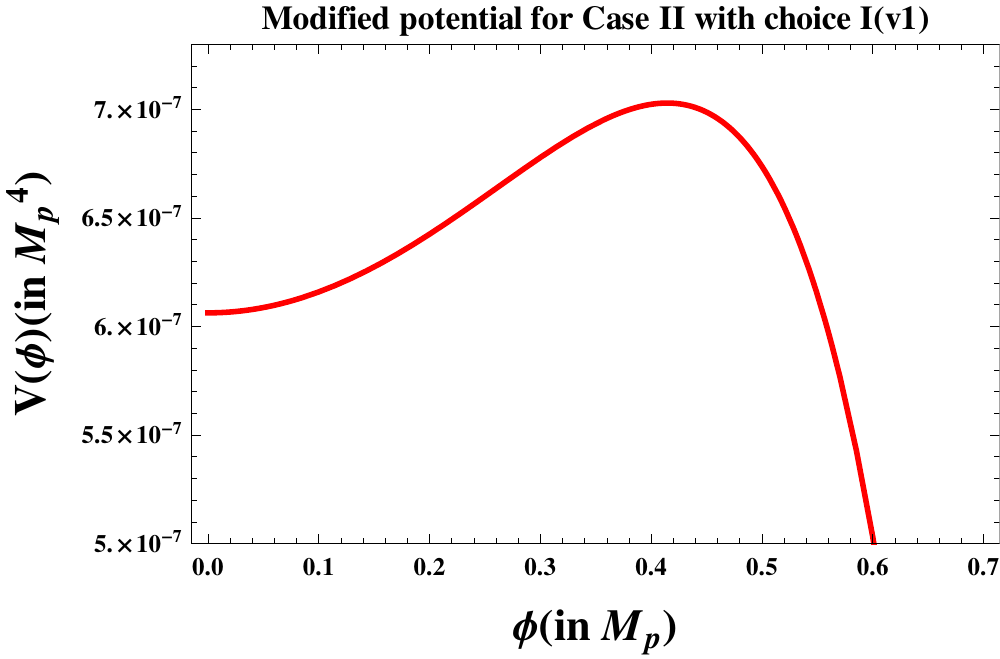}
    \label{fig3a}
}
\subfigure[$\textcolor{blue}{\bf \underline{Choice ~I(v2)}}:$ Modified potential from superstring theory with $\lambda>0$.]{
    \includegraphics[width=7.7cm,height=5cm] {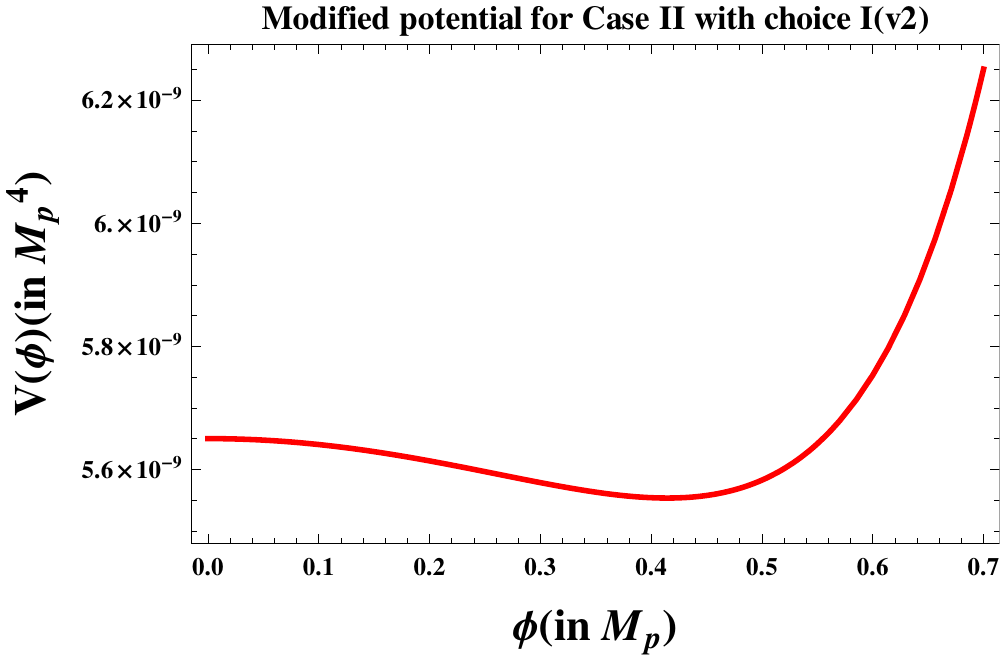}
    \label{fig3b}
}
\caption[Optional caption for list of figures]{Behaviour of the modified effective potential for case II with \ref{fig3a} $\textcolor{blue}{\bf \underline{Choice ~I(v1)}:}$ $V_0\neq0$, $\lambda<0$,  \ref{fig3b} $\textcolor{blue}{\bf \underline{Choice ~I(v2)}}:$ $V_0\neq0$, $\lambda>0$, where $M_p=2.43\times 10^{18}~GeV$.} 
\label{fzaqa}
\end{figure*}
 \item \textcolor{red}{\underline{\bf Choice II:}}\\
 It is possible to explain the reheating as well as the lite time cosmic acceleration once we switch on the effect of mass like quadratic term in the effective potential. In such a case the modified effective potential in Einstein frame
 can be written as:
  \bea\textcolor{red}{\bf v1:}~~~~
\tilde{W}(\phi,\Psi)
=\frac{M^{4}_{p}}{8\alpha}+\left(\frac{m^{2}_{c}}{2}\phi^2-\frac{\lambda}{4}\phi^{4}\right)e^{-\frac{2\sqrt{2}}{\sqrt{3}}\frac{\Psi}{M_p}}~~~~(\textcolor{red}{\bf for ~m^{2}_{c}>0,\lambda<0}),~~~~~~~\\
\textcolor{red}{\bf v2:}~~~~\tilde{W}(\phi,\Psi)
=\frac{M^{4}_{p}}{8\alpha}-\left(\frac{m^{2}_{c}}{2}\phi^2-\frac{\lambda}{4}\phi^{4}\right)e^{-\frac{2\sqrt{2}}{\sqrt{3}}\frac{\Psi}{M_p}}~~~~(\textcolor{red}{\bf for ~m^{2}_{c}<0,\lambda>0}).~~~~~~~\eea
Here to avoid any confusion we have taken out the signature of the coupling $\lambda$ outside in the expression for the effective potential for $\lambda<0$ case. In this context during inflation the inflaton field satisfies the constraint $\phi>>\sqrt{\frac{2}{|\lambda|}}|m_{c}|$. After inflation when
reheating starts then the field satisfies $\phi<<\sqrt{\frac{2}{|\lambda|}}|m_{c}|$.
Finally at the field value $\phi=\sqrt{\frac{2}{|\lambda|}}|m_{c}|$ the remnant energy $ V_0=M^4_{p}/8\alpha$
serves the purpose of explaining the late time acceleration of the universe. 
In the present context the field equations can be expressed as:
\bea
\textcolor{red}{{\bf For~ v1:}}~~~~~~~~~~~~~~~~~~~~~\label{eqs1nxvv1}
3\tilde{H}\frac{d\phi}{d\tilde{t}}+\left(m^{2}_{c}\phi-\lambda\phi^{3}\right)e^{-\frac{2\sqrt{2}}{\sqrt{3}}\frac{\Psi}{M_p}}=0,\,\\
\label{eqs2nxvv1}3\tilde{H}\frac{d\Psi}{d\tilde{t}}-\frac{2\sqrt{2}\left(\frac{m^{2}_{c}}{2}\phi^2-\frac{\lambda}{4}\phi^{4}\right)}{\sqrt{3} M_p}e^{-\frac{2\sqrt{2}}{\sqrt{3}}\frac{\Psi}{M_p}}=0,\,\\
\label{eqs3nxvv1}\tilde{H}^2=\frac{M^{2}_{p}}{24\alpha}+\frac{\left(\frac{m^{2}_{c}}{2}\phi^2-\frac{\lambda}{4}\phi^{4}\right)}{3M^2_p}e^{-\frac{2\sqrt{2}}{\sqrt{3}}\frac{\Psi}{M_p}}.
 \eea
 \bea
\textcolor{red}{{\bf For~ v2:}}~~~~~~~~~~~~~~~~~~~~~\label{eqs1nxvv2}
3\tilde{H}\frac{d\phi}{d\tilde{t}}-\left(m^{2}_{c}\phi-\lambda\phi^{3}\right)e^{-\frac{2\sqrt{2}}{\sqrt{3}}\frac{\Psi}{M_p}}=0,\,\\
\label{eqs2nxvv2}3\tilde{H}\frac{d\Psi}{d\tilde{t}}+\frac{2\sqrt{2}\left(\frac{m^{2}_{c}}{2}\phi^2-\frac{\lambda}{4}\phi^{4}\right)}{\sqrt{3} M_p}e^{-\frac{2\sqrt{2}}{\sqrt{3}}\frac{\Psi}{M_p}}=0,\,\\
\label{eqs3nxvv2}\tilde{H}^2=\frac{M^{2}_{p}}{24\alpha}-\frac{\left(\frac{m^{2}_{c}}{2}\phi^2-\frac{\lambda}{4}\phi^{4}\right)}{3M^2_p}e^{-\frac{2\sqrt{2}}{\sqrt{3}}\frac{\Psi}{M_p}}.
 \eea
 \begin{figure*}[htb]
 \centering
 \subfigure[$\textcolor{blue}{\bf \underline{Choice ~II(v1)}}:$ Modified potential with mass $m^{2}_{c}>0, \lambda<0$.]{
     \includegraphics[width=7.7cm,height=4.6cm] {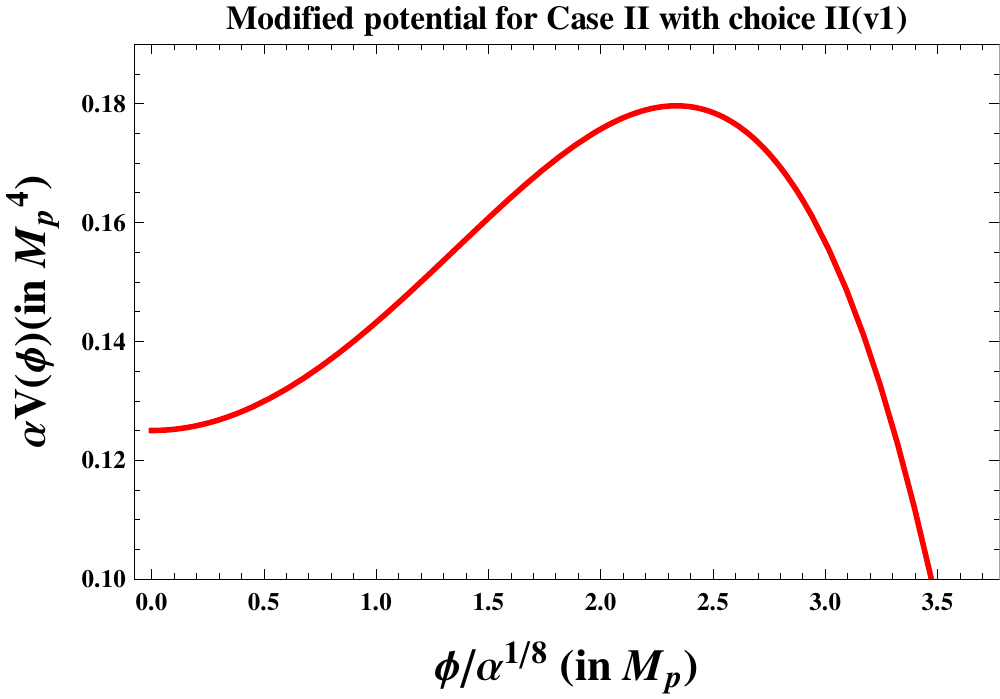}
     \label{fig4a}
 }
 \subfigure[$\textcolor{blue}{\bf \underline{Choice ~II(v2)}}:$ Modified potential with mass $m^{2}_{c}<0, \lambda>0$.]{
     \includegraphics[width=7.7cm,height=4.6cm] {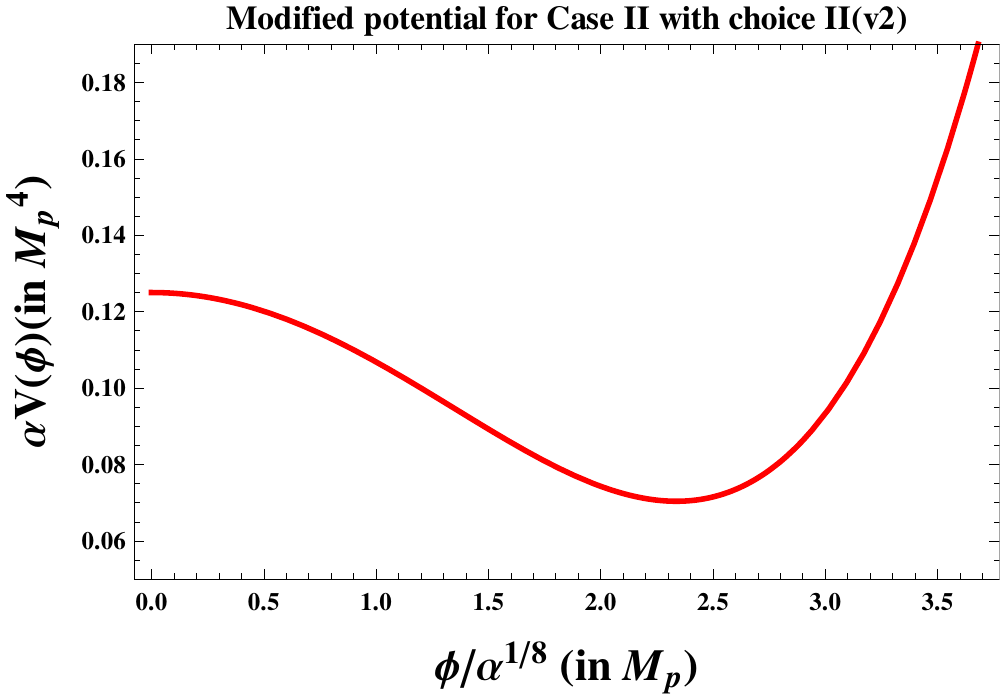}
     \label{fig4b}
 }
 \caption[Optional caption for list of figures]{Behaviour of the modified effective potential for case II with 
 \ref{fig4a} $\textcolor{blue}{\bf \underline{Choice ~II(v1)}}:$ $V_0\neq0$, $\lambda<0$, $m^{2}_{c}>0$ and $\phi<<\sqrt{\frac{2}{|\lambda|}}|m_{c}|$, \ref{fig4b} $\textcolor{blue}{\bf \underline{Choice ~II(v2)}}:$ $V_0\neq0$, $\lambda>0$, $m^{2}_{c}<0$ and $\phi<<\sqrt{\frac{2}{|\lambda|}}|m_{c}|$, where $M_p=2.43\times 10^{18}~GeV$.} 
 \label{fzaq}
 \end{figure*}
The solutions of Eq.~(\ref{eqs1nxvv1}-\ref{eqs3nxvv2}) are given by:
\bea\label{povvv1}
\textcolor{red}{\underline{\bf Choice~II(v1)}}~~~\nonumber\\
\Psi-\Psi_0&\approx&\frac{2\sqrt{2}M_p}{\sqrt{3}}\ln\left(\frac{a}{a_0}\right)
=\frac{M^{2}_{p}}{3\sqrt{\alpha}}\left(t-t_0\right) 
 =-\frac{1}{2\sqrt{6}M_p}\left[\left(\phi^2 -\phi^2_0\right)+\frac{m^{2}_{c}}{\lambda}\ln\left(\frac{m^2_{c}-\lambda\phi^{2}}{m^2_{c}-\lambda\phi^{2}_0}\right)\right],~~~~~~~~~\\
\label{po2vvv1} a&\approx&a_0~\exp\left[\frac{M_p}{2\sqrt{6\alpha}}\left(t-t_0\right)\right],~~~~~~~~~\\
\label{po3vvv1}{\cal N}(\phi)-{\cal N}(\phi_0)&=& \frac{M^4_p}{16m^{2}_{c}(\Psi)\alpha}\ln\left(\frac{\phi^2\left(m^2_{c}-\lambda\phi^2_0\right)}{\phi^2_0\left(m^2_{c}-\lambda\phi^2\right)}\right)\approx
\frac{M^4_p}{16m^{2}_{c}(\Psi)\alpha}\ln\left(\frac{\left[1-\frac{8M^2_p}{\phi^2_0}\ln\left(\frac{a}{a_0}\right)\right]\left(m^2_{c}-\lambda\phi^2_0\right)}{\left(m^2_{c}-\lambda\phi^2_0
\left[1-\frac{8M^2_p}{\phi^2_0}\ln\left(\frac{a}{a_0}\right)\right]\right)}\right),~~~~~~~~~~
\label{po4vvv1}
\eea \bea
\textcolor{red}{\underline{\bf Choice~II(v2)}}~~~\nonumber\\
\label{povvv2}\Psi-\Psi_0&\approx&\frac{2\sqrt{2}M_p}{\sqrt{3}}\ln\left(\frac{a}{a_0}\right)
=\frac{M^{2}_{p}}{3\sqrt{\alpha}}\left(t-t_0\right) 
 =-\frac{1}{2\sqrt{6}M_p}\left[\left(\phi^2 -\phi^2_0\right)+\frac{m^{2}_{c}}{\lambda}\ln\left(\frac{m^2_{c}-\lambda\phi^{2}}{m^2_{c}-\lambda\phi^{2}_0}\right)\right],~~~~~~~~~\\
\label{po2vvv2} a&\approx&a_0~\exp\left[\frac{M_p}{2\sqrt{6\alpha}}\left(t-t_0\right)\right],~~~~~~~~~\\
\label{po3vvv2} {\cal N}(\phi)-{\cal N}(\phi_0)&=&\frac{M^4_p}{16m^{2}_{c}(\Psi)\alpha}\ln\left(\frac{\phi^2_0\left(m^2_{c}-\lambda\phi^2\right)}{\phi^2\left(m^2_{c}-\lambda\phi^2_0\right)}\right)\approx
\frac{M^4_p}{16m^{2}_{c}(\Psi)\alpha}\ln\left(\frac{\left(m^2_{c}-\lambda\phi^2_0
\left[1-\frac{8M^2_p}{\phi^2_0}\ln\left(\frac{a}{a_0}\right)\right]\right)}{\left[1-\frac{8M^2_p}{\phi^2_0}\ln\left(\frac{a}{a_0}\right)\right]\left(m^2_{c}-\lambda\phi^2_0\right)}\right),~~~~~~~~~~
\label{po4vvv2}
\eea
The behavior of the effective 
potential in Einstein frame is plotted in fig.~(\ref{fig4a}) and fig.~(\ref{fig4b}), where the inflaton field is rolling down from a large field to lower value or the lower to 
larger field value and
after inflation take part in particle production and reheating. Here both of the situations are completely equivalent to the previous choice of the effective potentials as discussed earlier.
Here the only
difference is the scale of inflation, which are surely different compared to the previously mentioned scientific scenario. Additionally
it is note that
for both the cases the effective potential can be able to generate VEV at the field value,
$\phi=2.5~M_p$ which will finally take part to explain
the particle production and reheating mechanism. In both of the situations the lower bound on the
parameter $\alpha$ of the scale free gravity is fixed at, $\alpha\geq 2.51\times 10^{7}$, which is perfectly consistent
with Planck 2015 data and other available joint constraints \cite{Ade:2015lrj,Ade:2015ava,Ade:2015xua}.\\

\item \textcolor{red}{\underline{\bf Choice III:}}\\
In third option it is also possible to explain the reheating as well as the late time cosmic acceleration once we switch on the effect of non-minimal coupling between, $f(R)=\alpha R^2$ gravity sector and the matter field
sector. In that case the total effective action is modified in Jordan frame as:
\bea 
S&=&\int d^{4}x\sqrt{-g}\left[\frac{\alpha}{2}\left(1+\xi\phi^2\right)R^2+\frac{g^{\mu\nu}}{2}
\left(\partial_{\mu}\phi\right)\left(\partial_{\nu}\phi\right)-\frac{\lambda}{4}\left(\phi^2 -\phi^2_{V}\right)^2\right]~~~~~~~~~~
\eea
where $\xi$ represents the non-minimal coupling parameter and $\phi_{V}$ represents the VEV of the field $\phi$ in this context. After performing conformal transformation, the
effective action in the Einstein frame can be written as:
\bea\label{eq31}
S~~\underrightarrow{C.T.}~~ \tilde{S}
&=&\int d^{4}x \sqrt{-\tilde{g}}\left[\frac{M^{2}_{p}}{2}\tilde{R}+\frac{\tilde{g}^{\mu\nu}}{2}\tilde{\partial}_{\mu}\Psi\tilde{\partial}_{\nu}\Psi
+\frac{\tilde{g}^{\mu\nu}}{2}\tilde{\partial}_{\mu}\phi\tilde{\partial}_{\nu}\phi-\tilde{W}(\phi,\Psi)\right]~~~~~~~~~~~
 \eea
where after applying C.T.
the total modified effective action can be written as:
\be 
\tilde{W}(\phi,\Psi)=\frac{M^{4}_{p}}{8\alpha}+\frac{\frac{\lambda}{4}\left(\phi^2 -\phi^2_{V}\right)^2}{\left(1+\xi\phi^2\right)^2}e^{-\frac{2\sqrt{2}}{\sqrt{3}}\frac{\Psi}{M_p}}.
\ee
\begin{figure*}[htb]
\centering
{
    \includegraphics[width=12.2cm,height=5cm] {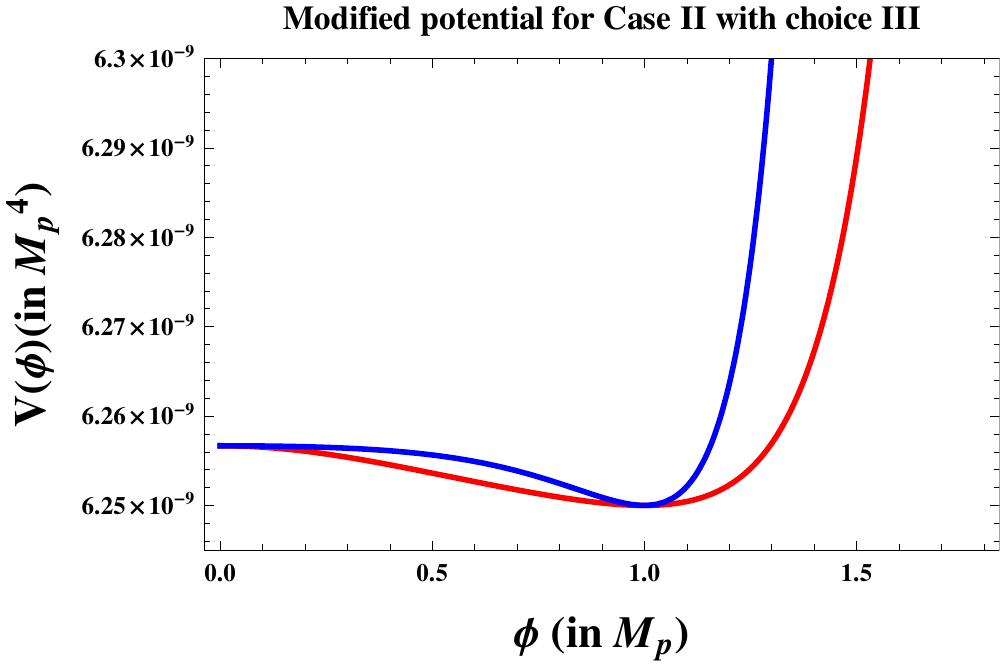}
    \label{fig5}
}
\caption[Optional caption for list of figures]{Behaviour of the modified effective potential in presence of non-minimal coupling in $R^2$ gravity for case II with $\textcolor{blue}{\bf \underline{Choice ~III}}:$ $V_0\neq0$, $\lambda>0$, $\xi=M^{-2}_{p} (\textcolor{red}{red}), 10^{-8}~M^{-2}_{p} (\textcolor{blue}{blue})$, where $M_p=2.43\times 10^{18}~GeV$.} 
\end{figure*}
In the present context the field equations can be expressed as:
\bea
\label{eqs1nxvvv1}
3\tilde{H}\frac{d\phi}{d\tilde{t}}+\frac{\lambda\phi\left(1+\xi\phi^2_{V}\right)\left(\phi^2-\phi^2_{V}\right)}{\left(1+\xi\phi^2\right)^3}e^{-\frac{2\sqrt{2}}{\sqrt{3}}\frac{\Psi}{M_p}}=0,\,\\
\label{eqs2nxvvv1}3\tilde{H}\frac{d\Psi}{d\tilde{t}}-\frac{\lambda\left(\phi^2 -\phi^2_{V}\right)^2}{\sqrt{6} M_p\left(1+\xi\phi^2\right)^2}e^{-\frac{2\sqrt{2}}{\sqrt{3}}\frac{\Psi}{M_p}}=0,\,\\
\label{eqs3nxvvv1}\tilde{H}^2=\frac{M^{2}_{p}}{24\alpha}+\frac{\frac{\lambda}{4}\left(\phi^2 -\phi^2_{V}\right)^2}{3M^2_p\left(1+\xi\phi^2\right)^2}e^{-\frac{2\sqrt{2}}{\sqrt{3}}\frac{\Psi}{M_p}}.
 \eea
 The solutions of Eq.~(\ref{eqs1nxvvv1}-\ref{eqs3nxvvv1}) are given by:
 \bea\label{povc1}
\textcolor{red}{\underline{\bf{\bf Choice~III}}}~~~\nonumber\\
\Psi-\Psi_0&\approx&\frac{2\sqrt{2}M_p}{\sqrt{3}}\ln\left(\frac{a}{a_0}\right)=\frac{M^{2}_{p}}{3\sqrt{\alpha}}\left(t-t_0\right)\nonumber\\
 &&~~~~~~~~~~~~~~~~~~~~~~=-\frac{1}{2\sqrt{6}M_p}\frac{\left[\left(\phi^2 -\phi^2_0 \right)\left(1+\frac{\xi}{2}\left(\phi^2 +\phi^2_0 -2\phi^2_{V}\right)\right)+2\phi^2_{V}\ln\left(\frac{\phi}{\phi_0}\right)\right]}{\left(1+\xi\phi^2_{V}\right)},~~~~~~~~~\\
\label{povc2} a&\approx& a_0~\exp\left[\frac{M_p}{2\sqrt{6\alpha}}\left(t-t_0\right)\right],~~~~~~~~~\\
\label{povc3} {\cal N}(\phi)-{\cal N}(\phi_0)&=& \frac{1}{M^{2}_{p}}\int^{\phi}_{\phi_0}d\phi \frac{\tilde{V}(\phi)}{\partial_{\phi}\tilde{V}(\phi)} 
=\frac{M^2_p}{16\phi^2_{V}\alpha\lambda(\Psi)\left(1+\xi\phi^2_{V}\right)}\ln\left(\frac{\phi^2_{0}\left(\phi^{2}-\phi^{2}_{V}\right)}{\phi^2\left(\phi^{2}_{0}-\phi^{2}_{V}\right)}\right)\nonumber\\&&~~~~~~~~~~~~~~~~~~~~~~~~~=
\frac{M^2_p}{16\phi^2_{V}\alpha\lambda(\Psi)\left(1+\xi\phi^2_{V}\right)}\ln\left(\frac{\left(\phi^2_0
\left[1-\frac{8M^2_p}{\phi^2_0}\ln\left(\frac{a}{a_0}\right)\right]-\phi^{2}_{V}\right)}{
\left[1-\frac{8M^2_p}{\phi^2_0}\ln\left(\frac{a}{a_0}\right)\right]\left(\phi^{2}_{0}-\phi^{2}_{V}\right)}\right),~~~~~~~~~~~~
\label{povc4}
\eea
In fig.~(\ref{fig5}), we have shown the behavior of the effective potential with respect to inflaton field in presence of non-minimal coupling parameter, $\xi=M^{-2}_{p}$ and $ \xi=10^{-8}~M^{-2}_{p}$ depicted by \textcolor{red}{red} and 
\textcolor{blue}{blue} colored curves respectively. For both of the cases we have taken the self interacting coupling parameter $\lambda>0$. Also it is important to mention here that, if we decrease the strength of 
the non-minimal coupling parameter then the effective potential become more steeper. For both the situations the inflaton field can roll-down from higher to lower or lower to higher field values and finally settle down to a local minimum
at $\phi_{V}=M_p$.
\end{itemize}

\section{\textcolor{blue}{Constraints on inflation with soft attractors}}
\label{s3}
Here we require the following constraints to study inflationary paradigm in the attractor regime:
\subsection{Number of e-foldings}
\label{s3a}
To get sufficient amount of inflation from the proposed setup (for both the \textcolor{red}{\underline{\bf Case I}} and \textcolor{red}{\underline{\bf Case II}}) it necessarily requires: 
\bea 
|{\cal N}(\phi_0)-{\cal N}(\phi_f)|\approx\left|\ln\left(\frac{a_f}{a_0}\right)\right|\gtrsim 50-70.
\eea
which is a necessary quantity that can able to solve horizon problem associated with standard big-bang cosmology. The subscripts `f' and `0' physically signify the final and initial values of the inflationary epoch. Further using Eq~(\ref{po3aa}) and Eq~(\ref{po6}) the field value at the end of inflation can be explicitly computed for the above mentioned two cases as:
\be\begin{array}{lll}\label{rd}
 \displaystyle\phi_f\displaystyle \sim\left\{\begin{array}{ll}
                    \displaystyle  \phi_{0} \left[1-\frac{480M^{2}_{p}}{\phi^{2}_{0}}\right]^{1/2} &
 \mbox{\small {\bf for \textcolor{red}{\underline{Case I}}}}  
\\ 
         \displaystyle  \frac{\phi_0}{\left[1+\frac{960\alpha\lambda(\Psi_f)\phi^2_0}{M^{2}_p}\right]^{1/2}}.~~~~~~~~~~ & \mbox{\small {\bf for \textcolor{red}{\underline{Case II}}}}.
          \end{array}
\right.
\end{array}\ee
Here it is important to mention the following facts:
\begin{itemize}
\item For the \textcolor{red}{\underline{\bf Case~I}} the expression for the field associated with the end of inflation $\phi_{f}$ is completely fixed by the value initial field value $\phi_0$. Here no information for the field dependent coupling $\lambda(\psi_f)=\lambda(\Psi=\Psi_f)$ is required for this case as the expression for $\phi_f$ is independent of the dilaton field dependent coupling.

\item For the \textcolor{red}{\underline{\bf Case~II}} the expression for the field associated with the end of inflation $\phi_{f}$ is fixed by the value initial field value $\phi_0$ as well as by the field dependent coupling $\lambda(\psi_f)=\lambda(\Psi=\Psi_f)$.
\end{itemize}  
\subsection{Primordial density perturbation}
\label{s3b}
\subsubsection{Two point function}
\label{s3b1}
 The next observational constraint comes from the imprints of density perturbations through scalar fluctuations.
Such fluctuations in CMB map directly implies that~\footnote{Here one equivalent notation for the amplitude of the scalar perturbation used as $\sqrt{{\cal P}_{cmb}}=\sqrt{{\cal P}({\cal N}_{cmb})}$ which we have used in the non attracor case.}: \be \frac{\delta\rho}{\rho}<\left(\frac{\delta\rho}{\rho}\right)_{cr}=\sqrt{A_S}\sim 10^{-5}\ee measured on the horizon crossing scales, where
$\delta\rho$ is the perturbation in the density $\rho$. Additionally it is important to note that, $A_S$, represents the amplitude of the scalar power spectrum.
Also in the present context for both the cases one can write:
\bea 
\left[\sigma\frac{\delta\rho}{\rho}\right]_{t_1}&=&\left[\sigma\frac{\delta\rho}{\rho}\right]_{t_2}
\eea
where the parameter $\sigma$ is the parameter in the present context, which can be expressed in terms of equation parameter as, $\sigma= 1+\frac{2}{3(1+w)},~~~ 
w=\frac{p}{\rho}$.
It is important to note that, $(t_{1},t_{2})$ represent the times when the perturbation first left and re-entered the
horizon, respectively. At time $t_{1}$, Eq~(\ref{con1a}) and Eq~(\ref{con1b}) perfectly hold good in the present context. On the other hand at time $t=t_2$ the representative parameter $\sigma$ take the value, 
$\sigma=3/2$ and $\sigma=5/3$ during radiation and matter dominated
epoch respectively. For the potential dominated inflationary epoch, $w\approx -1$ and consequently one can write the following constraint condition: 
\bea 
\left(\frac{\delta\rho}{\rho}\right)_{t_{2}}\approx\left(\left(1-\frac{1}{\sigma}\right)\frac{\delta\rho}{\rho}\right)_{t_{1}}.
\eea
Further using Eq~(\ref{con1a}) and Eq~(\ref{con1b}) and approximated equation of motion in slow-roll regime of fluctuation in the total energy density or equivalently in the scalar modes can be written as:
\bea 
\delta\rho &=& \dot{\phi}\delta{\dot{\phi}}+\dot{\Psi}\delta{\dot{\Psi}}-3\tilde{H}\left(\dot{\phi}\delta{\phi}+\dot{\Psi}\delta{\Psi}\right)
\approx -2\tilde{H}\left(\dot{\phi}\delta{\phi}+\dot{\Psi}\delta{\Psi}\right).
\eea
where we use the symbol as, $\dot~~\equiv d/d\tilde{t}$ and one can write down, $\delta\dot{\phi}\approx \tilde{H}\delta\phi$, $\delta\dot{\Psi}\approx \tilde{H}\delta\Psi$, $\delta\phi\approx \tilde{H}$,$\delta\Psi\approx \tilde{H}$, and finally the fractional density contrast 
can be expressed as:
\bea 
\left(\frac{\delta \rho}{\rho}\right)_{t_{2}}= \left(\frac{\tilde{H}^2 \left(|\dot{\phi}|+|\dot{\Psi}|\right)}{\dot{\phi}^2 +\dot{\Psi}^2}{\cal C}\right)_{t_{1}}
\eea
with the following constraint on the parameter ${\cal C}$ as given by, ${\cal C}\sim{\cal O}(1)$ and it serves the purpose of a normalization constant in this context. Then we get the two physically acceptable situations for both of the cases which can be written as:
\bea 
\textcolor{red}{\underline{\bf Region ~I}:}~~~~~~~~~|\dot{\phi}|<|\dot{\Psi}|&\Rightarrow& \frac{\delta\rho}{\rho}\approx \frac{\tilde{H}^2}{|\dot{\Psi}|}\approx \frac{\sqrt{\tilde{W}_{h}}}{2\sqrt{2}M^2_p}
,\\
\textcolor{red}{\underline{\bf Region~ II}:}~~~~~~~~|\dot{\phi}|>|\dot{\Psi}|&\Rightarrow& \frac{\delta\rho}{\rho}\approx \frac{\tilde{H}^2}{|\dot{\phi}|}\approx 
\frac{\tilde{W}^{3/2}_{h}}{M^3_p\left(\partial_{\phi}\tilde{W}\right)_h }
.
\eea
Here one can interpret the results as:
\begin{itemize}
\item In the \textcolor{red}{\underline{\bf Region~I}}, the amplitude of the density fluctuation at the horizon crossing is only controlled by the scale of inflation and the magnitude of the dilaton dependent effective coupling parameter $\lambda(\Phi_h)$.

\item In the \textcolor{red}{\underline{\bf Region~II}}, the amplitude of the density fluctuation at the horizon crossing is given by:
\be \left(\frac{\delta \rho}{\rho}\right)_{\textcolor{red}{\rm \bf Region~II}}=\frac{2}{\left(\sqrt{\epsilon_{\tilde{W}}}\right)_h}\left(\frac{\delta \rho}{\rho}\right)_{\textcolor{red}{\rm \bf Region~I}}.\ee
This implies that that contribution from the first slow roll parameter as given by, 
$\epsilon_{\tilde{W}}=\frac{M^2_p}{2}\left(\frac{\partial_{\phi}\tilde{W}}{\tilde{W}}\right)$,
controls the magnitude of the amplitude of density perturbation apart from the effect from the scale of inflation and  the magnitude of the dilaton dependent effective coupling parameter $\lambda(\Phi_h)$.

\end{itemize}
\subsubsection{Present observables}
\label{s3b2}
Further using the approximate equations of motion the fractional density contrast for the above mentioned two cases can be written as:
\be\begin{array}{lll}\label{rk9fin}
 \displaystyle\textcolor{red}{\underline{\bf Case~ I}}:\frac{\delta\rho}{\rho}\displaystyle \sim\left\{\begin{array}{ll}
                    \displaystyle   \frac{\phi^2_0}{4M^2_p}\sqrt{\frac{\lambda(\Psi_h)}{2}}\left[1-\frac{2\sqrt{6}M_p}{9\phi^2_0}\left(\Psi_h - \Psi_0\right)\right] &
 \mbox{\small {\bf for \textcolor{red}{\underline{Region I}}}}  
\\ 
         \displaystyle  \frac{\phi^3_0\sqrt{\lambda(\Psi_h)}}{8M^3_p} \left[1-\frac{2\sqrt{6}M_p}{9\phi^2_0}\left(\Psi_h - \Psi_0\right)\right]^{3/2} & \mbox{\small {\bf for \textcolor{red}{\underline{Region II}}}}.
          \end{array}
\right.
\end{array}\ee

\be\begin{array}{lll}\label{rk9finn}
 \displaystyle\textcolor{red}{\underline{\bf Case~ II}}:\frac{\delta\rho}{\rho}\displaystyle \sim\left\{\begin{array}{ll}
                    \displaystyle  \frac{1}{8\sqrt{\alpha}}\left[1+\frac{2\phi^2_0\alpha\lambda(\Psi_h)}{M^4_p}\left\{1-\frac{2\sqrt{6}M_p}{\phi^2_0}\left(\Psi_h-\Psi_0\right)\right\}^2\right]^{1/2} &
 \mbox{\small {\bf for \textcolor{red}{\underline{Region I}}}}  
\\ 
         \displaystyle \frac{M^3_p}{\lambda(\Psi_h)(8\alpha)^{3/2}\phi^3_0\left[1-\frac{2\sqrt{6}M_p}{\phi^2_0}\left(\Psi_h - \Psi_0\right)\right]^{3/2}} & \mbox{\small {\bf for \textcolor{red}{\underline{Region II}}}}.
          \end{array}
\right.
\end{array}\ee
Here one can interpret the results as:
\begin{itemize}
\item In the \textcolor{red}{\underline{\bf Region~I}} and \textcolor{red}{\underline{\bf Region~II}} of \textcolor{red}{\underline{\bf Case~I}}, the amplitude of the density fluctuation at the horizon crossing are related as:
\bea \left(\frac{\delta \rho}{\rho}\right)_{\textcolor{red}{\rm \bf Region~II}}&=&\frac{\phi_0}{\sqrt{2}M_p}\left(\frac{\delta \rho}{\rho}\right)_{\textcolor{red}{\rm \bf Region~I}}\left[1-\frac{2\sqrt{6}M_p}{9\phi^2_0}\left(\Psi_h - \Psi_0\right)\right]^{1/2}
\approx\frac{\phi_0}{\sqrt{2}M_p}\left(\frac{\delta \rho}{\rho}\right)_{\textcolor{red}{\rm \bf Region~I}}.~~~~\eea
This imples that if we know the field value at the starting point of inflation then one can directly quantify the amplitude of density perturbation. Most importantly, if inflation starts from the vicinity of the Planck scale i.e. $\phi_0 \sim \sqrt{2}M_p\sim {\cal O}(M_p)$ then by evaluating the amplitude of the density perturbation in the \textcolor{red}{\underline{\bf Region~I}} one can easily quantify the amplitude of the density perturbation in the \textcolor{red}{\underline{\bf Region~II}}.
In this setup within the range $50<{\cal N}_{f/h}<70$, we get:
\bea \left(\frac{\delta \rho}{\rho}\right)_{\textcolor{red}{\rm \bf Region~I}}&\sim& \left(\frac{\delta \rho}{\rho}\right)_{\textcolor{red}{\rm \bf Region~II}}\sim 2.2\times 10^{-9},\eea
which is consistent with Planck 2015 data.
But if inflation starts at the following field value,
 $\phi_0=\sqrt{2}\Delta~M_p$,
where the parameter $\Delta\gtrless 1$ then one ca write the following relationship between the amplitude of the density perturbation in the \textcolor{red}{\underline{\bf Region~I}} and \textcolor{red}{\underline{\bf Region~II}} as:
\bea \left(\frac{\delta \rho}{\rho}\right)_{\textcolor{red}{\rm \bf Region~II}}&=&\Delta\left(\frac{\delta \rho}{\rho}\right)_{\textcolor{red}{\rm \bf Region~I}}\left[1-\frac{\sqrt{6}}{9\Delta^2M_p}\left(\Psi_h - \Psi_0\right)\right]^{1/2}\approx\Delta\left(\frac{\delta \rho}{\rho}\right)_{\textcolor{red}{\rm \bf Region~I}}.\eea
This implies that for $\Delta\gtrless 1$ we get:
\bea \left(\frac{\delta \rho}{\rho}\right)_{\textcolor{red}{\rm \bf Region~II}}
&\gtrless&\left(\frac{\delta \rho}{\rho}\right)_{\textcolor{red}{\rm \bf Region~I}}.\eea
In this case for the \textcolor{red}{\underline{\bf Region~I}} we get:
\bea \left(\frac{\delta \rho}{\rho}\right)_{\textcolor{red}{\rm \bf Region~I}}&\sim& 2.2\times 10^{-9},\eea
then for the \textcolor{red}{\underline{\bf Region~II}} we get:
\bea \left(\frac{\delta \rho}{\rho}\right)_{\textcolor{red}{\rm \bf Region~I}}&\gtrless& 2.2\times 10^{-9}.\eea
This implies that for $\Delta\gtrless 1$ in \textcolor{red}{\underline{\bf Region~II}} we get tightly constrained result for the amplitude for the density perturbation.
\item In the \textcolor{red}{\underline{\bf Region~I}} and \textcolor{red}{\underline{\bf Region~II}} of \textcolor{red}{\underline{\bf Case~II}}, the amplitude of the density fluctuation at the horizon crossing are related as:
\be \left(\frac{\delta \rho}{\rho}\right)_{\textcolor{red}{\rm \bf Region~II}}\approx\frac{M^3_p}{\sqrt{8}\alpha\lambda(\Psi_h)\phi^3_0}\left(\frac{\delta \rho}{\rho}\right)_{\textcolor{red}{\rm \bf Region~I}}.\ee
This implies that if we know the field value at the starting point of inflation, the dilaton field dependent coupling at the horizon crossing $\lambda(\Psi_h)$ and the coupling of scale free gravity $\alpha$, then one can directly quantify the amplitude of density perturbation. Most importantly, if inflation starts from the vicinity of the Planck scale i.e. $\phi_0 \sim {\cal O}(M_p)$ 
and we have an additional constraint:
\be \lambda(\Psi_h)\sim \frac{1}{\sqrt{8}\alpha},\ee
then by evaluating the amplitude of the density perturbation in the \textcolor{red}{\underline{\bf Region~I}} one can easily quantify the amplitude of the density perturbation in the \textcolor{red}{\underline{\bf Region~II}}. Here one can also consider an equivalent constraint:
\be \phi_0 \sim \left(\frac{1}{\sqrt{8}\alpha\lambda(\Psi_h)}\right)^{1/3}~M_p.\ee
For both the situations in the present setup within the range $50<{\cal N}_{f/h}<70$, we get:
\bea \left(\frac{\delta \rho}{\rho}\right)_{\textcolor{red}{\rm \bf Region~I}}&\sim& \left(\frac{\delta \rho}{\rho}\right)_{\textcolor{red}{\rm \bf Region~II}}\sim 2.3\times 10^{-9},\eea
which is also consistent with Planck 2015 data. But if inflation starts at the following field value, $\phi_0=\Delta~M_p$,
where the parameter $\Delta\gtrless 1$ and we define:
\be \Gamma = \left(\frac{1}{\sqrt{8}\alpha\lambda(\Psi_h)\Delta^3}\right),\ee
where the parameter $\Gamma\gtrless 1$ and 
then one can write the following relationship between the amplitude of the density perturbation in the \textcolor{red}{\underline{\bf Region~I}} and \textcolor{red}{\underline{\bf Region~II}} as:
\be \left(\frac{\delta \rho}{\rho}\right)_{\textcolor{red}{\rm \bf Region~II}}=\Gamma\left(\frac{\delta \rho}{\rho}\right)_{\textcolor{red}{\rm \bf Region~I}}.\ee
This implies that for $\Delta\gtrless 1$ and $\Gamma\gtrless 1$ we get:
\bea \left(\frac{\delta \rho}{\rho}\right)_{\textcolor{red}{\rm \bf Region~II}}
&\gtrless&\left(\frac{\delta \rho}{\rho}\right)_{\textcolor{red}{\rm \bf Region~I}}.\eea
In this case for the \textcolor{red}{\underline{\bf Region~I}} we get:
\bea \left(\frac{\delta \rho}{\rho}\right)_{\textcolor{red}{\rm \bf Region~I}}&\sim& 2.3\times 10^{-9},\eea
then for the \textcolor{red}{\underline{\bf Region~II}} we get:
\bea \left(\frac{\delta \rho}{\rho}\right)_{\textcolor{red}{\rm \bf Region~I}}&\gtrless& 2.3\times 10^{-9}.\eea
This implies that for $\Delta\gtrless 1$ and $\Gamma\gtrless 1$ in \textcolor{red}{\underline{\bf Region~II}} we get tightly constrained result for the amplitude for the density perturbation.
\end{itemize}
\begin{figure*}[htb]
\centering
{
    \includegraphics[width=12.2cm,height=5cm] {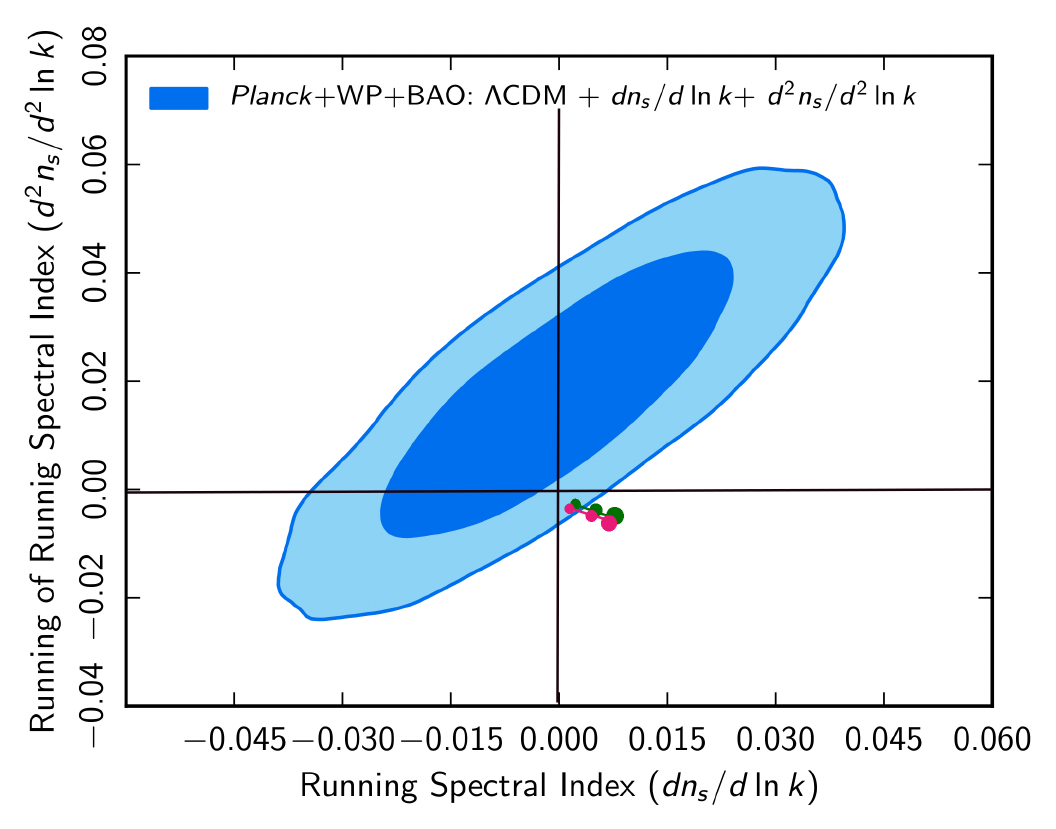}
    \label{figab1}
}
\caption[Optional caption for list of figures]{Plot for running of the running of spectral index $\kappa_{S}=d^2n_{S}/d^2\ln k$ vs running of the spectral index $\beta_{S}=dn_{S}/d\ln k$ for scalar modes. Here for \textcolor{red}{\underline{\bf Case~I}}  and 
\textcolor{red}{\underline{\bf Case~II}} we have drawn green and pink colored lines. We also draw the background of confidence contours obtained from  various joint constraints \cite{Ade:2015lrj,Ade:2015ava,Ade:2015xua}.} 
\end{figure*}
In this context the scalar spectral tilt can be written at the horizon crossing as~\footnote{Here we use a new symbol ${\cal N}_{f/h}$, which is defined as,
\be {\cal N}_{f/h}=\left|\ln\left(\frac{a_f}{a_h}\right)\right|=|{\cal N}(\phi_h)-{\cal N}(\phi_f)|\sim 50-70.\ee }:
\be\begin{array}{lll}\label{rk9finns}
 \displaystyle n_{S}-1=\left(\frac{d\ln A_S}{df}\right)_h \approx\left\{\begin{array}{ll}
                    \displaystyle   -\frac{3}{\left({\cal N}_{f/h}+1\right)} &
 \mbox{\small {\bf for \textcolor{red}{\underline{Case I}}}}  
\\ 
         \displaystyle -\frac{3}{{\cal N}_{f/h}} & \mbox{\small {\bf for \textcolor{red}{\underline{Case II}}}}.
          \end{array}
\right.
\end{array}\ee
Further using Eq~(\ref{rk9finns}) the running and running of the running of scalar spectral tilt can be computed as:
\be\begin{array}{lll}\label{run1}
 \displaystyle \beta_{S}=\left(\frac{dn_S}{df}\right)_h \approx\left\{\begin{array}{ll}
                    \displaystyle   \frac{3}{\left({\cal N}_{f/h}+1\right)^2} ~~~~&
 \mbox{\small {\bf for \textcolor{red}{\underline{Case I}}}}  
\\ 
         \displaystyle \frac{3}{{\cal N}^{2}_{f/h}} ~~~~& \mbox{\small {\bf for \textcolor{red}{\underline{Case II}}}}
          \end{array}
\right.
\end{array}\ee
and 
\be\begin{array}{lll}\label{run2}
 \displaystyle \kappa_{S}=\left(\frac{d\beta_S}{df}\right)_h \approx\left\{\begin{array}{ll}
                    \displaystyle   -\frac{6}{\left({\cal N}_{f/h}+1\right)^3} ~~~~&
 \mbox{\small {\bf for \textcolor{red}{\underline{Case I}}}}  
\\ 
         \displaystyle -\frac{6}{{\cal N}^{3}_{f/h}} ~~~~& \mbox{\small {\bf for \textcolor{red}{\underline{Case II}}}}.
          \end{array}
\right.
\end{array}\ee
Finally combining Eq~(\ref{rk9finns}), Eq~(\ref{run1}) and Eq~(\ref{run2}) we get the following consistency relation for both \textcolor{red}{\underline{\bf Case~I}} and \textcolor{red}{\underline{\bf Case~II}} we get:
\bea\label{run3}
\beta_{S}&=& \frac{\left(n_{S}-1\right)^2}{3}=3\left(-\frac{\kappa_{S}}{6}\right)^{2/3}.\eea
This is obviously a new a consistency relation for the present Higgsotic model of inflation and is also consistent with Planck 2015 data \cite{Ade:2015lrj,Ade:2015ava,Ade:2015xua}. In table~(\ref{tab2a}) we have shown the numerical estimations of the inflationary observables for the Higgsotic attractors depicted in \textcolor{red}{\underline{\bf Case~I}} and \textcolor{red}{\underline{\bf Case~II}} within the range $50<{\cal N}_{f/h}<70$.

In~fig.~(\ref{figab1}), we have plotted running of the running of spectral tilt for scalar perturbation ($\kappa_{S}=d^2n_{S}/d^2\ln k$) vs spectral tilt for scalar perturbation ($n_{S}$) in the light of Planck 2015 data along with various joint constraints. Here it is important to note that, for \textcolor{red}{\underline{\bf Case~I}} and \textcolor{red}{\underline{\bf Case~II}} the Higgsotic models are shown by the green and pink colored lines. Also the big circle, intermediate size circle and small circle represent the representative points in $(\kappa_S,n_{S})$ 2D plane for the number of e-foldings, ${\cal N}_{f/h}=70$, ${\cal N}_{f/h}=60$ and ${\cal N}_{f/h}=50$ respectively. To represent the present status as well as statistical significance of the Higgsotic model for the dynamical attractors as depicted in \textcolor{red}{\underline{\bf Case~I}} and \textcolor{red}{\underline{\bf Case~II}}, we have drawn the $1\sigma$ and $2\sigma$ confidence contours from Planck+WMAP+BAO 2015 joint data sets \cite{Ade:2015lrj,Ade:2015ava,Ade:2015xua}. It is clearly visualized from the fig.~(\ref{figan}) that, for \textcolor{red}{\underline{\bf Case~I}} we cover the range, $0.59\times 10^{-3}<\beta_{S}=\frac{dn_S}{d\ln k}<1.16\times 10^{-3}$ and $-1.65\times 10^{-5}>\kappa_{S}=\frac{d^2n_S}{d^2\ln k}>-4.56\times 10^{-5}$
in the $(\kappa_S,\beta_{S})$ 2D plane. Similarly for \textcolor{red}{\underline{\bf Case~II}} we cover the range, $0.62\times 10^{-3}<\beta_{S}=\frac{dn_S}{d\ln k}<1.20\times 10^{-3}$ and $-1.78\times 10^{-5}>\kappa_{S}=\frac{d^2n_S}{d^2\ln k}>-4.80\times 10^{-5}$
in the $(\kappa_S,\beta_{S})$ 2D plane.
\begin{table*}
\centering
\tiny\begin{tabular}{|c|c|c|c|c|c|c|c|c|c|c|}
\hline
\hline
 ${\cal N}_{f/h}$ & $A_{S}$&$n_{S}$& $\beta_{S}$ &
 $\kappa_{S}$  \\
  &  ($\times 10^{-9}$)& & ($ \times 10^{-3}$)  & ($ \times 10^{-5}$)    \\
  & \textcolor{red}{\underline{\bf Case~I}}~~~~~\textcolor{red}{\underline{\bf Case~II}}& \textcolor{red}{\underline{\bf Case~I}}~~~~~\textcolor{red}{\underline{\bf Case~II}}&\textcolor{red}{\underline{\bf Case~I}}~~~~~\textcolor{red}{\underline{\bf Case~II}}& \textcolor{red}{\underline{\bf Case~I}}~~~~~\textcolor{red}{\underline{\bf Case~II}}\\
\hline\hline\hline
 50 &  & 0.941~~~~~~~~0.940 & 1.16~~~~~~~~~~~~1.20& -4.56~~~~~~~~~~~~-4.80 
\\
 60 & 2.2~~~~~~~~~2.3 & 0.951~~~~~~~~0.950 & 0.80~~~~~~~~~~~~0.83 & -2.61~~~~~~~~~~~~-2.76  
\\
 70 &  & 0.958~~~~~~~~0.957  & 0.59~~~~~~~~~~~~0.62 &-1.65~~~~~~~~~~~~-1.78   
\\ \hline
\hline
\end{tabular}
\caption{Inflationary observables and model constraints in the light of Planck 2015 data \cite{Ade:2015lrj,Ade:2015ava,Ade:2015xua} for the dynamical attractors considered in \textcolor{red}{\underline{\bf Case~I}} and \textcolor{red}{\underline{\bf Case~II}}.}\label{tab2a}
\vspace{.4cm}
\end{table*}

\begin{figure*}[htb]
\centering
\subfigure[$\textcolor{blue}{\bf \underline{Case ~I}:}$ $r$ vs $\lambda(\Psi_{h})$.]{
    \includegraphics[width=7.7cm,height=5cm] {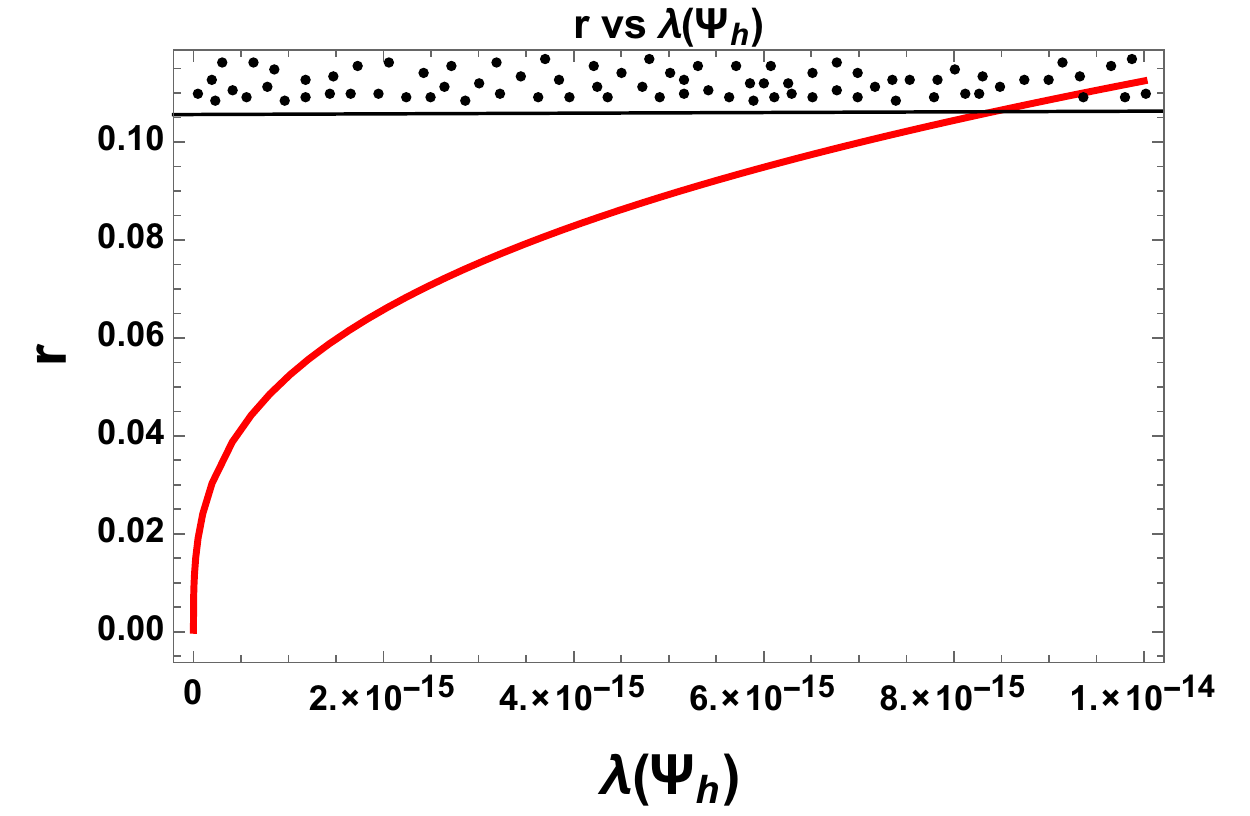}
    \label{figlm1}
}
\subfigure[$\textcolor{blue}{\bf \underline{Case ~II}}:$ $r$ vs $\alpha$.]{
    \includegraphics[width=7.7cm,height=5cm] {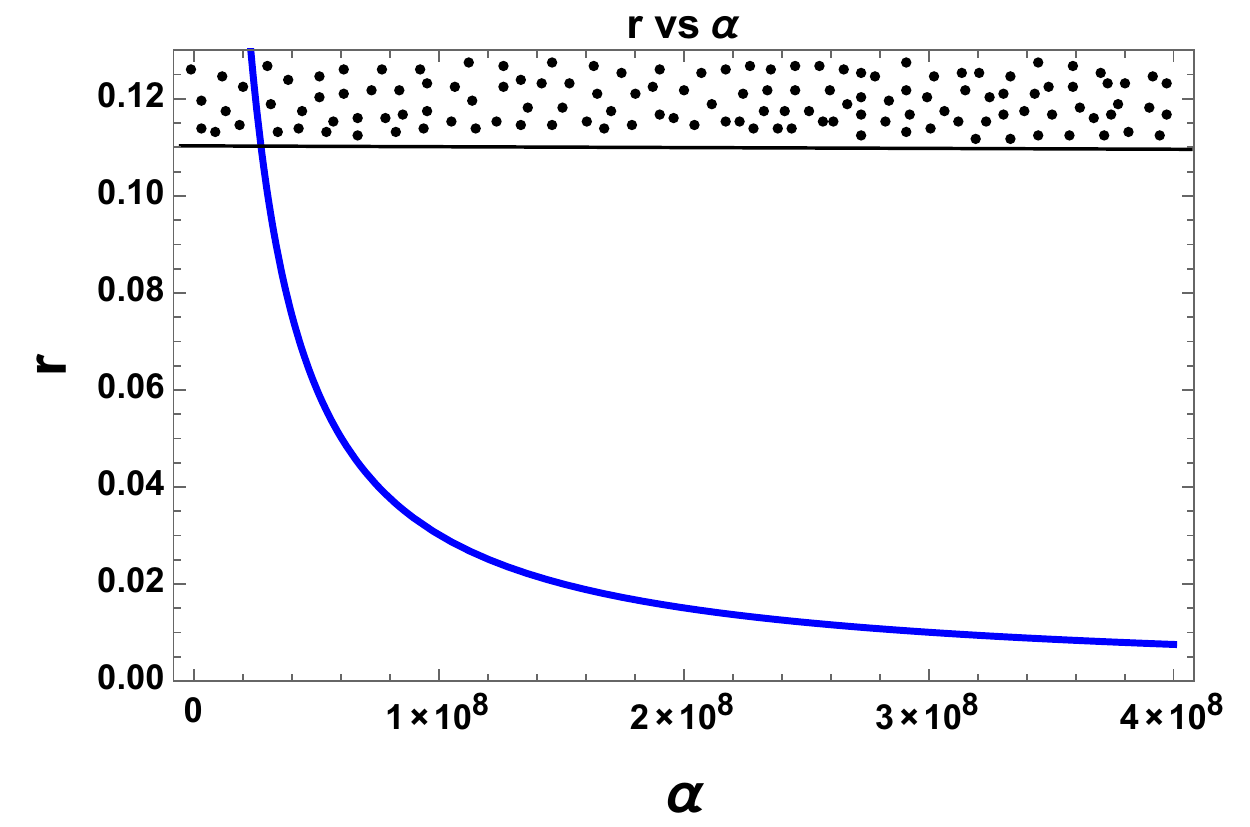}
    \label{figlm2}
}
\caption[Optional caption for list of figures]{Variation of tensor-to-scalar ratio $r$ with respect to \ref{figlm1} coupling parameter $\lambda(\Psi_{h})$ (\textcolor{red}{Case I}) and \ref{figlm2}
scale free parameter $\alpha$ (\textcolor{red}{Case II}). For both the plots dotted region is disfavoured by Planck 2015 data along with BICEP2+Keck Array joint constraint \cite{Ade:2015lrj,Ade:2015ava,Ade:2015xua}. } 
\label{fza}
\end{figure*}
\begin{figure*}[htb]
\centering
{
    \includegraphics[width=12.2cm,height=5cm] {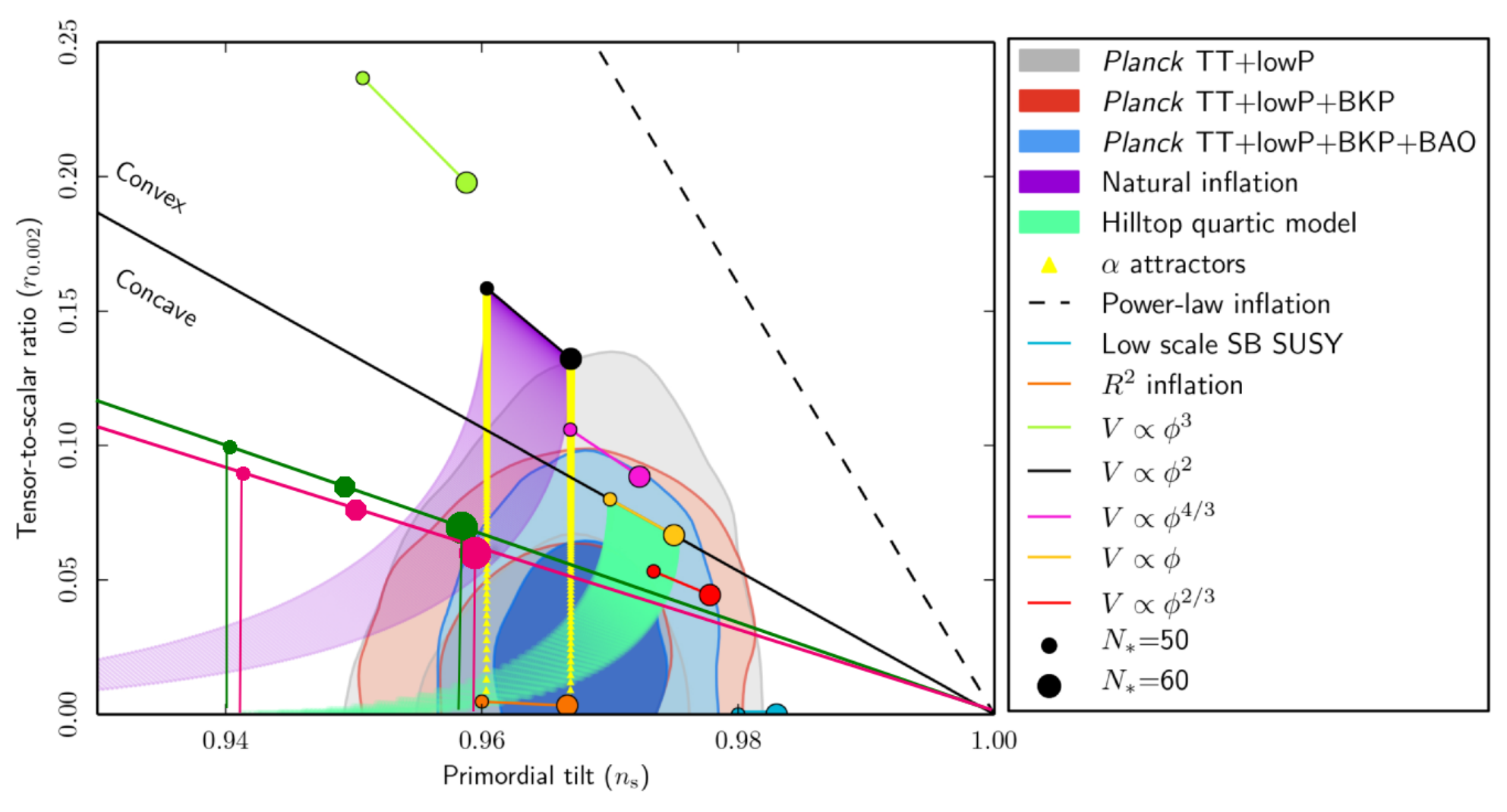}
    \label{figv1}
}
\caption[Optional caption for list of figures]{$r$ vs $n_{S}$ polt for \textcolor{red}{\underline{{\bf Case~I}}} and \textcolor{red}{\underline{{\bf Case~II}}} in the background of confidence contours obtained from  Planck TT+low P, Planck TT+low P+BKP, Planck TT+low P+BKP+BAO joint data sets.} 
\end{figure*}
\subsection{Primordial tensor modes and future observables}
\label{s3c}
In terms of the number of e-foldings (${\cal N}$) the the most useful parametrization of the primordial scalar and tensor 
power spectrum or equivalently for tensor-to-scalar ratio can be written near the horizon crossing ${\cal N}_h={\cal N}(\phi_h)$ as:
\bea 
r({\cal N})&=&\frac{8}{M^{2}_{p}}\left(\frac{d\phi}{d{\cal N}}\right)^2=r({\cal N}_h)e^{\left({\cal N}-{\cal N}_h\right)\left\{A_h+B_h\left({\cal N}-{\cal N}_h\right)\right\}}~~~~~
\eea
where in the slow-roll regime of inflation the tensor-to-scalar ratio $r({\cal N}_h)$ can be written in terms of the inflationary potential as:
\be\begin{array}{lll}\label{re}
 \displaystyle r=r({\cal N}_h)\displaystyle \approx8M^{2}_{p}\left(\frac{V^{'}_h}{V_h}\right)^2=\left\{\begin{array}{ll}
                    \displaystyle \frac{128M^{2}_{p}}{\phi^2_h} &
 \mbox{\small {\bf for \textcolor{red}{\underline{Case I}}}}  
\\ 
         \displaystyle \frac{512\alpha^2 \lambda^{2}(\Psi_h)\phi^{6}_{h}}{M^6_p}~~~~~~~~~~ & \mbox{\small {\bf for \textcolor{red}{\underline{Case II}}}}.
          \end{array}
\right.
\end{array}\ee
and the symbols $A_h$, $B_h$ and $C_h$ are expressed in terms of the inflationary observables at horizon crossing as, 
$A_h=n_{T}-n_{S}+1,~~
B_h=\frac{1}{2}\left(\beta_{T}-\beta_{S}\right)$.
In the above parametrization $ A_h>>B_h$ i.e.
$\beta_S-2(n_S-1)>>\beta_T-2n_T$, is always required for convergence of the Taylor expansion.  
Using this assumption the relationship between field excursion, $\Delta\phi=\phi_h -\phi_f$ and tensor-to-scalar ratio $r({\cal N}_h)$ can be 
computed as:
\bea \label{Lythx}
\frac{|\Delta\phi|}{M_p}\approx \sqrt{\frac{r({\cal N}_h)}{8}} e^{-\frac{A^{2}_{h}}{2B_h}}\sqrt{\frac{2\pi}{B_h}}\left|{\rm erfi}\left(\frac{A_h}{\sqrt{2B_h}}
\right)-{\rm erfi}\left(\frac{A_h}{\sqrt{2B_h}}
-\sqrt{\frac{B_h}{8}}{\cal N}_{f/h}\right)\right|.~~~~~~~~~~
\eea
Now the scale of infation is connected with the tensor-to-scalar ratio in the following fashion:
\bea\label{scale}
{V^{1/4}_h}&=& \left(\frac{3}{2}\pi^2 A_S r(f_h)\right)^{1/4}M_p\sim 7.9\times10^{-3}~M_p \times\left(\frac{r(f_h)}{0.11}\right)^{1/4}.
\eea
Substituting Eq~(\ref{scale}) in Eq~(\ref{Lythx}) we compute the relationship between field excursion and the scale of inflation as:
\bea \label{Lyth}
\frac{|\Delta\phi|}{M_p}\approx \sqrt{\frac{V_h}{6\pi M^{4}_{p}A_S B_h}} e^{-\frac{A^{2}_{h}}{2B_h}}\left|{\rm erfi}\left(\frac{A_h}{\sqrt{2B_h}}
\right)-{\rm erfi}\left(\frac{A_h}{\sqrt{2B_h}}
-\sqrt{\frac{B_h}{8}}{\cal N}_{f/h}\right)\right|.~~~~~~~~~~
\eea
Also using Eq~(\ref{scale}) the tensor-to-scalar ratio can be written as:
\be\begin{array}{lll}\label{rinflax}
 \displaystyle r=r({\cal N}_h)\displaystyle =\left\{\begin{array}{ll}
                    \displaystyle \frac{\lambda(\Psi_h)\phi^4_h}{\left(2\times 10^{-2} M_p\right)^4} &
 \mbox{\small {\bf for \textcolor{red}{\underline{Case I}}}}  
\\ 
         \displaystyle  \frac{1}{\alpha\left(2.4\times 10^{-2}\right)^4}.~~~~~~~~~~ & \mbox{\small {\bf for \textcolor{red}{\underline{Case II}}}}.
          \end{array}
\right.
\end{array}\ee
Further using Eq~(\ref{re}) and Eq~(\ref{rinflax}) we get the following constraints from primordial tensor perturbation:
\be\begin{array}{lll}\label{rinflasd}
 \displaystyle \phi_h\displaystyle =\left\{\begin{array}{ll}
                    \displaystyle \frac{0.17 M_p}{\sqrt[6]{\lambda(\Psi_h)}} &
 \mbox{\small {\bf for \textcolor{red}{\underline{Case I}}}}  
\\ 
         \displaystyle  \frac{4.25 M_p}{\alpha^{1/2}\sqrt[3]{\lambda(\Psi_h)}}.~~~~~~~~~~ & \mbox{\small {\bf for \textcolor{red}{\underline{Case II}}}}.
          \end{array}
\right.
\end{array}\ee
Consequently the model parameters of the prescribed theory can be recast in terms of tensor-to-scalar ratio as:
\bea
 \displaystyle \label{w1f}\lambda(\Psi_h)&=&9.358\times 10^{-15}\times \left(\frac{r}{0.11}\right)^3 ~~~~~~~~~~~
 \mbox{\small {\bf for \textcolor{red}{\underline{Case I}}}}  
\\ 
 \label{w2}\alpha&=&2.740\times 10^{7}\times \left(\frac{r}{0.11}\right)^{-1}.~~~~~~~~~~  \mbox{\small {\bf for \textcolor{red}{\underline{Case II}}}}.
\eea
To satisfy the upeer bound of tensor-to-scalar ratio as obtained from Planck 2015+ BICEP2 + Keck Array i.e. $r\sim 0.11$ \cite{Ade:2015lrj,Ade:2015ava,Ade:2015xua}, Eq~(\ref{w1f}) and Eq~(\ref{w2}), 
gives the upper bound of the model parameters $\lambda(\Psi_h)$ and $\alpha$ respectively.

In~fig.~(\ref{figlm1}) and figu.~(\ref{figlm2}), we have shown the variation of tensor-to-scalar ratio $r$ with respect to field dependent  coupling parameter $\lambda(\Psi_{h})$ for \textcolor{red}{\underline{\bf Case I}} and 
scale free parameter $\alpha$ for \textcolor{red}{\underline{\bf Case II}}. For both the plots dotted region is disfavoured by Planck 2015 data along with BICEP2+Keck Array joint constraint.

In~fig.~(\ref{figv1}), we have plotted tensor-to-scalar ratio ($r$) vs spectral tilt for scalar perturbation ($n_{S}$) in the light of Planck data along with various joint constraints. Here it is important to note that, for \textcolor{red}{\underline{\bf Case~I}} and \textcolor{red}{\underline{\bf Case~II}} the Higgsotic models are shown by the green and pink colored lines. Also the big circle, intermediate size circle and small circle represent the representative points in $(r,n_{S})$ 2D plane for the number of e-foldings, ${\cal N}_{f/h}=70$, ${\cal N}_{f/h}=60$ and ${\cal N}_{f/h}=50$ respectively. To represent the present status as well as statistical significance of the Higgsotic model for the dynamical attractors as depicted in \textcolor{red}{\underline{\bf Case~I}} and \textcolor{red}{\underline{\bf Case~II}}, we have drawn the $1\sigma$ and $2\sigma$ confidence contours from Planck TT+low P, Planck TT+low P+BKP, Planck TT+low P+BKP+BAO joint data sets \cite{Ade:2015lrj,Ade:2015ava,Ade:2015xua} in fig.~(\ref{figv1}). It is clearly visualized from the fig.~(\ref{figv1}) that, for \textcolor{red}{\underline{\bf Case~I}} we cover the range, $ 0.941<n_{S}<0.958$ and $0.07<r<0.11$
in the $(r,n_{S})$ 2D plane for the effective field dependent coupling constrained within the window, $5.956\times 10^{-15}<\lambda(\Psi_h)<9.358\times 10^{-15}$. Similarly for \textcolor{red}{\underline{\bf Case~II}} we cover the range, $0.940<n_{S}<0.957$ and $ 0.056<r<0.09$,
in the $(r,n_{S})$ 2D plane for the effective scale of the Higgsotic potential constrained within the window, $5.382\times 10^{7}>\alpha>3.349\times 10^{7}$.
Additionally it is important to mention here that,
the area bounded by the parallel vertical green lines and the pink lines represent the allowed parameter space in the $(r,n_{S})$ 2D plane for the two Higgsotic dynamical attractors as depicted in \textcolor{red}{\underline{\bf Case~I}} and \textcolor{red}{\underline{\bf Case~II}}.
\subsection{Reheating}
\label{s3d}
To get successful amount of reheating from the proposed setup we consider the fact that reheating to commence at the end of slow rolling of the inflaton, $\ddot{\phi}\approx 3\tilde{H}\dot{\phi}$. This condition 
translates into the following physical constraint for \textcolor{red}{\underline{\bf Case~I}} and \textcolor{red}{\underline{\bf Case~II}} as given by:
\bea 
\partial_{\phi\phi}\left(\sqrt{\tilde{W}}\right)=\frac{3}{2M^{2}_{p}}\sqrt{\tilde{W}}.
\eea
Further using this constraint in Eq~(\ref{rd}) the inflaton field value at the end of inflation can be computed for \textcolor{red}{\underline{\bf Case~I}} and \textcolor{red}{\underline{\bf Case~II}} as:
\be\begin{array}{lll}\label{rdgx}
 \displaystyle\phi_f\displaystyle \sim\left\{\begin{array}{ll}
                    \displaystyle \sqrt{\frac{6}{5}}~M_p &
 \mbox{\small {\bf for \textcolor{red}{\underline{Case I}}}}  
\\ 
         \displaystyle  \frac{\phi_0}{\left[1+\frac{270\phi^2_0}{M^{2}_p}\right]^{1/2}}.~~~~~~~~~~ & \mbox{\small {\bf for \textcolor{red}{\underline{Case II}}}}.
          \end{array}
\right.
\end{array}\ee
Additionally it is important to mention here that the slow-roll approximation in Eq~(\ref{eqs1n}), Eq~(\ref{eqs2n}) and Eq~(\ref{eqs3n}) is only valid when the following constraint is satisfied, $\tilde{W}\geq \frac{\partial_{\phi}\tilde{W}}{\sqrt{6}}M_p$.
For our present setup this condition translates into the following constraint for \textcolor{red}{\underline{\bf Case~I}} and \textcolor{red}{\underline{\bf Case~II}}:
\bea 
\textcolor{red}{\underline{\bf Case ~I}:}~~~~~~~~~\phi_{inf}&\geq& \frac{2\sqrt{6}}{3}~M_p,\\
\textcolor{red}{\underline{\bf Case ~II}:}~~~~~~~~~\phi_{inf}&\leq& \left(\frac{4\sqrt{2}}{3\sqrt{3}}\right)^{1/3}~M_p,
\eea
where $\phi_{inf}$ represent the field value of the inflaton during inflation. Here it is important to note that, once the $\dot{\phi}$ term become dominant, then the slow-roll condition is not valid. In such case we need to solve the equation of motions with large inflaton kinetic terms where, $|\ddot{\phi}/3\tilde{H}\dot{\phi}|\sim {\cal O}(1)$. This also implies that when $\dot{\phi}^2$ contribution is dominant in the energy density, the slow-roll approximation for inflaton field completely breaks down
and reheating starts. 

Further we assume that to occur successful reheating in the proposed framework it is important to convert energy from the potential
energy density to radiation. Consequently one can write the following expression for reheating temperature as given by:
\bea\label{reh}
T_{RH}=\left(\frac{30\rho_f}{\pi^2 g_*}\right)^{1/4}\approx\left(\frac{30\tilde{W}_f}{\pi^2 g_*}\right)^{1/4}>T_{RH,min}\sim {\cal E}~{\rm GeV}
\eea
where $g_*$ is the effective number of particle species and ${\cal E}$ is a parameter which is different for different models of inflation. In this context successful reheating does not require instant transition to a radiation-dominated universe after inflation. The inflaton decay can occur much later than the end of inflation, and formation of a fully thermalized radiation bath can take even longer. This implies that very large value of ${\cal E}$ is not allowed in the present context, which is consistent with various supersymmetric models of inflation. Equivalently Eq~(\ref{reh}) can be translated for \textcolor{red}{\underline{\bf Case~I}} and \textcolor{red}{\underline{\bf Case~II}} as:
\bea 
\textcolor{red}{\underline{\bf Case ~I}:}~~~~~~~~~\lambda(\Psi_f)=\lambda e^{-\frac{2\sqrt{2}}{\sqrt{3}}\frac{\Psi_f}{M_p}}&>&\left(\frac{5\pi^2 g_*{\cal E}^4}{54}\right)\times (4.12\times 10^{-19})^4,\\
\textcolor{red}{\underline{\bf Case ~II}:}~~~~~~~~~~~~~~~~~~~~~~~~~~~~~~\alpha&<&\left(\frac{15}{4\pi^2 g_*{\cal E}^4}\right)\times (4.12\times 10^{-19})^{-4}.
\eea
One can interpret the results obtained in this section as:
\begin{itemize}
\item For \textcolor{red}{\underline{Case~I}} we get a lower bound on the field dependent coupling $\lambda(\Psi_f)$ which is expressed in terms of the effective number of particle $g_*$ and  the parameter ${\cal E}$.  
\item Similarly for \textcolor{red}{\underline{Case~II}} we get a upper bound on the scale free gravity coupling $\alpha$ which is also expressed in terms of the effective number of particle $g_*$ and the parameter ${\cal E}$.  
\item For a given value of $g_*$ and ${\cal E}$ one can explicitly determine the respective bounds on the coupling parameters. For an example one can  fix $g_* \sim 100$ in the present context.

\end{itemize} 
\section{\textcolor{blue}{Cosmological solutions from soft attractors}}
\label{s4}
\subsection{Solutions for inflaton}
\label{s4a}
To study the inflationary constraints and the cosmological consequences from our proposed setup here we first express the the value of the inflaton field at the onset of inflation, the horizon, reheating 
and including the density perturbation conditions as given by:
\bea\label{po1a}
\textcolor{red}{\underline{\bf Case ~I}}~~~~\nonumber\\
\phi_0>\phi_h&=&\sqrt{\frac{6}{5}}~M_p \left[1+\frac{5}{27}{\cal N}_{f/h}\right]^{1/2},\\
\label{po2a} \phi_0<\phi_{RH}&=&T_{RH,min}\times \left(\frac{\pi^2 g_* \lambda(\Psi_0)}{120}\right)^{1/4},\\
\label{po3a}  \phi_0>\phi_{D}&=&\left\{\begin{array}{ll}
                    \displaystyle 2M_p\left[\left(\frac{\delta\rho}{\rho}\right)_{cr}\sqrt{\frac{2}{\lambda(\Psi_h)}}+
\left(\frac{\phi_0}{2M_p}\right)^2-\frac{3}{10}-\frac{20}{9}{\cal N}_{f/h}\right]^{1/2}
~~~~\mbox{\small {\bf for \textcolor{red}{\underline{Region I}}}}
\\ 
         \displaystyle  2M_p\left[\left(\frac{\delta\rho}{\rho}\right)^{2/3}_{cr}\frac{1}{(\lambda(\Psi_h))^{3/2}}+
\left(\frac{\phi_0}{2M_p}\right)^2-\frac{3}{10}-\frac{20}{9}{\cal N}_{f/h}\right]^{1/2}
 \mbox{\small {\bf for \textcolor{red}{\underline{Region II}}}}.
          \end{array}
\right.
\eea
\bea\label{po1b}
\textcolor{red}{\underline{\bf Case ~II}}~~~~\nonumber\\
\phi_0>\phi_h&=&M_p \left[\frac{\left(\frac{\phi_0}{M_p}\right)^2}{1+270\left(\frac{\phi_0}{M_p}\right)^2}+8{\cal N}_{f/h}\right]^{1/2},\\
\label{po2b} \phi_0>\phi_{RH}&=&M_p~\sqrt[4]{\frac{4}{\lambda(\Psi_0)}}\left[\frac{\pi^2 g_*}{30}\left(\frac{T_{RH,min}}{M_p}\right)^4 -\frac{1}{8\alpha}\right]^{1/4},\\
\label{po3b}  \phi_0>\phi_{D}&=&\left\{\begin{array}{ll}
                    \displaystyle M_p\left[\frac{1}{\sqrt{2\alpha\lambda(\Psi_h)}}\left\{64\alpha\left(\frac{\delta\rho}{\rho}\right)^2_{cr}-1\right\}^{1/2}
                    +\frac{270\left(\frac{\phi_0}{M_p}\right)^{4}}{1+270\left(\frac{\phi_0}{M_p}\right)^{2}}-16
                    {\cal N}_{f/h}\right]^{1/2}
~~~~\mbox{\small {\bf for \textcolor{red}{\underline{Region I}}}}
\\ 
         \displaystyle  M_p\left[\frac{1}{(8\alpha)^{3/2}\lambda(\Psi_h)\left(\frac{\phi_0}{M_p}\right)}\left(\frac{\delta\rho}{\rho}\right)^{-1}_{cr}
                    +\frac{270\left(\frac{\phi_0}{M_p}\right)^{4}}{1+270\left(\frac{\phi_0}{M_p}\right)^{2}}-16{\cal N}_{f/h}\right]^{1/2}
~~~~
 \mbox{\small {\bf for \textcolor{red}{\underline{Region II}}}}.
          \end{array}
\right.
\eea
The physical interpretation of the obtained results are given bellow:
\begin{itemize}
\item For \textcolor{red}{\underline{\bf Case~I}} the field value at the horizon crossing is completely specified by the number of e-foldings, which is lying within the window, $50<{\cal N}_{f/h}={\cal N}_{cmb}<70 $. On the other hand for \textcolor{red}{\underline{\bf Case~II}} field value at the horizon crossing is specified by two parameters- A. number of e-foldings and B. the field value at the starting point of inflation (initial condition).
\item For \textcolor{red}{\underline{\bf Case~I}} the field value during the time of reheating is specified by three parameters- A. minimum value of the reheating temperature, B. value of the field dependent coupling parameter at the starting point of inflation, and C. the effective number of degrees of freedom $g_*$. For \textcolor{red}{\underline{\bf Case~II}} to determine the field value at the time of reheating we need to know additionally the numerical value of the scale free gravity parameter $\alpha$. 
\item For \textcolor{red}{\underline{\bf Case~I}} the field value during the density perturbation is specified by four parameters- A. value of the density contrast or more precisely the amplitude of the scalar perturbation, B. value of the field dependent coupling parameter at the horizon crossing and at the starting point of inflation, and C. number of e-foldings. For \textcolor{red}{\underline{\bf Case~II}} to determine the field value during the density perturbation we need to know additionally the numerical value of the scale free gravity parameter $\alpha$.  For the \textcolor{red}{\underline{Case~I}} one can express the solution for \textcolor{red}{\underline{Region~II}} in terms of \textcolor{red}{\underline{Region~I}} as:
\bea \left(\phi_{D}\right)_{\textcolor{red}{\bf Region~II}}=2M_p\left[\frac{1}{(\lambda(\Psi_h))^{3/2}}\left(\frac{\phi_0}{\sqrt{2} M_p}\right)^{2/3}\left(\frac{\delta\rho}{\rho}\right)^{2/3}_{\textcolor{red}{\bf Region~I}}+\left(\frac{(\phi_D)_{\textcolor{red}{\bf Region~I}}}{2M_p}\right)^2\right]^{1/2}.~~~~~\eea
Similarly for the \textcolor{red}{\underline{Case~II}} one can express the solution for \textcolor{red}{\underline{Region~II}} in terms of \textcolor{red}{\underline{Region~I}} as:
\bea \left(\phi_{D}\right)_{\textcolor{red}{\bf Region~II}}&=&2M_p\left[\frac{1}{8\alpha^{1/2}}\frac{\phi^2_0}{M^2_p}\left(\frac{\delta\rho}{\rho}\right)^{-1}_{\textcolor{red}{\bf Region~I}}\right.\nonumber\\&& \left.~~~~~~~+
\left(\frac{(\phi_D)_{\textcolor{red}{\bf Region~I}}}{M_p}\right)^2-\frac{1}{\sqrt{2\alpha\lambda(\Psi_h)}}\left\{64\alpha\left(\frac{\delta\rho}{\rho}\right)^{2}_{\textcolor{red}{\bf Region~I}}-1\right\}\right]^{1/2}.~~~~~~~~\eea
\end{itemize}
\subsection{Solutions for field dependent coupling $\lambda(\Psi)$}
\label{s4b}
Now let us describe the behaviour of the running or the scale dependence of the field dependent coupling $\lambda(\Psi)$
in the above mentioned two cases as:
\bea\label{po1c}
\textcolor{red}{\underline{\bf Case ~I}}~~~~~~~~~~~~~~\quad\quad\quad~~~~~~~~~~~~~~~~~\nonumber\\
\lambda(\Psi)= \lambda(\Psi_0) e^{-\frac{2\sqrt{2}}{\sqrt{3}M_p}\left(\Psi-\Psi_0\right)}&=&\lambda(\Psi_0) 
e^{\frac{1}{3M^2_p}\left(\phi^2-\phi^2_0\right)}=\lambda(\Psi_0) \left(\frac{t_0}{t}\right)^2,~~~~~~~~~~\\
\textcolor{red}{\underline{\bf Case ~II}}~~~~~~~~~~~~~~\quad\quad\quad~~~~~~~~~~~~~~~~~\nonumber\\
\label{po2v}\lambda(\Psi)= \lambda(\Psi_0) e^{-\frac{2\sqrt{2}}{\sqrt{3}M_p}\left(\Psi-\Psi_0\right)}&=&\lambda(\Psi_0) 
e^{\frac{3}{M^2_p}\left(\phi^2-\phi^2_0\right)}=\lambda(\Psi_0) e^{-\frac{2\sqrt{2}M_p}{3\sqrt{3\alpha}}(t-t_0)}.~~~~~~~~~~
\eea
Further using Eq~(\ref{rinflasd}), Eq~(\ref{po1a}) and Eq~(\ref{po1b}) we get the following constraint on the field dependent coupling $\lambda(\Psi_h)$ at 
horizon crossing:
\be\begin{array}{lll}\label{rdg}
 \displaystyle\lambda(\Psi_h)\displaystyle =\left\{\begin{array}{ll}
                    \displaystyle \lambda e^{-\frac{2\sqrt{2}}{\sqrt{3}}\frac{\Psi_h}{M_p}}=\frac{1.4\times 10^{-5}}{\left[1+\frac{5}{27}\alpha_{f/h}\right]^{3}} &
 \mbox{\small {\bf for \textcolor{red}{\underline{\bf Case I}}}}  
\\ 
         \displaystyle  -\lambda e^{-\frac{2\sqrt{2}}{\sqrt{3}}\frac{\Psi_h}{M_p}}=-\frac{77~M_p}{\alpha^{3/2}\left[\frac{\left(\frac{\phi_0}{M_p}\right)^2}{1+270\left(\frac{\phi_0}{M_p}\right)^2}+8\alpha_{f/h}\right]^{3/2}}.~~~~~~~~~~ & \mbox{\small {\bf for \textcolor{red}{\underline{\bf Case II}}}}.
          \end{array}
\right.
\end{array}\ee
Similarly using Eq~(\ref{rdg}) the field dependent coupling $\lambda(\Psi_0)$ can be expressed in terms of the number of e-foldings as:
\be\begin{array}{lll}\label{rdgasw}
 \displaystyle\lambda(\Psi_0)\displaystyle\displaystyle =\left\{\begin{array}{ll}
                    \displaystyle \lambda e^{-\frac{2\sqrt{2}}{\sqrt{3}}\frac{\Psi_0}{M_p}}=\frac{1.4\times 10^{-5}\times e^{-\frac{2}{5}\left[1+\frac{5}{27}\alpha_{f/h}\right]}}{\left[1+\frac{5}{27}\alpha_{f/h}\right]^{3}}e^{\frac{\phi^2_0}{3M^2_p}} &
 \mbox{\small {\bf for \textcolor{red}{\underline{Case I}}}}  
\\ 
         \displaystyle  -\lambda e^{-\frac{2\sqrt{2}}{\sqrt{3}}\frac{\Psi_0}{M_p}}=-\frac{77~M_p\times e^{-\frac{1}{3}\left[\frac{\left(\frac{\phi_0}{M_p}\right)^2}{1+270\left(\frac{\phi_0}{M_p}\right)^2}+8\alpha_{f/h}\right]}}{\alpha^{3/2}\left[\frac{\left(\frac{\phi_0}{M_p}\right)^2}{1+270\left(\frac{\phi_0}{M_p}\right)^2}+8\alpha_{f/h}\right]^{3/2}}e^{\frac{\phi^2_0}{3M^2_p}}.~~~~~~~~~~ & \mbox{\small {\bf for \textcolor{red}{\underline{Case II}}}}.
          \end{array}
\right.
\end{array}\ee
Additionally, we get the following constraint condition on the ratio of the couplings at the horizon crossing and at the starting point of inflation as given by:
\be\begin{array}{lll}\label{rdgxxc}
 \displaystyle\frac{\lambda(\Psi_h)}{\lambda(\Psi_0)}e^{\frac{\phi^2_0}{3M^2_p}}\displaystyle =\left\{\begin{array}{ll}
                    \displaystyle e^{\frac{2}{5}\left[1+\frac{5}{27}\alpha_{f/h}\right]} &
 \mbox{\small {\bf for \textcolor{red}{\underline{Case I}}}}  
\\ 
         \displaystyle   e^{\frac{1}{3}\left[\frac{\left(\frac{\phi_0}{M_p}\right)^2}{1+270\left(\frac{\phi_0}{M_p}\right)^2}+8\alpha_{f/h}\right]}.~~~~~~~~~~ & \mbox{\small {\bf for \textcolor{red}{\underline{Case II}}}}.
          \end{array}
\right.
\end{array}\ee
\section{\textcolor{blue}{Beyond soft attractor: A single field approach}}
\label{s5}
In this section our prime objective is to analyze the non attractor phase of inflation. To serve this purpose let us start with
the Klien-Gordon field equations for inflaton field $\phi$ and dilaton field $\Psi$, which can be written in the flat ($k=0$) FLRW background as:
\bea
\label{eqs1}
\frac{d^{2}\phi}{d\tilde{t}^{2}}+3\tilde{H}\frac{d\phi}{d\tilde{t}}+\partial_{\phi}\tilde{W}(\phi,\Psi)&=&0\Rightarrow\frac{d^{2}\phi}{d\tilde{t}^{2}}+3\tilde{H}\frac{d\phi}{d\tilde{t}}+\lambda(\Psi)\phi^3 =0\,\\
\label{eqs2}\frac{d^{2}\Psi}{d\tilde{t}^{2}}+3\tilde{H}\frac{d\Psi}{d\tilde{t}}+\partial_{\Psi}\tilde{W}(\phi,\Psi)&=&0\Rightarrow\frac{d^{2}\Psi}{d\tilde{t}^{2}}+3\tilde{H}\frac{d\Psi}{d\tilde{t}}-\frac{\lambda(\Psi)\phi^4}{\sqrt{6} M_p}=0.\,
\eea
Now in the slow-roll approximated regime the field equations are approximated as:
\bea
\label{eqs1n}
3\tilde{H}\frac{d\phi}{d\tilde{t}}+\lambda(\Psi)\phi^3=0\,\\
\label{eqs2n}3\tilde{H}\frac{d\Psi}{d\tilde{t}}-\frac{\lambda(\Psi)\phi^4}{\sqrt{6} M_p}=0,\,\\
\label{eqs3n}\tilde{H}^2=\frac{\tilde{W}(\phi,\Psi)}{3M^2_p}=\frac{V_0}{3M^2_p}\left[1+\frac{2\alpha\lambda(\Psi)}{M^{4}_{p}}\phi^{4}\right],
\eea
During the non-attractor phase of inflation we assume that the $\phi$ field is the only dynamical field controlling the scenario and at that time the $\Psi$ field freezes at the Planck scale. On the other hand, at late times the dynamical contribution comes from the $\Psi$ field and the inflaton field $\phi$ freezes at Planck scale. Assuming this fact the general behavior during inflationary epoch are governed by:
\bea
\label{po2} a&=&a_i\exp\left[-\frac{1}{20M^5_p}\left(\phi^5 -\phi^5_b\right)-\frac{\sqrt{3\pi}}{16\alpha\bar{\lambda}}\left\{{\rm Erf}\left[\frac{\phi}{\sqrt{3}M_p}
\right]-{\rm Erf}\left[\frac{\phi_i}{\sqrt{3}M_p}\right]\right\}\right],~~~~~~~~~\\
\label{po3} t-t_i&\approx& -\frac{3}{2\bar{\lambda}\sqrt{2\alpha}}\left[\frac{\sqrt{3\pi}}{2}\left\{{\rm Erf}\left[\frac{\phi}{\sqrt{3}M_p}
\right]-{\rm Erf}\left[\frac{\phi_i}{\sqrt{3}M_p}\right]\right\}+\frac{\alpha \bar{\lambda}}{5M^5_p}\left(\phi^5-\phi^5_b\right)\right].~~~~~~~
\eea
where $`i'$ subscript is used to describe the boundary/initial condition within the prescribed setup.
It is important to note that in Eq~(\ref{po2},\ref{po3}) we introduce new symbol $\bar{\lambda}$, which signifies the value
 of the self coupling at the freezing value of dilaton field
 ${\bf \Psi}\sim {\cal O}(M_p)$ during inflation i.e. 
\be
\bar{\lambda}=\lambda({{\bf \Psi}})=\lambda \exp\left[-\frac{2\sqrt{2}}{\sqrt{3}}\right].
\ee 
On the other hand at late time inflaton field get its VEV at $\phi\sim {\cal O}(M_p)$ and correspondingly the self coupling $\hat{\lambda}$ at late time is defined as:
\bea 
\hat{\lambda}\equiv\frac{\lambda}{4}\hat{\bf \phi}^{4}\sim\frac{\lambda}{4}M^{4}_{p}.
\eea
Here the obtained results can be interpreted as following:
\begin{itemize}
\item Solution for the scale factor $a(t)$ and inflaton field $\phi(t)$ admits quasi de-Sitter behaviour in presence of additional contribution coming from error functions.
\item For the large value of the product $\alpha\lambda$ one can further neglect the contributions from the error function. In that case one can get back the exact de-Sitter behaviour in the present context.
\end{itemize}
For further analysis let us introduce the following Hubble flow functions in Einstein frame:
\bea\label{hubx}
\tilde{\epsilon}_{H}&=&-\frac{1}{\tilde{H}^2}\frac{d\tilde{H}}{d\tilde{t}},~~~~
\tilde{\eta}_{H}=-\frac{1}{\tilde{H}}\left(\frac{\frac{d^{2}\phi}{d\tilde{t}^2}}{\frac{d\phi}{d\tilde{t}}}\right).
\eea 
The flow functions in the Einstein frame can be expressed in terms of the Jordan frame Hubble flow functions as:
\bea\label{hubflow}
\displaystyle \tilde{\epsilon}_{H}&=&\epsilon_{H}\frac{\left[1-\frac{1}{2H^{2}\epsilon_{H}}
\frac{d^{2}\ln \Omega^{2}}{dt^{2}}+\frac{1}{H\epsilon_{H}}\frac{d\ln \Omega^2}{dt}
-\frac{1}{4}\left(\frac{d\ln \Omega^{2}}{dt}\right)^{2}\right]}{\left[1+\frac{1}{2H}\frac{d\ln \Omega^{2}}{dt}\right]^{2}},~~~
\tilde{\eta}_{H}= \eta_{H}\frac{\left[1+\frac{1}{2H\eta_{H}}\frac{d\ln \Omega^2}{dt}\right]}{\left[1+\frac{1}{2H}\frac{d\ln \Omega^2}{dt}\right]}.
\eea

Further we introduce potential flow-functions in Einstein frame for the Higgs field $\phi$ as:
\begin{eqnarray}\label{ra1}
    \tilde{\epsilon}_{\tilde{W}}&=&\frac{M^{2}_{p}}{2}\left(\partial_{\phi} \ln\tilde{W}(\phi,{\bf\Psi})\right)^{2}\,
=\frac{32\alpha^{2}\bar{\lambda}^{2}\phi^{6}}{M^{6}_{p}\left[1+\frac{2\alpha\bar{\lambda}}{M^{4}_{p}}\phi^{4}\right]^{2}},\\
    \label{ra2} \tilde{\eta}_{\tilde{W}}&=&{M^{2}_{p}}\frac{\partial_{\phi\phi}\tilde{W}(\phi,{\bf\Psi})}{\tilde{W}(\phi,{\bf\Psi})}\,
=\frac{24\alpha\bar{\lambda}\phi^{2}}{M^{2}_{p}\left[1+\frac{2\alpha\bar{\lambda}}{M^{4}_{p}}\phi^{4}\right]},\\
\label{ra3}\tilde{\xi}^{2}_{\tilde{W}}&=&M^{4}_{p}\frac{\left(\partial_{\phi}\ln\tilde{W}(\phi,{\bf\Psi})\right)\left(\partial_{\phi\phi\phi}
\tilde{W}(\phi,{\bf\Psi})\right)}{\left(\tilde{W}(\phi,{\bf\Psi})\right)^{2}}\,=\frac{384\alpha^{2}\bar{\lambda}^{2}\phi^{4}}{M^{4}_{p}
\left[1+\frac{2\alpha\bar{\lambda}}{M^{4}_{p}}\phi^{4}\right]^{2}}\\
\label{ra4}\tilde{\sigma}^{3}_{\tilde{W}}&=&M^{6}_{p}\frac{\left(\partial_{\phi}\ln\tilde{W}(\phi,{\bf\Psi})\right)^{2}\left(\partial_{\phi\phi\phi\phi}
\tilde{W}(\phi,{\bf\Psi})\right)}{\left(\tilde{W}(\phi,{\bf\Psi})\right)^{3}}\,=\frac{3072\alpha^{3}\bar{\lambda}^{3}\phi^{6}}{M^{6}_{p}
\left[1+\frac{2\alpha\bar{\lambda}}{M^{4}_{p}}\phi^{4}\right]^{3}}.
   \end{eqnarray}
where we assume that the dilaton field $\Psi$ freezes at the field value ${\bf\Psi}$ during inflation. For further numerical estimation during inflation we fix the freezing 
value of dilaton field ${\bf \Psi}\sim {\cal O}(M_p)$. During inflation potential is characterized by:
\bea\label{eq33ainfla}
\tilde{W}(\phi,{\bf \Psi})&=&\tilde{U}({\bf \Psi})+\tilde{V}(\phi)=V_0 \left[1+\frac{2\alpha\lambda({\bf \Psi})}{M^{4}_{p}}\phi^{4}\right]=V_0+\frac{\bar{\lambda}}{4}\phi^{4}.
\eea
On the other hand in Einstein frame the Potential and Hubble flow functions are connected through the following relations: 
\bea\label{hub}
\tilde{\epsilon}_{H}&\approx&\tilde{\epsilon}_{\tilde{W}}
=\frac{32\alpha^{2}\bar{\lambda}^{2}\phi^{6}}{M^{6}_{p}\left[1+\frac{2\alpha\bar{\lambda}}{M^{4}_{p}}\phi^{4}\right]^{2}},~~~
\tilde{\eta}_{H}\approx\tilde{\eta}_{\tilde{W}}-\tilde{\epsilon}_{\tilde{W}}
=\frac{24\alpha\bar{\lambda}\phi^{2}}{M^{2}_{p}\left[1+\frac{2\alpha\bar{\lambda}}{M^{4}_{p}}\phi^{4}\right]}-\frac{32\alpha^{2}\bar{\lambda}^{2}
\phi^{6}}{M^{6}_{p}\left[1+\frac{2\alpha\bar{\lambda}}{M^{4}_{p}}\phi^{4}\right]^{2}}.
\eea 

\section{\textcolor{blue}{Constraints on inflation beyond soft attractor}}
\label{s6}
\subsection{Number of e-foldings}
\label{s6a}
In the present context the total number of e-foldings is defined as:
\bea
\label{cx1}{\cal N}_{total}={\cal N}(t_{e},t_{i})=\int^{t_{e}}_{t_{i}}H ~dt={\cal N}_{cmb}+\Delta {\cal N}
\eea
where ${\cal N}_{cmb}$ and $\Delta {\cal N}$ are defined as:
\bea
\label{cx2}{\cal N}_{cmb}={\cal N}(t_{e},t_{cmb})=\int^{t_{e}}_{t_{cmb}}H ~dt,~~~~
\label{cx3}\Delta{\cal N}={\cal N}(t_{cmb},t_{i})=\int^{t_{cmb}}_{t_{i}}H ~dt.
\eea

Further substituting the explicit form of the potential preseneted in this paper, the number of e-foldings can be recast as:
\bea
\label{nq1b}{\cal N}_{total}&\approx&-\frac{1}{M^2_p}\int^{\phi_{e}}_{\phi_{i}}\frac{\tilde{W}(\phi,{\bf\Psi})}{\partial_{\phi}\tilde{W}(\phi,{\bf\Psi})} ~d\phi=
\frac{M^{2}_{p}}{16\alpha\bar{\lambda}}\left\{\frac{1}{\phi^2_e}-\frac{1}{\phi^2_i}\right\}-\frac{1}{8M^2_p}\left(\phi^2_e-\phi^2_i\right),\\
\label{nq1a}{\cal N}_{cmb}&\approx&-\frac{1}{M^2_p}\int^{\phi_{e}}_{\phi_{cmb}}\frac{\tilde{W}(\phi,{\bf\Psi})}{\partial_{\phi}\tilde{W}(\phi,{\bf\Psi})} ~d\phi=
\frac{M^{2}_{p}}{16\alpha\bar{\lambda}}\left\{\frac{1}{\phi^2_e}-\frac{1}{\phi^2_{cmb}}\right\}-\frac{1}{8M^2_p}\left(\phi^2_e-\phi^2_{cmb}\right),~~~~
\eea
where superscript $e$, $cmb$ and $i$ denote the values of the inflaton field evaluated at the end of inflation, horizon crossing and starting point of inflation respectively.

In the present context the field value of the inflaton at the inflation is determined from the following condition, $\epsilon_{\tilde{W}}(\phi_{e})=1=|\eta_{\tilde{W}}(\phi_{e})|$.
Further substituting Eq~(\ref{ra1},\ref{ra2}) in the inflaton field value at the end of inflation can be computed as,
$\phi_{e}=3\sqrt{2}~M_{p}$. Now the expression for the inflaton field value at the horizon crossing is given by:
\bea \label{new1}
\phi_{cmb}&=& \frac{M_{p}}{2}\sqrt{\frac{{\cal A}_{cmb}}{\alpha\bar{\lambda}}}
\left[1+\sqrt{1+\frac{8\alpha\bar{\lambda}}{{\cal A}^{2}_{cmb}}}\right]^{\frac{1}{2}} 
\approx \frac{M_p}{2}\sqrt{\frac{2{\cal A}_{cmb}}{\alpha\bar{\lambda}}}\approx 2M_{p}\sqrt{{\cal N}_{cmb}}~~~~~~~
\eea
where we use the following constraint condition, 
$\sqrt{1+\frac{8\alpha\bar{\lambda}}{{\cal A}^{2}_{cmb}}}\approx 1+\frac{4\alpha\bar{\lambda}}{{\cal A}^{2}_{cmb}}+\cdots\sim {\cal O}(1)$
as $\alpha\bar{\lambda}/{\cal A}^2_{cmb}<<1$. Here the parameter ${\cal A}_{cmb}$ is defined as:
\bea \label{new2}
{\cal A}_{cmb}&=& \alpha\bar{\lambda}\left(16{\cal N}_{cmb}+36\right)= 16\alpha\bar{\lambda}{\cal N}_{cmb}\left[1+\frac{9}{4{\cal N}_{cmb}}\right]\approx 16\alpha\bar{\lambda}{\cal N}_{cmb}.
\eea
Similarly using Eq~(\ref{new1}, \ref{new2}) in Eq~(\ref{nq1b}) the expression for the inflaton field value at the starting point of inflation is given by:
\bea \label{new3}
\phi_{i}&=& \frac{M_{p}}{2}\sqrt{\frac{{\cal A}_{total}}{\alpha\bar{\lambda}}}
\left[1+\sqrt{1+\frac{8\alpha\bar{\lambda}}{{\cal A}^{2}_{total}}}\right]^{\frac{1}{2}}
\approx \frac{M_{p}}{2}\sqrt{\frac{{\cal A}_{total}}{\alpha\bar{\lambda}}}\approx 2M_p\sqrt{{\cal N}_{total}}~~~~~~~~~~~
\eea
where we use the similar constraint as mentioned above i.e.
$\sqrt{1+\frac{8\alpha\bar{\lambda}}{{\cal A}^{2}_{total}}}\approx 1+\frac{4\alpha\bar{\lambda}}{{\cal A}^{2}_{total}}+\cdots\sim {\cal O}(1)$
as $\alpha\bar{\lambda}/{\cal A}^2_{total}<<1$. Here the parameter ${\cal A}_{total}$ is defined as:
\bea \label{new4}
{\cal A}_{total}&=& \alpha\bar{\lambda}\left(16{\cal N}_{total}+36\right)= 16\alpha\bar{\lambda}{\cal N}_{total}\left[1+\frac{9}{4{\cal N}_{total}}\right]\approx 16\alpha\bar{\lambda}{\cal N}_{total}.
\eea
In table~(\ref{tab1}) we have given an estimate of the inflaton field value at horizon crossing ($\phi_{cmb}$) and starting point of inflation ($\phi_i$) 
for different values of ${\cal N}_{cmb}$ within the window $50\leq {\cal N}_{cmb}\leq 70$. 

\begin{table*}
\centering
\tiny\begin{tabular}{|c|c|c|c|c|c|c|c|c|c|}
\hline
\hline
 ${\cal N}_{cmb}$ & ${\cal N}_{total}$& ${\cal N}_{reh}$ &$\Delta{\cal N}$&$\overline{\Delta{\cal N}}$& $\phi_{e}$ &  $\phi_{cmb}$ & $\phi_{i}$& $\phi_{reh}$ & $|\Delta\phi|=|\phi_{cmb}-\phi_{i}|$\\
 & & && & (in $M_p$) & (in $M_p$) &  (in $M_p$) & (in $M_p$) & (in $M_p$)\\
\hline\hline\hline
 50 & 60& 51 & & & & 14.1 & 15.5 &  & 1.4
\\
 60 & 70& 61 &10& 9 & $3\sqrt{2}$ & 15.5  & 16.7& 2 & 1.2
\\
 70 & 80&71 & & & & 16.7& 17.9&  & 1.2
\\ \hline
\hline
\end{tabular}
\caption{Inflaton field value at the end, horizon crossing and starting point of inflation.}\label{tab1}
\vspace{.4cm}
\end{table*}

\subsection{Primordial density perturbation}
\label{s6b}

\begin{table*}
\centering
\tiny\begin{tabular}{|c|c|c|c|c|c|c|c|c|c|}
\hline
\hline
 ${\cal N}_{cmb}$ & ${P}_{S}({\cal N}_{cmb})$ &$n_{S}({\cal N}_{cmb})$& $\beta_{S}({\cal N}_{cmb})$ &
 $\kappa_{S}({\cal N}_{cmb})$ & $r({\cal N}_{cmb})$ & $\bar{\lambda}$ & $\alpha$& $T_{reh}$\\
 &(in $ 10^{-9}$) & & (in $ 10^{-3}$)  & (in $ 10^{-5}$) &   &(in $ 10^{-13}$) & (in $10^{7}$)& (in $10^{-3}~{ M_p}$)\\
\hline\hline\hline
 50 &  & 0.94 & 1.2& -4.8 &  & $\lesssim 1.49$ & $\gtrsim 3.48$ & $\lesssim 3.17$
\\
 60 & 2.207  &0.95 & 0.8 & -2.8  & $\lesssim 0.11$ & $\lesssim 1.13$ &$\gtrsim 3.48$ & $\lesssim 3.17$
\\
 70 &  & 0.96 & 0.6 &-1.8 &  &$\lesssim 0.89$  &$\gtrsim 3.48$ & $\lesssim 3.17$
\\ \hline
\hline
\end{tabular}
\caption{Inflationary observables and model constraints in the light of Planck 2015 data.}\label{tab2}
\vspace{.4cm}
\end{table*}
In the present context it is important to note that the perturbations to the homogeneous FLRW metric is described by the well known ADM
formalism. The line element in the ADM formalism after cosmological perturbation takes the following simplified form:
\be\label{gr1} ds^2= −N^2dt^2 + g_{ij} (dx^i + N^i dt)(dx^j + N^j dt),\ee 
where $g_{ij}$ is the induced spatial metric on the three surface labeled by time coordinate $t$, and $N$, $N_i$ are the time dependent lapse and shift functions, respectively. To do further analysis in the present computation one needs to make a proper choice of gauge to fix the diffeomorphism invariance of the theory of soft inflation originated from extended theories of gravity. A convenient choice is the synchronous gauge, defined by imposing the  conditions, $N=1,~N^i=0$.
The perturbed metric in synchronous gauge has the mathematical structure,
$g_{ij}=a^2(t)\left[(1+2\zeta(t,{\bf x}))\delta_{ij}+\gamma_{ij}\right]$,$
\gamma_{ii}=0$, where $\zeta(t,{\bf x})$ and $\gamma_{ij}$ are the scalar and tensor perturbations in the metric, respectively. Here to proceed further we need to make a specific gauge choice to fix the diffeomorphism invariance of the inflationary theory. Additionally, it is important to mention here that, for the computation the inflationary
correlation functions, it is more convenient to choose the gauge, $\delta\phi(t,{\bf x})= 0$, where the inflaton is homogeneous and the scalar perturbations are also appearing in the metric. Our focus will be on computing the two and three point functions
for the scalar fluctuation at late time, when the modes of interest have exited the horizon.
\subsubsection{Two point function}
\label{s6b1}
To compute the two point function for the scalar fluctuation we start with the second order action for the curvature perturbation as given by:
		    \bea 
		    S^{(2)}_{\zeta}&\approx&\int d^{4}x ~a^3~M^2_p\epsilon\left[\dot{\zeta}^2
		      -
		      \frac{1}{a^2}(\partial_{i}\zeta)^2
		      \right].\eea
		     Next we introduce a new variable $v(\eta,{\bf x})$ which is defined as,
		      $v(\eta,{\bf x})=z~\zeta(\eta,{\bf x})~M_p$.
		      In general the parameter $z$
		      is defined as,
$z =a\sqrt{2\epsilon}$.		      
		      Now in terms of $v(\eta,{\bf x})$ the second order action for the curvature perturbation can be recast as:
		      \bea S^{(2)}_{\zeta}&\approx&\int d^{3}x~ d\eta ~\left[v^{'2}
		      		   -(\partial_{i}v)^2
		      		   \frac{1}{a^2}(\partial_{i}\zeta)^2-m^{2}_{eff}(\eta)v^2
		      		   \right],~~~~~~~~~~~~\eea
		      		   where the effective mass parameter $m_{eff}(\eta)$ is defined as,
		      		    $m^{2}_{eff}(\eta)=-\frac{1}{z}\frac{d^2z}{d\eta^2}$.
		      		    During inflation the scale factor and the parameter $z$ can be expressed in terms of the conformal time $\eta$ as,
		      		   $a(\eta)=-\frac{1}{H\eta}$
		      		   and 
		      		   $\displaystyle z =a\sqrt{2\epsilon}= -\sqrt{2\epsilon}/H\eta$
		      		   where $\epsilon$ is the Hubble slow-roll parameter defined earlier. But for simplicity one can neglect the contribution from $\epsilon$ in the 
		      		  leading order for quasi de-Sitter case as it is sufficiently small in the slow-roll regime. 
		               Now further taking the Fourier transform:
		      		   \be v(\eta,{\bfx})=\int \frac{d^3k}{(2\pi)^3}~v_{\bf k}(\eta)~e^{i{\bf k}.{\bf x}}\ee
		      		   one can write down the equation of motion for scalar fluctuation as:
		      		   \be v^{''}_{\bf k}+\left(k^2+m^{2}_{eff}(\eta)\right)v_{\bf k}=0.\ee
		      		   Here it is important to note that for de Sitter and quasi de Sitter case the effective mass parametrer can be expressed as, 
		      		   $m^{2}_{eff}(\eta) =  -\frac{2}{\eta^2}$.
		      		Finally, considering the behaviour of the mode function in the \textcolor{blue}{subhorizon regime} and \textcolor{blue}{superhorizon regime} one can write the expression in de Sitter with Bunch-Davies vacuum as:
		      		\bea\label{yu2zxzxa}
 \displaystyle \zeta(\eta,{\bf k}) &=& \frac{v_{\bf k}(\eta)}{z~M_p}=
  \frac{iH}{2~M_p\sqrt{\epsilon_{H}}~k^{\frac{3}{2}}}~e^{- i(k\eta+\pi)}
                    \left(1+ik\eta\right).\eea	                    	At the horizon crossing taking the late time limit one can write down the
following expression for the two point function for scalar fluctuation as:
\bea	\langle \zeta({\bf k}) \zeta({\bf q})\rangle
&=&(2\pi)^3\delta^{(3)}({\bf k}+{\bf q})P_{\zeta}(k_*)\frac{1}{k^3},\eea
where $P_{\zeta}(k_*)$ is known as the power spectrum at the horizon crossing for scalar fluctuations and it is defined as:
\bea
										 \displaystyle P_{\zeta}(k_*)
										  &=&\frac{H^2}{4~M^2_p\epsilon^{*}_H}.\eea 
For simplicity one can further define amplitude of the power spectrum ${\cal P}_{S}({\cal N}_{cmb})$ at the horizon crossing as:
							\bea\label{powas}
								 \displaystyle {\cal P}_{S}({\cal N}_{cmb})&=&\frac{1}{2\pi^2}P_{\zeta}(k_*)=\frac{H^2}{8\pi^2~M^2_p\epsilon^*_H}.\eea										  
\subsubsection{Present observables}
\label{s6b2}
\begin{figure*}[htb]
\centering
{
    \includegraphics[width=12.2cm,height=5cm] {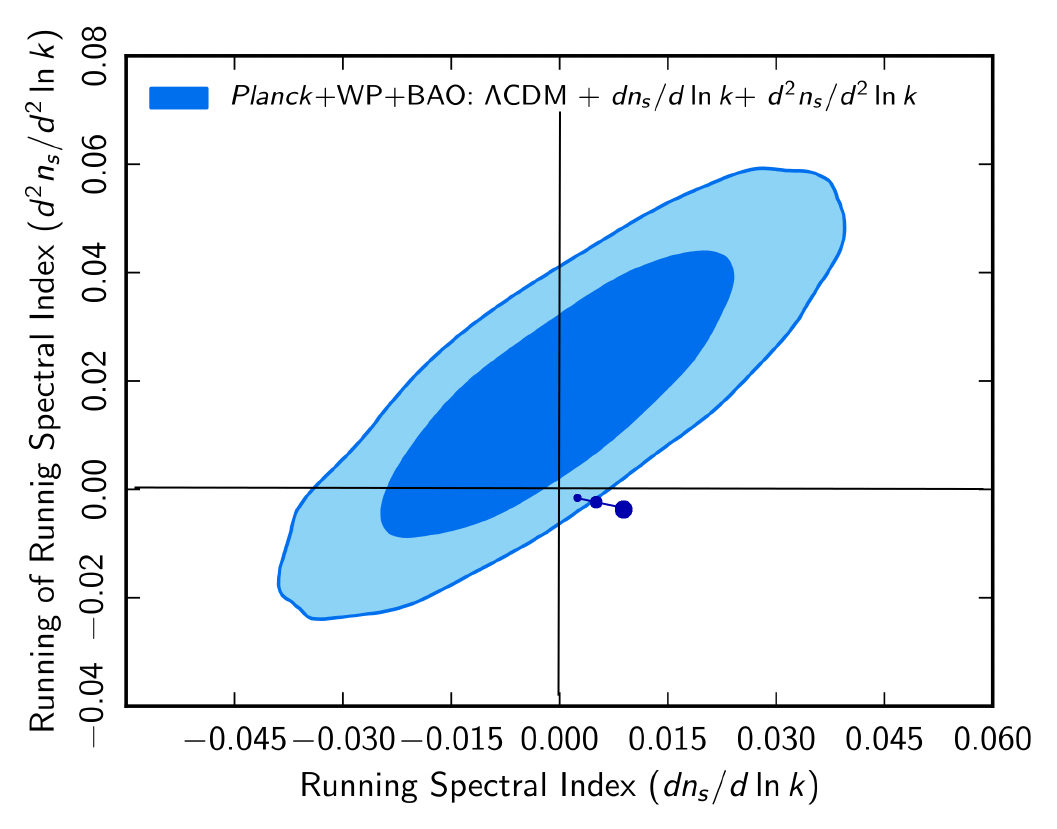}
    \label{figab}
}
\caption[Optional caption for list of figures]{Plot for running of the running of spectral index $\kappa_{S}=d^2n_{S}/d^2\ln k$ vs running of the spectral index $\beta_{S}=dn_{S}/d\ln k$ for scalar modes. Here for non attractor phase of inflation we have drawn blue colored line. We also draw the background of confidence contours obtained from  various joint constraints \cite{Ade:2015lrj,Ade:2015ava,Ade:2015xua}.} 
\end{figure*}
Applying slow-roll approximation in the present context the inflationary observables i.e. power spectrum, spectral tilt, running and running of the running of the tilt  
for scalar modes 
from our model at horizon crossing (${\cal N}={\cal N}_{cmb}$) can be computed as: 
\bea
\label{pu1}{\cal P}_{S}({\cal N}_{cmb})&=&\frac{\tilde{W}(\phi_{cmb},{\bf \Psi})}{24\pi^{2}M^{4}_{p}\epsilon^{*}_{\tilde{W}}}=
\frac{1}{393216\pi^{2}\alpha^{3}\bar{\lambda}^{2}{\cal N}^{3}_{cmb}}\left[1+32\alpha\bar{\lambda}{\cal N}^{2}_{cmb}\right]^{3},\\
\label{pu2}n_{S}({\cal N}_{cmb})&=&1+\left(\frac{d\ln {\cal P}_{S}}{d{\cal N}}\right)_{{\cal N}_{cmb}}
\approx 1-\frac{3}{{\cal N}_{cmb}},\\
\label{pu3}\alpha_{S}({\cal N}_{cmb})&=&\left(\frac{dn_{S}}{d{\cal N}}\right)_{{\cal N}_{cmb}}
\approx \frac{3}{{\cal N}^{2}_{cmb}},~~~
\label{pu4}\kappa_{S}({\cal N}_{cmb})=\left(\frac{d\alpha_{S}}{d{\cal N}}\right)_{{\cal N}_{cmb}}
\approx -\frac{6}{{\cal N}^{3}_{cmb}}.
\eea
where ${\bf \Psi}$ represents the freezing value of $\Psi$ field at Planck scale.

In~fig.~(\ref{figab}), we have plotted running of the running of spectral tilt for scalar perturbation ($\kappa_{S}=d^2n_{S}/d^2\ln k$) vs running of the spectral index $\beta_{S}=dn_{S}/d\ln k$ in the light of Planck data along with various joint constraints. Here it is important to note that, for non attractor phase of inflation the Higgsotic models are shown by the blue colored line. Also the big circle, intermediate size circle and small circle represent the representative points in $(\kappa_S,\beta_{S})$ 2D plane for the number of e-foldings, ${\cal N}_{cmb}=70$, ${\cal N}_{cmb}=60$ and ${\cal N}_{cmb}=50$ respectively. To represent the present status as well as statistical significance of the Higgsotic model for for non attractor phase of inflation, we have drawn the $1\sigma$ and $2\sigma$ confidence contours from Planck+WMAP+BAO joint data sets \cite{Ade:2015lrj,Ade:2015ava,Ade:2015xua}. It is clearly visualized from the fig.~(\ref{figan}) that, for non attractor phase of inflation we cover the range, $0.6\times 10^{-3}<\beta_{S}=\frac{dn_S}{d\ln k}<1.2\times 10^{-3}$ and $-1.8\times 10^{-5}>\kappa_{S}=\frac{d^2n_S}{d^2\ln k}>-4.8\times 10^{-5}$
in the $(\kappa_S,\beta_{S})$ 2D plane.
\subsection{Primordial tensor modes}
\label{s6c}
\subsubsection{Two point function}
\label{s6c1}
To compute the expression for the two point function for the tensor fluctuation here we start with the second order action for the spin-2 graviton as given by:
			 \bea S^{(2)}_{\gamma}&\approx&\int d^{4}x ~a^3~\frac{M^2_p}{8}\left[\dot{\gamma}_{ij}\dot{\gamma}_{ij}
			   -
			   \frac{1}{a^2}(\partial_{m}\gamma_{ij})^2
			   \right].\eea
			  In terms of conformal time in the present context second order action for tensor fluctuation can be recast as:
			   \bea S^{(2)}_{\gamma}&\approx&\int d^{3}x~d\eta ~a^2~\frac{M^2_p}{8}\left[\gamma^{'2}_{ij}
			   			   -
			   			  (\partial_{m}\gamma_{ij})^2
			   			   \right].\eea
			   In Fourier space one can further decompose the graviton $\gamma_{ij}(\eta,{\bf x})$ as:
			   \be \gamma_{ij}(\eta,{\bf x})=\sum_{\lambda=\times,+}\int\frac{d^3 k}{(2\pi)^{\frac{3}{2}}}\epsilon^{\lambda}_{ij}(k)~\gamma_{\lambda}(\eta,{\bf k})~e^{i{\bf k}.{\bf x}},\ee
			      where $\epsilon^{\lambda}_{ij}$ is the polarization tensor which satisfies the following property, 
			      $\epsilon^{\lambda}_{ii}=k^{i}\epsilon^{\lambda}_{ij}=0,~
			      \sum_{i,j}\epsilon^{\lambda}_{ij}\epsilon^{\lambda^{'}}_{ij}=2\delta_{\lambda\lambda^{'}}$.
			      Similar like scalar fluctuation here we also define a new variable $u_{\lambda}(\eta,{\bf k})$ in Fourier space as,
			      $u_{\lambda}(\eta,{\bf k}) =\frac{a}{\sqrt{2}}~M_p~\gamma_{\lambda}(\eta,{\bf k})=  -\frac{1}{\sqrt{2}H\eta}~M_p~\gamma_{\lambda}(\eta,{\bf k})$.
			      		      		   Using $u_{\lambda}(\eta,{\bf k})$ one can further 
			      		      		   write down the the second order action for the graviton as:
\bea S^{(2)}_{\gamma}&\approx&\int d^{3}x~d\eta ~a^2~\frac{M^2_p}{4}\left[u^{'2}_{\lambda}(\eta,{\bf k})
  -\left(k^2-\frac{a^{''}}{a}\right) (u_{\lambda}(\eta,{\bf k}))^2
			   			   \right].\eea		
			   			   From this action one can find out the mode equation for tensor fluctuation as:
			   			   \bea 
	u^{''}_{\lambda}(\eta,{\bf k})+\left(k^2-\frac{a^{''}}{a}\right)u_{\lambda}(\eta,{\bf k})=0.\eea
	It is important to mention here that for de-Sitter case we have,
	$\frac{a^{''}}{a}=\frac{2}{\eta^2}$.
Further considering the behaviour of the solution in \textcolor{blue}{superhorizon} and \textcolor{blue}{subhorizon} regime for Bunch-Davies vacuum we get:
	\bea u_{\lambda}(\eta,{\bf k})&=&\frac{1}{i\eta \sqrt{2}~ k^{\frac{3}{2}}}~e^{-i(k\eta+\pi)}~(1+ik\eta).\eea
		At the horizon crossing taking the late time limit one can write down the
following expression for the two point function for scalar fluctuation as:
\bea	\langle h(\eta,{\bf k}) h(\eta,{\bf q})\rangle=\sum_{\lambda,\lambda^{'}} \langle h_{\lambda}(\eta,{\bf k})h_{\lambda^{'}}(\eta,{\bf q})\rangle
			      		 &=&(2\pi)^3\delta^{(3)}({\bf k}+{\bf q})P_{h}(k,\eta),\eea
where $P_{h}(k_*)$ is known as the power spectrum at the horizon crossing for tensor fluctuations and it is defined as:
\bea
										 \displaystyle P_{h}(k_*)
										  &=&\frac{4H^2}{M^2_p}.\eea 
For simplicity one can further define amplitude of the power spectrum ${\cal P}_{T}({\cal N}_{cmb})$ at the horizon crossing as:
							\bea\label{powas2}
								 \displaystyle {\cal P}_{T}({\cal N}_{cmb})&=&\frac{1}{2\pi^2}P_{h}(k_*)=\frac{2H^2}{\pi^2~M^2_p}.\eea				
\subsubsection{Future observables}
\label{s6c2}
Applying slow-roll approximation in the present context the future inflationary observables i.e. power spectrum, spectral tilt, running and running of the running of the tilt  
for tensor modes from our model at horizon crossing (${\cal N}={\cal N}_{cmb}$) can be computed as: 
\bea
\label{pu1t}{\cal P}_{T}({\cal N}_{cmb})&=&\frac{2\tilde{W}(\phi_{cmb},{\bf \Psi})}{3\pi^{2}M^{4}_{p}}=
\frac{1}{12\pi^{2}\alpha}\left[1+32\alpha\bar{\lambda}{\cal N}^{2}_{cmb}\right],\\
\label{pu2t}n_{T}({\cal N}_{cmb})&=&\left(\frac{d\ln {\cal P}_{T}}{d{\cal N}}\right)_{{\cal N}_{cmb}}
= \frac{64\alpha\bar{\lambda}{\cal N}_{cmb}}{\left[1+32\alpha\bar{\lambda}{\cal N}^{2}_{cmb}\right]},~~~
\label{pu3t}\alpha_{T}({\cal N}_{cmb})=\left(\frac{dn_{T}}{d{\cal N}}\right)_{{\cal N}_{cmb}}
= \frac{64\alpha\bar{\lambda}\left[1-32\alpha\bar{\lambda}{\cal N}^{2}_{cmb}\right]}{\left[1+32\alpha\bar{\lambda}{\cal N}^{2}_{cmb}\right]^2},~~~~~~~~\\
\label{pu4z}\kappa_{T}({\cal N}_{cmb})&=&\left(\frac{d\alpha_{T}}{d{\cal N}}\right)_{{\cal N}_{cmb}}= -2048\alpha^2 \bar{\lambda}^2 {\cal N}_{cmb}\left[\frac{{\cal N}_{cmb}\left[1+32\alpha\bar{\lambda}{\cal N}^{2}_{cmb}\right]^2+4 
\left[1-1024\alpha^2 \bar{\lambda}^2 {\cal N}^{4}_{cmb}\right]}{\left[1+32\alpha\bar{\lambda}{\cal N}^{2}_{cmb}\right]^4}\right],~~~~~~~\\
\label{pu5}r({\cal N}_{cmb}) &=&\frac{{\cal P}_{T}({\cal N}_{cmb})}{{\cal P}_{S}({\cal N}_{cmb})}= 16\epsilon^{*}_{\tilde{W}}
= \frac{32768\alpha^{2}\bar{\lambda}^{2}{\cal N}^{3}_{cmb}}{\left[1+32\alpha\bar{\lambda}{\cal N}^{2}_{cmb}\right]^{2}}.
\eea
\begin{figure*}[htb]
\centering
{
    \includegraphics[width=12.2cm,height=5cm] {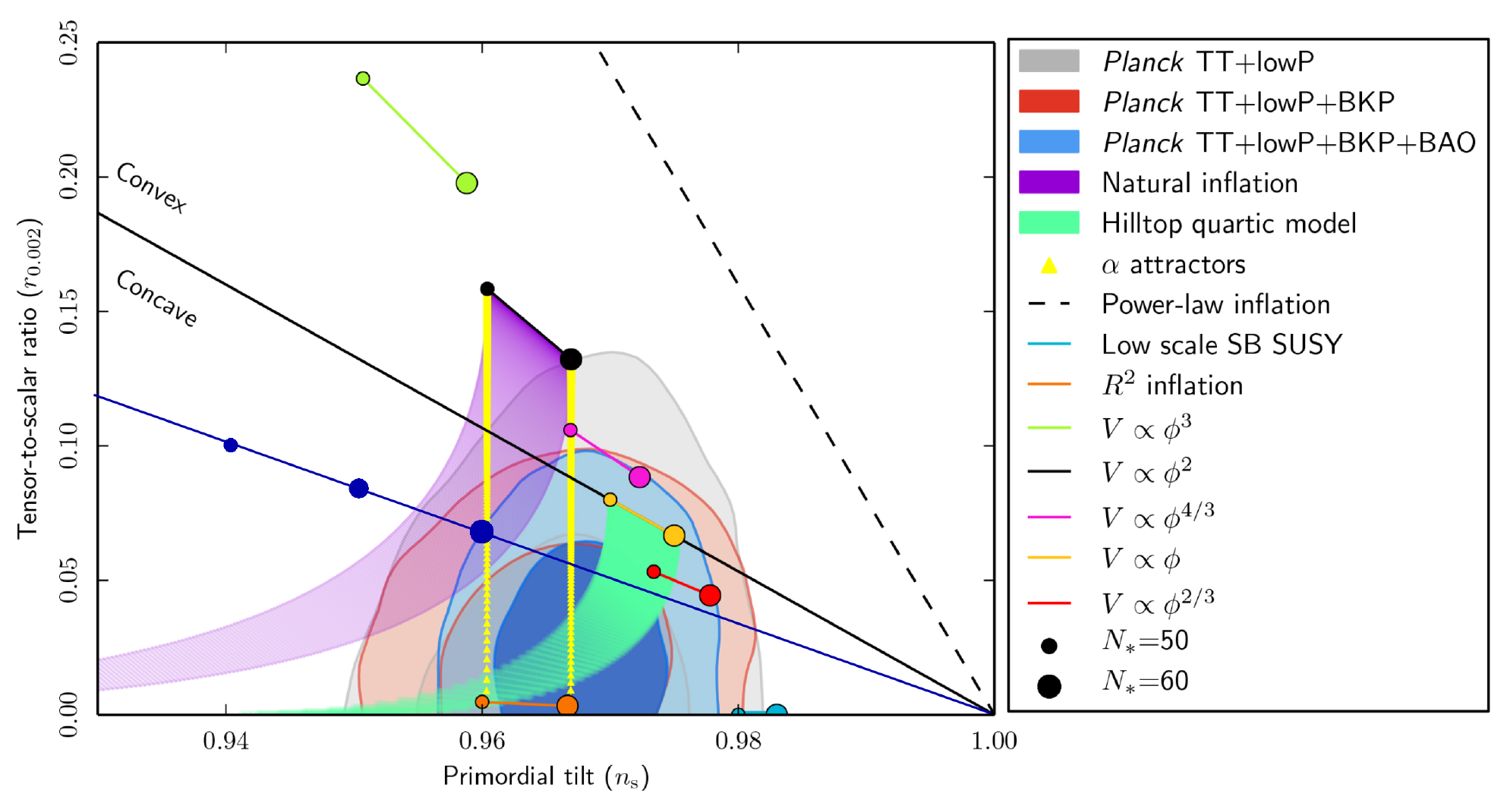}
    \label{figvns1}
}
\caption[Optional caption for list of figures]{$r$ vs $n_{S}$ polt for non attractor phase of inflation for Higgsotic model in the background of confidence contours obtained from  Planck TT+low P, Planck TT+low P+BKP, Planck TT+low P+BKP+BAO joint data sets.} 
\end{figure*}
In~fig.~(\ref{figvns1}), we have plotted tensor-to-scalar ratio ($r$) vs spectral tilt for scalar perturbation ($n_{S}$) in the light of Planck data alongwith various joint constraints. Here it is important to note that, non attractor phase of inflation for Higgsotic model is shown by the green and pink colored lines. Also the big circle, intermediate size circle and small circle represent the representative points in $(r,n_{S})$ 2D plane for the number of e-foldings, ${\cal N}_{cmb}=70$, ${\cal N}_{cmb}=60$ and ${\cal N}_{cmb}=50$ respectively. To represent the present status as well as statistical significance of the Higgsotic model in its non attractor phase, we have drawn the $1\sigma$ and $2\sigma$ confidence contours from Planck TT+low P, Planck TT+low P+BKP, Planck TT+low P+BKP+BAO joint data sets \cite{Ade:2015lrj,Ade:2015ava,Ade:2015xua} in fig.~(\ref{figvns1}). It is clearly visualized from the fig.~(\ref{figvns1}) that, we cover the range, $ 0.94<n_{S}<0.96$ and $0.06<r<0.11$
in the $(r,n_{S})$ 2D plane. 

Now to derive the constraints on the model parameters $\alpha$ and $\bar{\lambda}$ we use Eq~(\ref{pu1}) and Eq~(\ref{pu5}), which can be recast as:
\bea\label{alw1}\alpha&\approx& \frac{1}{(96\pi^{2}{\cal P}_{S}({\cal N}_{cmb}))^{1/3}\bar{\lambda}^{2/3}\times 16{\cal N}_{cmb}},\\
\label{alw2}\bar{\lambda}&\approx&(96\pi^{2}{\cal P}_{S}({\cal N}_{cmb}))\left[\frac{r({\cal N}_{cmb})}{128{\cal N}_{cmb}}\right]^{3/2}.\eea
Further substituting Eq~(\ref{alw2}) on Eq~(\ref{alw1}) finally we get:
\bea\label{alw23}
\alpha &\approx&\frac{1}{12\pi^{2}{\cal P}_{S}({\cal N}_{cmb})r({\cal N}_{cmb})}.
\eea
In terms of the number of e-foldings (${\cal N}$) the the most useful parametrization of the primordial scalar and tensor 
power spectrum or equivalently for tensor-to-scalar ratio can be written near the horizon crossing ${\cal N}={\cal N}_{cmb}$ as:
\bea 
r({\cal N})&=&\frac{{\cal P}_{T}({\cal N})}{{\cal P}_{S}({\cal N})}=r({\cal N}_{cmb})\exp\left[\left({\cal N}-{\cal N}_{cmb}\right)\left\{A({\cal N}_{cmb})+B({\cal N}_{cmb})\left({\cal N}-{\cal N}_{cmb}\right) 
+C({\cal N}_{cmb})\left({\cal N}-{\cal N}_{cmb}\right)^2\right\}\right]~~~~~~~~~~
\eea
where the symbols $A({\cal N}_{cmb})$, $B({\cal N}_{cmb})$ and $C({\cal N}_{cmb})$ are expressed in terms of the inflationary observables at horizon crossing as, 
$A({\cal N}_{cmb})=n_{T}({\cal N}_{cmb})-n_{S}({\cal N}_{cmb})+1$, 
$B({\cal N}_{cmb})=\frac{1}{2}\left(\alpha_{T}({\cal N}_{cmb})-\alpha_{S}({\cal N}_{cmb})\right)$,
$C({\cal N}_{cmb})=\frac{1}{6}\left(\kappa_{T}({\cal N}_{cmb})-\kappa_{S}({\cal N}_{cmb})\right)$.
In the above parametrization, $A({\cal N}_{cmb})>>B({\cal N}_{cmb})>>C({\cal N}_{cmb})$ is always required for convergence of the Taylor expansion. 
For the time being to make the computation simpler let us assume that the term involving the 
co-efficient of the quadratic term $B({\cal N}_{cmb})$ and cubic term $C({\cal N}_{cmb})$
 is negligibly small compared to the leading order term $A({\cal N}_{cmb})$ as appearing in the exponent of the above mentioned parametrization.  
Using this assumption the relation between field excursion and tensor-to-scalar ratio can be computed as:
\bea \label{Lythh}
\frac{|\Delta\phi|}{M_p}&\approx& \frac{2}{A({\cal N}_{cmb})}\sqrt{\frac{r({\cal N}_{cmb})}{8}}
\left[1-e^{-\Delta {\cal N}\left(\frac{A({\cal N}_{cmb})}{2}\right)}\right]\approx\sqrt{\frac{r({\cal N}_{cmb})}{8}}\Delta{\cal N}\eea 
and finally using the above relation from our $R^2$ gravity model we get:
\bea\label{Lyth2} r({\cal N}_{cmb})&\approx& 8\left(\frac{|\Delta\phi|}{M_p \Delta{\cal N}}\right)^{2}
\approx 32\left(\frac{\left[\sqrt{{\cal N}_{total}}-\sqrt{{\cal N}_{cmb}}\right]}{\Delta{\cal N}}\right)^{2}=\frac{32}{\left[\sqrt{{\cal N}_{total}}+\sqrt{{\cal N}_{cmb}}\right]^2}.~~~~
~~~~~\eea
For our prescribed model $|\Delta\phi|\approx 1.2 ~{\rm M}_{p}$ and $\Delta{\cal N}=10$ is fixed by Planck 2015 observation. 
Substituting these values in the relation stated in 
Eq~(\ref{Lyth2}), the upper bound of tensor-to-scalar ratio at the scale of horizon crossing computed from our setup as,
$r({\cal N}_{cmb})\lesssim 0.11$.
Now in the present context using Eq~(\ref{pu2}) we can express the number of e-foldings at the horizon crossing as:
\bea 
\label{con1} {\cal N}_{cmb}=\frac{3}{1-n_{S}({\cal N}_{cmb})}
\eea
and substituing in Eq~(\ref{pu3},\ref{pu4}) we get the following sets of consistency relations for scalar modes from our analysis:
 \bea 
\label{tr1}\alpha_{S}({\cal N}_{cmb}) &\approx& \frac{\left(1-n_{S}({\cal N}_{cmb})\right)^{2}}{3},\\
\label{tr2} \kappa_{S}({\cal N}_{cmb}) &\approx& -\frac{2\left(1-n_{S}({\cal N}_{cmb})\right)^{3}}{9}
\eea
and combining Eq~(\ref{tr1}) and Eq~(\ref{tr2}) we finally get:
\bea
\label{tr3} 1-n_{S}({\cal N}_{cmb})+\frac{3\kappa_{S}({\cal N}_{cmb})}{2\alpha_{S}({\cal N}_{cmb})}&\approx&0.
\eea
Similarly from tensor modes we get the following sets of consistency relations from our model:
\bea
\label{trx3} r({\cal N}_{cmb}) \approx16\epsilon^{*}_{\tilde{W}}&=&\frac{884736\alpha^{2}\bar{\lambda}^{2}}{\left(1-n_{S}({\cal N}_{cmb})\right)^{3}
\left[1+\frac{288\alpha\bar{\lambda}}{\left(1-n_{S}({\cal N}_{cmb})\right)^2}\right]^2}=\frac{884736\alpha^{2}\bar{\lambda}^{2}}{\left(3\alpha_{S}({\cal N}_{cmb})
\right)^{3/2}\left[1+\frac{96\alpha\bar{\lambda}}{\alpha_{S}({\cal N}_{cmb})}\right]^2} \nonumber \\
&=&-\frac{196608\alpha^{2}\bar{\lambda}^{2}}{\kappa_{S}({\cal N}_{cmb})\left[1+\frac{288\alpha\bar{\lambda}}{\left\{-\frac{9}{2}
\kappa_{S}({\cal N}_{cmb})\right\}^{2/3}}\right]^2}=\frac{24n^{2}_{T}({\cal N}_{cmb})}{1-n_{S}({\cal N}_{cmb})}.
\eea
It is important mention here that in the resent context the usual consistency relation for single field slow-roll inflation,  \be r({\cal N}_{cmb})=-8n_{T}({\cal N}_{cmb})\ee violates and after doing the 
anaysis we found a completely new consistency relation as presented in Eq~(\ref{trx3}). In case of usual slow-roll single field inflationary setup 
the tensor spectral tilt, $n_{T}({\cal N}_{cmb})<0$ always. But for prescribed setup Eq~(\ref{pu2t}), Eq~(\ref{alw2}) and  Eq~(\ref{alw23}) suggests that,
 $\bar{\lambda}>0,\alpha>0$ always imples $ n_{T}({\cal N}_{cmb})>0$. Further using Eq~(\ref{Lyth2}) in Eq~(\ref{trx3}) we get the following constraint relationship:
\bea 
\frac{|\Delta\phi|}{M_p}
\approx 2\left[\sqrt{{\cal N}_{total}}-\sqrt{{\cal N}_{cmb}}\right]=\sqrt{\frac{3}{1-n_{S}({\cal N}_{cmb})}}n_{T}({\cal N}_{cmb})\left[{\cal N}_{total}
-\frac{3}{1-n_{S}({\cal N}_{cmb})}\right].~~~~~~~~
\eea
\subsection{Reheating}
\label{s6d}
The above results provide limits on the reheating temperature $T_{reh}$, defined as the initial temperature of
the homogeneous radiation dominated universe. In general, the reheating temperature $T_{reh}$ is related to energy density $\rho_{reh}$ through the following expression:
\bea \label{retem}
\rho_{reh}=\frac{\pi^{2}}{30}g_{\rm eff}(T_{reh})T^{4}_{reh} \Rightarrow T_{reh}&=&\left(\frac{30}{\pi^{2}g_{\rm eff}(T_{reh})}\right)^{1/4}\rho^{1/4}_{reh}\approx \left(\frac{30}{\pi^{2}g_{\rm eff}(T_{reh})}\right)^{1/4}V^{1/4}_{reh}
\eea
where $g_{\rm eff}(T_{reh})$ is the effective number of relativistic degrees of freedom present in the thermal bath at
the temperature $T=T_{reh}$ and $V_{reh}$ represents the scale of reheating at $\phi=\phi_{reh}$ given by the expression, $V_{reh}=V(\phi=\phi_{reh})=V_0\left[1+\frac{2\alpha\bar{\lambda}}{M^{4}_{p}}\phi^{4}_{reh}\right]$.
Counting all degrees of freedom of the Standard Model and the dilaton degrees of freedom, one has $g_{\rm eff}(T_{reh})=107.75$. To find the reheating constraint from our prescribed setup let us introduce the number of e-foldings at the time of reheating defined as:
\bea \label{ne1}
{\cal N}_{reh}&=&\int^{t_{e}}_{t_{reh}} H~dt={\cal N}_{total}-\overline{\Delta{\cal N}}\approx -\frac{1}{M^2_p}\int^{\phi_{e}}_{\phi_{reh}}\frac{\tilde{W}(\phi,{\bf\Psi})}{\partial_{\phi}\tilde{W}(\phi,{\bf\Psi})} ~d\phi=\frac{M^{2}_{p}}{16\alpha\bar{\lambda}}\left\{\frac{1}{\phi^2_e}-\frac{1}{\phi^2_{reh}}\right\}-\frac{1}{8M^2_p}\left(\phi^2_e-\phi^2_{reh}\right).~~~~~~~~~~
\eea
For the sake of clarity let us express the interval $\overline{\Delta{\cal N}}$ as:
\bea \label{sde}
\overline{\Delta{\cal N}}&=&\int^{t_{e}}_{t_{i}} H~dt-\int^{t_{e}}_{t_{reh}} H~dt=\int^{t_{reh}}_{t_{i}} H~dt ={\cal N}_{total}-{\cal N}_{reh}=\Delta{\cal N}-\left({\cal N}_{reh}-{\cal N}_{cmb}\right)\nonumber\\
\Rightarrow\Delta{\cal N}-\overline{\Delta{\cal N}}&=&{\cal N}_{reh}-{\cal N}_{cmb}=\frac{M^{2}_{p}}{16\alpha\bar{\lambda}}
\left\{\frac{1}{\phi^2_{cmb}}-\frac{1}{\phi^2_{reh}}\right\}-\frac{1}{8M^2_p}\left(\phi^2_{cmb}-\phi^2_{reh}\right).~~~~~~
\eea
Finally using the last step of Eq~(\ref{sde}) the field value during reheating can be expressed as:
\bea \label{new1c}
\phi_{reh}= \frac{M_{p}}{2}\sqrt{\frac{{\cal M}_{reh}}{\alpha\bar{\lambda}}}
\left[1+\sqrt{1+\frac{8\alpha\bar{\lambda}}{{\cal M}^{2}_{reh}}}\right]^{\frac{1}{2}} 
\approx \frac{M_p}{2}\sqrt{\frac{{\cal M}_{reh}}{\alpha\bar{\lambda}}}\approx
 2M_{p}\sqrt{\left(\Delta {\cal N}-\overline{\Delta {\cal N}}\right)}\approx
 2M_{p}\sqrt{{\cal N}_{reh}-{\cal N}_{cmb}}~~~~~~~~~~~
\eea
where 
\bea \label{new2c}
{\cal M}_{reh}&=& \alpha\bar{\lambda}\left[16\left(\Delta {\cal N}-\overline{\Delta {\cal N}}\right)
+8{\cal N}_{cmb}\right]-\frac{1}{4{\cal N}_{cmb}}\approx 16\alpha\bar{\lambda}\left(\Delta {\cal N}-\overline{\Delta {\cal N}}\right).
\eea
Using Eq~(\ref{new2c}) finally we get the scale of reheating in terms of the number of e-foldings as:
\bea\label{uytwe1x}
V_{reh}\approx V_0\left[1+32\alpha\bar{\lambda}\left(\Delta {\cal N}-\overline{\Delta {\cal N}}\right)^{2}\right]=V_0\left[1+32\alpha\bar{\lambda}\left({\cal N}_{reh}-{\cal N}_{cmb}\right)^{2}\right].
\eea
Further substituting Eq~(\ref{uytwe1x}) in Eq~(\ref{retem}) the reheating temperature can be expressed in terms of the number of e-foldings and scale of inflation 
in the context of our proposed
model as:
\bea \label{retemc1}
 T_{reh}&\approx& \left(\frac{30}{\pi^{2}g_{\rm eff}(T_{reh})}\right)^{1/4}
V^{1/4}_{0}\left[1+32\alpha\bar{\lambda}\left(\Delta {\cal N}-\overline{\Delta {\cal N}}\right)^{2}\right]^{1/4}\nonumber\\
&=&\left(\frac{30}{\pi^{2}g_{\rm eff}(T_{reh})}\right)^{1/4}
\left(\frac{1}{8\alpha}+4\bar{\lambda}\left({\cal N}_{reh}-{\cal N}_{cmb}\right)^{2}\right)^{1/4}~M_p\nonumber\\
&=&\left[\frac{45\pi^{2}{\cal P}_{S}({\cal N}_{cmb})r({\cal N}_{cmb})}{g_{\rm eff}(T_{reh})}
\left\{1+2\left[\frac{r({\cal N}_{cmb})}{128{\cal N}^{3}_{cmb}}\right]^{1/2}\left({\cal N}_{reh}-{\cal N}_{cmb}\right)^{2}\right\}\right]^{1/4}M_p.~~~~~~~~~~
\eea
\section{\textcolor{blue}{Future probe:~Primordial Non-Gaussianity}}
\label{s7}
\subsection{Three point function}
\label{s7a}
\subsubsection{Using In-In formalism}
\label{s7a1}
Here we discuss about the constraint on the primordial three point scalar correlation function in the non attractor regime of soft inflation. In general one can write down the following expressions for the three point function of the scalar fluctuation as \cite{Baumann:2009ds,Maldacena:2002vr,Arkani-Hamed:2015bza,Choudhury:2015yna,Chen:2010xka,Bartolo:2004if,Chen:2013aj,Chen:2009zp,Creminelli:2003iq,Babich:2004gb,Creminelli:2004yq,Shukla:2016bnu,Kundu:2014gxa,Mata:2012bx}:
\bea \langle \zeta({\bf k}_1)\zeta({\bf k}_2)\zeta({\bf k}_3)\rangle&=&(2\pi)^3\delta^{(3)}({\bf k}_1+{\bf k}_2+{\bf k}_3)B(k_1,k_2,k_3).\eea
In our computation we choose Bunch-davies vacuum state and for single field soft slow-roll inflation we get the following expression for the bispectrum: 
\bea B(k_1,k_2,k_3)
&\approx&\frac{\tilde{W}^2(\phi_{cmb},{\bf \Psi})}{288(\epsilon^*_{\tilde{W}})^2M^6_p(k_1k_2k_3)^3}\left[2(3\epsilon^*_{\tilde{W}}-\eta^{*}_{\tilde{W}})\sum^{3}_{i=1}k^3_i 
+\epsilon^*_{\tilde{W}}\left(-\sum^{3}_{i=1}k^3_i+\sum^{3}_{i,j=1,i\neq j}k_i k^2_j+\frac{8}{K}\sum^{3}_{i,j=1,i> j}k^2_i k^2_j\right)\right],~~~~~~\eea
where $ K=k_1+k_2+k_3=\sum^{3}_{i=1}k_i$,and the potential flow-functions in Einstein frame can be expressed in terms of number of e-foldings ${\cal N}_{cmb}$ as:
\bea
    \tilde{\epsilon}_{\tilde{W}}&=&\left[\frac{M^{2}_{p}}{2}\left(\partial_{\phi} \ln\tilde{W}(\phi,{\bf\Psi})\right)^{2}\right]_{\phi=\phi_{cmb}}\,=\frac{32\alpha^{2}\bar{\lambda}^{2}\phi^{6}_{cmb}}{M^{6}_{p}\left[1+\frac{2\alpha\bar{\lambda}}{M^{4}_{p}}\phi^{4}_{cmb}\right]^{2}}=\frac{2048\alpha^{2}\bar{\lambda}^{2}{\cal N}^3_{cmb}}{\left[1+32\alpha\bar{\lambda}{\cal N}^2_{cmb}\right]^{2}},\\
    \tilde{\eta}_{\tilde{W}}&=&\left[{M^{2}_{p}}\frac{\partial_{\phi\phi}\tilde{W}(\phi,{\bf\Psi})}{\tilde{W}(\phi,{\bf\Psi})}\right]_{\phi=\phi_{cmb}}\,=\frac{24\alpha\bar{\lambda}\phi^{2}_{cmb}}{M^{2}_{p}\left[1+\frac{2\alpha\bar{\lambda}}{M^{4}_{p}}\phi^{4}_{cmb}\right]}=\frac{96\alpha\bar{\lambda}{\cal N}_{cmb}}{\left[1+32\alpha\bar{\lambda}{\cal N}^2_{cmb}\right]}.\eea
In the present context one can parameterize non-Gaussianity phenomenologically via a non-linear correction to a Gaussian perturbation $\zeta_{g}$ in position space as \cite{Baumann:2009ds}:
		   	   \bea \zeta({\bf x})&=&\zeta_{g}({\bf x})+\frac{3}{5}f^{loc}_{NL}\left[\zeta^2_{g}({\bf x})-\langle\zeta^2_{g}({\bf x})\rangle\right]+\cdots,\eea
		   	   where $\cdots$ represent higher order non-Gaussian contributions. This definition is local in real space
		   	   and therefore called local non-Gaussianity. In case local non-Gaussianity amplitude of the bispectrum from the three point function is defined as \cite{Baumann:2009ds}:
		   	   \bea f^{loc}_{NL}(k_1,k_2,k_3)&=&\frac{5}{6}\frac{B(k_1,k_2,k_3)}{\left[P_{\zeta}(k_1)P_{\zeta}(k_2)+P_{\zeta}(k_2)P_{\zeta}(k_3)+P_{\zeta}(k_1)P_{\zeta}(k_3)\right]}.\eea
		   	   Further substituting the expression for bispectrum and power spectrum the non-Gaussianity amplitude can be expressed as:
\bea f^{loc}_{NL}(k_1,k_2,k_3)
&\approx&\frac{5}{12}\frac{1}{\sum^{3}_{i=1}k^3_i} \left[2(3\epsilon^*_{\tilde{W}}-\eta^{*}_{\tilde{W}})\sum^{3}_{i=1}k^3_i 
+\epsilon^*_{\tilde{W}}\left(-\sum^{3}_{i=1}k^3_i+\sum^{3}_{i,j=1,i\neq j}k_i k^2_j+\frac{8}{K}\sum^{3}_{i,j=1,i> j}k^2_i k^2_j\right)\right].~~~~~~\eea
     \begin{figure*}[htb]
     \centering
     \subfigure[ $S$ channel diagram.]{
         \includegraphics[width=5.0cm,height=4.5cm] {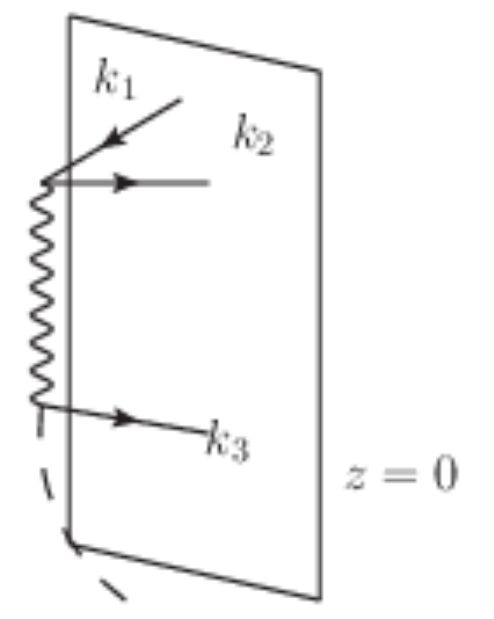}
         \label{figv1bbb}
     }
     \subfigure[$T$ channel diagram.]{
         \includegraphics[width=5.0cm,height=4.5cm] {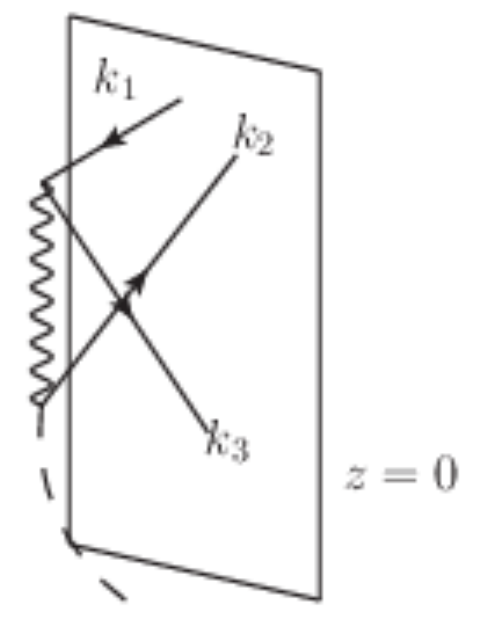}
         \label{figv2bbb}
     }
     \subfigure[$U$ channel diagram.]{
              \includegraphics[width=5.0cm,height=4.5cm] {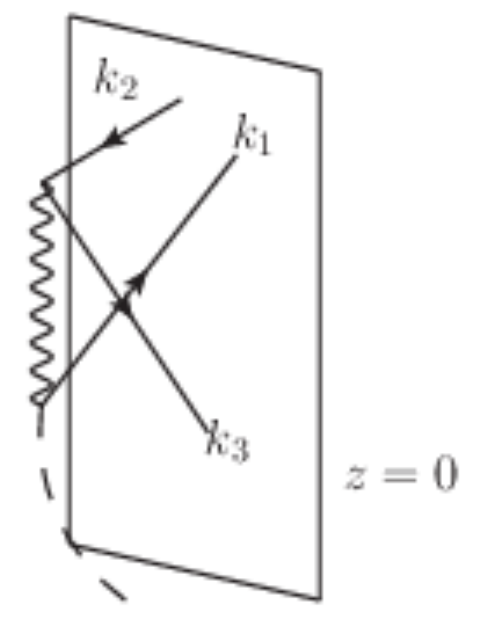}
              \label{figv3bbb}
          }
     \caption[Optional caption for list of figures]{Representative $S,T$ and $U$ channel Feynman-Witten diagram for bulk interpretation of three point scalar correlation function in presence of graviton exchange contribution. In all the diagrams graviton is propagating on the bulk and the end point of scalars are attached with the boundary at $z=0$. More precisely the wavy line denotes
          the bulk-to-bulk graviton propagator, the solid lines represent the bulk-to-boundary propagators
          for the scalar field and the dashed line denotes background represented by, $\partial_{z}{\bar{\phi}}$. } 
     \label{bulk1}
     \end{figure*}
     To give the bulk interpretation of the obtained results for scalar three point correlation function here we start with the graviton propagator which can be computed from the secornd order fluctuation in $\delta g_{\mu\nu}$ for the canonical scalar field action minimally coupled with Einstein gravity. In this context we choose a guage $\delta g_{zz}=0=\delta g_{zi}$ which is equivalent to choosing $N^i =0, N=1$ in ADM formalism. After choosing this gauge we get:
     \bea G_{ij;kl}(z_1,{\bf y}_{1};z_2,{\bf y}_{2})&=& \int\frac{d^3{\bf k}}{(2\pi)^3}e^{i{\bf k}.({\bf y}_1-{\bf y}_2)}\int^{\infty}_{0}qdq\frac{J_{\frac{3}{2}}(qz_1)J_{\frac{3}{2}}(qz_2)}{2\sqrt{z_1z_2}}\Delta_{ijkl},\eea
     where $\Delta_{ijkl}$ is defined as,
     $\Delta_{ijkl}=P_{ik}P_{jl}+P_{il}P_{jk}-P_{ij}P_{kl}$.
     	Here $P_{ij}$ is the projection operator in momentum space, which is defined as,
     	$P_{ij}=\left(\delta_{ij}+k_i k_j/q^2\right)$.
     	Further one can also write down the expression for the transverse part of the graviton propagator from this compution:
     	\bea \bar{G}_{ij;kl}(z_1,{\bf y}_{1};z_2,{\bf y}_{2})&=& \int\frac{d^3{\bf k}}{(2\pi)^3}e^{i{\bf k}.({\bf y}_1-{\bf y}_2)}\int^{\infty}_{0}qdq\frac{J_{\frac{3}{2}}(qz_1)J_{\frac{3}{2}}(qz_2)}{2\sqrt{z_1z_2}}\bar{\Delta}_{ijkl},\eea
     	where $\bar{\Delta}_{ijkl}$ is defined as,
     	     $\bar{\Delta}_{ijkl}=\bar{P}_{ik}\bar{P}_{jl}+\bar{P}_{il}\bar{P}_{jk}-\bar{P}_{ij}\bar{P}_{kl}$.
     	     	Here $\bar{P}_{ij}$ is the transverse part of the projection operator in momentum space, which is defined as, 
     	     	$\bar{P}_{ij}=\left(\delta_{ij}-k_i k_j/q^2\right)$.
     	     	Similarly the longitudinal part of the graviton propagator can be expressed as:
     	    \bea \hat{G}_{ij;kl}(z_1,{\bf y}_{1};z_2,{\bf y}_{2})&=&G_{ij;kl}(z_1,{\bf y}_{1};z_2,{\bf y}_{2})-\bar{G}_{ij;kl}(z_1,{\bf y}_{1};z_2,{\bf y}_{2}).\eea 
     	    Consequently in the bulk the onshell action can be written as a sum of transverse and longitudinal contribution as:
     	    \bea S^{\rm Bulk}_{\rm on-shell}&=&-\frac{3M^2_p}{2\Lambda}(Z_{tr}+Z_{long}),\eea	
     	    where $\Lambda$ is the cosmological constant. In this context the transverse contribution $Z_{tr}$ and longitudinal contribution $Z_{long}$ are defined as:
     	    \bea Z_{tr}&=&\int dz_{1}dz_{2}d^{3}{\bf y}_1 d^{3}{\bf y}_2 T_{mn}(z_1,{\bf y}_1)\bar{G}_{mn;kl}(z_1,{\bf y}_{1};z_2,{\bf y}_{2})T_{kl}(z_1,{\bf y}_1),\\
     	    Z_{long}&=&-\int \frac{dz}{z^2}\int d^3{\bf y}\left[T_{zk}(z,{\bf y})\partial^{-2}T_{zk}(z,{\bf y})+\frac{z}{2}\partial_{k}T_{zk}(z,{\bf y})\partial^{-2}T_{zz}(z,{\bf y}) 
     	    +\frac{1}{4}\partial_{k}T_{zk}(z,{\bf y})(\partial^{-2})^2\partial_{m}T_{zm}(z,{\bf y})\right].~~~~~~~~\eea
     	    Finally one can write down the follwing simplified expression:
     	    \bea S^{\rm Bulk}_{\rm on-shell}&=&-\frac{3M^2_p}{2\Lambda}\sqrt{2\epsilon}(2\pi)^3\delta^{(3)}({\bf k}_1+{\bf k}_2+{\bf k}_3)\prod^{3}_{n=1}\phi_{0}(k_{n}) 
     	    \left[-\sum^{3}_{i=1}k^3_i+\sum^{3}_{i,j=1,i\neq j}k_i k^2_j+\frac{8}{K}\sum^{3}_{i,j=1,i> j}k^2_i k^2_j\right].~~~~~~~~\eea
     	    further taking the derivatives with respect to the background field value $\phi_0$ and choosing the following gauge $\zeta=-H\delta\phi/\dot{\phi}$,
     	    one can write down the following expression for the scalr three point function:
     	    \bea \langle \zeta({\bf k}_1)\zeta({\bf k}_2)\zeta({\bf k}_3)\rangle&=&(2\pi)^3\delta^{(3)}({\bf k}_1+{\bf k}_2+{\bf k}_3)\frac{H^4}{32(\epsilon^*_{H})^2M^4_p}\frac{1}{(k_1k_2k_3)^3}\left[2(2\epsilon^*_{H}-\eta^*_{H})\sum^{3}_{i=1}k^3_i\right.\nonumber\\ &&\left.+\epsilon^*_{H}\left(-\sum^{3}_{i=1}k^3_i+\sum^{3}_{i,j=1,i\neq j}k_i k^2_j+\frac{8}{K}\sum^{3}_{i,j=1,i> j}k^2_i k^2_j\right)\right],\eea
     which can be expressed interms of the effective poteials using Friedman equations and using the relation between Hubble and potentail dependent slow-roll parameters.	    
      The representative $S$, $T$ and $U$ channel Feynman Witten diagrams for bulk interpretation of the three point scalar correlation function in presence of graviton exchange is shown in shown in fig.~(\ref{figv1bbb}), fig.~(\ref{figv2bbb}) and fig.~(\ref{figv3bbb}). In these diagrams we have explicitly shown that, graviton is propagating on the bulk and the end point of scalars are attached with the boundary at $z=0$. Additionally it is important to note that, the dashed line represents background denoted by, $\partial_{z}{\bar{\phi}}$ in all of the representative diagrams. Here $\bar{\phi}$ is the background field value. More precisely the wavy line denotes
          the bulk-to-bulk graviton propagator, the solid lines represent the bulk-to-boundary propagators
          for the scalar field. In our computation all the representative diagrams are important to explain the total three point scalar correlation function.

To analyze the shape of the bispectrum here we further consider two limiting configurations- \textcolor{blue}{equilateral limit} and \textcolor{blue}{squeezed limit}. In these limits, the final simplified results are appended below:
     \begin{figure*}[htb]
     \centering
     \subfigure[Range~I.]{
         \includegraphics[width=7.6cm,height=4.5cm] {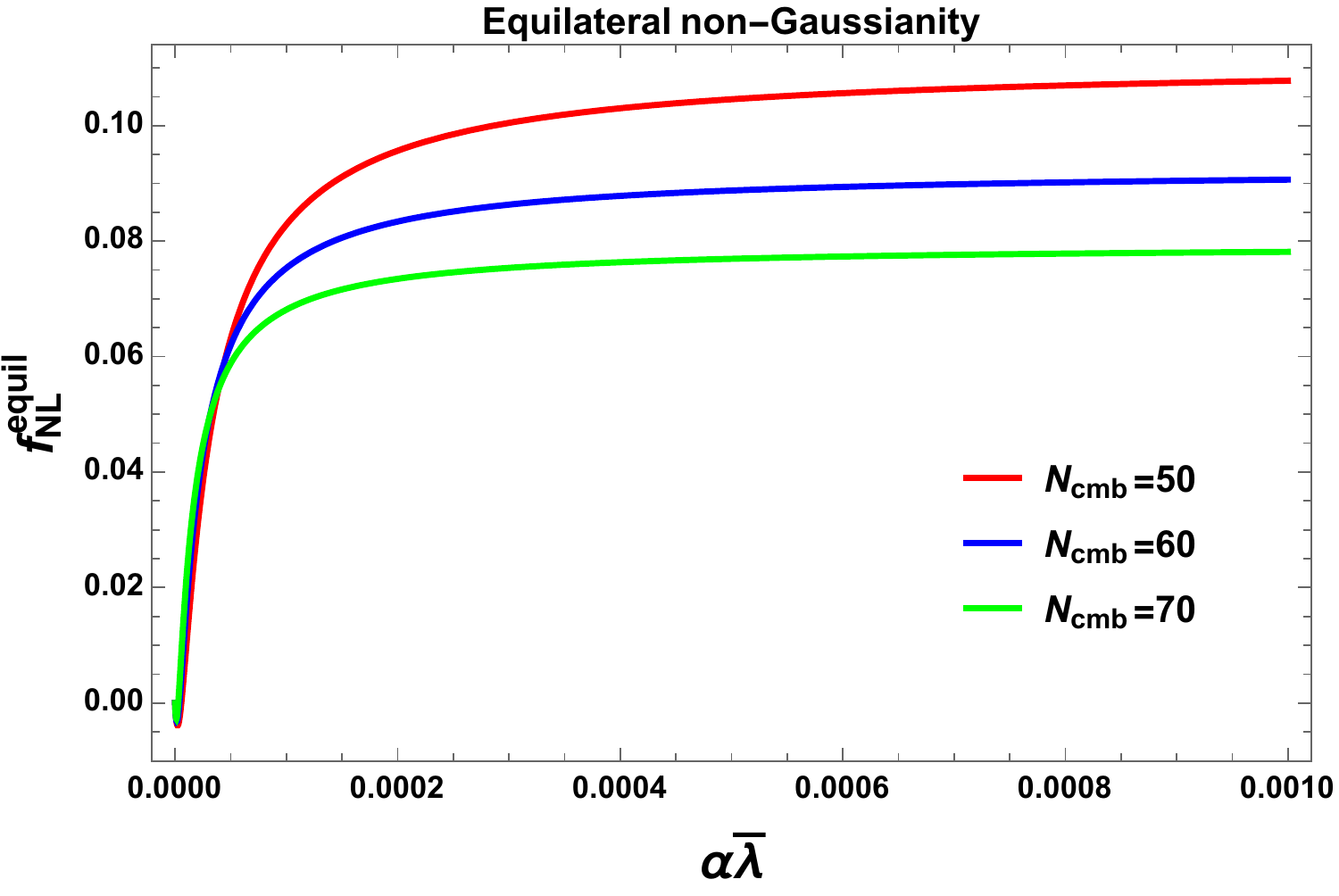}
         \label{fige1bb}
     }
     \subfigure[Range~II.]{
         \includegraphics[width=7.6cm,height=4.5cm] {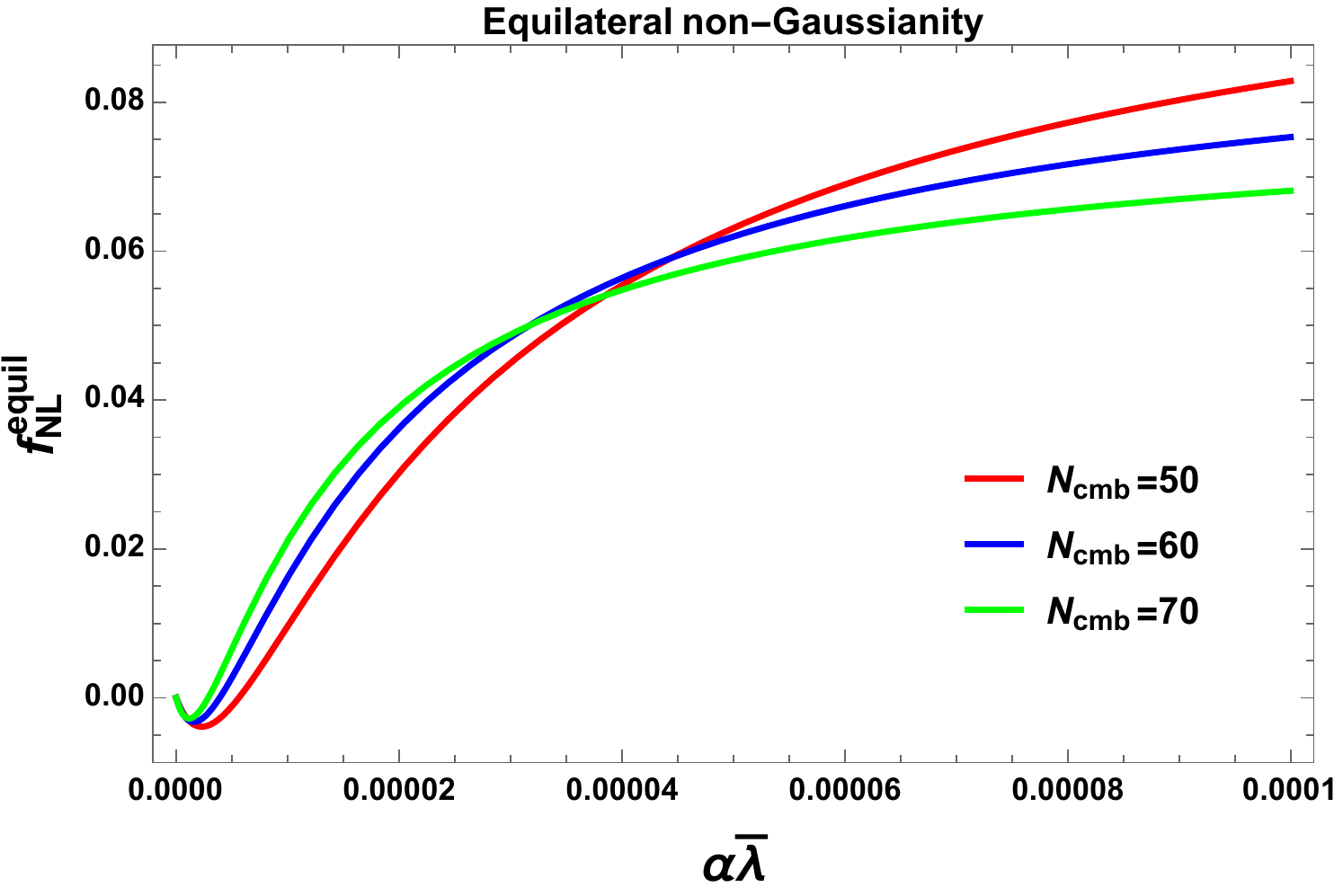}
         \label{fige2bb}
     }
     \subfigure[Rangle~III.]{
              \includegraphics[width=7.6cm,height=4.5cm] {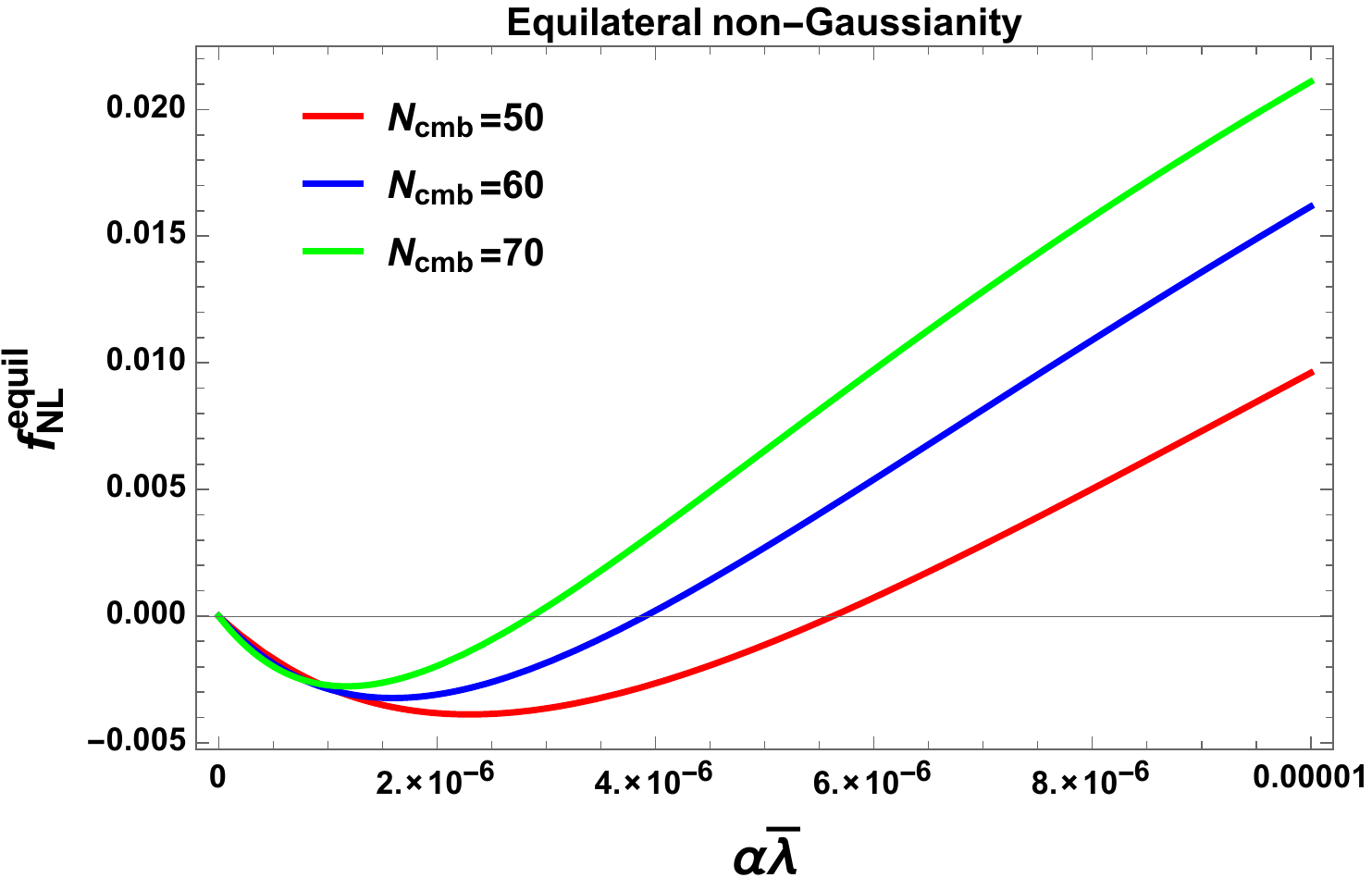}
              \label{fige3bb}
          }
          \subfigure[Range~IV.]{
                        \includegraphics[width=7.6cm,height=4.5cm] {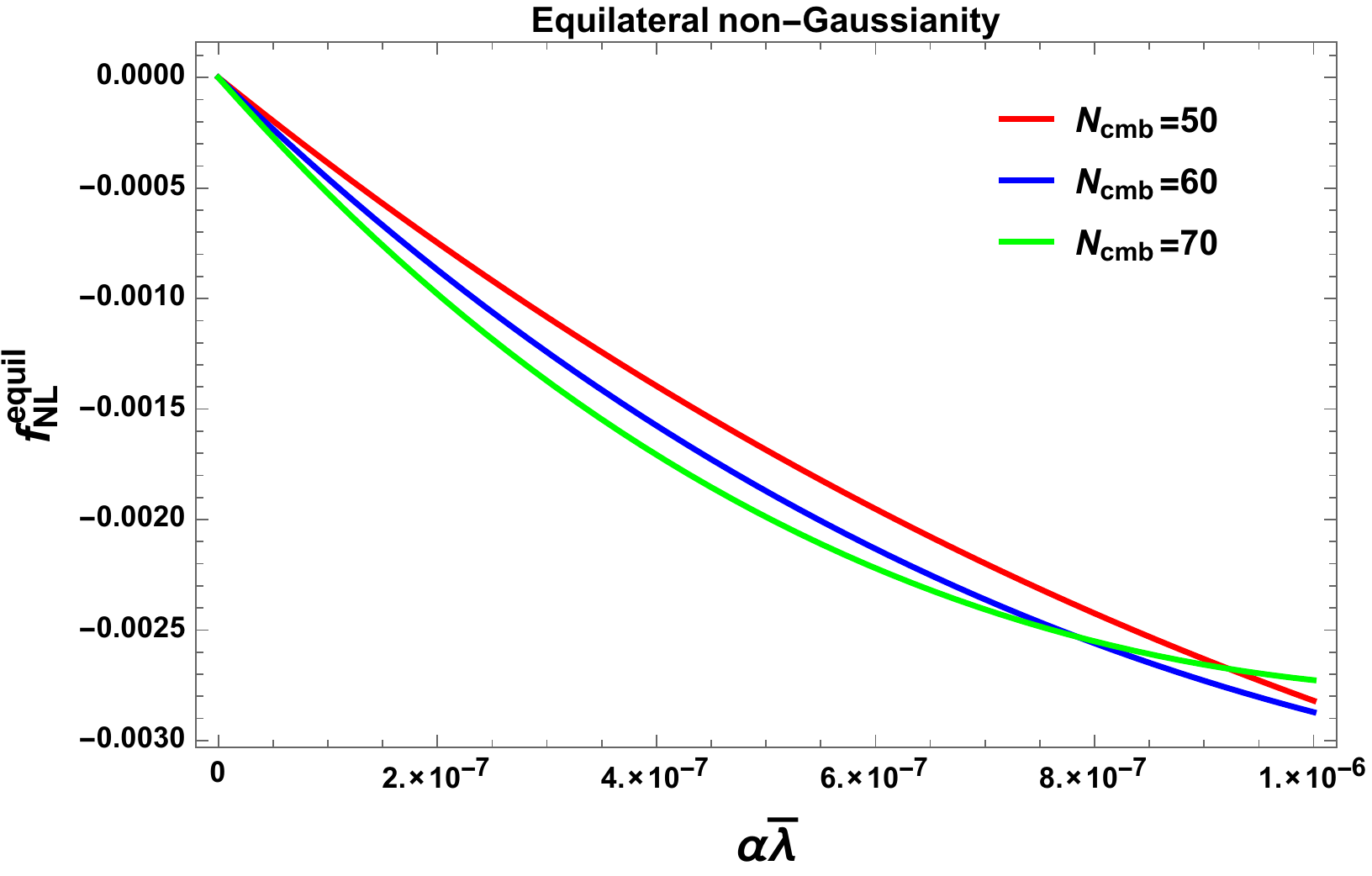}
                        \label{fige4bb}
                    }
     \caption[Optional caption for list of figures]{Representative diagram for equilateral non-Gaussian three point amplitude vs product of the parameters $\alpha\bar{\lambda}$ in four different region for ${\cal N}_{cmb}=50$ (red), ${\cal N}_{cmb}=60$ (blue) and ${\cal N}_{cmb}=70$ (green).} 
     \label{fnleeq}
     \end{figure*}
          \begin{figure*}[htb]
          \centering
          \subfigure[Angle~I.]{
              \includegraphics[width=7.6cm,height=4.5cm] {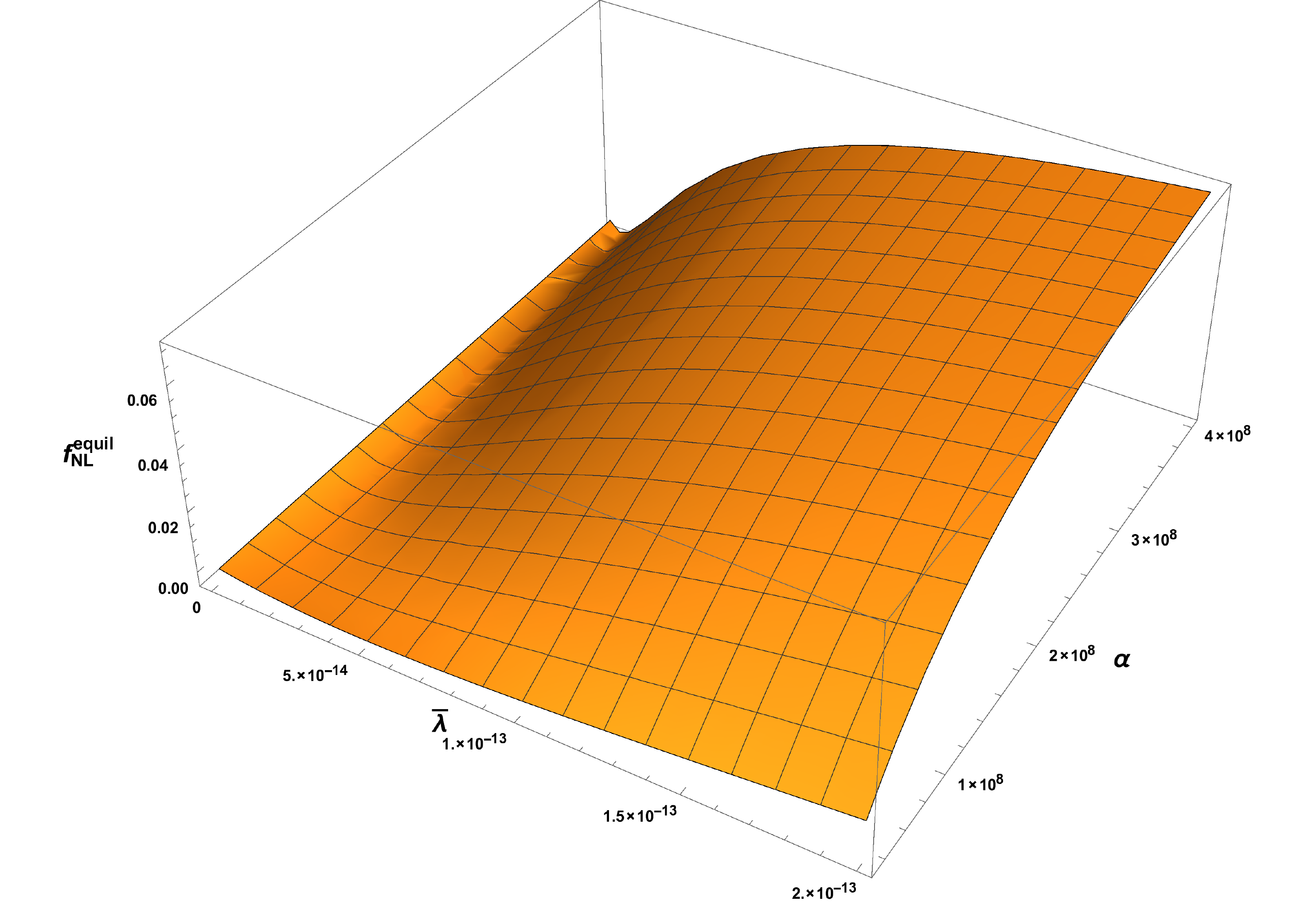}
              \label{fnl3d1}
          }
          \subfigure[Angle~II.]{
              \includegraphics[width=7.6cm,height=4.5cm] {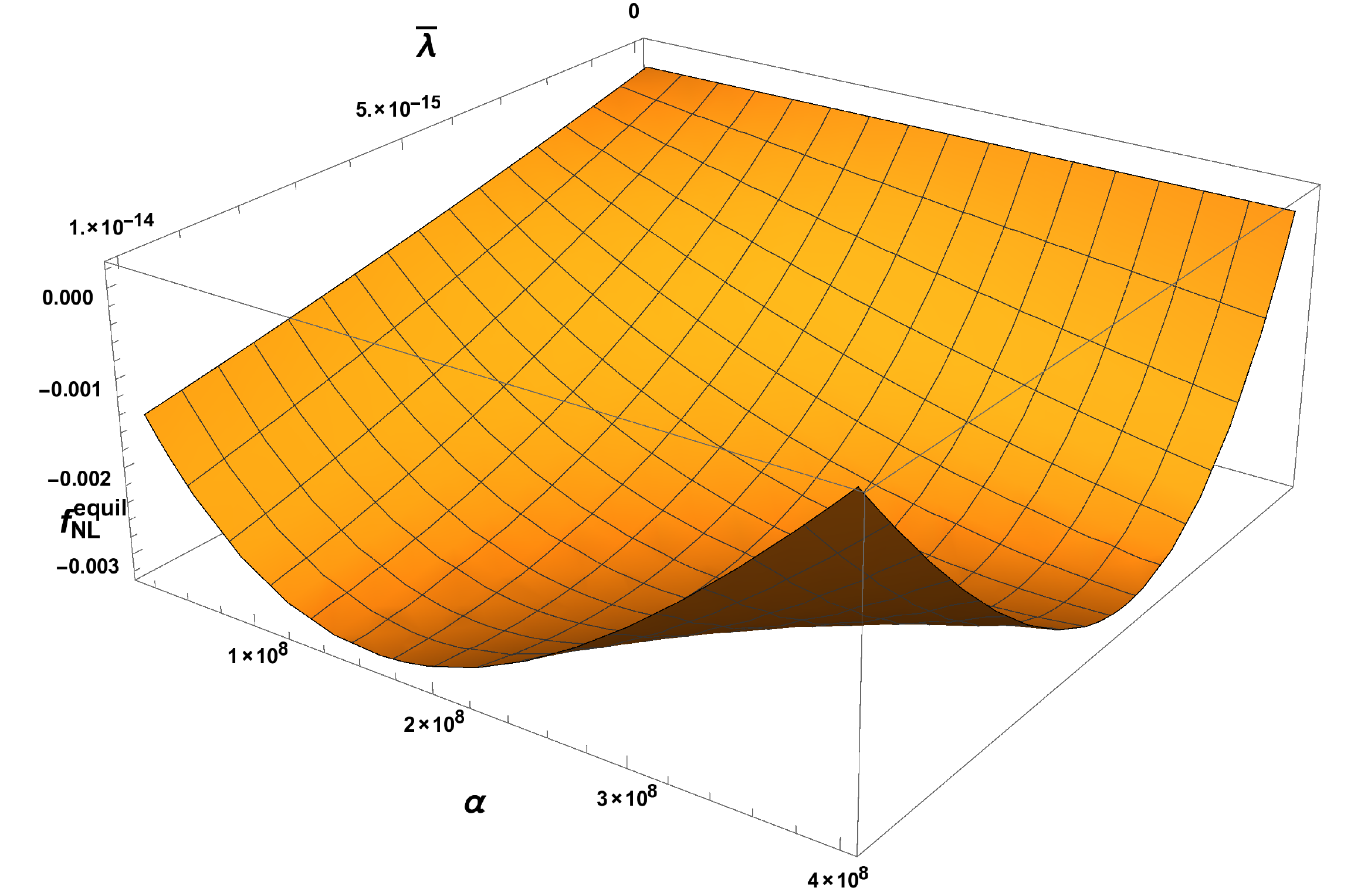}
              \label{fnl3d2}
          }
          \caption[Optional caption for list of figures]{Representative 3D diagram for equilateral non-Gaussian three point amplitude vs the model parameters $\alpha$ and $\bar{\lambda}$ for  ${\cal N}_{cmb}=60$ in two differenent angular views.} 
          \label{fnl3}
          \end{figure*}
               \begin{figure*}[htb]
               \centering
               \subfigure[Range~I.]{
                   \includegraphics[width=7.6cm,height=4.5cm] {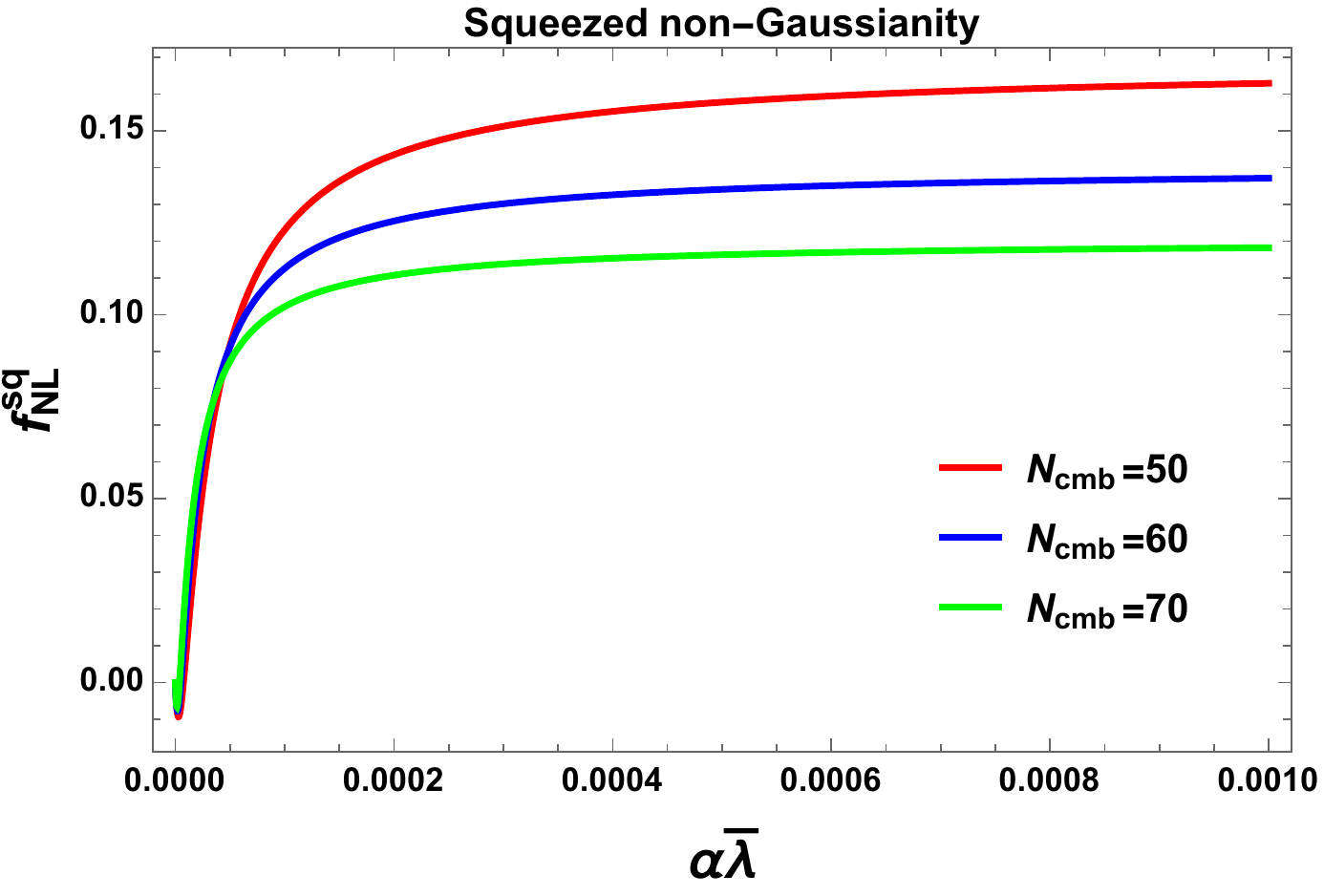}
                   \label{figs1bb}
               }
               \subfigure[Range~II.]{
                   \includegraphics[width=7.6cm,height=4.5cm] {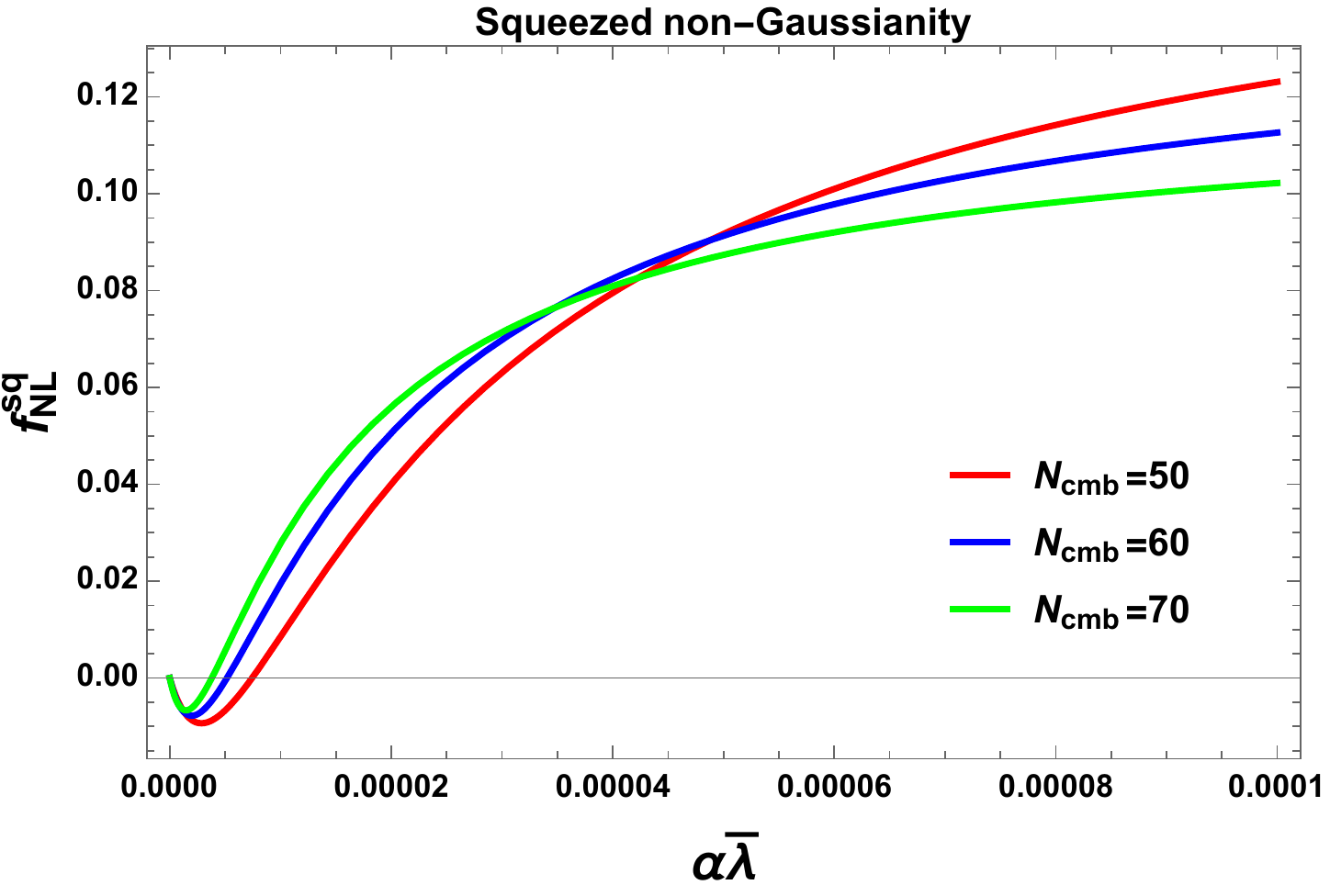}
                   \label{figs2bb}
               }
               \subfigure[Rangle~III.]{
                        \includegraphics[width=7.6cm,height=4.5cm] {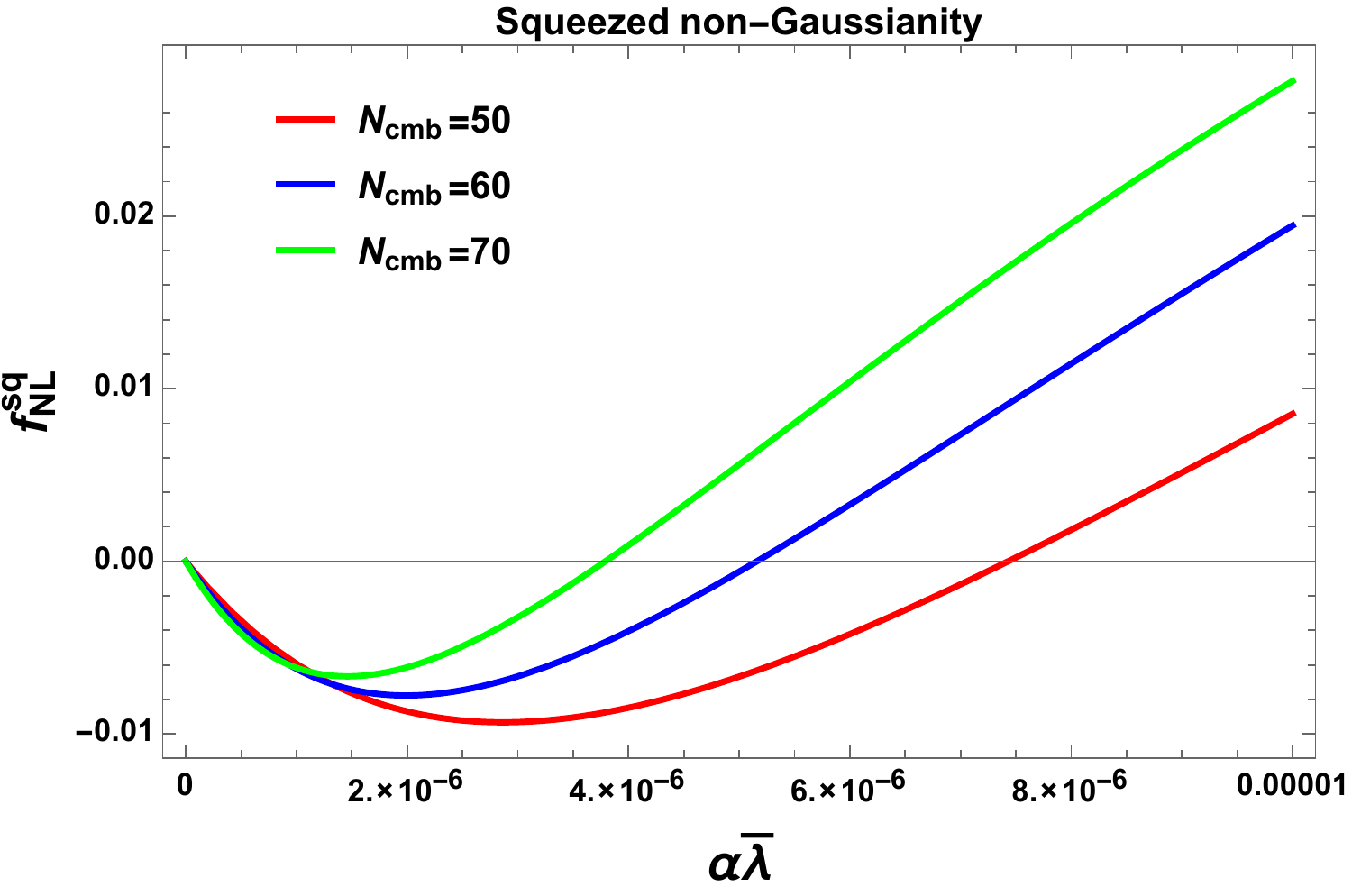}
                        \label{figs3bb}
                    }
                    \subfigure[Range~IV.]{
                                  \includegraphics[width=7.6cm,height=4.5cm] {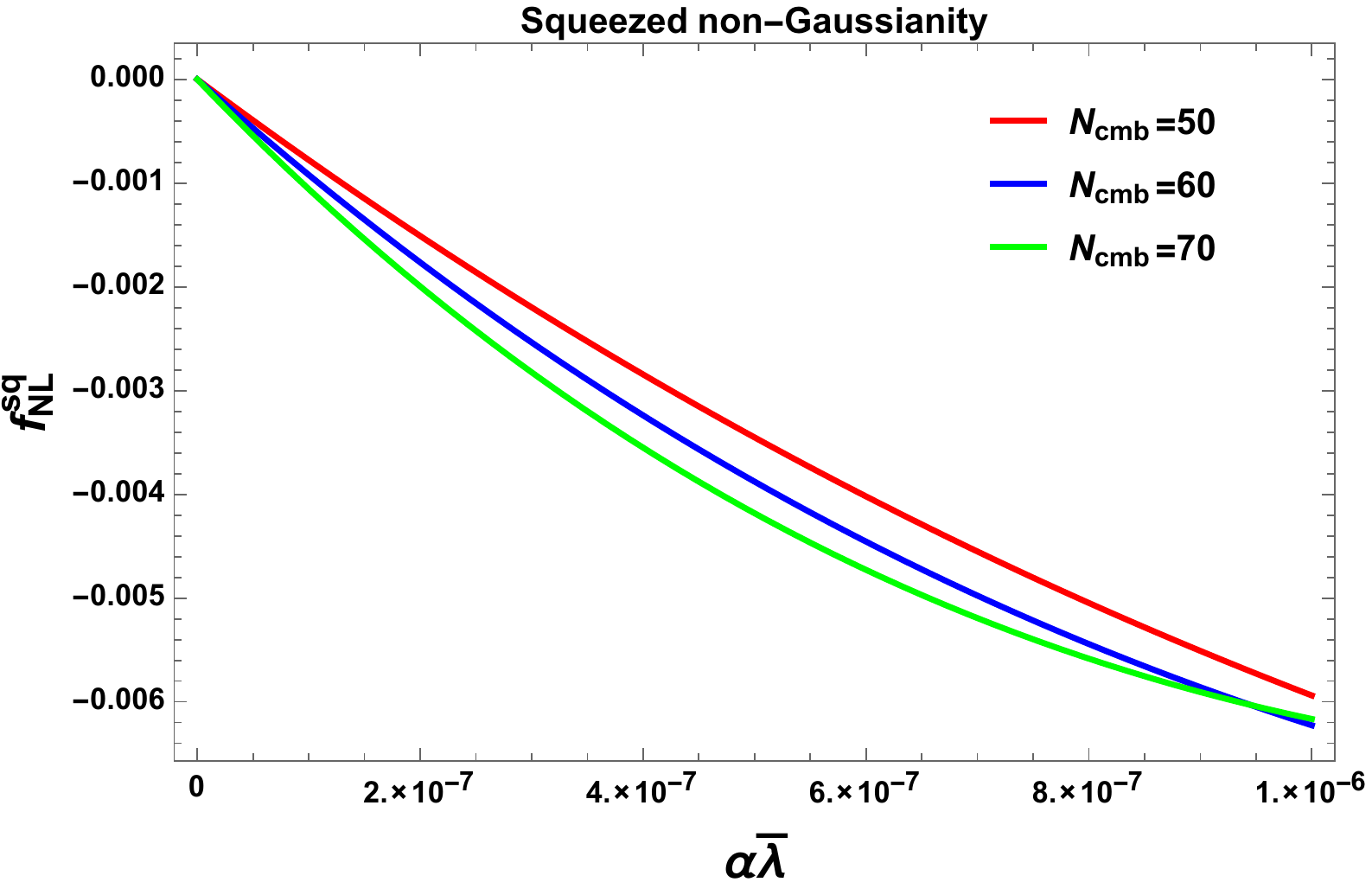}
                                  \label{figs4bb}
                              }
               \caption[Optional caption for list of figures]{Representative diagram for squeezed non-Gaussian three point amplitude vs product of the parameters $\alpha\bar{\lambda}$ in four different region for ${\cal N}_{cmb}=50$ (red), ${\cal N}_{cmb}=60$ (blue) and ${\cal N}_{cmb}=70$ (green).} 
               \label{fnls}
               \end{figure*}
                    \begin{figure*}[htb]
                    \centering
                    \subfigure[Angle~I.]{
                        \includegraphics[width=7.6cm,height=4.5cm] {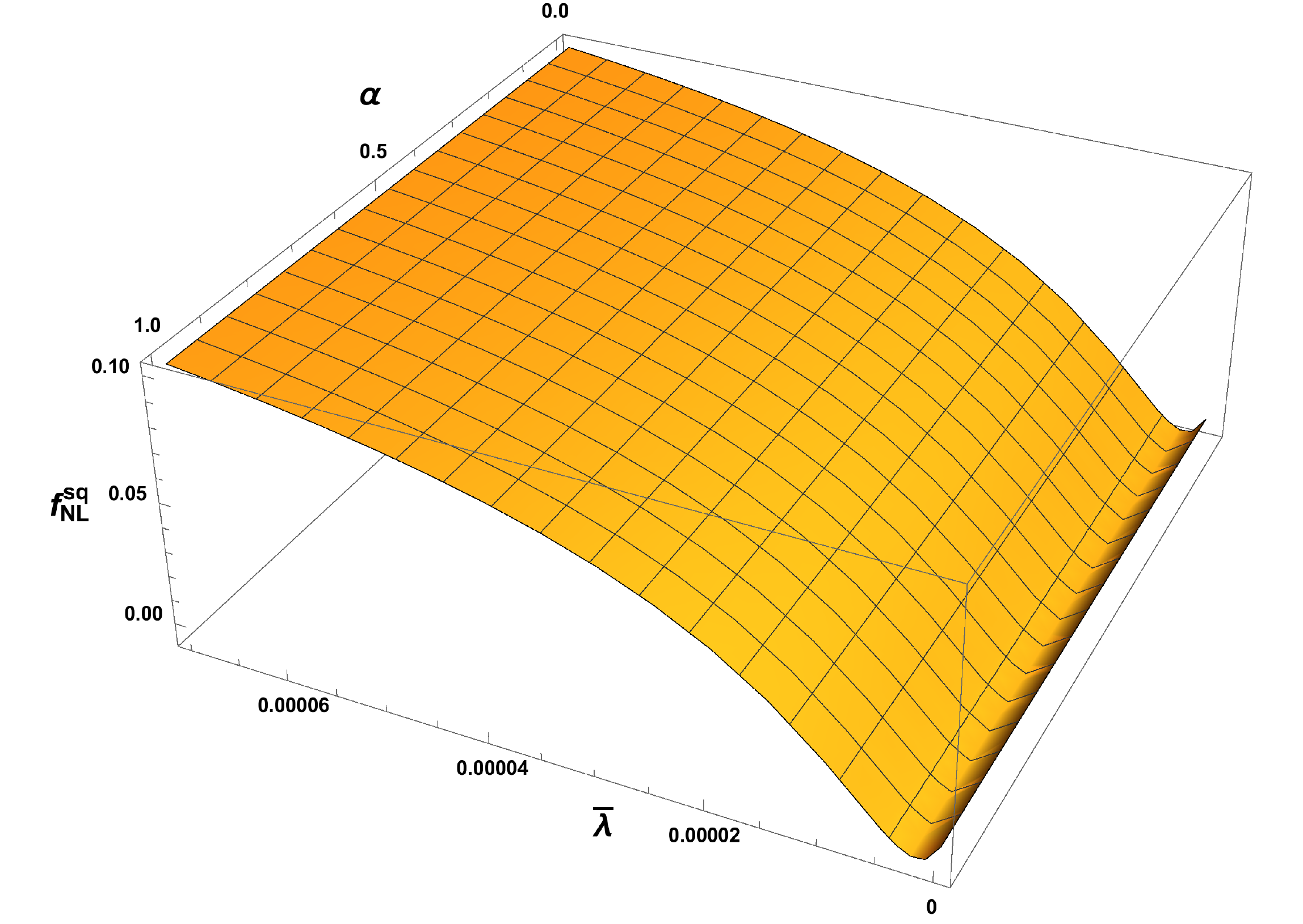}
                        \label{fnl3sd1}
                    }
                    \subfigure[Angle~II.]{
                        \includegraphics[width=7.6cm,height=4.5cm] {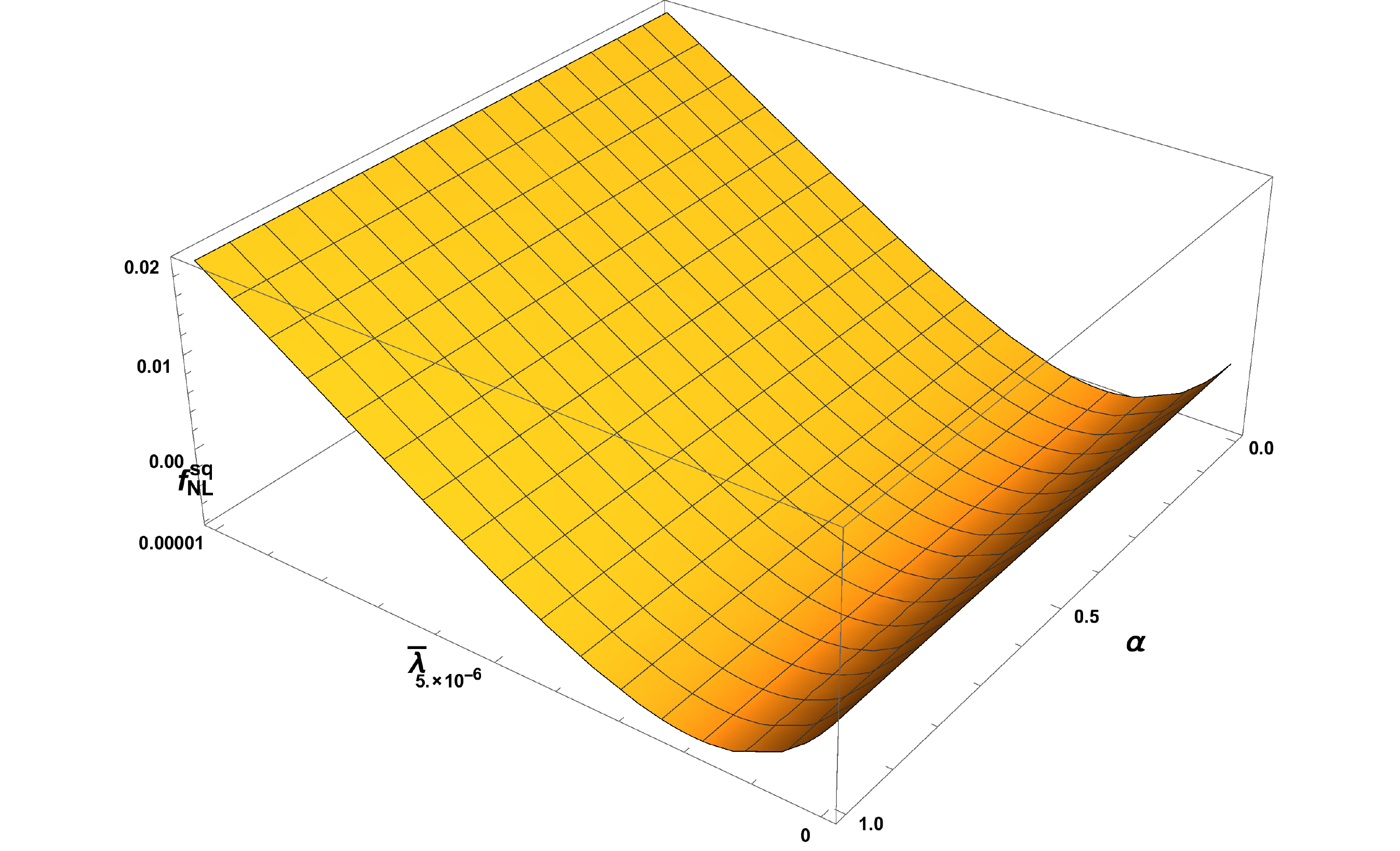}
                        \label{fnl3sd2}
                    }
                    \caption[Optional caption for list of figures]{Representative 3D diagram for squeezed non-Gaussian three point amplitude vs the model parameters $\alpha$ and $\bar{\lambda}$ for  ${\cal N}_{cmb}=60$ in two differenent angular views.} 
                    \label{fnl3s}
                    \end{figure*}
\begin{enumerate}
\item \underline{\textcolor{blue}{Equilateral limit configuration}:}\\
For this case we have $|{\bf k}_1|=|{\bf k}_2|=|{\bf k}_3|=k$ and the bispectrum for scalar fluctuation can be written as:
\bea B(k,k,k) 
\approx \frac{\tilde{W}^2(\phi_{cmb},{\bf \Psi})}{288(\epsilon^*_{\tilde{W}})^2M^6_p}\frac{1}{k^6}\left[29\epsilon^*_{\tilde{W}}-6\eta^*_{\tilde{W}} 
\right].~~~~~~~~\eea
In this case the non-Gaussian amplitude for bispectrum can be expressed as:
\bea f^{equil}_{NL}=f^{loc}_{NL}(k,k,k) 
\approx\frac{5}{36} \left[29\epsilon^*_{\tilde{W}}-6\eta^*_{\tilde{W}} 
\right].\eea	
\item \underline{\textcolor{blue}{Squeezed limit configuration}:}\\
For this case we have $k_{1}\approx k_{2}(=k_{L})>>k_{3}(=k_{S})$,
where $k_{i}=|{\bf k_{i}}|\forall i=1,2,3$. Here $k_{L}$ and $k_{S}$ represent momentum for long and short modes respectively. Consequently the bispectrum for scalar fluctuation for arbitrary vacuum can be expressed as:
\bea B(k_L,k_L,k_S)
&\approx&\frac{\tilde{W}^2(\phi_{cmb},{\bf \Psi})}{288(\epsilon^*_{\tilde{W}})^2M^6_p}\frac{1}{k^3_Lk^3_S}\left[4(4\epsilon^*_{\tilde{W}}-\eta^*_{\tilde{W}}) 
+10\epsilon^*_{\tilde{W}}\left(\frac{k_S}{k_L}\right)^2-(2\eta^*_{\tilde{W}}-\epsilon^*_{\tilde{W}})\left(\frac{k_S}{k_L}\right)^3\right].~~~~~~~~~~~~~ \eea
In this case the non-Gaussian amplitude for bispectrum can be expressed as:
\bea f^{sq}_{NL}=f^{loc}_{NL}(k_L,k_L,k_S)
&\approx&\frac{5}{12}\left[4(4\epsilon^*_{\tilde{W}}-\eta^*_{\tilde{W}}) 
+10\epsilon^*_{\tilde{W}}\left(\frac{k_S}{k_L}\right)^2-(2\eta^*_{\tilde{W}}-\epsilon^*_{\tilde{W}})\left(\frac{k_S}{k_L}\right)^3\right].~~~~~~~~~~~~~ \eea
\end{enumerate}
  \begin{table*}
      \centering
       \tiny
      \begin{tabular}{|c|c|c|c|}
     \hline
      \hline
       \textcolor{red}{Scanning Region} & \textcolor{red}{Bound on} $\alpha\bar{\lambda}$&$f^{equil}_{NL}$ &$f^{sq}_{NL}$\\
       & & & \\
      \hline\hline\hline
       \textcolor{red}{ I} & $0.0001<\alpha\bar{\lambda}<0.001$ &  $0.06<f^{equil}_{NL}<0.11$ & $0.09<f^{sq}_{NL}<0.16$
           \\
      \hline\hline
       \textcolor{red}{ II} & $0.00001<\alpha\bar{\lambda}<0.0001$ & $0.01<f^{equil}_{NL}<0.08$ & $0.01<f^{sq}_{NL}<0.12$
           \\
      \hline\hline
      \textcolor{red}{ III} & $0.000001<\alpha\bar{\lambda}<0.00001$ & $-0.004<f^{equil}_{NL}<0.02$ & $-0.01<f^{sq}_{NL}<0.03$
           \\ 
      \hline\hline
      \textcolor{red}{ IV} & $0.0000001<\alpha\bar{\lambda}<0.000001$ & $-0.0001<f^{equil}_{NL}<-0.0029$ & $-0.0005<f^{sq}_{NL}<-0.006$
           \\ 
      \hline\hline
           \textcolor{red}{ I+II+III+IV} & $0.0000001<\alpha\bar{\lambda}<0.001$ & $-0.0001<f^{equil}_{NL}<0.11$ & $-0.0005<f^{sq}_{NL}<0.16$
                     \\ \hline\hline
      \hline
      \end{tabular}
      \caption{Contraint on scalar three point non-Gaussian amplitude from equilateral and squeezed configuration.}\label{tab11}
      \vspace{.7cm}
      \end{table*}
      In table.~(\ref{tab11}), we give the numerical estimates and constraints on the three point non-Gaussian amplitude from equilateral and squeezed configuration. Here all the obtained results are consistent with the two point constraints as well as with the Planck 2015 data.
      
     In fig.~(\ref{fnleeq}), we have shown the features of non-Gaussian amplitude from three point function in equilateral limit configuration in four different scanning region of product of the two parameters $\alpha\bar{\lambda}$ in the $(f^{equil}_{NL},\alpha\bar{\lambda})$ 2D plane for the number of e-foldings $50<{\cal N}_{cmb}<70$. Physical explanation of the obtained features are appended following:-
     \begin{itemize}
     \item \textcolor{red}{\underline{Region~I}:} \\
     Here for the parameter space $0.0001<\alpha\bar{\lambda}<0.001$ the non-Gaussian amplitude lying within the window $0.06<f^{equil}_{NL}<0.11.$ Further if we increase the numerical value of $\alpha\bar{\lambda}$, then the magnitude of the non-Gaussian amplitude saturates and we get maximum value for ${\cal N}_{cmb}=50$, 
     $|f^{equil}_{NL}|_{max}\sim 0.11$.
     \item \textcolor{red}{\underline{Region~II}}\\
          Here for the parameter space $0.00001<\alpha\bar{\lambda}<0.0001$ the non-Gaussian amplitude lying within the window $0.01<f^{equil}_{NL}<0.08$. In this region we get maximum value for ${\cal N}_{cmb}=50$,
          $|f^{equil}_{NL}|_{max}\sim 0.08$.
          Additionally it is important to note that, in this case for $\alpha\bar{\lambda}=0.00004$ the lines obtained for ${\cal N}_{cmb}=50$, ${\cal N}_{cmb}=60$ and ${\cal N}_{cmb}=70$ cross each other.
     \item \textcolor{red}{\underline{Region~III}}\\
               Here for the parameter space $0.000001<\alpha\bar{\lambda}<0.00001$ the non-Gaussian amplitude lying within the window $ -0.004<f^{equil}_{NL}<0.02$. In this region we get maximum value for ${\cal N}_{cmb}=70$,
               $|f^{equil}_{NL}|_{max}\sim 0.02$.
               Additionally it is important to note that, in this case for $0.000003\leq \alpha\bar{\lambda}\leq 0.000006$ the lines obtained for ${\cal N}_{cmb}=50$, ${\cal N}_{cmb}=60$ and ${\cal N}_{cmb}=70$ cross the zero line of non-Gaussian amplitude and transition takes place from negative to positive values of $f^{equil}_{NL}$.
     \item \textcolor{red}{\underline{Region~IV}}\\
                    Here for the parameter space $0.0000001<\alpha\bar{\lambda}<0.000001$ the non-Gaussian amplitude lying within the window $ -0.0001<f^{equil}_{NL}<-0.0029$. In this region we get maximum value for ${\cal N}_{cmb}=60$,
                    $|f^{equil}_{NL}|_{max}\sim 0.0029$.
     \end{itemize} 
     Further combining the contribution from \textcolor{red}{\underline{Region~I}}, \textcolor{red}{\underline{Region~II}}, \textcolor{red}{\underline{Region~III}} and   \textcolor{red}{\underline{Region~IV}} we finally get the following constraint on the three point non-Gaussian amplitude in the equilateral limit configuration:
     
     \bea \textcolor{red}{\underline{\rm Region~I}}+\textcolor{red}{\underline{\rm Region~II}}+\textcolor{red}{\underline{\rm Region~III}}+\textcolor{red}{\underline{\rm Region~IV}:}~~-0.0001<f^{equil}_{NL}<0.11
     \eea
     for the following parameter space:
     \bea \textcolor{red}{\underline{\rm Region~I}}+\textcolor{red}{\underline{\rm Region~II}}+\textcolor{red}{\underline{\rm Region~III}}+\textcolor{red}{\underline{\rm Region~IV}:}~~~~~0.0000001<\alpha\bar{\lambda}<0.001.~~~~~~~
          \eea
          In this analysis we get the following maximum value of the three point non-Gaussian amplitude in the equilateral limit configuration as given by:
     \be |f^{equil}_{NL}|_{max}\sim 0.11.\ee
     To visualize these constraints more clearly we have also presented $(f^{equil}_{NL},\alpha,\bar{\lambda})$ 3D plot in fig.~(\ref{fnl3d1}) and fig.~(\ref{fnl3d2}), for two different angular orientations as given by \textcolor{red}{\underline{Angle~I}} and \textcolor{red}{\underline{Angle~II}}. From the the representative surfaces it is clearly observed the behavior of three point non-Gaussian amplitude in the equilateral limit for the variation of two fold parameter $\alpha$ and $\bar{\lambda}$ and the results are consistent with the obtained constraints in 2D analysis. Here all the obtained results are consistent with the two point constraints and the Planck 2015 data.

     In fig.~(\ref{fnls}), we have shown the features of non-Gaussian amplitude from three point function in squeezed limit configuration in four different scanning region of product of the two parameters $\alpha\bar{\lambda}$ in the $(f^{sq}_{NL},\alpha\bar{\lambda})$ 2D plane for the number of e-foldings $50<{\cal N}_{cmb}<70$. Physical explanation of the obtained features are appended following:-
          \begin{itemize}
          \item \textcolor{red}{\underline{Region~I}:} \\
          Here for the parameter space $0.0001<\alpha\bar{\lambda}<0.001$ the non-Gaussian amplitude lying within the window $0.09<f^{sq}_{NL}<0.16$. Further if we increase the numerical value of $\alpha\bar{\lambda}$, then the magnitude of the non-Gaussian amplitude saturates and we get maximum value for ${\cal N}_{cmb}=50$, 
          $|f^{sq}_{NL}|_{max}\sim 0.16$.
          \item \textcolor{red}{\underline{Region~II}}\\
               Here for the parameter space $ 0.00001<\alpha\bar{\lambda}<0.0001$ the non-Gausiian amplitude lying within the window $ 0.01<f^{sq}_{NL}<0.12$. In this region we get maximum value for ${\cal N}_{cmb}=50$,
               $|f^{sq}_{NL}|_{max}\sim 0.12$.
               Additionally it is important to note that, in this case for $\alpha\bar{\lambda}=0.00004$ the lines obtained for ${\cal N}_{cmb}=50$, ${\cal N}_{cmb}=60$ and ${\cal N}_{cmb}=70$ cross each other.
          \item \textcolor{red}{\underline{Region~III}}\\
                    Here for the parameter space $0.000001<\alpha\bar{\lambda}<0.00001$ the non-Gaussian amplitude lying within the window $ -0.01<f^{sq}_{NL}<0.03$. In this region we get maximum value for ${\cal N}_{cmb}=70$,
                    $|f^{sq}_{NL}|_{max}\sim 0.03$.
                    Additionally it is important to note that, in this case for $ 0.000003\leq \alpha\bar{\lambda}\leq 0.000006$ the lines obtained for ${\cal N}_{cmb}=50$, ${\cal N}_{cmb}=60$ and ${\cal N}_{cmb}=70$ cross the zero line of non-Gaussian amplitude and transition takes place from negative to positive values of $f^{sq}_{NL}$.
          \item \textcolor{red}{\underline{Region~IV}}\\
                         Here for the parameter space $ 0.0000001<\alpha\bar{\lambda}<0.000001$ the non-Gaussian amplitude lying within the window $ -0.0005<f^{sq}_{NL}<-0.006.$ In this region we get maximum value for ${\cal N}_{cmb}=60$,
                         $|f^{sq}_{NL}|_{max}\sim 0.006$.
          \end{itemize} 
          Further combining the contribution from \textcolor{red}{\underline{Region~I}}, \textcolor{red}{\underline{Region~II}}, \textcolor{red}{\underline{Region~III}} and   \textcolor{red}{\underline{Region~IV}} we finally get the following constraint on the three point non-Gaussian amplitude in the squeezed limit configuration:
          
          \bea \textcolor{red}{\underline{\rm Region~I}}+\textcolor{red}{\underline{\rm Region~II}}+\textcolor{red}{\underline{\rm Region~III}}+\textcolor{red}{\underline{\rm Region~IV}:}~~-0.0005<f^{sq}_{NL}<0.16
          \eea
          for the following parameter space:
          \bea \textcolor{red}{\underline{\rm Region~I}}+\textcolor{red}{\underline{\rm Region~II}}+\textcolor{red}{\underline{\rm Region~III}}+\textcolor{red}{\underline{\rm Region~IV}:}~~~~~0.0000001<\alpha\bar{\lambda}<0.001.~~~~~~~
               \eea
     To visualize these constraints more clearly we have also presented $(f^{sq}_{NL},\alpha,\bar{\lambda})$ 3D plot in fig.~(\ref{fnl3sd1}) and fig.~(\ref{fnl3sd2}), for two different angular orientations as given by \textcolor{red}{\underline{Angle~I}} and \textcolor{red}{\underline{Angle~II}}. From the the representative surfaces it is clearly observed the behavior of three point non-Gaussian amplitude in the squeezed limit for the variation of two fold parameter $\alpha$ and $\bar{\lambda}$ and the results are consistent with the obtained constraints in 2D analysis. Here all the obtained results are consistent with the two point constraints and the Planck 2015 data.
    \subsubsection{Using $\delta {\cal N}$ formalism}
\label{s7a2}
\underline{\textcolor{blue}{\bf A. Basic methodology:}}

\begin{figure}[ht]
{\centerline{\includegraphics[width=12.2cm, height=5cm] {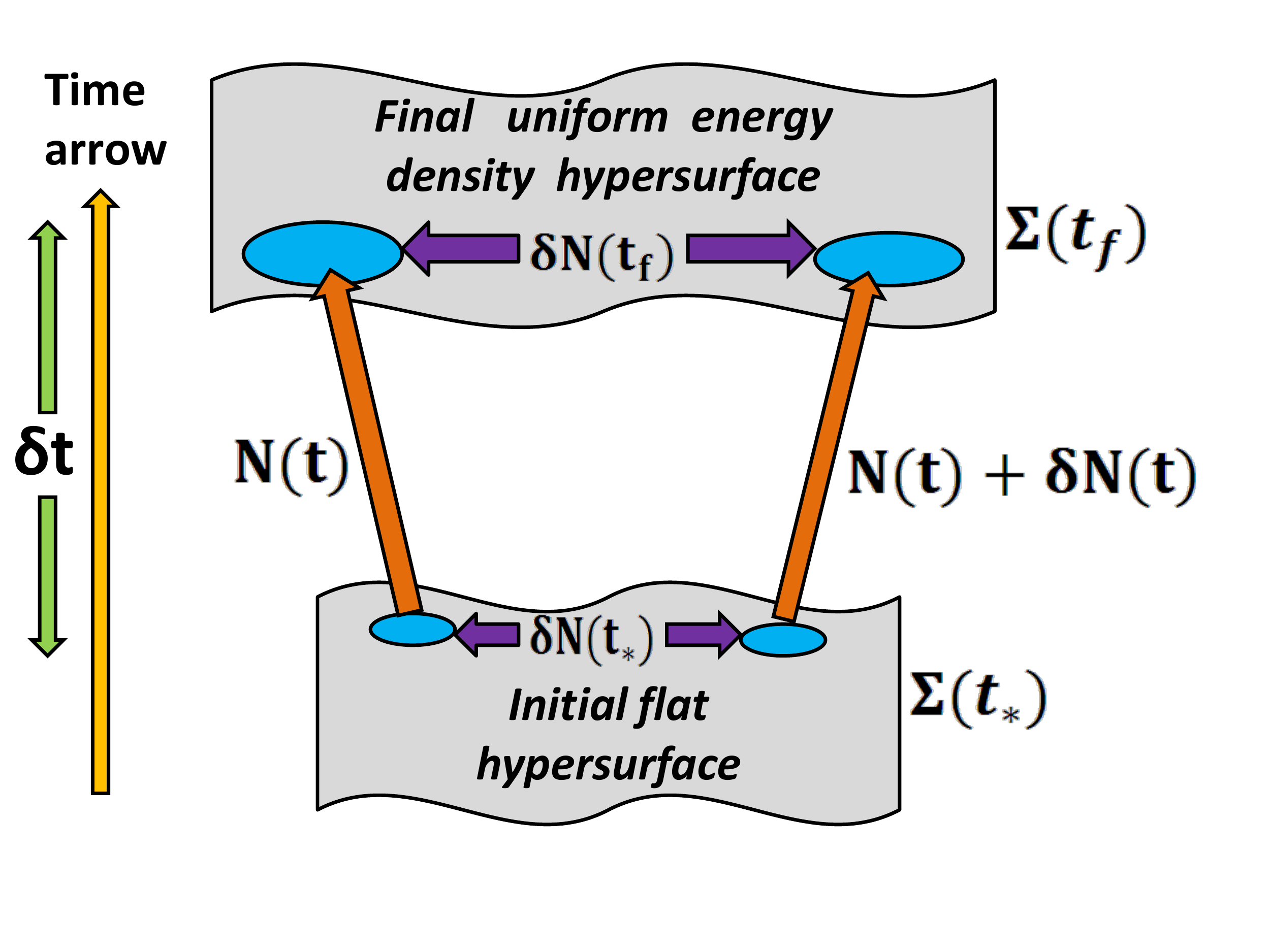}}}
\caption{Diagrammatic representation of $\delta N$ formalism.} \label{figdn}
\end{figure}

In this section our prime objective is to use $\delta {\cal N}$ formalism 
to compute the three point and four point correlation functions in the attractor regime. Here ${\cal N}$ signifies the number of e-foldings as we have defined earlier.
In this formalism the dominant contribution comes from only on the
perturbations of the scalar field 
trajectories with respect to the field value at the initial
hypersurface $\phi, \Psi$ and the velocity $\dot \phi, \dot \Psi$. This can be realized by providing two initial conditions on both of them on the initial hypersurface. 
More specifically, in the present context, we have assumed that the evolution of the universe is governed by a unique fashion
after the value of the scalar field achieved at $\phi=\phi_*$ and $\Psi=\Psi_*$, where it is mimicking the role of standard clock in inflationary cosmology. Here
the value of its velocity $\dot \phi_*$ and $\dot \Psi_*$ is completely insignificant. 
Let us mention that only in this case $\delta {\cal N}$ is equal to
the final value of the comoving curvature perturbation $\zeta$ which
is conserved at the epoch $t\geq t_*$. In figure~(\ref{figdn}), we have shown the schematic diagram of $\delta {\cal N}$ formalism.

For further computation we assume that on large scales the dynamical behaviour which permit us to ignore time
derivatives appearing in the cosmological perturbation theory, the horizon volume will evolve in such a way that it were a perfectly self contained universe. As a result the scalar curvature perturbation can be expressed
beyond liner order in cosmological perturbation theory as:
\bea \zeta &=& \delta {\cal N}= \left[{\cal N}_{,\phi}\delta \phi+{\cal N}_{,\Psi}\delta \Psi\right]+\frac{1}{2!}\left[{\cal N}_{,\phi\phi}\delta \phi\delta \phi+\left({\cal N}_{,\phi\Psi}+{\cal N}_{,\Psi\phi}\right)\delta \phi \delta \Psi+N_{,\Psi\Psi}\delta \Psi\delta \Psi\right]\nonumber\\&&~~~~~~~~~~~~~~~~~~~+\frac{1}{3!}\left[{\cal N}_{,\phi\phi\phi}\delta \phi\delta \phi\delta \phi+\left({\cal N}_{,\phi\Psi\Psi}+{\cal N}_{,\Psi\phi\Psi}+{\cal N}_{,\Psi\Psi\phi}\right)\delta \phi\delta \Psi\delta \Psi\right.\nonumber\\ 
&&\left.~~~~~~~~~~~~~~+\left({\cal N}_{,\phi\phi\Psi}+N_{,\phi\Psi\phi}+{\cal N}_{,\Psi\phi\phi}\right)\delta \phi\delta \phi\delta \Psi+{\cal N}_{,\Psi\Psi\Psi}\delta \Psi\delta \Psi\delta \Psi\right]+\cdots,~~~~~~~~~~~~~~~~ \eea
where we use the following notations for simplicity:
\bea {\cal N}_{,\phi}&=&\partial_{\phi}{\cal N},~~
{\cal N}_{,\Psi}=\partial_{\Psi}{\cal N},\\
{\cal N}_{,\phi\phi}&=&\partial^2_{\phi}{\cal N},~~
{\cal N}_{,\Psi\Psi}=\partial^2_{\Psi}{\cal N},~~
N_{,\phi\Psi}=\partial_{\phi}\partial_{\Psi}{\cal N},~~
{\cal N}_{,\Psi\phi}=\partial_{\Psi}\partial_{\phi}{\cal N},\\
{\cal N}_{,\phi\phi\phi}&=&\partial^3_{\phi}{\cal N},~~
{\cal N}_{,\Psi\Psi\Psi}=\partial^3_{\Psi}{\cal N},~~
{\cal N}_{,\phi\phi\Psi}=\partial_{\phi}\partial_{\phi}\partial_{\Psi}{\cal N},~~
{\cal N}_{,\phi\Psi\phi}=\partial_{\phi}\partial_{\Psi}\partial_{\phi}{\cal N},\\
{\cal N}_{,\Psi\phi\phi}&=&\partial_{\Psi}\partial_{\phi}\partial_{\phi}{\cal N},~~
{\cal N}_{,\phi\Psi\Psi}=\partial_{\phi}\partial_{\Psi}\partial_{\Psi}{\cal N},\\
{\cal N}_{,\Psi\phi\Psi}&=&\partial_{\Psi}\partial_{\phi}\partial_{\Psi}{\cal N},~~
{\cal N}_{,\Psi\Psi\phi}=\partial_{\Psi}\partial_{\Psi}\partial_{\phi}{\cal N}.~~~~~~~
\eea
Here we use the notation, $\partial_{\phi}=\partial/\partial\phi$ and $\partial_{\Psi}=\partial/\partial\Psi$ to denote the partial derivatives. 
But here we have to point that in attractor regime both the fields $\phi$ and $\Psi$ are connected with each other, which we have already pointed earlier in this paper. 

Further the curvature perturbation can be recast as:
\bea \label{xcv}\zeta &=& \delta {\cal N}= 2{\cal N}_{,\phi}\delta \phi+\left\{2{\cal N}_{,\phi\phi}-\frac{{\cal V}^{'}(\phi)}{{\cal V}(\phi)}~{\cal N}_{,\phi}\right\}\delta \phi\delta\phi\nonumber\\
&&~~~~~~~~~~~~+\left\{\frac{4}{3}{\cal N}_{,\phi\phi\phi}-2\frac{{\cal V}^{'}(\phi)}{{\cal V}(\phi)}{\cal N}_{,\phi\phi}+\left(\frac{5}{3}\frac{{\cal V}^{'2}(\phi)}{{\cal V}^2(\phi)}-\frac{1}{6}\frac{{\cal V}^{''}(\phi)}{{\cal V}(\phi)}\right){\cal N}_{,\phi}\right\}\delta\phi\delta\phi\delta\phi+\cdots,~~~~~~~~~~ \eea
which implies that if we compute ${\cal N}_{,\phi}$, ${\cal N}_{,\phi\phi}$ and ${\cal N}_{,\phi\phi\phi}$, then one can determine the curvature perturbation and also compute the three and four point functions using Eq~(\ref{xcv}). The detalied computation of the field derivatives of ${\cal N}$ are explicitly given in the Appendix for all the derived effective potentials.\\ \\
\underline{\textcolor{blue}{\bf B. Generalized convention for field solution:}}\\ 
In $\delta N$ formalism to compute ${\cal N}_{,\phi}$, ${\cal N}_{,\phi\phi}$ and ${\cal N}_{,\phi\phi\phi}$ we start with the background equation of motion for the $\phi$ field:
\bea \label{sdf}\ddot{\phi}+3\tilde{H}\dot{\phi}+\partial_{\phi}\tilde{W}(\phi,\Psi)=0,\eea
where the effective potential $\tilde{W}(\phi,\Psi)$ is given by:
\be\footnotesize\begin{array}{lll}\label{rinfla1}
 \displaystyle \tilde{W}(\phi,\Psi)\displaystyle =\left\{\begin{array}{ll}
                    \displaystyle \frac{\lambda}{4}e^{-\frac{2\sqrt{2}}{\sqrt{3}}\frac{\Psi}{M_p}}\phi^{4}~~~~~ &
 \mbox{\small \textcolor{red}{\bf for \underline{Case I}}}  
\\ 
         \displaystyle \frac{M^{4}_{p}}{8\alpha}-\frac{\lambda}{4}e^{-\frac{2\sqrt{2}}{\sqrt{3}}\frac{\Psi}{M_p}}\phi^{4}.~~~~ & \mbox{\small \textcolor{red}{\bf for \underline{Case II}}}  
         \\ 
                  \displaystyle \frac{M^{4}_{p}}{8\alpha}-\frac{\lambda}{4}e^{-\frac{2\sqrt{2}}{\sqrt{3}}\frac{\Psi}{M_p}}\left(\phi^4-\phi^{4}_{V}\right).~~~~ & \mbox{\small \textcolor{red}{\bf for \underline{Case ~II+Choice~I(v1)}}}\\ 
                  \displaystyle \frac{M^{4}_{p}}{8\alpha}+\frac{\lambda}{4}e^{-\frac{2\sqrt{2}}{\sqrt{3}}\frac{\Psi}{M_p}}\left(\phi^4-\phi^{4}_{V}\right).~~~~ & \mbox{\small \textcolor{red}{\bf for \underline{Case ~II+Choice~I(v2)}}}\\ 
                                    \displaystyle \frac{M^{4}_{p}}{8\alpha}+\left(\frac{m^{2}_{c}}{2}\phi^2-\frac{\lambda}{4}\phi^{4}\right)e^{-\frac{2\sqrt{2}}{\sqrt{3}}\frac{\Psi}{M_p}}.~~~~ & \mbox{\small \textcolor{red}{\bf for \underline{Case ~II+Choice~II(v1)}}}\\ 
                                    \displaystyle \frac{M^{4}_{p}}{8\alpha}-\left(\frac{m^{2}_{c}}{2}\phi^2-\frac{\lambda}{4}\phi^{4}\right)e^{-\frac{2\sqrt{2}}{\sqrt{3}}\frac{\Psi}{M_p}}.~~~~ & \mbox{\small \textcolor{red}{\bf for \underline{Case ~II+Choice~II(v2)}}}\\ 
                                                      \displaystyle \frac{M^{4}_{p}}{8\alpha}+\frac{\frac{\lambda}{4}\left(\phi^2 -\phi^2_{V}\right)^2}{\left(1+\xi\phi^2\right)^2}e^{-\frac{2\sqrt{2}}{\sqrt{3}}\frac{\Psi}{M_p}}.~~~~ & \mbox{\small \textcolor{red}{\bf for \underline{Case ~II+Choice~III}}}.
          \end{array}
\right.
\end{array}\ee
Here it is important to note that the exact connecting relations between the $\Psi$ field and the inflaton field $\phi$ is given by:
\be\footnotesize\begin{array}{lll}\label{rinfla2}
 \displaystyle \Psi-\Psi_{0}\displaystyle =- \frac{1}{2\sqrt{6}M_p}\times\left\{\begin{array}{ll}
                    \displaystyle 9\left(\phi^2 -\phi^2_0\right). &
 \mbox{\small \textcolor{red}{\bf for \underline{Case I}}}  
\\ 
         \displaystyle  \left(\phi^2 -\phi^2_0\right). & \mbox{\small \textcolor{red}{\bf for \underline{Case II}}}  
         \\ 
                  \displaystyle \left[\left(\phi^2 -\phi^2_0\right)+\phi^4_V\left(\frac{1}{\phi^2}-\frac{1}{\phi^2_0}\right)\right]. & \mbox{\small \textcolor{red}{\bf for \underline{Case ~II+Choice~I(v1)}}}\\ 
                  \displaystyle \left[\left(\phi^2 -\phi^2_0\right)+\phi^4_V\left(\frac{1}{\phi^2}-\frac{1}{\phi^2_0}\right)\right]. & \mbox{\small \textcolor{red}{\bf for \underline{Case ~II+Choice~I(v2)}}}\\ 
                                    \displaystyle \left[\left(\phi^2 -\phi^2_0\right)+\frac{m^{2}_{c}}{\lambda}\ln\left(\frac{m^2_{c}-\lambda\phi^{2}}{m^2_{c}-\lambda\phi^{2}_0}\right)\right]. & \mbox{\small \textcolor{red}{\bf for \underline{Case ~II+Choice~II(v1)}}}\\ 
                                    \displaystyle\left[\left(\phi^2 -\phi^2_0\right)+\frac{m^{2}_{c}}{\lambda}\ln\left(\frac{m^2_{c}-\lambda\phi^{2}}{m^2_{c}-\lambda\phi^{2}_0}\right)\right]. & \mbox{\small \textcolor{red}{\bf for \underline{Case ~II+Choice~II(v2)}}}\\ 
                                                      \displaystyle \frac{1}{\left(1+\xi\phi^2_{V}\right)}\left[\left(\phi^2 -\phi^2_0 \right)\left(1+\frac{\xi}{2}\left(\phi^2 +\phi^2_0 -2\phi^2_{V}\right)\right)\right.\\
                                                      \displaystyle \left.~~~~~~~~~~~~~~~+2\phi^2_{V}\ln\left(\frac{\phi}{\phi_0}\right)\right]. & \mbox{\small \textcolor{red}{\bf for \underline{Case ~II+Choice~III}}}.
          \end{array}
\right.
\end{array}\ee
It is obvious from the structural form of the effective potential for all of these cases that the general analytical solution for the inflaton field $\phi$ is too much complicated. To simplify the job here we consider a particular solution of the following form:
\bea \phi=\phi_L\propto \exp({\cal Y}Ht)~~~(i.e.~~\phi=\phi_{L}({\cal N})=\phi_{*}\exp(-{\cal Y}{\cal N})).\eea
Here we assume that ${\cal Y}$ is a time independent quantity.
Further our prime motivation is to obtain a more generalized version of the solution for FLRW cosmological background up to the consistent second order in cosmological perturbations around the prescribed particular solution. During our computation we also assume that the boundary between the attractor phase and the non attractor phase is determined by the field value $\phi=\phi_{*}=\phi({\cal N}_{cmb})=\phi_{cmb}$, which in cosmological literature identified to be the field value associated at the pivot scale.

To proceed further here we define a theoretical perturbative parameter which accounts the deviation from the actual inflaton field value compared to the field value after perturbation:
\bea \Delta_{part}\equiv \phi-\phi_0 -\phi_L=\sum^{\infty}_{n=1}\Delta_{n},\eea
where in general $\phi_0$ is the VEV of the inflaton field $\phi$. Here we assume that the parameter can take into account the difference between the true FLRW background solution and the proposed reference solution to solve the background Eq~(\ref{sdf}) in the physical domian where cosmological perturbation theory is valid. Additionally we claim that to validate the cosmological perturbation theory in the preferred physical domain, the infinite series sum should be convergent. Consequently in the present context we only look into $\Delta_{1}$ and $\Delta_{2}$, which are the general linearized and second order solution within cosmological perturbation theory for the background field equations. We also neglect all the higher order contribution in the perturbative regime of solution as they are very small. 
\\ \\
\underline{\textcolor{blue}{\bf C. Linearized perturbative solution:}}\\ \\
Before proceed further in this section let us clearly mention that, here we use the following ansatz to derive the results for linearized solution in the perturbative regime. 

In this case we assume that at the equation of motion level in the linear regime of perturbation theory there is no contribution from effective potential which contains quadratic structure or more complicated than that in terms of field $\phi$. In our calculation we treat all such contributions to be the back reactions and in the linear perturbative regime of solution our claim is such effects are small and largely suppressed. In this paper the derived effective potentials for all of these cases are also complicated and to get a preferred analytical solution in the linearized perturbative regime 
we use this \textcolor{red}{\bf Ansatz}. Here we get:
\be \left[\partial_{\phi}\tilde{W}\right]_{\phi=\phi_0+\phi_L+\Delta_1}\approx 0,\ee
which is valid for all types of derived potentials in the present context. Now let us consider the linearized perturbative solution $\Delta_{1}$ in this section. Consequently in the leading order of cosmological perturbation the background linearized version of the equation of motion takes the following form using the prescribed \textcolor{red}{\bf Ansatz}:
\bea\label{asdf2} \ddot{\Delta}_{1}+3H\dot{\Delta}_{1}+\phi_{L}\left\{{\cal Y}^2H^2\left(1-2tH\epsilon_{H}\right)-2{\cal Y}H^2\epsilon_{H}+3H^2{\cal Y}(1-tH\epsilon_{H})\right\}=0,\eea
where $\epsilon_{H}$ is the Hubble slow-roll parameter, 
$\epsilon_{H}=-\dot{H}/H^2$. The exact analytical solution of the Eq~(\ref{asdf2}) is given by:
\bea\label{dft2}
\displaystyle \Delta_{1}={\bf D}_{2}-\frac{1}{3H}{\bf D}_{1}e^{-3Ht}+\frac{1}{{\cal Y}(3+{\cal Y})^2}\phi_* e^{H{\cal Y}t}\left[-{\cal Y}(3+{\cal Y})^2 
\displaystyle +\epsilon_{H}\left(-9+{\cal Y}(3+{\cal Y})\left\{-2+H(3+2{\cal Y})t\right\}\right)\right].~~~~~
\eea
Here ${\bf D}_{1}$ and ${\bf D}_{2}$ are dimensionful arbitrary integration constants which can be determined by imposing the appropriate boundary condition. Additionally it is important to note that, in the present context this solution is valid in case quasi de-Sitter case also where the Hubble parameter $H$ is not exactly constant. 

\underline{\textcolor{blue}{\bf D. Second-order perturbative solution:}}\\ \\
Here we have considered the effect from the second order cosmological perturbation, $\Delta_2$.
It is important to note that during the computation here we also follow the same \textcolor{red}{\bf Ansatz}, which we have already introduced in the last section.
As a result including the contribution from slow-roll correction, the perturbative second order background equation of motion takes the following simplified form:
\bea\displaystyle \label{asdf2ffff} \ddot{\Delta}_{2}+3H\dot{\Delta}_{2}+\phi_{L}\left\{{\cal Y}^2H^2\left(1-2tH\epsilon_{H}\right)-2{\cal Y}H^2\epsilon_{H}+3H^2{\cal Y}(1-tH\epsilon_{H})\right\}=\Sigma_{S}.\eea
Here it is important to note that in  Eq~(\ref{asdf2ffff}), $\Sigma_{S}$ is the source contribution which is commping from the linear order perturbation $\Delta_1$. In this paper $\Sigma_S$ can be expressed for all derived effective potentials as:
\be\footnotesize\begin{array}{lll}\label{rinfla1cxc}
 \displaystyle \Sigma_{S}\displaystyle =\left\{\begin{array}{ll}
                    \displaystyle \Lambda_{c}e^{\frac{3\left(\left(\Delta_1+\phi_L\right)^2 -\phi^2_0\right)}{M^2_p}}\left(\Delta_1+\phi_L\right)^{4}~~~~~ &
 \mbox{\small \textcolor{red}{\bf for \underline{Case I}}}  
\\ 
         \displaystyle \frac{M^{3}_{p}}{8\alpha}-\Lambda_{c}e^{\frac{\left(\left(\Delta_1+\phi_L\right)^2 -\phi^2_0\right)}{3M^2_p}}\left(\Delta_1+\phi_L\right)^{4}.~~~~ & \mbox{\small \textcolor{red}{\bf for \underline{Case II}}}  
         \\ 
                  \displaystyle \frac{M^{3}_{p}}{8\alpha}-\Lambda_{c}e^{\frac{\left[\left(\left(\Delta_1+\phi_L\right)^2 -\phi^2_0\right)+\phi^4_V\left(\frac{1}{\left(\Delta_1+\phi_L\right)^2}-\frac{1}{\phi^2_0}\right)\right]}{3M^2_p}}\left(\left(\Delta_1+\phi_L\right)^4-\phi^{4}_{V}\right).~~~~ & \mbox{\small \textcolor{red}{\bf for \underline{Case ~II+Choice~I(v1)}}}\\ 
                  \displaystyle \frac{M^{3}_{p}}{8\alpha}+\Lambda_{c}e^{\frac{\left[\left(\left(\Delta_1+\phi_L\right)^2 -\phi^2_0\right)+\phi^4_V\left(\frac{1}{\left(\Delta_1+\phi_L\right)^2}-\frac{1}{\phi^2_0}\right)\right]}{3M^2_p}}\left(\left(\Delta_1+\phi_L\right)^4-\phi^{4}_{V}\right).~~~~ & \mbox{\small \textcolor{red}{\bf for \underline{Case ~II+Choice~I(v2)}}}\\ 
                                    \displaystyle \frac{M^{3}_{p}}{8\alpha}+\left(\frac{m^{2}_{c}}{2M_p}\left(\Delta_1+\phi_L\right)^2\right.\\ \left. \displaystyle~~~~~~~~-\Lambda_{c}\left(\Delta_1+\phi_L\right)^{4}\right)e^{\frac{\left[\left(\left(\Delta_1+\phi_L\right)^2 -\phi^2_0\right)+\frac{m^{2}_{c}}{\lambda}\ln\left(\frac{m^2_{c}-\lambda\left(\Delta_1+\phi_L\right)^{2}}{m^2_{c}-\lambda\phi^{2}_0}\right)\right]}{3M^2_p}}.~~~~ & \mbox{\small \textcolor{red}{\bf for \underline{Case ~II+Choice~II(v1)}}}\\ 
                                    \displaystyle \frac{M^{3}_{p}}{8\alpha}-\left(\frac{m^{2}_{c}}{2 M_p}\left(\Delta_1+\phi_L\right)^2\right.\\ \left. \displaystyle~~~~~~~~-\Lambda_{c}\left(\Delta_1+\phi_L\right)^{4}\right)e^{\frac{\left[\left(\left(\Delta_1+\phi_L\right)^2 -\phi^2_0\right)+\frac{m^{2}_{c}}{\lambda}\ln\left(\frac{m^2_{c}-\lambda\left(\Delta_1+\phi_L\right)^{2}}{m^2_{c}-\lambda\phi^{2}_0}\right)\right]}{3M^2_p}}.~~~~ & \mbox{\small \textcolor{red}{\bf for \underline{Case ~II+Choice~II(v2)}}}\\ 
                                                      \displaystyle \frac{M^{3}_{p}}{8\alpha}+\frac{\Lambda_{c}\left(\left(\Delta_1+\phi_L\right)^2 -\phi^2_{V}\right)^2}{\left(1+\xi\left(\Delta_1+\phi_L\right)^2\right)^2}\\
                                                      \displaystyle~~~\times e^{\frac{\left[\left(\left(\Delta_1+\phi_L\right)^2 -\phi^2_0 \right)\left(1+\frac{\xi}{2}\left(\left(\Delta_1+\phi_L\right)^2 +\phi^2_0 -2\phi^2_{V}\right)\right)+2\phi^2_{V}\ln\left(\frac{\left(\Delta_1+\phi_L\right)}{\phi_0}\right)\right]}{3M^2_p\left(1+\xi\phi^2_{V}\right)}}.~~~~ & \mbox{\small \textcolor{red}{\bf for \underline{Case ~II+Choice~III}}}.
          \end{array}
\right.
\end{array}\ee
where we define a new parameter:
\be \Lambda_{c}=\frac{\lambda}{4M_p}e^{-\frac{2\sqrt{2}}{\sqrt{3}}\frac{\Psi_0}{M_p}}.\ee
From the complicated mathematical structure of the source function $\Sigma_{S}$ it is clear that using it it is not possible to solve second order perturbation equations. To solve this problem one can simplify the the source function in the following way:
\be\footnotesize\begin{array}{lll}\label{rinfla1cxcc}
 \displaystyle \Sigma_{S}\displaystyle \approx\left\{\begin{array}{ll}
                    \displaystyle \Lambda_{c}\phi^4_L\left(1+4\frac{\Delta_1}{\phi_L}\right)~~~~~ &
 \mbox{\small \textcolor{red}{\bf for \underline{Case I}}}  
\\ 
         \displaystyle \beta-\Lambda_{c}\phi^4_L\left(1+4\frac{\Delta_1}{\phi_L}\right).~~~~ & \mbox{\small \textcolor{red}{\bf for \underline{Case II}}}  
         \\ 
                  \displaystyle \beta-\Lambda_{c}\left(\phi^4_L\left(1+4\frac{\Delta_1}{\phi_L}\right)-\phi^{4}_{V}\right).~~~~ & \mbox{\small \textcolor{red}{\bf for \underline{Case ~II+Choice~I(v1)}}}\\ 
                  \displaystyle \beta+\Lambda_{c}\left(\phi^4_L\left(1+4\frac{\Delta_1}{\phi_L}\right)-\phi^{4}_{V}\right).~~~~ & \mbox{\small \textcolor{red}{\bf for \underline{Case ~II+Choice~I(v2)}}}\\ 
                                    \displaystyle \beta+\left(\frac{M_{c}}{2}\phi^2_L\left(1+2\frac{\Delta_1}{\phi_L}\right)-\Lambda_{c}\phi^4_L\left(1+4\frac{\Delta_1}{\phi_L}\right)\right).~~~~ & \mbox{\small \textcolor{red}{\bf for \underline{Case ~II+Choice~II(v1)}}}\\ 
                                    \displaystyle \beta-\left(\frac{M_{c}}{2}\phi^2_L\left(1+2\frac{\Delta_1}{\phi_L}\right)-\Lambda_{c}\phi^4_L\left(1+4\frac{\Delta_1}{\phi_L}\right)\right).~~~~ & \mbox{\small \textcolor{red}{\bf for \underline{Case ~II+Choice~II(v2)}}}\\ 
                                                      \displaystyle \beta+\Gamma_\xi\left\{1+\Theta_{\xi}\frac{\Delta_1}{\phi_L}\right\}.~~~~ & \mbox{\small \textcolor{red}{\bf for \underline{Case ~II+Choice~III}}}.
          \end{array}
\right.
\end{array}\ee
where $\beta,M_c$ and $\Gamma_c$ is defined as:
\bea \beta&=&\frac{M^{3}_{p}}{8\alpha},~~
M_c=\frac{m^2_c}{M_p},~~
\Gamma_\xi=\Lambda_{c}\left(\phi^2_L-\phi^2_V\right)^2\left(1+2\xi\phi^2_L\right),~~
\Theta_{\xi}=4\phi^2_L\left(\xi+\frac{1}{\phi^2_L-\phi^2_V}\right).\eea
The representative solutions of Eq~(\ref{asdffffx}) for various sources are given in the Appendix.\\ \\
\underline{\textcolor{blue}{\bf D. Implementation of $\delta {\cal N}$ at the final hypersurface:}}\\ 
Using the results derived in the previous two sections here our prime objective is to explicitly compute the expression for the cosmological scalar perturbations in terms of the number of e-folds, $\delta {\cal N}$, which we have already introduced earlier. In the present conetxt the truncated version of the background solution of the inflaton field $\phi$ corrected upto the second order cosmological perturbations around the reference trajectory, $\phi_L\propto e^{-{\cal Y}{\cal N}}$ or $\phi_{L}\propto e^{{\cal Y}Ht}$, is generically given by for all the various physical cases are:
\bea\label{cfr1}
 \phi({\cal N})&=&\phi_0+\frac{\phi_*}{1+\hat{\Delta}_{1}({\cal N}=0)+\hat{\Delta}_{2}({\cal N}=0)}\left(e^{-{\cal Y}{\cal N}}+\hat{\Delta}_{1}({\cal N})+\hat{\Delta}_{2}({\cal N})\right),\eea
 or equivalently one can write:
 \bea\label{cfr2}
  \phi(t)&=&\phi_0+\frac{\phi_*}{1+\hat{\Delta}_{1}(t=0)+\hat{\Delta}_{2}(t=0)}\left(e^{{\cal Y}Ht}+\hat{\Delta}_{1}(t)+\hat{\Delta}_{2}(t)\right).\eea
  But for the sake of simplicity we use Eq~(\ref{cfr1}) as we want to implement the methodology of $\delta {\cal N}$ formalism. Additionally, it is important to mention that the symbol $\hat{}$ is introduced in the present context to rescale the integration constants and the perturbative solutions by the field value $\phi_*$ i.e.${\Delta}_1=\phi_* \hat{\Delta}_1$,$
  {\Delta}_2=\phi_* \hat{\Delta}_2$.
  Expresssions for the perturbative solutions $\Delta_{1}({\cal N}=0)$ and $\Delta_{2}({\cal N}=0)$ are explicitly written in the Appendix.

In the present context all of the sets of scaled integration constants parameterizes different trajectories and for our computation we set:
$ \phi(0,\hat{\bf W}_{k})=\phi_*,$
where $\hat{\bf W}_{k}$ is defined as the collection of all integration constants in a specific situation as defined as, 
$\hat{\bf W}_{k}\displaystyle =[\hat{\bf W}_{1},\hat{\bf W}_{2},\hat{\bf W}_{3},\hat{\bf W}_{4}]=[\hat{\bf D}_1, \hat{\bf D}_2, \hat{\bf D}_3, \hat{\bf D}_4]$.
Further inverting Eq~(\ref{cfr1}), for a specified set of values of the constants $\hat{\bf W}_{k}$, we have obtained the following simplified expression for $\delta {\cal N}$ as a implicit function of the inflaton field $\phi$, additional field $\Psi$ and $\hat{\bf W}_{k}$ as:
\bea\label{opopxx} \delta {\cal N}(\phi,\Psi,\hat{\bf W}_{k})&=& {\cal N}(\phi+\delta\phi,\Psi+\delta\Psi,\hat{\bf W}_{k})-{\cal N}(\phi,\Psi, 0)=\sum^{2}_{\alpha=1}\sum^{4}_{k=0}\sum_{n,m}\frac{1}{n!m!}\partial^{n}_{\phi^n_{\alpha}}\partial^{m}_{\hat{\bf W}^m_{k}}\left\{{\cal N}(\phi_{\alpha},0)\right\}~\delta\phi^n_{\alpha}~\hat{\bf W}^m_{k}.~~~~~~~~\eea

For this computation we have introduced the shift of the inflaton field $\phi$, additional field  $\Psi$ and the number of e-foldings ${\cal N}$ as,
$\phi\rightarrow \phi+\delta\phi$,$
\Psi\rightarrow\Psi+\delta\Psi$,
${\cal N}\rightarrow{\cal N}+\delta {\cal N}$,
in both the sides of Eq~(\ref{cfr1}), to compute the analytical expression for $\delta {\cal N}$ in a iterative way from our present setup. Additionally it is important to note that in this present context $\phi$ field and $\Psi$ field are not independent. They are related via Eq~(\ref{rinfla2}), as we have already mentioned earlier. In the present setup, we have already obtained the second order perturbative solutions of the scalar inflaton field trajectories around the particular reference solution, $\phi_L=\phi_* e^{{\cal Y}Ht}=\phi_* e^{-{\cal Y}{\cal N}}$, as we have already pointed earlier. Additionally important to note that if we neglect the sub dominant contribution of the form $\Delta_{1}\propto e^{{\cal Y}Ht}$, then the analysis only holds good only at thsufficiently late time epochs. This directly implies that in this computation if we use such assumption then we choose the initial time in such a way that it is very close to the final time for the number of e-folds ${\cal N}\leq 1$. To serve this purpose the simplest possibility is to choose the initial time epoch is infinitesimally close to the time scale at $\phi=\phi_{*}=\phi({\cal N}_{cmb})=\phi_{cmb}$.

For the sake of simplicity one can further assume that the final expression for curvature perturbation in $\delta {\cal N} $ formalism is independent of the coefficients $\hat{\bf W}_{k}$ at ${\cal N}=0$ for which the following constraints holds good perfectly, 
$\partial^{m}_{\hat{\bf W}_{k}}{\cal N}=0~~~\forall m=1,.......,\infty$.
Consequently we get the following simplified expression:
\bea\label{opopopc} \zeta=\delta {\cal N}
=\sum^{2}_{\alpha=1}\sum_{n}\frac{1}{n!}\partial^{n}_{\phi^n_{\alpha}}\left\{{\cal N}(\phi_{\alpha},0)\right\}~\delta\phi^n_{\alpha}&=&2{\cal N}_{,\phi}\delta \phi+\left\{2{\cal N}_{,\phi\phi}-\frac{{\cal V}^{'}(\phi)}{{\cal V}(\phi)}~{\cal N}_{,\phi}\right\}\delta \phi\delta\phi\nonumber\\
&&+\left\{\frac{4}{3}{\cal N}_{,\phi\phi\phi}-2\frac{{\cal V}^{''}(\phi)}{{\cal V}(\phi)}{\cal N}_{,\phi\phi}+\left(\frac{5}{3}\frac{{\cal V}^{'2}(\phi)}{{\cal V}^2(\phi)}-\frac{1}{6}\frac{{\cal V}^{''}(\phi)}{{\cal V}(\phi)}\right){\cal N}_{,\phi}\right\}\delta\phi\delta\phi\delta\phi+\cdots,\nonumber\\
&&\eea
where the function ${\cal V}(\phi)$ we have explicitly defined earlier for all the derived effective potentials. Here $\cdots$ corresponds to the higher order contributions, which are very very small compared to the leading order contributions appearing from cosmological perturbation theory for scalar fluctuations.

Next we take the derivatives of both sides of Eq~(\ref{cfr1}) and further set the following two constraints, 
${\cal N}=0,~~~
\hat{\bf W}_k=0 ~~\forall k$, at the final stage of the calculation. 
Our next task is to derive the analytical expression for inflaton fluctuation $\delta\phi_*$ and the coefficients $\hat{\bf W}_{k}$, which are generated via quantum fluctuations on the flat slice of $\delta\phi$. To implement this computational technique let us consider the evolution of fluctuation in the inflaton field $\delta\phi$ on super horizon scales. The field fluctuation or more precisely the shift in the inflaton field $\phi$ can be expressed as:
\bea \delta\phi({\cal N})&=&\sum^{2}_{i=1}\delta\phi_{i}({\cal N})=\phi_* \sum^{2}_{i=1}\hat{\Delta}_{i}({\cal N}),\eea
where the subscript $"1"$ and $"2"$ signify the linear and second order solution appearing from cosmological perturbation. Additionally, it is important to note that both the solutions $\hat{\Delta}_{1}({\cal N})$ and $\hat{\Delta}_{2}({\cal N})$, contain the growing and decaying mode characteristics. Further imposing the appropriate boundary condition from the end of the non-attractor region, where the number of e-folds ${\cal N}=0$, we get the
following expression for the shift in the inflaton field from linear order and second order cosmological perturbation at $\phi=\phi_*$ as:
    \bea \delta\phi_{*}=\delta\phi(0)&=&\sum^{2}_{i=1}\delta\phi_{i*}=\phi_* \sum^{2}_{i=1}\hat{\Delta}_{i}(0)=\phi_*\left(\hat{\Delta}_{1}(0)+\hat{\Delta}_{2}(0)\right).~~~~~\eea     
    See Appendix for more details. Now in the present context as we have started our computation from the reference solution $\phi\propto e^{-{\cal Y}{\cal N}}$, then using this relationship one can write down the explicit expression for the number of e-folds in terms of the inflaton field value as, 
    ${\cal N}(\phi)=\frac{1}{{\cal Y}}\ln\left(\frac{\phi_*}{\phi}\right)$,   
    which is consistent with the boundary condition that at $\phi=\phi_*$ the number of e-folds is ${\cal N}=0$ in the present case.
    Using these result at $\phi=\phi_{*}$ one can write down following expression for the curvature perturbation using $\delta {\cal N}$ formalism as:
   \bea\label{opopopp} \zeta&=&\delta {\cal N}
   ={\cal A}(\phi_*)\delta \phi_*+{\cal B}(\phi_*)\delta \phi_*\delta\phi_*+{\cal C}(\phi_*)\delta\phi_*\delta\phi_*\delta\phi_*+\cdots,\eea
      where ${\cal A}(\phi_*)$, ${\cal B}(\phi_*)$ and ${\cal C}(\phi_*)$ is defined as:
      \bea {\cal A}(\phi_*)&=&-\frac{2}{{\cal Y}\phi_*},\\ 
      {\cal B}(\phi_*)&=&\left\{\frac{2}{{\cal Y} \phi^2_*}+\left(\frac{{\cal V}^{'}(\phi)}{{\cal V}(\phi)}\right)_{*}~\frac{1}{{\cal Y}\phi_*}\right\},\\ 
      {\cal C}(\phi_*)&=&\left\{-\frac{2}{{\cal Y} \phi^3_*}\frac{4}{3}-2\left(\frac{{\cal V}^{''}(\phi)}{{\cal V}(\phi)}\right)_{*}\frac{1}{{\cal Y} \phi^2_*} 
      -\left(\frac{5}{3}\left(\frac{{\cal V}^{'2}(\phi)}{{\cal V}^2(\phi)}\right)_*-\frac{1}{6}\left(\frac{{\cal V}^{''}(\phi)}{{\cal V}(\phi)}\right)_*\right)\frac{1}{{\cal Y}\phi_*}\right\}.\eea
      Explicit forms of ${\cal B}(\phi_*)$ and ${\cal C}(\phi_*)$ are written in the Appendix for all the derived effective potentials.

            Next we decompose the product of the fluctuation in the inflaton field $\delta \phi_*\delta \phi_*$ and $\delta \phi_*\delta \phi_*\delta \phi_*$ into two parts which comes from linear and second order cosmological perturbation in the following way:
            \bea \delta\phi_*\delta\phi_*&=&\delta\phi(0)\delta\phi(0)=\sum^{2}_{i=1}\sum^{2}_{j=1}\delta\phi_{i*}\delta\phi_{j*}=\phi^2_{*}\sum^{2}_{i=1}\sum^{2}_{j=1}\hat{\Delta}_{i}(0)\hat{\Delta}_{j}(0),\\
            \delta\phi_*\delta\phi_*\delta\phi_*&=&\delta\phi(0)\delta\phi(0)\delta\phi(0)=\sum^{2}_{i=1}\sum^{2}_{j=1}\sum^{2}_{k=1}\delta\phi_{i*}\delta\phi_{j*}\delta\phi_{k*}=\phi^3_{*}\sum^{2}_{i=1}\sum^{2}_{j=1}\sum^{2}_{k=1}\hat{\Delta}_{i}(0)\hat{\Delta}_{j}(0)\hat{\Delta}_{k}(0).\eea
   an write down following expression for the curvature perturbation using $\delta {\cal N}$ formalism as:
      \bea\label{opopopx} \zeta=\delta {\cal N}&=&\phi_*{\cal A}(\phi_*)\sum^{2}_{i=1}\hat{\Delta}_{i}(0)+\phi^2_*{\cal B}(\phi_*)\sum^{2}_{i=1}\sum^{2}_{j=1}\hat{\Delta}_{i}(0)\hat{\Delta}_{j}(0)+\phi^3_*{\cal C}(\phi_*)\sum^{2}_{i=1}\sum^{2}_{j=1}\sum^{2}_{k=1}\hat{\Delta}_{i}(0)\hat{\Delta}_{j}(0)\hat{\Delta}_{k}(0)+\cdots\nonumber\\&=&\phi_*{\cal A}(\phi_*)\left(\hat{\Delta}_{1}(0)+\hat{\Delta}_{2}(0)\right)+\phi^2_*{\cal B}(\phi_*)\left(\hat{\Delta}^2_{1}(0)+\hat{\Delta}^2_{2}(0)+2\hat{\Delta}_{1}(0)\hat{\Delta}_{2}(0)\right)\nonumber\\
     &&~~~~~~~~~~~~~~~~~~~                                                                                           +\phi^3_*{\cal C}(\phi_*)\left(\hat{\Delta}^3_{1}(0)+\hat{\Delta}^3_{2}(0)+3\hat{\Delta}^2_{1}(0)\hat{\Delta}_{2}(0)+3\hat{\Delta}_{1}(0)\hat{\Delta}^2_{2}(0)\right)+\cdots,\eea 
 Further using local configuration in momentum space one can define the non-Gaussian amplitude associated with the three point function using $\delta {\cal N}$ formalism as \cite{Sugiyama:2012tj}:
      \bea f^{loc}_{NL}&=&\frac{5}{6}\frac{B(k_1,k_2,k_3)}{\left[P_{\zeta}(k_1)P(k_2)+P_{\zeta}(k_2)P_{\zeta}(k_3)+P_{\zeta}(k_3)P_{\zeta}(k_1)\right]}=\frac{5}{6}\frac{{\cal N}_{,IJ}{\cal N}_{,I}{\cal N}_{,J}}{({\cal N}_{,K}{\cal N}_{,K})^2},~~~~~~~~~~~~\eea      
      Here $B(k_1,k_2,k_3)$ is the bispectrum and $P_{\zeta}(k)$ is the power spectrum for scalar perturbations. Here $I,J,K$ are the field configuration indices i.e. $I,J,K=\phi,\Psi$. In terms of inflaton field $\Phi$ and additional field $\Psi$ we get the following simplified expression for the non-Gaussian amplitude associated with the three point function:
      \bea f^{loc}_{NL}&=&\frac{5}{6}\left[\frac{{\cal N}_{,\phi\phi}{\cal N}_{,\phi}{\cal N}_{,\phi}
      +{\cal N}_{,\Psi\Psi}{\cal N}_{,\Psi}{\cal N}_{,\Psi}+\left({\cal N}_{,\phi\Psi}+{\cal N}_{,\Psi\phi}\right){\cal N}_{,\phi}{\cal N}_{,\Psi}}{({\cal N}_{,\phi}{\cal N}_{,\phi}+{\cal N}_{,\Psi}{\cal N}_{,\Psi})^2}\right]_{*}.\eea  
      Now as the $\Psi$ field can be expressed in therms of $\phi$ field, using this crucial fact we get the following result of non-Gaussian amplitude in the attractor regime as:
       \bea f^{loc}_{NL}&=&\frac{5}{6}\left[\frac{\left(1+\frac{2}{{\cal V}^2(\phi)}+\frac{1}{{\cal V}^4(\phi)}\right)}{\left(1+\frac{1}{{\cal V}^2(\phi)}\right)^2 }\frac{{\cal N}_{,\phi\phi}}{{\cal N}^2_{,\phi}}
                          -\frac{\frac{{\cal V}^{'}(\phi)}{{\cal V}^3(\phi)}}{\left(1+\frac{1}{{\cal V}^2(\phi)}\right)}\frac{1}{{\cal N}_{,\phi}}\right]_{*}.\eea
  Further substituting the explicit form of the function ${\cal V}(\phi)$ and ${\cal N}_{,\phi}$, ${\cal N}_{,\phi\phi}$ for all derived effective potentials at $\phi=\phi_*$ we get:
   \bea f^{loc}_{NL}&=&\frac{5{\cal Y}}{6}\left[{\cal G}_{1}(\phi_*)+{\cal G}_{2}(\phi_*)\phi_*\right],\eea
                            where the functions ${\cal G}_{1}(\phi_*)$ and ${\cal G}_{2}(\phi_*)$ are defined in the Appendix.

  Here it is important to note that the exact momentum dependence will not be calculable using the semi classical techniques used in $\delta {\cal N}$ formalism in the attractor regime of cosmological perturbations. But to know the exact momentum dependence of the non-Gaussian amplitude obtained from the three point function of the scalar curvature fluctuation it is always useful to follow exact quantum mechanical techniques used in in-in formalism as discussed earlier part of this section. In case of in-in formalism we freeze the the value of the additional field $\Psi$ at the Planck scale and perform the calculation in the non-attractor regime of perturbation theory. But to get the correct estimate one can claim that the results obtained using both of the techniques should match at the horizon crossing iff we freeze the value of the $\Psi$ field at the Planck scale in $\delta {\cal N}$ formalism. This is also a strong information from the observational point of view, as Planck and the other future observation trying to probe the value of non-Gaussianity at this scale. In this work, we have done both the calculations for three point function for scalar curvature fluctuation by following semi classical and quantum mechanical techniques. In case of in-in formalism we have computed the results we use two physical shape configurations or templates- \textcolor{blue}{equilateral} and \textcolor{blue}{squeezed} to analyse the non-Gaussian amplitude obtained from the three point function for scalar curvature fluctuation by freezing the value of the additional $\Psi$ field at the Planck scale. Now to implement the equality between two results at the horizon crossing we have to fix the value of the additional field $\Psi$ in the $\delta {\cal N}$ formalism also. After freezing the value of $\Psi$ in the all derived effective potentials we get the following result for curvature perturbation in terms of $\delta {\cal N}$ at $\phi=\phi_*$:
  \bea\label{opopop} \zeta=\delta {\cal N}
        &=&\phi_*{\cal D}(\phi_*)\sum^{2}_{i=1}\hat{\Delta}_{i}(0)+\phi^2_*{\cal E}(\phi_*)\sum^{2}_{i=1}\sum^{2}_{j=1}\hat{\Delta}_{i}(0)\hat{\Delta}_{j}(0)+\phi^3_*{\cal F}(\phi_*)\sum^{2}_{i=1}\sum^{2}_{j=1}\sum^{2}_{k=1}\hat{\Delta}_{i}(0)\hat{\Delta}_{j}(0)\hat{\Delta}_{k}(0)+\cdots\nonumber\\&=&\phi_*{\cal D}(\phi_*)\left(\hat{\Delta}_{1}(0)+\hat{\Delta}_{2}(0)\right)+\phi^2_*{\cal E}(\phi_*)\left(\hat{\Delta}^2_{1}(0)+\hat{\Delta}^2_{2}(0)+2\hat{\Delta}_{1}(0)\hat{\Delta}_{2}(0)\right)\nonumber\\
       &&~~~~~~~~~~~~~~~~~~~                                                                                           +\phi^3_*{\cal F}(\phi_*)\left(\hat{\Delta}^3_{1}(0)+\hat{\Delta}^3_{2}(0)+3\hat{\Delta}^2_{1}(0)\hat{\Delta}_{2}(0)+3\hat{\Delta}_{1}(0)\hat{\Delta}^2_{2}(0)\right)+\cdots,\eea
       where the new functions ${\cal D}(\phi_*)$, ${\cal E}(\phi_*)$ and ${\cal F}(\phi_*)$ are defined as:
       \bea {\cal D}(\phi_*)&=&\left({\cal N}_{,\phi}\right)_*=-\frac{1}{{\cal Y}\phi_*},~~ 
             {\cal E}(\phi_*)=\frac{1}{2}\left({\cal N}_{,\phi\phi}\right)_*=\frac{1}{2{\cal Y} \phi^2_*},~~ 
             {\cal F}(\phi_*)=\frac{1}{6}\left({\cal N}_{,\phi\phi\phi}\right)_*=-\frac{1}{3{\cal Y} \phi^3_*}.\eea
             After freezing the value $\Psi$ in the Planck scale in the non attractor regime of cosmological perturbation theory we get the following expression for the non-Gaussian amplitude from three point scalar curvature fluctuation as:
             \bea\label{p23} f^{loc}_{NL}&=&\frac{5}{6}\left[\frac{{\cal N}_{,\phi\phi}}{{\cal N}_{,\phi}}\right]_{*}=\frac{5}{6}{\cal Y}.\eea
   Now further we use the general momentum dependent result at the horizon crossing and also use two different templates to equate with the results obtained from $\delta{\cal N}$ formalism and finally we get folowwing expression for the unknown factor ${\cal Y}$ as:
  \bea {\cal Y}
&\approx&\frac{1}{2\sum^{3}_{i=1}k^3_i} \left[2(3\epsilon^*_{\tilde{W}}-\eta^{*}_{\tilde{W}})\sum^{3}_{i=1}k^3_i 
  +\epsilon^*_{\tilde{W}}\left(-\sum^{3}_{i=1}k^3_i+\sum^{3}_{i,j=1,i\neq j}k_i k^2_j+\frac{8}{K}\sum^{3}_{i,j=1,i> j}k^2_i k^2_j\right)\right].\eea
  However it is crucial to note that, without freezing the value of the addition field $\Psi$ in the Planck scale in the non attractor regime of cosmological perturbation theory one can perform the exact quantum mechanical in-in calculation where solution of the $\Psi$ field is related to the inflaton field $\phi$ and finally match with the results obtained from the $\delta {\cal N}$ formalism. In this paper we have not computed this in case in in formalism and we also hope to generalize this methodology in the attarctor regime as well in near future. 
  
  Next we use use the two physical templates for the shape configurations-\textcolor{blue}{equilateral} and \textcolor{blue}{squeezed} to determine the 
  functional form of the unknown factor ${\cal Y}$ which is appearing in $\delta{\cal N}$ formalism. In this context we get:
  \begin{enumerate}
  \item  \underline{\textcolor{blue}{Equilateral limit configuration}:}
  \bea {\cal Y}&\approx&\frac{1}{6} \left[29\epsilon^*_{\tilde{W}}-6\eta^*_{\tilde{W}} 
      \right].\eea
  \item \underline{\textcolor{blue}{Squeezed limit configuration}:}
  \bea {\cal Y}
    &\approx&\frac{1}{2}\left[4(4\epsilon^*_{\tilde{W}}-\eta^*_{\tilde{W}}) 
    +10\epsilon^*_{\tilde{W}}\left(\frac{k_S}{k_L}\right)^2-(2\eta^*_{\tilde{W}}-\epsilon^*_{\tilde{W}})\left(\frac{k_S}{k_L}\right)^3\right].\eea
  \end{enumerate}
  which are correct results of the unknown factor ${\cal Y}$ at the level of three point function computed from scalar curvature perturbation.

\subsection{Four point function}
\label{s7b}
\subsubsection{Using In-In formalism}
\label{s7b1}
Here we discuss about the constraint on the primordial four point scalar correlation function in the non attractor regime of soft inflation. In general one can write down the following expressions for the four point function of the scalar fluctuation as \cite{Chen:2009bc,Ghosh:2014kba,Kundu:2015xta,Arroja:2008ga,Seery:2006js,Seery:2008ax}:
\bea \langle \zeta({\bf k}_1)\zeta({\bf k}_2)\zeta({\bf k}_3)\zeta({\bf k}_4)\rangle&=&(2\pi)^3\delta^{(3)}({\bf k}_1+{\bf k}_2+{\bf k}_3+{\bf k}_4)T(k_1,k_2,k_3,k_4).\eea
In our computation we choose Bunch-Davies vacuum state and for single field soft slow-roll inflation we get the following expression for the trispectrum: 
\bea T(k_1,k_2,k_3,k_4)
&\approx&\frac{\tilde{W}^3(\phi_{cmb},{\bf \Psi})}{216M^{12}_p (\epsilon^*_{\tilde{W}})^2}\frac{1}{(k_1 k_2 k_3k_4)^3}\left[\hat{G}^{S}({\bf k}_1,{\bf k}_2,{\bf k}_3,{\bf k}_4)+\hat{G}^{S}({\bf k}_1,{\bf k}_3,{\bf k}_2,{\bf k}_4)\right.\nonumber\\ 
&&\left.~~~~~+\hat{G}^{S}({\bf k}_1,{\bf k}_4,{\bf k}_3,{\bf k}_2)-\hat{W}^{S}({\bf k}_1,{\bf k}_2,{\bf k}_3,{\bf k}_4) 
-\hat{W}^{S}({\bf k}_1,{\bf k}_3,{\bf k}_2,{\bf k}_4)-\hat{W}^{S}({\bf k}_1,{\bf k}_4,{\bf k}_3,{\bf k}_2)\right.\nonumber\\ 
&&\left.~~~~-2\left\{\hat{R}^{S}({\bf k}_1,{\bf k}_2,{\bf k}_3,{\bf k}_4)+\hat{R}^{S}({\bf k}_1,{\bf k}_3,{\bf k}_2,{\bf k}_4)+\hat{R}^{S}({\bf k}_1,{\bf k}_4,{\bf k}_3,{\bf k}_2)\right\}\right],~~~~~~~~~~~~\eea
where the momentum dependent functions $\hat{G}^{S}({\bf k}_1,{\bf k}_2,{\bf k}_3,{\bf k}_4)$, $\hat{W}^{S}({\bf k}_1,{\bf k}_2,{\bf k}_3,{\bf k}_4)$ and $\hat{R}^{S}({\bf k}_1,{\bf k}_2,{\bf k}_3,{\bf k}_4)$ are defined in the Appendix.

Here it is important to mention that, our derived result is consists of three following parts:
\begin{enumerate}
\item First of all, we have the contribution from {\it contact interaction} term $\hat{R}^{S}$, which appears due to the longitudinal graviton $S$-channel propagator as given by:
\bea R^{S}({\bf k}_1,{\bf k}_2,{\bf k}_3,{\bf k}_4)&=&16(2\pi)^3\delta^{(3)}({\bf k}_1+{\bf k}_2+{\bf k}_3+{\bf k}_4)\left[\prod^{4}_{I=1}\phi({\bf k}_{I})\right]\hat{R}^{S}({\bf k}_1,{\bf k}_2,{\bf k}_3,{\bf k}_4).~~~~~~~~~~~~~\eea

\item Next we have the contribution from the terms like $\hat{W}^{S}$, which comes from the contribution which appears due to the transverse graviton propagator as given by:
\bea \widetilde{W}&=&\int dz_1 d^3{\bf x}_1\int dz_2 d^3{\bf x}_2T_{i^{'}j^{'}}(z_1,{\bf x}_1)\delta^{i^{'}i }\delta^{j^{'}j }\tilde{G}_{ij,kl}(z_1,{\bf x}_1;z_2,{\bf x}_2)\delta^{kk^{''}}\delta^{ll^{''}}T_{k^{''}l^{''}}(z_2,{\bf x}_2),~~~~~~~~~~~~~\eea
where the transverse graviton Green's function $\tilde{G}_{ij,kl}(z_1,{\bf x}_1;z_2,{\bf x}_2)$ is given by:
\bea \tilde{G}_{ij,kl}(z_1,{\bf x}_1;z_2,{\bf x}_2)&=&\int\frac{d^3{\bf k }}{(2\pi)^3}~e^{i{\bf k}.({\bf x}_1-{\bf x}_2)} 
\int^{\infty}_{0}dp^2\frac{1}{4}\left[\frac{J_{\frac{3}{2}}(pz_1)J_{\frac{3}{2}}(pz_2)}{\sqrt{z_1z_2}({\bf k}^2+p^2)}\left(\widetilde{P}_{ik}\widetilde{P}_{jl}+\widetilde{P}_{il}\widetilde{P}_{jk}-\widetilde{P}_{ij}\widetilde{P}_{kl}\right)\right].~~~~~~~\eea
     \begin{figure*}[htb]
     \centering
     \subfigure[ $S$ channel diagram.]{
         \includegraphics[width=5.0cm,height=4.5cm] {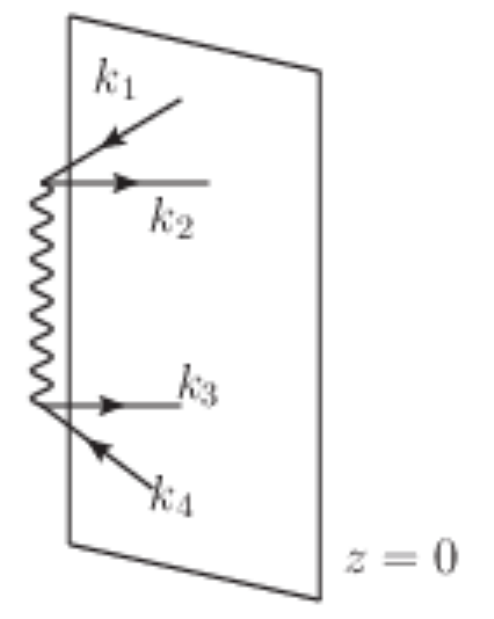}
         \label{figv1bb}
     }
     \subfigure[$T$ channel diagram.]{
         \includegraphics[width=5.0cm,height=4.5cm] {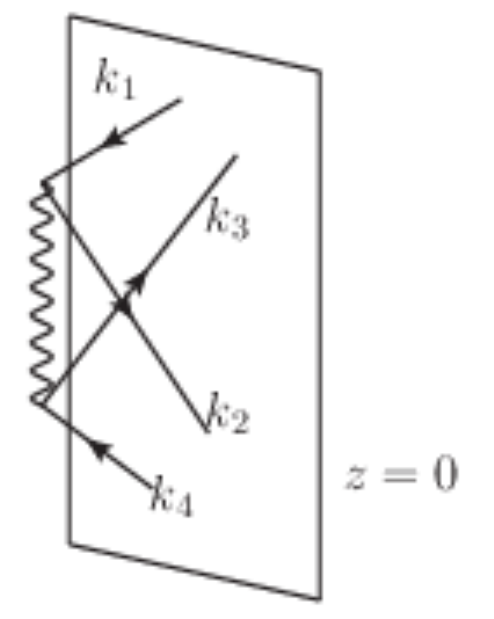}
         \label{figv2bb}
     }
     \subfigure[$U$ channel diagram.]{
              \includegraphics[width=5.0cm,height=4.5cm] {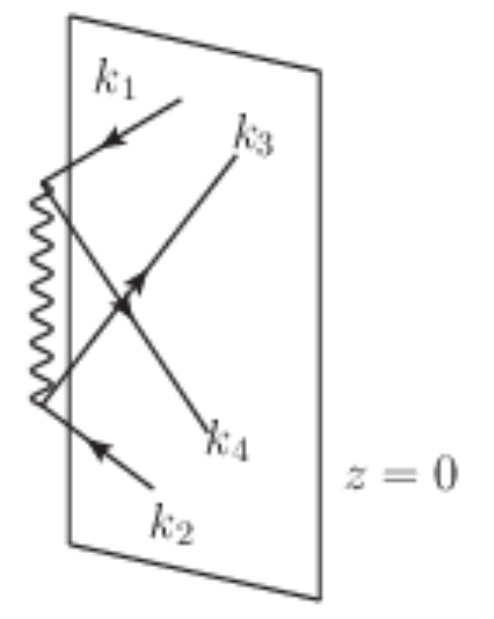}
              \label{figv3bb}
          }
     \caption[Optional caption for list of figures]{Representative $S,T$ and $U$ channel diagram for bulk interpretation of four point scalar correlation function in presence of graviton exchange contribution. In all the diagrams graviton is propagating on the bulk and the end point of scalars are attached with the boundary at $z=0$.} 
     \label{bulk2}
     \end{figure*}
Here $\widetilde{P}_{ij}$ is the which is the
transverse traceless projector onto the directions perpendicular to ${\bf k}$ as given by,
$\widetilde{P}_{ij}=\left(\delta_{ij}-\frac{k_ik_j}{k^2}\right)$,
and $J_{\frac{3}{2}}(x)$ is the Bessel function with characteristic index $3/2$, which can be expressed in terms of the following simplified form:
\bea J_{\frac{3}{2}}(x)&=&\sqrt{\frac{2}{\pi x}}\left(\frac{\sin x}{x}-\cos x\right)=\sqrt{\frac{2}{\pi x}}\frac{(1-ix)~e^{ix}-(1+ix)~e^{-ix}}{2ix}.\eea
Additionally in the present context the expression for stress tensor $T_{ij}(z,{\bf x})$ in terms of scalar field inflaton fluctuation $\delta\phi(z,{\bf x})$ is given by the following simplified expression:
\bea T_{ij}(z,{\bf x})&=&2(\partial_{i}\delta\phi)(\partial_{j}\delta\phi)-\delta_{ij}\left[(\partial_{z}\delta\phi)^2+\eta^{kl}(\partial_{k}\delta\phi)(\partial_{l}\delta\phi)\right].\eea 
Here it is important to mention that, two different insertions of the stress tensor corresponds to two different values of the radial variable $z=(z_1,z_2)$ which we finally integrate out. Finally, for $S$-channel contribution substituting for $\delta\phi$ in Fourier space we get the following expression for the transverse graviton propagator:
\bea \label{uiop}\widetilde{W}^{S}({\bf k}_1,{\bf k}_2,{\bf k}_3,{\bf k}_4)&=&16(2\pi)^3\delta^{(3)}({\bf k}_1+{\bf k}_2+{\bf k}_3+{\bf k}_4)\left[\prod^{4}_{I=1}\phi({\bf k}_{I})\right]\nonumber\\
&&~~~~\times k^{i}_{1}k^{j}_{2}k^{k}_{3}k^{l}_{4}\left(\widetilde{P}_{ik}\widetilde{P}_{jl}+\widetilde{P}_{il}\widetilde{P}_{jk}-\widetilde{P}_{ij}\widetilde{P}_{kl}\right)\Theta(k_1,k_2,k_3,k_4),~~~~~~~~~~~~~\eea
where $\Theta(k_1,k_2,k_3,k_4)$ and the transverse projector along with appropriate index contraction in momentum direction are defined as:
\bea \Theta(k_1,k_2,k_3,k_4)
&=&-\frac{2k_1k_2(k_1+k_2)^2((k_1+k_2)^2-k^2_3-k^2_4-4k_3k_4)}{(\hat{K}-2(k_3+k_4))^2\hat{K}^2((k_1+k_2)^2-K^2_s)}\nonumber\\
&&~~~~\times\left(\frac{3}{2(k_1+k_2)}-\frac{1}{\hat{K}}-\frac{1}{\hat{K}-2(k_3+k_4)}-\frac{k_1+k_2}{2k_1k_2}\right.\nonumber\\
&& \left.~~~~~~~~~+\frac{k_1+k_2}{K^2_s-(k_1+k_2)^2}-\frac{k_1+k_2}{k^2_3+k^2_4+4k_3k_4-(k_1+k_2)^2}\right)+(1,2\leftrightarrow 3,4)\nonumber\\
&&~~~~~+\frac{K^3_s(K^2_s-k^2_1-k^2_2-4k_1k_2)(K^2_s-k^2_3-k^2_4-4k_3k_4)}{(K^2_s-k^2_1-k^2_2-2k_1k_2)^2(K^2_s-k^2_3-k^2_4-2k_3k_4)^2}.\eea
\bea k^{i}_{1}k^{j}_{2}k^{k}_{3}k^{l}_{4}\left(\widetilde{P}_{ik}\widetilde{P}_{jl}+\widetilde{P}_{il}\widetilde{P}_{jk}-\widetilde{P}_{ij}\widetilde{P}_{kl}\right)&=&\left[{\bf k}_1.{\bf k}_3+\frac{({\bf k}_1.({\bf k}_1+{\bf k}_2))({\bf k}_3.({\bf k}_3+{\bf k}_4))}{|{\bf k}_1+{\bf k}_2|^2}\right]\nonumber\\
&&\left[{\bf k}_2.{\bf k}_4+\frac{({\bf k}_2.({\bf k}_1+{\bf k}_2))({\bf k}_4.({\bf k}_3+{\bf k}_4))}{|{\bf k}_1+{\bf k}_2|^2}\right]\nonumber\\
&&+\left[{\bf k}_1.{\bf k}_4+\frac{({\bf k}_1.({\bf k}_1+{\bf k}_2))({\bf k}_4.({\bf k}_3+{\bf k}_4))}{|{\bf k}_1+{\bf k}_2|^2}\right]\nonumber\\
&&\left[{\bf k}_2.{\bf k}_3+\frac{({\bf k}_2.({\bf k}_1+{\bf k}_2))({\bf k}_3.({\bf k}_3+{\bf k}_4))}{|{\bf k}_1+{\bf k}_2|^2}\right]\nonumber\\
&&-\left[{\bf k}_1.{\bf k}_2-\frac{({\bf k}_1.({\bf k}_1+{\bf k}_2))({\bf k}_2.({\bf k}_3+{\bf k}_4))}{|{\bf k}_1+{\bf k}_2|^2}\right]\nonumber\\
&&\left[{\bf k}_3.{\bf k}_4-\frac{({\bf k}_3.({\bf k}_1+{\bf k}_2))({\bf k}_4.({\bf k}_3+{\bf k}_4))}{|{\bf k}_1+{\bf k}_2|^2}\right].~~~~~~~~~~~~\eea
where $K_s$ is defined as the norm of the total momentum required for graviton exchange in $S$-channel, 
$K_s=|{\bf k}_1+{\bf k}_2|=|-({\bf k}_3+{\bf k}_4)|=|{\bf k}_3+{\bf k}_4|$.

Finally substituting all of these expressions in Eq~(\ref{uiop}) we get the following simplified expression for the $S$-channel contribution in transverse graviton propagator:
\bea \widetilde{W}^{S}({\bf k}_1,{\bf k}_2,{\bf k}_3,{\bf k}_4)&=&16(2\pi)^3\delta^{(3)}({\bf k}_1+{\bf k}_2+{\bf k}_3+{\bf k}_4)\left[\prod^{4}_{I=1}\phi({\bf k}_{I})\right]\hat{W}^{S}({\bf k}_1,{\bf k}_2,{\bf k}_3,{\bf k}_4),~~~~~~~~~~~~~\eea
where $\hat{W}^{S}({\bf k}_1,{\bf k}_2,{\bf k}_3,{\bf k}_4)$ is defined as:
\bea \hat{W}^{S}({\bf k}_1,{\bf k}_2,{\bf k}_3,{\bf k}_4) &=&k^{i}_{1}k^{j}_{2}k^{k}_{3}k^{l}_{4}\left(\widetilde{P}_{ik}\widetilde{P}_{jl}+\widetilde{P}_{il}\widetilde{P}_{jk}-\widetilde{P}_{ij}\widetilde{P}_{kl}\right)\Theta(k_1,k_2,k_3,k_4).~~~~~~~~~\eea
Here it is important to note that, the contribution from the $T$ and $U$-channel can be obtained by replacing the following momenta:
\bea T-{\rm channel}:~~~~~{\bf k}_{2}&\leftrightarrow&{\bf k}_{3},\\
     U-{\rm channel}:~~~~~{\bf k}_{2}&\leftrightarrow&{\bf k}_{4}.\eea
     The representative $S$, $T$ and $U$ channel diagrams for bulk interpretation of the four point scalar correlation function in presence of graviton exchange is shown in shown in fig.~(\ref{figv1bb}), fig.~(\ref{figv2bb}) and fig.~(\ref{figv3bb}). In these diagrams we have explicitly shown that, graviton is propagating on the bulk and the end point of scalars are attached with the boundary at $z=0$. In or computation all the representative diagrams are important to explain the total four point scalar correlation function.

\item Additionally, the extra contributions $\hat{G}^{S}$ appears due to integrating
out the metric perturbation.
\end{enumerate}
In the present context, including the contribution from four point function one can parameterize non-Gaussianity phenomenologically via a non-linear correction to a Gaussian perturbation $\zeta_{g}$ in position space as:
		   	   \bea \zeta({\bf x})&=&\zeta_{g}({\bf x})+\frac{3}{5}f^{loc}_{NL}\left[\zeta^2_{g}({\bf x})-\langle\zeta^2_{g}({\bf x})\rangle\right]+\frac{9}{25}g^{loc}_{NL}\zeta^3_{g}({\bf x})+\cdots,~~~~~~\eea
		   	   where $\cdots$ represent higher order non-Gaussian contributions. In case local non-Gaussianity amplitude of the bispectrum from the three point function is defined as \cite{Choudhury:2015yna,Chen:2009bc}:
\bea \label{cv2} T(k_1,k_2,k_3,k_4)&=&\left[\tau^{loc}_{NL}\sum_{j<p,i\neq j,p}P_{\zeta}(k_{ij})P_{\zeta}(k_j)P_{\zeta}(k_p)+\frac{54}{25}g^{loc}_{NL}\sum_{i<j<p}P_{\zeta}(k_{i})P_{\zeta}(k_j)P_{\zeta}(k_p)\right],~~~~~\nonumber\\
&&\eea		   	  
where $\tau^{loc}_{NL}=\tau^{loc}_{NL}(k_1,k_2,k_3,k_4)$ and
     $g^{loc}_{NL}=g^{loc}_{NL}(k_1,k_2,k_3,k_4)$.
Here additionally it is important to note that the connecting relation between the non-Gaussian parameter $\tau^{loc}_{NL}$ and $g^{loc}_{NL}$ can be expressed as:
\bea \label{cv1} g^{loc}_{NL}&=&{\cal N}_{NORM}\tau^{loc}_{NL},~~~~~~~~~~~~~\eea
where ${\cal N}_{NORM}$ is defined as the appropriate normalization factor. In general the values of the normalization factor is different in different shape configurations.
Further using Eq~(\ref{cv1}) in Eq~(\ref{cv2}) we get the following simplified expression for the non-Gaussian parameter $\tau^{loc}_{NL}$ and $g^{loc}_{NL}$ as obtained from the four point scalar function:
\bea \tau^{loc}_{NL}&=&\frac{T(k_1,k_2,k_3,k_4)}{\left[\sum_{j<p,i\neq j,p}P_{\zeta}(k_{ij})P_{\zeta}(k_j)P_{\zeta}(k_p)+\frac{54}{25}{\cal N}_{NORM}\sum_{i<j<p}P_{\zeta}(k_{i})P_{\zeta}(k_j)P_{\zeta}(k_p)\right]},~~~~~~\\
g^{loc}_{NL}&=&\frac{{\cal N}_{NORM}T(k_1,k_2,k_3,k_4)}{\left[\sum_{j<p,i\neq j,p}P_{\zeta}(k_{ij})P_{\zeta}(k_j)P_{\zeta}(k_p)+\frac{54}{25}{\cal N}_{NORM}\sum_{i<j<p}P_{\zeta}(k_{i})P_{\zeta}(k_j)P_{\zeta}(k_p)\right]}.~~~~~~ \eea
Additionally it is important to note that, in the non attractor regime of soft inflation the model exactly similar to the single field slow roll model of inflation, where it is a well known fact that the non-Gaussian parameter $\tau^{loc}_{NL}$
and $f^{loc}_{NL}$ are connected via the following constraint relationship:
\bea \tau^{loc}_{NL}=\frac{36}{25}(f^{loc}_{NL})^2,\eea
which is commonly known as \textcolor{blue}{\it Suyama-Yamaguchi} consistency relation. If this relation perfectly holds good in the present context, then one can easily 
get:
\bea \tau^{loc}_{NL}
&\approx&\frac{36}{144}\frac{1}{\left(\sum^{3}_{i=1}k^3_i\right)^2} \left[2(3\epsilon^*_{\tilde{W}}-\eta^{*}_{\tilde{W}})\sum^{3}_{i=1}k^3_i 
+\epsilon^*_{\tilde{W}}\left(-\sum^{3}_{i=1}k^3_i+\sum^{3}_{i,j=1,i\neq j}k_i k^2_j+\frac{8}{K}\sum^{3}_{i,j=1,i> j}k^2_i k^2_j\right)\right]^2.~~~~~~~\eea		  
from which one can find out the following expression for the normalization factor: 
\bea {\cal N}_{NORM}&=&\frac{25}{54}\frac{\left[\frac{25T(k_1,k_2,k_3,k_4)}{36(f^{loc}_{NL})^2}-\sum_{j<p,i\neq j,p}P_{\zeta}(k_{ij})P_{\zeta}(k_j)P_{\zeta}(k_p)\right]}{\sum_{i<j<p}P_{\zeta}(k_{i})P_{\zeta}(k_j)P_{\zeta}(k_p)},\eea
Here it is very easy to observe that the normalization factor is different for different shapes.

But in general always the connecting relationship between the non-Gaussian parameters 
$\tau^{loc}_{NL}$ and $f^{loc}_{NL}$ or more precisely the \textcolor{blue}{\it Suyama-Yamaguchi} consistency relation is not perfectly holds good as the cosmological perturbation during inflationary epoch is subject to quantum mechanical interference
effects at the time of horizon crossing and such prescriptions does not satisfy a simple type of parameterization in terms of momentum independent-coefficients in Fourier space. In that specific case one can write down the connecting relation between the non-Gaussian parameter $\tau^{loc}_{NL}$ and $g^{loc}_{NL}$ as:
\bea \label{cv3} g^{loc}_{NL}&=&f(k_1,k_2,k_3,k_4,k_{12},k_{14},k_{13})\tau^{loc}_{NL},~~~~~~~~~~~~~\eea
where one can choose the momentum dependent function $f(k_1,k_2,k_3,k_4,k_{12},k_{14},k_{13})$ as:
\bea f(k_1,k_2,k_3,k_4,k_{12},k_{14},k_{13})
&=&\frac{64}{\hat{K}^3}\sum_{i<j,m\neq i,j}k^3_ik^3_j\left(\frac{1}{k^3_{im}}+\frac{1}{k^3_{jm}}\right),\eea
which is motivated from the choice of the shape function for trispectrum. 

In this situation we get the following simplified expression for the non-Gaussian parameter $\tau^{loc}_{NL}$ and $g^{loc}_{NL}$ as obtained from the four point scalar function:
\bea \tau^{loc}_{NL}&=& \frac{T(k_1,k_2,k_3,k_4)}{\left[\sum_{j<p,i\neq j,p}P_{\zeta}(k_{ij})P_{\zeta}(k_j)P_{\zeta}(k_p)+\frac{54}{25}f(k_1,k_2,k_3,k_4,k_{12},k_{14},k_{13})\sum_{i<j<p}P_{\zeta}(k_{i})P_{\zeta}(k_j)P_{\zeta}(k_p)\right]},\nonumber\\ \\ g^{loc}_{NL}&=& \frac{f(k_1,k_2,k_3,k_4,k_{12},k_{14},k_{13})T(k_1,k_2,k_3,k_4)}{\left[\sum_{j<p,i\neq j,p}P_{\zeta}(k_{ij})P_{\zeta}(k_j)P_{\zeta}(k_p)+\frac{54}{25}f(k_1,k_2,k_3,k_4,k_{12},k_{14},k_{13})\sum_{i<j<p}P_{\zeta}(k_{i})P_{\zeta}(k_j)P_{\zeta}(k_p)\right]},\nonumber\\ \eea

In this context we denote the angle between two momentum vectors as:
\bea \cos\theta_{12}=\cos\theta_{34}\equiv\cos\theta_3,\\
\cos\theta_{23}=\cos\theta_{14}\equiv\cos\theta_1,\\ 
\cos\theta_{13}=\cos\theta_{24}\equiv\cos\theta_2, \eea
which satisfies the costraint condition, $\sum^{4}_{i<j=1}\cos\theta_{ij}=-2$
and can be equivalently written as, 
$\sum^{3}_{\alpha=1}\cos\theta_{\alpha}=\cos\theta_1 + \cos\theta_2 + \cos\theta_3 =-1$.
This comes as an outcome of conservation of momentum. Additionally here,
\bea k_{14}=k_{23}=|{\bf k}_{1}+{\bf k}_{4}|=|{\bf k}_{2}+{\bf k}_{3}|&=&\sqrt{k^2_1+k^2_4+2k_1k_4\cos\theta_1}=\sqrt{k^2_2+k^2_3+2k_2k_3\cos\theta_1}, \\ k_{24}=k_{13}=|{\bf k}_{2}+{\bf k}_{4}|=|{\bf k}_{1}+{\bf k}_{3}|&=&\sqrt{k^2_2+k^2_4+2k_2k_4\cos\theta_2}=\sqrt{k^2_1+k^2_3+2k_1k_3\cos\theta_2}, \\  k_{34}=k_{12}=|{\bf k}_{3}+{\bf k}_{4}|=|{\bf k}_{1}+{\bf k}_{2}|&=&\sqrt{k^2_3+k^2_4+2k_3k_4\cos\theta_3}=\sqrt{k^2_1+k^2_2+2k_1k_2\cos\theta_3}, \eea
Let us now concentrate on the following limiting configurations for the trispectrum to analyze the shape properly from the obtained results:
\begin{enumerate}
\item \underline{\textcolor{blue}{Equilateral limit configuration:}}\\
For this case we have \be |{\bf k}_1|=|{\bf k}_2|=|{\bf k}_3|=k=k_i~\forall i=1,2,3,4,\ee and  
 this implies:
\bea k_{ij}=|k_{i}+k_{j}|=\sqrt{2}k\sqrt{1+\cos\theta_{ij}}=\sqrt{2}k\sqrt{1+\cos\theta_{\alpha}},\forall (i,j=1,2,3,4)~{\rm with}~i<j, \alpha=1,2,3.~~~~~~~~~\eea
Additionally in the equilateral limit configuration, $\theta=\theta_{\alpha}\forall \alpha=1,2,3$.
further using the constraint condition as stated in Eq~(\ref{ w2}), we get:
\be \cos\theta=\cos\theta_{i}=-\frac{1}{3}~~~\forall i=1,2,3.\ee
and further using these results trispectrum for scalar fluctuation can be written as:
\bea \label{ww} T(k,k,k,k)
&\approx&\frac{3H^6}{8M^6_p (\epsilon^*_{\tilde{W}})^2}\frac{1}{k^{12}}\left[\hat{G}^{S}({\bf k},{\bf k},{\bf k},{\bf k})-\hat{W}^{S}({\bf k},{\bf k},{\bf k},{\bf k})-2\hat{R}^{S}({\bf k},{\bf k},{\bf k},{\bf k})\right],~~~~~~~~~~~~\eea
where the momentum dependent functions $\hat{G}^{S}({\bf k},{\bf k},{\bf k},{\bf k})$, $\hat{W}^{S}({\bf k},{\bf k},{\bf k},{\bf k})$ and $\hat{R}^{S}({\bf k},{\bf k},{\bf k},{\bf k})$ in the equilateral limit configuration are defined as:
\bea\label{w1} \hat{G}^{S}({\bf k},{\bf k},{\bf k},{\bf k})&=&0, \\ \label{ w2}\hat{R}^{S}({\bf k},{\bf k},{\bf k},{\bf k})
&=&-\frac{1}{48}k^3,\\
\label{w3} \hat{W}^{S}({\bf k},{\bf k},{\bf k},{\bf k})
&=& 0\eea
where we use $\hat{K}=4k$.
Also the momentum dependent functions $A_{1}({\bf k},{\bf k},{\bf k},{\bf k})$, $A_{2}({\bf k},{\bf k},{\bf k},{\bf k})$ and $A_{3}({\bf k},{\bf k},{\bf k},{\bf k})$ are defined as:
\bea A_{1}({\bf k},{\bf k},{\bf k},{\bf k})&=&-\frac{7}{24}k^4,~~
 A_{2}({\bf k},{\bf k},{\bf k},{\bf k})=\frac{1}{4}k^5,~~
 A_{3}({\bf k},{\bf k},{\bf k},{\bf k})=\frac{7}{3}k^6.\eea
Substituting Eq.~(\ref{w1}) and Eq.~(\ref{w2}) and Eq.~(\ref{w3}) in Eq~(\ref{ww}), we get the following simplified expression for the trispectrum for scalar fluctuation:
\bea \label{w} T(k,k,k,k)&=&\frac{H^6}{64M^6_p (\epsilon^*_H)^2}\frac{1}{k^9}\approx\frac{\tilde{W}^3(\phi_{cmb},{\bf \Psi})}{1728M^{12}_p (\epsilon^*_{\tilde{W}})^2}\frac{1}{k^9},~~~~~~~~~~~~\eea
     \begin{figure*}[htb]
     \centering
     \subfigure[Range~I.]{
         \includegraphics[width=7.6cm,height=4.5cm] {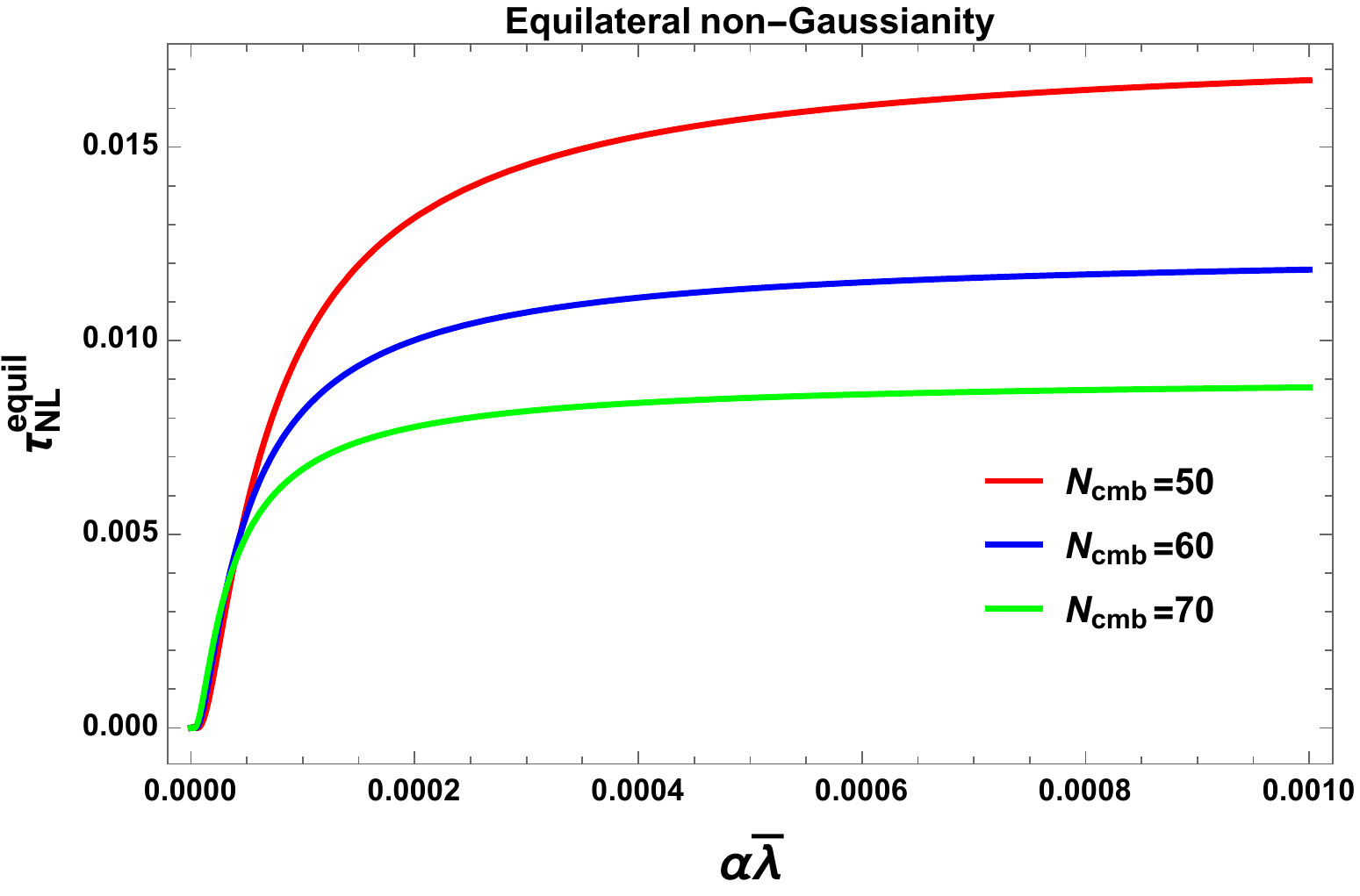}
         \label{fige1bbb}
     }
     \subfigure[Range~II.]{
         \includegraphics[width=7.6cm,height=4.5cm] {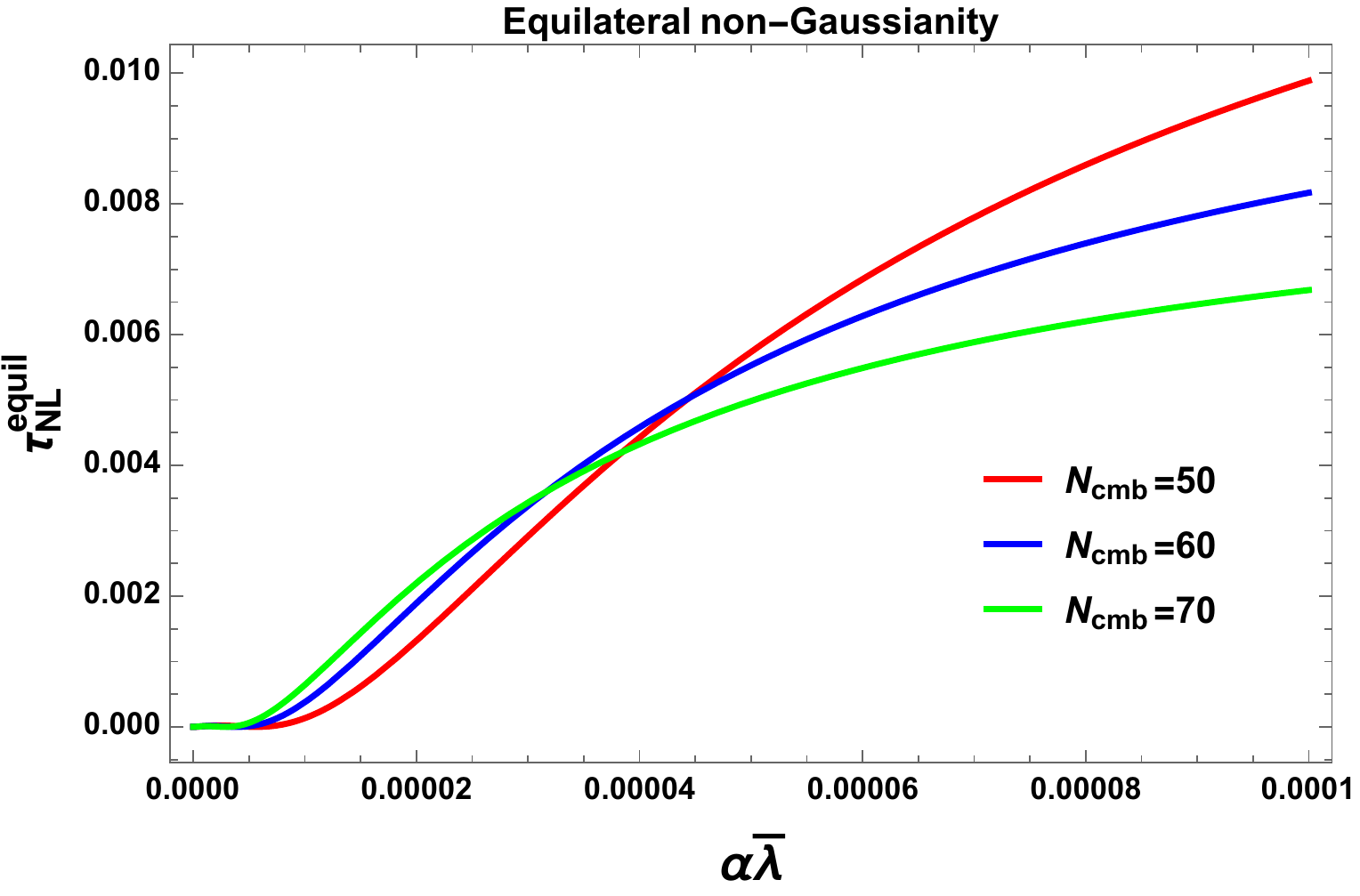}
         \label{fige2bbb}
     }
     \subfigure[Rangle~III.]{
              \includegraphics[width=7.6cm,height=4.5cm] {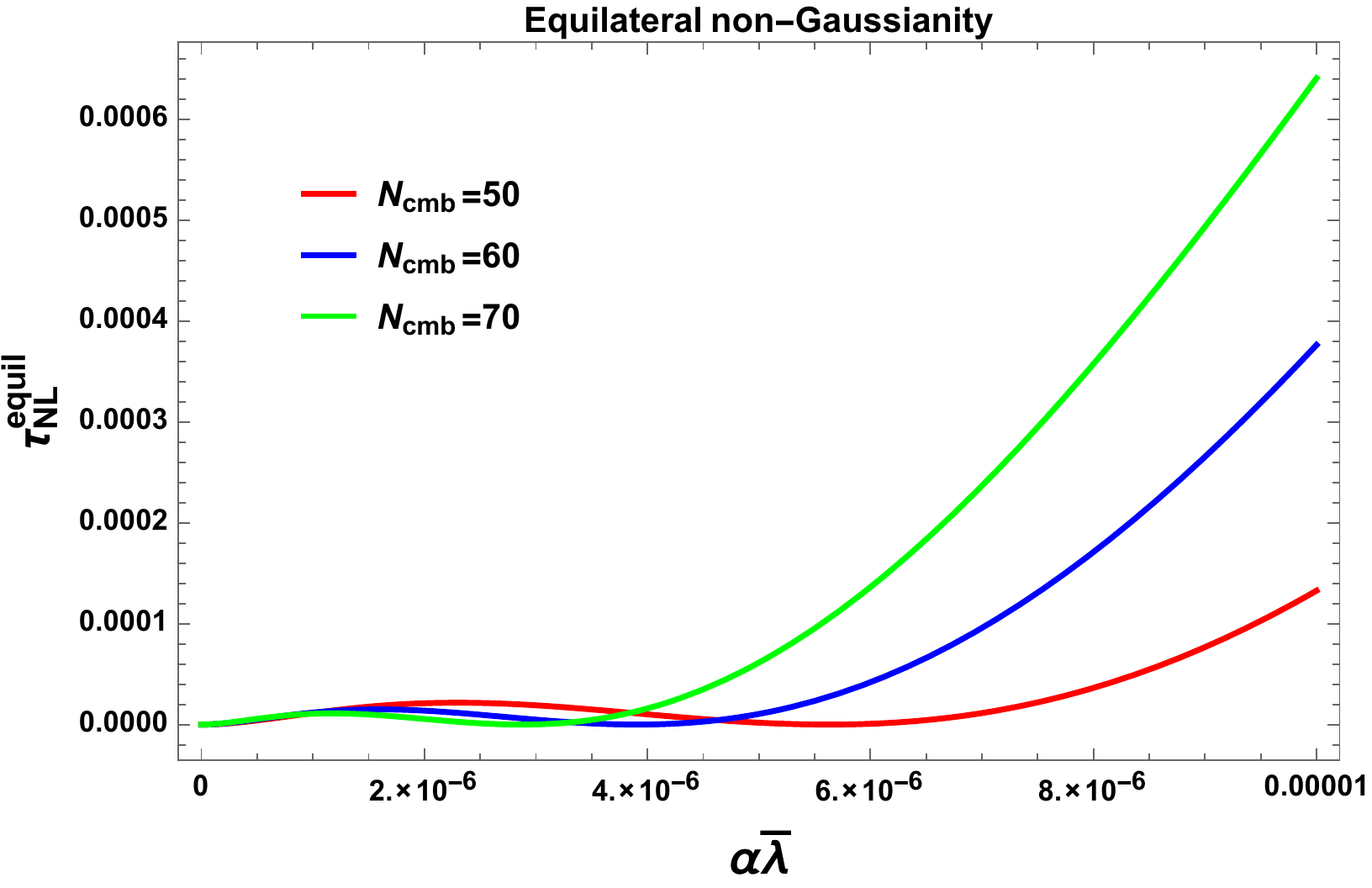}
              \label{fige3bbb}
          }
          \subfigure[Range~IV.]{
                        \includegraphics[width=7.6cm,height=4.5cm] {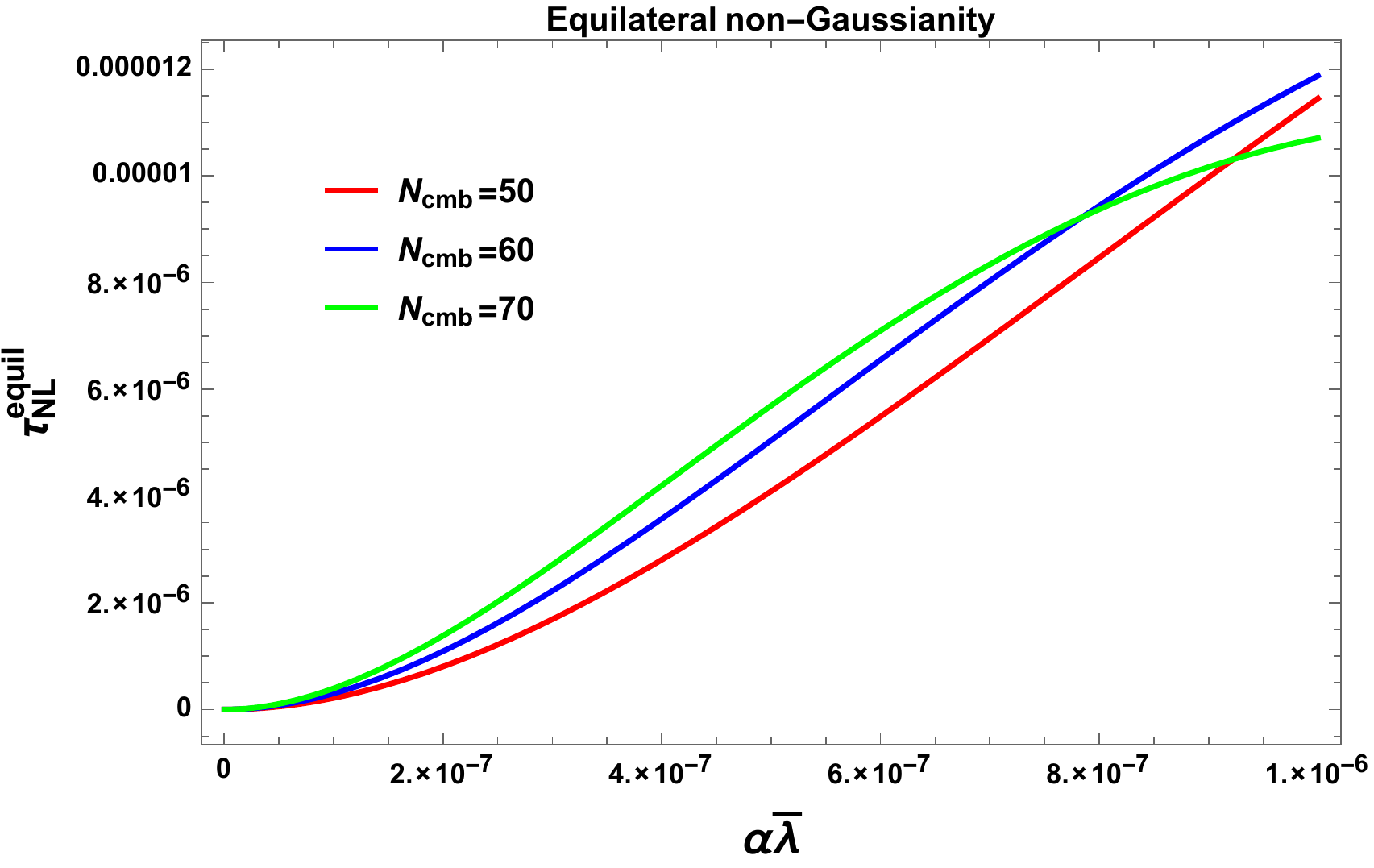}
                        \label{fige4bbb}
                    }
     \caption[Optional caption for list of figures]{Representative diagram for equilateral non-Gaussian three point amplitude vs product of the parameters $\alpha\bar{\lambda}$ in four different region for ${\cal N}_{cmb}=50$ (red), ${\cal N}_{cmb}=60$ (blue) and ${\cal N}_{cmb}=70$ (green).} 
     \label{fnle}
     \end{figure*}
          \begin{figure*}[htb]
          \centering
          \subfigure[Angle~I.]{
              \includegraphics[width=7.6cm,height=4.5cm] {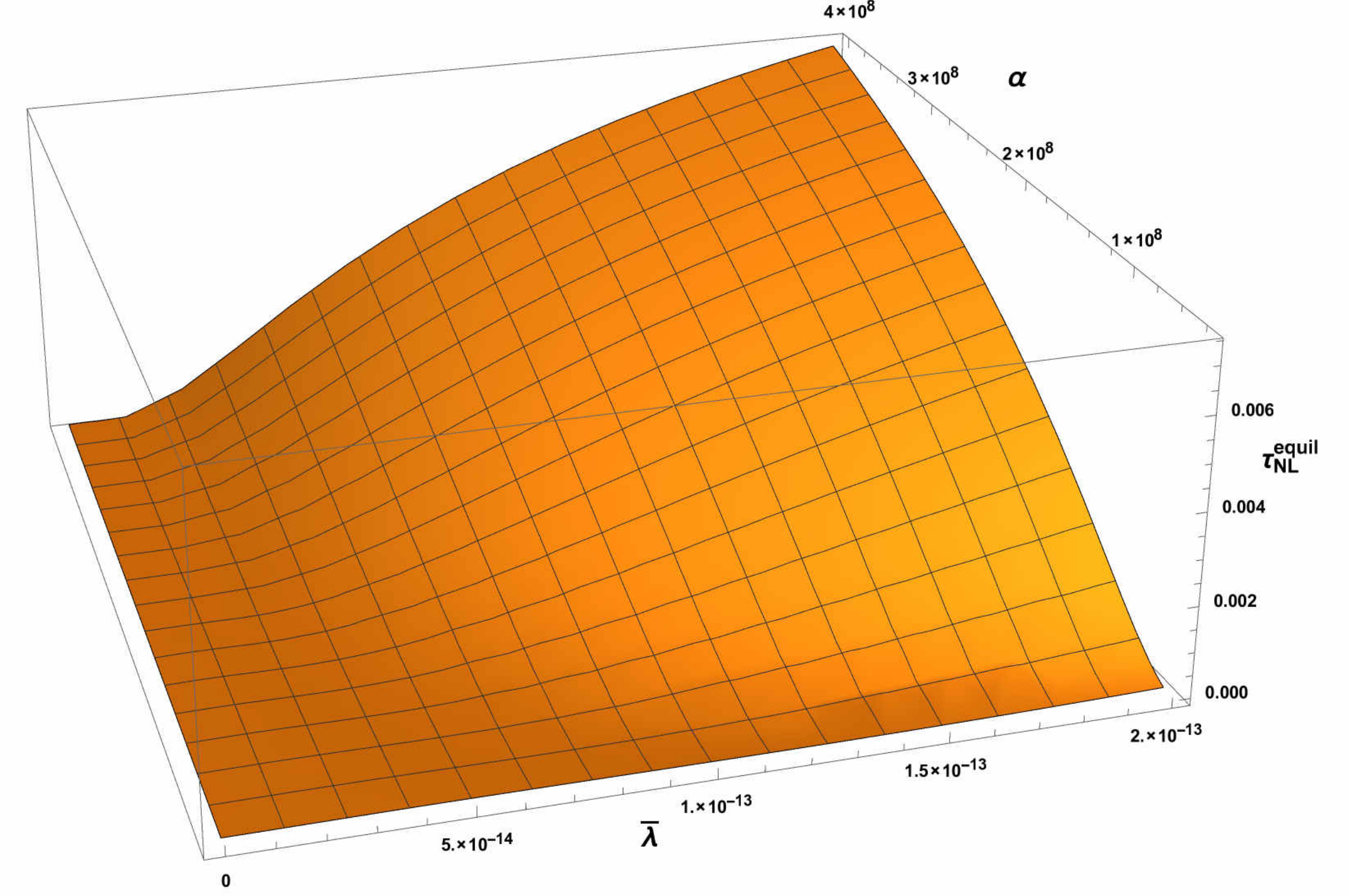}
              \label{tnl3d1}
          }
          \subfigure[Angle~II.]{
              \includegraphics[width=7.6cm,height=4.5cm] {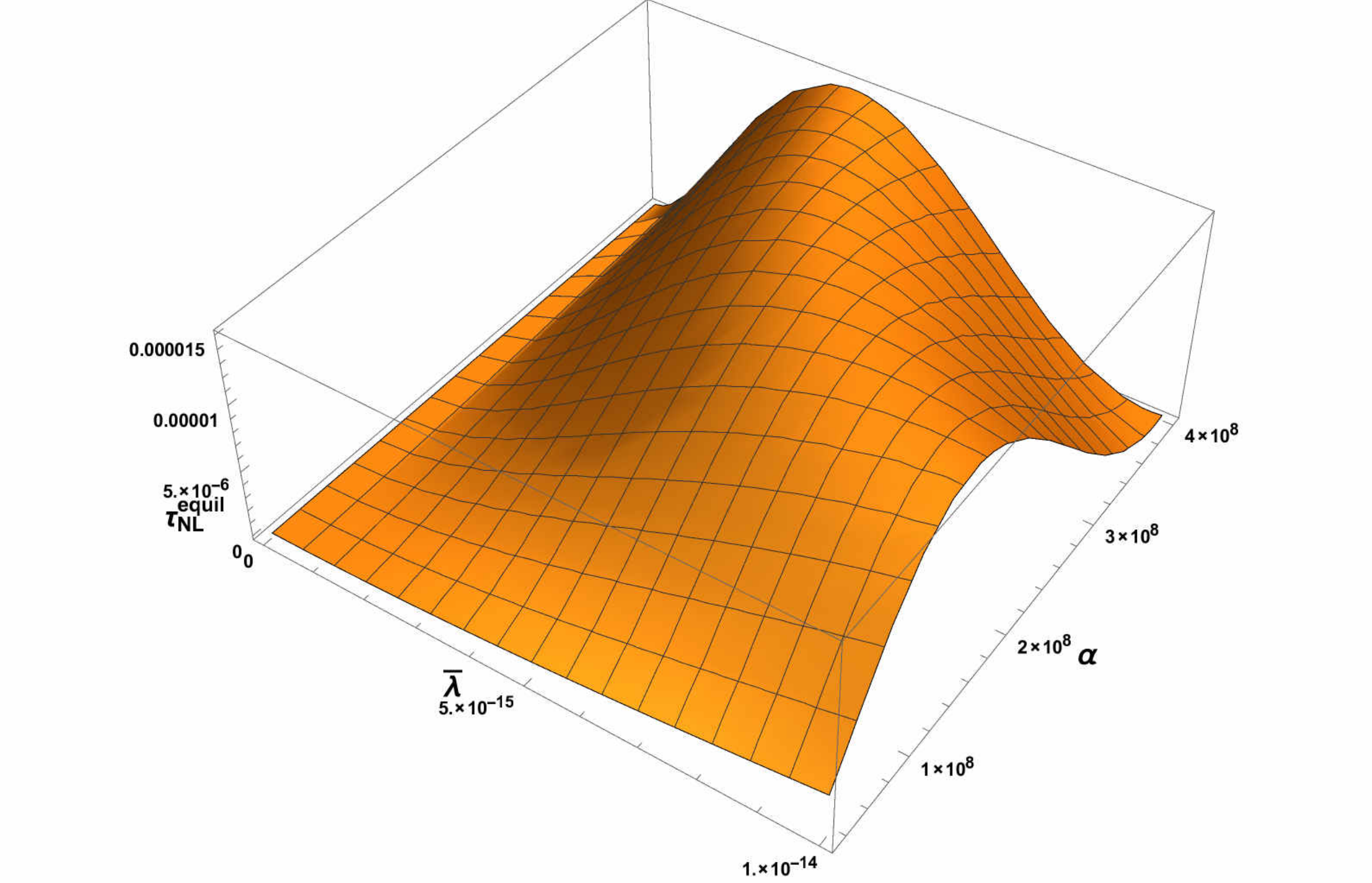}
              \label{tnl3d2}
          }
          \caption[Optional caption for list of figures]{Representative 3D diagram for equilateral non-Gaussian three point amplitude vs the model parameters $\alpha$ and $\bar{\lambda}$ for  ${\cal N}_{cmb}=60$ in two differenent angular views.} 
          \label{tnl3}
          \end{figure*}
       \begin{figure*}[htb]
       \centering
       \subfigure[Range~I.]{
           \includegraphics[width=7.6cm,height=4.5cm] {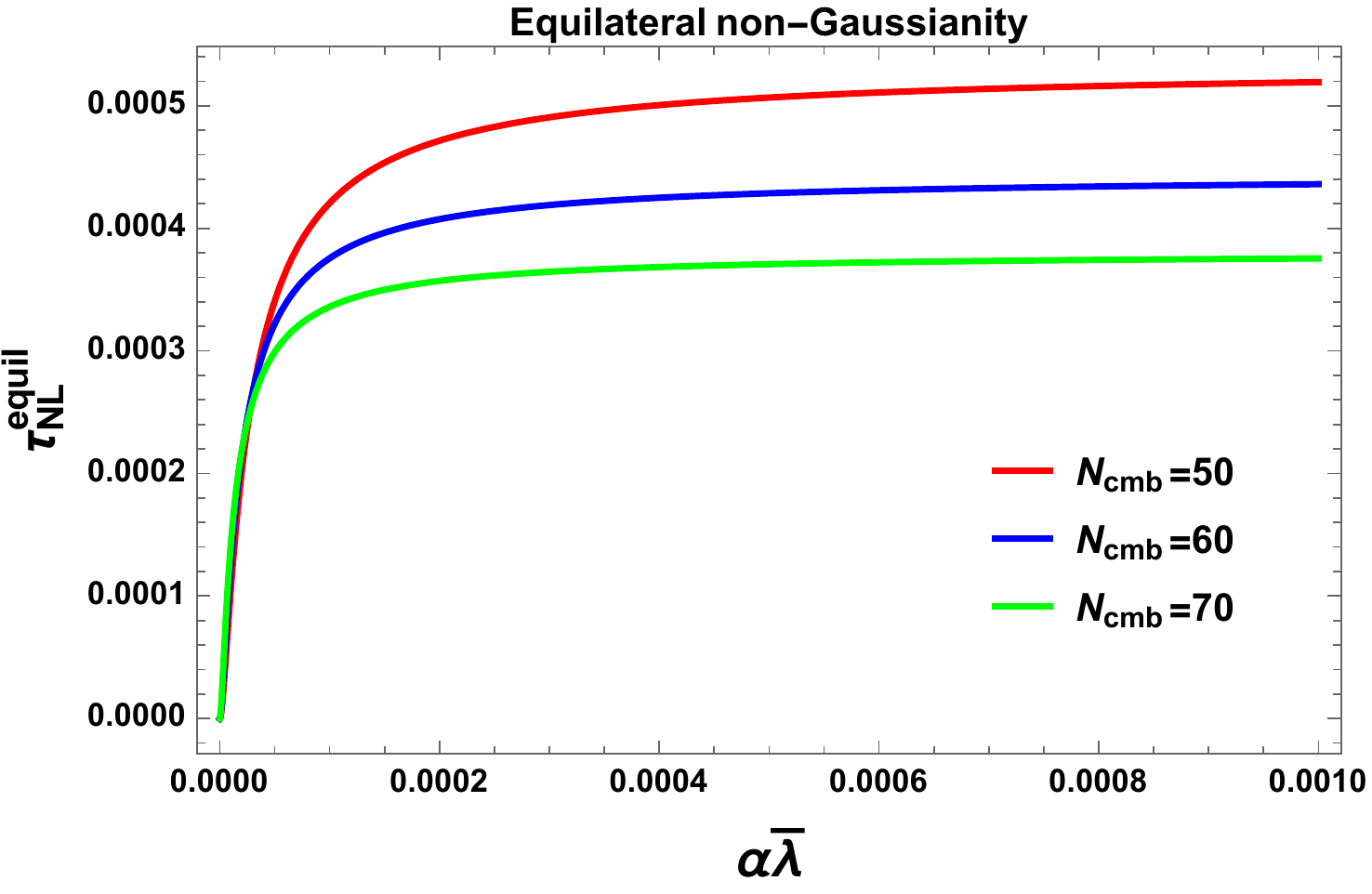}
           \label{fige1bbbv2}
       }
       \subfigure[Range~II.]{
           \includegraphics[width=7.6cm,height=4.5cm] {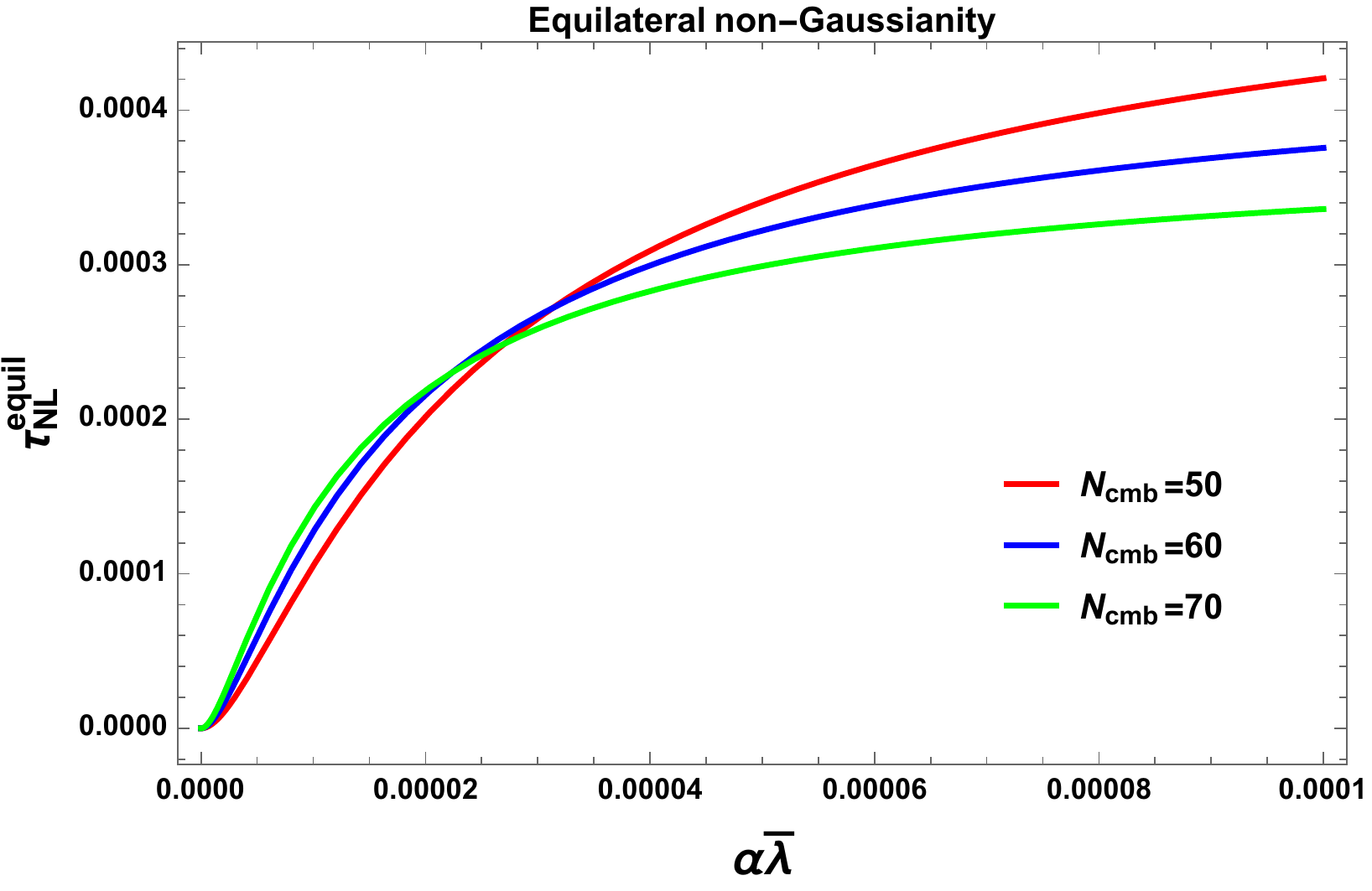}
           \label{fige2bbbv2}
       }
       \subfigure[Rangle~III.]{
                \includegraphics[width=7.6cm,height=4.5cm] {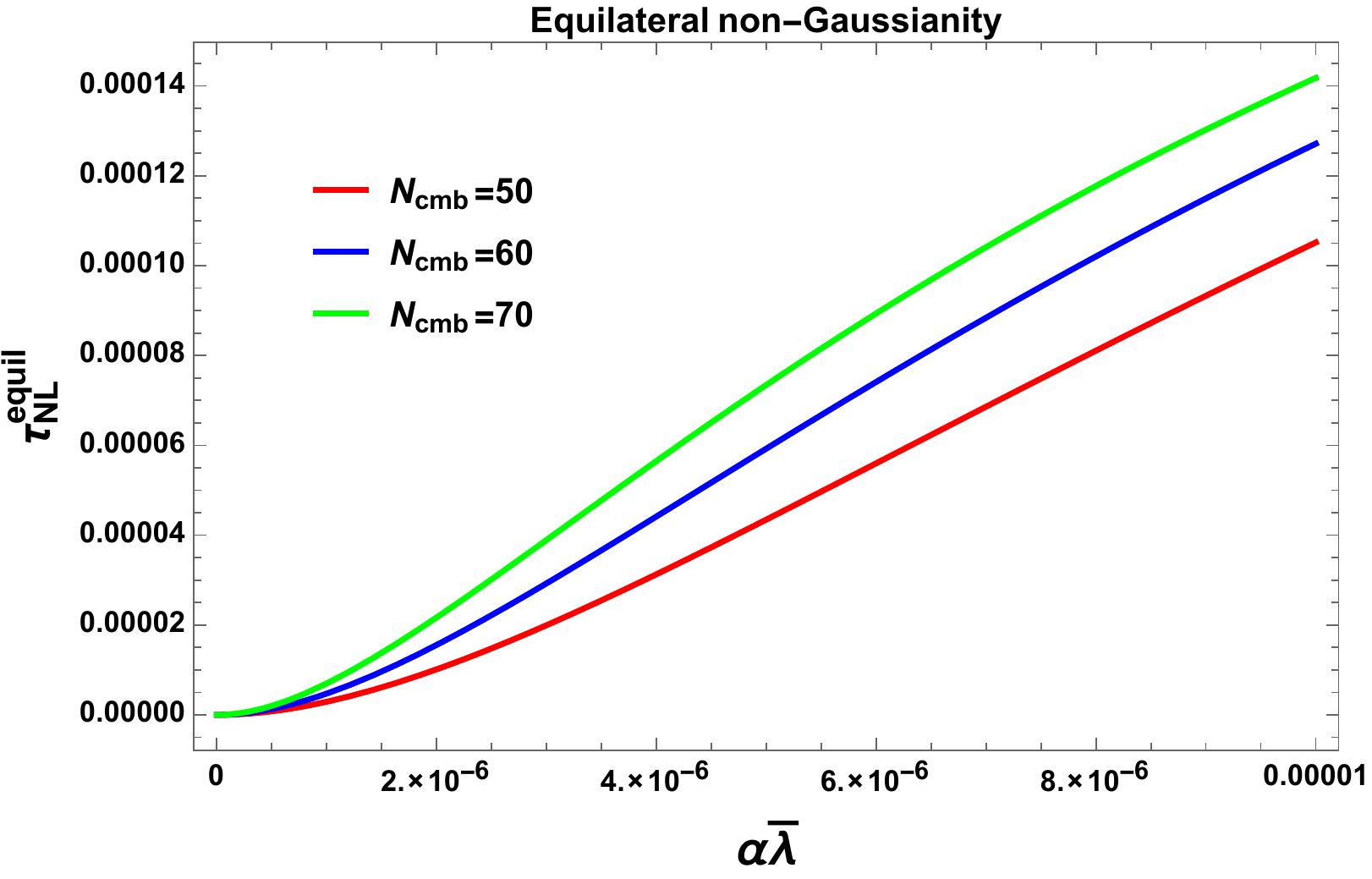}
                \label{fige3bbbv2}
            }
            \subfigure[Range~IV.]{
                          \includegraphics[width=7.6cm,height=4.5cm] {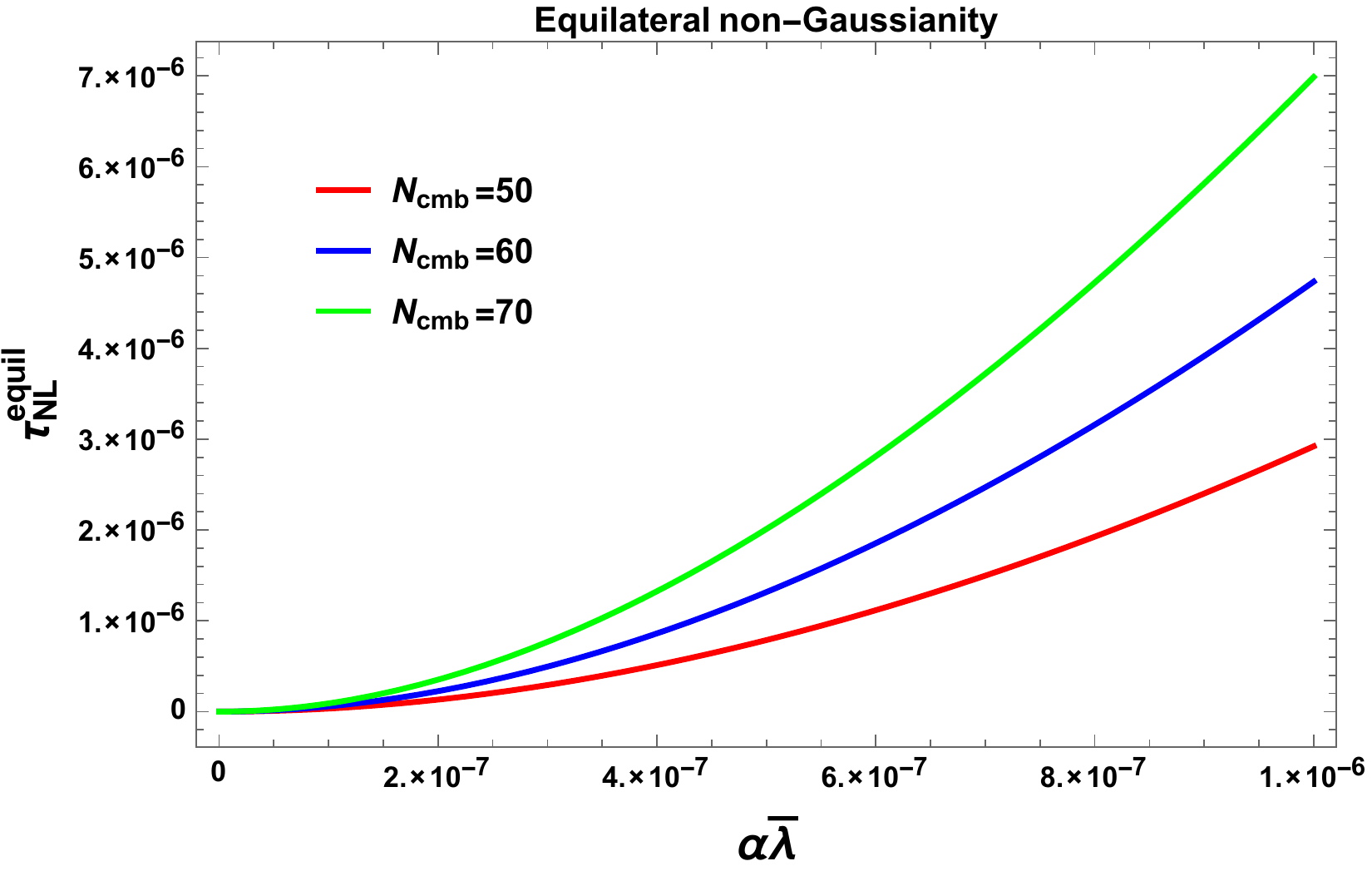}
                          \label{fige4bbbv2}
                      }
       \caption[Optional caption for list of figures]{Representative diagram for equilateral non-Gaussian three point amplitude vs product of the parameters $\alpha\bar{\lambda}$ in four different region for ${\cal N}_{cmb}=50$ (red), ${\cal N}_{cmb}=60$ (blue) and ${\cal N}_{cmb}=70$ (green).} 
       \label{fnlea}
       \end{figure*}
            \begin{figure*}[htb]
            \centering
            \subfigure[Angle~I.]{
                \includegraphics[width=7.6cm,height=4.5cm] {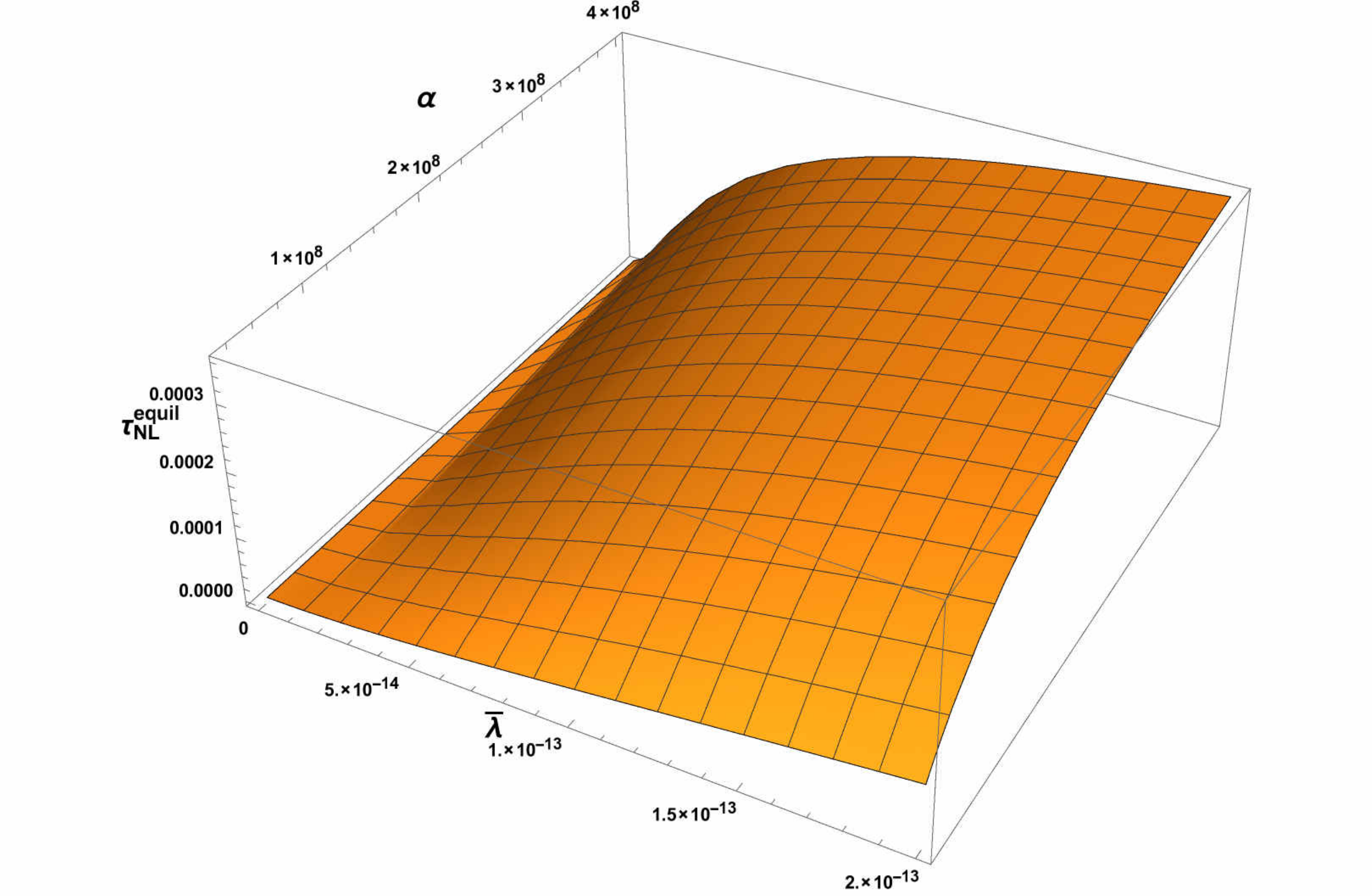}
                \label{tnl3d1a}
            }
            \subfigure[Angle~II.]{
                \includegraphics[width=7.6cm,height=4.5cm] {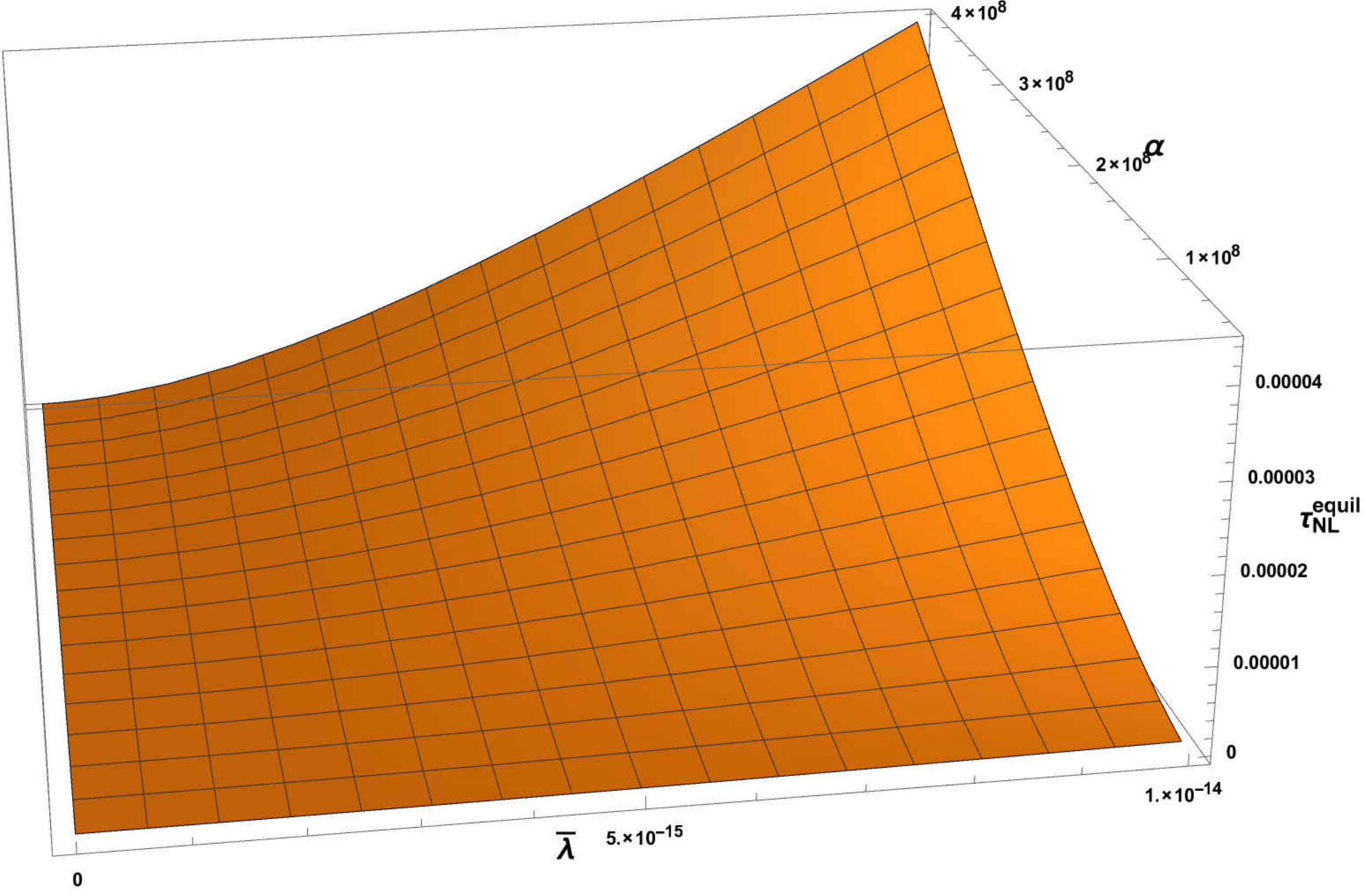}
                \label{tnl3d2a}
            }
            \caption[Optional caption for list of figures]{Representative 3D diagram for equilateral non-Gaussian three point amplitude vs the model parameters $\alpha$ and $\bar{\lambda}$ for  ${\cal N}_{cmb}=60$ in two different angular views.} 
            \label{tnl3a}
            \end{figure*}
                 \begin{figure*}[htb]
                 \centering
                 \subfigure[Range~I.]{
                     \includegraphics[width=7.6cm,height=4.5cm] {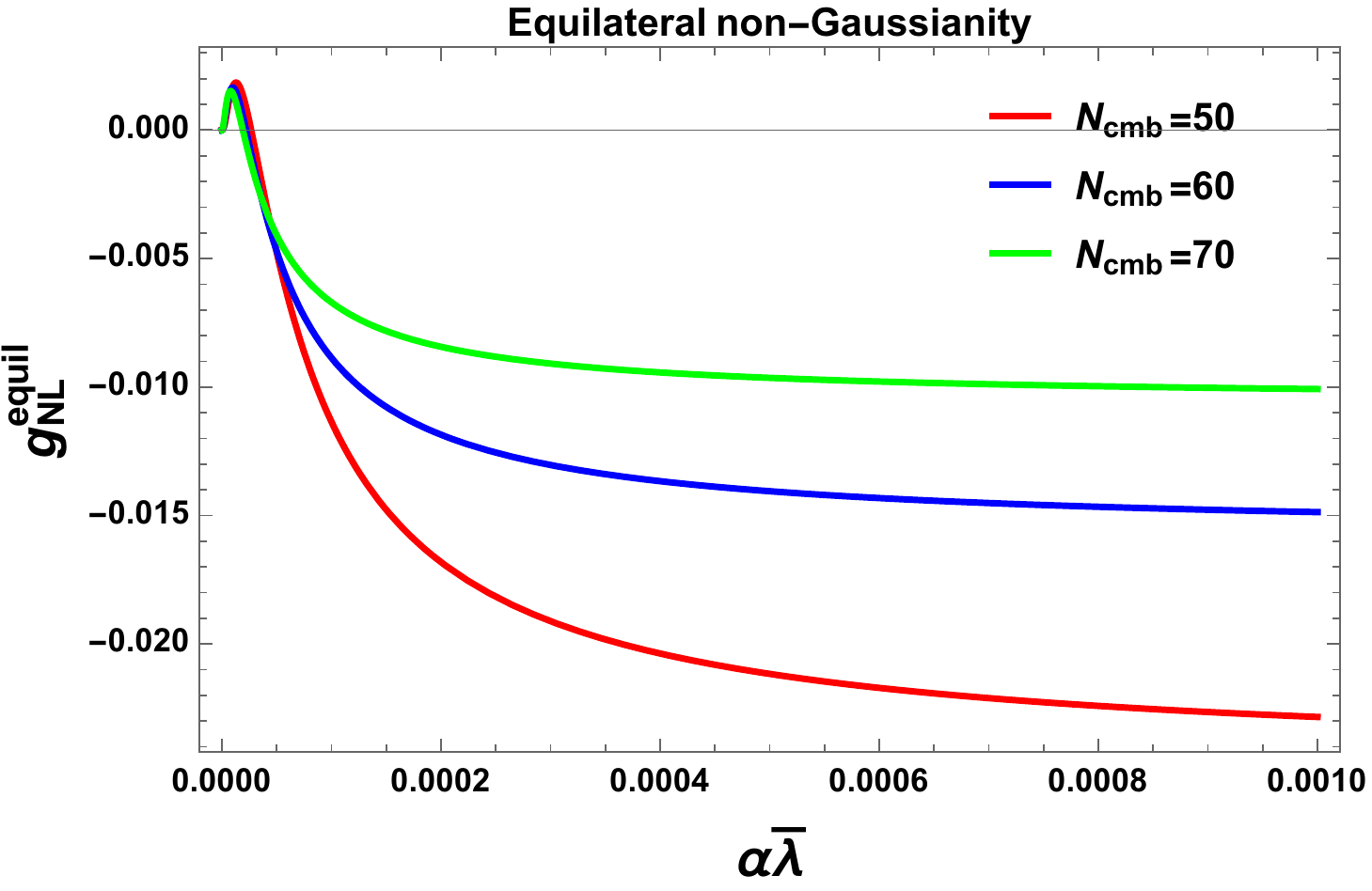}
                     \label{fige1bbbb}
                 }
                 \subfigure[Range~II.]{
                     \includegraphics[width=7.6cm,height=4.5cm] {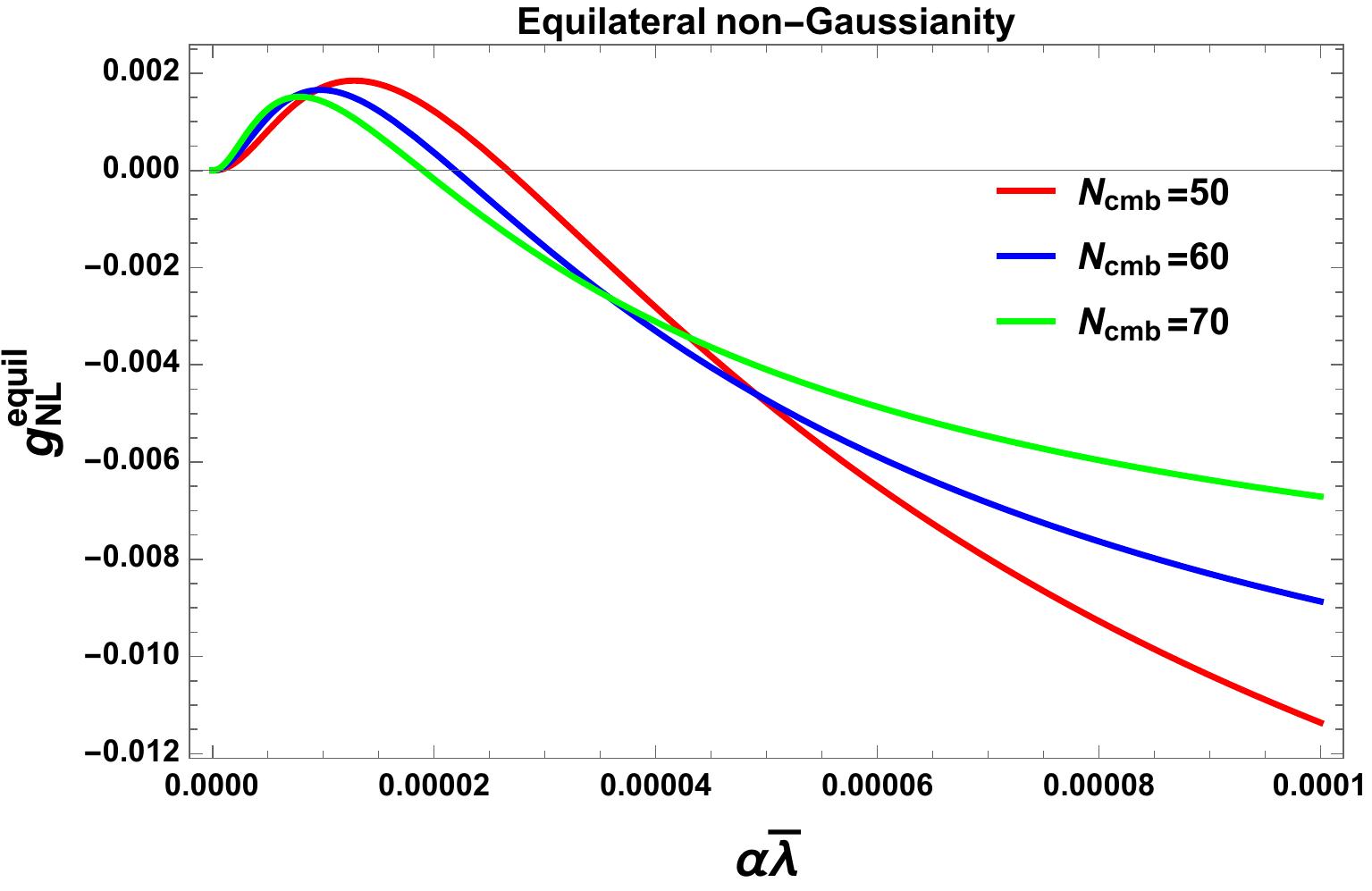}
                     \label{fige2bbbb}
                 }
                 \subfigure[Rangle~III.]{
                          \includegraphics[width=7.6cm,height=4.5cm] {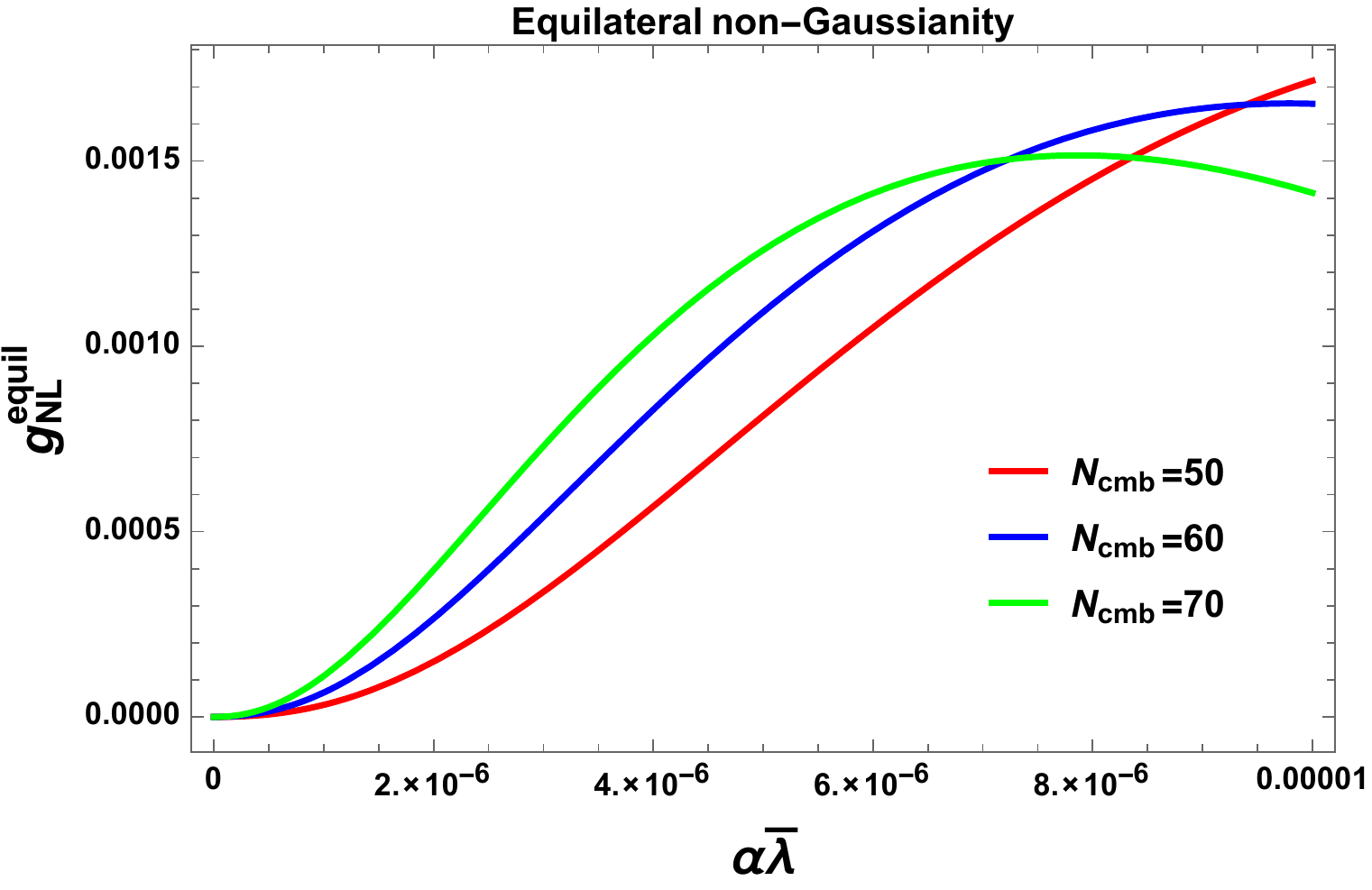}
                          \label{fige3bbbb}
                      }
                      \subfigure[Range~IV.]{
                                    \includegraphics[width=7.6cm,height=4.5cm] {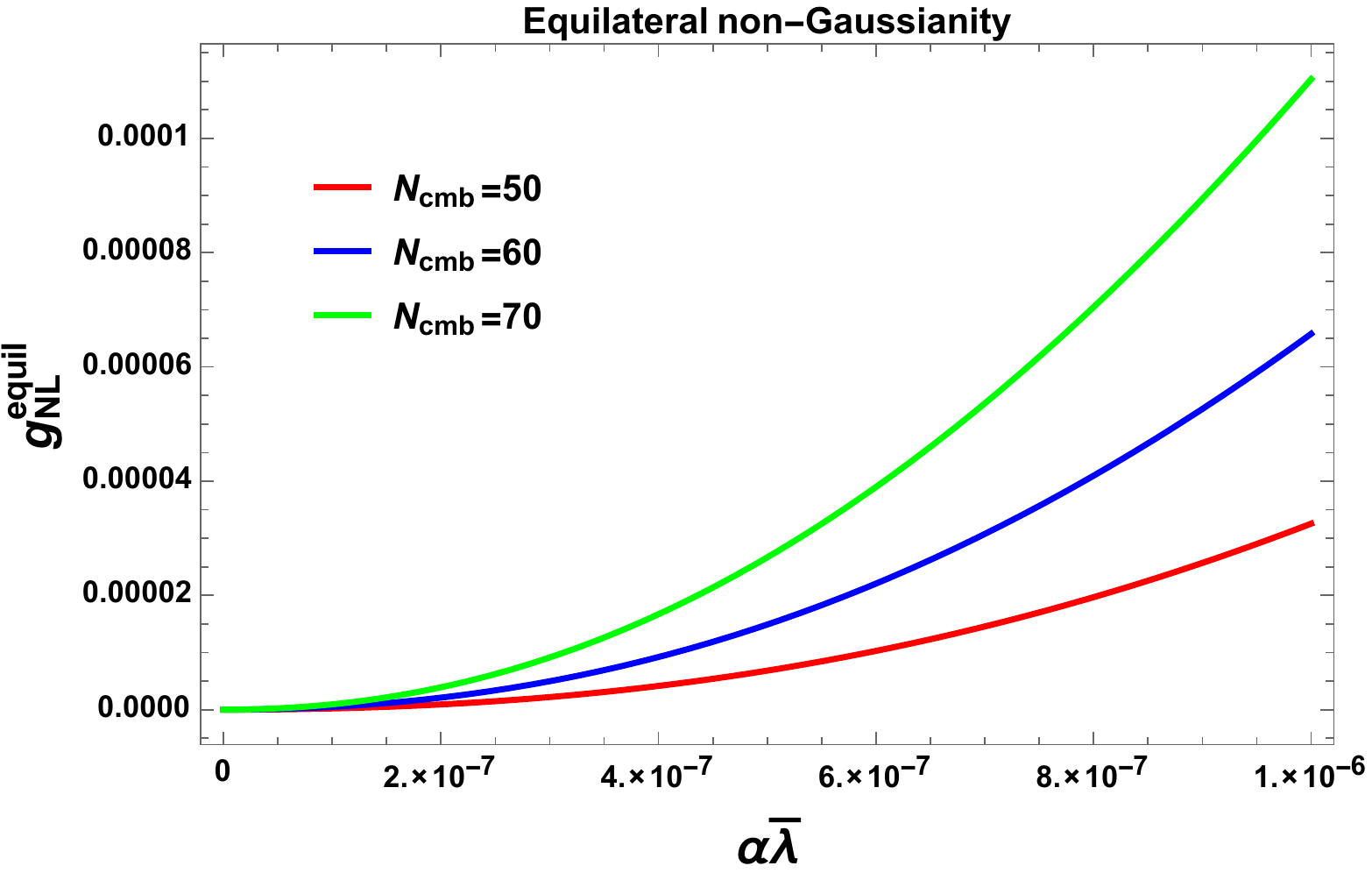}
                                    \label{fige4bbbb}
                                }
                 \caption[Optional caption for list of figures]{Representative diagram for equilateral non-Gaussian three point amplitude vs product of the parameters $\alpha\bar{\lambda}$ in four different region for ${\cal N}_{cmb}=50$ (red), ${\cal N}_{cmb}=60$ (blue) and ${\cal N}_{cmb}=70$ (green).}
                 \label{fnleaa}
                 \end{figure*}
                      \begin{figure*}[htb]
                      \centering
                      \subfigure[Angle~I.]{
                          \includegraphics[width=7.6cm,height=4.5cm] {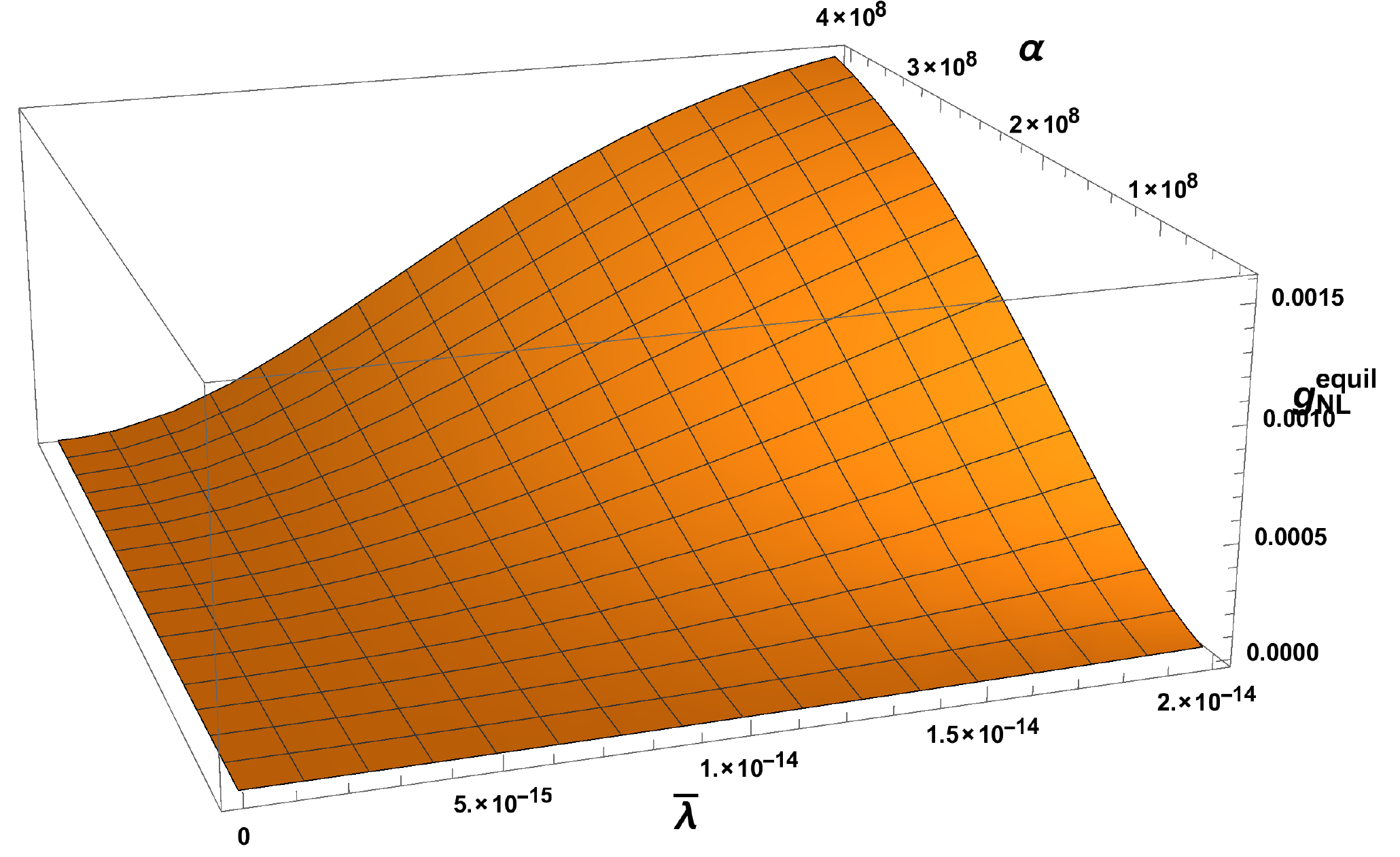}
                          \label{gnl3da}
                      }
                      \subfigure[Angle~II.]{
                          \includegraphics[width=7.6cm,height=4.5cm] {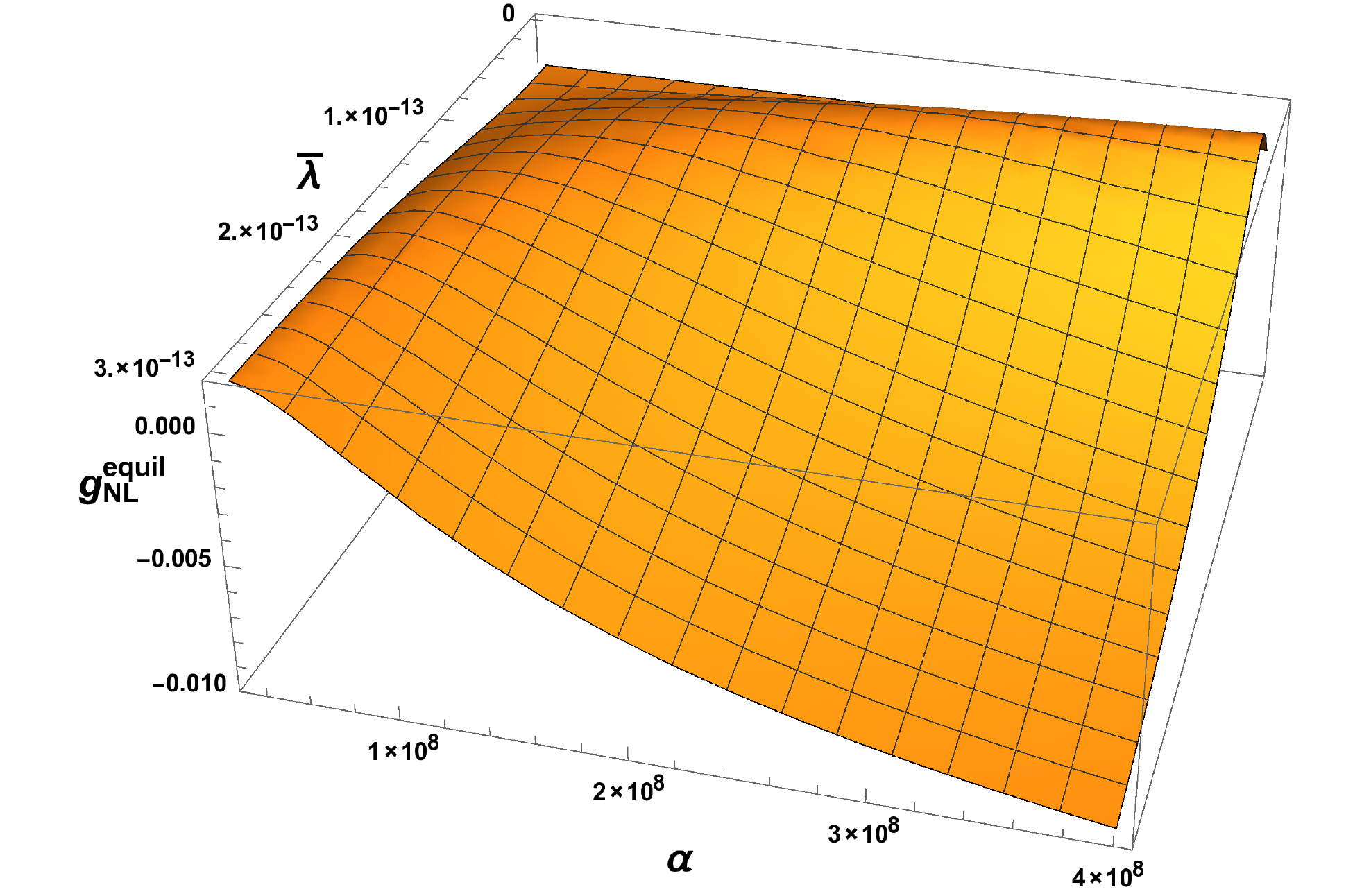}
                          \label{gnl3db}
                      }
                      \caption[Optional caption for list of figures]{Representative 3D diagram for equilateral non-Gaussian three point amplitude vs the model parameters $\alpha$ and $\bar{\lambda}$ for  ${\cal N}_{cmb}=60$ in two different angular views.} 
                      \label{gnl3aa}
                      \end{figure*}
                 \begin{figure*}[htb]
                 \centering
                 \subfigure[Range~I.]{
                     \includegraphics[width=7.6cm,height=4.5cm] {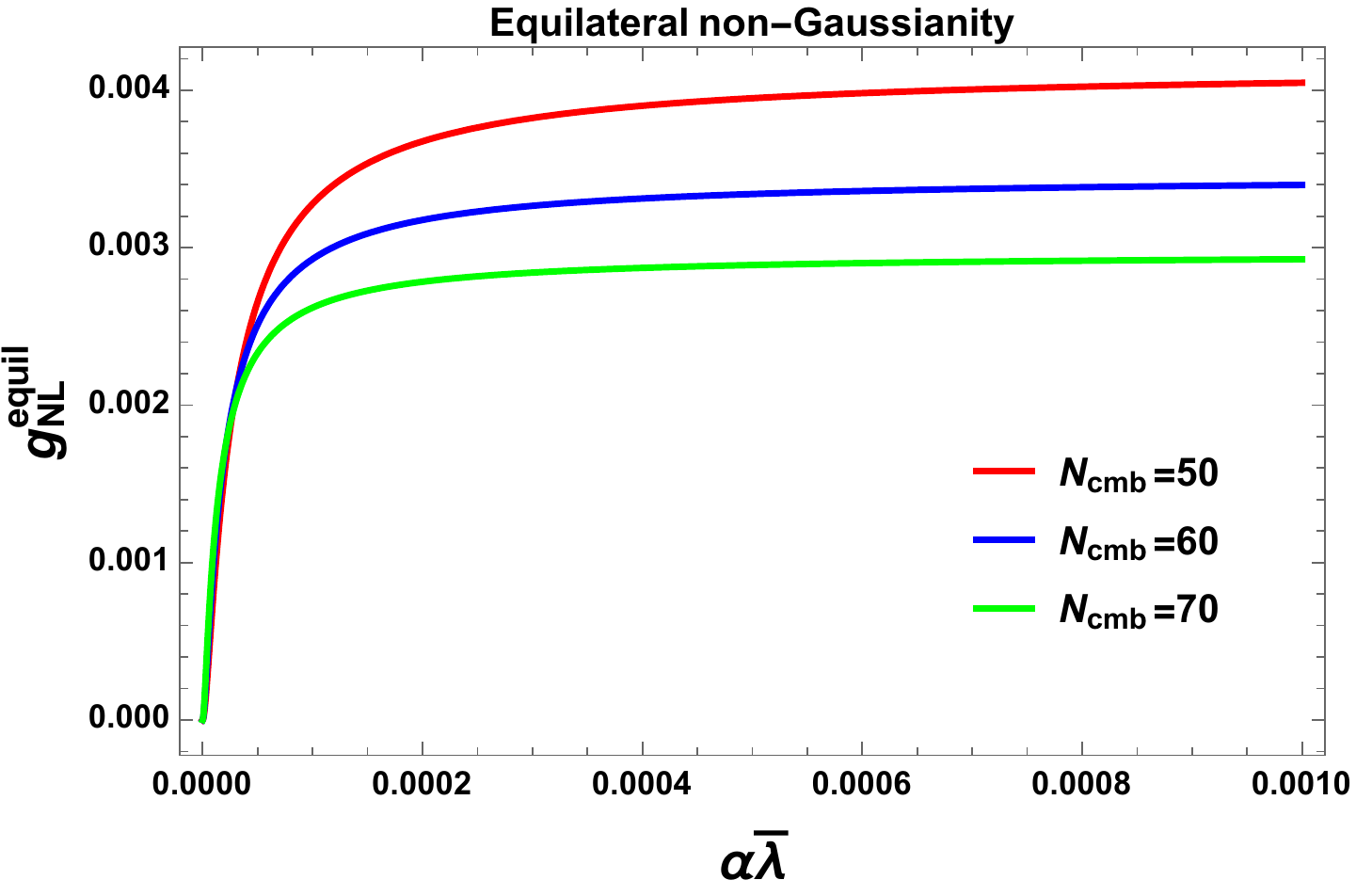}
                     \label{fige1bbbbb}
                 }
                 \subfigure[Range~II.]{
                     \includegraphics[width=7.6cm,height=4.5cm] {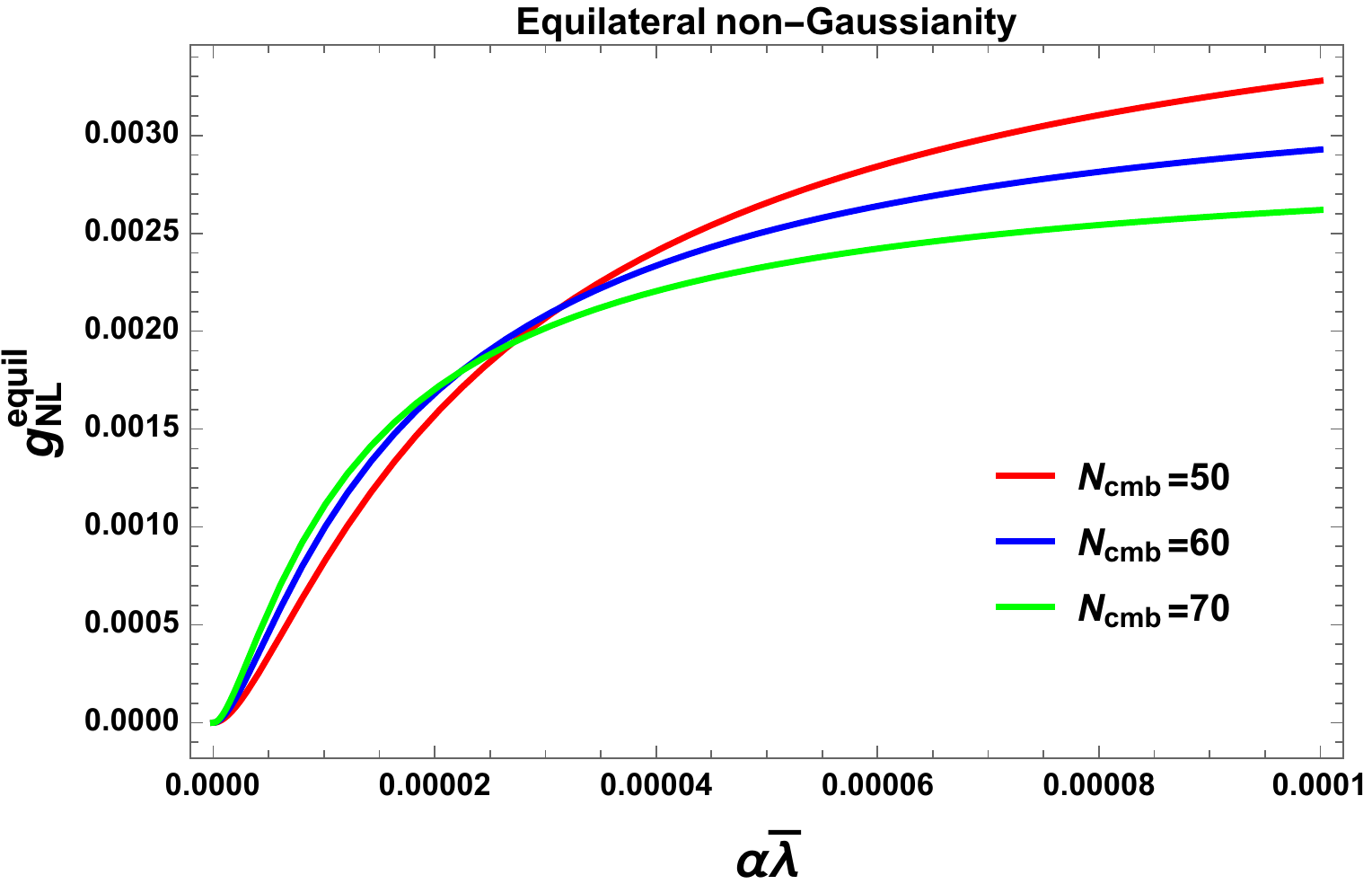}
                     \label{fige2bbbbb}
                 }
                 \subfigure[Rangle~III.]{
                          \includegraphics[width=7.6cm,height=4.5cm] {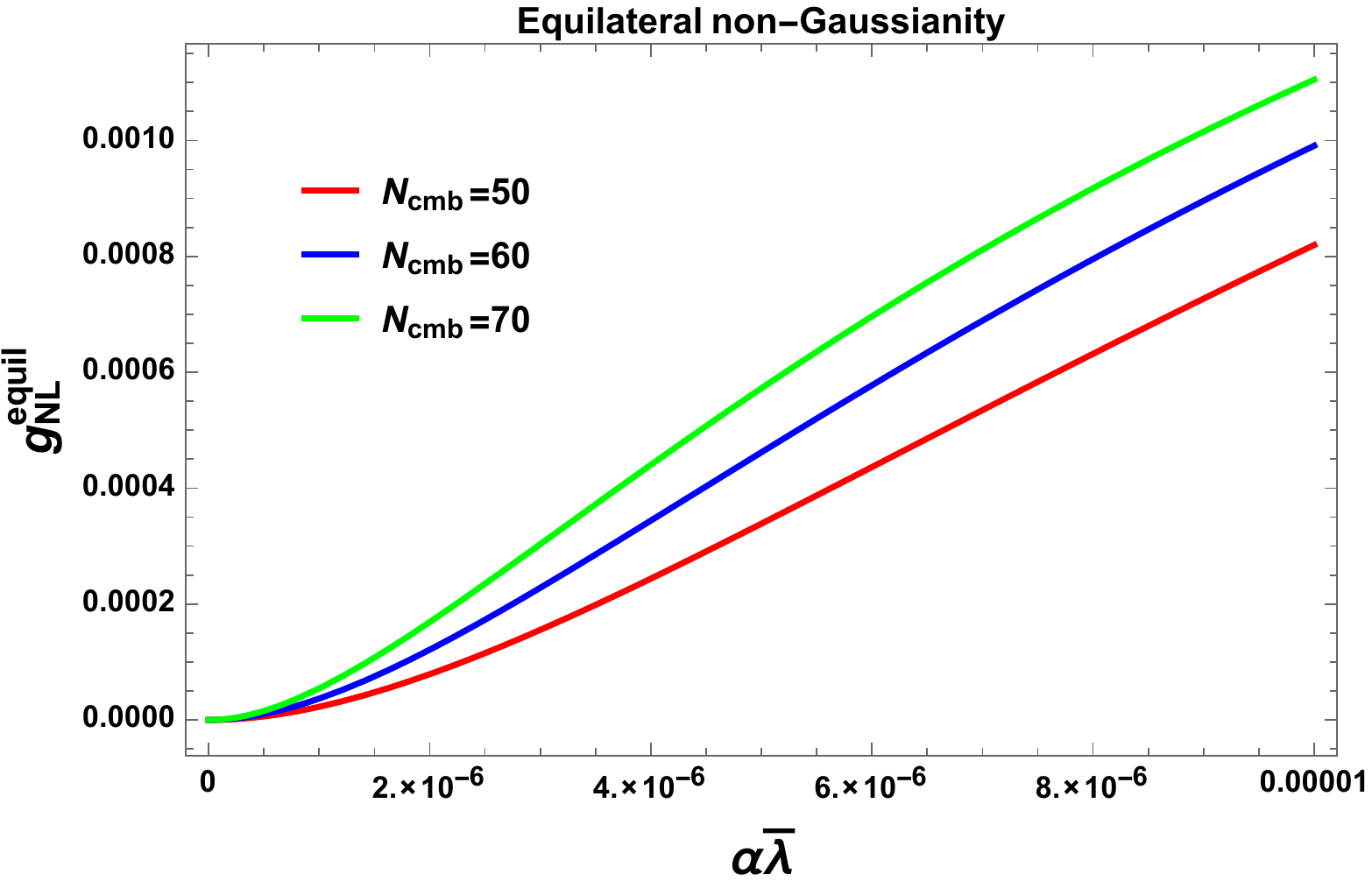}
                          \label{fige3bbbbb}
                      }
                      \subfigure[Range~IV.]{
                                    \includegraphics[width=7.6cm,height=4.5cm] {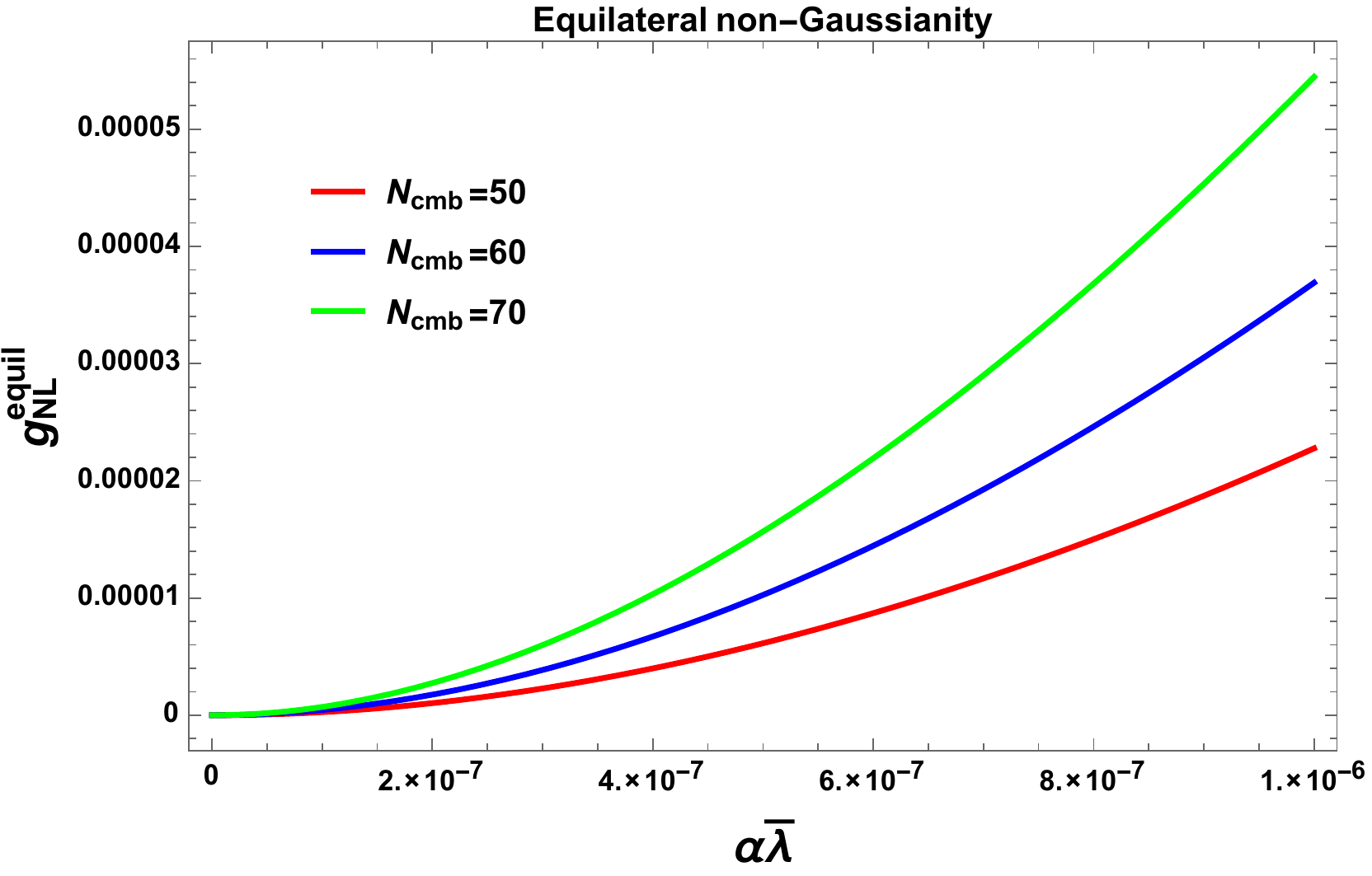}
                                    \label{fige4bbbbb}
                                }
                 \caption[Optional caption for list of figures]{Representative diagram for equilateral non-Gaussian three point amplitude vs product of the parameters $\alpha\bar{\lambda}$ in four different region for ${\cal N}_{cmb}=50$ (red), ${\cal N}_{cmb}=60$ (blue) and ${\cal N}_{cmb}=70$ (green).}
                 \label{fnleaaa}
                 \end{figure*}
                      \begin{figure*}[htb]
                      \centering
                      \subfigure[Angle~I.]{
                          \includegraphics[width=7.6cm,height=4.5cm] {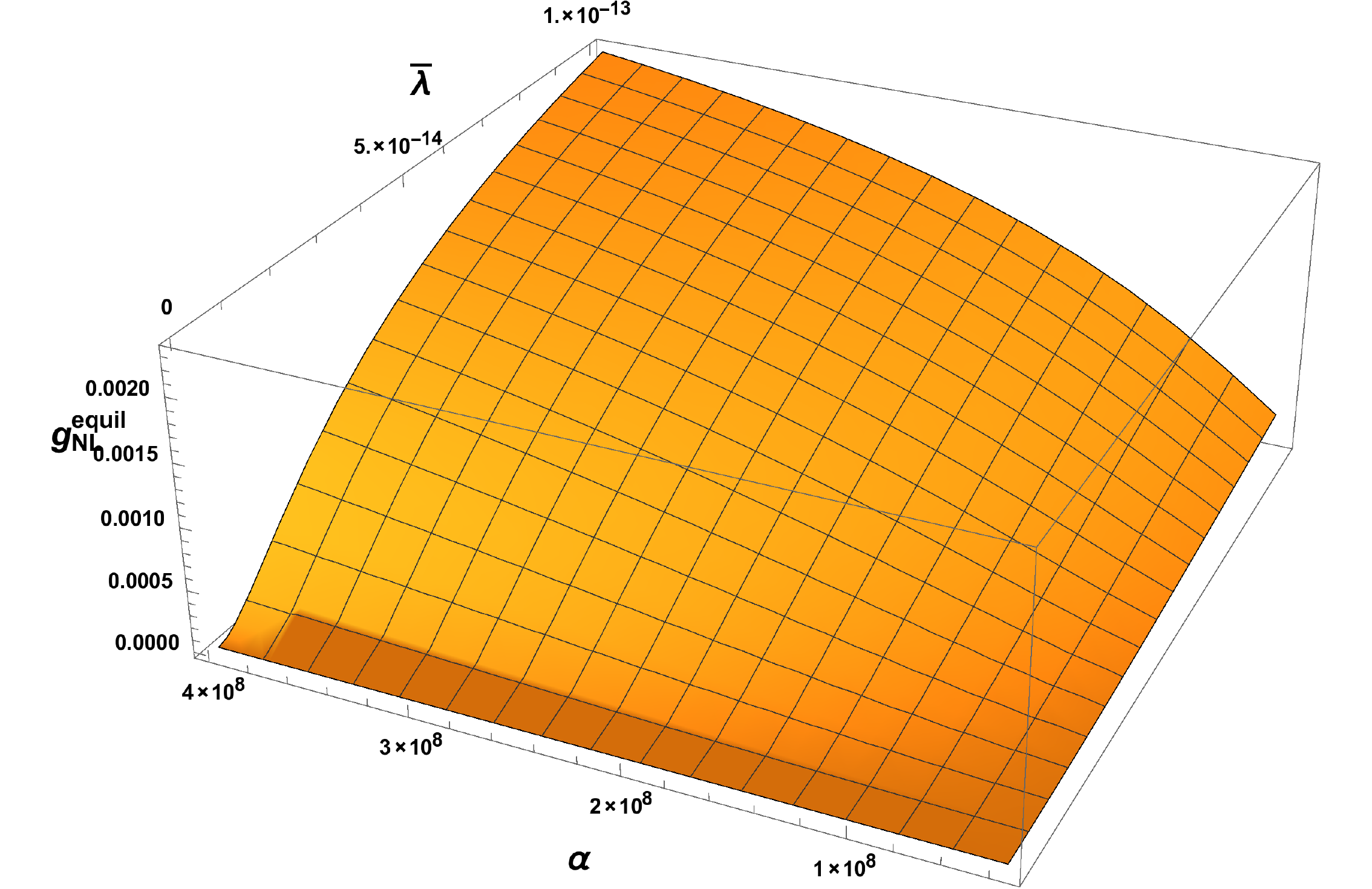}
                          \label{gnl3daa}
                      }
                      \subfigure[Angle~II.]{
                          \includegraphics[width=7.6cm,height=4.5cm] {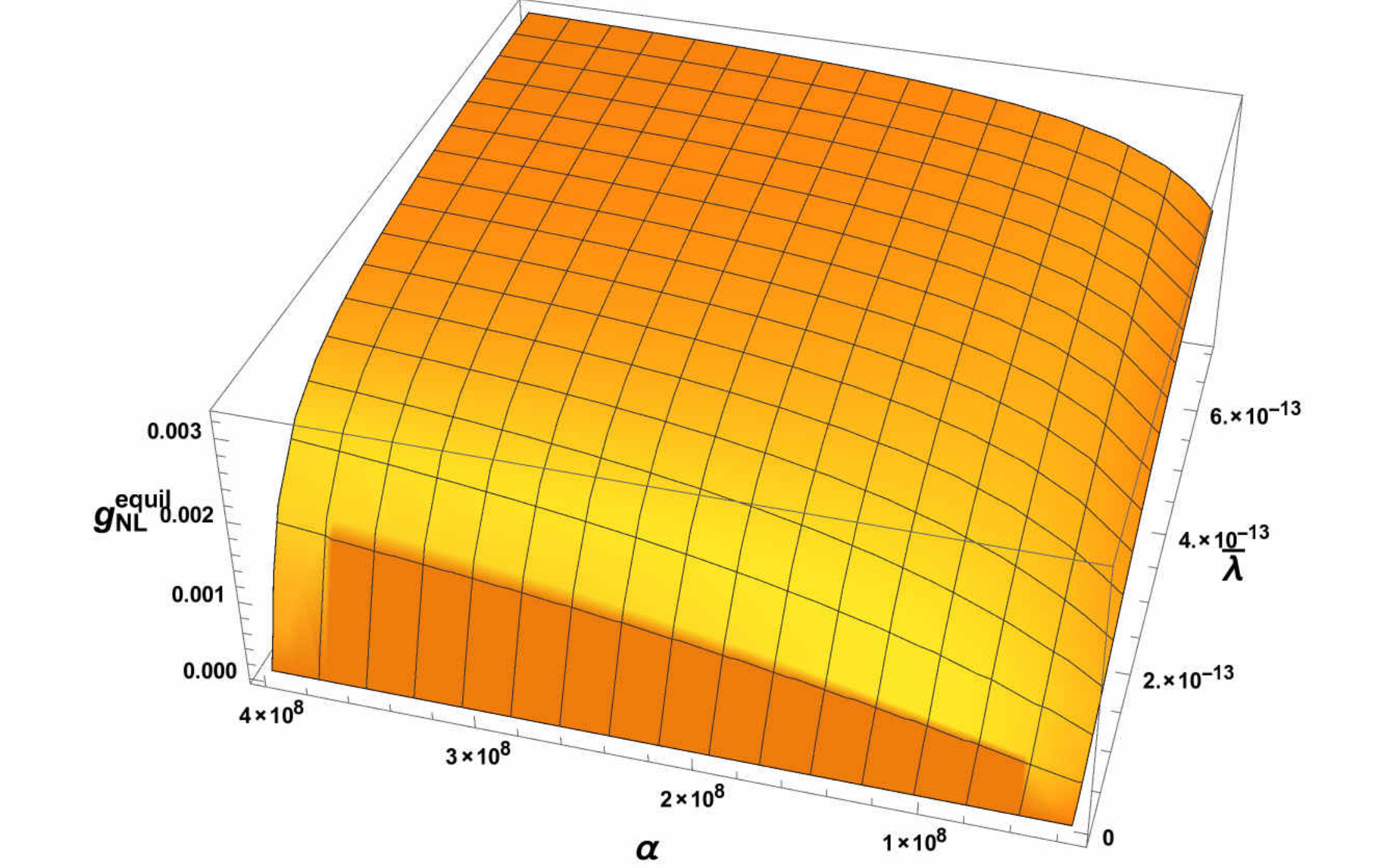}
                          \label{gnl3dbb}
                      }
                      \caption[Optional caption for list of figures]{Representative 3D diagram for equilateral non-Gaussian three point amplitude vs the model parameters $\alpha$ and $\bar{\lambda}$ for  ${\cal N}_{cmb}=60$ in two different angular views.} 
                      \label{gnl3zx}
                      \end{figure*}
Now if we assume that the non-Gaussian parameter $\tau^{equil}_{NL}$ and $f^{equil}_{NL}$ are connected through \textcolor{blue}{\it Suyama Yamaguchi} consistency relation, then we in the \textcolor{blue}{equlilateral limiting configuration} we get the following expression for the four point non-Gaussian parameter:
\bea \tau^{equil}_{NL}&\approx&\frac{1}{36} \left[29\epsilon^*_{\tilde{W}}-6\eta^*_{\tilde{W}} 
\right]^2.\eea 
 In this limiting configuration the normalization factor ${\cal N}_{NORM}$ that connects the two non-Gaussian parameters $\tau^{loc}_{NL}$ and $g^{loc}_{NL}$ computed from four point function as:
 \bea  {\cal N}_{NORM}&\approx& \left\{\frac{9\epsilon^*_H}{ \left[29\epsilon^*_{\tilde{W}}-6\eta^*_{\tilde{W}}
 \right]^2}-\frac{9\sqrt{3}}{8}\right\}. \eea   
 Consequently the non-Gaussian parameter $g^{loc}_{NL}$ can be express as:
 \bea \label{sd1} g^{equil}_{NL}&\approx&\frac{\left\{\frac{9\epsilon^*_{\tilde{W}}}{\left[29\epsilon^*_{\tilde{W}}-6\eta^*_{\tilde{W}}
     \right]^2}-\frac{9\sqrt{3}}{8}\right\}\epsilon^*_H}{\left[\frac{9\sqrt{3}}{8}+\frac{162}{25}\left\{\frac{9\epsilon^*_H}{\left[29\epsilon^*_{\tilde{W}}-6\eta^*_{\tilde{W}}
       \right]^2}-\frac{9\sqrt{3}}{8}\right\}\right]},\eea   
       or equivalently one can write the expression for non-Gaussian parameter $g^{loc}_{NL}$ as:
        \bea\label{sd2} g^{equil}_{NL}&\approx&\frac{1}{36}\left\{9\epsilon^*_{\tilde{W}}-\frac{9\sqrt{3}}{8}\left[29\epsilon^*_{\tilde{W}}-6\eta^*_{\tilde{W}}
                  \right]^2\right\},\eea 
       as we have assumed \textcolor{blue}{\it Suyama Yamaguchi} consistency relation
       perfectly holds good. Here it is important to mention that the, consistency of 
       the results obtained from Eq~(\ref{sd1}) and Eq~(\ref{sd2}) is perfectly consistent as in the leading order both of them predicts similar magnitude, which is proportional to the slow roll parameter $\epsilon^*_H$ or $\epsilon^*_{\tilde{W}}$.
       
          In fig.~(\ref{fnle}) and fig.~(\ref{fnleaa}), we have shown the features of non-Gaussian amplitude from four point scalar function $\tau^{equil}_{NL}$ and $g^{equil}_{NL}$ in equilateral limit configuration in four different scanning region of product of the two parameters $\alpha\bar{\lambda}$ in the $(\tau^{equil}_{NL},\alpha\bar{\lambda})$ and $(g^{equil}_{NL},\alpha\bar{\lambda})$ 2D plane for the number of e-foldings $50<{\cal N}_{cmb}<70$. Physical explanation of the obtained features are appended following:-
            \begin{itemize}
            \item \textcolor{red}{\underline{Region~I}:} \\
            Here for the parameter space $0.0001<\alpha\bar{\lambda}<0.001$ the non-Gaussian amplitude lying within the window $0.006<\tau^{equil}_{NL}<0.016$,$-0.004<g^{equil}_{NL}<-0.023$. Further if we increase the numerical value of $\alpha\bar{\lambda}$, then the magnitude of the non-Gaussian amplitude saturates and we get maximum value for ${\cal N}_{cmb}=50$, 
           $|\tau^{equil}_{NL}|_{max}\sim 0.016$, 
             $|g^{equil}_{NL}|_{max}\sim 0.023$.
            \item \textcolor{red}{\underline{Region~II}}\\
                 Here for the parameter space $0.00001<\alpha\bar{\lambda}<0.0001$ the non-Gaussian amplitude lying within the window $ 0.001<\tau^{equil}_{NL}<0.009$, $
                 0.002<g^{equil}_{NL}<-0.011$. In this region we get maximum value for ${\cal N}_{cmb}=50$,
                 $|\tau^{equil}_{NL}|_{max}\sim 0.009$,
                 $|g^{equil}_{NL}|_{max}\sim 0.011$.
                 Additionally it is important to note that, in this case for $\alpha\bar{\lambda}=0.00004$ the lines obtained for ${\cal N}_{cmb}=50$, ${\cal N}_{cmb}=60$ and ${\cal N}_{cmb}=70$ cross each other.
            \item \textcolor{red}{\underline{Region~III}}\\
                      Here for the parameter space $ 0.000001<\alpha\bar{\lambda}<0.00001$ the non-Gaussian amplitude lying within the window $ 0.00002<\tau^{equil}_{NL}<0.00062$, $
                      0.0001<g^{equil}_{NL}<0.0017$. In this region we get maximum value for ${\cal N}_{cmb}=70$,
                      $|\tau^{equil}_{NL}|_{max}\sim 0.00062$,
                      $|g^{equil}_{NL}|_{max}\sim 0.0017$.
                      Additionally it is important to note that, in this case for $0.000003\leq \alpha\bar{\lambda}\leq 0.000006$ the lines obtained for ${\cal N}_{cmb}=50$, ${\cal N}_{cmb}=60$ and ${\cal N}_{cmb}=70$ cross cross each other and then show increasing behaviour.
            \item \textcolor{red}{\underline{Region~IV}}\\
                           Here for the parameter space $ 0.0000001<\alpha\bar{\lambda}<0.000001$ the non-Gaussian amplitude lying within the window $10^{-6}<\tau^{equil}_{NL}<0.000012$, 
                          ~$2\times 10^{-6}<g^{equil}_{NL}<0.00011$. In this region we get maximum value for ${\cal N}_{cmb}=60$, 
                           $|\tau^{equil}_{NL}|_{max}\sim 0.000012$,
                           $|g^{equil}_{NL}|_{max}\sim 0.00011$.
            \end{itemize} 
            Further combining the contribution from \textcolor{red}{\underline{Region~I}}, \textcolor{red}{\underline{Region~II}}, \textcolor{red}{\underline{Region~III}} and   \textcolor{red}{\underline{Region~IV}} we finally get the following constraint on the four point non-Gaussian amplitude in the equilateral limit configuration:
            \bea \textcolor{red}{\underline{\rm Region~I}}+\textcolor{red}{\underline{\rm Region~II}}+\textcolor{red}{\underline{\rm Region~III}}+\textcolor{red}{\underline{\rm Region~IV}:}~~10^{-6}<\tau^{equil}_{NL}<0.016,~~
            -0.023<g^{equil}_{NL}<0.002~~~~~~~~~~
            \eea
            for the following parameter space:
            \bea \textcolor{red}{\underline{\rm Region~I}}+\textcolor{red}{\underline{\rm Region~II}}+\textcolor{red}{\underline{\rm Region~III}}+\textcolor{red}{\underline{\rm Region~IV}:}~~~~~0.0000001<\alpha\bar{\lambda}<0.001.~~~~~~~
                 \eea
                 In this analysis we get the following maximum value of the three point non-Gaussian amplitude in the equilateral limit configuration as given by:
            \bea |\tau^{equil}_{NL}|_{max}\sim 0.016,~~~ |g^{equil}_{NL}|_{max}\sim 0.002.\eea
            To visualize these constraints more clearly we have also presented $(\tau^{equil}_{NL},\alpha,\bar{\lambda})$ and $(g^{equil}_{NL},\alpha,\bar{\lambda})$ 3D plot in fig.~(\ref{tnl3d1}),  fig.~(\ref{tnl3d1}), fig.~(\ref{gnl3da}) and  fig.~(\ref{gnl3db}) for two different angular orientations as given by \textcolor{red}{\underline{Angle~I}} and \textcolor{red}{\underline{Angle~II}}. From the the representative surfaces it is clearly observed the behavior of three point non-Gaussian amplitude in the equilateral limit for the variation of two fold parameter $\alpha$ and $\bar{\lambda}$ and the results are consistent with the obtained constarints in 2D analysis. Here all the obtained results are consistent with the two point and three point constaints as well as with the Planck 2015 data \cite{Ade:2015lrj,Ade:2015ava,Ade:2015xua}.

       But as we have already pointed that if we relax the assumption of holding the \textcolor{blue}{\it Suyama Yamaguchi} consistency relation in the present context of discussion, then using Eq~(\ref{ w2}) one can write down the expression for momentum dependent function $f(k,k,k,k,\frac{2}{\sqrt{3}}k,\frac{2}{\sqrt{3}}k,\frac{2}{\sqrt{3}}k)$ in the  \textcolor{blue}{equlilateral limiting configuration} as,  
  $f\left(k,k,k,k,\frac{2}{\sqrt{3}}k,\frac{2}{\sqrt{3}}k,\frac{2}{\sqrt{3}}k\right)=\frac{9\sqrt{3}}{2}$, 
 using which we get the following simplified expression for the non-Gaussian parameter $\tau^{equil}_{NL}$ and $g^{equil}_{NL}$ as obtained from the four point scalar function
 in \textcolor{blue}{equlilateral limiting configuration} as: 
  \bea \tau^{equil}_{NL}&=& \frac{50\sqrt{3}}{6507}\epsilon^*_H\approx \frac{50\sqrt{3}}{6507}\epsilon^*_{\tilde{W}},~~~~ g^{equil}_{NL}= \frac{25}{241}\epsilon^*_H\approx  \frac{25}{241}\epsilon^*_{\tilde{W}}. \eea  
    In fig.~(\ref{fnlea}) and fig.~(\ref{fnleaaa}), without assuming the \textcolor{blue}{\it Suyama Yamaguchi} consistency relation we have shown the features of non-Gaussian amplitude from four point scalar function $\tau^{equil}_{NL}$ and $g^{equil}_{NL}$ in equilateral limit configuration in four different scanning region of product of the two parameters $\alpha\bar{\lambda}$ in the $(\tau^{equil}_{NL},\alpha\bar{\lambda})$ and $(g^{equil}_{NL},\alpha\bar{\lambda})$ 2D plane for the number of e-foldings $50<{\cal N}_{cmb}<70$. Physical explanation of the obtained features are appended following:-
                 \begin{itemize}
                 \item \textcolor{red}{\underline{Region~I}:} \\
                 Here for the parameter space $0.0001<\alpha\bar{\lambda}<0.001$ the non-Gaussian amplitude lying within the window $ 0.00028<\tau^{equil}_{NL}<0.00052$,~$
                 0.0022<g^{equil}_{NL}<0.004$. Further if we increase the numerical value of $\alpha\bar{\lambda}$, then the magnitude of the non-Gaussian amplitude saturates and we get maximum value for ${\cal N}_{cmb}=50$,
                 $|\tau^{equil}_{NL}|_{max}\sim 0.00052$,$|g^{equil}_{NL}|_{max}\sim 0.004$.
                 \item \textcolor{red}{\underline{Region~II}}\\
                      Here for the parameter space $0.00001<\alpha\bar{\lambda}<0.0001$ the non-Gaussian amplitude lying within the window $ 0.00005<\tau^{equil}_{NL}<0.00042$,~$
                      0.0005<g^{equil}_{NL}<0.0033$. In this region we get maximum value for ${\cal N}_{cmb}=50$,
                      $|\tau^{equil}_{NL}|_{max}\sim 0.00042$,$|g^{equil}_{NL}|_{max}\sim 0.0033$.
                      Additionally it is important to note that, in this case for $\alpha\bar{\lambda}=0.00004$ the lines obtained for ${\cal N}_{cmb}=50$, ${\cal N}_{cmb}=60$ and ${\cal N}_{cmb}=70$ cross each other.
                 \item \textcolor{red}{\underline{Region~III}}\\
                           Here for the parameter space $ 0.000001<\alpha\bar{\lambda}<0.00001$ the non-Gaussian amplitude lying within the window $ 0.00001<\tau^{equil}_{NL}<0.00014$,~$
                           0.00008<g^{equil}_{NL}<0.0014$. In this region we get maximum value for ${\cal N}_{cmb}=70$,
                           $|\tau^{equil}_{NL}|_{max}\sim 0.00014$,$
                           |g^{equil}_{NL}|_{max}\sim 0.0014$.
                           Additionally it is important to note that, in this case for $0.000003\leq \alpha\bar{\lambda}\leq 0.000006$ the lines obtained for ${\cal N}_{cmb}=50$, ${\cal N}_{cmb}=60$ and ${\cal N}_{cmb}=70$ cross cross each other and then show increasing behaviour.
                 \item \textcolor{red}{\underline{Region~IV}}\\
                                Here for the parameter space $ 0.0000001<\alpha\bar{\lambda}<0.000001$ the non-Gaussian amplitude lying within the window $ 10^{-7}<\tau^{equil}_{NL}<7\times 10^{-6}$,~
                               $5\times 10^{-8}<g^{equil}_{NL}<0.000052$. In this region we get maximum value for ${\cal N}_{cmb}=60$,
                               $|\tau^{equil}_{NL}|_{max}\sim 7\times 10^{-6}$,$
                                |g^{equil}_{NL}|_{max}\sim 0.000052$.
                 \end{itemize} 
                 Further combining the contribution from \textcolor{red}{\underline{Region~I}}, \textcolor{red}{\underline{Region~II}}, \textcolor{red}{\underline{Region~III}} and   \textcolor{red}{\underline{Region~IV}} we finally get the following constraint on the four point non-Gaussian amplitude in the equilateral limit configuration:
                 \bea \textcolor{red}{\underline{\rm Region~I}}+\textcolor{red}{\underline{\rm Region~II}}+\textcolor{red}{\underline{\rm Region~III}}+\textcolor{red}{\underline{\rm Region~IV}:}~~10^{-7}<\tau^{equil}_{NL}<0.00052,~~~
                 5\times 10^{-8}<g^{equil}_{NL}<0.004~~~~~~~~~~
                 \eea
                 for the following parameter space:
                 \bea \textcolor{red}{\underline{\rm Region~I}}+\textcolor{red}{\underline{\rm Region~II}}+\textcolor{red}{\underline{\rm Region~III}}+\textcolor{red}{\underline{\rm Region~IV}:}~~~~~0.0000001<\alpha\bar{\lambda}<0.001.~~~~~~~
                      \eea
                      In this analysis we get the following maximum value of the three point non-Gaussian amplitude in the equilateral limit configuration as given by:
                 \bea |\tau^{equil}_{NL}|_{max}\sim 0.00052,~~~ |g^{equil}_{NL}|_{max}\sim 0.004.\eea
                 To visualize these constraints more clearly we have also presented $(\tau^{equil}_{NL},\alpha,\bar{\lambda})$ and $(g^{equil}_{NL},\alpha,\bar{\lambda})$ 3D plot in fig.~(\ref{tnl3d1a}),  fig.~(\ref{tnl3d1a}), fig.~(\ref{gnl3daa}) and  fig.~(\ref{gnl3dbb}) for two different angular orientations as given by \textcolor{red}{\underline{Angle~I}} and \textcolor{red}{\underline{Angle~II}}. From the the representative surfaces it is clearly observed the behavior of three point non-Gaussian amplitude in the equilateral limit for the variation of two fold parameter $\alpha$ and $\bar{\lambda}$ and the results are consistent with the obtained constarints in 2D analysis. Here all the obtained results are consistent with the two point and three point constaints as well as with the Planck 2015 data \cite{Ade:2015lrj,Ade:2015ava,Ade:2015xua}.
                                 
\item \underline{\textcolor{blue}{Counter-collinier or folded kite limiting configuration:}}\\
For this case we have the situation where the magnitude of the sum of the two momenta is taken to zero, which implies:
\bea k_{ij}=|k_{i}+k_{j}|=\sqrt{k^2_i+k^2_j+2k_ik_j\cos\theta_{ij}}\rightarrow 0,\forall (i,j=1,2,3,4)~{\rm with}~i<j,~~~~~\eea
which implies,$ \cos\theta_{ij}\rightarrow -\frac{(k^2_i+k^2_j)}{2k_ik_j},\forall (i,j=1,2,3,4)~{\rm with}~i<j,$
and this satisfies, 
$\sum^{4}_{i<j=1}\frac{(k^2_i+k^2_j)}{k_ik_j}\rightarrow 4$.
In the present context of discussion, we identify this situation as
the \textcolor{blue}{counter-collinear}  limiting configuration as is this case for each momentum there is another associated momentum which have equal
magnitude along with opposite direction. In this specific case one can construct a  quadrilateral which is formed by the
momentum vectors participating in this limit using two fold ways. From the analysis it is observed that if our choice on the momenta are of the same
order in magnitude then the counter-collinear configurations are adjacent. Sometimes in literature this identified as the \textcolor{blue}{folded kite}  limiting configuration. On the contrary, here one can also choose the momenta in such a way, where the counter-collinear configurations are on the opposite sides
of the quadrilateral formed in the present context. But both the situations are  specifically dual configurations to each other. Consequently, the mathematical structure of the local trispectrum for scalar fluctuation simplifies in the
\textcolor{blue}{counter-collinear} or \textcolor{blue}{folded kite} limiting configuration. In this limit here we actually take:
\bea k_{12}<<k_{1}\approx k_{2},k_{3}\approx k_{4}.\eea
Consequently we have, 
$\cos\theta_1=\cos\theta_2,$ 
$\cos\theta_3=-1.$
In this case the trispectrum for scalar fluctuation can be written as:
\bea T(k_1,k_1,k_3,k_3)
&\approx&\frac{\tilde{W}^3(\phi_{cmb},{\bf \Psi})}{216M^{12}_p (\epsilon^*_{\tilde{W}})^2}\frac{1}{(k_1 k_3 )^6}\left[\frac{9}{4}\frac{k^3_1k^3_3}{k^3_{12}}\sin^2\alpha_1\sin^2\alpha_3\cos 2\chi_{12,34}+\cdots\right],~~~~~~~~~~~~\eea
where in the \textcolor{blue}{counter-collinear} or \textcolor{blue}{folded kite} limiting configuration contribution from the momentum dependent functions $\hat{W}^{S}({\bf k}_1,{\bf k}_2,{\bf k}_3,{\bf k}_4)$ and $\hat{R}^{S}({\bf k}_1,{\bf k}_2,{\bf k}_3,{\bf k}_4)$ are finite but the contributions are sub dominant for which one can easily neglect this part compared to the graviton exchange contribution. Here the graviton exchange contribution in \textcolor{blue}{counter-collinear} or \textcolor{blue}{folded kite} limiting configuration defined as:
\bea \hat{G}^{S}({\bf k}_1,{\bf k}_2,{\bf k}_3,{\bf k}_4)&=&\left[\frac{9}{4}\frac{k^3_1k^3_3}{k^3_{12}}\sin^2\alpha_1\sin^2\alpha_3\cos 2\chi_{12,34}+\cdots\right]\eea
where in \textcolor{blue}{counter-collinear} or \textcolor{blue}{folded kite} limiting configuration we have used additionally the following results:
\bea S({\bf k}_1,{\bf k}_2)&\approx&\frac{3}{2}k_1,~~~~
S({\bf k}_3,{\bf k}_4)\approx\frac{3}{2}k_3,\eea
and for the polarization sum we use:
\bea
\sum_{s=+,\times}\epsilon^{s}_{ij}({\bf k}_{12})\epsilon^{s}_{lm}({\bf k}_{34})k^{i}_{1}k^{j}_{2}k^{l}_{3}k^{m}_{4}
&=&k^2_1k^2_3\sin^2\alpha_1\sin^2\alpha_3\cos 2\chi_{12,34}.
\eea
Here we have to mention that:
\bea \epsilon^{+}_{ij}&=&{\bf a}_{i}{\bf a}_{j}-\bar{\bf a}_{i}\bar{\bf a}_{j},~~ 
\epsilon^{\times}_{ij}={\bf a}_{i}{\bf a}_{j}+\bar{\bf a}_{i}\bar{\bf a}_{j},\\ 
\epsilon^{+}_{ij}({\bf k}_{12})k^{i}_{1}k^{j}_{2}&=&k_{1}k_{2}\sin\alpha_{1}\sin\alpha_{2}\cos\left(\beta_{1}+\beta_{2}\right),\\
\epsilon^{+}_{ij}({\bf k}_{34})k^{l}_{3}k^{m}_{4}&=&k_{3}k_{4}\sin\alpha_{3}\sin\alpha_{4}\cos\left(\beta_{1}+\beta_{2}\right)\\
\epsilon^{\times}_{ij}({\bf k}_{12})k^{i}_{1}k^{j}_{2}&=&k_{1}k_{2}\sin\alpha_{1}\sin\alpha_{2}\sin\left(\beta_{1}+\beta_{2}\right),\\
\epsilon^{\times}_{ij}({\bf k}_{34})k^{l}_{3}k^{m}_{4}&=&k_{3}k_{4}\sin\alpha_{3}\sin\alpha_{4}\sin\left(\beta_{1}+\beta_{2}\right),\\
\frac{\sin\alpha_{2}}{\sin\alpha_{1}}&=&\frac{k_1}{k_2}\approx 1,~~
\frac{\sin\alpha_{4}}{\sin\alpha_{3}}=\frac{k_3}{k_4}\approx 1,\\ 
\beta_{2}-\beta_{1}&=&\pi,~~
\beta_{4}-\beta_{3}=\pi,~~
\beta_{1}-\beta_{3}=\chi_{12,34},
\eea
and we use the following coordinate to parameterize the momentum vector:
\bea {\bf k}_{i}&=& k_{i}\left(\sin\alpha_{i}\cos\beta_{i},\sin\alpha_{i}\sin\beta_{i},\cos\alpha_{i}\right)\forall i=(1,2,3,4),\eea 
where
\bea \alpha_{i}&\equiv&\cos^{-1}\left(\hat{{\bf k}}_{i}.\hat{{\bf k}}_{12}\right)\forall i=(1,2,3,4),\\
\beta_{i}&\equiv&\cos^{-1}\left(\hat{{\bf k}}_{i}.{\bf a}\right)\forall i=(1,2,3,4).\eea
Now if we assume that the non-Gaussian parameter $\tau^{loc}_{NL}$ and $f^{loc}_{NL}$ are connected through \textcolor{blue}{\it Suyama Yamaguchi} consistency relation, then in the \textcolor{blue}{counter-collinear} or \textcolor{blue}{folded kite} limiting configuration we get the following expression for the four point non-Gaussian parameter:
\bea \tau^{foldkite}_{NL}
&\approx&\frac{36}{144}\frac{1}{\left(2k^3_1+k^3_3\right)^2} \left[2(3\epsilon^*_{\tilde{W}}-\eta^{*}_{\tilde{W}})\left(2k^3_1+k^3_3\right)\right.\nonumber\\ &&\left.+\epsilon^*_{\tilde{W}}\left(-\left(2k^3_1+k^3_3\right)+2\left(k^3_1+k_1k_3(k_1+k_3)\right)+\frac{8k^2_1}{(2k_1+k_3)}(k^2_1+2k^2_3)\right)\right]^2.~~~~~~~~~~~\eea
                  \begin{figure*}[htb]
                  \centering
                  \subfigure[Range~I.]{
                      \includegraphics[width=7.6cm,height=4.5cm] {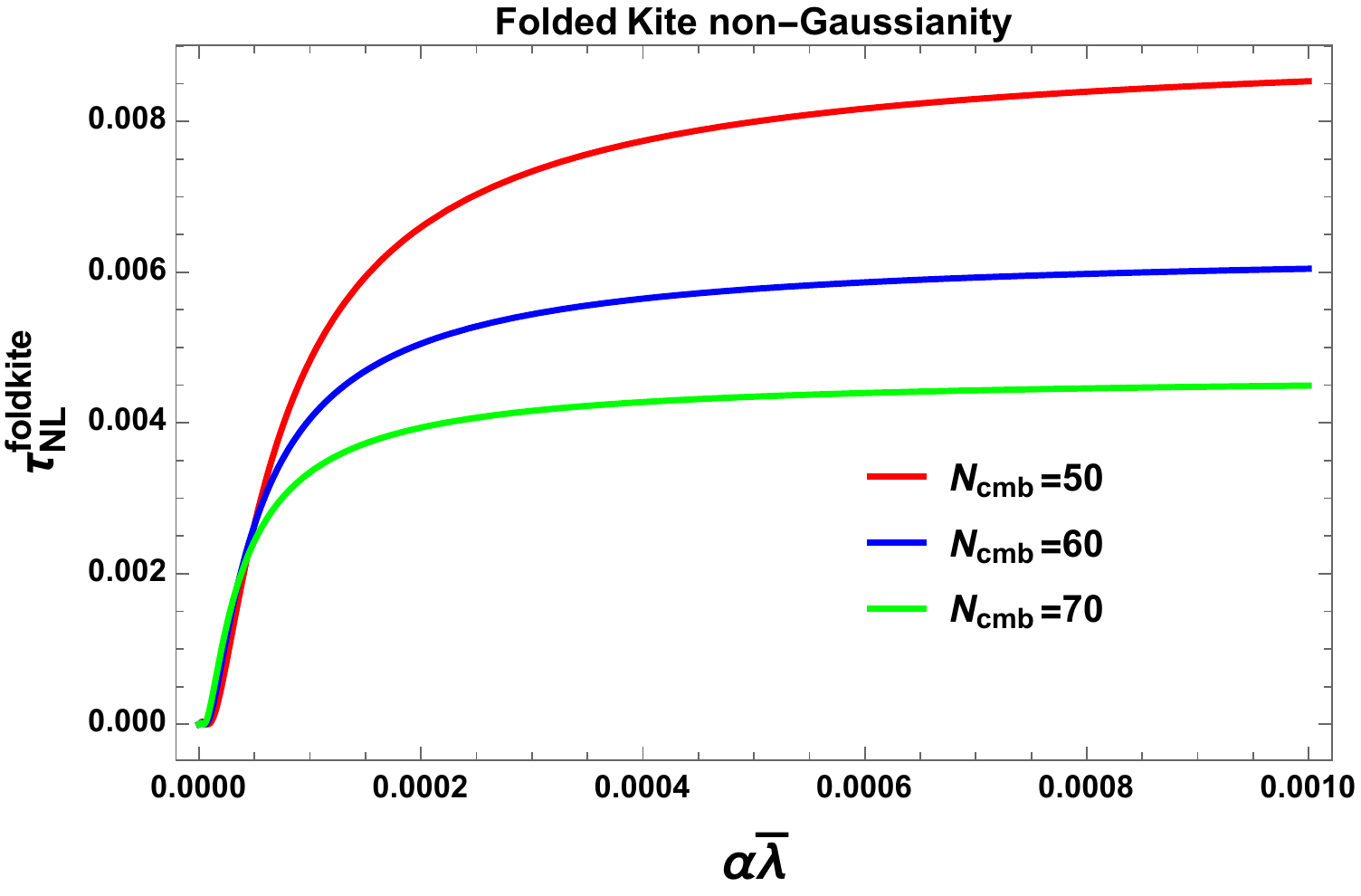}
                      \label{fige1bbbbbbvv}
                  }
                  \subfigure[Range~II.]{
                      \includegraphics[width=7.6cm,height=4.5cm] {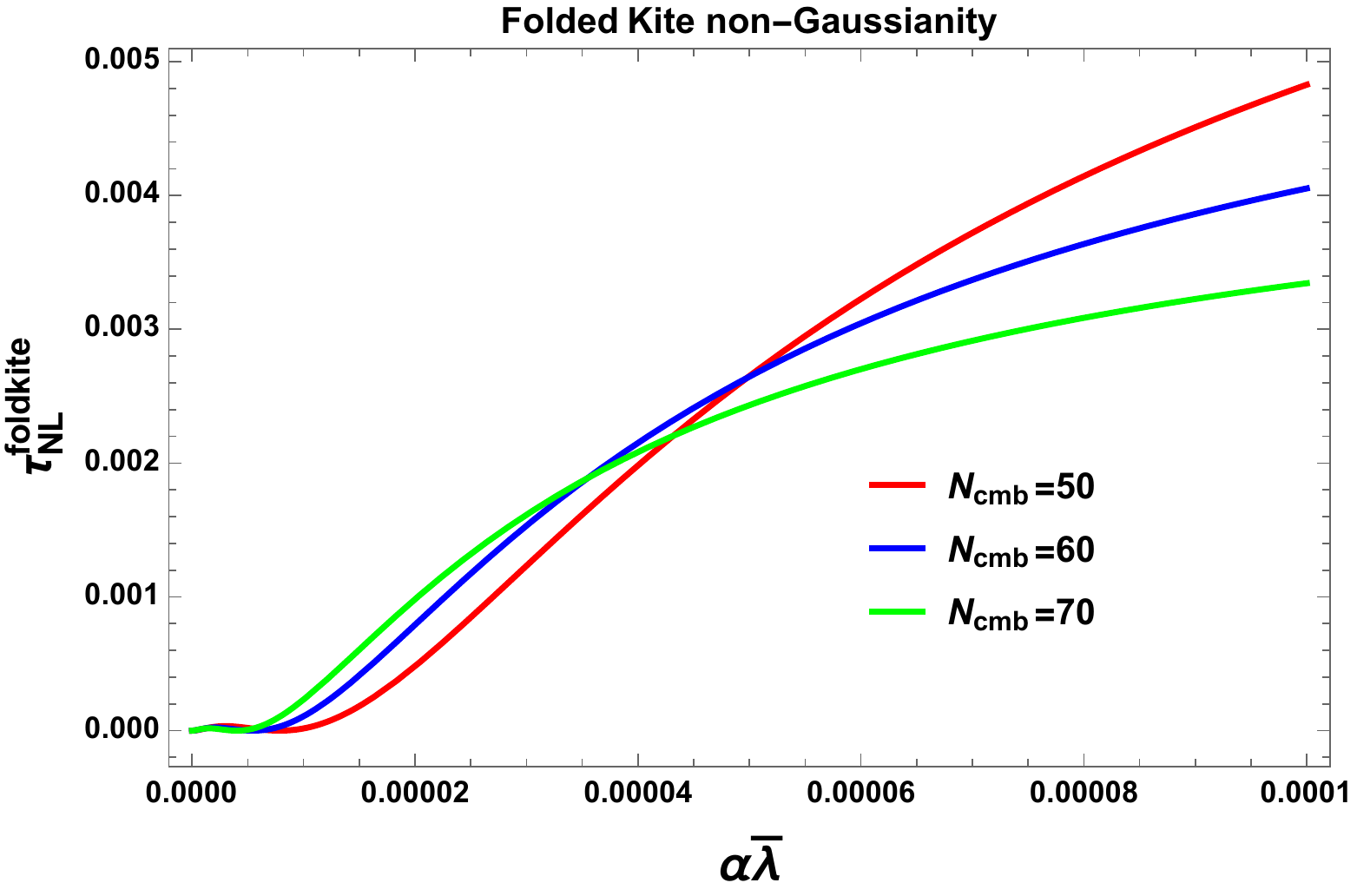}
                      \label{fige2bbbbbbvv}
                  }
                  \subfigure[Rangle~III.]{
                           \includegraphics[width=7.6cm,height=4.5cm] {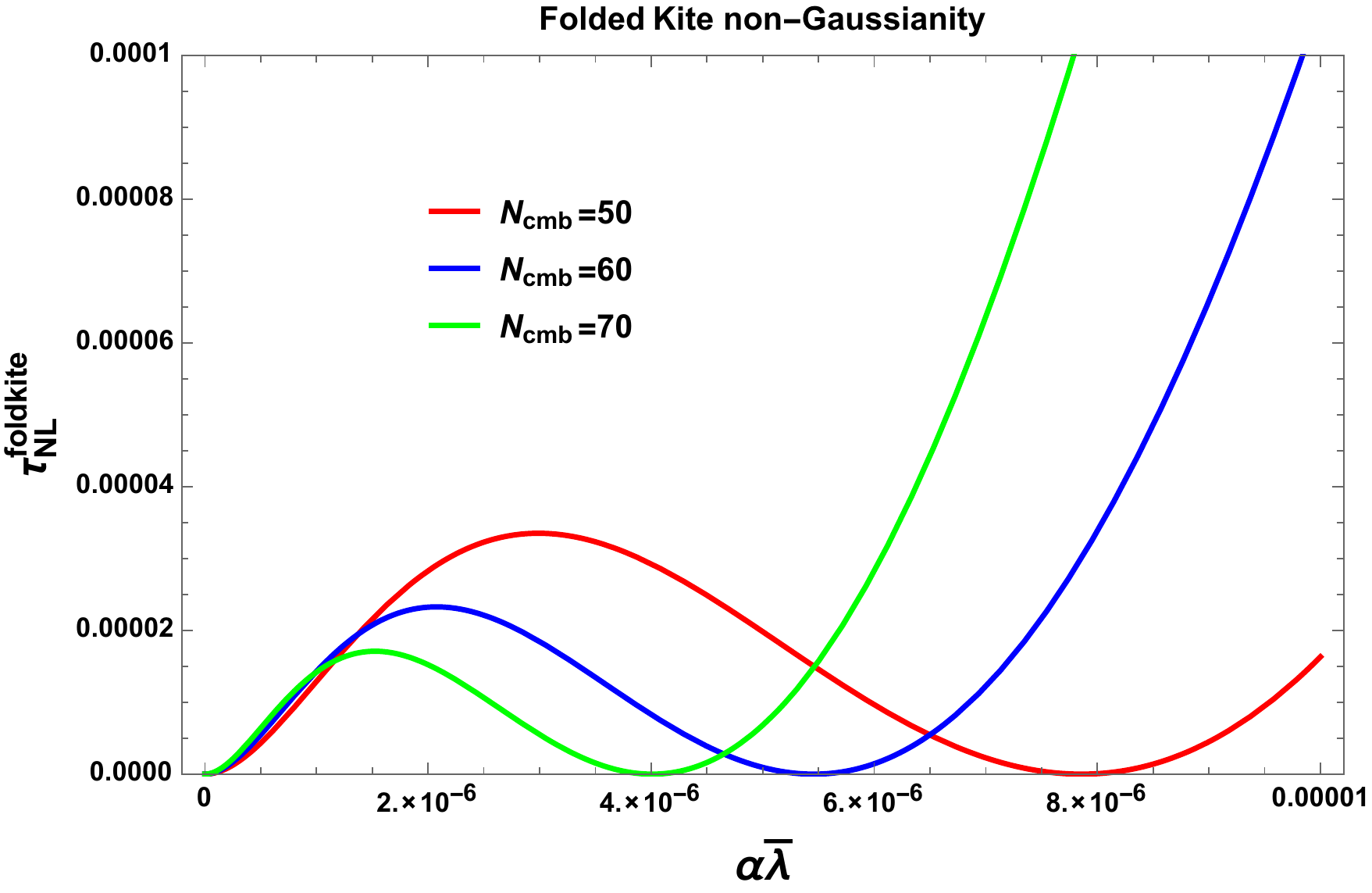}
                           \label{fige3bbbbbbvv}
                       }
                       \subfigure[Range~IV.]{
                                     \includegraphics[width=7.6cm,height=4.5cm] {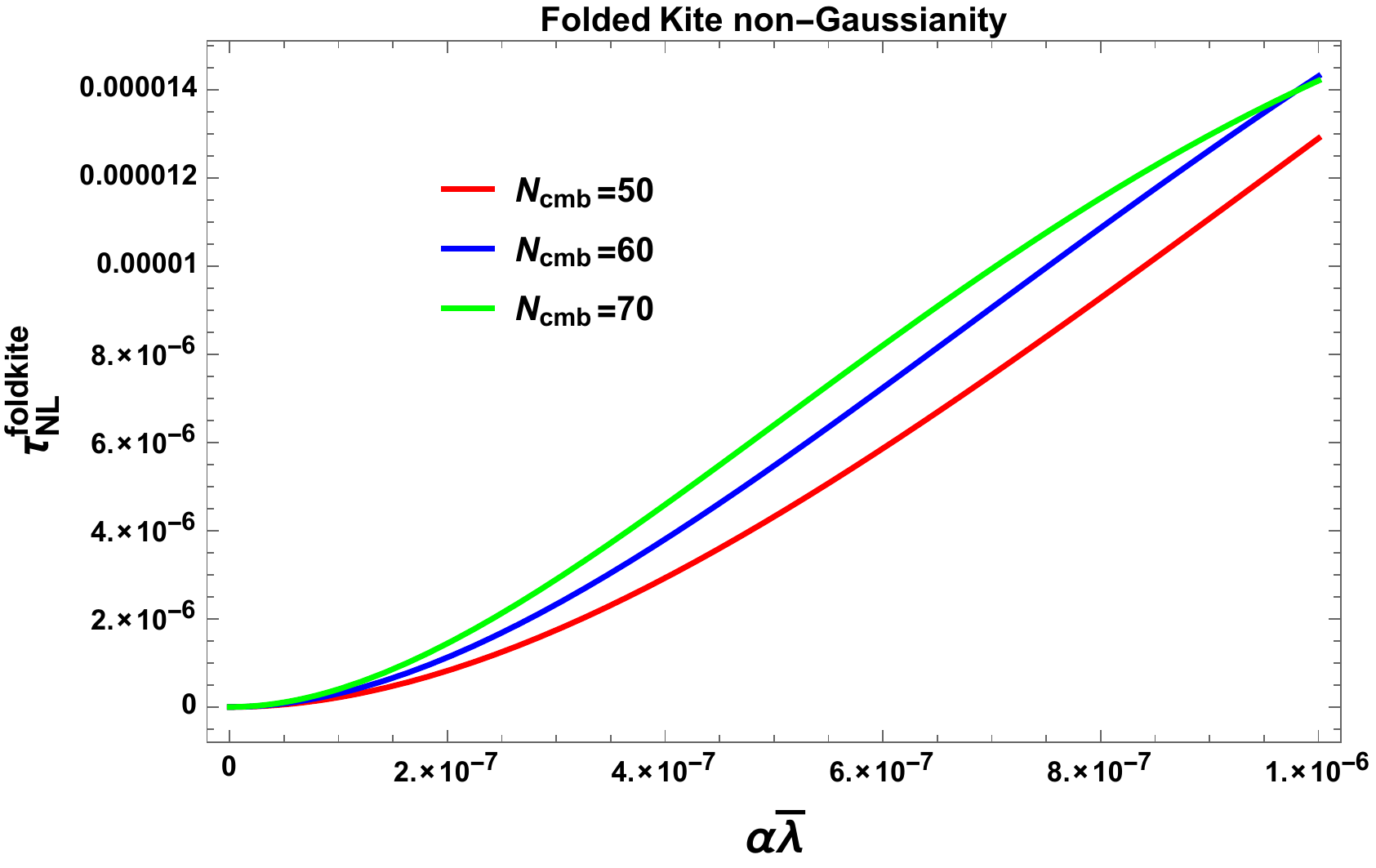}
                                     \label{fige4bbbbbbvv}
                                 }
                  \caption[Optional caption for list of figures]{Representative diagram for equilateral non-Gaussian three point amplitude vs product of the parameters $\alpha\bar{\lambda}$ in four different region for ${\cal N}_{cmb}=50$ (red), ${\cal N}_{cmb}=60$ (blue) and ${\cal N}_{cmb}=70$ (green).}
                  \label{tnleaaaacv}
                  \end{figure*}
                       \begin{figure*}[htb]
                       \centering
                       \subfigure[Angle~I.]{
                           \includegraphics[width=7.6cm,height=4.5cm] {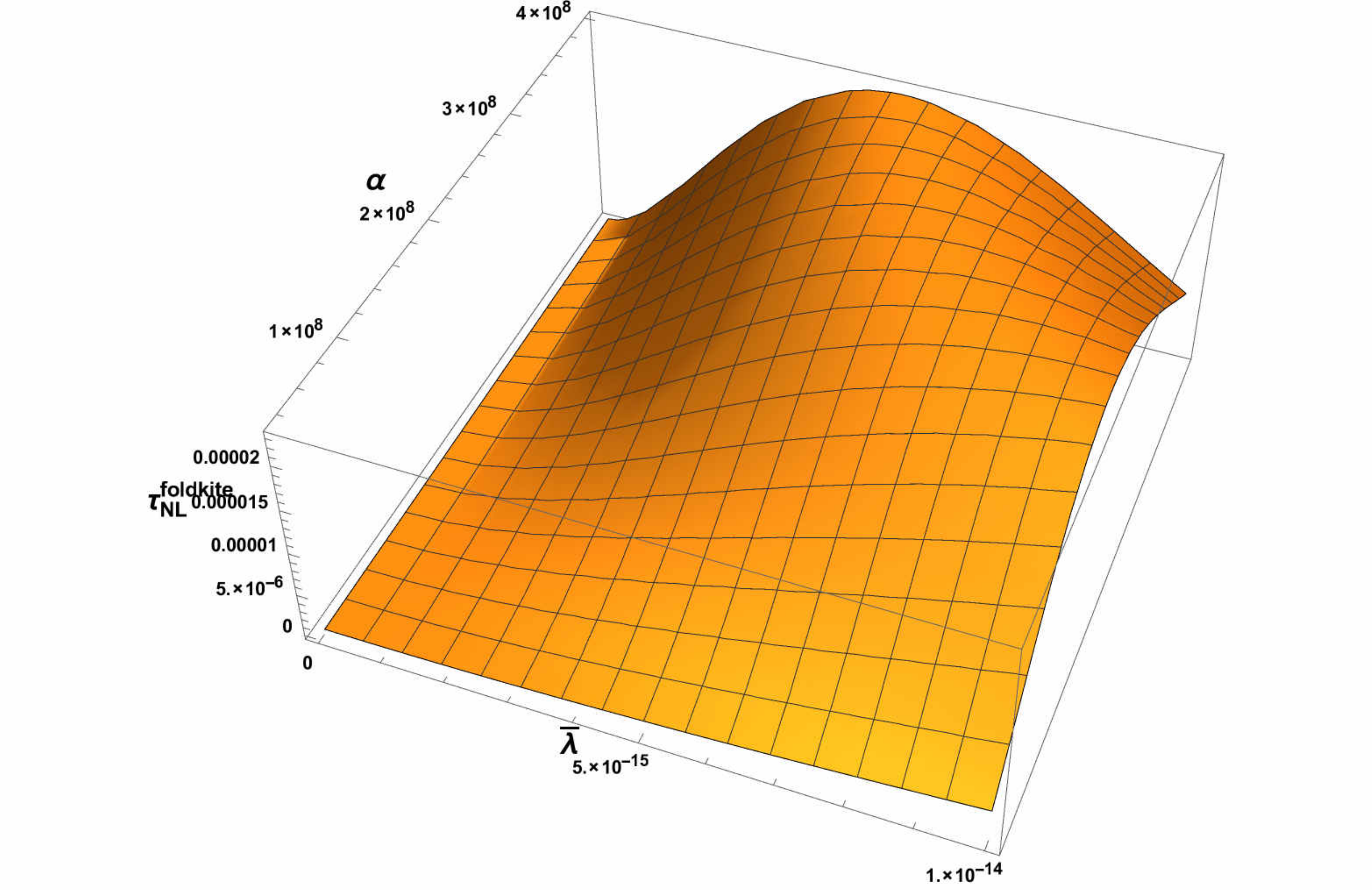}
                           \label{tnl3daaax}
                       }
                       \subfigure[Angle~II.]{
                           \includegraphics[width=7.6cm,height=4.5cm] {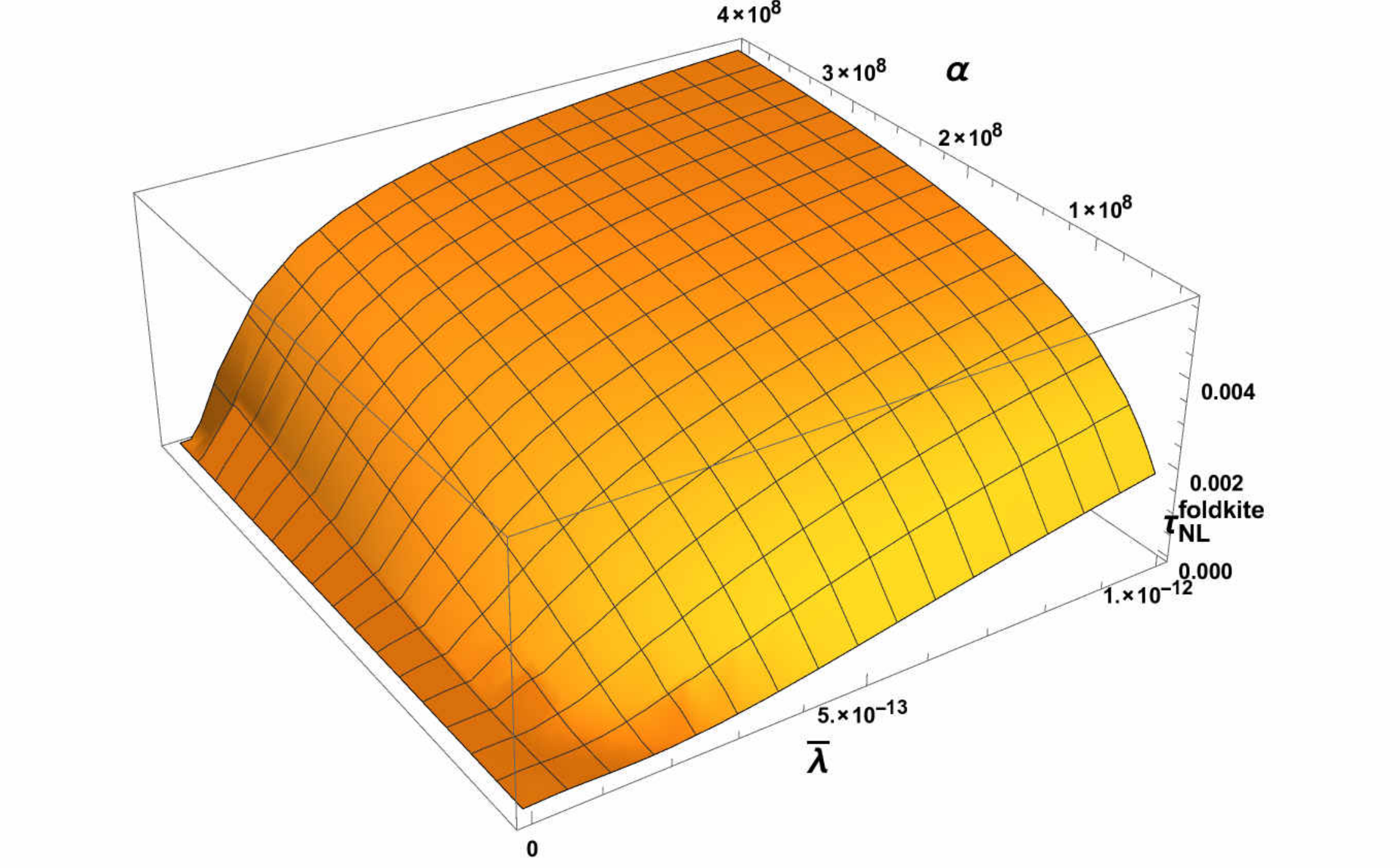}
                           \label{tnl3dbbbx}
                       }
                       \caption[Optional caption for list of figures]{Representative 3D diagram for equilateral non-Gaussian three point amplitude vs the model parameters $\alpha$ and $\bar{\lambda}$ for  ${\cal N}_{cmb}=60$ in two different angular views.} 
                       \label{gnl3vv}
                       \end{figure*}
                  \begin{figure*}[htb]
                  \centering
                  \subfigure[Range~I.]{
                      \includegraphics[width=7.6cm,height=4.5cm] {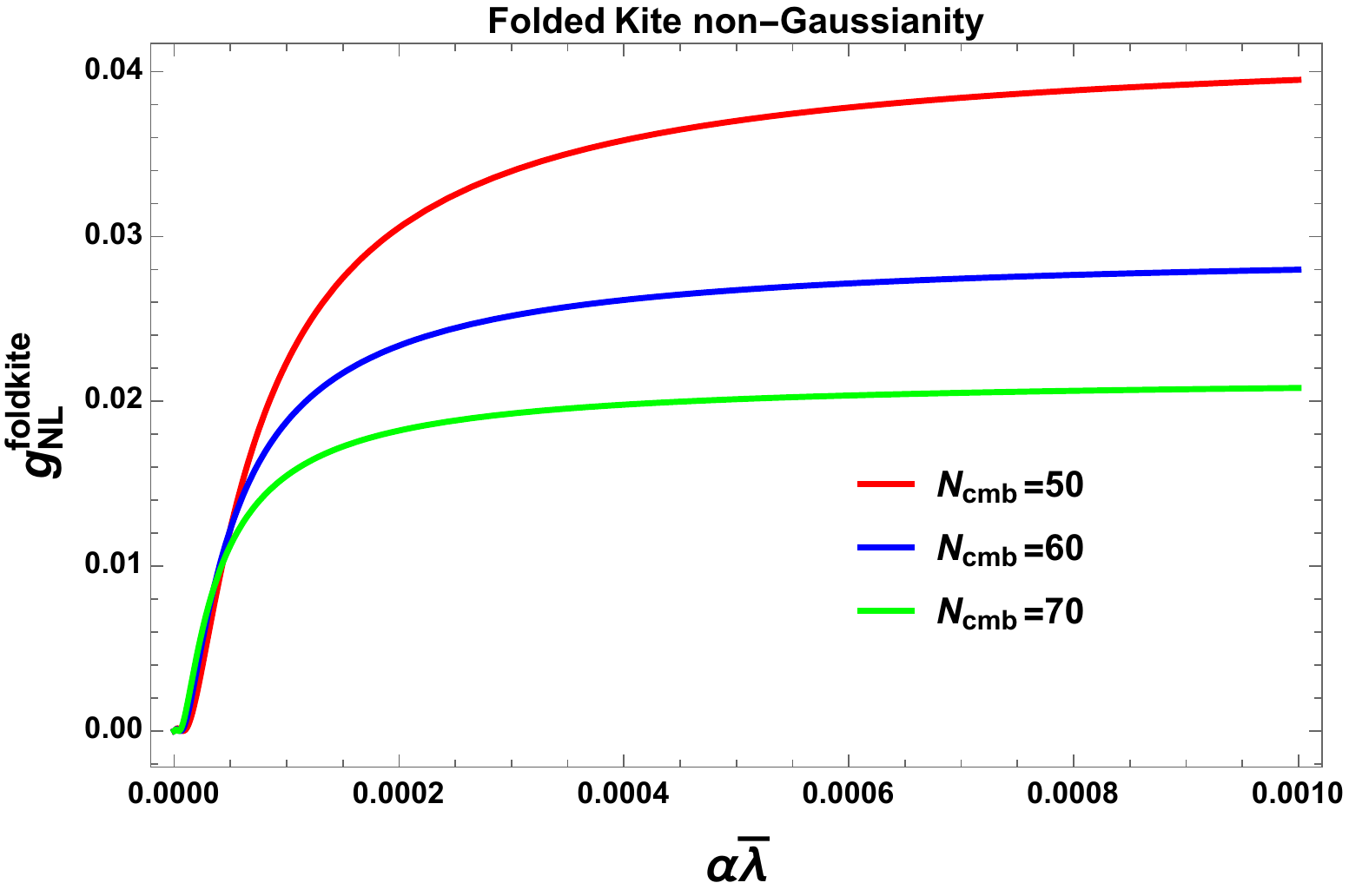}
                      \label{fige1bbbbbb}
                  }
                  \subfigure[Range~II.]{
                      \includegraphics[width=7.6cm,height=4.5cm] {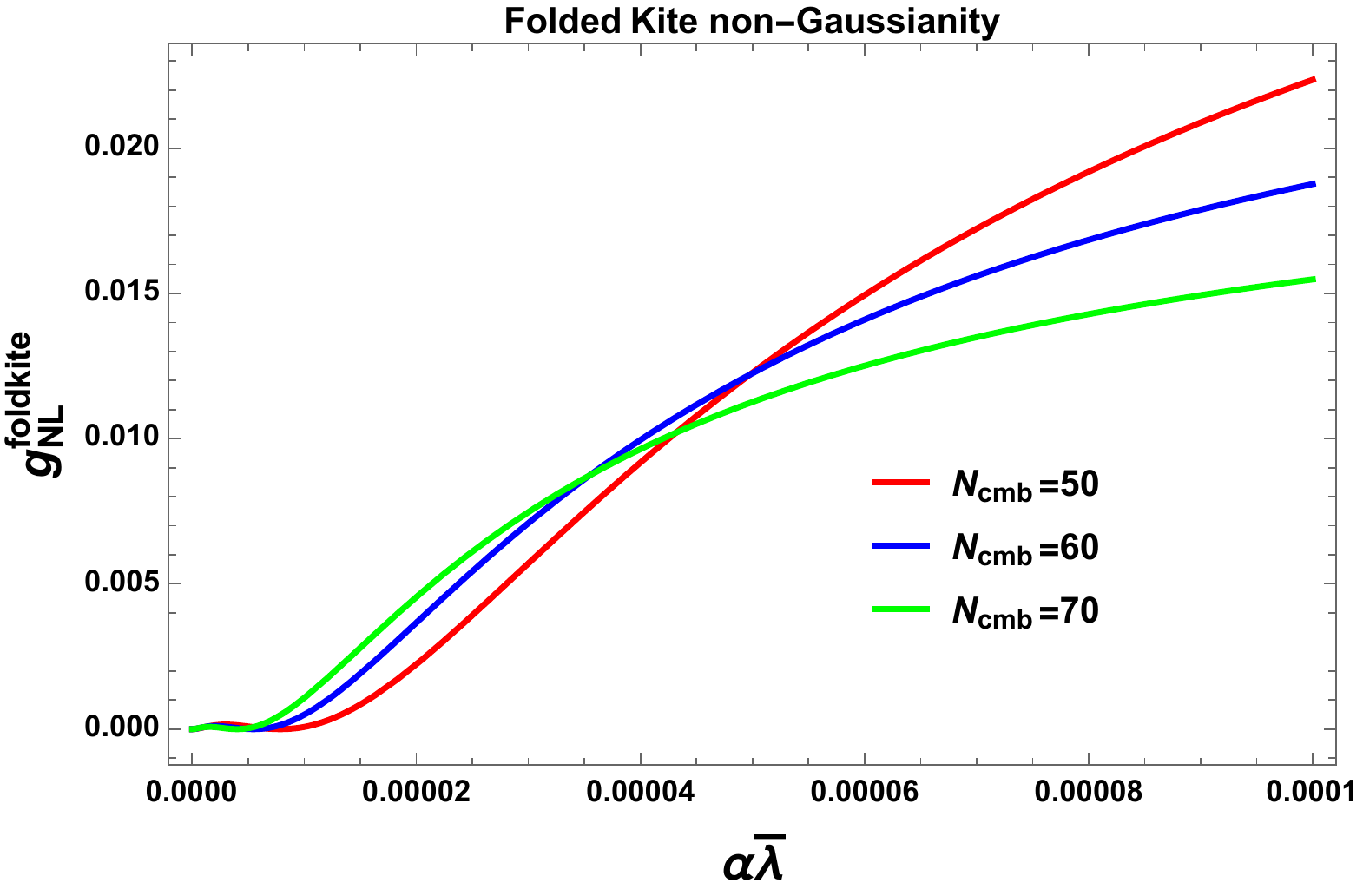}
                      \label{fige2bbbbbb}
                  }
                  \subfigure[Rangle~III.]{
                           \includegraphics[width=7.6cm,height=4.5cm] {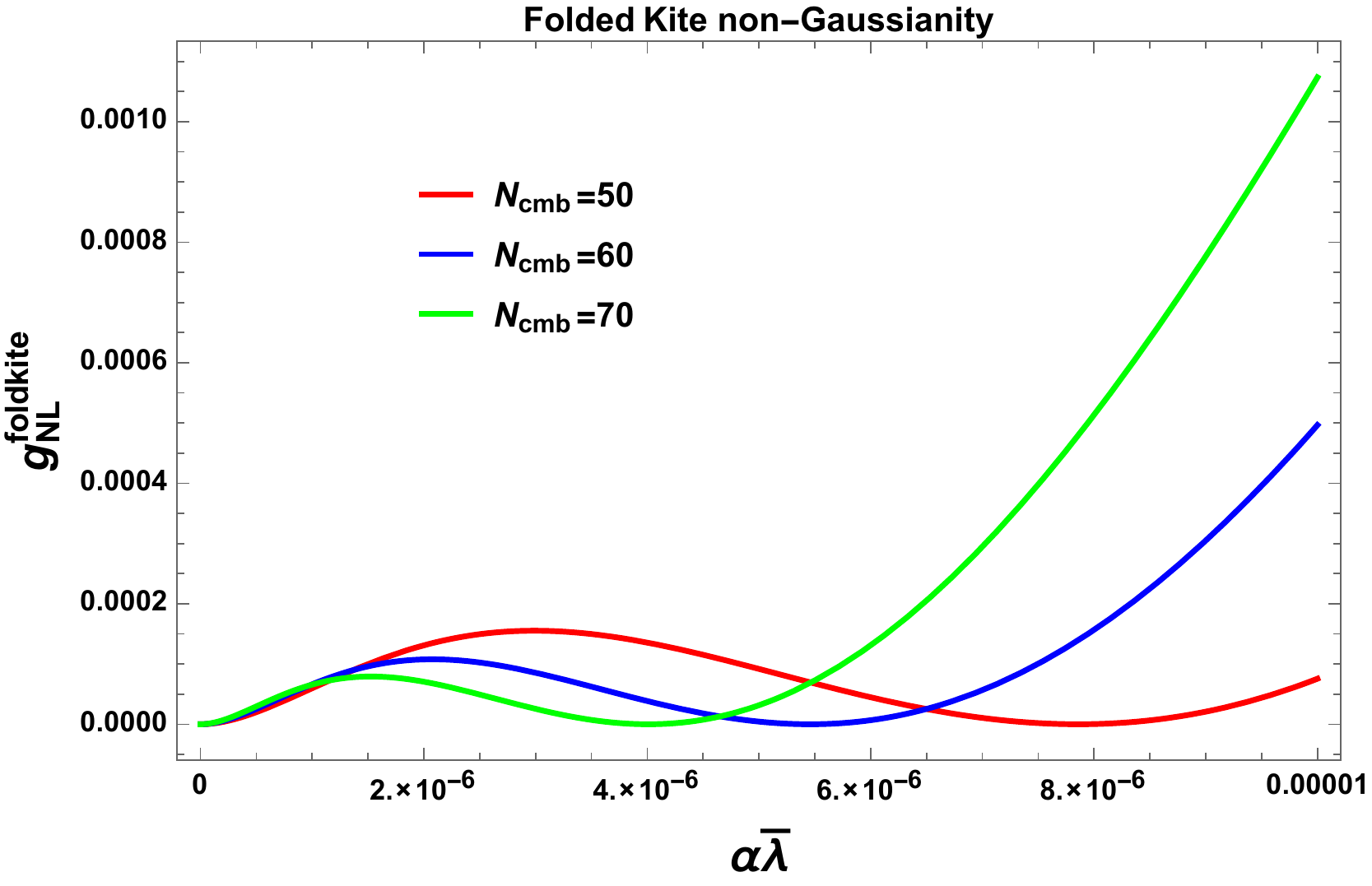}
                           \label{fige3bbbbbb}
                       }
                       \subfigure[Range~IV.]{
                                     \includegraphics[width=7.6cm,height=5cm] {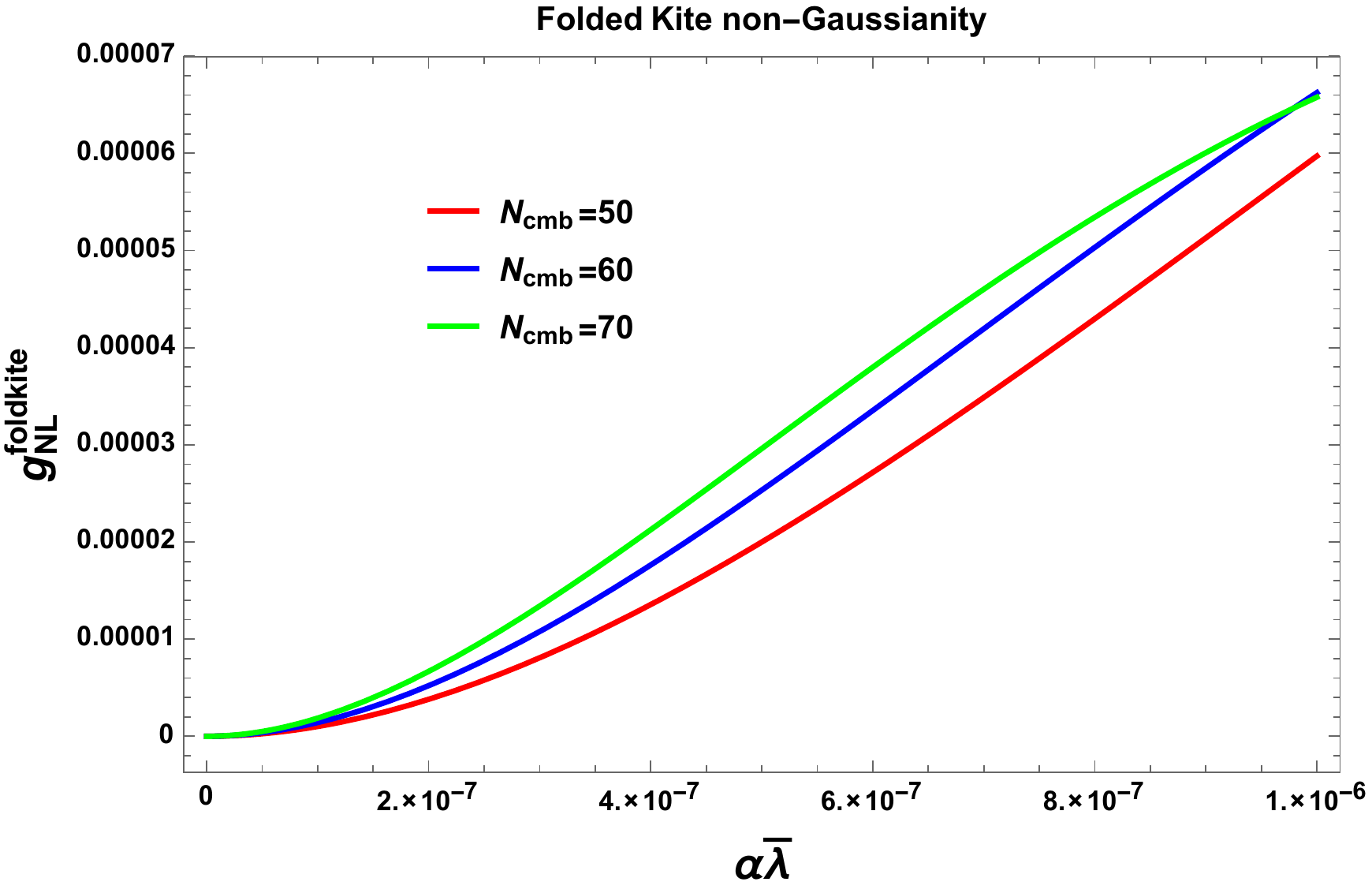}
                                     \label{fige4bbbbbb}
                                 }
                  \caption[Optional caption for list of figures]{Representative diagram for equilateral non-Gaussian three point amplitude vs product of the parameters $\alpha\bar{\lambda}$ in four different region for ${\cal N}_{cmb}=50$ (red), ${\cal N}_{cmb}=60$ (blue) and ${\cal N}_{cmb}=70$ (green).}
                  \label{tnleaaaa}
                  \end{figure*}
                       \begin{figure*}[htb]
                       \centering
                       \subfigure[Angle~I.]{
                           \includegraphics[width=7.6cm,height=4.5cm] {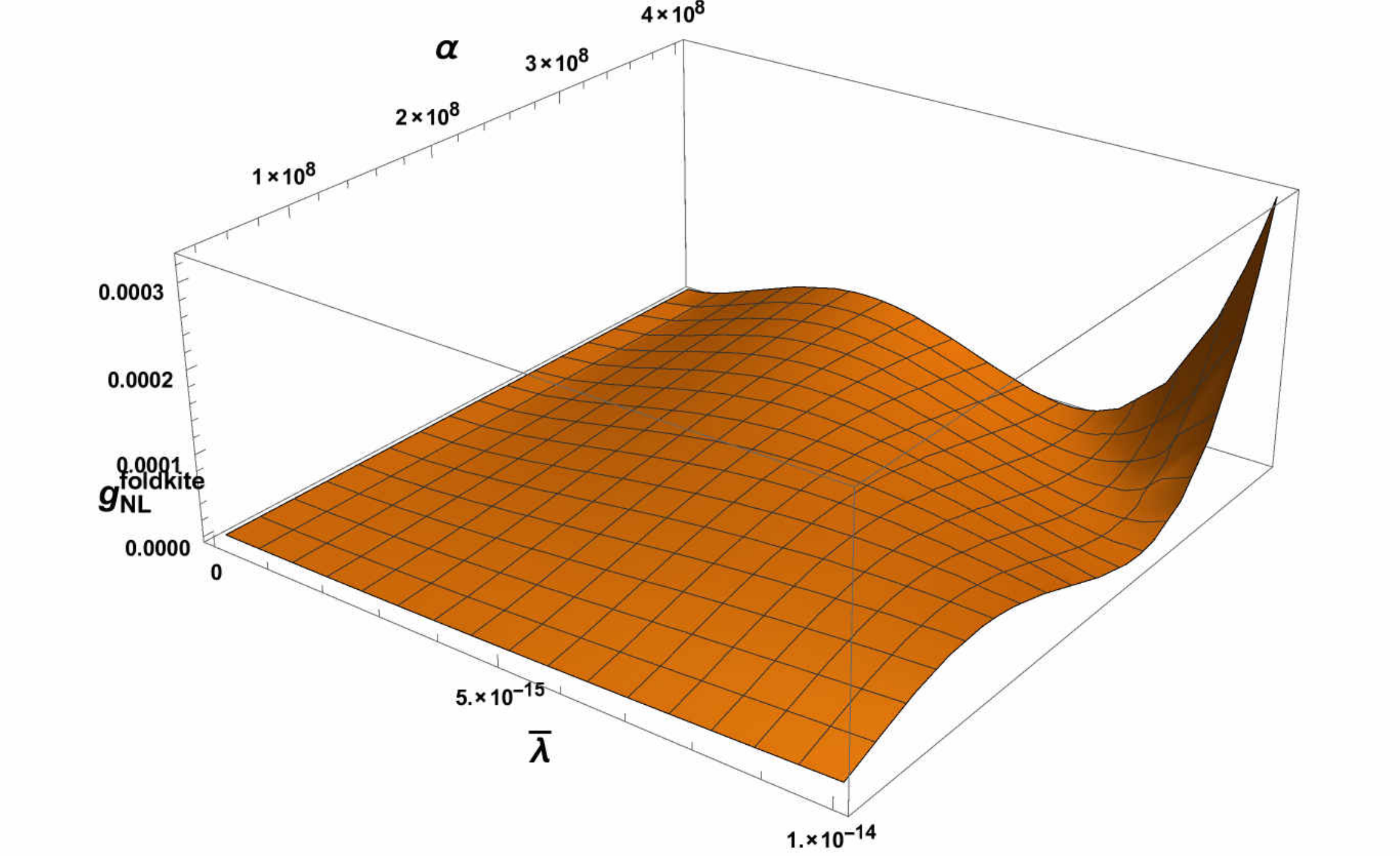}
                           \label{tnl3daaa}
                       }
                       \subfigure[Angle~II.]{
                           \includegraphics[width=7.6cm,height=4.5cm] {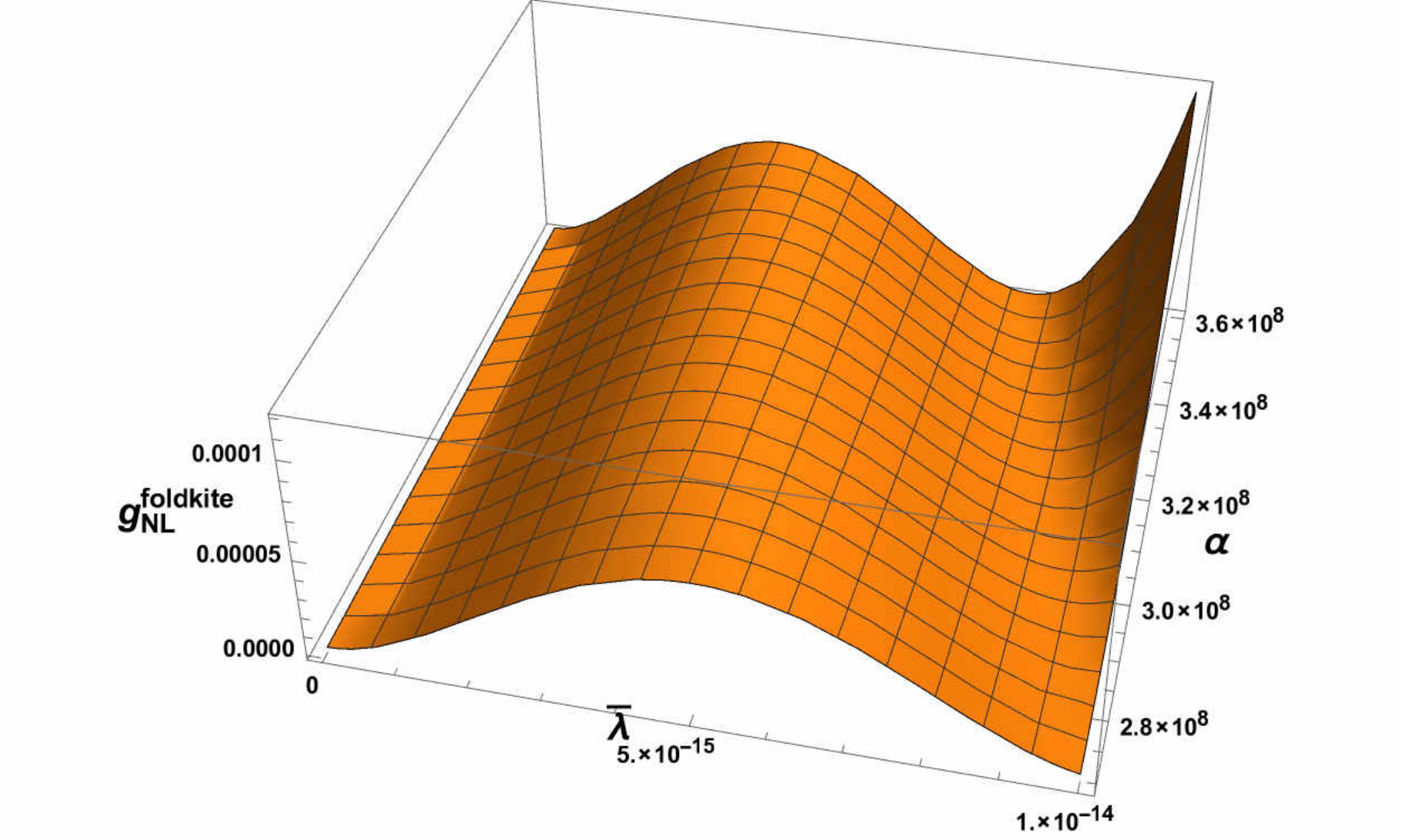}
                           \label{tnl3dbbb}
                       }
                       \caption[Optional caption for list of figures]{Representative 3D diagram for equilateral non-Gaussian three point amplitude vs the model parameters $\alpha$ and $\bar{\lambda}$ for  ${\cal N}_{cmb}=60$ in two different angular views.} 
                       \label{gnl3}
                       \end{figure*}
 In this limiting configuration the normalization factor ${\cal N}_{NORM}$ that connects the two non-Gaussian parameters $\tau^{foldkite}_{NL}$ and $g^{foldkite}_{NL}$ computed from frour point function as:
\bea {\cal N}_{NORM}&=&\frac{25}{54}\left[\frac{\Delta_{1}-\Delta_{2}}{\Delta_{3}}\right],\eea
where the momentum dependent factors $\Delta_{1}$, $\Delta_{2}$ and $\Delta_{3}$ are defined as:
\bea \Delta_{1}
&\approx&\frac{\frac{\tilde{W}^3(\phi_{cmb},{\bf \Psi})}{54M^{12}_p (\epsilon^*_{\tilde{W}})^2}\frac{(2k^3_1+k^3_3)^2}{(k_1 k_3 )^6}\left[\frac{9}{4}\frac{k^3_1k^3_3}{k^3_{12}}\sin^2\alpha_1\sin^2\alpha_3\cos 2\chi_{12,34}+\cdots\right]}{\left[2(3\epsilon^*_{\tilde{W}}-\eta^{*}_{\tilde{W}})\left(2k^3_1+k^3_3\right)+\epsilon^*_{\tilde{W}}\left(-\left(2k^3_1+k^3_3\right)+2\left(k^3_1+k_1k_3(k_1+k_3)\right)+\frac{8k^2_1(k^2_1+2k^2_3)}{(2k_1+k_3)}\right)\right]^{2}},~~~~~~~~~
\\
\Delta_{2}&=&\sum_{j<p,i\neq j,p}P_{\zeta}(k_{ij})P_{\zeta}(k_j)P_{\zeta}(k_p)\approx\frac{\tilde{W}^3(\phi_{cmb},{\bf \Psi})}{1728 M^{12}_{p}(\epsilon^*_{\tilde{W}})^3}\left[\frac{4}{k^{3}_{12}k^{3}_{1}k^{3}_{3}}+\frac{1}{k^{3}_{13}k^{3}_{3}}\left(\frac{3}{k^3_1}+\frac{1}{k^3_3}\right)+\frac{1}{k^{3}_{23}}\left(\frac{1}{k^3_1}+\frac{1}{k^3_3}\right)^2\right],~~~~~~~~~\eea
\bea
\Delta_{3}&=&\sum_{i<j<p}P_{\zeta}(k_{i})P_{\zeta}(k_j)P_{\zeta}(k_p)\approx\frac{\tilde{W}^3(\phi_{cmb},{\bf \Psi})}{864 M^{12}_{p}(\epsilon^*_{\tilde{W}})^3}\frac{1}{k^3_1k^3_3}\left(\frac{1}{k^3_1}+\frac{1}{k^3_3}\right),\eea  
 Consequently the non-Gaussian parameter $g^{loc}_{NL}$ can be express as:
 \bea g^{foldkite}_{NL}
 &\approx&\frac{25}{216}\left[\frac{\Delta_{1}-\Delta_{2}}{\Delta_{3}}\right]\frac{1}{\left(2k^3_1+k^3_3\right)^2} \left[2(3\epsilon^*_{\tilde{W}}-\eta^{*}_{\tilde{W}})\left(2k^3_1+k^3_3\right)\right.\nonumber\\ &&\left.+\epsilon^*_{\tilde{W}}\left(-\left(2k^3_1+k^3_3\right)+2\left(k^3_1+k_1k_3(k_1+k_3)\right)+\frac{8k^2_1}{(2k_1+k_3)}(k^2_1+2k^2_3)\right)\right]^2.~~~~~~~~~~~\eea
  In fig.~(\ref{tnleaaaacv}) and fig.~(\ref{tnleaaaa}), we have shown the features of non-Gaussian amplitude from four point scalar function $\tau^{foldkite}_{NL}$ and $g^{foldkite}_{NL}$ in folded kite limit configuration in four different scanning region of product of the two parameters $\alpha\bar{\lambda}$ in the $(\tau^{foldkite}_{NL},\alpha\bar{\lambda})$ and $(g^{foldkite}_{NL},\alpha\bar{\lambda})$ 2D plane for the number of e-foldings $50<{\cal N}_{cmb}<70$. Physical explanation of the obtained features are appended following:-
             \begin{itemize}
             \item \textcolor{red}{\underline{Region~I}:} \\
             Here for the parameter space $0.0001<\alpha\bar{\lambda}<0.001$ the non-Gaussian amplitude lying within the window $0.0025<\tau^{foldkite}_{NL}<0.0085$,
             $0.014<g^{foldkite}_{NL}<0.038$. Further if we increase the numerical value of $\alpha\bar{\lambda}$, then the magnitude of the non-Gausiian amplitude saturates and we get maximum value for ${\cal N}_{cmb}=50$,
             $|\tau^{foldkite}_{NL}|_{max}\sim 0.0085$,$|g^{foldkite}_{NL}|_{max}\sim 0.038$.
             \item \textcolor{red}{\underline{Region~II}}\\
                  Here for the parameter space $0.00001<\alpha\bar{\lambda}<0.0001$ the non-Gaussian amplitude lying within the window $ 0.0002<\tau^{foldkite}_{NL}<0.0048$,~
                  $0.001<g^{foldkite}_{NL}<0.023$. In this region we get maximum value for ${\cal N}_{cmb}=50$, 
                  $|\tau^{foldkite}_{NL}|_{max}\sim 0.0048$,$
                  |g^{foldkite}_{NL}|_{max}\sim 0.023$.
                  Additionally it is important to note that, in this case for $\alpha\bar{\lambda}=0.00004$ the lines obtained for ${\cal N}_{cmb}=50$, ${\cal N}_{cmb}=60$ and ${\cal N}_{cmb}=70$ cross each other.
             \item \textcolor{red}{\underline{Region~III}}\\
                       Here for the parameter space $ 0.000001<\alpha\bar{\lambda}<0.00001$ the non-Gaussian amplitude lying within the window $ 0.000018<\tau^{foldkite}_{NL}<0.001$,~
                       $0.0001<g^{foldkite}_{NL}<0.001$. In this region we get maximum value for ${\cal N}_{cmb}=70$,
                       $|\tau^{foldkite}_{NL}|_{max}\sim 0.00062$,$|g^{foldkite}_{NL}|_{max}\sim 0.001$.
                       Additionally it is important to note that, in this case for $0.000001\leq \alpha\bar{\lambda}\leq 0.000006$ the lines obtained for ${\cal N}_{cmb}=50$, ${\cal N}_{cmb}=60$ and ${\cal N}_{cmb}=70$ show increasing, decreasing and further increasing behaviour.
             \item \textcolor{red}{\underline{Region~IV}}\\
                            Here for the parameter space $ 0.0000001<\alpha\bar{\lambda}<0.000001$ the non-Gaussian amplitude lying within the window $ 10^{-6}<\tau^{foldkite}_{NL}<0.000014,$~~
                           $2\times 10^{-6}<g^{foldkite}_{NL}<0.000066$. In this region we get maximum value for ${\cal N}_{cmb}=60$, 
                            $|\tau^{foldkite}_{NL}|_{max}\sim 0.000014$,$
                            |g^{foldkite}_{NL}|_{max}\sim 0.000066$.
             \end{itemize} 
             Further combaining the contribution from \textcolor{red}{\underline{Region~I}}, \textcolor{red}{\underline{Region~II}}, \textcolor{red}{\underline{Region~III}} and   \textcolor{red}{\underline{Region~IV}} we finally get the following contraint on the four point non-Gaussian amplitude in the equilateral limit configuration:
             \bea \textcolor{red}{\underline{\rm Region~I}}+\textcolor{red}{\underline{\rm Region~II}}+\textcolor{red}{\underline{\rm Region~III}}+\textcolor{red}{\underline{\rm Region~IV}:}~~10^{-6}<\tau^{foldkite}_{NL}<0.016,~~
             -0.023<g^{foldkite}_{NL}<0.002~~~~~~~~
             \eea
             for the following parameter space:
             \bea \textcolor{red}{\underline{\rm Region~I}}+\textcolor{red}{\underline{\rm Region~II}}+\textcolor{red}{\underline{\rm Region~III}}+\textcolor{red}{\underline{\rm Region~IV}:}~~~~~0.0000001<\alpha\bar{\lambda}<0.001.~~~~~~~
                  \eea
                  In this analysis we get the following maximum value of the three point non-Gaussian amplitude in the equilateral limit configuration as given by:
             \bea |\tau^{foldkite}_{NL}|_{max}\sim 0.016,~~~ |g^{foldkite}_{NL}|_{max}\sim 0.002.\eea
             To visualize these constraints more clearly we have also presented $(\tau^{foldkite}_{NL},\alpha,\bar{\lambda})$ and $(g^{foldkite}_{NL},\alpha,\bar{\lambda})$ 3D plot in fig.~(\ref{tnl3d1}),  fig.~(\ref{tnl3d1}), fig.~(\ref{gnl3da}) and  fig.~(\ref{gnl3db}) for two different angular orientations as given by \textcolor{red}{\underline{Angle~I}} and \textcolor{red}{\underline{Angle~II}}. From the the representative surfaces it is clearly observed the behavior of scalar four point non-Gaussian amplitude in the folded kite limit for the variation of two fold parameter $\alpha$ and $\bar{\lambda}$ and the results are consistent with the obtained constarints in 2D analysis. Here all the obtained results are consistent with the two point and three point constaints as well as with the Planck 2015 data \cite{Ade:2015lrj,Ade:2015ava,Ade:2015xua}.
             
       But as we have already pointed that if we relax the assumption of holding the \textcolor{blue}{\it Suyama Yamaguchi} consistency relation in the present context of discussion, then using Eq~(\ref{ w2}) one can write down the expression for momentum dependent function $f(k_1,k_1,k_3,k_3,k_{12},k_{13},k_{23})$ in the  \textcolor{blue}{equlilateral limiting configuration} as:  
  \bea f\left(k_1,k_1,k_3,k_3,k_{12},k_{13},k_{23}\right)&=&\frac{8}{(k_1+k_3)^3}\left[(k^6_1 +k^6_3)\left(\frac{1}{k^3_{13}}+\frac{1}{k^3_{23}}\right)+2k^3_1 k^3_3\left(\frac{2}{k^3_{12}}+\frac{1}{k^3_{13}}+\frac{1}{k^3_{23}}\right)\right],~~~~~~~~~\eea
 using which we get the following simplified expression for the non-Gaussian parameter $\tau^{foldkite}_{NL}$ and $g^{foldkite}_{NL}$ as obtained from the four point scalar function
 in \textcolor{blue}{equlilateral limiting configuration} as: 
  \bea \tau^{foldkite}_{NL}
  &\approx&\frac{\frac{\tilde{W}^3(\phi_{cmb},{\bf \Psi})}{216M^{12}_p (\epsilon^*_{\tilde{W}})^2}\frac{1}{(k_1 k_3 )^6}\left[\frac{9}{4}\frac{k^3_1k^3_3}{k^3_{12}}\sin^2\alpha_1\sin^2\alpha_3\cos 2\chi_{12,34}+\cdots\right]}{\left[\Delta_{2}+\frac{54}{25}\frac{8}{(k_1+k_3)^3}\left\{(k^6_1 +k^6_3)\left(\frac{1}{k^3_{13}}+\frac{1}{k^3_{23}}\right)+2k^3_1 k^3_3\left(\frac{2}{k^3_{12}}+\frac{1}{k^3_{13}}+\frac{1}{k^3_{23}}\right)\right\}\Delta_{3}\right]},~~~~~~~~~~~~~~~~~\\ g^{foldkite}_{NL}&\approx& \frac{\frac{\tilde{W}^3(\phi_{cmb},{\bf \Psi})}{216M^{12}_p (\epsilon^*_{\tilde{W}})^2}\frac{1}{(k_1 k_3 )^6}\left[\frac{9}{4}\frac{k^3_1k^3_3}{k^3_{12}}\sin^2\alpha_1\sin^2\alpha_3\cos 2\chi_{12,34}+\cdots\right]}{\left[\frac{\Delta_{2}}{\frac{8}{(k_1+k_3)^3}\left\{(k^6_1 +k^6_3)\left(\frac{1}{k^3_{13}}+\frac{1}{k^3_{23}}\right)+2k^3_1 k^3_3\left(\frac{2}{k^3_{12}}+\frac{1}{k^3_{13}}+\frac{1}{k^3_{23}}\right)\right\}}+\frac{54}{25}\Delta_{3}\right]}.\eea

    \item \underline{\textcolor{blue}{Squeezed limiting configuration:}}\\
    For this case we have $k_{1}\approx k_{2}\approx k_{3}(=k_{L})>>k_{4}(=k_{S})$,
    where $k_{i}=|{\bf k_{i}}|\forall i=1,2,3.$ Here $k_{L}$ and $k_{S}$ represent momentum for long and short modes respectively. 
 In this case one can write, 
   $\cos\theta_{i}=-\frac{1}{2}\left(1+\frac{k_{S}}{k_{L}}\right),~~~~\forall ~i=(1,2,3),$
   which gives an estimate of the factor $k_S/k_L$ in the \textcolor{blue}{squeezed limit} configuration and this estimate we will use for future computation.
    
     In this case the trispectrum for scalar fluctuation can be written as:
    \bea T(k_L,k_L,k_L,k_S)
    &\approx&\frac{\tilde{W}^3(\phi_{cmb},{\bf \Psi})}{216M^{12}_p (\epsilon^*_{\tilde{W}})^2}\frac{1}{k^6_Lk^3_S}\left[\frac{9}{4}\frac{1}{\left(1-\frac{k_S}{k_L}\right)^{\frac{3}{2}}}\sin^2\alpha_1\sin^2\alpha_3\cos 2\chi_{12,34}\right].~~~~~~~~~~~~\eea
                       \begin{figure*}[htb]
                       \centering
                       \subfigure[Range~I.]{
                           \includegraphics[width=7.6cm,height=4.5cm] {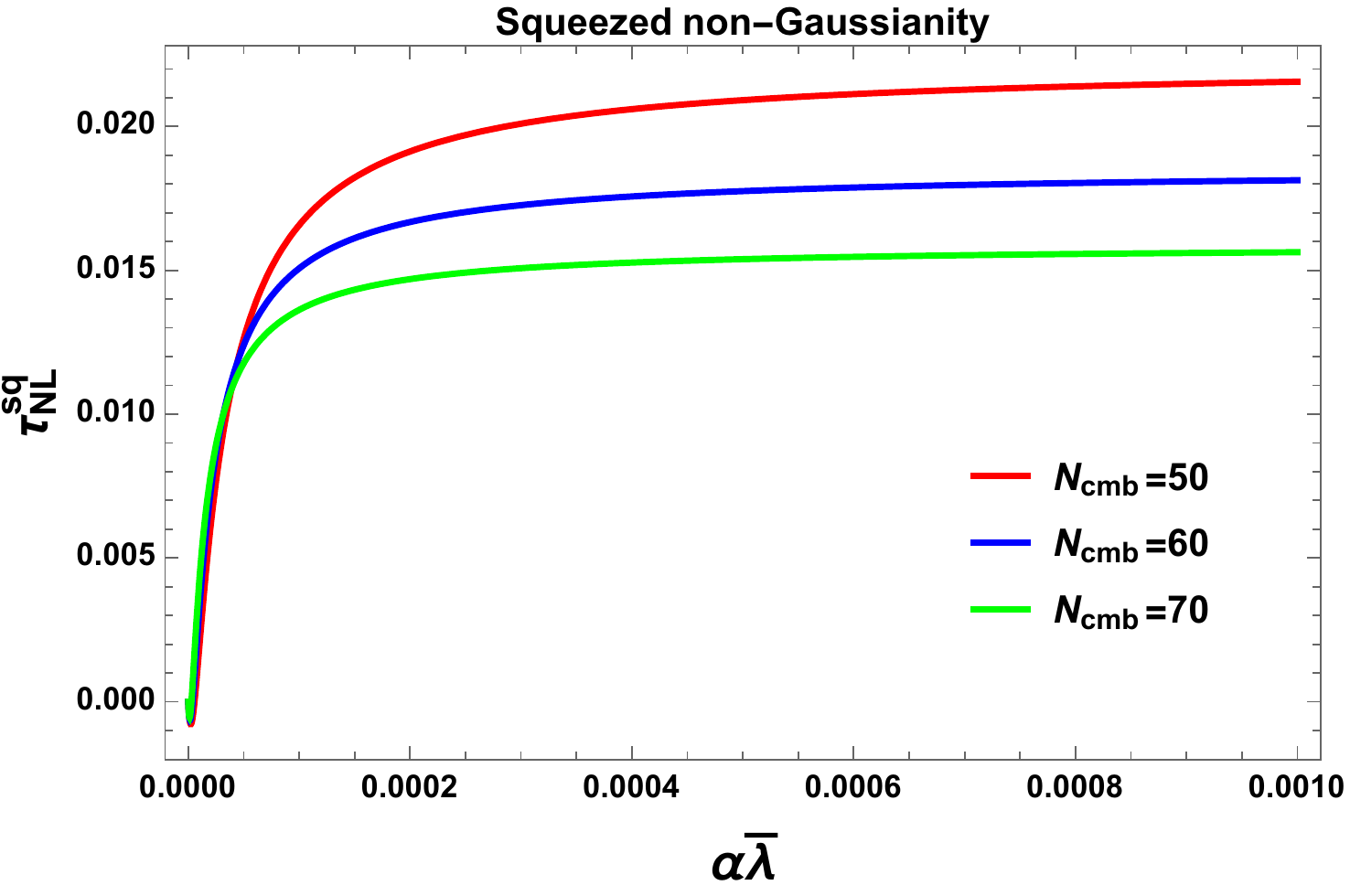}
                           \label{fige1bbbbbbb}
                       }
                       \subfigure[Range~II.]{
                           \includegraphics[width=7.6cm,height=4.5cm] {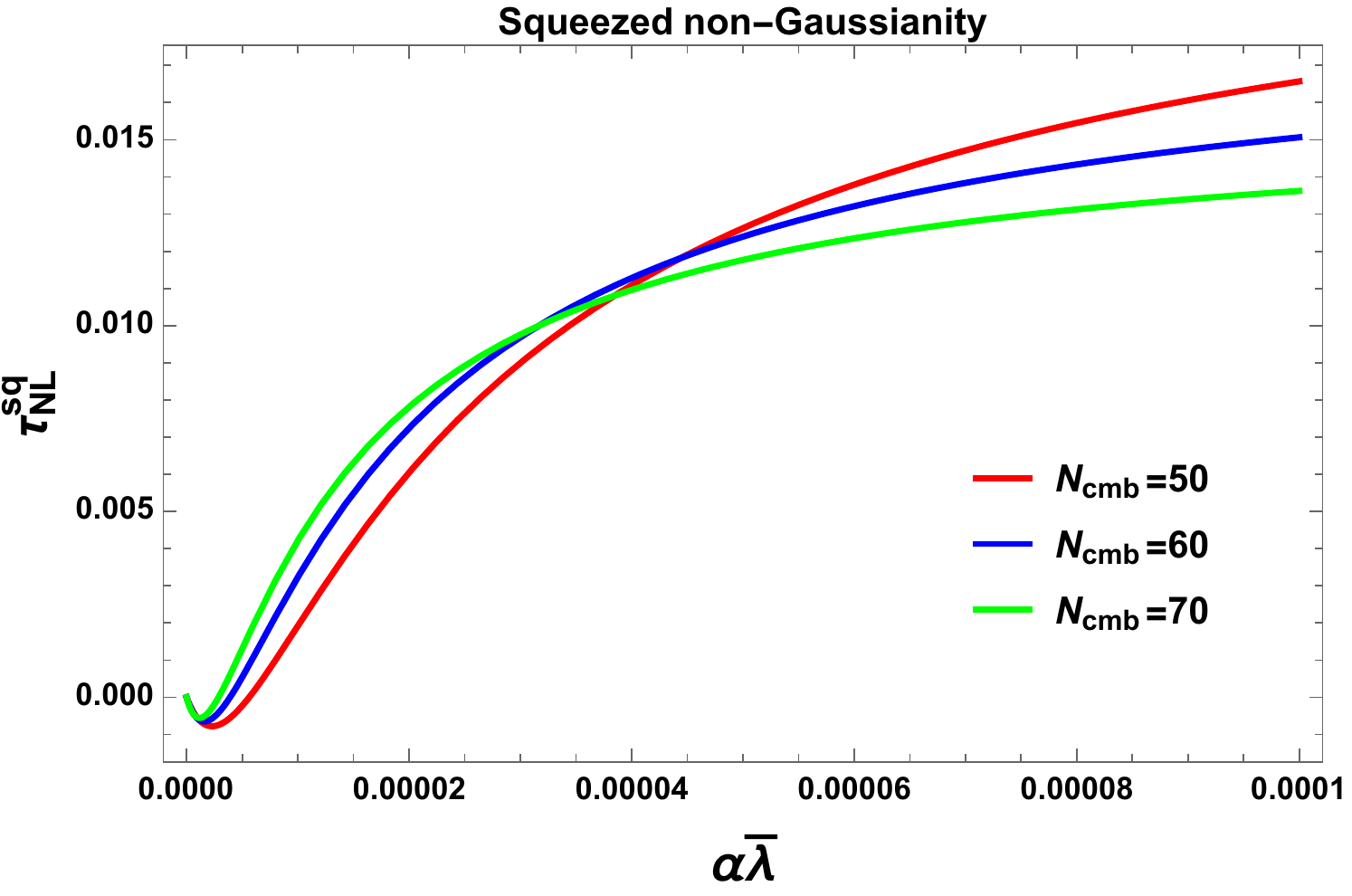}
                           \label{fige2bbbbbbb}
                       }
                       \subfigure[Rangle~III.]{
                                \includegraphics[width=7.6cm,height=4.5cm] {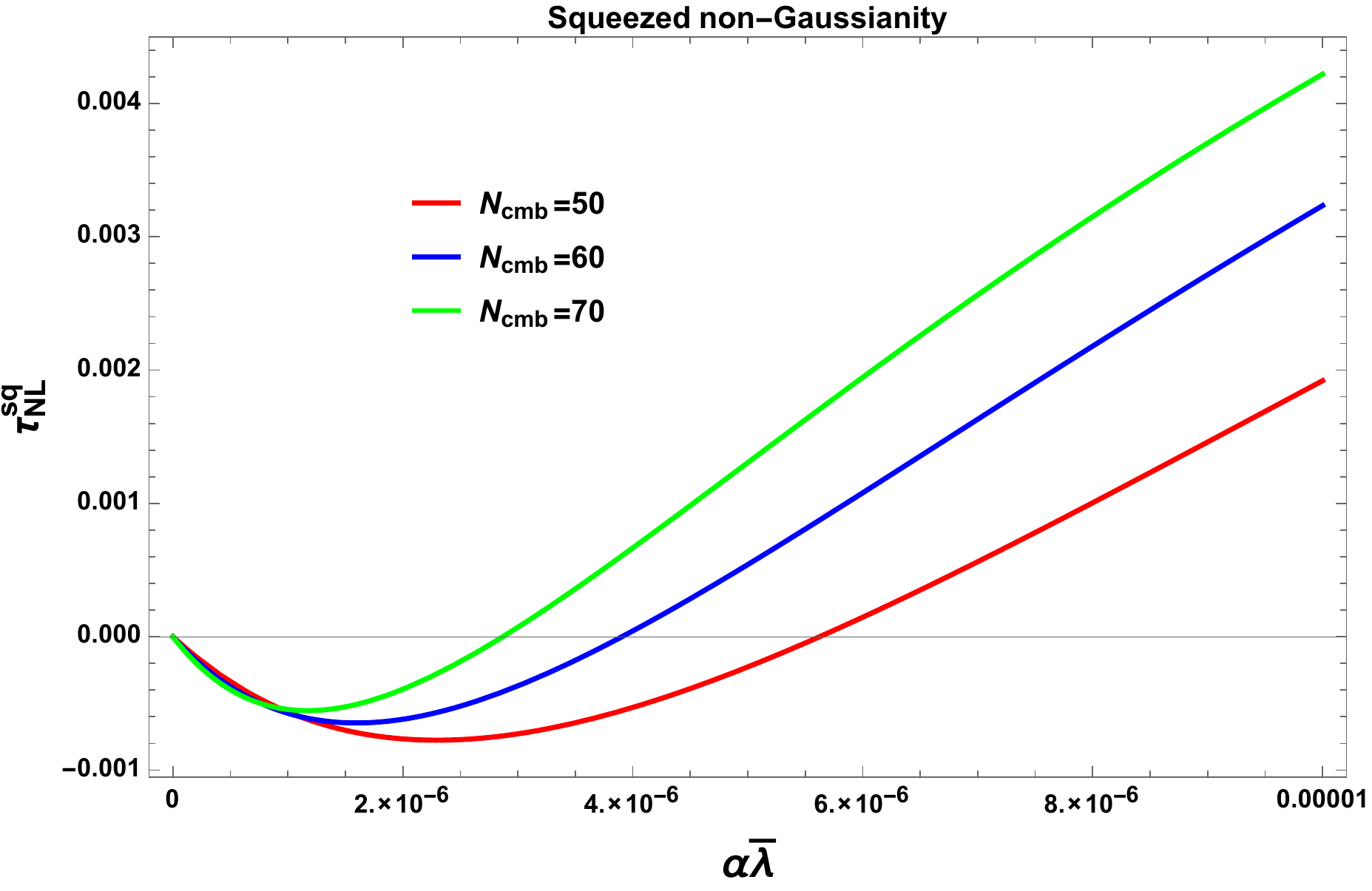}
                                \label{fige3bbbbbbb}
                            }
                            \subfigure[Range~IV.]{
                                          \includegraphics[width=7.6cm,height=4.5cm] {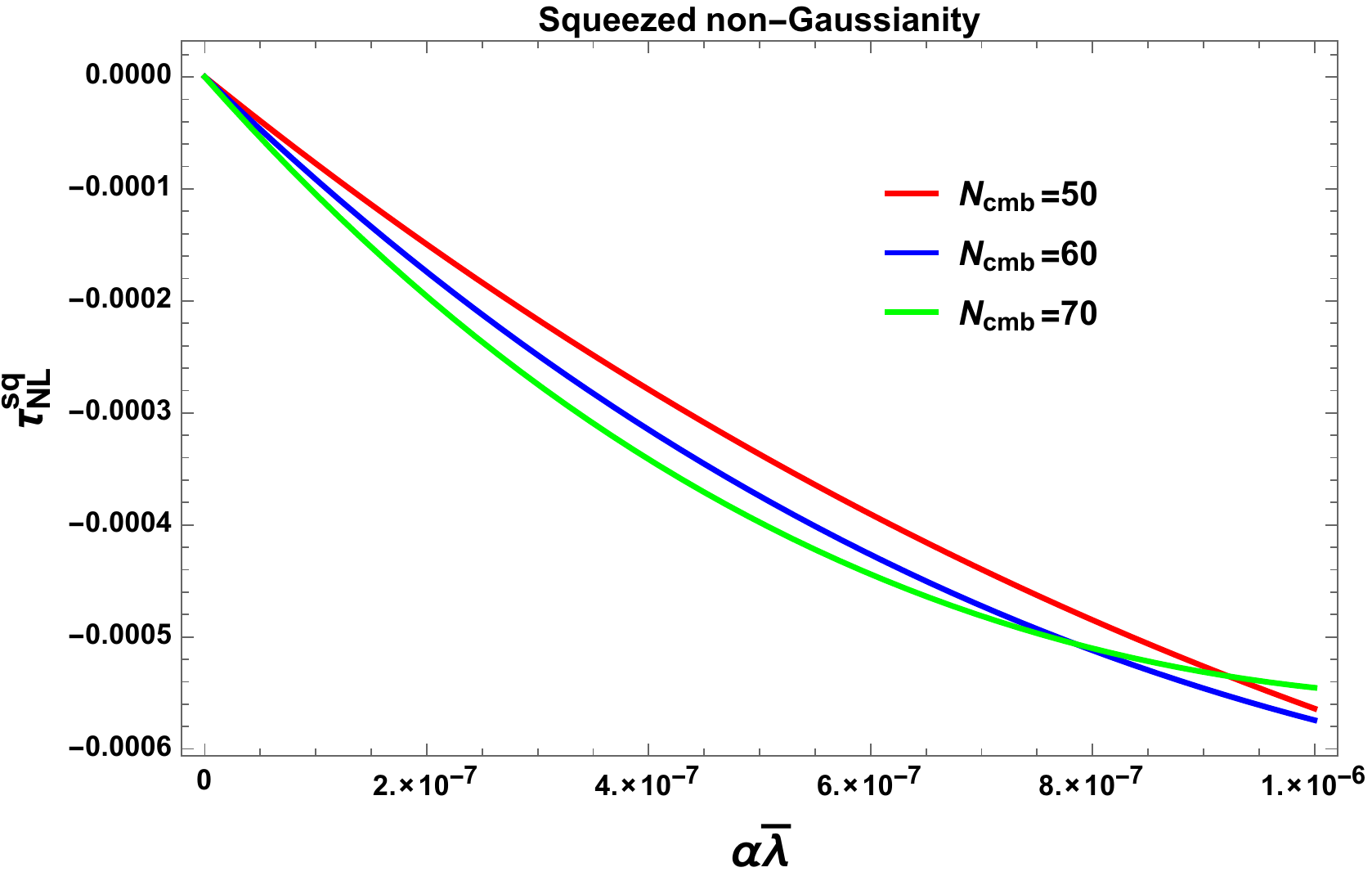}
                                          \label{fige4bbbbbbb}
                                      }
                       \caption[Optional caption for list of figures]{Representative diagram for equilateral non-Gaussian three point amplitude vs product of the parameters $\alpha\bar{\lambda}$ in four different region for ${\cal N}_{cmb}=50$ (red), ${\cal N}_{cmb}=60$ (blue) and ${\cal N}_{cmb}=70$ (green).}
                       \label{tnleaaaaa}
                       \end{figure*}
                            \begin{figure*}[htb]
                            \centering
                            \subfigure[Angle~I.]{
                                \includegraphics[width=7.6cm,height=4.5cm] {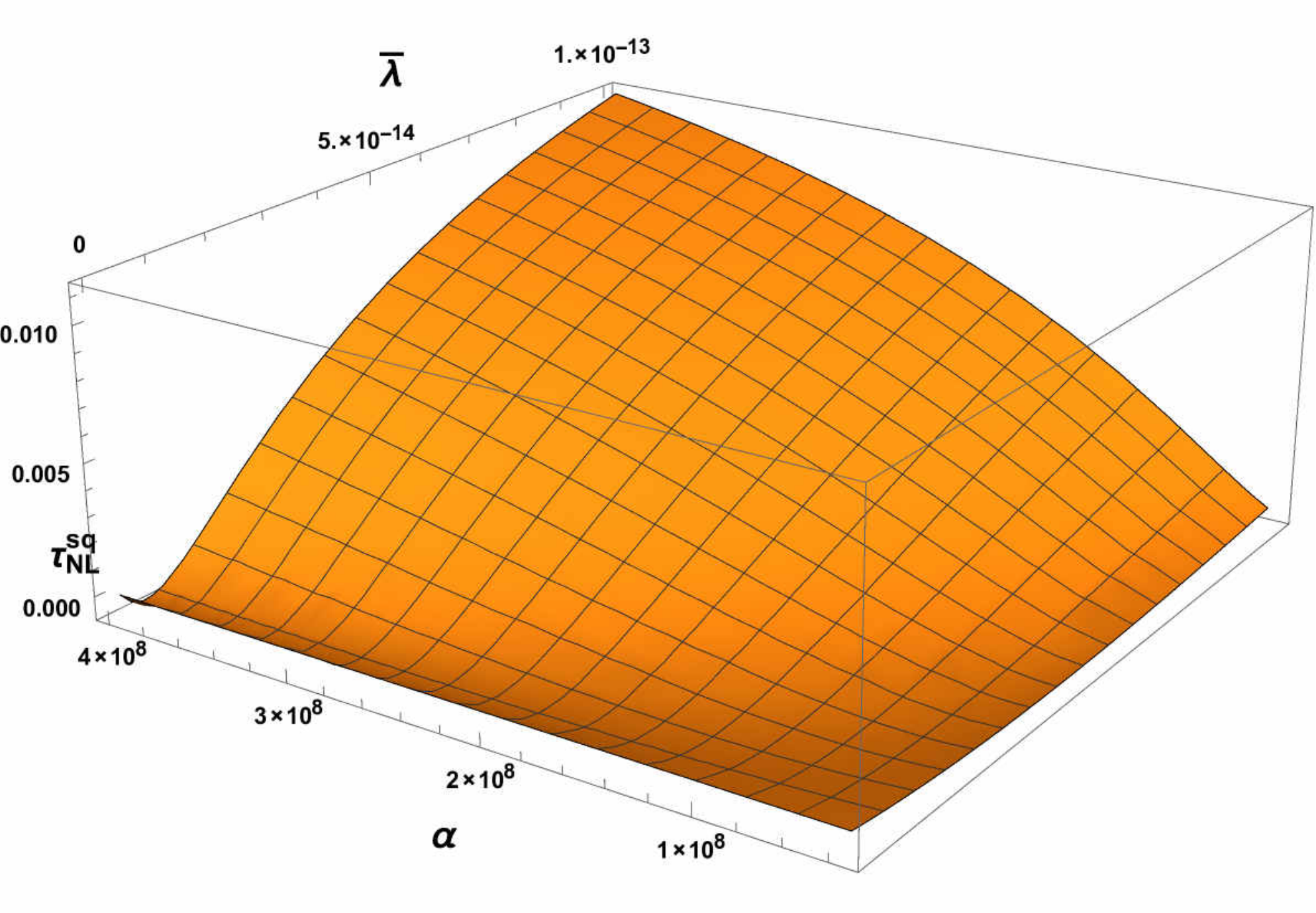}
                                \label{tnl3daaaa}
                            }
                            \subfigure[Angle~II.]{
                                \includegraphics[width=7.6cm,height=4.5cm] {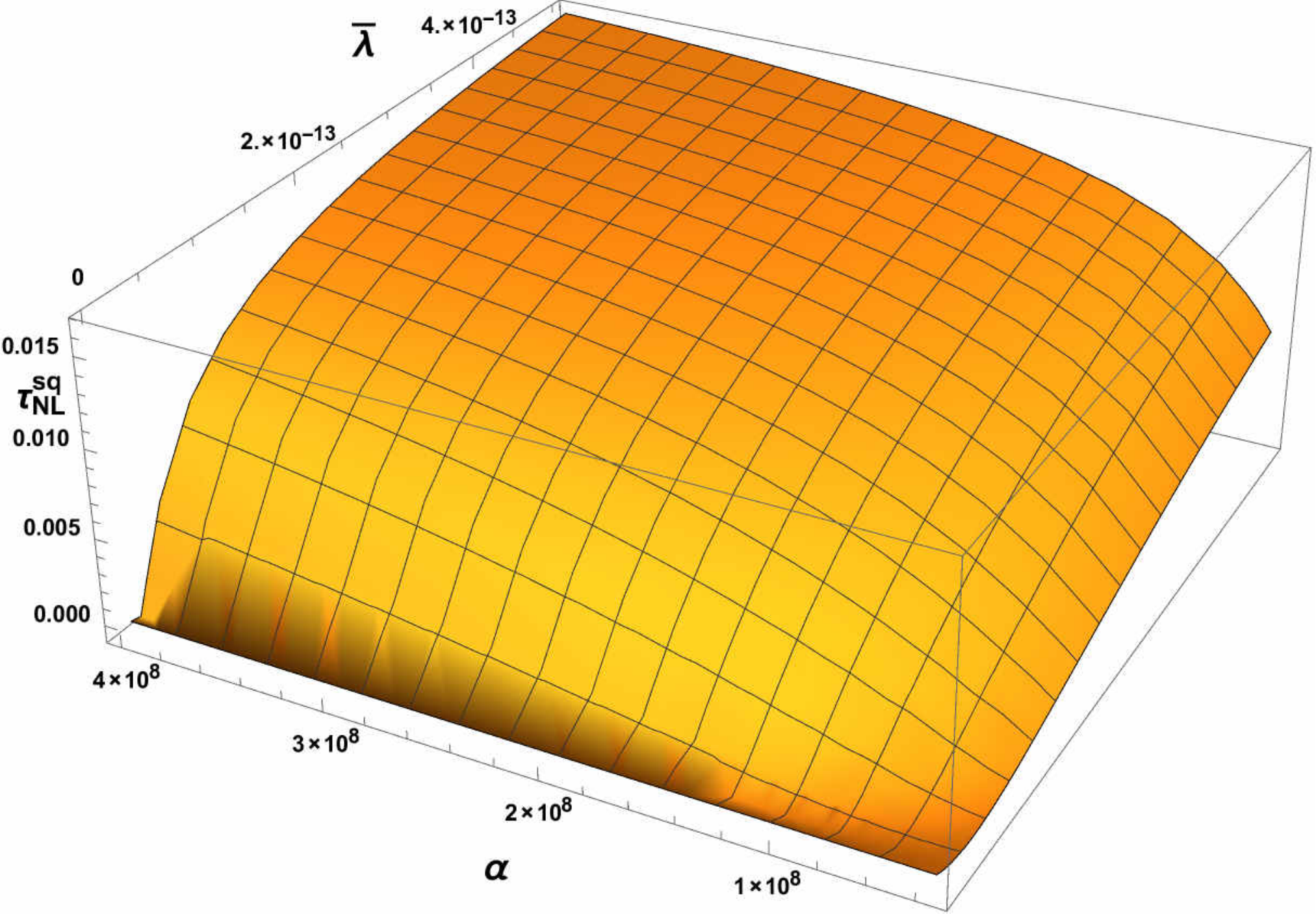}
                                \label{tnl3dbbbb}
                            }
                            \caption[Optional caption for list of figures]{Representative 3D diagram for equilateral non-Gaussian three point amplitude vs the model parameters $\alpha$ and $\bar{\lambda}$ for  ${\cal N}_{cmb}=60$ in two different angular views.} 
                            \label{tnl3bb}
                            \end{figure*}
                        \begin{figure*}[htb]
                        \centering
                        \subfigure[Range~I.]{
                            \includegraphics[width=7.6cm,height=4.5cm] {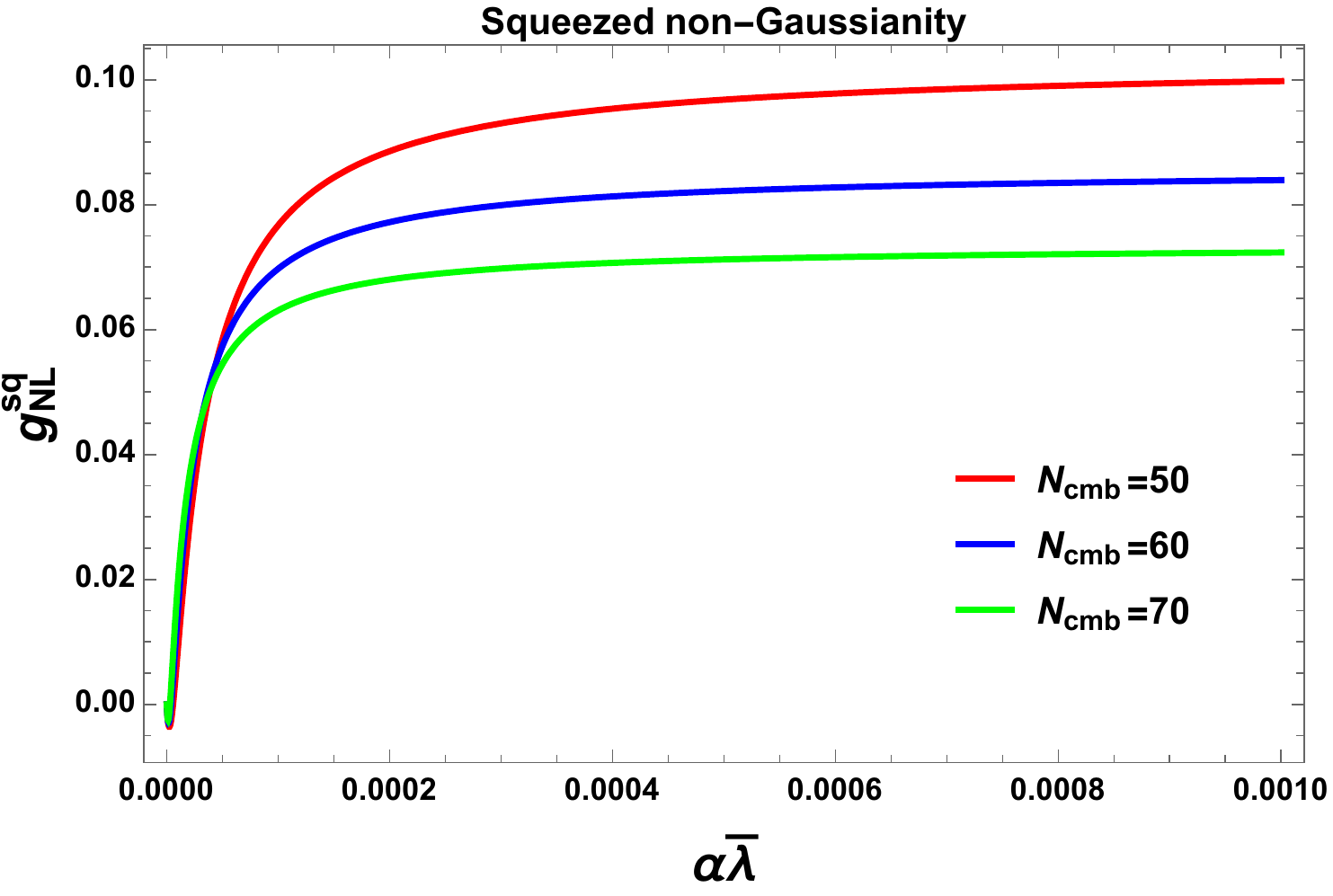}
                            \label{fige1bbbbbbbb}
                        }
                        \subfigure[Range~II.]{
                            \includegraphics[width=7.6cm,height=4.5cm] {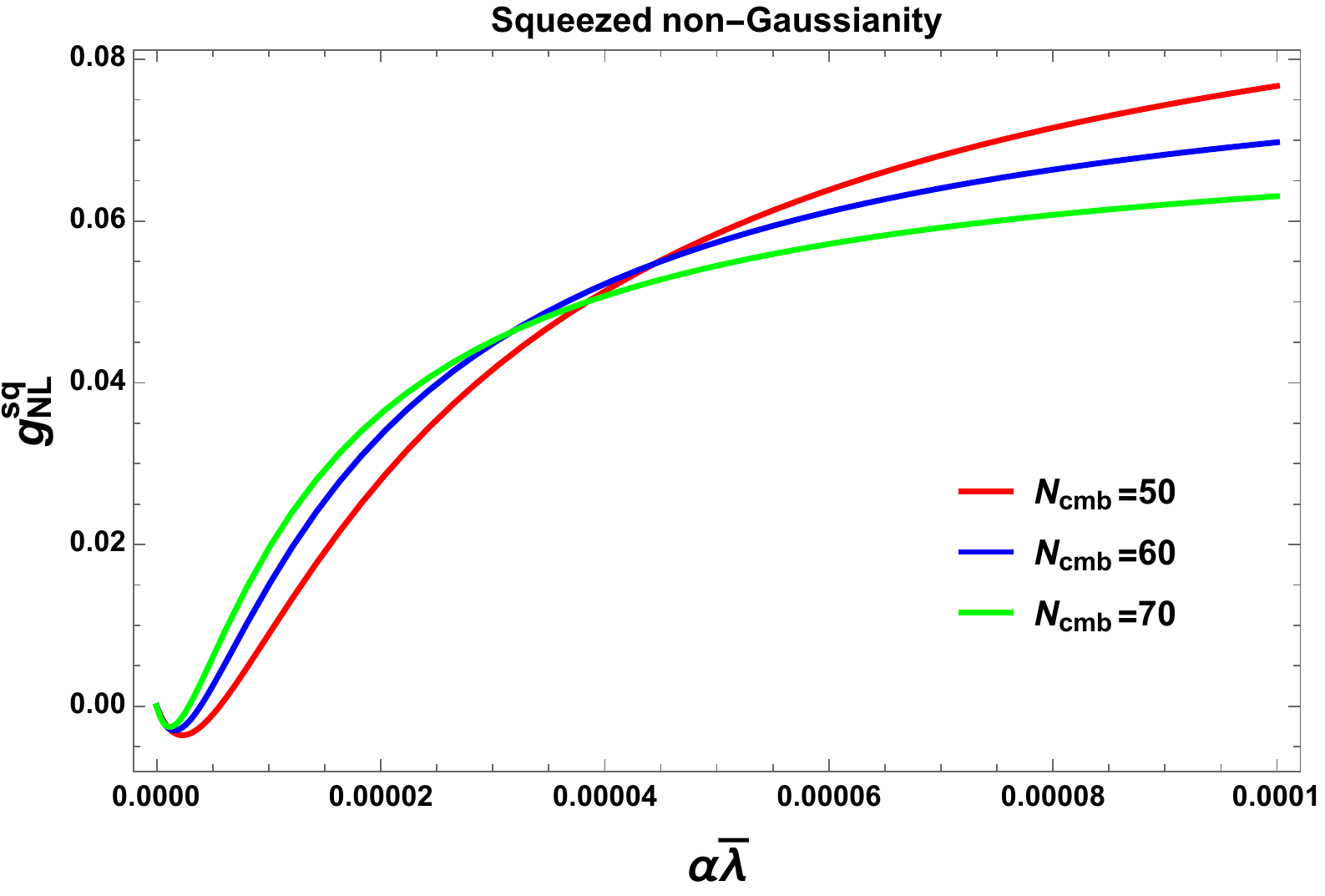}
                            \label{fige2bbbbbbbb}
                        }
                        \subfigure[Rangle~III.]{
                                 \includegraphics[width=7.6cm,height=4.5cm] {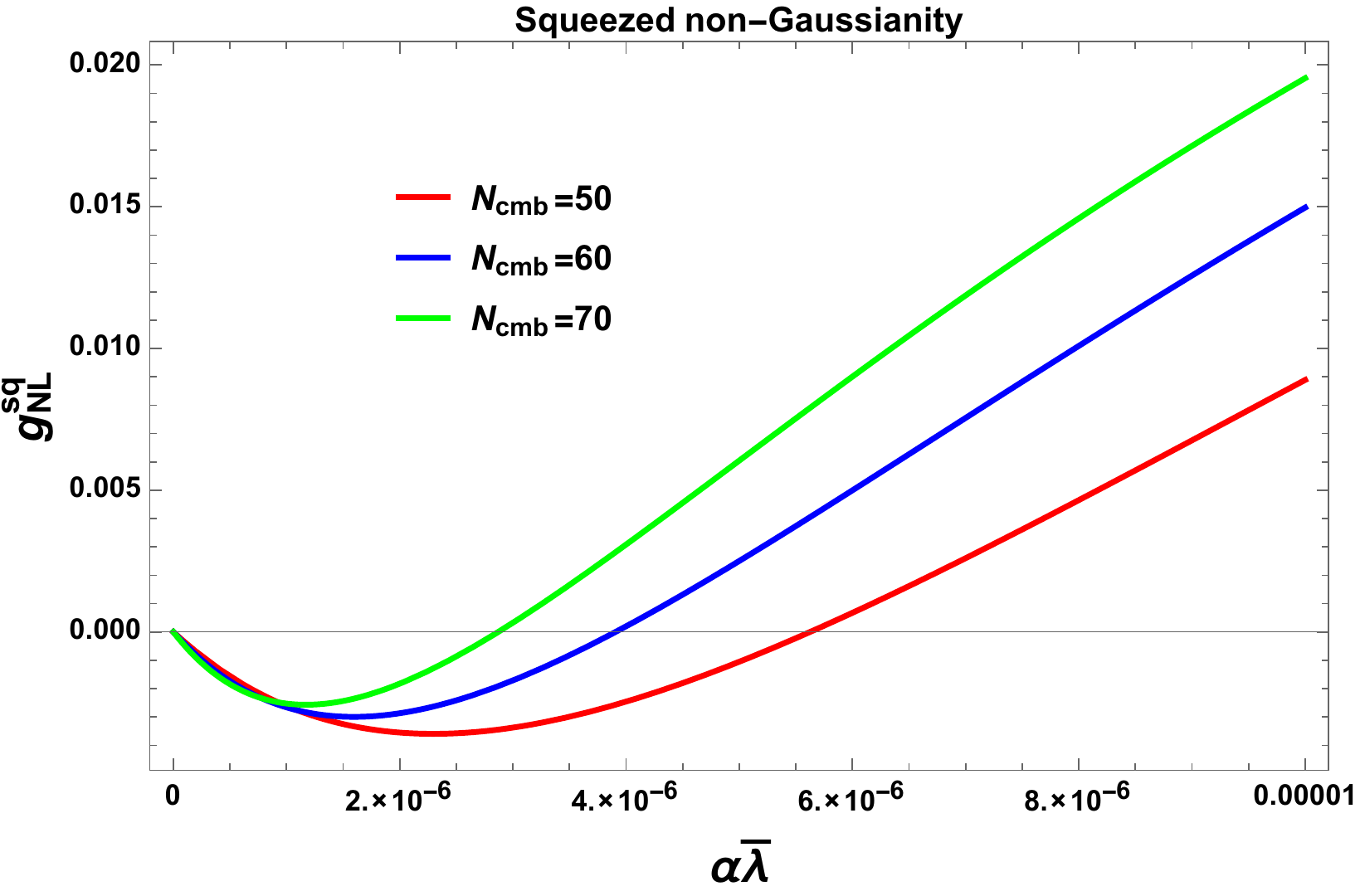}
                                 \label{fige3bbbbbbbb}
                             }
                             \subfigure[Range~IV.]{
                                           \includegraphics[width=7.6cm,height=4.5cm] {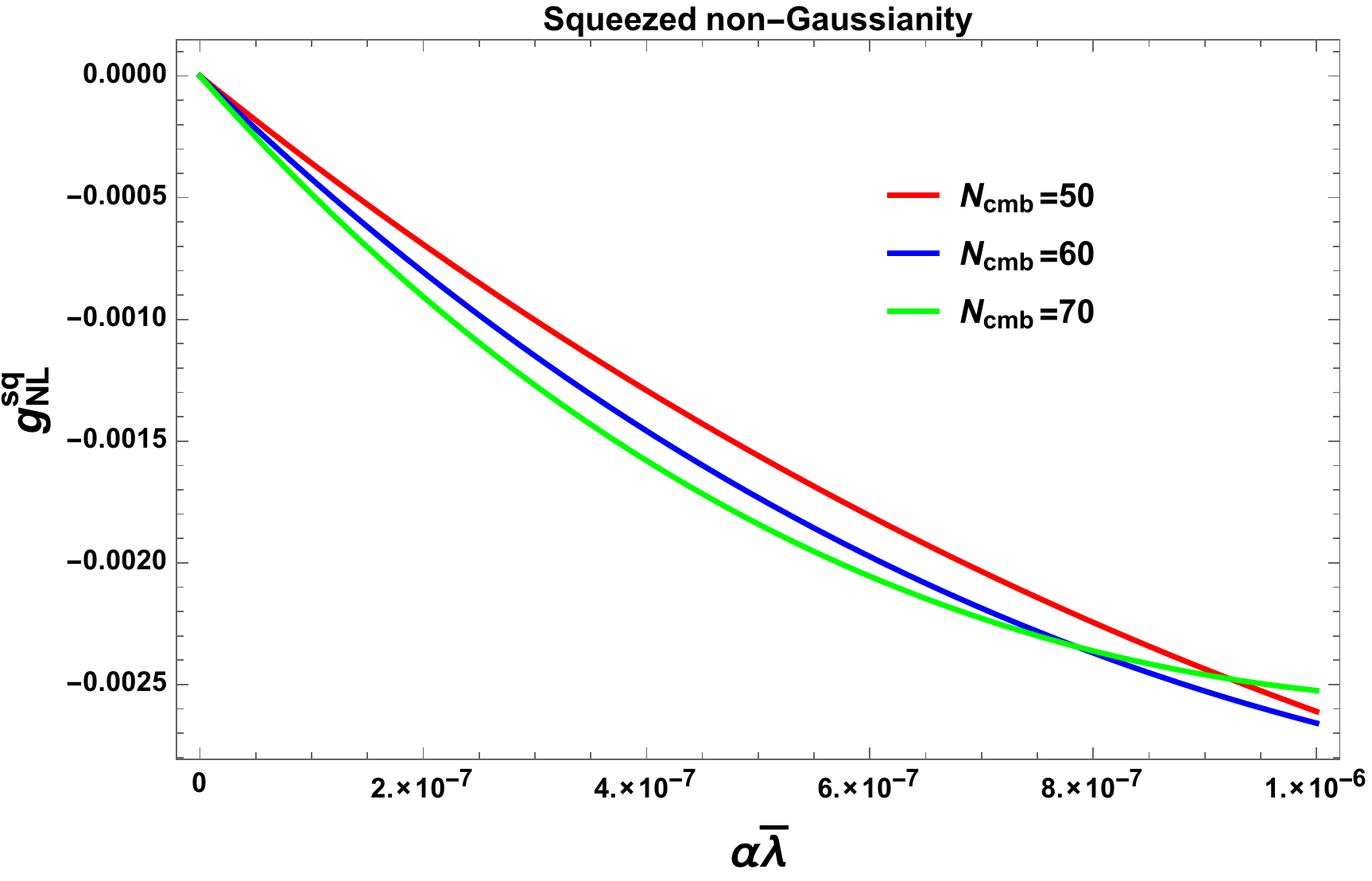}
                                           \label{fige4bbbbbbbb}
                                       }
                        \caption[Optional caption for list of figures]{Representative diagram for equilateral non-Gaussian three point amplitude vs product of the parameters $\alpha\bar{\lambda}$ in four different region for ${\cal N}_{cmb}=50$ (red), ${\cal N}_{cmb}=60$ (blue) and ${\cal N}_{cmb}=70$ (green).}
                        \label{gnleaaaaa}
                        \end{figure*}
                             \begin{figure*}[htb]
                             \centering
                             \subfigure[Angle~I.]{
                                 \includegraphics[width=7.6cm,height=4.5cm] {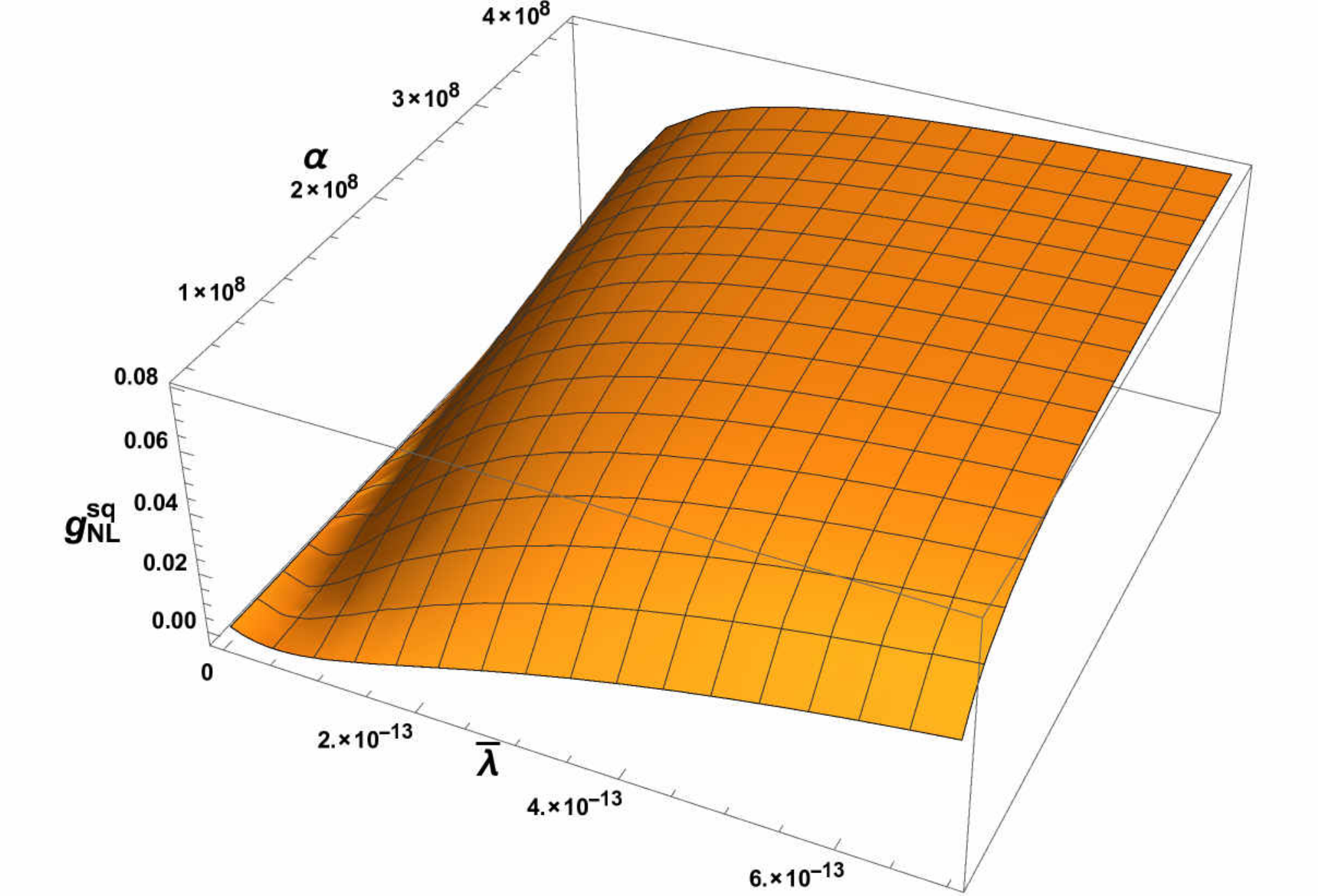}
                                 \label{gnl3daaaa}
                             }
                             \subfigure[Angle~II.]{
                                 \includegraphics[width=7.6cm,height=4.5cm] {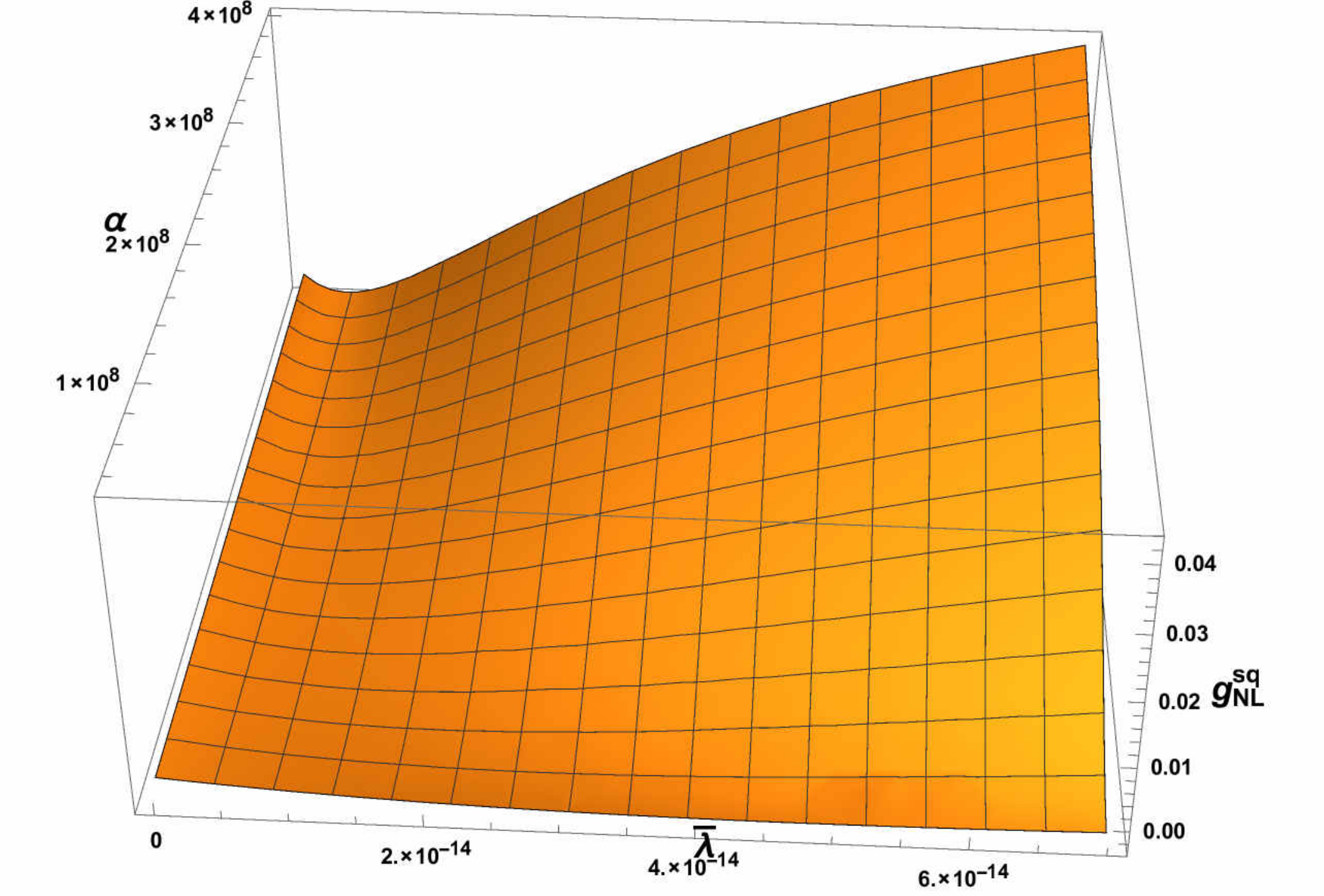}
                                 \label{gnl3dbbbb}
                             }
                             \caption[Optional caption for list of figures]{Representative 3D diagram for equilateral non-Gaussian three point amplitude vs the model parameters $\alpha$ and $\bar{\lambda}$ for  ${\cal N}_{cmb}=60$ in two different angular views.} 
                             \label{gnl3bbb}
                             \end{figure*}
    where in the \textcolor{blue}{squeezed} limiting configuration contribution from the momentum dependent functions $\hat{W}^{S}({\bf k}_1,{\bf k}_2,{\bf k}_3,{\bf k}_4)$ and $\hat{R}^{S}({\bf k}_1,{\bf k}_2,{\bf k}_3,{\bf k}_4)$ vanishes in leading order in slow-roll and negligibly small but finite contribution comes from the graviton exchange term. Here the graviton exchange contribution in \textcolor{blue}{squeezed} limiting configuration defined as:
    \bea \hat{G}^{S}({\bf k}_1,{\bf k}_2,{\bf k}_3,{\bf k}_4)
    &=&\frac{\frac{9}{4}k^6_L}{\left(1-\frac{k_S}{k_L}\right)^{\frac{3}{2}}}\sin^2\alpha_1\sin^2\alpha_3\cos 2\chi_{12,34},~~~~~~~~~~~\eea
    where in \textcolor{blue}{squeezed} limiting configuration we have used additionally the following results:
    \bea S({\bf k}_1,{\bf k}_2)&\approx&\frac{3}{2}k_L\approx S({\bf k}_3,{\bf k}_4),\eea
    and for the polarization sum we use the same results that we have used in case of \textcolor{blue}{counter collinear} limit. Additionally here we have:
    \bea \frac{\sin\alpha_{2}}{\sin\alpha_{1}}&=&\frac{k_L}{k_L}\approx 1,~~~
    \frac{\sin\alpha_{4}}{\sin\alpha_{3}}=\frac{k_L}{k_S}>> 1.
    \eea
    
    Now if we assume that the non-Gaussian parameter $\tau^{loc}_{NL}$ and $f^{loc}_{NL}$ are connected through \textcolor{blue}{\it Suyama Yamaguchi} consistency relation, then in the case where $k_{1}\sim k_{2} \sim k_{3}\approx k_L$, we get the following expression for the four point non-Gaussian parameter:
    \bea \label{opwe}\tau^{sq}_{NL}
    &\approx&\frac{1}{36} \left[29\epsilon^*_{\tilde{W}}-6\eta^{*}_{\tilde{W}}\right]^2.~~~~~~~~~~~\eea
     In this limiting configuration the normalization factor ${\cal N}_{NORM}$ that connects the two non-Gaussian parameters $\tau^{loc}_{NL}$ and $g^{loc}_{NL}$ computed from frour point function as:
   \bea {\cal N}_{NORM}&=&\frac{25}{54}\left[\frac{\Delta_{1}-\Delta_{2}}{\Delta_{3}}\right],\eea
   where the momentum dependent factors $\Delta_{1}$, $\Delta_{2}$ and $\Delta_{3}$ are defined as:
   \bea \Delta_{1}&=&\frac{25 T(k_1,k_1,k_3,k_3)}{36(f^{loc}_{NL})^2}\approx\frac{\frac{\tilde{W}^3(\phi_{cmb},{\bf \Psi})}{6M^{12}_p (\epsilon^*_{\tilde{W}})^2}\frac{1}{k^6_Lk^3_S}\left[\frac{9}{4}\frac{1}{\left(1-\frac{k_S}{k_L}\right)^{\frac{3}{2}}}\sin^2\alpha_1\sin^2\alpha_3\cos 2\chi_{12,34}\right]}{ \left[29\epsilon^*_{\tilde{W}}-6\eta^*_{\tilde{W}}\right]^2}\\
   \Delta_{2}&=&\sum_{j<p,i\neq j,p}P_{\zeta}(k_{ij})P_{\zeta}(k_j)P_{\zeta}(k_p)
   =\frac{\tilde{W}^3(\phi_{cmb},{\bf \Psi})}{1728 M^{12}_{p}(\epsilon^*_{\tilde{W}})^3}\frac{6}{k^6_L}\left(\frac{1}{k^3_L}+\frac{1}{k^3_S}\right)\frac{1}{\left(1-\frac{k_S}{k_L}\right)^{\frac{3}{2}}},~~~~~~~~~~~~~~\\
   \Delta_{3}&=&\sum_{i<j<p}P_{\zeta}(k_{i})P_{\zeta}(k_j)P_{\zeta}(k_p)=\frac{\tilde{W}^3(\phi_{cmb},{\bf \Psi})}{1728 M^{12}_{p}(\epsilon^*_{\tilde{W}})^3}\frac{1}{k^6_L}\left(\frac{1}{k^3_L}+\frac{3}{k^3_S}\right),\eea  
    Consequently the non-Gaussian parameter $g^{loc}_{NL}$ can be expressed as:
    \bea g^{sq}_{NL}&\approx&\frac{25}{1944}\left[\frac{\Delta_{1}-\Delta_{2}}{\Delta_{3}}\right] \left[29\epsilon^*_{\tilde{W}}-6\eta^{*}_{\tilde{W}}\right]^2.~~~~~~~~~~~\eea
    In fig.~(\ref{tnleaaaaa}) and fig.~(\ref{gnleaaaaa}), we have shown the features of non-Gaussian amplitude from four point scalar function $\tau^{sq}_{NL}$ and $g^{sq}_{NL}$ in squeezed limit configuration in four different scanning region of product of the two parameters $\alpha\bar{\lambda}$ in the $(\tau^{sq}_{NL},\alpha\bar{\lambda})$ and $(g^{sq}_{NL},\alpha\bar{\lambda})$ 2D plane for the number of e-foldings $50<{\cal N}_{cmb}<70$. Physical explanation of the obtained features are appended following:-
                 \begin{itemize}
                 \item \textcolor{red}{\underline{Region~I}:} \\
                 Here for the parameter space $0.0001<\alpha\bar{\lambda}<0.001$ the non-Gaussian amplitude lying within the window $ 0.01<\tau^{sq}_{NL}<0.021$,~~
                 $0.05<g^{sq}_{NL}<0.095$. Further if we increase the numerical value of $\alpha\bar{\lambda}$, then the magnitude of the non-Gaussian amplitude saturates and we get maximum value for ${\cal N}_{cmb}=50$,
                 $|\tau^{sq}_{NL}|_{max}\sim 0.021$,$
                  |g^{sq}_{NL}|_{max}\sim 0.095$.
                 \item \textcolor{red}{\underline{Region~II}}\\
                      Here for the parameter space $ 0.00001<\alpha\bar{\lambda}<0.0001$ the non-Gaussian amplitude lying within the window $ 0.002<\tau^{sq}_{NL}<0.017$,~~
                      $0.01<g^{sq}_{NL}<0.075$. In this region we get maximum value for ${\cal N}_{cmb}=50$,
                      $|\tau^{sq}_{NL}|_{max}\sim 0.017$,$
                      |g^{sq}_{NL}|_{max}\sim 0.075$.
                      Additionally it is important to note that, in this case for $\alpha\bar{\lambda}=0.00004$ the lines obtained for ${\cal N}_{cmb}=50$, ${\cal N}_{cmb}=60$ and ${\cal N}_{cmb}=70$ cross each other.
                 \item \textcolor{red}{\underline{Region~III}}\\
                           Here for the parameter space $ 0.000001<\alpha\bar{\lambda}<0.00001$ the non-Gaussian amplitude lying within the window $ -0.0008<\tau^{sq}_{NL}<0.0042$,~~
                           $-0.0008<g^{sq}_{NL}<0.02$. In this region we get maximum value for ${\cal N}_{cmb}=70$,
                           $|\tau^{sq}_{NL}|_{max}\sim 0.0042$,$
                           |g^{sq}_{NL}|_{max}\sim 0.02$.
                           Additionally it is important to note that, in this case for $0.000001\leq \alpha\bar{\lambda}\leq 0.000006$ the lines obtained for ${\cal N}_{cmb}=50$, ${\cal N}_{cmb}=60$ and ${\cal N}_{cmb}=70$ show increasing, decreasing and further increasing behaviour.
                 \item \textcolor{red}{\underline{Region~IV}}\\
                                Here for the parameter space $0.0000001<\alpha\bar{\lambda}<0.000001$ the non-Gaussian amplitude lying within the window $ -0.00005<\tau^{sq}_{NL}<-0.00058$,~
                               $-0.0001<g^{sq}_{NL}<-0.0027$. In this region we get maximum value for ${\cal N}_{cmb}=60$, 
                                $|\tau^{sq}_{NL}|_{max}\sim 0.00058$,$
                                |g^{sq}_{NL}|_{max}\sim 0.0027$.
                 \end{itemize} 
                 Further combining the contribution from \textcolor{red}{\underline{Region~I}}, \textcolor{red}{\underline{Region~II}}, \textcolor{red}{\underline{Region~III}} and   \textcolor{red}{\underline{Region~IV}} we finally get the following constraint on the four point non-Gaussian amplitude in the equilateral limit configuration:\bea \textcolor{red}{\underline{\rm Region~I}}+\textcolor{red}{\underline{\rm Region~II}}+\textcolor{red}{\underline{\rm Region~III}}+\textcolor{red}{\underline{\rm Region~IV}:}~~10^{-6}<\tau^{sq}_{NL}<0.016,~~
                 -0.023<g^{sq}_{NL}<0.002~~~~~~~~~~
                 \eea
                 for the following parameter space:
                 \bea \textcolor{red}{\underline{\rm Region~I}}+\textcolor{red}{\underline{\rm Region~II}}+\textcolor{red}{\underline{\rm Region~III}}+\textcolor{red}{\underline{\rm Region~IV}:}~~~~~0.0000001<\alpha\bar{\lambda}<0.001.~~~~~~~
                      \eea
                      In this analysis we get the following maximum value of the three point non-Gaussian amplitude in the equilateral limit configuration as given by:
                 \bea |\tau^{sq}_{NL}|_{max}\sim 0.021,~~~ |g^{sq}_{NL}|_{max}\sim 0.095.\eea
                 To visualize these constraints more clearly we have also presented $(\tau^{sq}_{NL},\alpha,\bar{\lambda})$ and $(g^{sq}_{NL},\alpha,\bar{\lambda})$ 3D plot in fig.~(\ref{tnl3daaaa}),  fig.~(\ref{tnl3dbbbb}), fig.~(\ref{gnl3daaaa}) and  fig.~(\ref{gnl3dbbbb}) for two different angular orientations as given by \textcolor{red}{\underline{Angle~I}} and \textcolor{red}{\underline{Angle~II}}. From the the representative surfaces it is clearly observed the behavior of scalar four point non-Gaussian amplitude in the squeezed limit for the variation of two fold parameter $\alpha$ and $\bar{\lambda}$ and the results are consistent with the obtained constraints in 2D analysis. Here all the obtained results are consistent with the two point and three point constaints as well as with the Planck 2015 data \cite{Ade:2015lrj,Ade:2015ava,Ade:2015xua}.
                 
    For the sake of simplicity one can further neglect all the contrubition from the very very small factor $k_S/k_L$ and finally write the following expression for the trispectrum in the \textcolor{blue}{squeezed} limiting configuration as:
      \bea\label{dfg} T(k_L,k_L,k_L,k_S)&\approx&3\left(\tau^{sq}_{NL}+\frac{54}{25}g^{sq}_{NL}\right)P_{\zeta}(k_S)P^2_{\zeta}(k_L),\eea
      which implies that in the \textcolor{blue}{squeezed} limiting configuration if we neglect the contribution from very small term $k_S/k_L$ then the trispectrum contributes equally to the four point non-Gaussian parameter $\tau^{loc}_{NL}$ and $g^{loc}_{NL}$. So if we now assume that the \textcolor{blue}{Suyma Yamguchi} relation holds good perfectly in the present context i.e. Eq~(\ref{opwe}) is completely correct then using Eq~(\ref{dfg}) one can find out the following expresseion for the four point non-Gaussian parameter $g^{loc}_{NL}$ as:
      \bea g^{sq}_{NL}&\approx& 6\epsilon^*_{\tilde{W}}\sin^2\alpha_1\sin^2\alpha_3\cos 2\chi_{12,34}-\frac{1}{36} \left[29\epsilon^*_{\tilde{W}}-6\eta^*_{\tilde{W}}\right]^2,\eea
       where we have used the approximation, $(1+\cos\theta_3)\approx \frac{1}{2}\left(1-\frac{k_S}{k_L}\right)\approx \frac{1}{2}$, due to the smallness of of the momentum ratio $k_S/k_L$ as it is much smaller than unity.
       
 But as we have already pointed that if we relax the assumption of holding the \textcolor{blue}{\it Suyama Yamaguchi} consistency relation in the present context of discussion, then using Eq~(\ref{ w2}) one can write down the expression for momentum dependent function in the  \textcolor{blue}{squeezed limiting configuration} as:  
      \bea f\left(k_L,k_L,k_L,k_S,k_L\sqrt{1-\frac{k_S}{k_L}},k_L\sqrt{1-\frac{k_S}{k_L}},k_L\sqrt{1-\frac{k_S}{k_L}}\right) \approx \frac{\frac{128}{9}\left(1+\frac{k^3_S}{k^3_L}\right)}{\left(1+\frac{k_S}{3k_L}\right)^3}\frac{1}{\left(1-\frac{k_S}{k_L}\right)^{\frac{3}{2}}},\eea
     using which we get the following simplified expression for the non-Gaussian parameter $\tau^{loc}_{NL}$ and $g^{loc}_{NL}$ as obtained from the four point scalar function
     in \textcolor{blue}{squeezed limiting configuration} as: 
      \bea \tau^{sq}_{NL}
      &\approx&\frac{\frac{\tilde{W}^3(\phi_{cmb},{\bf \Psi})}{216M^{12}_p (\epsilon^*_{\tilde{W}})^2}\frac{1}{k^9_L}\frac{\sin^3\alpha_4}{\sin^3\alpha_3}\left[\frac{9}{4}\frac{1}{\left(1-\frac{\sin\alpha_3}{\sin\alpha_4}\right)^{\frac{3}{2}}}\sin^2\alpha_1\sin^2\alpha_3\cos 2\chi_{12,34}\right]}{\left[\Delta_2+\frac{54}{25}\frac{128}{9}\left(1+\frac{\sin^3\alpha_3}{\sin^3\alpha_4}\right)\frac{1}{\left(1+\frac{\sin\alpha_3}{3\sin\alpha_4}\right)^3}\frac{1}{\left(1-\frac{\sin\alpha_3}{\sin\alpha_4}\right)^{\frac{3}{2}}}\Delta_3\right]},\eea
      \bea
       g^{sq}_{NL}
       &\approx& \frac{\frac{\tilde{W}^3(\phi_{cmb},{\bf \Psi})}{216M^{12}_p (\epsilon^*_{\tilde{W}})^2}\frac{1}{k^9_L}\frac{\sin^3\alpha_4}{\sin^3\alpha_3}\left[\frac{9}{4}\frac{1}{\left(1-\frac{\sin\alpha_3}{\sin\alpha_4}\right)^{\frac{3}{2}}}\sin^2\alpha_1\sin^2\alpha_3\cos 2\chi_{12,34}\right]}{\left[\frac{\Delta_2}{\frac{128}{9}\left(1+\frac{\sin^3\alpha_3}{\sin^3\alpha_4}\right)\frac{1}{\left(1+\frac{\sin\alpha_3}{3\sin\alpha_4}\right)^3}\frac{1}{\left(1-\frac{\sin\alpha_3}{\sin\alpha_4}\right)^{\frac{3}{2}}}}+\frac{54}{25}\Delta_3\right]}. \eea      
      Further if we neglect the contribution from very small term $k_S/k_L$ then the four point non-Gaussian parameter $\tau^{sq}_{NL}$ and $g^{sq}_{NL}$ can be expressed as:
      \bea \tau^{sq}_{NL}
            &\approx&\frac{\frac{\tilde{W}^3(\phi_{cmb},{\bf \Psi})}{216M^{12}_p (\epsilon^*_{\tilde{W}})^2}\frac{1}{k^9_L}\frac{\sin^3\alpha_4}{\sin^3\alpha_3}\left[\frac{9}{4}\sin^2\alpha_1\sin^2\alpha_3\cos 2\chi_{12,34}\right]}{\left[\Delta_2+\frac{54}{25}\frac{128}{9}\Delta_3\right]},\eea 
            \bea
             g^{sq}_{NL}
             &\approx& \frac{\frac{\tilde{W}^3(\phi_{cmb},{\bf \Psi})}{216M^{12}_p (\epsilon^*_{\tilde{W}})^2}\frac{1}{k^9_L}\frac{\sin^3\alpha_4}{\sin^3\alpha_3}\left[\frac{9}{4}\sin^2\alpha_1\sin^2\alpha_3\cos 2\chi_{12,34}\right]}{\left[\frac{9}{128}\Delta_2+\frac{54}{25}\Delta_3\right]}, \eea   
             where the momentum dependent factors can be approximated as:
              \bea f\left(k_L,k_L,k_L,k_S,k_L,k_L,k_L\right)
                         \displaystyle  &\approx&\frac{128}{9},\eea
             \bea
                \Delta_{2}
                &=&\frac{\tilde{W}^3(\phi_{cmb},{\bf \Psi})}{1728 M^{12}_{p}(\epsilon^*_{\tilde{W}})^3}\frac{6}{k^6_L}\left(\frac{1}{k^3_L}+\frac{1}{k^3_S}\right)\frac{1}{\left(1-\frac{k_S}{k_L}\right)^{\frac{3}{2}}},~~~~~~~~\\
                \Delta_{3}&=&\frac{\tilde{W}^3(\phi_{cmb},{\bf \Psi})}{1728 M^{12}_{p}(\epsilon^*_{\tilde{W}})^3}\frac{3}{k^6_Lk^3_S},\eea 
\end{enumerate}
 
 \begin{table*}
      \centering
       \tiny
      \begin{tabular}{|c|c|c|}
     \hline
      \hline
       \textcolor{red}{Scanning Region} &$\tau^{equil}_{NL}$ &$g^{equil}_{NL}$\\
      \hline\hline\hline
       \textcolor{red}{ I}  &  $0.006<\tau^{equil}_{NL}<0.016$ & $-0.004<g^{equil}_{NL}<-0.023$
           \\
      \hline\hline
       \textcolor{red}{ II}  & $0.001<\tau^{equil}_{NL}<0.009$ & $0.002<g^{equil}_{NL}<-0.011$
           \\
      \hline\hline
      \textcolor{red}{ III}  & $0.00002<\tau^{equil}_{NL}<0.00062$ & $0.0001<g^{equil}_{NL}<0.0017$
           \\ 
      \hline\hline
      \textcolor{red}{ IV}  & $10^{-6}<\tau^{equil}_{NL}<0.000012$ & $2\times 10^{-6}<g^{equil}_{NL}<0.00011$
           \\ 
      \hline\hline
           \textcolor{red}{ I+II+III+IV}  & $10^{-6}<\tau^{equil}_{NL}<0.016$ & $2\times -0.023<g^{equil}_{NL}<0.002$
                     \\ \hline\hline
      \hline
      \end{tabular}
       
      \begin{tabular}{|c|c|c|}
           \hline
            \hline
             \textcolor{red}{Scanning Region} &$\tau^{foldkite}_{NL}$ &$g^{foldkite}_{NL}$\\
            \hline\hline\hline
             \textcolor{red}{ I}  &  $0.0025<\tau^{foldkite}_{NL}<0.0085$ & $0.014<g^{foldkite}_{NL}<0.038$
                 \\
            \hline\hline
             \textcolor{red}{ II}  & $0.0002<\tau^{foldkite}_{NL}<0.0048$ & $0.001<g^{foldkite}_{NL}<0.023$
                 \\
            \hline\hline
            \textcolor{red}{ III}  & $0.000018<\tau^{foldkite}_{NL}<0.001$ & $0.0001<g^{foldkite}_{NL}<0.001$
                 \\ 
            \hline\hline
            \textcolor{red}{ IV}  & $10^{-6}<\tau^{foldkite}_{NL}<0.000014$ & $2\times 10^{-6}<g^{foldkite}_{NL}<0.000066$
                 \\ 
            \hline\hline
                 \textcolor{red}{ I+II+III+IV}  & $10^{-6}<\tau^{foldkite}_{NL}<0.016$ & $-0.023<g^{foldkite}_{NL}<0.002$
                           \\ \hline\hline
            \hline
            \end{tabular}
            \begin{tabular}{|c|c|c|}
                 \hline
                  \hline
                   \textcolor{red}{Scanning Region} &$\tau^{sq}_{NL}$ &$g^{sq}_{NL}$\\
                  \hline\hline\hline
                   \textcolor{red}{ I}  &  $0.00028<\tau^{equil}_{NL}<0.00052$ & $0.0022<g^{equil}_{NL}<0.004$
                       \\
                  \hline\hline
                   \textcolor{red}{ II}  & $0.00005<\tau^{equil}_{NL}<0.00042$ & $0.0005<g^{equil}_{NL}<0.0033$
                       \\
                  \hline\hline
                  \textcolor{red}{ III}  & $0.00001<\tau^{equil}_{NL}<0.00014$ & $0.00008<g^{equil}_{NL}<0.0014$
                       \\ 
                  \hline\hline
                  \textcolor{red}{ IV}  & $10^{-7}<\tau^{equil}_{NL}<7\times 10^{-6}$ & $5\times 10^{-8}<g^{equil}_{NL}<0.000052$
                       \\ 
                  \hline\hline
                       \textcolor{red}{ I+II+III+IV}  & $10^{-7}<\tau^{equil}_{NL}<0.00052$ & $5\times 10^{-8}<g^{equil}_{NL}<0.004$
                                 \\ \hline\hline
                  \hline
                  \end{tabular}
      \caption{Contraint on scalar four point non-Gaussian amplitude from equilateral, folded kite and squeezed configuration with assuming Suyama Yamguchi consistency relation.}\label{tab111}
      \vspace{.7cm}
      \end{table*}
      \begin{table*}
            \centering
             \tiny
            \begin{tabular}{|c|c|c|}
           \hline
            \hline
             \textcolor{red}{Scanning Region} &$\tau^{equil}_{NL}$ &$g^{equil}_{NL}$\\
            \hline\hline\hline
             \textcolor{red}{ I}  &  $0.00028<\tau^{equil}_{NL}<0.00052$ & $0.0022<g^{equil}_{NL}<0.004$
                 \\
            \hline\hline
             \textcolor{red}{ II}  & $0.00005<\tau^{equil}_{NL}<0.00042$ & $0.0005<g^{equil}_{NL}<0.0033$
                 \\
            \hline\hline
            \textcolor{red}{ III}  & $0.00001<\tau^{equil}_{NL}<0.00014$ & $0.00008<g^{equil}_{NL}<0.0014$
                 \\ 
            \hline\hline
            \textcolor{red}{ IV}  & $10^{-7}<\tau^{equil}_{NL}<7\times 10^{-6}$ & $5\times 10^{-8}<g^{equil}_{NL}<0.000052$
                 \\ 
            \hline\hline
                 \textcolor{red}{ I+II+III+IV}  & $10^{-7}<\tau^{equil}_{NL}<0.00052$ & $5\times 10^{-8}<g^{equil}_{NL}<0.004$
                           \\ \hline\hline
            \hline
            \end{tabular}
            \caption{Contraint on scalar four point non-Gaussian amplitude from equilateral configuration without assuming Suyama Yamguchi consistency relation.}\label{tab111b}
            \vspace{.7cm}
            \end{table*}
      In table.~(\ref{tab111}), we give the numerical estimates and constraints on the four point non-Gaussian amplitude from equilateral configuration with assuming Suyama Yamguchi consistency relation. Also in table.~(\ref{tab111b}), we give the numerical estimates and constarints on the four point non-Gaussian amplitude from equilateral configuration without assuming Suyama Yamguchi consistency relation. Here all the obtained results are consistent with the two point and three point constaints as well as with the Planck 2015 data \cite{Ade:2015lrj,Ade:2015ava,Ade:2015xua}.
\subsubsection{Using $\delta {\cal N}$ formalism}
\label{s7b2}
In this section using the prescription of $\delta {\cal N}$ formalism in the attractor regime of cosmological perturbation we derive the expression for the non-Gaussian amplitudes associated with the four point function of scalar curvature fluctuation as:
\bea \tau^{loc}_{NL}&=&\frac{{\cal N}_{,JI}{\cal N}_{,JK}{\cal N}_{,I}{\cal N}_{,K}}{\left({\cal N}_{,M}{\cal N}_{,M}\right)^3},\eea \bea 
g^{loc}_{NL}&=&\frac{25}{54}\frac{{\cal N}_{,IJK}{\cal N}_{,I}{\cal N}_{,J}{\cal N}_{,K}}{\left({\cal N}_{,M}{\cal N}_{,M}\right)^3}.
\eea
Further writing the expressions for the non-Gaussian amplitudes in terms of the inflaton field $\phi$ and the additional field $\Psi$ we get:
\bea \tau^{loc}_{NL}&=&\frac{1}{\left({\cal N}_{,\phi}{\cal N}_{,\phi}+{\cal N}_{,\Psi}{\cal N}_{,\Psi}\right)^3_{*}}\left[\left({\cal N}_{,\phi\phi}{\cal N}_{,\phi\phi}+{\cal N}_{,\Psi\phi}{\cal N}_{,\Psi\phi}\right){\cal N}_{,\phi}{\cal N}_{,\phi}\right.\nonumber\\  &&\left.~~~~~~~~~~~~~~~~~~~~~~~+\left({\cal N}_{,\Psi\Psi}{\cal N}_{,\Psi\Psi}+{\cal N}_{,\phi\Psi}{\cal N}_{,\phi\Psi}\right){\cal N}_{,\Psi}{\cal N}_{,\Psi}\right.\nonumber\\  &&\left.~~~~~~~~~~~~~~~~~~~~~~~+2\left({\cal N}_{,\phi\phi}{\cal N}_{,\phi\Psi}+{\cal N}_{,\Psi\Psi}{\cal N}_{,\Psi\phi}\right){\cal N}_{,\phi}{\cal N}_{,\Psi}\right]_{*},\\
g^{loc}_{NL}&=&\frac{25}{54}\frac{1}{\left({\cal N}_{,\phi}{\cal N}_{,\phi}+{\cal N}_{,\Psi}{\cal N}_{,\Psi}\right)^3_{*}}\left[{\cal N}_{,\phi\phi\phi}{\cal N}_{,\phi}{\cal N}_{,\phi}{\cal N}_{,\phi}+{\cal N}_{,\Psi\Psi\Psi}{\cal N}_{,\Psi}{\cal N}_{,\Psi}{\cal N}_{,\Psi}\right.\nonumber\\  &&\left.~~~~~~~~~~~~~~~~~~+\left({\cal N}_{,\phi\phi\Psi}+{\cal N}_{,\Psi\phi\phi}+{\cal N}_{,\phi\Psi\phi}\right){\cal N}_{,\phi}{\cal N}_{,\phi}{\cal N}_{,\Psi}\right.\nonumber\\  &&\left.~~~~~~~~~~~~~~~~~~+\left({\cal N}_{,\phi\Psi\Psi}+{\cal N}_{,\Psi\Psi\phi}+{\cal N}_{,\Psi\phi\Psi}\right){\cal N}_{,\phi}{\cal N}_{,\Psi}{\cal N}_{,\Psi}\right]_{*}.
\eea
Now we already know that in the attractor regime cosmological perturbation, solution for the additional field $\Psi$ can be expressed in terms odf the inflaton field $\phi$ and using this fact the expression for the non-Gaussian amplitudes associated with the four point function of scalar curvature fluctuation can be recast as:
\bea \tau^{loc}_{NL}&=&\left[X_{1}(\phi)~\frac{{\cal N}_{,\phi\phi}{\cal N}_{,\phi\phi}}{\left({\cal N}_{,\phi}{\cal N}_{,\phi}\right)^2}+X_2(\phi)~\frac{{\cal N}_{,\phi\phi}}{{\cal N}^3_{,\phi}}+X_3(\phi)~\frac{1}{{\cal N}_{,\phi}{\cal N}_{,\phi}}\right]_{*},\\
g^{loc}_{NL}&=&\frac{25}{54}\left[X_{4}(\phi)\frac{{\cal N}_{,\phi\phi\phi}}{{\cal N}_{,\phi}{\cal N}_{,\phi}{\cal N}_{,\phi}}+X_{5}(\phi)\frac{{\cal N}_{,\phi\phi}}{{\cal N}_{,\phi}{\cal N}_{,\phi}{\cal N}_{,\phi}}+X_{6}(\phi)\frac{1}{{\cal N}_{,\phi}{\cal N}_{,\phi}}\right]_{*}.~~~~~~
\eea 
where the new functions $X_{1}(\phi),\cdots,X_{6}(\phi)$ are defined as:
\bea X_{1}(\phi)&=& f(\phi)\left(1+\frac{2}{{\cal V}^2(\phi)}+\frac{2}{{\cal V}^4(\phi)}+\frac{1}{{\cal V}^6(\phi)}\right),~~~
X_{2}(\phi)=-3f(\phi)\left(\frac{{\cal V}^{'}(\phi)}{{\cal V}^3(\phi)}+\frac{{\cal V}^{'}(\phi)}{{\cal V}^5(\phi)}\right),\nonumber\\
X_{3}(\phi)&=& f(\phi)\left(\frac{{\cal V}^{'2}(\phi)}{{\cal V}^6(\phi)}+\frac{{\cal V}^{'2}(\phi)}{{\cal V}^4(\phi)}\right),~~~
X_{4}(\phi)=f(\phi)\left(1+\frac{3}{{\cal V}^2(\phi)}+\frac{3}{{\cal V}^4(\phi)}+\frac{1}{{\cal V}^6(\phi)}\right),\nonumber\\
X_{5}(\phi)&=&-3f(\phi)\left(\frac{{\cal V}^{'}(\phi)}{{\cal V}^7(\phi)}+2\frac{{\cal V}^{'}(\phi)}{{\cal V}^5(\phi)}+\frac{{\cal V}^{'}(\phi)}{{\cal V}^3(\phi)}\right),\nonumber\\
X_{6}(\phi)&=&-f(\phi)\left(\frac{{\cal V}^{''}(\phi)}{{\cal V}^3(\phi)}-2\frac{{\cal V}^{'2}(\phi)}{{\cal V}^4(\phi)}-3\frac{{\cal V}^{'2}(\phi)}{{\cal V}^8(\phi)}-5~\frac{{\cal V}^{'2}(\phi)}{{\cal V}^6(\phi)}\right). \eea
where $f(\phi)=\left(1+\frac{1}{{\cal V}^2(\phi)}\right)^{-3}$.
Further substituting the explicit form of the function ${\cal V}(\phi)$ and ${\cal N}_{,\phi}$, ${\cal N}_{,\phi\phi}$,  ${\cal N}_{,\phi\phi\phi}$ for all derived effective potentials at $\phi=\phi_*$ we get:
\bea\label{p1} \tau^{loc}_{NL}&=&{\cal Y}^2\left[X_{1}(\phi_{*})+X_2(\phi_{*})\phi_*+X_3(\phi_{*})\phi^2_*\right],\\
\label{p2} g^{loc}_{NL}&=&\frac{25}{54}{\cal Y}^2\left[2X_{4}(\phi_{*})+X_{5}(\phi_{*})\phi_*+X_{6}(\phi_{*})\phi^2_*\right].~~~~~~
\eea 
Now we comment on the consistency relation between the non-Gaussian parameters derived from four point and theree point scalar correlation function in the attractor regime of inflation.  To establish this connection we start with Eq.~(\ref{p23}),~Eq.~(\ref{p1}) and Eq.~(\ref{p2}) and finally getnew set of consistency relations:
\bea\label{p1111} \tau^{loc}_{NL}&=&\frac{36}{25}(f^{loc}_{NL})^2\left[X_{1}(\phi_{*})+X_2(\phi_{*})\phi_*+X_3(\phi_{*})\phi^2_*\right],\\
\label{p2111} g^{loc}_{NL}&=&\frac{10}{27}(f^{loc}_{NL})^2\left[2X_{4}(\phi_{*})+X_{5}(\phi_{*})\phi_*+X_{6}(\phi_{*})\phi^2_*\right].~~~~~~
\\ \label{p2111z} g^{loc}_{NL}&=&\frac{125}{486}\tau^{loc}_{NL}\frac{\left[2X_{4}(\phi_{*})+X_{5}(\phi_{*})\phi_*+X_{6}(\phi_{*})\phi^2_*\right]}{\left[2X_{4}(\phi_{*})+X_{5}(\phi_{*})\phi_*+X_{6}(\phi_{*})\phi^2_*\right]}.~~~~~~
\eea 
It is a very well known fact that in the non attractor regime, where the additional field $\Psi$ is freezed in the Planck scale {\it Suyama-Yamguchi} consistency relation \cite{Smith:2011if,Suyama:2010uj,Suyama:2007bg} holds good, which states:
\bea \tau^{loc}_{NL}=\frac{36}{25}(f^{loc}_{NL})^2.\eea
Further using this results one can estimate the devation in the {\it Suyama-Yamguchi} consistency relation if we go from attractor regime to non-attractor regime of cosmological peturbation as:
\bea |\Delta \tau^{loc}_{NL}|&=&|\left[\tau^{loc}_{NL}|_{non-attractor}-\tau^{loc}_{NL}|_{attractor}\right]=\frac{36}{25}(f^{loc}_{NL})^2|_{non-attractor}|\left\{1-Q_{corr}\right\}|,
~~~~~~~~~~~~\eea
where the correction factor $Q_{corr}$ can bw written as:
\bea Q_{corr}&=&\frac{(f^{loc}_{NL})^2|_{attractor}}{(f^{loc}_{NL})^2|_{non-attractor}}\left[X_{1}(\phi_{*})+X_2(\phi_{*})\phi_*+X_3(\phi_{*})\phi^2_*\right].\eea
Here we need to point few crucial issues as appended below:
\begin{itemize}
\item First of all, to estimate the magnitude of the deviation factor $Q_{corr}$ we need to concentrate on two physical situations,
\textcolor{red}{\bf I. Super planckian field regime} and
\textcolor{red}{\bf II. Sub planckian field regime}.
\item In the super planckian field regime the deviation factor $Q_{corr}$ can be expressed as:
\be\begin{array}{lll}\label{rinfla1cxccdfzzz}\tiny
    \displaystyle Q_{corr}\displaystyle=\Delta_{f}\times\left\{\begin{array}{ll}
                      \displaystyle \left(1-\frac{18M^2_p}{81\phi^2_*}-\frac{72M^4_p}{6561\phi^4_*}-\frac{3888M^8_p}{43046721\phi^8_*}+\cdots\right)
                                                 ~~~~~ &
    \mbox{\small \textcolor{red}{\bf for \underline{Case I}}}  
   \\ 
            \displaystyle \left(1-\frac{18M^2_p}{\phi^2_*}-\frac{72M^4_p}{\phi^4_*}-\frac{3888M^8_p}{\phi^8_*}+\cdots\right)& \mbox{\small \textcolor{red}{\bf for \underline{Case II}}}  
            \\  
                     \displaystyle
                     \left(1-\frac{18M^2_p}{\phi^2_*\left(1-\frac{\phi^4_V}{\phi^4_*}\right)^2}-\frac{72M^4_p}{\phi^4_*\left(1-\frac{\phi^4_V}{\phi^4_*}\right)^4} 
                     \displaystyle
                     -\frac{3888M^8_p}{\phi^8_*\left(1-\frac{\phi^4_V}{\phi^4_*}\right)^8}+\cdots\right) & \mbox{\small \textcolor{red}{\bf for \underline{Case ~II+Choice~I}}}\\ 
                                       \displaystyle
                                        \left(1-\frac{18M^2_p}{\phi^2_*\left(1-\frac{m^2_c}{m^2_c-\lambda\phi^2_*}\right)^2}-\frac{72M^4_p}{\phi^4_*\left(1-\frac{m^2_c}{m^2_c-\lambda\phi^2_*}\right)^4} 
                                         -\frac{3888M^8_p}{\phi^8_*\left(1-\frac{m^2_c}{m^2_c-\lambda\phi^2_*}\right)^8}+\cdots\right)
                                                                  & \mbox{\small \textcolor{red}{\bf for \underline{Case ~II+Choice~II}}}\\ 
                                                         \displaystyle
       \left(1-\frac{18M^2_p}{\phi^2_*\left(1+\xi(\phi^2_*-\phi^2_V)+\frac{\phi^2_V}{\phi^2_*}\right)^2}-\frac{72M^4_p}{\phi^4_*\left(1+\xi(\phi^2_*-\phi^2_V)+\frac{\phi^2_V}{\phi^2_*}\right)^4}\right.\\ \left.~~~~~~~~~~~~~~~~~~~~~~~~~~~~\displaystyle -\frac{3888M^8_p}{\phi^8_*\left(1+\xi(\phi^2_*-\phi^2_V)+\frac{\phi^2_V}{\phi^2_*}\right)^8}+\cdots\right)                                                   & \mbox{\small \textcolor{red}{\bf for \underline{Case ~II+Choice~III}}}.
             \end{array}
   \right.
   \end{array}\ee            
       where the factor $\Delta_{f}$ is defined as:
       \bea \Delta_{f}=\frac{(f^{loc}_{NL})^2|_{attractor}}{(f^{loc}_{NL})^2|_{non-attractor}}.\eea    
      Now to give a proper estimate of the deviation in the magnitude of the amplitude of non-Gaussian parameter computed from four point function in terms of the three point non-Gaussian amplitude for the time being we assume that the results obtained from the attractor and non-attractor formalism is almost at the same order of magnitude. In that case we have, $\Delta_f \sim {\cal O}(1)$.
      Consequently the deviation factor can be recast as:
      \bea Q_{corr}\sim \Delta_f(1-J_{corr})\sim 1-J_{corr},\eea 
      where the correction factor $J_{corr}<<1$ is highly suppressed in the super Planckian region of the perturbation theory, but those small corrections are important as precision measurement is concerned in the context of cosmology. In case of our derived effective potentials we get following approximated expressions for the correction factor:
      \be\begin{array}{lll}\label{rinfla1cxccdfzz}\tiny
          \displaystyle J_{corr}\displaystyle\sim\left\{\begin{array}{ll}
                            \displaystyle \left(\frac{18M^2_p}{81\phi^2_*}+\frac{72M^4_p}{6561\phi^4_*}+\frac{3888M^8_p}{43046721\phi^8_*}+\cdots\right)
                                                       ~~~~~ &
          \mbox{\small \textcolor{red}{\bf for \underline{Case I}}}  
         \\ 
                  \displaystyle \left(\frac{18M^2_p}{\phi^2_*}+\frac{72M^4_p}{\phi^4_*}+\frac{3888M^8_p}{\phi^8_*}+\cdots\right)& \mbox{\small \textcolor{red}{\bf for \underline{Case II}}}  
                  \\  
                           \displaystyle
                           \left(\frac{18M^2_p}{\phi^2_*\left(1-\frac{\phi^4_V}{\phi^4_*}\right)^2}+\frac{72M^4_p}{\phi^4_*\left(1-\frac{\phi^4_V}{\phi^4_*}\right)^4} 
                           \displaystyle
                           +\frac{3888M^8_p}{\phi^8_*\left(1-\frac{\phi^4_V}{\phi^4_*}\right)^8}+\cdots\right) & \mbox{\small \textcolor{red}{\bf for \underline{Case ~II+Choice~I}}}\\ 
                                             \displaystyle
                                              \left(\frac{18M^2_p}{\phi^2_*\left(1-\frac{m^2_c}{m^2_c-\lambda\phi^2_*}\right)^2}+\frac{72M^4_p}{\phi^4_*\left(1-\frac{m^2_c}{m^2_c-\lambda\phi^2_*}\right)^4} \right.\\ \left.~~~~~~~~~~~~~~~~~~~~~~~~~~~~\displaystyle
                                               +\frac{3888M^8_p}{\phi^8_*\left(1-\frac{m^2_c}{m^2_c-\lambda\phi^2_*}\right)^8}+\cdots\right)
                                                                        & \mbox{\small \textcolor{red}{\bf for \underline{Case ~II+Choice~II}}}\\ 
                                                               \displaystyle
             \left(\frac{18M^2_p}{\phi^2_*\left(1+\xi(\phi^2_*-\phi^2_V)+\frac{\phi^2_V}{\phi^2_*}\right)^2}+\frac{72M^4_p}{\phi^4_*\left(1+\xi(\phi^2_*-\phi^2_V)+\frac{\phi^2_V}{\phi^2_*}\right)^4}\right.\\ \left.~~~~~~~~~~~~~~~~~~~~~~~~~~~~\displaystyle +\frac{3888M^8_p}{\phi^8_*\left(1+\xi(\phi^2_*-\phi^2_V)+\frac{\phi^2_V}{\phi^2_*}\right)^8}+\cdots\right)                                                   & \mbox{\small \textcolor{red}{\bf for \underline{Case ~II+Choice~III}}}.
                   \end{array}
         \right.
         \end{array}\ee           
 Further using this results one can estimate the deviation in the {\it Suyama-Yamguchi} consistency relation if we go from attractor regime to non-attractor regime of cosmological perturbation as:
 \bea |\Delta \tau^{loc}_{NL}|&=& \frac{36}{25}(f^{loc}_{NL})^2|_{non-attractor}|\left\{1-\Delta_{f}(1-J_{corr})\right\}|\sim \frac{36}{25}(f^{loc}_{NL})^2|_{non-attractor}|J_{corr}|.
 ~~~~~~~~~~~~\eea  
 Also the fractional change can be expressed as:
 \bea \left|\frac{\Delta \tau^{loc}_{NL}}{(\tau^{loc}_{NL})_{non-attractor}}\right|_{\phi_*>>M_p}&=& |1-\Delta_{f}(1-J_{corr})|\sim |J_{corr}|.
  ~~~~~~~~~~~~\eea 
 So it is clear that that $|J_{corr}|$ captures the effect of the deviation in {\it Suyama Yamaguchi} consistency relation which are very small and highly suppressed in the super Planckian regime of inflation. But as far as precision cosmology is concerned, this small effect is also very useful to discriminate between all derived effective models considered in this paper. If in near future Planck or any other observational probe detect the signature of primordial non-Gaussianity with high statistical significance then one can also further comment on the significance of attractors and non-atttractors in the context of early universe cosmology.   
\item In the sub planckian field regime the deviation factor $Q_{corr}$ can be expressed as:
\be\begin{array}{lll}\label{rinfla1cxccdfz}\tiny
    \displaystyle Q_{corr}\displaystyle=\Delta_{f}\times\left\{\begin{array}{ll}
                      \displaystyle \left(1+\frac{19683}{8}\frac{\phi^6_*}{M^6_p}+\cdots\right)
                                                  &
    \mbox{\small \textcolor{red}{\bf for \underline{Case I}}}  
   \\ 
            \displaystyle \left(1+\frac{1}{216}\frac{\phi^6_*}{M^6_p}+\cdots\right)& \mbox{\small \textcolor{red}{\bf for \underline{Case II}}}  
            \\  
                     \displaystyle
                     \left(1+\frac{1}{216}\frac{\phi^6_*}{M^6_p}\left(1-\frac{\phi^4_V}{\phi^4_*}\right)^6+\cdots\right) & \mbox{\small \textcolor{red}{\bf for \underline{Case ~II+Choice~I}}}\\ 
                                       \displaystyle
                                       \left(1+\frac{1}{216}\frac{\phi^6_*}{M^6_p}\left(1-\frac{m^2_c}{m^2_c-\lambda\phi^2_*}\right)^6+\cdots\right) 
                                                                  & \mbox{\small \textcolor{red}{\bf for \underline{Case ~II+Choice~II}}}\\ 
                                                         \displaystyle \left(1+\frac{1}{216}\frac{\phi^6_*}{M^6_p}\left(1+\xi(\phi^2_*-\phi^2_V)+\frac{\phi^2_V}{\phi^2_*}\right)^6+\cdots\right)
                                                       & \mbox{\small \textcolor{red}{\bf for \underline{Case ~II+Choice~III}}}.
             \end{array}
   \right.
   \end{array}\ee            
       where the factor $\Delta_{f}$ is defined earlier, which is $\Delta_f \sim {\cal O}(1)$.
      Consequently the deviation factor can be recast as:
      \bea Q_{corr}\sim \Delta_f(1+C_{corr})\sim 1+C_{corr},\eea 
      where the correction factor $C_{corr}<<1$ is suppressed in the sub Planckian region of the perturbation theory, but those small corrections are important as precision measurement is concerned in the context of cosmology. In case of our derived effective potentials we get following approximated expressions for the correction factor:
      \be\begin{array}{lll}\label{rinfla1cxccdfzzfg}\tiny
          \displaystyle C_{corr}\displaystyle\sim\Delta_{f}\times\left\{\begin{array}{ll}
                                \displaystyle \left(\frac{19683}{8}\frac{\phi^6_*}{M^6_p}+\cdots\right)
                                                            &
              \mbox{\small \textcolor{red}{\bf for \underline{Case I}}}  
             \\ 
                      \displaystyle \left(\frac{1}{216}\frac{\phi^6_*}{M^6_p}+\cdots\right)& \mbox{\small \textcolor{red}{\bf for \underline{Case II}}}  
                      \\  
                               \displaystyle
                               \left(\frac{1}{216}\frac{\phi^6_*}{M^6_p}\left(1-\frac{\phi^4_V}{\phi^4_*}\right)^6+\cdots\right) & \mbox{\small \textcolor{red}{\bf for \underline{Case ~II+Choice~I}}}\\ 
                                                 \displaystyle
                                                 \left(1+\frac{1}{216}\frac{\phi^6_*}{M^6_p}\left(1-\frac{m^2_c}{m^2_c-\lambda\phi^2_*}\right)^6+\cdots\right) 
                                                                            & \mbox{\small \textcolor{red}{\bf for \underline{Case ~II+Choice~II}}}\\ 
                                                                   \displaystyle \left(\frac{1}{216}\frac{\phi^6_*}{M^6_p}\left(1+\xi(\phi^2_*-\phi^2_V)+\frac{\phi^2_V}{\phi^2_*}\right)^6+\cdots\right)
                                                                 & \mbox{\small \textcolor{red}{\bf for \underline{Case ~II+Choice~III}}}.
                       \end{array}
             \right.
             \end{array}\ee            
 Further using this results one can estimate the deviation in the {\it Suyama-Yamguchi} consistency relation if we go from attractor regime to non-attractor regime of cosmological perturbation as:
 \bea |\Delta \tau^{loc}_{NL}|&=& \frac{36}{25}(f^{loc}_{NL})^2|_{non-attractor}|\left\{1-\Delta_{f}(1+C_{corr})\right\}|\sim \frac{36}{25}(f^{loc}_{NL})^2|_{non-attractor}|C_{corr}|.
 ~~~~~~~~~~~~\eea  
 Also the fractional change can be expressed as:
 \bea \left|\frac{\Delta \tau^{loc}_{NL}}{(\tau^{loc}_{NL})_{non-attractor}}\right|_{\phi_*<<M_p}&=& |1-\Delta_{f}(1+C_{corr})|\sim |C_{corr}|.
  ~~~~~~~~~~~~\eea 
 So it is clear that that $|C_{corr}|$ captures the effect of the deviation in {\it Suyama Yamaguchi} consistency relation which are very small and suppressed in the sub Planckian regime of inflation.                    
\item From the study of sub Planckian and super Planckian regime it is evident that when $\Delta_{f}\sim {\cal O}(1)$ i.e. the non-Gaussian amplitude obtained from three point function in attractor and non-attractor regime for all the derived effective potentials are of the same order then deviation from {\it Suyama Yamaguchi} consistency relation is very small. The only difference is in sub Planckian case this correction is greater than unity and on the other hand in the super Planckian case this correction factor is less than unity. But since we are interested in the precision cosmological measurement, such small but distinctive corrections will play significant role to discriminate between the classes of effective models of inflation derived in this paper.
\item Finally, if we relax the assumption that the deviation factor, $\Delta_{f} \neq 1$, then one can consider the following two situations-
\begin{enumerate}
\item First we consider, $\Delta_{f}>>1$.
In this case in the super Planckian and sub Planckian regime we get the following results for the deviation in the {\it Suyama Yamaguchi} consistency relation:
 \bea |\Delta \tau^{loc}_{NL}|_{\phi_*>>M_p}&=& \frac{36}{25}(f^{loc}_{NL})^2|_{non-attractor}|\Delta_{f}(1-J_{corr})|.
 ~~~~~~~~~~~~\\ |\Delta \tau^{loc}_{NL}|_{\phi_*<<M_p}&=& \frac{36}{25}(f^{loc}_{NL})^2|_{non-attractor}|\Delta_{f}(1+C_{corr})|.
  ~~~~~~~~~~~~\eea 
  Also the fractional change in the {\it Suyama Yamaguchi} consistency relation can be expressed as:
  \bea \left|\frac{\Delta \tau^{loc}_{NL}}{(\tau^{loc}_{NL})_{non-attractor}}\right|_{\phi_*>>M_p}&=&|\Delta_{f}(1-J_{corr})|,~~\left|\frac{\Delta \tau^{loc}_{NL}}{(\tau^{loc}_{NL})_{non-attractor}}\right|_{\phi_*<<M_p}= |\Delta_{f}(1+C_{corr})|.
      ~~~~~~~~~~~~\eea 
      In this specific situation deviation factor is large and consequently one can achieve maximum amount of violation in {\it Suyama Yamaguchi} consistency relation. Here the results of the super Planckian and sub Planckian field value differs due to the presece of the correction factors $J_{corr}$ and $C_{corr}$. Here both $J_{corr}<1$ and $C_{corr}<1$, but for model discrimination such small corrects are significant as mentioned earlier.
\item Next we consider, $\Delta_{f}<<1$.
In this case in the super Planckian and sub Planckian regime we get the following results for the deviation in the {\it Suyama Yamaguchi} consistency relation:
 \bea |\Delta \tau^{loc}_{NL}|_{\phi_*>>M_p}&=& \frac{36}{25}(f^{loc}_{NL})^2|_{non-attractor}|1-\Delta_{f}|.
 ~~~~~~~~~~~~\\ |\Delta \tau^{loc}_{NL}|_{\phi_*<<M_p}&=& \frac{36}{25}(f^{loc}_{NL})^2|_{non-attractor}|1-\Delta_{f}|.
    ~~~~~~~~~~~~\eea 
    Also the fractional change in the {\it Suyama Yamaguchi} consistency relation can be expressed as:
      \bea \left|\frac{\Delta \tau^{loc}_{NL}}{(\tau^{loc}_{NL})_{non-attractor}}\right|_{\phi_*>>M_p}&=&|1-\Delta_{f}|,~~~~~~~ \left|\frac{\Delta \tau^{loc}_{NL}}{(\tau^{loc}_{NL})_{non-attractor}}\right|_{\phi_*<<M_p}=|1-\Delta_{f}|.
          ~~~~~~~~~~~~\eea 
          In this specific situation deviation factor is small and consequently one can achieve very small amount of violation in {\it Suyama Yamaguchi} consistency relation. Here the results of the super Planckian and sub Planckian field value are almost same as we have neglected the terms $\Delta_{f}J_{corr}<<1$ and $\Delta_{f}C_{corr}<<1$.
\end{enumerate} 
\end{itemize}
Now to derive the results of non-Gaussian amplitudes in the non-attractor regime using $\delta {\cal N}$ formalism we need to freeze the value of the additional field $\Psi$ in the Planck scale. If we do this job then the expression for the four point non-Gaussian amplitude computed from scalar fluctuation can be expressed as:
\bea \tau^{loc}_{NL}&=&\left[\frac{{\cal N}_{,\phi\phi}{\cal N}_{,\phi\phi}}{\left({\cal N}_{,\phi}{\cal N}_{,\phi}\right)^2}\right]_{*}={\cal Y}^2,~~~~
g^{loc}_{NL}=\frac{25}{54}\left[\frac{{\cal N}_{,\phi\phi\phi}}{{\cal N}_{,\phi}{\cal N}_{,\phi}{\cal N}_{,\phi}}\right]_{*}=\frac{25}{108}{\cal Y}^2.
\eea
In this case we also derive the modified consistency relation between the four point and three point non-Gaussian amplitude for scalar fluctuations as:
\bea g^{loc}_{NL}&=&\frac{25}{108}\tau^{loc}_{NL},~~~~
\tau^{loc}_{NL}=\frac{972}{625}(f^{loc}_{NL})^2.\eea
This implies that the well known {\it Suyama Ymaguchi  } consistency relation also violates in this context and the amount of violation is given by:
\bea |\Delta\tau^{loc}_{NL}|=|(\tau^{loc}_{NL})_{\delta {\cal N }}-(\tau^{loc}_{NL})_{In-In}|=\frac{36}{25}((f^{loc}_{NL})^2)_{In-In}\left|\frac{27}{25}W_f-1\right|,\eea
where the factor $W_{f}$ is defined as:
\be W_f = \frac{(f^{loc}_{NL})^2)_{\delta {\cal N }}}{((f^{loc}_{NL})^2)_{In-In}}.\ee
Also the factional change is given by:
\bea \left|\frac{\Delta\tau^{loc}_{NL}}{(\tau^{loc}_{NL})_{In-In}}\right|=\left|\frac{27}{25}W_f-1\right|.\eea 
Now if we claim that at the horizon crossing non-Gaussian amplitudes obtained from $\delta{\cal N}$ and In In formalism are of the same order then in that case we get, $W_f\sim {\cal O}(1)$. Consequently the deviation in {\it Suyama Ymaguchi  } consistency relation can be recast as:
\bea |\Delta\tau^{loc}_{NL}|=|(\tau^{loc}_{NL})_{\delta {\cal N }}-(\tau^{loc}_{NL})_{In-In}|\sim\frac{72}{625}((f^{loc}_{NL})^2)_{In-In}.\eea
Consequently the factional deviation is given by, $\left|\frac{\Delta\tau^{loc}_{NL}}{(\tau^{loc}_{NL})_{In-In}}\right|\sim \frac{2}{25}$. 
\section{\textcolor{blue}{Conclusion}}
\label{s8}
To summarize, in the present article, we have addressed the following points:
\begin{itemize}
\item Firstly we have started our discussion with a specific class of modified theory of gravity, {\it aka} $f(R)$ gravity where a single matter (scalar field) is minimally coupled with the gravity sector. For simplicity we consider the case where the matter field contains only canoniocal kinetic term. To build effective potential from this toy setup of modified gravity in 4D we choose $f(R)=\alpha R^2$ gravity.  

\item Next to start with in the matter sector we choose a very simple model of potential, $V(\phi)=\frac{\lambda}{4}\phi^4$, where $\phi$ is a real scalar field and $\lambda$ is a real parameter of the monomial model. This type of potential can be treated as a Higgs like potential as the structure of Higgs potential is given by, $V(H)=\frac{\lambda}{4}(H^{\dagger}H-V^2)$, where $\lambda$ is Yukawa coupling, $H$ is the Higgs SU(2) doublet and $\langle 0|H|0\rangle={\cal V}\sim 125~{\rm GeV}$ is the VEV of the Hiigs field. Now one can write the Higgs SU(2) doublet as, $H ^{\dagger}=(\phi~~ 0)$ and the corresponding Higgs potential can be recast as, $V(\phi)=\frac{\lambda}{4}(\phi^2-{\cal V}^2)^2$. Now at the scale of inflation, which is at ${\cal O}(10^{16}~{\rm GeV})$,  contribution from the VEV is almost negligible and consequently one can recast the Higgs potential in the monomial form, $V(\phi)\approx\frac{\lambda}{4}\phi^4$. The only difference is in case of Higgs $\lambda$ is Yukawa coupling and in case of general monomial model $\lambda$ is a free parameter of the theory. Due to the similar structural form of the potential we call the general $\phi^4$ monomial model as Higgsotic potential.

\item Further, we provide the field equations in spatially flat FLRW background, which are extremely complicated to solve for this setup. To simplify, next we perform a conformal transformation in the metric and write down the model action in the transformed Einstein frame. Next, we derive the field equations in spatially flat FLRW background and try to solve them for two dynamical attractor features as given by-
 I. Power law solution and II. Exponential solution.
However, the second case give rise to tachyonic behaviour which can be resolved by considering  non-BPS D-brane in superstring theory, considering 
  the effect of mass like quadratic term in the effective potential and considering the effect of non-minimal coupling between $f(R)=\alpha R^2$ scale free gravity sector and the matter field
sector. 
\item Next, using two dynamical attractors, \underline{Power law} and \underline{Exponential} solution we have studied the cosmological constraints in presence of two field in Einstein frame. We have studied the constraints from primordial density perturbation, by deriving the expressions for two point function and the present observables- amplitude of power spectrum for density perturbations, corresponding spectral tilt and associated running and running of the running for inflation. We have repeated the analysis for tensor modes and also comment on the future observables-amplitude of the tensor fluctuations, associated tilt and running, tensor-to-scalar ratio. We also provide a modified formula for the field excursion in terms of tensor-to-scalar ratio, scale of inflation and the number of e-foldings. Further, we have compared our model with Planck 2015 data and constrain the parameter $\alpha$ of the scale free gravity and non-minimal coupling parameter $\lambda(\Psi_h)$. Additionally, we have studied the constraint for reheating temperature. Finally, we derive the expression for inflaton and the coupling parameter at horizon crossing, during reheating and at the onset of inflation which are very useful to study the scale dependent behaviour. Most importantly, in the present context one can interpret such scale dependence as an outcome of RG flow in usual Quantum Field Theory.

\item Further, we have explored the cosmological solutions beyond attractor regime. We have shown that this possibility can be achieved if we freeze the field value of the dilaton field  in Einstein frame. This possibility can be treated as a single field model where a additaional field freezes at certain field value, which we fix at the reduced Planck scale. To serve this purpose we have used ADM formalism and compute two point function and associated present inflationary observables using Bunch Davies initial condition for scalar fluctuations. We have repeated the procedure for tensor fluctuations as well. In the non-attractor regime, we have also derived a modified version of field excursion formula in terms of tensor-to-scalar ratio, scale of inflation and the number of e-foldings. We have also derived few sets of consistency relations in this context which are different from the usual single field slow roll models. For an example, instead of getting $r=-8n_T$ here we get, $r=\frac{24n^2_T}{1-n_S}$ at horizon crossing scale. Further, we derive the constraints on reheating temperature in terms of inflationary observables and number of e-foldings. 

\item Next, as a future probe we have computed the expression for three point function and the bispectrum of scalar fluctuations using In-In formalism for non attractor case and $\delta {\cal N}$ formalism for the attractor case. Following the fact that the local ansatz for curvature perturbation holds good perfectly, we have derived the results for non-Gaussian amplitude $f^{loc}_{NL}$ for equilateral limit and squeezed limit triangular shape configuration. We also give a bulk interpretation of each of the momentum dependent terms appearing in the expression for the three point scalar correlation function in terms of $S$, $T$ and $U$ channel contributions. It is important to note that, in the attractor phase since we have started with various proposals of the effective potentials as mentioned earlier, we have found various non-trivial features upto second order perturbation in $\delta {\cal N}$ formalism. Further, for the coinsistency check we freeze the dilaton field in Planck scale and redo the analysis of $\delta {\cal N}$ formalism. By doing this we have found out that the expression for the three point non-Gaussian amplitude is slightly different as expected for single field case. Further, we compare the results obtained from In-In formalism and $\delta{\cal N}$ formalism for the non attractor phase, where the dilaton field is fixed in Planck scale. Here finally, we give a theoretical bound on the scalar three point non-Gaussian amplitude computed from equilateral and squeezed limit configurations. The obtained results are consistent with the Planck 2015 data.

\item Finally, as an additional future probe we have also computed the expression for four point function and as well as the trispectrum of scalar fluctuations using In-In formalism for non attractor case and $\delta {\cal N}$ formalism for the attractor case. We have derived the results for non-Gaussian amplitude $g^{loc}_{NL}$ and $\tau^{loc}_{NL}$ for equilateral limit, counter colliniear or folded kite limit and squeezed limit shape configuration from In-In formalism. Further we have given the bulk interpretation of each of the momentum dependent terms appearing in the expression for the four point scalar correlation function. We have identified the $S$, $T$ and $U$ channel contributions in momentum space from our computation. In our computation we have considered the contribution from contact interaction term, scalar and graviton exchange. In the attractor phase following the prescription of $\delta {\cal N}$ formalism we also derive the expressions for the four point non-Gaussian amplitude $g^{loc}_{NL}$ and $\tau^{loc}_{NL}$. Next we have shown that the consistency relation connecting three and four point non-Gaussian amplitude {\it aka} Suyama Yamaguchi relation is modified in attractor phase and further given an estimate of the amout of deviation. Further, for the coinsistency check we freeze the dilaton field in Planck scale and redo the analysis of $\delta {\cal N}$ formalism. By doing this we have found out that the expression for the four point non-Gaussian amplitude is slightly different as expected for single field case. Next we have also shown that the exact numerical deviation of the consistency relation is of the order of $2/25$ by assuming non-Gaussian three point amplitude for attractor and non-attractor phase are of the same order of magnitude. Further, we compare the results obtained from In-In formalism and $\delta{\cal N}$ formalism for the non attractor phase, where the dilaton field is fixed in Planck scale. Here finally, we give a theoretical bound on the scalar four point non-Gaussian amplitude computed from equilateral, folded kite and squeezed limit configurations. The obtained results are consistent with the Planck 2015 data.
 \end{itemize}
The future prospects of our work are appended below:
\begin{itemize}
 \item We have restricted our analysis up to monomial $\phi^4$ model and due to the structural similarity with Higgs potential at the scale of inflation we have identified monomial $\phi^4$ model as Higgsotic model in the present context. 
  
  \item To investigate the role of scale free theory of gravity, as an example we have used $\alpha R^2$ gravity. But the present analysis can be generalized to any class of $f(R)$ gravity models and other class of higher derivative gravity models.
  
  \item In the matter sector for completeness one can consider most generalized version of $P(X,\phi)$ models, where $X=-\frac{1}{2}g^{\mu\nu}\partial_{\mu}\phi\partial_{\nu}\phi$. DBI is one of the examples of $P(X,\phi)$ model which can be implemented in the matter sector instead of simple canonical kinetic contribution.

 \item In this work, we have not given any computation of three point and found point scalar correlation function and representative non-Gaussian amplitudes using In-In formalism in the attractor regime in presence of both the fields $\phi$ and $\Psi$ for all classes of Higgsotic models. In near future we are planning to present the detailed calculation on this important issue.
 
 \item Generation of primordial magnetic field through inflationary magnetogenesis is one of the important issues in the context of primordial cosmology, which we have not explored yet from our setup. One can consider such interactions by breaking conformal invariance of the $U(1)$ gauge field in presence of time dependent coupling $f(\phi(\eta))$ to study the features of primordial magnetic field through inflationary magnetogenesis. We have also a future plan to address this issue.
 
 \item In this work we have restricted our analysis within the class of Higgsotic models. For completeness in future we will extend this idea to all class of potentials allowed by the presently available observed Planck data. We will also include the effects of various types of non minimal and non canonical interactions in the present setup.
 
 \item In the same direction one can also carry forward the present analysis in the context of various types of higher derivative gravity set up and comment on the constraints on the primordial non-Gaussianity, reheating and generation of primordial magnetic field through inflationary magntogenesis for completeness. Also one can consider the possibility of non-minimal interaction between $\alpha R^2$ gravity and matter sector. In future we will investigate the possibility of appearing new consistency relations in presence of higher derivative gravity set up and will give proper estimate of the amount of violation from Suyama Yamaguchi consistency relation.
 
 \item During the compuation of correlation functions using semi classical method, via $\delta {\cal N}$ formalism, we have restricted upto second order contributions in the solution of the field equation in FLRW background and also neglected the contributions from the back reaction for all type of effective Higgsotic models derived in Einstein frame. For more completeness, one can relax these assumptions and redo the analysis by taking care of all such contributions. Additionally, we have a future plan to extend the semi classical computation of $\delta{\cal N}$ formalism of cosmological perturbation theory in a more sophisticated way and will redo the analysis in the present context.
 
 \item In this work, we also have not investigated the possibility of getting dark matter and dark energy  constraints from the present up. Most importantly the present structure of interactions in
 the Einstein frame shows that both the field $\phi$ and $\Psi$ are coupled and due to this fact if we want to explain the possibility of dark matter
 and dark energy together from this setup it is very clear that both of them are coupled. But this is not very clear at the level of analytics and detailed calculations.
 Here one can also investigate these possibilities from this setup. 
 
 \item In this work we have not investigated the contribution from the loop effects (radiative corrections) in all of the effective Higgsotic interactions (specifically in the self couplings) derived in the Einstein frame. After switching on all such effects one can investigate the specific numerical contribution of such terms and comment on the effects of such terms in precision cosmology measurement.

  \item Here one can generalize the results for $\alpha$ vacua and study its cosmological consequences for all types of derived potential in the present context.
  
  \item In the present context one can also study the quantum entanglement between the Bell pairs, which can be created through the Bell's inequality violation in cosmology \cite{Maldacena:2015bha,Choudhury:2016cso,Choudhury:2016pfr}.

\end{itemize}

\section*{\textcolor{blue}{Acknowledgments}}
SC would like to thank Department of Theoretical Physics, Tata Institute of Fundamental Research, Mumbai and specially the Quantum Structure of the Spactime Group
for providing me Visiting (Post-Doctoral) Research Fellowship. The work of SC was supported
in part by Infosys Endowment for the study of the Quantum Structure of Space Time. SC take this opportunity to thank sincerely to Ashok Das, Sudhakar Panda, Shiraz Minwalla,
Sandip P. Trivedi, Gautam Mandal and Varun Sahni for their constant support and inspiration. SC also thank the organizers of Indian String Meet 2016 and Advanced String School 2017 for providing the local hospitality during the work. SC also thanks Institute of Physics, Bhubaneswar for providing
the academic visit during the work. Last but not the
least, SC would all like to acknowledge our debt to the people of India for their generous and steady support
for research in natural sciences, especially for theoretical high energy physics, string theory and cosmology.

\section{\textcolor{blue}{Appendix}}
\subsection{Effective Higgsotic models for generalized $P(X,\phi)$ theory}
\label{df1}
In this section, to give a broad overview of the effective Higgsotic models let us start with a general $f(R)$ theory in the gravity sector and generalized  $P(X,\phi)$ theory in the matter sector. The representative actions in Jordan frame is given by:
\bea S=\int d^{4}x \sqrt{-g}\left[f(R)+P(X,\phi)\right],\eea
where $P(X,\phi)$ is a arbitrary function of single scalar field $\phi$ and the kinetic term $ X=-\frac{1}{2}g^{\mu\nu}\partial_{\mu}\phi\partial_{\nu}\phi$. 
In general $f(R)$ is any arbitrary function of $R$. But for our purpose we choose $f(R)=\alpha R^2$ to study the consequences from scale free gravity. From this representative action one can write down the field equations in spatially flat FLRW background as:
\bea
H^{2}&=&\left(\frac{\dot{a}}{a}\right)^2=\frac{\rho_{\phi}}{6\alpha R}+\frac{R}{2}-\left(\frac{\dot{R}}{R}\right)H,\\
\label{eq9}2\dot{H}+3H^{2}&=&2\left(\frac{\ddot{a}}{a}\right)+\left(\frac{\dot{a}}{a}\right)^{2}=-\frac{p_{\phi}}{2\alpha R}
-2\left(\frac{\dot{R}}{R}\right)H-\frac{\ddot{R}}{R}+\frac{R}{4}
\eea
where for generalized  $P(X,\phi)$ theory pressure $p_{\phi}$ and density $\rho_{\phi}$ can be written as:
\bea p_{\phi}&=&P(X,\phi),~~~~
\rho_{\phi}=2XP_{,X}(X,\phi)-P(X,\phi).\eea
Here effective speed of sound parameter $c_{S}$ is defined as:
\be c_{S}=\sqrt{\frac{P_{,X}(X,\phi)}{P_{,X}(X,\phi)+2XP_{,XX}(X,\phi)}}.\ee
If we choose the following functional form of $P(X,\phi)$ which is:
\be\label{xxc} P(X,\phi)=-\frac{1}{f(\phi)}\sqrt{1-2Xf(\phi)}+\frac{1}{f(\phi)}-V(\phi),\ee
as pointed earlier, then we get the following simplified expression for $p_{\phi}$ and density $\rho_{\phi}$ as given by:
\bea p&=&\frac{1}{f(\phi)}(1-c_S)-V(\phi),~~~~
\rho=\frac{1}{f(\phi)}\left(\frac{1}{c_S}-1\right)+V(\phi).\eea
Also one can consider any arbitrary slow-roll effective potential but for our purpose we choose monomial Higgsotic model, $V(\phi)=\frac{\lambda}{4}\phi^4$ in the Jordan frame.

In the present context let us introduce a scale dependent mode $\Psi$, which can be written in terms of a no scale dilaton mode $\Theta$ as, $\Theta = f^{'}(R)M^{-2}_{p}=2\alpha R~M^{-2}_{p}=e^{\sqrt{\frac{2}{3}}\frac{\Psi}{M_p}}=\Omega^2$,
which plays the role of a Lagrange multiplier and arises in the Jordan frame without space-time derivatives. Here $\Omega$ is the conformal factor of the conformal transformation that we perform from Jordan frame to Einstein frame.

In terms of the newly introduced no scale dilaton mode $\Theta$ the total action of the theory (see Eq~(\ref{eq1})) can be recast as:
\bea
S&=&\int d^{4}x \sqrt{-g}\left\{\frac{M^{2}_{p}}{2}\Theta R-\frac{M^{4}_{p}}{8\alpha}\Theta^{2}+P(X,\phi)\right\}.
\eea 
After doing C.T. the total action can be recast in the Einstein frame as:
\bea
S~~\underrightarrow{C.T.}~~ \tilde{S}
&=&\int d^{4}x \sqrt{-\tilde{g}}\left[\frac{M^{2}_{p}}{2}\tilde{R}+G(\tilde{X},\phi,\Psi)\right]
 \eea
where after applying C.T. the functional $G(\tilde{X},\phi,\Psi)$ is defined in Einstein frame as:
\be G(\tilde{X},\phi,\Psi)=\frac{1}{\Omega^4}\left[P(\tilde{X},\phi)-\frac{M^{4}_{p}}{8\alpha}e^{2\sqrt{\frac{2}{3}}\frac{\Psi}{M_p}}\right].\ee
Here $\tilde{X}$ is the kinetic term after conformal transformation, which is defined as,
$\tilde{X}=-\frac{1}{2}\tilde{g}^{\mu\nu}\tilde{\partial}_{\mu}\phi \tilde{\partial}_{\nu}\phi$.
Now in case of the specific form of $P(X,\phi)$ as stated in Eq~(\ref{xxc}) after conformal transformation we get:
\be G(\tilde{X},\phi,\Psi)=\frac{1}{\Omega^4}\left[-\frac{1}{f(\phi)}\sqrt{1-2\tilde{X}f(\phi)}+\frac{1}{f(\phi)}-V(\phi)-\frac{M^{4}_{p}}{8\alpha}e^{2\sqrt{\frac{2}{3}}\frac{\Psi}{M_p}}\right].\ee
Here the total potential can be recast as:
\bea
\tilde{W}(\phi,\Psi)&=&\frac{\frac{M^{4}_{p}}{8\alpha}e^{2\sqrt{\frac{2}{3}}\frac{\Psi}{M_p}}+V(\phi)+\frac{1}{f(\phi)}}{\Omega^4}=V_0 \left[1+\left(\frac{8\alpha}{M^{4}_{p}}V(\phi)+\frac{1}{f(\phi)}\right)e^{-\frac{2\sqrt{2}}{\sqrt{3}}\frac{\Psi}{M_p}}\right],~~~~~~~
\eea
where $V_0=M^{4}_{p}/8\alpha$,
exactly mimics the role of cosmological constant as mentioned earlier.

In case of Higgsotic model we can rewrite the total potential as:
\bea
\tilde{W}(\phi,\Psi)&=&\frac{\frac{M^{4}_{p}}{8\alpha}e^{2\sqrt{\frac{2}{3}}\frac{\Psi}{M_p}}+\frac{\lambda}{4}\phi^{4}+\frac{1}{f(\phi)}}{\Omega^4}= V_0 \left[1+\frac{2\alpha\lambda(\Psi)}{M^{4}_{p}}\phi^{4}+\frac{1}{f(\phi)}e^{-\frac{2\sqrt{2}}{\sqrt{3}}\frac{\Psi}{M_p}}\right].
\eea
Here the effective matter coupling ($\lambda(\Psi)$) in the potential sector is given by:
\bea\label{eq33}
\lambda(\Psi)&=&\frac{\lambda}{\Omega^4}=\lambda e^{-\frac{2\sqrt{2}}{\sqrt{3}}\frac{\Psi}{M_p}}. 
\eea 
Rest of the computation is exactly similar as we have performed earlier, only the structure of the total effective potential changes.

Here it is important to note that, apart from $f(R)$ gravity one can consider various other possibilities. To give a clear picture about various classes of two field attractor models one can consider the following 4D effective action in Einstein frame:
\bea S=\int d^{4}x\sqrt{-g}\left[\frac{M^2_p}{2}R+J(X,Y,\phi,\Psi)\right].\eea
where $J(X,Y,\phi,\Psi)$ is the general functional of the two field $\phi$ and $\Psi$ as given by the following specific mathematical structure:
\be J(X,Y,\phi,\Psi)=e^{-\frac{c_{1}\Psi}{M_p}}X+Y-W(\phi,\Psi).\ee
Here $c_{1}$ and $c_{2}$ characterize effective coupling cosnstant in 4D, which are different for various types of source theories. In EFT setup these are identified as the Wilson co-efficients. Additionally, it is important to note that the kinetic term for the $\phi$ and $\Psi$ field is defined as, 
$X= -\frac{g^{\mu\nu}}{2}\partial_{\mu}\phi\partial_{\nu}\phi$ and
$Y=-\frac{g^{\mu\nu}}{2}\partial_{\mu}\Psi\partial_{\nu}\Psi$.
Here $W(\phi,\Psi)$ is the 4D effective potential, which is given by the following expression:
\be W(\phi,\Psi)=e^{-\frac{c_{2}\Psi}{M_p}}V(\phi).\ee
This is non separable form of two field effective potential where one can treat $V(\phi)$ as usual inflaton field and $e^{-\frac{c_{2}\Psi}{M_p}}$ as the dilaton exponential coupling.

This type of effective theory can be derived from the following class of models:
\begin{enumerate}
\item \textcolor{red}{\underline{Type~I:}} Consider an action in Jordan frame where the scalar field $\Phi$ is non-minimally coupled with gravity sector as given by:
\bea S=\int d^4 x \sqrt{-g}\left[f_1(\Phi)R-f_2(\Phi)g^{\mu\nu}\partial_{\mu}\Phi\partial_{\nu}\Phi-U(\Phi)+X-V(\phi)\right].\eea
Here $f_{1}(\Phi)$ is the non-minimal coupling and $f_{2}(\Phi)$ is the non-canonical interaction. This type of theories include the following subclass of models:
\begin{itemize}
\item \textcolor{red}{\underline{Jordan Brans Dicke (JBD) theory:}}\\ 
In this case we have:
\bea f_1(\Phi)&=&\frac{\Phi}{16\pi},~~~ 
f_2(\Phi)=\frac{\omega}{16\pi\Phi},~~~
U(\Phi)= 0,~~~
c_{1}=\frac{c_2}{2},~~~
c_{2}=\sqrt{\frac{8}{2\omega+3}},\\
\Psi&=&M_p\sqrt{\omega+\frac{3}{2}}\ln\left(\frac{\Phi}{2M^2_p}\right).
\eea
Here $\omega$ is the JBD parameter and for power law inflation $\omega>1/2$.
\item \textcolor{red}{\underline{Induced gravity theory:}}\\
In this case we have:
\bea f_1(\Phi)&=&\frac{g_1}{2}\Phi^2,~~~ 
f_2(\Phi)=\frac{1}{2},~~~
U(\Phi)= \frac{\lambda}{8}(\Phi^2-g^2_2)^2,~~~
c_{1}=\frac{c_2}{2},~~~
c_{2}=\sqrt{\frac{16g_1}{6g_1+1}},\\
\Psi&=&M_p\sqrt{6+\frac{1}{g_1}}\ln\left(\frac{\sqrt{g_1}\Phi}{M_p}\right).~~~~
\eea
Here $g_1$ and $g_2$ are coupling constants. For power law inflation $g_{1}<1/2$.
\item \textcolor{red}{\underline{Nonminimally coupled theory:}}\\
In this case we have:
\bea f_1(\Phi)&=&\frac{M^2_p}{2}-\frac{\xi}{2}\Phi^2,~~~ 
f_2(\Phi)=\frac{1}{2},~~~
U(\Phi)= 0,~~~
c_{1}=\frac{c_2}{2},~~~
c_{2}=\sqrt{\frac{3\xi M^2_p}{2(6\xi M^2_p+1)}},~~~~~~~~~~~~\\
\Psi&=&\left\{\begin{array}{ll}
                      \displaystyle 
            \sqrt{6}M_p\left\{{\rm tan}^{-1}\left[\frac{\sqrt{6}\xi\Phi}{\xi\left(6\xi+\frac{1}{M^2_p}\right)\Phi^2-1}\right]\right.\\ \left.
            ~~~~~~~~~~~\displaystyle -\sqrt{1+\frac{1}{6\xi M^2_p}}~{\rm sin}^{-1}\left[\sqrt{\xi\left(6\xi+\frac{1}{M^2_p}\right)}\Phi\right]\right\}                                      &
    \mbox{\small \textcolor{red}{\bf for \underline{$\xi\neq 1/6$}}}  
   \\ 
            \displaystyle \frac{M_p}{\sqrt{6}}~{\rm sin}^{-1}\left[\sqrt{6}\frac{\Phi}{M^2_p}\right]& \mbox{\small \textcolor{red}{\bf for \underline{$\xi=1/6$}}}.
             \end{array}
   \right.~~~~~~~~
\eea
After doing conformal transformation in Einstein frame one can derive the required form of the effective action from all of these models.
        
\end{itemize}
\item \textcolor{red}{\underline{Type~II:}} Consider an action in Jordan frame where the scalar field $\Phi$ is minimally coupled with $f(R)$ gravity sector as given by:
\bea S=\int d^4 x \sqrt{-g}\left[f(R)+X-V(\phi)\right].\eea
Here $f(R)$ is the arbitrary functional of Ricci scalar $R$. After doing conformal transformation in Einstein frame one can derive the required form of the effective action.

\item \textcolor{red}{\underline{Type~III:}} Consider a $4+D$ dimensional Kaluza Klien theory with an additional scalar field. This type of theories include the following subclass of models:
\begin{itemize}
\item \textcolor{red}{\underline{Extra dimensional theory-I:}} In this case the inflaton is introduced in the 4D effective action in Jordan frame as given by:
\bea S&=&\int d^4x\sqrt{-g}\left[\frac{M^2_p}{2}\left\{\Phi^2R+4\left(1-\frac{1}{D}\right)g^{\mu\nu}\partial_{\mu}\Phi\partial_{\nu}\Phi\right\}-U(\Phi)+X-V(\phi)\right].~~~~~~~~~~~~~\eea
In this case we have:
\bea
c_{1}&=&\frac{c_2}{2},~~
c_{2}=\sqrt{\frac{8D}{D+2}},~~
\Psi=M_p\sqrt{2\left(1+\frac{2}{D}\right)}\ln\left(\Phi\right).
\eea
But from this type of model no power law inflationary solutions are possible.
\item \textcolor{red}{\underline{Extra dimensional theory-II:}} In this case the inflaton is introduced in the $4+D$ dimentional action in Jordan frame as given by:
\bea S&=&\int d^{4+D}x\sqrt{-g_{4+D}}\left[\frac{1}{2\kappa^2_{4+D}}R+X-V(\phi)\right].~~~~~~~~~~~~~\eea
Here $g_{4+D}$ is the determinant of the $4+D$ dimensional metric and $\kappa^2_{4+D}$ is the $4+D$ dimensional gravitational coupling constant. In this case also we have:
\bea
c_{1}&=&0,~~
c_{2}=\sqrt{\frac{2D}{D+2}},~~
\Psi=M_p\sqrt{2\left(1+\frac{2}{D}\right)}\ln\left(\Phi\right).
\eea
From this type of model power law inflationary solutions are possible for all extra $D$ dimensions.
\end{itemize}

\item \textcolor{red}{\underline{Type~IV:}} Consider an action in Jordan frame from superstring theory in $10$ dimension with fixed Kalb Ramond background. In this case the scalar field $\Phi$ is non-minimally coupled with gravity sector as given by:
\bea S=\int d^4 x \sqrt{-g}\left[e^{-2\Phi}R+4g^{\mu\nu}\partial_{\mu}\Phi\partial_{\nu}\Phi+X-V(\phi)\right].\eea
In this case $\Phi$ is known as the dilaton field. But from this type of model no power law inflationary solutions are possible. Here additionally we have two class of solutions:\\
\textcolor{red}{\underline{Class I:}} 
\bea
c_{1}&=&\frac{c_2}{2},~~
c_{2}=2\sqrt{2},~~
\Psi=\frac{M_p}{\sqrt{2}}\left(6\ln b-\frac{\Phi}{2}\right).
\eea
\textcolor{red}{\underline{Class II:}} 
\bea
c_{1}&=&c_2,~~
c_{2}=-\sqrt{6},~~
\Psi=\frac{M_p}{\sqrt{2}}\left(2\ln b+\frac{\Phi}{2}\right).
\eea
\end{enumerate}
\subsection{Dynamical dilaton at late times}
After the completion of the phase of reheating, the total system enters the radiation dominated stage, at the beginning of which the
total energy density is governed by Eq~(\ref{retem}). At that stage, the scalar inflaton fields have almost settled down
in one of the potential valleys of the derived EFT potentials and get its VEV for the proposed model in $R^2$ gravity setup in Einstein frame. To make the computation simpler we also assume that, at
 the level of perturbations the dilaton field $\Psi$
 is almost decoupled from the Standard Model fields and the only dynamical field present in the model at late times. Henceforth, we will treat $\Psi$ as a dynamical field
 minimally coupled to the $R^2$ gravity in a conformally transformed Einstein frame and also assume that the $\Psi$ field is non-interacting with other matter degrees of freedom and radiation content of the Universe at late times.
During this epoch the total potential is characterized by the following expression:
\bea\label{late}
\tilde{W}(\hat{ \phi},\Psi)&=&V_0 \left[1+\frac{2\alpha\lambda(\Psi)}{M^{4}_{p}}\hat{\bf \phi}^{4}\right]=V_0+
\underbrace{\hat{\lambda}
\exp\left[-\frac{2\sqrt{2}}{\sqrt{3}}\frac{\Psi}{M_p}\right]}_{\bf \rm Dominant~at~late ~time}.
\eea
where $V_0$ is defined as, $V_0=\frac{M^{4}_{p}}{8\alpha}$,
and the VEV of the inflaton field $\phi$ is denoted by the symbol $\hat{\phi}$. 
Here one can set $\hat{\phi}\sim {\cal O}(M_p)$ for the proposed model at late time scale. 

Once the contribution of the inflaton sclar field $\phi$ get its VEV the corresponding energy density $\rho_{m}\equiv \rho_{\phi}={\rm Constant}$. Now in the present context to characterize the features of late time acceleration of Universe let us introduce equation of state parameter $w_{{\bf  X}}(=w_{\Psi})$, which is defined as:
\bea 
\label{ooo}w_{\bf  X} &=&\frac{p_{{\bf  X}}}{\rho_{{\bf  X}}}=\frac{\left(\frac{d\Psi}{d\tilde{t}}\right)^2 -\tilde{W}(\hat{\phi},\Psi)}{\left(\frac{d\Psi}{d\tilde{t}}\right)^2 +\tilde{W}(\hat{\phi},\Psi)}
\eea
and the continuity equation in the present context can be written as:
\bea 
\frac{d\rho_{\bf  X}}{d\tilde{t}}+3\tilde{H}(1+w_{\bf  X})\rho_{\bf  X}=0.
\eea
For the qualitative analysis of the prescribed systemin Einstein frame and in order to compare with present day observations, we introduce the following sets of dimensionless density parameters and shifted equation of state parameter:
\bea 
\label{o11}\Omega_{\bf  X}&\equiv& \Omega_{\Psi}=\frac{\rho_{\bf  X}}{3\tilde{H}^{2}M^{2}_{p}},~~
\Omega_m \equiv \Omega_{\phi}=\frac{\rho_m}{3\tilde{H}^{2}M^{2}_{p}},~~
\Omega_r \equiv \frac{\rho_r}{3\tilde{H}^{2}M^{2}_{p}},\\
\label{o44}\Delta_{\bf  X} &\equiv& \Delta_{\Psi}=1+w_{\Psi}=1+w_{\bf  X},~~
\label{o55}\Delta_{  m} \equiv \Delta_{\phi}=1+w_{\phi}=1+w_{  m}.
\eea
In order to transform the cosmological equations into a simplified
autonomous system, we define the following dimensionless auxiliary variables for the study of present dynamical system at late time scale:
\bea  
x&\equiv& \frac{\dot{\Psi}}{\sqrt{6}\tilde{H}M_p},~~
y\equiv \frac{\dot{\tilde{W}}(\hat{\phi},\Psi)}{\sqrt{3}\tilde{H}M_p},~~
\Theta\equiv -M_p\partial_{\Psi}\ln\tilde{W}(\hat{\phi},\Psi),~~
\Sigma\equiv\frac{\tilde{W}(\hat{\phi},\Psi)\partial_{\Psi\Psi}\tilde{W}(\hat{\phi},\Psi)}{\left(\partial_{\Psi}\tilde{W}(\hat{\phi},\Psi)\right)^{2}}
\eea
which can be recast in the autonomous form as:
\bea
\frac{dx}{d{\cal N}} &=& \frac{x}{2}\left(\Omega_r-3y^2-3\right)+\frac{3x^3}{2}
 +\sqrt{\frac{3}{2}}y^2\Theta,~~~
 \frac{dy}{ d{\cal N}} =
\frac{y}{2}\left(3x^2-\sqrt{6}x\Theta+3+\Omega_r\right) 
 -\frac{3 y^3}{2},~~~\nonumber\\
 \hspace*{-1.5em}
 \frac{d \Theta}{d{\cal N}} &=&
 -\frac{\sqrt{6}}{2} \Theta^2 (\Sigma-1)x,~~
 \frac{d \Omega_r}{d{\cal N}} =
 -\Omega_r\left(1-3(x^2 -y^2)-\Omega_r\right),~~
 \frac{d \Omega_m}{d{\cal N}} =
 \Omega_m\left(3(x^2 -y^2)+\Omega_r\right),~~~~~~
\eea
together with an additional constraint condition,
$\Omega_X+\Omega_r+\Omega_m=x^2+y^2+\Omega_m +\Omega_r=1$.
Also using these dimensionless variables Eq~(\ref{ooo}) and Eq~(\ref{o11}) can be recast as: 
\bea
\label{wphiquin}
w_{X} &\equiv& \frac{p_{X}}
{\rho_{X}}=\frac{ x^2-y^2}
{ x^2+y^2}=\frac{x^2-y^2}{\Omega_X}=\frac{w_{\rm eff}-\frac{\Omega_r}{3}}{\Omega_X}, ~~~~
\Omega_{X} \equiv
\frac{\rho_{X}}{3\tilde{H}^2M^{2}_{p}}
= x^2+y^2.
\eea
One can also define the total effective equation of state as:
\bea
\label{weffquinddd}
w_{\rm eff} 
&\equiv& \frac{p_{\rm eff}}{\rho_{\rm eff}}=\frac{p_{\bf X}+p_m+p_r}
{\rho_{ X}+\rho_m+\rho_r} =\frac{p_{\bf \Psi}+p_{\phi}+p_r}
{\rho_{ \Psi}+\rho_{\phi}+\rho_r} = x^2 -y^2 +\frac{\Omega_r}{3}.
\eea
For an accelerated expansion effective equation of state satisfy the following constraint,
 $w_{\rm eff}<-1/3$. Using this methodology mentioned in this section one can study the constraints on the model from late time acceleration which is beyond the scope of our discussion in this paper.
 \subsection{Details of $\delta {\cal N}$ formalism:}
 \subsubsection{Useful field derivatives of ${\cal N}$}
 To simplify the calculation for $\delta {\cal N}$ let us consider the all of these possibilities to write down the infinitesimal change in $\Psi$ field in terms of the inflaton field $\phi$:
 \bea\label{po11}
 \textcolor{red}{\underline{\bf Case ~I}:}~~~~~~~~~
 \delta\Psi&=& -\frac{9\phi}{\sqrt{6}M_p}~\delta\phi,\\ 
 \textcolor{red}{\underline{\bf Case ~II}:}~~~~~~~~~
 \delta\Psi&=& -\frac{\phi}{\sqrt{6}M_p}~\delta\phi,\\ 
 \textcolor{red}{\underline{\bf Case ~II+Choice~I(v1\& v2)}:}~~~~~~~~~
 \delta\Psi&=& -\frac{\phi}{\sqrt{6}M_p}~\delta\phi\left[1-\frac{\phi^4_V}{\phi^4}\right],\\ 
 \textcolor{red}{\underline{\bf Case ~II+Choice~II(v1\& v2)}:}~~~~~~~~~
 \delta\Psi&=& -\frac{\phi}{\sqrt{6}M_p}~\delta\phi\left[1-\frac{m^2_c}{(m^2_c-\lambda\phi^2)}\right],\\
 \textcolor{red}{\underline{\bf Case ~II+Choice~III}:}~~~~~~~~~
 \delta\Psi&=& -\frac{\phi}{\sqrt{6}M_p}~\delta\phi\left[1+\frac{\xi}{2}(\phi^2+\phi^2_0-2\phi^2_V) 
 +\frac{\xi}{2}(\phi^2-\phi^2_0)+\frac{\phi^2_V}{\phi^2}\right].~~~~~~~~~~~~~
 \eea 
 Combining all of these possibilities one can write the following expression:
 \bea \delta\Psi&=& {\cal V}(\phi)~\delta\phi,\eea
 where we introduce a function ${\cal V}(\phi)$, which can be written as:
 \be\tiny\begin{array}{lll}\label{rinfla}
  \displaystyle {\cal V}(\phi)\displaystyle =-\frac{\phi}{\sqrt{6}M_p}\times\left\{\begin{array}{ll}
                     \displaystyle 9 &
  \mbox{\small \textcolor{red}{\bf for \underline{Case I}}}  
 \\ 
          \displaystyle 1 .~~~~ & \mbox{\small \textcolor{red}{\bf for \underline{Case II}}}  
          \\ 
                   \displaystyle \left[1-\frac{\phi^4_V}{\phi^4}\right].~~~~ & \mbox{\small \textcolor{red}{\bf for \underline{Case ~II+Choice~I(v1\& v2)}}}\\ 
                                     \displaystyle \left[1-\frac{m^2_c}{(m^2_c-\lambda\phi^2)}\right].~~~~ & \mbox{\small \textcolor{red}{\bf for \underline{Case ~II+Choice~II(v1\& v2)}}}\\ 
                                                       \displaystyle \left[1+\frac{\xi}{2}(\phi^2+\phi^2_0-2\phi^2_V) 
                                                       +\frac{\xi}{2}(\phi^2-\phi^2_0)+\frac{\phi^2_V}{\phi^2}\right] .~~~~ & \mbox{\small \textcolor{red}{\bf for \underline{Case ~II+Choice~III}}}.~~~~~~~~~~~~
           \end{array}
 \right.
 \end{array}\ee
 This additionally implies that one can write down the following differential operator for the $\Psi$ field:
 \bea \partial_{\Psi}&=&\frac{1}{{\cal V}(\phi)}~\partial_{\phi},~~~
 \partial^2_{\Psi}=\left[\frac{1}{{\cal V}^2(\phi)}~\partial^2_{\phi}-\frac{{\cal V}^{'}(\phi)}{{\cal V}^3(\phi)}\partial_{\phi}\right],~~
 \partial^3_{\Psi}=\left[\frac{1}{{\cal V}^3(\phi)}~\partial^3_{\phi}-3\frac{{\cal V}^{'}(\phi)}{{\cal V}^4(\phi)}~\partial^2_{\phi}+3\frac{{\cal V}^{'2}(\phi)}{{\cal V}^5(\phi)}~\partial_{\phi}\right],~~~~~~~~~~~~\\ \partial_{\phi}\partial_{\Psi}&=&\left[\frac{1}{{\cal V}(\phi)}~\partial^2_{\phi}-\frac{{\cal V}^{'}(\phi)}{{\cal V}^2(\phi)}~\partial_{\phi}\right],~~ \partial_{\Psi}\partial_{\phi}=\frac{1}{{\cal V}(\phi)}~\partial^2_{\phi},\\
  \partial_{\phi}\partial_{\phi}\partial_{\Psi}&=&\left[\frac{1}{{\cal V}(\phi)}~\partial^3_{\phi}-2\frac{{\cal V}^{'}(\phi)}{{\cal V}^2(\phi)}~\partial^2_{\phi}-\left(\frac{{\cal V}^{''}(\phi)}{{\cal V}^2(\phi)}-2\frac{{\cal V}^{'2}(\phi)}{{\cal V}^3(\phi)}\right)\partial_{\phi}\right],~~
 \partial_{\phi}\partial_{\Psi}\partial_{\phi}=\left[\frac{1}{{\cal V}(\phi)}~\partial^3_{\phi}-\frac{{\cal V}^{'}(\phi)}{{\cal V}^2(\phi)}~\partial^2_{\phi}\right],\\
 \partial_{\Psi}\partial_{\phi}\partial_{\phi}&=&\frac{1}{{\cal V}(\phi)}~\partial^3_{\phi},~~
 \partial_{\phi}\partial_{\Psi}\partial_{\Psi}=\left[\frac{1}{{\cal V}^2(\phi)}~\partial^{3}_{\phi}-3\frac{{\cal V}^{'}(\phi)}{{\cal V}^3(\phi)}~\partial^{2}_{\phi}+3~\frac{{\cal V}^{'2}(\phi)}{{\cal V}^4(\phi)}~\partial_{\phi}\right],\\
 \partial_{\Psi}\partial_{\phi}\partial_{\Psi}&=&\left[\frac{1}{{\cal V}^2(\phi)}~\partial^{3}_{\phi}-2\frac{{\cal V}^{'}(\phi)}{{\cal V}^3(\phi)}~\partial^{2}_{\phi}+2~\frac{{\cal V}^{'2}(\phi)}{{\cal V}^4(\phi)}~\partial_{\phi}\right],~~
 \partial_{\Psi}\partial_{\Psi}\partial_{\phi}=\left[\frac{1}{{\cal V}^2(\phi)}~\partial^{3}_{\phi}-\frac{{\cal V}^{'}(\phi)}{{\cal V}^3(\phi)}~\partial^{2}_{\phi}\right],\eea
 where $'$ is defined as the partial derivative with respect to the field $\phi$ i.e. $'=\partial_{\phi}$.
 
 Consequently one can write:
 \bea 
 {\cal N}_{,\Psi}&=&\frac{1}{{\cal V}(\phi)}~\partial_{\phi}{\cal N}=\frac{1}{{\cal V}(\phi)}~{\cal N}_{,\phi},~~
 {\cal N}_{,\Psi\Psi}=\left[\frac{1}{{\cal V}^2(\phi)}~\partial^2_{\phi}-\frac{{\cal V}^{'}(\phi)}{{\cal V}^3(\phi)}\partial_{\phi}\right]{\cal N}=\left[\frac{1}{{\cal V}^2(\phi)}~{\cal N}_{,\phi\phi}-\frac{{\cal V}^{'}(\phi)}{{\cal V}^3(\phi)}~{\cal N}_{,\phi}\right],~~~~~~~~~~~~\\
 {\cal N}_{,\phi\Psi}&=&\left[\frac{1}{{\cal V}(\phi)}~\partial^2_{\phi}-\frac{{\cal V}^{'}(\phi)}{{\cal V}^2(\phi)}~\partial_{\phi}\right]{\cal N}=\left[\frac{1}{{\cal V}(\phi)}~{\cal N}_{,\phi\phi}-\frac{{\cal V}^{'}(\phi)}{{\cal V}^2(\phi)}~{\cal N}_{,\phi}\right],~~ {\cal N}_{,\Psi\phi}=\frac{1}{{\cal V}(\phi)}~\partial^2_{\phi}~{\cal N}=\frac{1}{{\cal V}(\phi)}~{\cal N}_{,\phi\phi},\\
 {\cal N}_{,\Psi\Psi\Psi}&=&\left[\frac{1}{{\cal V}^3(\phi)}~\partial^3_{\phi}-3\frac{{\cal V}^{'}(\phi)}{{\cal V}^4(\phi)}~\partial^2_{\phi}+3\frac{{\cal V}^{'2}(\phi)}{{\cal V}^5(\phi)}~\partial_{\phi}\right]~{\cal N}=\left[\frac{1}{{\cal V}^3(\phi)}~{\cal N}_{,\phi\phi\phi}-3\frac{{\cal V}^{'}(\phi)}{{\cal V}^4(\phi)}~{\cal N}_{,\phi\phi}+3\frac{{\cal V}^{'2}(\phi)}{{\cal V}^5(\phi)}~{\cal N}_{,\phi}\right],\\
 {\cal N}_{,\phi\phi\Psi}&=&\left[\frac{1}{{\cal V}(\phi)}~\partial^3_{\phi}-2\frac{{\cal V}^{'}(\phi)}{{\cal V}^2(\phi)}~\partial^2_{\phi}-\left(\frac{{\cal V}^{''}(\phi)}{{\cal V}^2(\phi)}-2\frac{{\cal V}^{'2}(\phi)}{{\cal V}^3(\phi)}\right)\partial_{\phi}\right]~{\cal N},\nonumber\\
 &=&\left[\frac{1}{{\cal V}(\phi)}~{\cal N}_{,\phi\phi\phi}-2\frac{{\cal V}^{'}(\phi)}{{\cal V}^2(\phi)}~{\cal N}_{,\phi\phi}-\left(\frac{{\cal V}^{''}(\phi)}{{\cal V}^2(\phi)}-2\frac{{\cal V}^{'2}(\phi)}{{\cal V}^3(\phi)}\right)~{\cal N}_{,\phi}\right],\\
 {\cal N}_{,\phi\Psi\phi}&=&\left[\frac{1}{{\cal V}(\phi)}~\partial^3_{\phi}-\frac{{\cal V}^{'}(\phi)}{{\cal V}^2(\phi)}~\partial^2_{\phi}\right]~{\cal N}=\left[\frac{1}{{\cal V}(\phi)}~{\cal N}_{,\phi\phi\phi}-\frac{{\cal V}^{'}(\phi)}{{\cal V}^2(\phi)}~{\cal N}_{,\phi\phi}\right],\\
 N_{,\Psi\phi\phi}&=&\frac{1}{{\cal V}(\phi)}~\partial^3_{\phi}~{\cal N}=\frac{1}{{\cal V}(\phi)}~{\cal N}_{,\phi\phi\phi},\\
 {\cal N}_{,\phi\Psi\Psi}&=&\left[\frac{1}{{\cal V}^2(\phi)}~\partial^{3}_{\phi}-3\frac{{\cal V}^{'}(\phi)}{{\cal V}^3(\phi)}~\partial^{2}_{\phi}+3~\frac{{\cal V}^{'2}(\phi)}{{\cal V}^4(\phi)}~\partial_{\phi}\right]~{\cal N}=\left[\frac{1}{{\cal V}^2(\phi)}~{\cal N}_{,\phi\phi\phi}-3\frac{{\cal V}^{'}(\phi)}{{\cal V}^3(\phi)}~{\cal N}_{,\phi\phi}+3~\frac{{\cal V}^{'2}(\phi)}{{\cal V}^4(\phi)}~{\cal N}_{,\phi}\right],~~~~~~~~~~~~~~\\
 {\cal N}_{,\Psi\phi\Psi}&=&\left[\frac{1}{{\cal V}^2(\phi)}~\partial^{3}_{\phi}-2\frac{{\cal V}^{'}(\phi)}{{\cal V}^3(\phi)}~\partial^{2}_{\phi}+2~\frac{{\cal V}^{'2}(\phi)}{{\cal V}^4(\phi)}~\partial_{\phi}\right]~{\cal N}=\left[\frac{1}{{\cal V}^2(\phi)}~{\cal N}_{,\phi\phi\phi}-2\frac{{\cal V}^{'}(\phi)}{{\cal V}^3(\phi)}~{\cal N}_{,\phi\phi}+2~\frac{{\cal V}^{'2}(\phi)}{{\cal V}^4(\phi)}~{\cal N}_{,\phi}\right],\\
 {\cal N}_{,\Psi\Psi\phi}&=&\left[\frac{1}{{\cal V}^2(\phi)}~\partial^{3}_{\phi}-\frac{{\cal V}^{'}(\phi)}{{\cal V}^3(\phi)}~\partial^{2}_{\phi}\right]~{\cal N}=\left[\frac{1}{{\cal V}^2(\phi)}~{\cal N}_{,\phi\phi\phi}-\frac{{\cal V}^{'}(\phi)}{{\cal V}^3(\phi)}~{\cal N}_{,\phi\phi}\right].
 \eea
 \subsubsection{Second-order perturbative solution with various source}
 If we negelect the quadratic slow-roll corrections then the solution of Eq~(\ref{asdf2ffff}) takes the following form for the all different cases considered here:
 \be\small\begin{array}{llll}\label{er12}
 \displaystyle \textcolor{red}{\underline{\bf For~Case~I}:}
 \\
 \displaystyle \Delta_2={\bf D}_{4}+\frac{1}{27H^3}\left[\frac{27\phi_* H e^{H{\cal Y}t}}{{\cal Y}^2(3+{\cal Y})^3} 
 \left\{-{\cal Y}(3+{\cal Y})^2\left(4\Lambda_c \phi^3_L+H^2{\cal Y}(3+{\cal Y})\right)\right.\right.\\ \left.\left.\displaystyle~~~~~~~~~+\epsilon_H\left(4\Lambda_c \phi^3_L(-18+{\cal Y}(3+{\cal Y})(-6+Ht(3+2{\cal Y})))\right.\right.\right. \\ \left.\left.\left. \displaystyle~~~~~~~~~+H^2{\cal Y}(3+{\cal Y})(-9+{\cal Y}(3+{\cal Y})(-2+Ht(3+2{\cal Y})))\right)\right\}\right.\\ \left.\displaystyle~~~~~~~~+9H^2\Lambda_c\phi^3_L t\left(\phi_L+4{\bf D}_2\right)+e^{-3Ht}\left(4\Lambda_c \phi^3_L(1+3Ht){\bf D}_1-9H^2{\bf D}_3\right)\right]. \end{array}\ee
 \be\small\begin{array}{llll}\label{er22}
 \displaystyle \textcolor{red}{\underline{\bf For~Case~II}:}
 \\
 \displaystyle \Delta_2={\bf D}_{4}+\frac{1}{27H^3}\left[\frac{27\phi_* H e^{H{\cal Y}t}}{{\cal Y}^2(3+{\cal Y})^3} 
 \left\{-{\cal Y}(3+{\cal Y})^2\left(-4\Lambda_c \phi^3_L+H^2{\cal Y}(3+{\cal Y})\right)\right.\right.\\ \left.\left.\displaystyle~~~~~~~~~+\epsilon_H\left(-4\Lambda_c \phi^3_L(-18+{\cal Y}(3+{\cal Y})(-6+Ht(3+2{\cal Y})))\right.\right.\right. \\ \left.\left.\left. \displaystyle~~~~~~~~~+H^2{\cal Y}(3+{\cal Y})(-9+{\cal Y}(3+{\cal Y})(-2+Ht(3+2{\cal Y})))\right)\right\}\right.\\ \left.\displaystyle~~~~~~~~+9H^2 t\left(\beta-\Lambda_c\phi^3_L \left(\phi_L+4{\bf D}_2\right)\right)-e^{-3Ht}\left(4\Lambda_c \phi^3_L(1+3Ht){\bf D}_1+9H^2{\bf D}_3\right)\right].\end{array}\ee
 \be\small\begin{array}{llll}\label{er32}
 \displaystyle \textcolor{red}{\underline{\bf For~Case ~II+Choice~I(v1)}:}
 \\
 \displaystyle \Delta_2={\bf D}_{4}+\frac{1}{27H^3}\left[\frac{27\phi_* H e^{H{\cal Y}t}}{{\cal Y}^2(3+{\cal Y})^3} 
 \left\{-{\cal Y}(3+{\cal Y})^2\left(-4\Lambda_c \phi^3_L+H^2{\cal Y}(3+{\cal Y})\right)\right.\right.\\ \left.\left.\displaystyle~~~~~~~~~+\epsilon_H\left(-4\Lambda_c \phi^3_L(-18+{\cal Y}(3+{\cal Y})(-6+Ht(3+2{\cal Y})))\right.\right.\right. \\ \left.\left.\left. \displaystyle~~~~~~~~~+H^2{\cal Y}(3+{\cal Y})(-9+{\cal Y}(3+{\cal Y})(-2+Ht(3+2{\cal Y})))\right)\right\}\right.\\ \left.\displaystyle~~~~~~~~+9H^2 t\left(\beta+\Lambda_c\phi^4_V-\Lambda_c\phi^3_L \left(\phi_L+4{\bf D}_2\right)\right)-e^{-3Ht}\left(4\Lambda_c \phi^3_L(1+3Ht){\bf D}_1+9H^2{\bf D}_3\right)\right]. \end{array}\ee
 \be\small\begin{array}{llll}\label{er52}
 \displaystyle \textcolor{red}{\underline{\bf For~Case ~II+Choice~I(v2)}:}
 \\
 \displaystyle \Delta_2={\bf D}_{4}+\frac{1}{27H^3}\left[\frac{27\phi_* H e^{H{\cal Y}t}}{{\cal Y}^2(3+{\cal Y})^3} 
 \left\{-{\cal Y}(3+{\cal Y})^2\left(4\Lambda_c \phi^3_L+H^2{\cal Y}(3+{\cal Y})\right)\right.\right.\\ \left.\left.\displaystyle~~~~~~~~~+\epsilon_H\left(4\Lambda_c \phi^3_L(-18+{\cal Y}(3+{\cal Y})(-6+Ht(3+2{\cal Y})))\right.\right.\right. \\ \left.\left.\left. \displaystyle~~~~~~~~~+H^2{\cal Y}(3+{\cal Y})(-9+{\cal Y}(3+{\cal Y})(-2+Ht(3+2{\cal Y})))\right)\right\}\right.\\ \left.\displaystyle~~~~~~~~+9H^2 t\left(\beta-\Lambda_c\phi^4_V+\Lambda_c\phi^3_L \left(\phi_L+4{\bf D}_2\right)\right)+e^{-3Ht}\left(4\Lambda_c \phi^3_L(1+3Ht){\bf D}_1-9H^2{\bf D}_3\right)\right]. \end{array}\ee
 \be\small\begin{array}{llll}\label{er62}
 \displaystyle \textcolor{red}{\underline{\bf For~Case ~II+Choice~II(v1)}:}
 \\
 \displaystyle \Delta_2={\bf D}_{4}+\frac{1}{54H^3}\left[\frac{54\phi_* H e^{H{\cal Y}t}}{{\cal Y}^2(3+{\cal Y})^3} 
 \left\{-{\cal Y}(3+{\cal Y})^2\left(M_c\phi_L-4\Lambda_c\phi^3_L+H^2{\cal Y}(3+{\cal Y})\right)\right.\right.\\ \left.\left.\displaystyle 
 ~~~~~~~~~~~~+\epsilon_H \left((4\Lambda_c \phi^3_L-M_c\phi_L )(18-{\cal Y}(3+{\cal Y})(-6+Ht(3+2{\cal Y})))\right.\right.\right.\\ \left.\left.\left.\displaystyle~~~~~~~~~~~+H^2{\cal Y}(3+{\cal Y})(-9+{\cal Y}(3+{\cal Y})(-2+Ht(3+2{\cal Y})))\right)\right\}\right.\\ \left.~~~~~~~~~~~~~~~~~~
 +9H^2t\left(2\beta+\phi_L(M_c(\phi_L+2{\bf D}_2)-\Lambda_c\phi^2_L(\phi_L+4{\bf D}_2))\right)\right.\\ \left.\displaystyle~~~~~~~~~~~~~~~~~~+e^{-3Ht}\left(2\phi_L(M_c-4\Lambda_c\phi^2_L)(1+3Ht){\bf D}_1-18H^2{\bf D}_3\right)\right]. \end{array}\ee
 \be\small\begin{array}{llll}\label{er72}
 \displaystyle \textcolor{red}{\underline{\bf For~Case ~II+Choice~II(v2)}:}
 \\
 \displaystyle \Delta_2={\bf D}_{4}+\frac{1}{54H^3}\left[\frac{54\phi_* H e^{H{\cal Y}t}}{{\cal Y}^2(3+{\cal Y})^3} 
 \left\{-{\cal Y}(3+{\cal Y})^2\left(-M_c\phi_L+4\Lambda_c\phi^3_L+H^2{\cal Y}(3+{\cal Y})\right)\right.\right.\\ \left.\left.\displaystyle 
 ~~~~~~~~~~~~+\epsilon_H \left((-4\Lambda_c \phi^3_L+M_c\phi_L )(18-{\cal Y}(3+{\cal Y})(-6+Ht(3+2{\cal Y})))\right.\right.\right.\\ \left.\left.\left.\displaystyle~~~~~~~~~~~+H^2{\cal Y}(3+{\cal Y})(-9+{\cal Y}(3+{\cal Y})(-2+Ht(3+2{\cal Y})))\right)\right\}\right.\\ \left.~~~~~~~~~~~~~~~~~~
 +9H^2t\left(2\beta+\phi_L(-M_c(\phi_L+2{\bf D}_2)+\Lambda_c\phi^2_L(\phi_L+4{\bf D}_2))\right)\right.\\ \left.\displaystyle~~~~~~~~~~~~~~~~~~-e^{-3Ht}\left(2\phi_L(M_c-4\Lambda_c\phi^2_L)(1+3Ht){\bf D}_1+18H^2{\bf D}_3\right)\right]. \end{array}\ee
 \be\small\begin{array}{llll}\label{er82}
 \displaystyle \textcolor{red}{\underline{\bf For~Case~III}:}
 \\
 \displaystyle \Delta_2={\bf D}_{4}+\frac{1}{27H^3\phi_L}\left[\frac{27\phi_* H e^{H{\cal Y}t}}{{\cal Y}^2(3+{\cal Y})^3} 
 \left\{-{\cal Y}(3+{\cal Y})^2\left(\Gamma_{\xi}\Theta_{\xi}+H^2\phi_L{\cal Y}(3+{\cal Y})\right)\right.\right.\\ \left.\left.\displaystyle~~~~~~~~+\epsilon_H\left(\Gamma_{\xi}\Theta_{\xi}(-18+{\cal Y}(3+{\cal Y})(-6+Ht(3+2{\cal Y})))\right.\right.\right.\\ \left.\left.\left.\displaystyle~~~~~~~+H^2\phi_L{\cal Y}(3+{\cal Y})(-9+{\cal Y}(3+{\cal Y})(-2+Ht(3+2{\cal Y})))\right)\right\}\right.\\ \left.\displaystyle~~~~~~~+9H^2t\left(\phi_L(\beta+\Gamma_{\xi})+\Gamma_{\xi}\Theta_{\xi}{\bf D}_2\right)+e^{-3Ht}\left(\Gamma_{\xi}\Theta_{\xi}(1+3Ht){\bf D}_1-9H^2\phi_L {\bf D}_3\right)\right].\end{array}\ee
 Here ${\bf D}_{3}$ and ${\bf D}_{4}$ are dimensionful arbitrary integration constants which can be determined by imposing the appropriate boundary condition.

 \subsubsection{Expressions for perturbative solutions in final hypersurface}
 Negelecting the contribution from the qudratic slow roll term and taking upto linear order term in slow-roll we get the following result:
   \be\begin{array}{llll}\label{dft2v2}
   \displaystyle \Delta_{1}({\cal N}=0)={\bf D}_{2}-\frac{1}{3H}{\bf D}_{1}+\frac{1}{{\cal Y}(3+{\cal Y})^2}\phi_* \left[-{\cal Y}(3+{\cal Y})^2-2\epsilon_{H}\left(-9+{\cal Y}(3+{\cal Y})\right)\right].
   \end{array}\ee
   Similarly if we negelect the quadratic slow-roll corrections then the solution of $\Delta_{2}(N=0)$ takes the following form for the all different cases considered here:
   \be\small\begin{array}{llll}\label{er12v1q}
   \displaystyle \textcolor{red}{\underline{\bf For~Case~I}:}
   \\
   \displaystyle \Delta_2({\cal N}=0)={\bf D}_{4}+\frac{1}{27H^3}\left[\frac{27\phi_* H}{{\cal Y}^2(3+{\cal Y})^3} 
   \left\{-{\cal Y}(3+{\cal Y})^2\left(4\Lambda_c \phi^3_L+H^2{\cal Y}(3+{\cal Y})\right)\right.\right.\\ \left.\left.\displaystyle+\epsilon_H\left(-4\Lambda_c \phi^3_L(18+6{\cal Y}(3+{\cal Y}))-H^2{\cal Y}(3+{\cal Y})(9+2{\cal Y}(3+{\cal Y}))\right)\right\}+\left(4\Lambda_c \phi^3_L{\bf D}_1-9H^2{\bf D}_3\right)\right]. \end{array}\ee
   \be\small\begin{array}{llll}\label{er22v1q}
   \displaystyle \textcolor{red}{\underline{\bf For~Case~II}:}
   \\
   \displaystyle \Delta_2({\cal N}=0)={\bf D}_{4}+\frac{1}{27H^3}\left[\frac{27\phi_* H }{{\cal Y}^2(3+{\cal Y})^3}\left\{-{\cal Y}(3+{\cal Y})^2\left(-4\Lambda_c \phi^3_L+H^2{\cal Y}(3+{\cal Y})\right)\right.\right.\\ \left.\left.\displaystyle+\epsilon_H\left(4\Lambda_c \phi^3_L(18+6{\cal Y}(3+{\cal Y}))-H^2{\cal Y}(3+{\cal Y})(9+2{\cal Y}(3+{\cal Y}))\right)\right\}-\left(4\Lambda_c \phi^3_L{\bf D}_1+9H^2{\bf D}_3\right)\right].\end{array}\ee
   \be\small\begin{array}{llll}\label{er32v1c}
   \displaystyle \textcolor{red}{\underline{\bf For~Case ~II+Choice~I(v1)}:}
   \\
   \displaystyle \Delta_2({\cal N}=0)={\bf D}_{4}+\frac{1}{27H^3}\left[\frac{27\phi_* H }{{\cal Y}^2(3+{\cal Y})^3}\left\{-{\cal Y}(3+{\cal Y})^2\left(-4\Lambda_c \phi^3_L+H^2{\cal Y}(3+{\cal Y})\right)\right.\right.\\ \left.\left.\displaystyle+\epsilon_H\left(4\Lambda_c \phi^3_L(18+6{\cal Y}(3+{\cal Y}))-H^2{\cal Y}(3+{\cal Y})(9+2{\cal Y}(3+{\cal Y}))\right)\right\}-\left(4\Lambda_c \phi^3_L{\bf D}_1+9H^2{\bf D}_3\right)\right]. \end{array}\ee
   \be\small\begin{array}{llll}\label{er52v1q}
   \displaystyle \textcolor{red}{\underline{\bf For~Case ~II+Choice~I(v2)}:}
   \\
   \displaystyle \Delta_2({\cal N}=0)={\bf D}_{4}+\frac{1}{27H^3}\left[\frac{27\phi_* H }{{\cal Y}^2(3+{\cal Y})^3}\left\{-{\cal Y}(3+{\cal Y})^2\left(4\Lambda_c \phi^3_L+H^2{\cal Y}(3+{\cal Y})\right)\right.\right.\\ \left.\left.\displaystyle+\epsilon_H\left(-4\Lambda_c \phi^3_L(18+6{\cal Y}(3+{\cal Y}))-H^2{\cal Y}(3+{\cal Y})(9+2{\cal Y}(3+{\cal Y}))\right)\right\}+\left(4\Lambda_c \phi^3_L{\bf D}_1-9H^2{\bf D}_3\right)\right]. \end{array}\ee
   \be\small\begin{array}{llll}\label{er62v1q}
   \displaystyle \textcolor{red}{\underline{\bf For~Case ~II+Choice~II(v1)}:}
   \\
   \displaystyle \Delta_2({\cal N}=0)={\bf D}_{4}+\frac{1}{54H^3}\left[\frac{54\phi_* H }{{\cal Y}^2(3+{\cal Y})^3}\left\{-{\cal Y}(3+{\cal Y})^2\left(M_c\phi_L-4\Lambda_c\phi^3_L+H^2{\cal Y}(3+{\cal Y})\right)\right.\right.\\ \left.\left.\displaystyle 
   ~~~~~~~~~~~~+\epsilon_H \left((4\Lambda_c \phi^3_L-M_c\phi_L )(18+6{\cal Y}(3+{\cal Y}))\right.\right.\right.\\ \left.\left.\left.\displaystyle~~~~~~~~~~~-H^2{\cal Y}(3+{\cal Y})(9+2{\cal Y}(3+{\cal Y})))\right)\right\}
   +\left(2\phi_L(M_c-4\Lambda_c\phi^2_L){\bf D}_1-18H^2{\bf D}_3\right)\right]. \end{array}\ee
   \be\small\begin{array}{llll}\label{er72v1q}
   \displaystyle \textcolor{red}{\underline{\bf For~Case ~II+Choice~II(v2)}:}
   \\
   \displaystyle \Delta_2({\cal N}=0)={\bf D}_{4}+\frac{1}{54H^3}\left[\frac{54\phi_* H }{{\cal Y}^2(3+{\cal Y})^3}\left\{-{\cal Y}(3+{\cal Y})^2\left(-M_c\phi_L+4\Lambda_c\phi^3_L+H^2{\cal Y}(3+{\cal Y})\right)\right.\right.\\ \left.\left.\displaystyle 
   ~~~~~~~~~~~~+\epsilon_H \left((-4\Lambda_c \phi^3_L+M_c\phi_L )(18+6{\cal Y}(3+{\cal Y}))\right.\right.\right.\\ \left.\left.\left.\displaystyle~~~~~~~~~~~-H^2{\cal Y}(3+{\cal Y})(9+2{\cal Y}(3+{\cal Y}))\right)\right\}
   -\left(2\phi_L(M_c-4\Lambda_c\phi^2_L){\bf D}_1+18H^2{\bf D}_3\right)\right]. \end{array}\ee
   \be\small\begin{array}{llll}\label{er82v1q}
   \displaystyle \textcolor{red}{\underline{\bf For~Case~III}:}
   \\
   \displaystyle \Delta_2({\cal N}=0)={\bf D}_{4}+\frac{1}{27H^3\phi_L}\left[\frac{27\phi_* H }{{\cal Y}^2(3+{\cal Y})^3}\left\{-{\cal Y}(3+{\cal Y})^2\left(\Gamma_{\xi}\Theta_{\xi}+H^2\phi_L{\cal Y}(3+{\cal Y})\right)\right.\right.\\ \left.\left.\displaystyle~~~~~~~~+\epsilon_H\left(-\Gamma_{\xi}\Theta_{\xi}(18+6{\cal Y}(3+{\cal Y}))\right.\right.\right.\\ \left.\left.\left.\displaystyle~~~~~~~-H^2\phi_L{\cal Y}(3+{\cal Y})(9+2{\cal Y}(3+{\cal Y}))\right)\right\}+\left(\Gamma_{\xi}\Theta_{\xi}{\bf D}_1-9H^2\phi_L {\bf D}_3\right)\right].\end{array}\ee

   \subsubsection{Shift in the inflaton field due to $\delta{\cal N}$}
    Analytical expression for the shift in the inflaton field from linear order and second order cosmological perturbation theory can be written upto considering the contributions from the first order slow-roll contribution as:
     \be\begin{array}{llll}\label{dft23}
     \displaystyle \delta\phi_{1}({\cal N}=0)=\delta\phi_{1*}=\phi_* \hat{\Delta}_{1}({\cal N}=0)=\phi_*\hat{\bf D}_{2}-\frac{\phi_*}{3H}\hat{\bf D}_{1}+\frac{\phi_*}{{\cal Y}(3+{\cal Y})^2}\left[-{\cal Y}(3+{\cal Y})^2\right.\\ \left.~~~~~~~~~~~~~~~~~~~~~~~~~~~~~~~~~~~~~~~~~~~~~~~~~~~\displaystyle +\epsilon_{H}\left(-9+{\cal Y}(3+{\cal Y})\left\{-2+H(3+2{\cal Y})t\right\}\right)\right].
     \end{array}\ee
     \be\small\begin{array}{llll}\label{er12v1}
       \displaystyle \textcolor{red}{\underline{\bf For~Case~I}:}
       \\
        \displaystyle\delta\phi_{2}({\cal N}=0)=\delta\phi_{2*}=\phi_* \hat{\Delta}_{2}({\cal N}=0)\\
                   \displaystyle~~~~~~~=\phi_*\hat{\bf D}_{4}+\frac{\phi_*}{27H^3}\left[\frac{27H}{{\cal Y}^2(3+{\cal Y})^3} 
       \left\{-{\cal Y}(3+{\cal Y})^2\left(4\Lambda_c \phi^3_L+H^2{\cal Y}(3+{\cal Y})\right)\right.\right.\\ \left.\left.\displaystyle+\epsilon_H\left(-4\Lambda_c \phi^3_L(18+6{\cal Y}(3+{\cal Y}))-H^2{\cal Y}(3+{\cal Y})(9+2{\cal Y}(3+{\cal Y}))\right)\right\}+\left(4\Lambda_c \phi^3_L\hat{\bf D}_1-9H^2\hat{\bf D}_3\right)\right]. \end{array}\ee
       \be\small\begin{array}{llll}\label{er22v1}
       \displaystyle \textcolor{red}{\underline{\bf For~Case~II}:}
       \\
        \displaystyle\delta\phi_{2}({\cal N}=0)=\delta\phi_{2*}=\phi_* \hat{\Delta}_{2}({\cal N}=0)\\
                   \displaystyle~~~~~~~=\phi_*\hat{\bf D}_{4}+\frac{\phi_*}{27H^3}\left[\frac{27H }{{\cal Y}^2(3+{\cal Y})^3}\left\{-{\cal Y}(3+{\cal Y})^2\left(-4\Lambda_c \phi^3_L+H^2{\cal Y}(3+{\cal Y})\right)\right.\right.\\ \left.\left.\displaystyle+\epsilon_H\left(4\Lambda_c \phi^3_L(18+6{\cal Y}(3+{\cal Y}))-H^2{\cal Y}(3+{\cal Y})(9+2{\cal Y}(3+{\cal Y}))\right)\right\}-\left(4\Lambda_c \phi^3_L\hat{\bf D}_1+9H^2\hat{\bf D}_3\right)\right].\end{array}\ee
       \be\small\begin{array}{llll}\label{er32v1}
       \displaystyle \textcolor{red}{\underline{\bf For~Case ~II+Choice~I(v1)}:}
       \\
        \displaystyle\delta\phi_{2}({\cal N}=0)=\delta\phi_{2*}=\phi_* \hat{\Delta}_{2}({\cal N}=0)\\
                   \displaystyle~~~~~~~=\phi_*\hat{\bf D}_{4}+\frac{\phi_*}{27H^3}\left[\frac{27H }{{\cal Y}^2(3+{\cal Y})^3}\left\{-{\cal Y}(3+{\cal Y})^2\left(-4\Lambda_c \phi^3_L+H^2{\cal Y}(3+{\cal Y})\right)\right.\right.\\ \left.\left.\displaystyle+\epsilon_H\left(4\Lambda_c \phi^3_L(18+6{\cal Y}(3+{\cal Y}))-H^2{\cal Y}(3+{\cal Y})(9+2{\cal Y}(3+{\cal Y}))\right)\right\}-\left(4\Lambda_c \phi^3_L\hat{\bf D}_1+9H^2\hat{\bf D}_3\right)\right]. \end{array}\ee
       \be\small\begin{array}{llll}\label{er52v1}
       \displaystyle \textcolor{red}{\underline{\bf For~Case ~II+Choice~I(v2)}:}
       \\
        \displaystyle\delta\phi_{2}({\cal N}=0)=\delta\phi_{2*}=\phi_* \hat{\Delta}_{2}({\cal N}=0)\\
                   \displaystyle~~~~~~~=\phi_*\hat{\bf D}_{4}+\frac{\phi_*}{27H^3}\left[\frac{27H }{{\cal Y}^2(3+{\cal Y})^3}\left\{-{\cal Y}(3+{\cal Y})^2\left(4\Lambda_c \phi^3_L+H^2{\cal Y}(3+{\cal Y})\right)\right.\right.\\ \left.\left.\displaystyle+\epsilon_H\left(-4\Lambda_c \phi^3_L(18+6{\cal Y}(3+{\cal Y}))-H^2{\cal Y}(3+{\cal Y})(9+2{\cal Y}(3+{\cal Y}))\right)\right\}+\left(4\Lambda_c \phi^3_L\hat{\bf D}_1-9H^2\hat{\bf D}_3\right)\right]. \end{array}\ee
       \be\begin{array}{llll}\label{er62v1}
       \displaystyle \textcolor{red}{\underline{\bf For~Case ~II+Choice~II(v1)}:}
       \\
       \displaystyle \Delta_2({\cal N}=0)=\phi_*\hat{\bf D}_{4}+\frac{\phi_*}{54H^3}\left[\frac{54 H }{{\cal Y}^2(3+{\cal Y})^3}\left\{-{\cal Y}(3+{\cal Y})^2\left(M_c\phi_L-4\Lambda_c\phi^3_L+H^2{\cal Y}(3+{\cal Y})\right)\right.\right.\\ \left.\left.\displaystyle 
       ~~~~~~~~~~~~+\epsilon_H \left((4\Lambda_c \phi^3_L-M_c\phi_L )(18+6{\cal Y}(3+{\cal Y}))\right.\right.\right.\\ \left.\left.\left.\displaystyle~~~~~~~~~~~-H^2{\cal Y}(3+{\cal Y})(9+2{\cal Y}(3+{\cal Y})))\right)\right\}
       +\left(2\phi_L(M_c-4\Lambda_c\phi^2_L)\hat{\bf D}_1-18H^2\hat{\bf D}_3\right)\right]. \end{array}\ee
       \be\small\begin{array}{llll}\label{er72v1}
       \displaystyle \textcolor{red}{\underline{\bf For~Case ~II+Choice~II(v2)}:}
       \\
        \displaystyle\delta\phi_{2}({\cal N}=0)=\delta\phi_{2*}=\phi_* \hat{\Delta}_{2}({\cal N}=0)\\
                   \displaystyle~~~~~~~=\phi_*\hat{\bf D}_{4}+\frac{\phi_*}{54H^3}\left[\frac{54 H }{{\cal Y}^2(3+{\cal Y})^3}\left\{-{\cal Y}(3+{\cal Y})^2\left(-M_c\phi_L+4\Lambda_c\phi^3_L+H^2{\cal Y}(3+{\cal Y})\right)\right.\right.\\ \left.\left.\displaystyle 
       ~~~~~~~~~~~~+\epsilon_H \left((-4\Lambda_c \phi^3_L+M_c\phi_L )(18+6{\cal Y}(3+{\cal Y}))\right.\right.\right.\\ \left.\left.\left.\displaystyle~~~~~~~~~~~-H^2{\cal Y}(3+{\cal Y})(9+2{\cal Y}(3+{\cal Y}))\right)\right\}
       -\left(2\phi_L(M_c-4\Lambda_c\phi^2_L)\hat{\bf D}_1+18H^2\hat{\bf D}_3\right)\right]. \end{array}\ee
       \be\small\begin{array}{llll}\label{er82v1}
       \displaystyle \textcolor{red}{\underline{\bf For~Case~III}:}
       \\
        \displaystyle\delta\phi_{2}({\cal N}=0)=\delta\phi_{2*}=\phi_* \hat{\Delta}_{2}({\cal N}=0)\\
                   \displaystyle~~~~~~~=\phi_*\hat{\bf D}_{4}+\frac{\phi_*}{27H^3\phi_L}\left[\frac{27 H }{{\cal Y}^2(3+{\cal Y})^3}\left\{-{\cal Y}(3+{\cal Y})^2\left(\Gamma_{\xi}\Theta_{\xi}+H^2\phi_L{\cal Y}(3+{\cal Y})\right)\right.\right.\\ \left.\left.\displaystyle~~~+\epsilon_H\left(-\Gamma_{\xi}\Theta_{\xi}(18+6{\cal Y}(3+{\cal Y})) 
                   -H^2\phi_L{\cal Y}(3+{\cal Y})(9+2{\cal Y}(3+{\cal Y}))\right)\right\}+\left(\Gamma_{\xi}\Theta_{\xi}\hat{\bf D}_1-9H^2\phi_L \hat{\bf D}_3\right)\right].\end{array}\ee
                   
 \subsubsection{Various useful constants for $\delta{\cal N}$}
  For the derived effective potentials ${\cal B}(\phi_*)$ and ${\cal C}(\phi_*)$ can be recast as:
        \be\tiny\begin{array}{lll}\label{rinfla1cxcvvv}
         \displaystyle {\cal B}(\phi_*)\displaystyle =\left\{\begin{array}{ll}
                            \displaystyle \frac{3}{{\cal Y} \phi^2_*}~~~~~ &
         \mbox{\small \textcolor{red}{\bf for \underline{Case I \& II}}} 
                 \\ 
                          \displaystyle \frac{1}{{\cal Y} \phi^2_*}\frac{\left(3+\frac{\phi^4_V}{\phi^4_*}\right)}{\left(1-\frac{\phi^4_V}{\phi^4_*}\right)}~~~~ & \mbox{\small \textcolor{red}{\bf for \underline{Case ~II+Choice~I(v1\& v2)}}}\\
                                            \displaystyle \frac{1}{{\cal Y} \phi^2_*}\left\{3-\frac{2\lambda m^2_c \phi^2_*}{(m^2_c-\lambda\phi^2_*)^2\left(1-\frac{m^2_c}{(m^2_c-\lambda\phi^2_*)}\right)}\right\}~~~~ & \mbox{\small \textcolor{red}{\bf for \underline{Case ~II+Choice~II(v1\& v2)}}}\\
                                                              \displaystyle \frac{2}{{\cal Y} \phi^2_*}\left\{2+\frac{1+\xi(3\phi^2_*-\phi^2_V)-\frac{\phi^2_V}{\phi^2_*}}{1+\xi(\phi^2_*-\phi^2_V)+\frac{\phi^2_V}{\phi^2_*}}\right\}~~~~ & \mbox{\small \textcolor{red}{\bf for \underline{Case ~II+Choice~III}}}.
                  \end{array}
        \right.
        \end{array}\ee
        \be\tiny\begin{array}{lll}\label{rinfla1cxcvv}
               \displaystyle {\cal C}(\phi_*)\displaystyle \approx\left\{\begin{array}{ll}
                                  \displaystyle -\frac{19}{3}\frac{1}{{\cal Y} \phi^3_*}~~~~~ &
               \mbox{\small \textcolor{red}{\bf for \underline{Case I \& II}}} 
                       \\ 
                                \displaystyle-\frac{1}{{\cal Y} \phi^3_*}\left\{\frac{8}{3}+\frac{2\left(1+2\frac{\phi^4_V}{\phi^4_*}\right)}{\left(1-\frac{\phi^4_V}{\phi^4_*}\right)}-\frac{5}{3}\frac{\left(1+3\frac{\phi^4_V}{\phi^4_*}\right)^2}{\left(1-\frac{\phi^4_V}{\phi^4_*}\right)^2}\right\} & \mbox{\small \textcolor{red}{\bf for \underline{Case ~II+Choice~I(v1\& v2)}}}\\ 
                                                  \displaystyle -\frac{1}{{\cal Y} \phi^3_*}\left\{\frac{8}{3}+\frac{2\left(1-\frac{m^2_c}{m^2_c-\lambda\phi^2_*}-\frac{2\lambda m^2_c\phi^2_*}{(m^2_c-\lambda\phi^2_*)^2}\right)}{\left(1-\frac{m^2_c}{m^2_c-\lambda\phi^2_*}\right)}\right.\\ \left.\displaystyle~~~ +\left(\frac{5}{3}\frac{\left(1-\frac{m^2_c}{m^2_c-\lambda\phi^2_*}-\frac{2\lambda m^2_c\phi^2_*}{(m^2_c-\lambda\phi^2_*)^2}\right)^2}{\left(1-\frac{m^2_c}{m^2_c-\lambda\phi^2_*}\right)^2} 
                                                  +\frac{1}{6}\frac{\left(\frac{6m^2_c\lambda\phi_*}{(m^2_c-\lambda\phi^2_*)^2}+\frac{8\lambda^2 m^2_c\phi^3_*}{(m^2_c-\lambda\phi^2_*)^3}\right)}{\left(1-\frac{m^2_c}{m^2_c-\lambda\phi^2_*}\right)}\right)\right\} & \mbox{\small \textcolor{red}{\bf for \underline{Case ~II+Choice~II(v1\&v2)}}}\\ 
                                                                    \displaystyle -\frac{1}{{\cal Y} \phi^3_*}\left\{\frac{8}{3}+\frac{2\left(1+\xi(3\phi^2_*-\phi^2_V)-\frac{\phi^2_V}{\phi^2_*}\right)}{1+\xi(\phi^2_*-\phi^2_V)+\frac{\phi^2_V}{\phi^2_*}}\right.\\ \left.\displaystyle~~~~~~~~~~+\left(\frac{5}{3}\frac{\left(1+\xi(3\phi^2_*-\phi^2_V)-\frac{\phi^2_V}{\phi^2_*}\right)^2}{(1+\xi(\phi^2_*-\phi^2_V)+\frac{\phi^2_V}{\phi^2_*})^2} 
                                                                    -\frac{1}{6}\frac{\left(6\xi\phi^2_*+2\frac{\phi^2_V}{\phi^2_*}\right)}{\left(1+\xi(\phi^2_*-\phi^2_V)+\frac{\phi^2_V}{\phi^2_*}\right)}\right)\right\} & \mbox{\small \textcolor{red}{\bf for \underline{Case ~II+Choice~III}}}.
                        \end{array}
              \right.
              \end{array}\ee
   Additionally the constants ${\cal G}_{1}(\phi_*)$ and ${\cal G}_{2}(\phi_*)$, as appearing  in the expression for $f^{loc}_{NL}$ are defined as:      
 \be\tiny\begin{array}{lll}\label{rinfla1cxccdfx}
    \displaystyle {\cal G}_{1}(\phi_*)\displaystyle=\left\{\begin{array}{ll}
                       \displaystyle \left(1+\frac{6M^2_p}{81\phi^2_*}\right)^{-2} \left(1+\frac{12M^2_p}{81\phi^2_*}+\frac{36M^4_p}{6561\phi^4_*}\right)
                                                 ~~~~~ &
    \mbox{\small \textcolor{red}{\bf for \underline{Case I}}}  
   \\ 
            \displaystyle \left(1+\frac{6M^2_p}{\phi^2_*}\right)^{-2} \left(1+\frac{12M^2_p}{\phi^2_*}+\frac{36M^4_p}{\phi^4_*}\right)& \mbox{\small \textcolor{red}{\bf for \underline{Case II}}}  
            \\ 
                     \displaystyle
                     \left(1+\frac{6M^2_p}{\phi^2_*\left(1-\frac{\phi^4_V}{\phi^4_*}\right)^2}\right)^{-2} \left(1+\frac{12M^2_p}{\phi^2_*\left(1-\frac{\phi^4_V}{\phi^4_*}\right)^2}+\frac{36M^4_p}{\phi^4_*\left(1-\frac{\phi^4_V}{\phi^4_*}\right)^4}\right) & \mbox{\small \textcolor{red}{\bf for \underline{Case ~II+Choice~I(v1\&v2)}}}\\
                                       \displaystyle
                                        \left(1+\frac{6M^2_p}{\phi^2_*\left(\frac{m^2_c}{m^2_c-\lambda\phi^2_*}\right)^2}\right)^{-2} \left(1+\frac{12M^2_p}{\phi^2_*\left(1-\frac{m^2_c}{m^2_c-\lambda\phi^2_*}\right)^2}+\frac{36M^4_p}{\phi^4_*\left(1-\frac{m^2_c}{m^2_c-\lambda\phi^2_*}\right)^4}\right)
                                                                  & \mbox{\small\textcolor{red}{\bf for \underline{Case ~II+Choice~II(v1\&v2)}}}\\
                                                         \displaystyle
                                                         \left(1+\frac{6M^2_p}{\phi^2_*\left(1+\xi(\phi^2_*-\phi^2_V)+\frac{\phi^2_V}{\phi^2_*}\right)^2}\right)^{-2} \left(1+\frac{12M^2_p}{\phi^2_*\left(1+\xi(\phi^2_*-\phi^2_V)+\frac{\phi^2_V}{\phi^2_*}\right)^2}+\frac{36M^4_p}{\phi^4_*\left(1+\xi(\phi^2_*-\phi^2_V)+\frac{\phi^2_V}{\phi^2_*}\right)^4}\right) & \mbox{\small \textcolor{red}{\bf for \underline{Case ~II+Choice~III}}}.
             \end{array}
   \right.
   \end{array}\ee 
   and                           
  \be\tiny\begin{array}{lll}\label{rinfla1cxccas}
   \displaystyle {\cal G}_{2}(\phi_*)\displaystyle=\left\{\begin{array}{ll}
                      \displaystyle \left(1+\frac{6M^2_p}{81\phi^2_*}\right)^{-1}\frac{6M^2_p}{81\phi^3_*}~~~~~ &
   \mbox{\small \textcolor{red}{\bf for \underline{Case I}}}  
  \\ 
           \displaystyle \left(1+\frac{6M^2_p}{\phi^2_*}\right)^{-1}\frac{6M^2_p}{\phi^3_*}~~~~~& \mbox{\small \textcolor{red}{\bf for \underline{Case II}}}  
           \\ 
                    \displaystyle
                   \left(1+\frac{6M^2_p}{\phi^2_*\left(1-\frac{\phi^4_V}{\phi^4_*}\right)^2}\right)^{-1}\frac{6M^2_p\left(1+3\frac{\phi^4_V}{\phi^4_*}\right)}{\phi^3_*\left(1-\frac{\phi^4_V}{\phi^4_*}\right)^3}~~~~~ & \mbox{\small \textcolor{red}{\bf for \underline{Case ~II+Choice~I(v1\&v2)}}}\\
                                      \displaystyle
                                       \left(1+\frac{6M^2_p}{\phi^2_*\left(\frac{m^2_c}{m^2_c-\lambda\phi^2_*}\right)^2}\right)^{-1}\frac{6M^2_p\left(1-\frac{m^2_c}{m^2_c-\lambda\phi^2_*}-\frac{2\lambda m^2_c\phi^2_*}{(m^2_c-\lambda\phi^2_*)^2}\right)}{\phi^3_*\left(\frac{m^2_c}{m^2_c-\lambda\phi^2_*}\right)^3}~~~~~ & \mbox{\small \textcolor{red}{\bf for \underline{Case ~II+Choice~II(v1\&v2)}}}\\
                                                        \displaystyle
                                                         \left(1+\frac{6M^2_p}{\phi^2_*\left(1+\xi(\phi^2_*-\phi^2_V)+\frac{\phi^2_V}{\phi^2_*}\right)^2}\right)^{-1}\frac{6M^2_p\left(1+\xi(3\phi^2_*-\phi^2_V)-\frac{\phi^2_V}{\phi^2_*}\right)}{\phi^3_*\left(1+\xi(\phi^2_*-\phi^2_V)+\frac{\phi^2_V}{\phi^2_*}\right)^3}~~~~~~~~~~ & \mbox{\small \textcolor{red}{\bf for \underline{Case ~II+Choice~III}}}.
            \end{array}
  \right.
  \end{array}\ee
 \subsection{Momentum dependent functions in four point function}   
 Momentum dependent functions $\hat{G}^{S}({\bf k}_1,{\bf k}_2,{\bf k}_3,{\bf k}_4)$, $\hat{W}^{S}({\bf k}_1,{\bf k}_2,{\bf k}_3,{\bf k}_4)$ and $\hat{R}^{S}({\bf k}_1,{\bf k}_2,{\bf k}_3,{\bf k}_4)$ as appearing in four point function are defined as:
  \bea \hat{G}^{S}({\bf k}_1,{\bf k}_2,{\bf k}_3,{\bf k}_4)&=&\frac{S(\tilde{\bf k},{\bf k}_1,{\bf k}_2)S(\tilde{\bf k},{\bf k}_3,{\bf k}_4)}{|{\bf k}_1+{\bf k}_2|^3}\times\nonumber\\
  &&\left[\left\{{\bf k}_1.{\bf k}_3+\frac{\left[{\bf k}_1.({\bf k}_1+{\bf k}_2)\right]\left[{\bf k}_3.({\bf k}_3+{\bf k}_4)\right]}{|{\bf k}_1+{\bf k}_2|^2}\right\}\right.\nonumber\\
  &&\left.~~~~~~~~~~~~~~~~~~~~\times\left\{{\bf k}_2.{\bf k}_4+\frac{\left[{\bf k}_2.({\bf k}_1+{\bf k}_2)\right]\left[{\bf k}_4.({\bf k}_3+{\bf k}_4)\right]}{|{\bf k}_1+{\bf k}_2|^2}\right\}\right.\nonumber\\&& \left.+\left\{{\bf k}_1.{\bf k}_4+\frac{\left[{\bf k}_1.({\bf k}_1+{\bf k}_2)\right]\left[{\bf k}_4.({\bf k}_3+{\bf k}_4)\right]}{|{\bf k}_1+{\bf k}_2|^2}\right\}\right.\nonumber\\
  &&\left.~~~~~~~~~~~~~~~~~~~~\times\left\{{\bf k}_2.{\bf k}_3+\frac{\left[{\bf k}_2.({\bf k}_1+{\bf k}_2)\right]\left[{\bf k}_3.({\bf k}_3+{\bf k}_4)\right]}{|{\bf k}_1+{\bf k}_2|^2}\right\}\right.\nonumber\\&& \left.-\left\{{\bf k}_1.{\bf k}_2+\frac{\left[{\bf k}_1.({\bf k}_1+{\bf k}_2)\right]\left[{\bf k}_2.({\bf k}_3+{\bf k}_4)\right]}{|{\bf k}_1+{\bf k}_2|^2}\right\}\right.\nonumber\\
  &&\left.~~~~~~~~~~~~~~~~~~~~\times\left\{{\bf k}_3.{\bf k}_4+\frac{\left[{\bf k}_3.({\bf k}_3-{\bf k}_4)\right]\left[{\bf k}_4.({\bf k}_3-{\bf k}_4)\right]}{|{\bf k}_1+{\bf k}_2|^2}\right\}\right]\nonumber\\
  && \eea
  with 
  \bea S(\tilde{\bf k},{\bf k}_1,{\bf k}_2)&=&\left[K-\frac{1}{K}\sum_{i>j}k_ik_j-\frac{1}{K^2}\prod^{3}_{i=1}k_i\right]_{\tilde{\bf k}=-({\bf k}_1+{\bf k}_2)},\eea
  and 
  \bea \hat{R}^{S}({\bf k}_1,{\bf k}_2,{\bf k}_3,{\bf k}_4)
  &=&\sum^{3}_{n=1}\frac{1}{\hat{K}^{n}}A_{n}({\bf k}_1,{\bf k}_2,{\bf k}_3,{\bf k}_4),\eea
  where 
  \be \hat{K}=k_1+k_2+k_3+k_4=\sum^{4}_{i=1}k_i=K+k_4,\ee
  and the momentum dependent functions $A_{1}({\bf k}_1,{\bf k}_2,{\bf k}_3,{\bf k}_4)$, $A_{2}({\bf k}_1,{\bf k}_2,{\bf k}_3,{\bf k}_4)$ and $A_{3}({\bf k}_1,{\bf k}_2,{\bf k}_3,{\bf k}_4)$ are defined as:
  \bea A_{1}({\bf k}_1,{\bf k}_2,{\bf k}_3,{\bf k}_4)&=&\left[\frac{({\bf k}_3.{\bf k}_4)(({\bf k}_1.{\bf k}_2)(k^2_1+k^2_2)+2k^2_1k^2_2)}{8|{\bf k}_1+{\bf k}_2|^2}+(1,2\leftrightarrow 3,4)\right]\nonumber\\
  && -\frac{(k^2_1k^2_4({\bf k}_2.{\bf k}_3)+k^2_1k^2_3({\bf k}_2.{\bf k}_4)+k^2_2k^2_4({\bf k}_1.{\bf k}_3)+k^2_1k^2_4({\bf k}_2.{\bf k}_3))}{2|{\bf k}_1+{\bf k}_2|^2}\nonumber\\
  &&-\frac{(({\bf k}_1.{\bf k}_2)(k^2_1+k^2_2)+2k^2_1k^2_2)(({\bf k}_3.{\bf k}_4)(k^2_3+k^2_4)+2k^2_3k^2_4)}{8|{\bf k}_1+{\bf k}_2|^4}.~~~~~\eea
  \bea A_{2}({\bf k}_1,{\bf k}_2,{\bf k}_3,{\bf k}_4)&=&-\frac{\left[k_3k_4(k_3+k_4)(({\bf k}_1.{\bf k}_2)(k^2_1+k^2_2)+2k^2_1k^2_2)(k_3k_4+{\bf k}_3.{\bf k}_4)+(3,4\leftrightarrow 1,2)
  \right]}{8|{\bf k}_{1}+{\bf k}_{2}|^4}\nonumber\\
  &&-\frac{1}{2|{\bf k}_1+{\bf k}_2|^2}\left[k^2_1k^2_4({\bf k}_2.{\bf k}_3)(k_2+k_3)+k^2_1k^2_3({\bf k}_2.{\bf k}_4)(k_2+k_4)\right.\nonumber\\&&\left.~~~~~~~~~+k^2_2k^2_4({\bf k}_1.{\bf k}_3)(k_1+k_3)+k^2_2k^2_3({\bf k}_1.{\bf k}_4)(k_1+k_4)\right]\nonumber\\
  &&~~~~+\left\{\frac{({\bf k}_1.{\bf k}_2)}{8|{\bf k}_1+{\bf k}_2|^2}\left[((k_1+k_2)(({\bf k}_3.{\bf k}_4)(k^2_3+k^2_4)+2k^2_3k^2_4)\right.\right.\nonumber\\&&\left.\left.~~~~~~~~~+k_3k_4(k_3+k_4)(k_3k_4+{\bf k}_3.{\bf k}_4))\right]+(1,2\leftrightarrow 3,4)\right\}\eea
  \bea A_{3}({\bf k}_1,{\bf k}_2,{\bf k}_3,{\bf k}_4)&=&-\frac{k_1k_2k_3k_4(k_1+k_2)(k_3+k_4)(k_1k_2+{\bf k}_1.{\bf k}_2)(k_3k_4+{\bf k}_3.{\bf k}_4)}{4|{\bf k}_1+{\bf k}_2|^4}\nonumber\\
  &&-\frac{k_1k_2k_3k_4(k_1k_4({\bf k}_2.{\bf k}_3)+k_1k_3({\bf k}_2.{\bf k}_4)+k_2k_4({\bf k}_1,{\bf k}_3)+k_2k_3({\bf k}_1,{\bf k}_4))}{|{\bf k}_1+{\bf k}_2|^2}\nonumber\\
  &&+\frac{\left[k_3k_4(k_3+k_4)(({\bf k}_1.{\bf k}_2)(k^2_1+k^2_2)+2k^2_1k^2_2)(k_3k_4+{\bf k}_3.{\bf k}_4)+(3,4\leftrightarrow 1,2)
  \right]}{2|{\bf k}_{1}+{\bf k}_{2}|^2}\nonumber\\
  &&+\frac{3k_1k_2k_3k_4(k_1k_2+{\bf k}_1.{\bf k}_2)(k_3k_4+{\bf k}_3.{\bf k}_4)}{4|{\bf k}_1+{\bf k}_2|^2}.\eea
  \bea 
  \hat{W}^{S}({\bf k}_1,{\bf k}_3,{\bf k}_2,{\bf k}_4)
  &=&-\frac{2|{\bf k}_1+{\bf k}_2|^3\hat{G}^{S}({\bf k}_1,{\bf k}_3,{\bf k}_2,{\bf k}_4)}{S(\tilde{\bf k},{\bf k}_1,{\bf k}_2)S(\tilde{\bf k},{\bf k}_3,{\bf k}_4)}
  \left[\left\{\frac{k_1k_2(k_1+k_2)^2((k_1+k_2)^2-k^2_3-k^2_4-k_3k_4)}{(\hat{K}-2(k_3+k_4))^2\hat{K}^2((k_1+k_2)^2-|{\bf k}_1+{\bf k}_2|^2)}\right.\right.\nonumber \\ &&\left.\left. \times\left(-\frac{k_1+k_2}{2k_1k_2}-\frac{k_1+k_2}{k^2_3+k^2_4+4k_3k_4-(k_1+k_2)^2}+\frac{k_1+k_2}{|{\bf k}_1+{\bf k}_2|^2-(k_1+k_2)^2}\right.\right.\right.\nonumber \\ &&\left.\left.\left.+\frac{1}{\hat{K}-2(k_1+k_2)}-\frac{1}{\hat{K}}+\frac{3}{2(k_1+k_2)}\right)+(1,2\leftrightarrow 3,4)\right\}\right.\nonumber\\
  &&\left.-\frac{|{\bf k}_1+{\bf k}_2|^3(|{\bf k}_1+{\bf k}_2|^2-k^2_1-k^2_2-4k_1k_2)(|{\bf k}_1+{\bf k}_2|^2-k^2_3-k^2_4-4k_3k_4)}{2(|{\bf k}_1+{\bf k}_2|^2-k^2_1-k^2_2-2k_1k_2)(|{\bf k}_1+{\bf k}_2|^2-k^2_3-k^2_4-2k_3k_4)}\right]\nonumber\\
  &&\eea              


\begin{thebibliography}{}

\bibitem{Guth:1980zm}
\textsf{A.~H.~Guth},
\textcolor{red}{\textsf{The Inflationary Universe: A Possible Solution to the Horizon and Flatness Problems}},
\textcolor{violet}{{Phys.\ Rev.\ D {\bf 23} (1981) 347.}}


\bibitem{Baumann:2009ds}
\textsf{D.~Baumann},
\textcolor{red}{\textsf{TASI lectures on Inflation}},
\textcolor{violet}{{ arXiv:0907.5424 [hep-th].}}

\bibitem{Senatore:2016aui}
\textsf{L.~Senatore},
\textcolor{red}{\textsf{Lectures on Inflation}},
\textcolor{violet}{{1609.00716 [hep-th].}}

\bibitem{Liddle:1999mq}
\textsf{A.~R.~Liddle},
\textcolor{red}{\textsf{An Introduction to cosmological inflation}},
\textcolor{violet}{{astro-ph/9901124.}}

\bibitem{Langlois:2010xc}
\textsf{D.~Langlois},
\textcolor{red}{\textsf{Lectures on inflation and cosmological perturbations}},
\textcolor{violet}{{Lect.\ Notes Phys.\  {\bf 800} (2010) 1 [arXiv:1001.5259 [astro-ph.CO]].}}

\bibitem{Riotto:2002yw}
\textsf{A.~Riotto},
\textcolor{red}{\textsf{Inflation and the theory of cosmological perturbations}},
\textcolor{violet}{{hep-ph/0210162.}}

\bibitem{Lyth:1998xn}
\textsf{D.~H.~Lyth and A.~Riotto},
\textcolor{red}{\textsf{Particle physics models of inflation and the cosmological density perturbation}},
\textcolor{violet}{{Phys.\ Rept.\  {\bf 314} (1999) 1 [hep-ph/9807278].}}

\bibitem{Lyth:2007qh}
\textsf{D.~H.~Lyth},
\textcolor{red}{\textsf{Particle physics models of inflation}},
\textcolor{violet}{{Lect.\ Notes Phys.\  {\bf 738} (2008) 81 [hep-th/0702128].}}

\bibitem{Weinberg:2003sw}
\textsf{S.~Weinberg},
\textcolor{red}{\textsf{Adiabatic modes in cosmology}},
\textcolor{violet}{{Phys.\ Rev.\ D {\bf 67} (2003) 123504 [astro-ph/0302326].}}

\bibitem{Weinberg:2008hq}
\textsf{S.~Weinberg},
\textcolor{red}{\textsf{Effective Field Theory for Inflation}},
\textcolor{violet}{{Phys.\ Rev.\ D {\bf 77} (2008) 123541 [arXiv:0804.4291 [hep-th]].}}

\bibitem{Cheung:2007st}
\textsf{C.~Cheung, P.~Creminelli, A.~L.~Fitzpatrick, J.~Kaplan and L.~Senatore},
\textcolor{red}{\textsf{The Effective Field Theory of Inflation}},
\textcolor{violet}{{JHEP {\bf 0803} (2008) 014 [arXiv:0709.0293 [hep-th]].}}

\bibitem{Bardeen:1980kt}
\textsf{J.~M.~Bardeen},
\textcolor{red}{\textsf{Gauge Invariant Cosmological Perturbations}},
\textcolor{violet}{{Phys.\ Rev.\ D {\bf 22} (1980) 1882.}}

\bibitem{Choudhury:2011sq}
\textsf{S.~Choudhury and S.~Pal},
\textcolor{red}{\textsf{Brane inflation in background supergravity}},
\textcolor{violet}{{Phys.\ Rev.\ D {\bf 85} (2012) 043529 [arXiv:1102.4206 [hep-th]].}}

\bibitem{Choudhury:2011jt}
\textsf{S.~Choudhury and S.~Pal},
\textcolor{red}{\textsf{Fourth level MSSM inflation from new flat directions}},
\textcolor{violet}{{JCAP {\bf 1204} (2012) 018 [arXiv:1111.3441 [hep-ph]].}}

\bibitem{Choudhury:2012yh}
\textsf{S.~Choudhury and S.~Pal},
\textcolor{red}{\textsf{DBI Galileon inflation in background SUGRA}},
\textcolor{violet}{{Nucl.\ Phys.\ B {\bf 874} (2013) 85 [arXiv:1208.4433 [hep-th]].}}

\bibitem{Choudhury:2012ib}
\textsf{S.~Choudhury and S.~Pal},
\textcolor{red}{\textsf{Brane inflation: A field theory approach in background supergravity}},
\textcolor{violet}{{J.\ Phys.\ Conf.\ Ser.\  {\bf 405} (2012) 012009 [arXiv:1209.5883 [hep-th]].}}

\bibitem{Choudhury:2013zna}
\textsf{S.~Choudhury, T.~Chakraborty and S.~Pal},
\textcolor{red}{\textsf{Higgs inflation from new Kähler potential}},
\textcolor{violet}{{Nucl.\ Phys.\ B {\bf 880} (2014) 155 [arXiv:1305.0981 [hep-th]].}}

\bibitem{Choudhury:2013jya}
\textsf{S.~Choudhury, A.~Mazumdar and S.~Pal},
\textcolor{red}{\textsf{Low \& High scale MSSM inflation, gravitational waves and constraints from Planck}},
\textcolor{violet}{{JCAP {\bf 1307} (2013) 041 [arXiv:1305.6398 [hep-ph]].}}

\bibitem{Choudhury:2014sxa}
\textsf{S.~Choudhury, A.~Mazumdar and E.~Pukartas},
\textcolor{red}{\textsf{Constraining ${\cal N}=1$ supergravity inflationary framework with non-minimal K$\ddot{a}$hler operators}},
\textcolor{violet}{{JHEP {\bf 1404} (2014) 077 [arXiv:1402.1227 [hep-th]].}}

\bibitem{Choudhury:2015pqa}
\textsf{S.~Choudhury},
\textcolor{red}{\textsf{Reconstructing inflationary paradigm within Effective Field Theory framework}},
\textcolor{violet}{{Phys.\ Dark Univ.\  {\bf 11} (2016) 16 [arXiv:1508.00269 [astro-ph.CO]].}}

\bibitem{Choudhury:2014hja}
\textsf{S.~Choudhury, B.~K.~Pal, B.~Basu and P.~Bandyopadhyay},
\textcolor{red}{\textsf{Quantum Gravity Effect in Torsion Driven Inflation and CP violation}},
\textcolor{violet}{{JHEP {\bf 1510} (2015) 194 [arXiv:1409.6036 [hep-th]].}}

\bibitem{Choudhury:2016wlj}
\textsf{S.~Choudhury},
\textcolor{red}{\textsf{Field Theoretic Approaches To Early Universe}},
\textcolor{violet}{{arXiv:1603.08306 [hep-th].}}


\bibitem{Choudhury:2015hvr}
\textsf{S.~Choudhury and S.~Panda},
\textcolor{red}{\textsf{COSMOS-e’-GTachyon from string theory}},
\textcolor{violet}{{Eur.\ Phys.\ J.\ C {\bf 76} (2016) no.5,  278 [arXiv:1511.05734 [hep-th]].}}

\bibitem{Bhattacharjee:2014toa}
\textsf{A.~Bhattacharjee, A.~Deshamukhya and S.~Panda},
\textcolor{red}{\textsf{A note on low energy effective theory of chromo-natural inflation in the light of BICEP2 results}},
\textcolor{violet}{{Mod.\ Phys.\ Lett.\ A {\bf 30} (2015) no.11,  1550040 [arXiv:1406.5858 [astro-ph.CO]].}}

\bibitem{Deshamukhya:2009wc}
\textsf{A.~Deshamukhya and S.~Panda},
\textcolor{red}{\textsf{Warm tachyonic inflation in warped background}},
\textcolor{violet}{{Int.\ J.\ Mod.\ Phys.\ D {\bf 18} (2009) 2093 [arXiv:0901.0471 [hep-th]].}}

\bibitem{Kachru:2003sx}
\textsf{S.~Kachru, R.~Kallosh, A.~D.~Linde, J.~M.~Maldacena, L.~P.~McAllister and S.~P.~Trivedi},
\textcolor{red}{\textsf{Towards inflation in string theory}},
\textcolor{violet}{{JCAP {\bf 0310} (2003) 013 [hep-th/0308055].}}

\bibitem{Kachru:2003aw}
\textsf{S.~Kachru, R.~Kallosh, A.~D.~Linde and S.~P.~Trivedi},
\textcolor{red}{\textsf{De Sitter vacua in string theory}},
\textcolor{violet}{{Phys.\ Rev.\ D {\bf 68} (2003) 046005 [hep-th/0301240].}}

\bibitem{Iizuka:2004ct}
\textsf{N.~Iizuka and S.~P.~Trivedi},
\textcolor{red}{\textsf{An Inflationary model in string theory}},
\textcolor{violet}{{Phys.\ Rev.\ D {\bf 70} (2004) 043519 [hep-th/0403203].}}

\bibitem{Choudhury:2014kma}
\textsf{S.~Choudhury and A.~Mazumdar},
\textcolor{red}{\textsf{Reconstructing inflationary potential from BICEP2 and running of tensor modes}},
\textcolor{violet}{{arXiv:1403.5549 [hep-th].}}

\bibitem{Choudhury:2013iaa}
  \textsf{S.~Choudhury and A.~Mazumdar},
  \textcolor{red}{\textsf{An accurate bound on tensor-to-scalar ratio and the scale of inflation}},
  \textcolor{violet}{{Nucl.\ Phys.\ B {\bf 882} (2014) 386
  [arXiv:1306.4496 [hep-ph]]}}.
  
  \bibitem{Choudhury:2013woa}
     \textsf{S.~Choudhury and A.~Mazumdar},
    \textcolor{red}{\textsf{Primordial blackholes and gravitational waves for an inflection-point model of inflation}},
    \textcolor{violet}{{Phys.\ Lett.\ B {\bf 733} (2014) 270
    [arXiv:1307.5119 [astro-ph.CO]]}}.
    
    \bibitem{Choudhury:2014sua}
      \textsf{S.~Choudhury},
      \textcolor{red}{\textsf{Can Effective Field Theory of inflation generate large tensor-to-scalar ratio within Randall–Sundrum single braneworld?}},
       \textcolor{violet}{{Nucl.\ Phys.\ B {\bf 894} (2015) 29
      [arXiv:1406.7618 [hep-th]]}}.

\bibitem{Choudhury:2014wsa}
\textsf{S.~Choudhury and A.~Mazumdar},
\textcolor{red}{\textsf{Sub-Planckian inflation \& large tensor to scalar ratio with $r\geq 0.1$}},
\textcolor{violet}{{arXiv:1404.3398 [hep-th].}}

\bibitem{BuenoSanchez:2006rze}
\textsf{J.~C.~Bueno Sanchez, K.~Dimopoulos and D.~H.~Lyth},
\textcolor{red}{\textsf{A-term inflation and the MSSM}},
\textcolor{violet}{{JCAP {\bf 0701} (2007) 015 [hep-ph/0608299].}}

\bibitem{Allahverdi:2006iq}
\textsf{R.~Allahverdi, K.~Enqvist, J.~Garcia-Bellido and A.~Mazumdar},
\textcolor{red}{\textsf{Gauge invariant MSSM inflaton}},
\textcolor{violet}{{Phys.\ Rev.\ Lett.\  {\bf 97} (2006) 191304 [hep-ph/0605035].}}

\bibitem{Ross:1995dq}
\textsf{G.~G.~Ross and S.~Sarkar},
\textcolor{red}{\textsf{Successful supersymmetric inflation}},
\textcolor{violet}{{Nucl.\ Phys.\ B {\bf 461} (1996) 597 [hep-ph/9506283].}}

\bibitem{Allahverdi:2006rt}
\textsf{R.~Allahverdi},
\textcolor{red}{\textsf{MSSM flat direction inflation}},
\textcolor{violet}{{eConf C {\bf 0605151} (2006) 0020 [hep-ph/0610180].}}

\bibitem{Enqvist:2003gh}
\textsf{K.~Enqvist and A.~Mazumdar},
\textcolor{red}{\textsf{Cosmological consequences of MSSM flat directions}},
\textcolor{violet}{{Phys.\ Rept.\  {\bf 380} (2003) 99 [hep-ph/0209244].}}

\bibitem{Allahverdi:2006we}
\textsf{R.~Allahverdi, K.~Enqvist, J.~Garcia-Bellido, A.~Jokinen and A.~Mazumdar},
\textcolor{red}{\textsf{MSSM flat direction inflation: Slow roll, stability, fine tunning and reheating}},
\textcolor{violet}{{JCAP {\bf 0706} (2007) 019 [hep-ph/0610134].}}

\bibitem{Coleman:1973jx}
\textsf{S.~R.~Coleman and E.~J.~Weinberg},
\textcolor{red}{\textsf{Radiative Corrections as the Origin of Spontaneous Symmetry Breaking}},
\textcolor{violet}{{Phys.\ Rev.\ D {\bf 7} (1973) 1888.}}


\bibitem{Barenboim:2013wra}
\textsf{G.~Barenboim, E.~J.~Chun and H.~M.~Lee},
\textcolor{red}{\textsf{Coleman-Weinberg Inflation in light of Planck}},
\textcolor{violet}{{Phys.\ Lett.\ B {\bf 730} (2014) 81 [arXiv:1309.1695 [hep-ph]].}}

\bibitem{Fodor:1994bs}
\textsf{Z.~Fodor and A.~Hebecker},
\textcolor{red}{\textsf{Finite temperature effective potential to order $g^4$, $\lambda^2$ and the electroweak phase transition}},
\textcolor{violet}{{Nucl.\ Phys.\ B {\bf 432} (1994) 127 [hep-ph/9403219].}}

\bibitem{Quiros:1999jp}
\textsf{M.~Quiros},
\textcolor{red}{\textsf{Finite temperature field theory and phase transitions}},
\textcolor{violet}{{hep-ph/9901312.}}

\bibitem{Ade:2015lrj}
\textsf{P.~A.~R.~Ade {\it et al.} [Planck Collaboration]},
\textcolor{red}{\textsf{Planck 2015 results. XX. Constraints on inflation}},
\textcolor{violet}{{Astron.\ Astrophys.\  {\bf 594} (2016) A20 [arXiv:1502.02114 [astro-ph.CO]].}}


\bibitem{Ade:2015ava}
\textsf{P.~A.~R.~Ade {\it et al.} [Planck Collaboration]},
\textcolor{red}{\textsf{Planck 2015 results. XVII. Constraints on primordial non-Gaussianity}},
\textcolor{violet}{{Astron.\ Astrophys.\  {\bf 594} (2016) A17 [arXiv:1502.01592 [astro-ph.CO]].}}

\bibitem{Ade:2015xua}
\textsf{P.~A.~R.~Ade {\it et al.} [Planck Collaboration]},
\textcolor{red}{\textsf{Planck 2015 results. XIII. Cosmological parameters}},
\textcolor{violet}{{Astron.\ Astrophys.\  {\bf 594} (2016) A13 [arXiv:1502.01589 [astro-ph.CO]].}}

\bibitem{Ade:2015tva}
\textsf{P.~A.~R.~Ade {\it et al.} [BICEP2 and Planck Collaborations]},
\textcolor{red}{\textsf{Joint Analysis of BICEP2/$Keck  Array$ and $Planck$ Data}},
\textcolor{violet}{{Phys.\ Rev.\ Lett.\  {\bf 114} (2015) 101301 [arXiv:1502.00612 [astro-ph.CO]].}}

\bibitem{Yamaguchi:2011kg}
\textsf{M.~Yamaguchi},
\textcolor{red}{\textsf{Supergravity based inflation models: a review}},
\textcolor{violet}{{Phys.\ Rev.\ Lett.\  {\bf 114} (2015) 101301 [arXiv:1502.00612 [astro-ph.CO]].}}

\bibitem{Stewart:1994ts}
\textsf{E.~D.~Stewart},
\textcolor{red}{\textsf{Inflation, supergravity and superstrings}},
\textcolor{violet}{{Phys.\ Rev.\ D {\bf 51} (1995) 6847 [hep-ph/9405389].}}

\bibitem{McAllister:2007bg}
\textsf{L.~McAllister and E.~Silverstein},
\textcolor{red}{\textsf{String Cosmology: A Review}},
\textcolor{violet}{{Gen.\ Rel.\ Grav.\  {\bf 40} (2008) 565 [arXiv:0710.2951 [hep-th]].}}

\bibitem{Baumann:2009ni}
\textsf{D.~Baumann and L.~McAllister},
\textcolor{red}{\textsf{Advances in Inflation in String Theory}},
\textcolor{violet}{{Ann.\ Rev.\ Nucl.\ Part.\ Sci.\  {\bf 59} (2009) 67 [arXiv:0901.0265 [hep-th]].}}


\bibitem{Nilles:1983ge}
\textsf{H.~P.~Nilles},
\textcolor{red}{\textsf{Supersymmetry, Supergravity and Particle Physics}},
\textcolor{violet}{{Phys.\ Rept.\  {\bf 110} (1984) 1.}}

\bibitem{Linde:1997sj}
\textsf{A.~D.~Linde and A.~Riotto},
\textcolor{red}{\textsf{Hybrid inflation in supergravity}},
\textcolor{violet}{{Phys.\ Rev.\ D {\bf 56} (1997) R1841 [hep-ph/9703209].}}

\bibitem{BasteroGil:2006cm}
\textsf{M.~Bastero-Gil, S.~F.~King and Q.~Shafi},
\textcolor{red}{\textsf{Supersymmetric Hybrid Inflation with Non-Minimal Kahler potential}},
\textcolor{violet}{{Phys.\ Lett.\ B {\bf 651} (2007) 345 [hep-ph/0604198].}}

\bibitem{Choudhury:2011rz}
\textsf{S.~Choudhury and S.~Pal},
\textcolor{red}{\textsf{Reheating and leptogenesis in a SUGRA inspired brane inflation}},
\textcolor{violet}{{Nucl.\ Phys.\ B {\bf 857} (2012) 85 [arXiv:1108.5676 [hep-ph]].}}

\bibitem{Alishahiha:2004eh}
\textsf{M.~Alishahiha, E.~Silverstein and D.~Tong},
\textcolor{red}{\textsf{DBI in the sky}},
\textcolor{violet}{{Phys.\ Rev.\ D {\bf 70} (2004) 123505 [hep-th/0404084].}}

\bibitem{Silverstein:2016ggb}
\textsf{E.~Silverstein},
\textcolor{red}{\textsf{TASI lectures on cosmological observables and string theory}},
\textcolor{violet}{{arXiv:1606.03640 [hep-th].}}

\bibitem{Flauger:2014ana}
\textsf{R.~Flauger, L.~McAllister, E.~Silverstein and A.~Westphal},
\textcolor{red}{\textsf{Drifting Oscillations in Axion Monodromy}},
\textcolor{violet}{{arXiv:1412.1814 [hep-th].}}

\bibitem{McAllister:2014mpa}
\textsf{L.~McAllister, E.~Silverstein, A.~Westphal and T.~Wrase},
\textcolor{red}{\textsf{The Powers of Monodromy}},
\textcolor{violet}{{JHEP {\bf 1409} (2014) 123 [arXiv:1405.3652 [hep-th]].}}

\bibitem{Silverstein:2013wua}
\textsf{E.~Silverstein},
\textcolor{red}{\textsf{Les Houches lectures on inflationary observables and string theory}},
\textcolor{violet}{{arXiv:1311.2312 [hep-th].}}

\bibitem{Silverstein:2008sg}
\textsf{E.~Silverstein and A.~Westphal},
\textcolor{red}{\textsf{Monodromy in the CMB: Gravity Waves and String Inflation}},
\textcolor{violet}{{Phys.\ Rev.\ D {\bf 78} (2008) 106003 [arXiv:0803.3085 [hep-th]].}}

\bibitem{Panda:2010uq}
\textsf{S.~Panda, Y.~Sumitomo and S.~P.~Trivedi},
\textcolor{red}{\textsf{Axions as Quintessence in String Theory}},
\textcolor{violet}{{Phys.\ Rev.\ D {\bf 83} (2011) 083506 [arXiv:1011.5877 [hep-th]].}}

\bibitem{Panda:2007ie}
\textsf{S.~Panda, M.~Sami and S.~Tsujikawa},
\textcolor{red}{\textsf{Prospects of inflation in delicate D-brane cosmology}},
\textcolor{violet}{{Phys.\ Rev.\ D {\bf 76} (2007) 103512 [arXiv:0707.2848 [hep-th]].}}

\bibitem{Mazumdar:2001mm}
\textsf{A.~Mazumdar, S.~Panda and A.~Perez-Lorenzana},
\textcolor{red}{\textsf{Assisted inflation via tachyon condensation}},
\textcolor{violet}{{Nucl.\ Phys.\ B {\bf 614} (2001) 101 [hep-ph/0107058].}}

\bibitem{Choudhury:2002xu}
\textsf{D.~Choudhury, D.~Ghoshal, D.~P.~Jatkar and S.~Panda},
\textcolor{red}{\textsf{On the cosmological relevance of the tachyon}},
\textcolor{violet}{{Phys.\ Lett.\ B {\bf 544} (2002) 231 [hep-th/0204204].}}

\bibitem{Choudhury:2003vr}
\textsf{D.~Choudhury, D.~Ghoshal, D.~P.~Jatkar and S.~Panda},
\textcolor{red}{\textsf{Hybrid inflation and brane - anti-brane system}},
\textcolor{violet}{{JCAP {\bf 0307} (2003) 009 [hep-th/0305104].}}

\bibitem{Choudhury:2015baa}
\textsf{S.~Choudhury and S.~Banerjee},
\textcolor{red}{\textsf{Hysteresis in the Sky}},
\textcolor{violet}{{Astropart.\ Phys.\  {\bf 80} (2016) 34 [arXiv:1506.02260 [hep-th]].}}

\bibitem{Choudhury:2015fzb}
\textsf{S.~Choudhury and S.~Banerjee},
\textcolor{red}{\textsf{Cosmic Hysteresis}},
\textcolor{violet}{{arXiv:1512.08360 [hep-th].}}

\bibitem{Choudhury:2016rtp}
\textsf{S.~Choudhury and S.~Banerjee},
\textcolor{red}{\textsf{Cosmological Hysteresis in Cyclic Universe from Membrane Paradigm}},
\textcolor{violet}{{arXiv:1603.02805 [hep-th].}}

\bibitem{Maharana:1997cz}
\textsf{J.~Maharana, S.~Mukherji and S.~Panda},
\textcolor{red}{\textsf{Notes on axion, inflation and graceful exit in stringy cosmology}},
\textcolor{violet}{{Mod.\ Phys.\ Lett.\ A {\bf 12} (1997) 447 [hep-th/9701115].}}

\bibitem{Headrick:2004hz}
\textsf{M.~Headrick, S.~Minwalla and T.~Takayanagi},
\textcolor{red}{\textsf{Closed string tachyon condensation: An Overview}},
\textcolor{violet}{{Class.\ Quant.\ Grav.\  {\bf 21} (2004) S1539 [hep-th/0405064].}}

\bibitem{Minwalla:2003hj}
\textsf{S.~Minwalla and T.~Takayanagi},
\textcolor{red}{\textsf{Evolution of D branes under closed string tachyon condensation}},
\textcolor{violet}{{JHEP {\bf 0309} (2003) 011 [hep-th/0307248].}}

\bibitem{David:2001vm}
\textsf{J.~R.~David, M.~Gutperle, M.~Headrick and S.~Minwalla},
\textcolor{red}{\textsf{Closed string tachyon condensation on twisted circles}},
\textcolor{violet}{{JHEP {\bf 0202} (2002) 041 [hep-th/0111212].}}

\bibitem{Gopakumar:2000rw}
\textsf{R.~Gopakumar, S.~Minwalla and A.~Strominger},
\textcolor{red}{\textsf{Symmetry restoration and tachyon condensation in open string theory}},
\textcolor{violet}{{JHEP {\bf 0104} (2001) 018 [hep-th/0007226].}}

\bibitem{Sen:2000kd}
\textsf{A.~Sen},
\textcolor{red}{\textsf{Fundamental strings in open string theory at the tachyonic vacuum}},
\textcolor{violet}{{J.\ Math.\ Phys.\  {\bf 42} (2001) 2844 [hep-th/0010240].}}

\bibitem{Rastelli:2000hv}
\textsf{L.~Rastelli, A.~Sen and B.~Zwiebach},
\textcolor{red}{\textsf{String field theory around the tachyon vacuum}},
\textcolor{violet}{{Adv.\ Theor.\ Math.\ Phys.\  {\bf 5} (2002) 353 [hep-th/0012251].}}

\bibitem{Sen:2002qa}
\textsf{A.~Sen},
\textcolor{red}{\textsf{Time and tachyon}},
\textcolor{violet}{{Int.\ J.\ Mod.\ Phys.\ A {\bf 18} (2003) 4869 [hep-th/0209122].}}

\bibitem{Sen:2002in}
\textsf{A.~Sen},
\textcolor{red}{\textsf{Tachyon matter}},
\textcolor{violet}{{JHEP {\bf 0207} (2002) 065 [hep-th/0203265].}}

\bibitem{Sen:2002an}
\textsf{A.~Sen},
\textcolor{red}{\textsf{Field theory of tachyon matter}},
\textcolor{violet}{{Mod.\ Phys.\ Lett.\ A {\bf 17} (2002) 1797 [hep-th/0204143].}}

\bibitem{Sen:1999xm}
\textsf{A.~Sen},
\textcolor{red}{\textsf{Universality of the tachyon potential}},
\textcolor{violet}{{JHEP {\bf 9912} (1999) 027 [hep-th/9911116].}}

\bibitem{Sen:2002nu}
\textsf{A.~Sen},
\textcolor{red}{\textsf{Rolling tachyon}},
\textcolor{violet}{{JHEP {\bf 0204} (2002) 048 [hep-th/0203211].}}

\bibitem{Mandal:2003tj}
\textsf{G.~Mandal and S.~R.~Wadia},
\textcolor{red}{\textsf{Rolling tachyon solution of two-dimensional string theory}},
\textcolor{violet}{{JHEP {\bf 0405} (2004) 038 [hep-th/0312192].}}


\bibitem{Starobinsky:1979ty}
\textsf{A.~A.~Starobinsky},
\textcolor{red}{\textsf{Spectrum of relict gravitational radiation and the early state of the universe}},
\textcolor{violet}{{JETP Lett.\  {\bf 30} (1979) 682 [Pisma Zh.\ Eksp.\ Teor.\ Fiz.\  {\bf 30} (1979) 719].}}

\bibitem{Sebastiani:2015kfa}
\textsf{L.~Sebastiani and R.~Myrzakulov},
\textcolor{red}{\textsf{F(R) gravity and inflation}},
\textcolor{violet}{{Int.\ J.\ Geom.\ Meth.\ Mod.\ Phys.\  {\bf 12} (2015) no.9,  1530003 [arXiv:1506.05330 [gr-qc]].}}

\bibitem{DeFelice:2010aj}
\textsf{A.~De Felice and S.~Tsujikawa},
\textcolor{red}{\textsf{f(R) theories}},
\textcolor{violet}{{Living Rev.\ Rel.\  {\bf 13} (2010) 3 [arXiv:1002.4928 [gr-qc]].}}

\bibitem{Sotiriou:2008rp}
\textsf{T.~P.~Sotiriou and V.~Faraoni},
\textcolor{red}{\textsf{f(R) Theories Of Gravity}},
\textcolor{violet}{{Rev.\ Mod.\ Phys.\  {\bf 82} (2010) 451 [arXiv:0805.1726 [gr-qc]].}}

\bibitem{Kanti:2015pda}
\textsf{P.~Kanti, R.~Gannouji and N.~Dadhich},
\textcolor{red}{\textsf{Gauss-Bonnet Inflation}},
\textcolor{violet}{{Phys.\ Rev.\ D {\bf 92} (2015) no.4,  041302 [arXiv:1503.01579 [hep-th]].}}

\bibitem{Nozari:2008ny}
\textsf{K.~Nozari and B.~Fazlpour},
\textcolor{red}{\textsf{Gauss-Bonnet Cosmology with Induced Gravity and Non-Minimally Coupled Scalar Field on the Brane}},
\textcolor{violet}{{JCAP {\bf 0806} (2008) 032 [arXiv:0805.1537 [hep-th]].}}

\bibitem{Bezrukov:2007ep}
\textsf{F.~L.~Bezrukov and M.~Shaposhnikov},
\textcolor{red}{\textsf{The Standard Model Higgs boson as the inflaton}},
\textcolor{violet}{{Phys.\ Lett.\ B {\bf 659} (2008) 703 [arXiv:0710.3755 [hep-th]].}}

\bibitem{Hertzberg:2010dc}
\textsf{M.~P.~Hertzberg},
\textcolor{red}{\textsf{On Inflation with Non-minimal Coupling}},
\textcolor{violet}{{JHEP {\bf 1011} (2010) 023 [arXiv:1002.2995 [hep-ph]].}}


\bibitem{Pallis:2010wt}
\textsf{C.~Pallis},
\textcolor{red}{\textsf{Non-Minimally Gravity-Coupled Inflationary Models}},
\textcolor{violet}{{Phys.\ Lett.\ B {\bf 692} (2010) 287 [arXiv:1002.4765 [astro-ph.CO]].}}

\bibitem{Budhi:2017yjd}
\textsf{R.~H.~S.~Budhi},
\textcolor{red}{\textsf{Inflation due to non-minimal coupling of $f(R)$ gravity to a scalar field}},
\textcolor{violet}{{arXiv:1701.03814 [hep-ph].}}

\bibitem{Modesto:2011kw}
  \textsf{L.~Modesto},
  \textcolor{red}{\textsf{Super-renormalizable Quantum Gravity}},
  \textcolor{violet}{{Phys.\ Rev.\ D {\bf 86} (2012) 044005
  [arXiv:1107.2403 [hep-th]]}}.

\bibitem{Biswas:2011ar}
\textsf{T.~Biswas, E.~Gerwick, T.~Koivisto and A.~Mazumdar},
\textcolor{red}{\textsf{Towards singularity and ghost free theories of gravity}},
\textcolor{violet}{{Phys.\ Rev.\ Lett.\  {\bf 108} (2012) 031101 [arXiv:1110.5249 [gr-qc]].}}

\bibitem{Biswas:2012ka}
\textsf{T.~Biswas, J.~Kapusta and A.~Reddy},
\textcolor{red}{\textsf{Thermodynamics of String Field Theory Motivated Nonlocal Models}},
\textcolor{violet}{{JHEP {\bf 1212} (2012) 008 [arXiv:1201.1580 [hep-th]].}}

\bibitem{Biswas:2012bp}
\textsf{T.~Biswas, A.~S.~Koshelev, A.~Mazumdar and S.~Y.~Vernov},
\textcolor{red}{\textsf{Stable bounce and inflation in non-local higher derivative cosmology}},
\textcolor{violet}{{JCAP {\bf 1208} (2012) 024 [arXiv:1206.6374 [astro-ph.CO]].}}

\bibitem{Biswas:2013cha}
\textsf{T.~Biswas, A.~Conroy, A.~S.~Koshelev and A.~Mazumdar},
\textcolor{red}{\textsf{Generalized ghost-free quadratic curvature gravity}},
\textcolor{violet}{{Class.\ Quant.\ Grav.\  {\bf 31} (2014) 015022 Erratum: [Class.\ Quant.\ Grav.\  {\bf 31} (2014) 159501] [arXiv:1308.2319 [hep-th]].}}

\bibitem{Talaganis:2014ida}
\textsf{S.~Talaganis, T.~Biswas and A.~Mazumdar},
\textcolor{red}{\textsf{Towards understanding the ultraviolet behavior of quantum loops in infinite-derivative theories of gravity}},
\textcolor{violet}{{Class.\ Quant.\ Grav.\  {\bf 32} (2015) no.21,  215017 [arXiv:1412.3467 [hep-th]].}}

\bibitem{Biswas:2014tua}
\textsf{T.~Biswas and S.~Talaganis},
\textcolor{red}{\textsf{String-Inspired Infinite-Derivative Theories of Gravity: A Brief Overview}},
\textcolor{violet}{{Mod.\ Phys.\ Lett.\ A {\bf 30} (2015) no.03n04,  1540009 [arXiv:1412.4256 [gr-qc]].}}

\bibitem{Modesto:2014lga}
  \textsf{L.~Modesto and L.~Rachwal},
  \textcolor{red}{\textsf{Super-renormalizable and finite gravitational theories}},
  \textcolor{violet}{{Nucl.\ Phys.\ B {\bf 889} (2014) 228
  [arXiv:1407.8036 [hep-th]]}}.
  
  \bibitem{Dona:2015tra}
    \textsf{ P.~Donà, S.~Giaccari, L.~Modesto, L.~Rachwal and Y.~Zhu},
   \textcolor{red}{\textsf{Scattering amplitudes in super-renormalizable gravity}},
    \textcolor{violet}{{JHEP {\bf 1508} (2015) 038
    [arXiv:1506.04589 [hep-th]]}}.
    
    \bibitem{Koshelev:2016xqb}
      \textsf{A.~S.~Koshelev, L.~Modesto, L.~Rachwal and A.~A.~Starobinsky},
     \textcolor{red}{\textsf{Occurrence of exact $R^2$ inflation in non-local UV-complete gravity}},
      \textcolor{violet}{{JHEP {\bf 1611} (2016) 067
      [arXiv:1604.03127 [hep-th]]}}.

\bibitem{Brans:1961sx}
\textsf{C.~Brans and R.~H.~Dicke},
\textcolor{red}{\textsf{Mach's principle and a relativistic theory of gravitation}},
\textcolor{violet}{{Phys.\ Rev.\  {\bf 124} (1961) 925.}}

\bibitem{Brans:1962zz}
\textsf{C.~H.~Brans},
\textcolor{red}{\textsf{Mach's Principle and a Relativistic Theory of Gravitation. II}},
\textcolor{violet}{{Phys.\ Rev.\  {\bf 125} (1962) 2194.}}

\bibitem{Chen:2006nt}
\textsf{X.~Chen, M.~x.~Huang, S.~Kachru and G.~Shiu},
\textcolor{red}{\textsf{Observational signatures and non-Gaussianities of general single field inflation}},
\textcolor{violet}{{JCAP {\bf 0701} (2007) 002 [hep-th/0605045].}}

\bibitem{Khoury:2010gb}
\textsf{J.~Khoury, J.~L.~Lehners and B.~Ovrut},
\textcolor{red}{\textsf{Supersymmetric P(X,$\phi$) and the Ghost Condensate}},
\textcolor{violet}{{Phys.\ Rev.\ D {\bf 83} (2011) 125031 [arXiv:1012.3748 [hep-th]].}}

\bibitem{Berkin:1990ju}
\textsf{A.~L.~Berkin, K.~i.~Maeda and J.~Yokoyama},
\textcolor{red}{\textsf{Soft Inflation}},
\textcolor{violet}{{Phys.\ Rev.\ Lett.\  {\bf 65} (1990) 141.}}

\bibitem{Berkin:1991nm}
\textsf{A.~L.~Berkin and K.~I.~Maeda},
\textcolor{red}{\textsf{Inflation in generalized Einstein theories}},
\textcolor{violet}{{Phys.\ Rev.\ D {\bf 44} (1991) 1691.}}

\bibitem{Bars:1992sr}
\textsf{I.~Bars and K.~Sfetsos},
\textcolor{red}{\textsf{Conformally exact metric and dilaton in string theory on curved space-time}},
\textcolor{violet}{{Phys.\ Rev.\ D {\bf 46} (1992) 4510 [hep-th/9206006].}}

\bibitem{Callan:1985ia}
\textsf{C.~G.~Callan, Jr., E.~J.~Martinec, M.~J.~Perry and D.~Friedan},
\textcolor{red}{\textsf{Strings in Background Fields}},
\textcolor{violet}{{Nucl.\ Phys.\ B {\bf 262} (1985) 593.}}

\bibitem{Bars:1990rb}
\textsf{I.~Bars and D.~Nemeschansky},
\textcolor{red}{\textsf{String Propagation in Backgrounds With Curved Space-time}},
\textcolor{violet}{{Nucl.\ Phys.\ B {\bf 348} (1991) 89.}}

\bibitem{Choudhury:2013yg}
\textsf{S.~Choudhury and S.~Sengupta},
\textcolor{red}{\textsf{Features of warped geometry in presence of Gauss-Bonnet coupling}},
\textcolor{violet}{{JHEP {\bf 1302} (2013) 136 [arXiv:1301.0918 [hep-th]].}}

\bibitem{Choudhury:2013aqa}
\textsf{S.~Choudhury and S.~SenGupta},
\textcolor{red}{\textsf{A step toward exploring the features of Gravidilaton sector in Randall–Sundrum scenario via lightest Kaluza–Klein graviton mass}},
\textcolor{violet}{{Eur.\ Phys.\ J.\ C {\bf 74} (2014) no.11,  3159 [arXiv:1311.0730 [hep-ph]].}}

\bibitem{Choudhury:2014hna}
\textsf{S.~Choudhury, J.~Mitra and S.~SenGupta},
\textcolor{red}{\textsf{ Modulus stabilization in higher curvature dilaton gravity}},
\textcolor{violet}{{JHEP {\bf 1408} (2014) 004 [arXiv:1405.6826 [hep-th]].}}

\bibitem{Choudhury:2015wfa}
\textsf{S.~Choudhury, J.~Mitra and S.~SenGupta},
\textcolor{red}{\textsf{Fermion localization and flavour hierarchy in higher curvature spacetime}},
\textcolor{violet}{{arXiv:1503.07287 [hep-th].}}

\bibitem{Derendinger:1985kk}
\textsf{J.~P.~Derendinger, L.~E.~Ibanez and H.~P.~Nilles},
\textcolor{red}{\textsf{On the Low-Energy d = 4, N=1 Supergravity Theory Extracted from the d = 10, N=1 Superstring}},
\textcolor{violet}{{Phys.\ Lett.\  {\bf 155B} (1985) 65.}}

\bibitem{Ellis:1986zt}
\textsf{J.~R.~Ellis, D.~V.~Nanopoulos and M.~Quiros},
\textcolor{red}{\textsf{On the Axion, Dilaton, Polonyi, Gravitino and Shadow Matter Problems in Supergravity and Superstring Models}},
\textcolor{violet}{{Phys.\ Lett.\ B {\bf 174} (1986) 176.}}

\bibitem{Pilch:2000ue}
\textsf{K.~Pilch and N.~P.~Warner},
\textcolor{red}{\textsf{N=2 supersymmetric RG flows and the IIB dilaton}},
\textcolor{violet}{{Nucl.\ Phys.\ B {\bf 594} (2001) 209 [hep-th/0004063].}}

\bibitem{Kallosh:2013yoa}
 \textsf{ R.~Kallosh, A.~Linde and D.~Roest},
  \textcolor{red}{\textsf{Superconformal Inflationary $\alpha$-Attractors}},
  \textcolor{violet}{{JHEP {\bf 1311} (2013) 198
  doi:10.1007/JHEP11(2013)198
  [arXiv:1311.0472 [hep-th]]}}.
  
  \bibitem{Kallosh:2014rga}
    \textsf{R.~Kallosh, A.~Linde and D.~Roest},
    \textcolor{red}{\textsf{Large field inflation and double $\alpha$-attractors}},
     \textcolor{violet}{{JHEP {\bf 1408} (2014) 052
    [arXiv:1405.3646 [hep-th]]}}.
    
    \bibitem{Galante:2014ifa}
      \textsf{M.~Galante, R.~Kallosh, A.~Linde and D.~Roest},
      \textcolor{red}{\textsf{Unity of Cosmological Inflation Attractors}},
      \textcolor{violet}{{Phys.\ Rev.\ Lett.\  {\bf 114} (2015) no.14,  141302
      [arXiv:1412.3797 [hep-th]]}}.
      
      \bibitem{Kallosh:2015lwa}
        \textsf{R.~Kallosh and A.~Linde},
        \textcolor{red}{\textsf{Planck, LHC, and $\alpha$-attractors}},
        \textcolor{violet}{{Phys.\ Rev.\ D {\bf 91} (2015) 083528
        [arXiv:1502.07733 [astro-ph.CO]]}}.
        
        \bibitem{Linde:2015uga}
          \textsf{A.~Linde},
          \textcolor{red}{\textsf{Single-field $\alpha$-attractors}},
           \textcolor{violet}{{JCAP {\bf 1505} (2015) 003
          [arXiv:1504.00663 [hep-th]]}}.
          
          \bibitem{Carrasco:2015uma}
            \textsf{J.~J.~M.~Carrasco, R.~Kallosh, A.~Linde and D.~Roest},
            \textcolor{red}{\textsf{Hyperbolic geometry of cosmological attractors}},
            \textcolor{violet}{{Phys.\ Rev.\ D {\bf 92} (2015) no.4,  041301
            [arXiv:1504.05557 [hep-th]]}}.
            
            \bibitem{Carrasco:2015rva}
              \textsf{J.~J.~M.~Carrasco, R.~Kallosh and A.~Linde},
              \textcolor{red}{\textsf{Cosmological Attractors and Initial Conditions for Inflation}},
              \textcolor{violet}{{Phys.\ Rev.\ D {\bf 92} (2015) no.6,  063519
              [arXiv:1506.00936 [hep-th]]}}.
              
              \bibitem{Carrasco:2015pla}
                \textsf{J.~J.~M.~Carrasco, R.~Kallosh and A.~Linde},
               \textcolor{red}{\textsf{$\alpha $-Attractors: Planck, LHC and Dark Energy}},
                \textcolor{violet}{{JHEP {\bf 1510} (2015) 147
                [arXiv:1506.01708 [hep-th]]}}.
                
                \bibitem{Kallosh:2016gqp}
                  \textsf{R.~Kallosh and A.~Linde},
                  \textcolor{red}{\textsf{Cosmological Attractors and Asymptotic Freedom of the Inflaton Field}},
                  \textcolor{violet}{{JCAP {\bf 1606} (2016) no.06,  047
                  [arXiv:1604.00444 [hep-th]]}}.
                  
                  \bibitem{Kallosh:2016sej}
                    \textsf{R.~Kallosh, A.~Linde, D.~Roest and T.~Wrase},
                    \textcolor{red}{\textsf{Sneutrino inflation with $\alpha$-attractors}},
                    \textcolor{violet}{{JCAP {\bf 1611} (2016) 046
                    [arXiv:1607.08854 [hep-th]]}}.
                    
  \bibitem{Ferrara:2016fwe}
    \textsf{S.~Ferrara and R.~Kallosh},
    \textcolor{red}{\textsf{Seven-disk manifold, $\alpha$-attractors, and $B$ modes}},
    \textcolor{violet}{{Phys.\ Rev.\ D {\bf 94} (2016) no.12,  126015
    [arXiv:1610.04163 [hep-th]]}}.                  

\bibitem{Felder:2001kt}
\textsf{G.~N.~Felder, L.~Kofman and A.~D.~Linde},
\textcolor{red}{\textsf{Tachyonic instability and dynamics of spontaneous symmetry breaking}},
\textcolor{violet}{{Phys.\ Rev.\ D {\bf 64} (2001) 123517 [hep-th/0106179].}}

\bibitem{Felder:2002jk}
\textsf{G.~N.~Felder, A.~V.~Frolov, L.~Kofman and A.~D.~Linde},
\textcolor{red}{\textsf{Cosmology with negative potentials}},
\textcolor{violet}{{Phys.\ Rev.\ D {\bf 66} (2002) 023507 [hep-th/0202017].}}

\bibitem{Gumrukcuoglu:2016jbh}
\textsf{A.~E.~Gümrükçüoğlu, S.~Mukohyama and T.~P.~Sotiriou},
\textcolor{red}{\textsf{Low energy ghosts and the Jeans’ instability}},
\textcolor{violet}{{Phys.\ Rev.\ D {\bf 94} (2016) no.6,  064001 [arXiv:1606.00618 [hep-th]].}}

\bibitem{Saitou:2011hv}
\textsf{R.~Saitou and S.~Nojiri},
\textcolor{red}{\textsf{The unification of inflation and late-time acceleration in the frame of $k$-essence}},
\textcolor{violet}{{Eur.\ Phys.\ J.\ C {\bf 71} (2011) 1712 [arXiv:1104.0558 [hep-th]].}}

\bibitem{Tsujikawa:2010zza}
\textsf{S.~Tsujikawa},
\textcolor{red}{\textsf{Modified gravity models of dark energy}},
\textcolor{violet}{{Lect.\ Notes Phys.\  {\bf 800} (2010) 99 [arXiv:1101.0191 [gr-qc]].}}

\bibitem{Sen:1999mg}
\textsf{A.~Sen},
\textcolor{red}{\textsf{NonBPS states and Branes in string theory}},
\textcolor{violet}{{hep-th/9904207.}}

\bibitem{Frau:1999qs}
\textsf{M.~Frau, L.~Gallot, A.~Lerda and P.~Strigazzi},
\textcolor{red}{\textsf{Stable nonBPS D-branes in type I string theory}},
\textcolor{violet}{{Nucl.\ Phys.\ B {\bf 564} (2000) 60 [hep-th/9903123].}}

\bibitem{Eyras:2000my}
\textsf{E.~Eyras and S.~Panda},
\textcolor{red}{\textsf{NonBPS branes in a type I orbifold}},
\textcolor{violet}{{JHEP {\bf 0105} (2001) 056 [hep-th/0009224].}}

\bibitem{Bergshoeff:2000dq}
\textsf{E.~A.~Bergshoeff, M.~de Roo, T.~C.~de Wit, E.~Eyras and S.~Panda},
\textcolor{red}{\textsf{T duality and actions for nonBPS D-branes}},
\textcolor{violet}{{JHEP {\bf 0005} (2000) 009 [hep-th/0003221].}}

\bibitem{Eyras:2000ig}
\textsf{E.~Eyras and S.~Panda},
\textcolor{red}{\textsf{The Space-time life of a nonBPS D particle}},
\textcolor{violet}{{Nucl.\ Phys.\ B {\bf 584} (2000) 251 [hep-th/0003033].}}

\bibitem{Brax:2000cf}
\textsf{P.~Brax, G.~Mandal and Y.~Oz},
\textcolor{red}{\textsf{Supergravity description of nonBPS branes}},
\textcolor{violet}{{Phys.\ Rev.\ D {\bf 63} (2001) 064008 [hep-th/0005242].}}

\bibitem{Sen:2004nf}
\textsf{A.~Sen},
\textcolor{red}{\textsf{Tachyon dynamics in open string theory}},
\textcolor{violet}{{Int.\ J.\ Mod.\ Phys.\ A {\bf 20} (2005) 5513 [hep-th/0410103].}}

\bibitem{Sugiyama:2012tj}
\textsf{N.~S.~Sugiyama, E.~Komatsu and T.~Futamase},
\textcolor{red}{\textsf{$\delta$N formalism}},
\textcolor{violet}{{Phys.\ Rev.\ D {\bf 87} (2013) no.2,  023530 [arXiv:1208.1073 [gr-qc]].}}

\bibitem{Choudhury:2014uxa}
\textsf{S.~Choudhury},
\textcolor{red}{\textsf{Constraining ${\cal N} = 1$ supergravity inflation with non-minimal Kaehler operators using $\delta$N formalism}},
\textcolor{violet}{{JHEP {\bf 1404} (2014) 105 [arXiv:1402.1251 [hep-th]].}}

\bibitem{Lee:2005bb}
\textsf{H.~C.~Lee, M.~Sasaki, E.~D.~Stewart, T.~Tanaka and S.~Yokoyama},
\textcolor{red}{\textsf{A New $\delta$N formalism for multi-component inflation}},
\textcolor{violet}{{JCAP {\bf 0510} (2005) 004 [astro-ph/0506262].}}

\bibitem{Domenech:2016zxn}
\textsf{G.~Domenech, J.~O.~Gong and M.~Sasaki},
\textcolor{red}{\textsf{Consistency relation and inflaton field redefinition in the $\delta$N formalism}},
\textcolor{violet}{{arXiv:1606.03343 [astro-ph.CO].}}

\bibitem{Chen:2013eea}
\textsf{X.~Chen, H.~Firouzjahi, E.~Komatsu, M.~H.~Namjoo and M.~Sasaki},
\textcolor{red}{\textsf{In-in and $\delta$N calculations of the bispectrum from non-attractor single-field inflation}},
\textcolor{violet}{{JCAP {\bf 1312} (2013) 039 [arXiv:1308.5341 [astro-ph.CO]].}}

\bibitem{Choudhury:2015jaa}
\textsf{S.~Choudhury},
\textcolor{red}{\textsf{Constraining brane inflationary magnetic field from cosmoparticle physics after Planck}},
\textcolor{violet}{{JHEP {\bf 1510} (2015) 095 [arXiv:1504.08206 [astro-ph.CO]].}}

\bibitem{Choudhury:2014hua}
\textsf{S.~Choudhury},
\textcolor{red}{\textsf{Inflamagnetogenesis redux: Unzipping sub-Planckian inflation via various cosmoparticle probes}},
\textcolor{violet}{{Phys.\ Lett.\ B {\bf 735} (2014) 138 [arXiv:1403.0676 [hep-th]].}}

\bibitem{Subramanian:2015lua}
\textsf{K.~Subramanian},
\textcolor{red}{\textsf{The origin, evolution and signatures of primordial magnetic fields}},
\textcolor{violet}{{Rept.\ Prog.\ Phys.\  {\bf 79} (2016) no.7,  076901 [arXiv:1504.02311 [astro-ph.CO]].}}

\bibitem{Choudhury:2015zlc}
\textsf{S.~Choudhury, M.~Sen and S.~Sadhukhan},
\textcolor{red}{\textsf{Can Dark Matter be an artifact of extended theories of gravity?}},
\textcolor{violet}{{Eur.\ Phys.\ J.\ C {\bf 76} (2016) no.9,  494 [arXiv:1512.08176 [hep-ph]].}}

\bibitem{Choudhury:2016tzz}
\textsf{S.~Choudhury, M.~Sen and S.~Sadhukhan},
\textcolor{red}{\textsf{From Extended theories of Gravity to Dark Matter}},
\textcolor{violet}{{Acta Phys.\ Polon.\ Supp.\  {\bf 9} (2016) 789 [arXiv:1605.04043 [hep-th]].}}

\bibitem{Peterson:2010np}
\textsf{C.~M.~Peterson and M.~Tegmark},
\textcolor{red}{\textsf{Testing Two-Field Inflation}},
\textcolor{violet}{{Phys.\ Rev.\ D {\bf 83} (2011) 023522 [arXiv:1005.4056 [astro-ph.CO]].}}

\bibitem{Wands:2002bn}
\textsf{D.~Wands, N.~Bartolo, S.~Matarrese and A.~Riotto},
\textcolor{red}{\textsf{An Observational test of two-field inflation}},
\textcolor{violet}{{Phys.\ Rev.\ D {\bf 66} (2002) 043520 [astro-ph/0205253].}}

\bibitem{GarciaBellido:1995qq}
\textsf{J.~Garcia-Bellido and D.~Wands},
\textcolor{red}{\textsf{Metric perturbations in two field inflation}},
\textcolor{violet}{{Phys.\ Rev.\ D {\bf 53} (1996) 5437 [astro-ph/9511029].}}

\bibitem{Kaiser:2013sna}
\textsf{D.~I.~Kaiser and E.~I.~Sfakianakis},
\textcolor{red}{\textsf{Multifield Inflation after Planck: The Case for Nonminimal Couplings}},
\textcolor{violet}{{Phys.\ Rev.\ Lett.\  {\bf 112} (2014) no.1,  011302 [arXiv:1304.0363 [astro-ph.CO]].}}

\bibitem{Wands:2007bd}
\textsf{D.~Wands},
\textcolor{red}{\textsf{Multiple field inflation}},
\textcolor{violet}{{Lect.\ Notes Phys.\  {\bf 738} (2008) 275 [astro-ph/0702187 [ASTRO-PH]].}}

\bibitem{Maldacena:2002vr}
\textsf{J.~M.~Maldacena},
\textcolor{red}{\textsf{Non-Gaussian features of primordial fluctuations in single field inflationary models}},
\textcolor{violet}{{JHEP {\bf 0305} (2003) 013 [astro-ph/0210603].}}

\bibitem{Arkani-Hamed:2015bza}
\textsf{N.~Arkani-Hamed and J.~Maldacena},
\textcolor{red}{\textsf{Cosmological Collider Physics}},
\textcolor{violet}{{arXiv:1503.08043 [hep-th].}}

\bibitem{Choudhury:2015yna}
\textsf{S.~Choudhury and S.~Pal},
\textcolor{red}{\textsf{Primordial non-Gaussian features from DBI Galileon inflation}},
\textcolor{violet}{{Eur.\ Phys.\ J.\ C {\bf 75}, no. 6, 241 (2015) [arXiv:1210.4478 [hep-th]].}}

\bibitem{Chen:2010xka}
\textsf{X.~Chen},
\textcolor{red}{\textsf{Primordial Non-Gaussianities from Inflation Models}},
\textcolor{violet}{{Adv.\ Astron.\  {\bf 2010} (2010) 638979 [arXiv:1002.1416 [astro-ph.CO]].}}

\bibitem{Bartolo:2004if}
\textsf{N.~Bartolo, E.~Komatsu, S.~Matarrese and A.~Riotto},
\textcolor{red}{\textsf{Non-Gaussianity from inflation: Theory and observations}},
\textcolor{violet}{{Phys.\ Rept.\  {\bf 402} (2004) 103 [astro-ph/0406398].}}

\bibitem{Chen:2013aj}
\textsf{X.~Chen, H.~Firouzjahi, M.~H.~Namjoo and M.~Sasaki},
\textcolor{red}{\textsf{A Single Field Inflation Model with Large Local Non-Gaussianity}},
\textcolor{violet}{{Europhys.\ Lett.\  {\bf 102} (2013) 59001 [arXiv:1301.5699 [hep-th]].}}

\bibitem{Chen:2009zp}
\textsf{X.~Chen and Y.~Wang},
\textcolor{red}{\textsf{Quasi-Single Field Inflation and Non-Gaussianities}},
\textcolor{violet}{{JCAP {\bf 1004} (2010) 027 [arXiv:0911.3380 [hep-th]].}}

\bibitem{Creminelli:2003iq}
\textsf{P.~Creminelli},
\textcolor{red}{\textsf{On non-Gaussianities in single-field inflation}},
\textcolor{violet}{{JCAP {\bf 0310} (2003) 003 [astro-ph/0306122].}}

\bibitem{Babich:2004gb}
\textsf{D.~Babich, P.~Creminelli and M.~Zaldarriaga},
\textcolor{red}{\textsf{The Shape of non-Gaussianities}},
\textcolor{violet}{{JCAP {\bf 0408} (2004) 009 [astro-ph/0405356].}}

\bibitem{Creminelli:2004yq}
\textsf{P.~Creminelli and M.~Zaldarriaga},
\textcolor{red}{\textsf{Single field consistency relation for the 3-point function}},
\textcolor{violet}{{JCAP {\bf 0410} (2004) 006 [astro-ph/0407059].}}

\bibitem{Shukla:2016bnu}
\textsf{A.~Shukla, S.~P.~Trivedi and V.~Vishal},
\textcolor{red}{\textsf{Symmetry constraints in inflation, $\alpha$-vacua, and the three point function}},
\textcolor{violet}{{JHEP {\bf 1612} (2016) 102 [arXiv:1607.08636 [hep-th]].}}

\bibitem{Kundu:2014gxa}
\textsf{N.~Kundu, A.~Shukla and S.~P.~Trivedi},
\textcolor{red}{\textsf{Constraints from Conformal Symmetry on the Three Point Scalar Correlator in Inflation}},
\textcolor{violet}{{JHEP {\bf 1504} (2015) 061 [arXiv:1410.2606 [hep-th]].}}

\bibitem{Mata:2012bx}
\textsf{I.~Mata, S.~Raju and S.~P.~Trivedi},
\textcolor{red}{\textsf{CMB from CFT}},
\textcolor{violet}{{JHEP {\bf 1307} (2013) 015 [arXiv:1211.5482 [hep-th]].}}

\bibitem{Chen:2009bc}
\textsf{X.~Chen, B.~Hu, M.~x.~Huang, G.~Shiu and Y.~Wang},
\textcolor{red}{\textsf{Large Primordial Trispectra in General Single Field Inflation}},
\textcolor{violet}{{JCAP {\bf 0908} (2009) 008 [arXiv:0905.3494 [astro-ph.CO]].}}

\bibitem{Ghosh:2014kba}
\textsf{A.~Ghosh, N.~Kundu, S.~Raju and S.~P.~Trivedi},
\textcolor{red}{\textsf{Conformal Invariance and the Four Point Scalar Correlator in Slow-Roll Inflation}},
\textcolor{violet}{{JHEP {\bf 1407} (2014) 011 [arXiv:1401.1426 [hep-th]].}}

\bibitem{Kundu:2015xta}
\textsf{N.~Kundu, A.~Shukla and S.~P.~Trivedi},
\textcolor{red}{\textsf{Ward Identities for Scale and Special Conformal Transformations in Inflation}},
\textcolor{violet}{{JHEP {\bf 1601} (2016) 046 [arXiv:1507.06017 [hep-th]].}}

\bibitem{Arroja:2008ga}
\textsf{F.~Arroja and K.~Koyama},
\textcolor{red}{\textsf{Non-gaussianity from the trispectrum in general single field inflation}},
\textcolor{violet}{{Phys.\ Rev.\ D {\bf 77} (2008) 083517 [arXiv:0802.1167 [hep-th]].}}

\bibitem{Seery:2006js}
\textsf{D.~Seery and J.~E.~Lidsey},
\textcolor{red}{\textsf{Non-Gaussianity from the inflationary trispectrum}},
\textcolor{violet}{{JCAP {\bf 0701} (2007) 008 [astro-ph/0611034].}}

\bibitem{Seery:2008ax}
\textsf{D.~Seery, M.~S.~Sloth and F.~Vernizzi},
\textcolor{red}{\textsf{Inflationary trispectrum from graviton exchange}},
\textcolor{violet}{{JCAP {\bf 0903} (2009) 018 [arXiv:0811.3934 [astro-ph]].}}

\bibitem{Smith:2011if}
\textsf{K.~M.~Smith, M.~LoVerde and M.~Zaldarriaga},
\textcolor{red}{\textsf{A universal bound on N-point correlations from inflation}},
\textcolor{violet}{{Phys.\ Rev.\ Lett.\  {\bf 107} (2011) 191301 [arXiv:1108.1805 [astro-ph.CO]].}}

\bibitem{Suyama:2010uj}
\textsf{T.~Suyama, T.~Takahashi, M.~Yamaguchi and S.~Yokoyama},
\textcolor{red}{\textsf{On Classification of Models of Large Local-Type Non-Gaussianity}},
\textcolor{violet}{{JCAP {\bf 1012} (2010) 030 [arXiv:1009.1979 [astro-ph.CO]].}}

\bibitem{Suyama:2007bg}
\textsf{T.~Suyama and M.~Yamaguchi},
\textcolor{red}{\textsf{Non-Gaussianity in the modulated reheating scenario}},
\textcolor{violet}{{Phys.\ Rev.\ D {\bf 77} (2008) 023505 [arXiv:0709.2545 [astro-ph]].}}

\bibitem{Maldacena:2015bha}
\textsf{J.~Maldacena},
\textcolor{red}{\textsf{A model with cosmological Bell inequalities}},
\textcolor{violet}{{Fortsch.\ Phys.\  {\bf 64} (2016) 10 [arXiv:1508.01082 [hep-th]].}}

\bibitem{Choudhury:2016cso}
\textsf{S.~Choudhury, S.~Panda and R.~Singh},
\textcolor{red}{\textsf{Bell violation in the Sky}},
\textcolor{violet}{{Eur.\ Phys.\ J.\ C {\bf 77} (2017) no.2,  60 [arXiv:1607.00237 [hep-th]].}}

\bibitem{Choudhury:2016pfr}
\textsf{S.~Choudhury, S.~Panda and R.~Singh},
\textcolor{red}{\textsf{Bell violation in primordial cosmology}},
\textcolor{violet}{{Universe {\bf 3} (2017) 13 [arXiv:1612.09445 [hep-th]].}}

\end{thebibliography}
\end{document}